\documentclass[11pt]{cernrep}
 
\usepackage{epsfig, url, psfrag, color}
\usepackage{fancyhdr, multind}
\usepackage[square,comma,numbers]{natbib}
\usepackage{dcolumn, bm, mathrsfs, xspace, axodraw, graphicx,subfigure}
\usepackage{array,amsfonts,amssymb,amsmath,longtable}

\usepackage{tev4lhc}

\renewcommand{\theequation}{\thesubsection.\arabic{equation}}
\renewcommand{\thefigure}{\thesubsection.\arabic{figure}}
\renewcommand{\thetable}{\thesubsection.\arabic{table}}
\begin{document}



\date{\today}
\documentlabel{hep-ph/0612172 \\ FERMILAB-CONF-06-467-E-T}


\title{\bf Tevatron-for-LHC Report: Higgs}







\author{
(TeV4LHC Higgs Working Group)
\mbox{U. Aglietti,$^{1,2}$}
\mbox{A. Belyaev,$^{3}$}
\mbox{S. Berge,$^{4}$}
\mbox{A. Blum,$^{3}$}
\mbox{R. Bonciani,$^{5}$}
\mbox{J. Cammin,$^{6}$}
\mbox{M. Carena,$^{7}$}
\mbox{S. Chivukula,$^{3}$}
\mbox{H. Davoudiasl,$^{8}$}
\mbox{S. Dawson,$^{9}$}
\mbox{G. Degrassi,$^{1,10}$}
\mbox{A. Dominguez,$^{11}$}
\mbox{J. Donini,$^{1,12}$}
\mbox{T. Dorigo,$^{1,12}$}
\mbox{B.J. Field,$^{13}$}
\mbox{T. Hahn,$^{14}$}
\mbox{T. Han,$^{8}$}
\mbox{S. Heinemeyer,$^{15,16}$}
\mbox{S. Hesselbach,$^{17}$}
\mbox{G.-Y. Huang,$^{8}$}
\mbox{I. Iashvilli,$^{18}$}
\mbox{C.B. Jackson,$^{9}$}
\mbox{T. Junk,$^{19}$}
\mbox{S.-W. Lee,$^{20}$}
\mbox{H.E. Logan,$^{21}$}
\mbox{F. Maltoni,$^{22}$}
\mbox{B. Mellado,$^{8}$}
\mbox{S. Moretti,$^{17}$}
\mbox{S. Mrenna,$^{7}$}
\mbox{P.M. Nadolsky,$^{23}$}
\mbox{F.I. Olness,$^{4}$}
\mbox{W. Quayle,$^{8}$}
\mbox{J. Rathsman,$^{24}$}
\mbox{L. Reina,$^{13}$}
\mbox{E.H. Simmons,$^{3}$}
\mbox{A. Sopczak,$^{25}$}
\mbox{A. Vicini,$^{1,26}$}
\mbox{D. Wackeroth,$^{18}$}
\mbox{C.E.M. Wagner,$^{23,27}$}
\mbox{G. Weiglein,$^{28}$}
\mbox{G. Weiglein,$^{28}$}
\mbox{S. Willenbrock,$^{19}$}
\mbox{S.L. Wu,$^{8}$}
\mbox{C.P. Yuan$^{3}$}
}
\institute{
\mbox{$^{1}$ INFN},
\mbox{$^{2}$ Universit\`a di Roma ``La Sapienza''},
\mbox{$^{3}$ Michigan State University},
\mbox{$^{4}$ Southern Methodist University},
\mbox{$^{5}$ Universitat de Valencia},
\mbox{$^{6}$ University of Rochester},
\mbox{$^{7}$ Fermilab},
\mbox{$^{8}$ University of Wisconsin -- Madison},
\mbox{$^{9}$ Brookhaven National Laboratory},
\mbox{$^{10}$ Universit\`a di Roma Tre},
\mbox{$^{11}$ University of Nebraska -- Lincoln},
\mbox{$^{12}$ Sezione di Padova-Trento},
\mbox{$^{13}$ Florida State University},
\mbox{$^{14}$ Max Planck Institut f\"ur Physik},
\mbox{$^{15}$ CERN},
\mbox{$^{16}$ Universidad de Zaragoza},
\mbox{$^{17}$ Southampton University},
\mbox{$^{18}$ State University of New York at Buffalo},
\mbox{$^{19}$ University of Illinois at Urbana-Champaign},
\mbox{$^{20}$ Texas Tech University},
\mbox{$^{21}$ Carleton University},
\mbox{$^{22}$ Universit\'e Catholique de Louvain},
\mbox{$^{23}$ Argonne National Laboratory},
\mbox{$^{24}$ Uppsala University},
\mbox{$^{25}$ Lancaster University},
\mbox{$^{26}$ Universit\`a degli Studi di Milano},
\mbox{$^{27}$ University of Chicago},
\mbox{$^{28}$ University of Durham}
}

\maketitle
\begin{abstract}
  The search for Higgs bosons in both the standard model and its
  extensions is well under way at the Tevatron.  As the integrated
  luminosity collected increases into the multiple inverse femptobarn
  range, these searches are becoming very interesting indeed.
  Meanwhile, the construction of the Large Hadron Collider (LHC) and
  its associated experiments at CERN are nearing completion.  In this
  TeV4LHC workshop, it was realized that any experience at the
  Tevatron with respect to backgrounds, experimental techniques and
  theoretical calculations that can be verified at the Tevatron which
  have relevance for future measurements at the LHC were important.
  Studies and contributions to these efforts were made in three broad
  categories: theoretical calculations of Higgs production and decay
  mechanisms; theoretical calculations and discussions pertaining to
  non-standard model Higgs bosons; and experimental reviews, analyses
  and developments at both the Tevatron and the upcoming LHC
  experiments.  All of these contributions represent real progress
  towards the elucidation of the mechanism of electroweak symmetry
  breaking.
\end{abstract}
\vfill
\makebox[4in]{\hrulefill}\\
\noindent $^{\P{}}$ Convenors of the Higgs Working Group\\
$^\dagger$ Organizers of the TeV4LHC Workshop


\pagestyle{plain}
\clearpage
\tableofcontents

\label{chap:higgs}


\clearpage
\section{Introduction}
\label{sec:higgsintro}
\textbf{Contributed by: S. Willenbrock, A. Dominguez, I. Iashvilli}
\vspace{0.25in}

The Fermilab Tevatron, which has been colliding protons and
antiprotons for over twenty years, was not designed to search for
the Higgs boson.  However, the advent of high-efficiency $b$
tagging, developed in the context of the search for the top quark,
made it possible to consider searching for the Higgs boson, produced
in association with a weak boson, via the decay $h\to b\bar b$
\cite{Stange:1993ya}.  It was realized that this would require very
high luminosity, and that other discovery modes, such as $h\to
W^+W^- \to \ell^+\ell^-\nu\bar\nu$, might also become viable with
sufficient integrated luminosity \cite{Han:1998ma}.  The strategy
for the Standard Model Higgs search was developed in the TeV2000
workshop \cite{Amidei:1996dt}, and was further refined, along with
the case of the supersymmetric Higgs, in the SUSY/Higgs workshop
\cite{Carena:2000yx}.

The search for a Higgs boson, both standard and supersymmetric, is
in full swing at the Tevatron, and is becoming increasingly
interesting as the integrated luminosity mounts.  Meanwhile, the
construction of the CERN Large Hadron Collider (LHC) is nearing
completion.  At this workshop, dubbed TeV4LHC, the Higgs working
group used the first meeting to decide what ``TeV4LHC'' means in the
context of the Higgs boson.  We decided that anything having to do
with the Higgs at the Tevatron was relevant to the workshop, since
this experience will surely be valuable at the LHC.  Any experience
at the Tevatron with backgrounds to Higgs searches is also relevant
to the workshop. Finally, any experimental techniques being
developed for the Higgs search at the Tevatron or the LHC should
also be included in the workshop.

The proceedings of the Higgs working group comprises a large number
of contributions on a wide variety of topics.  Roughly speaking, the
contributions fall into one of three categories.

The first category is theoretical calculations of Higgs production
and decay processes, including higher-order corrections and
resummation to all orders.  There is an overview of Higgs total
cross sections, both in the Standard Model and with supersymmetry.
There is a review of calculations of Higgs production in association
with heavy quarks, either bottom or top.  In the case of Higgs
production in association with bottom quarks, there is a discussion
of the Higgs transverse momentum distribution, including the
resummation of soft gluons, for both inclusive Higgs production as
well as production in association with a high-\PT $b$ jet.  These
calculations make use of the $b$ distribution function in the
proton, and there is a contribution regarding sets of parton
distribution functions with no heavy quarks, with only $c$ quarks,
or with both $c$ and $b$ quarks, at next-to-next-to-leading order in
QCD.  Finally, there is a calculation of the electroweak corrections
to Higgs production via $gg\to h$, which is the dominant production
mechanism.

The second category is non-standard Higgs bosons, either with or
without supersymmetry.  There is a discussion of the impact of
radiative corrections on the search for supersymmetric Higgs bosons
at the Tevatron and the LHC.  There is an analysis of the search for
a Higgs decaying via $h\to aa\to b\bar b\tau^+\tau^-$ at the
Tevatron, where $a$ is also a Higgs scalar (or pseudoscalar). There
is a discussion on how to use the processes $b\bar b\to h$, $h\to
\tau^+\tau^-$, and $h\to \gamma\gamma$ to disentangle the nature of
electroweak symmetry breaking.  Methods to search for a Higgs boson
that decays invisibly are proposed.  Finally, there is a discussion
of the search for charged Higgs bosons at hadron colliders.

The third category is experimental reviews, analyses, and
developments.  There are reviews from both CDF and D0 on the status
and prospects for Higgs searches at the Tevatron.  There are studies
on $b$ jets, one on $Z\to b\bar b$ and the other on improving the
$b$-jet resolution.  There are studies on $h\to W^+W^-\to
\ell^+\ell^-\nu\bar\nu$ and $h\to \tau^+\tau^-$ at the LHC. There is
a discussion of the diphoton background at the Tevatron, which is
relevant to the search for the Higgs via $h\to \gamma\gamma$ at the
LHC.

All of these contributions represent real progress towards the
elucidation of the mechanism of electroweak symmetry breaking.  It
will require the best efforts of us all to extract the maximal
information from the data coming from the Tevatron and the LHC.

\subsection*{Acknowledgment}
This material is based upon work supported by the National Science
Foundation under Grant Number 0547780.

\clearpage
\section{SM and MSSM Higgs Boson Production Cross Sections}
\noindent\textbf{Contributed by: T. Hahn, S. Heinemeyer, F. Maltoni, S. Willenbrock}

\vspace*{0.25in}

We present the SM and MSSM Higgs-boson production cross sections at
the Tevatron and the LHC. The SM cross sections are a compilation of
state-of-the-art theoretical predictions. The MSSM cross
sections are obtained from the SM ones by means of an effective
coupling approximation, as implemented in {\tt FeynHiggs}.  Numerical
results have been obtained in four benchmark scenarios for two values
of $\tan\beta$, $\tan\beta = 5, 40$.





\subsection{Introduction}

Deciphering the mechanism of electroweak symmetry breaking (EWSB) is
one of the main quests of the high energy physics community.
Electroweak precision data in combination with the direct top-quark 
mass measurement at the Tevatron have
strongly constrained the range of possible scenarios and hinted to the
existence of a light scalar particle~\cite{Group:2005di}. 
Both in the standard model (SM) and in its minimal 
supersymmetric extensions (MSSM), the $W$ and $Z$
bosons and fermions acquire masses by coupling to the vacuum
expectation value(s) of scalar SU(2) doublet(s), via the so-called
Higgs mechanism. The common prediction of such models is the
existence of at least one scalar state, the Higgs boson.  Within the
SM, LEP has put a lower bound on the Higgs mass, $\mh >
114$ GeV~\cite{LEPHiggsSM}, and has contributed to the
indirect evidence that the Higgs boson should be relatively light 
with a 95\% probability for its mass to be below 186 GeV~\cite{Group:2005di}.  
In the MSSM the experimental lower bound for the mass of the lightest
state is somewhat  
weaker, and internal consistency of the theory predicts an upper bound
of 135 GeV~\cite{mhiggslong,mhiggsAEC,Allanach:2004rh}.

If the Higgs sector is realized as implemented in the SM or the MSSM,
at least one Higgs boson
should be discovered at the Tevatron and/or at the LHC. Depending on the mass,
there are various channels available where Higgs searches can be performed.
The power of each signature depends on the production
cross section, $\sigma$, and the Higgs branching ratio into final
state particles, such as leptons or $b$-jets, 
the total yield of events being proportional to $\sigma \cdot$ BR. 
In some golden channels, such as $gg\to h \to Z^{(*)}Z \to 4\mu$,  a
discovery will be straightfoward and mostly independent from our
ability to predict signal and/or backgrounds. On the other hand,
for coupling measurements or for searches in more difficult channels, 
such as $t\bar t h\to t \bar t b \bar b$ associated production, 
precise predictions for both signal and backgrounds are mandatory. 
Within the MSSM such precise predictions for signal and backgrounds
are necessary in order to relate the experimental results to the
underlying SUSY parameters.

The aim of this note is to collect up-to-date predictions for 
the most relevant signal cross sections, for both the SM and the MSSM. 
In Section~\ref{sec:SM} we collect the results 
of state-of-the-art calculations for the SM cross sections as a function
of the Higgs mass.  In Section~\ref{sec:MSSM} we present the MSSM 
cross sections for the neutral Higgs-bosons in four benchmark scenarios.
These results are obtained by rescaling the SM cross sections presented
in the previous sections, using an effective coupling approximation.


\subsection{SM Higgs production cross sections}
\label{sec:SM}

\begin{figure}[t]
\begin{center}
\includegraphics[angle=-90,width=15cm]{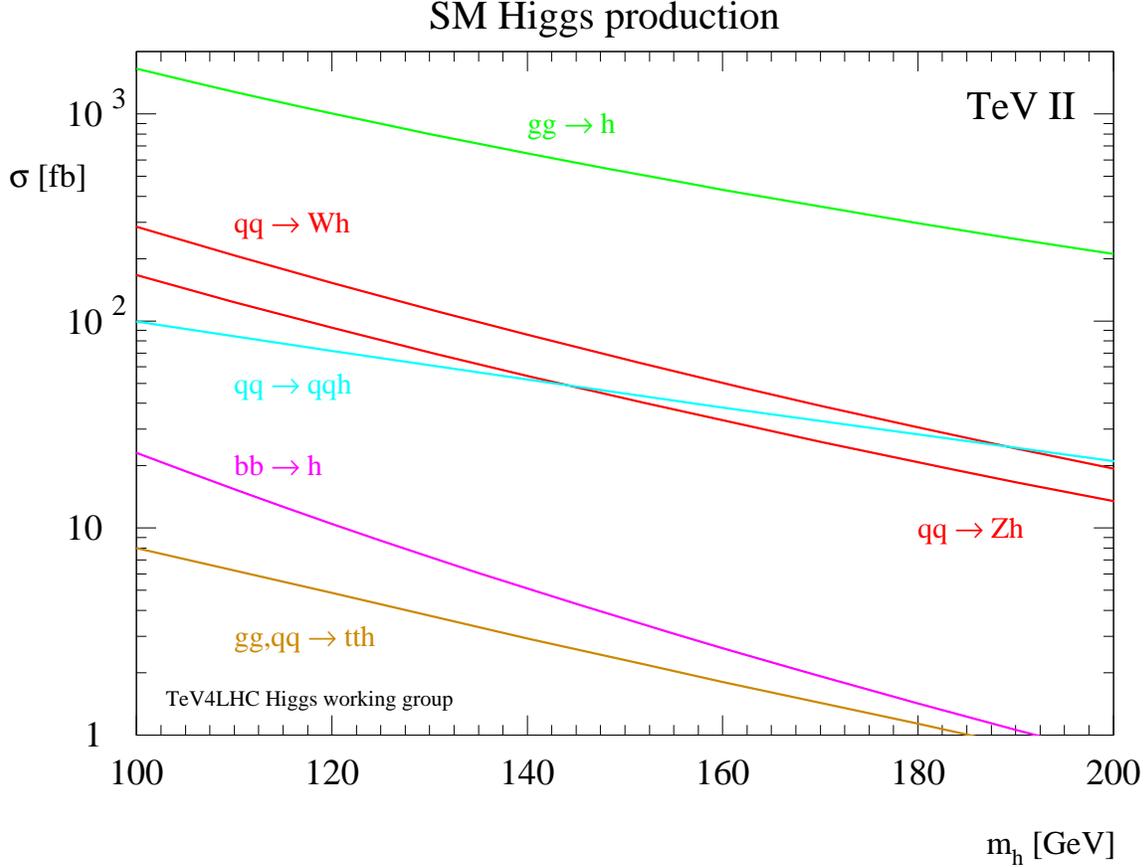}
\end{center}
\caption{Higgs-boson production cross sections (fb) at the Tevatron
  ($\sqrt{s}=1.96$ TeV) for the most relevant production mechanims as
  a function of the Higgs-boson mass. Results for $gg \to h$, $q\bar
  q\to Vh$, $b\bar b \to h$ are at NNLO in the QCD expansion.  Weak
  boson fusion ($qq \to qq h$) and $t\bar t $ associated
  production are at NLO accuracy.}
\label{fig:tev}
\end{figure}

\begin{figure}[t]
\begin{center}
\includegraphics[angle=-90,width=15cm]{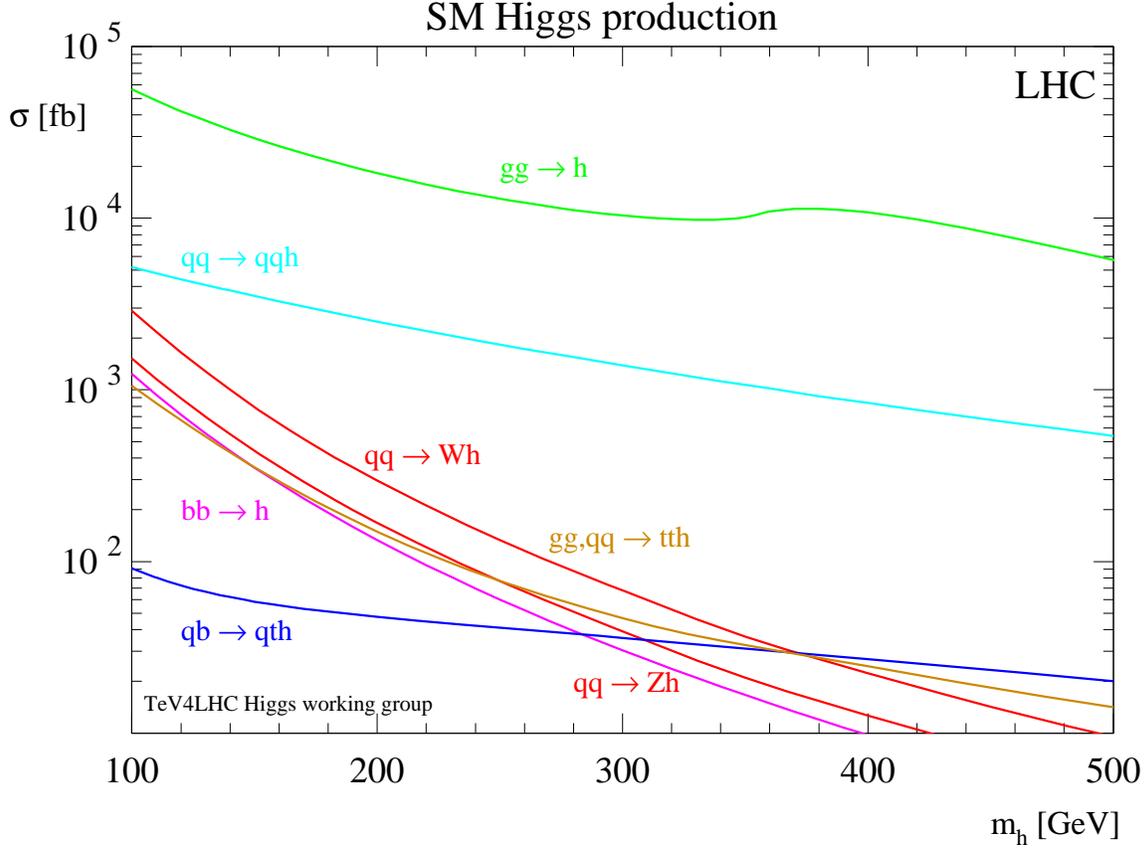}
\end{center}
\caption{Higgs-boson production cross sections (fb) at the LHC
  ($\sqrt{s}=14$ TeV) for the most relevant production mechanims as a
  function of the Higgs-boson mass. Results for $gg \to h$, $q\bar
  q\to Vh$, $b\bar b \to h$ are at NNLO in the QCD expansion.  Weak
  boson fusion ($qq \to qq h$) and $t\bar t $ associated
  production are at NLO accuracy. Single-top associated production
  ($qb \to qth$) is at LO.} 
\label{fig:lhc}
\end{figure}

In this section we collect the predictions for the most important
SM Higgs production processes at the Tevatron and at the LHC. The
relevant cross sections are presented in Figs.~\ref{fig:tev} and 
\ref{fig:lhc} as function of the Higgs mass. The results refer to
fully inclusive cross sections. No acceptance cuts or branching ratios
are applied%
\footnote{
More details and data files can be found at 
{\tt maltoni.web.cern.ch/maltoni/TeV4LHC} .
}%
. We do not consider here diffractive Higgs production,
$pp \to p \oplus H \oplus p$~\cite{Albrow:2000na,Khoze:2001xm,DeRoeck:2002hk,Cox:2004rv,Forshaw:2005qp}. For the discussion of this
channel in the MSSM we refer to \citere{diffHMSSM}.

We do not aim here at a detailed discussion of the importance of each
signature at the Tevatron or the LHC, but only at providing the most
accurate and up-to-date theoretical predictions. To gauge the progress
made in the last years, it is interesting to compare the accuracy of
the results available in the year 2000, at the time of the Tevatron Higgs
Working Group ~\cite{Carena:2000yx}, with those shown here.  All
relevant cross sections are now known at least one order better in the
strong-coupling expansion, and in some cases also electroweak
corrections are available.

\begin{itemize}
\item $gg\to h + X$: gluon fusion \\

This process is known at NNLO in
QCD~\cite{Harlander:2002wh,Anastasiou:2002yz,Ravindran:2003um} (in the 
large top-mass limit) and at NLO in QCD for a quark of an arbitrary
mass circulating in the loop~\cite{Graudenz:1992pv,QCDg2}.
Some N$^3$LO results have recently been obtained in 
\citeres{Moch:2005ky,Laenen:2005uz}.
The NNLO results plotted here are from Ref.~\cite{Catani:2003zt} and
include soft-gluon resummation effects at NNLL. MRST2002 at NNLO has
been used~\cite{Martin:2002aw}, with the renormalization and
factorization scales set 
equal to the Higgs-boson mass. The overall residual theoretical
uncertainty is estimated to be around 10\%. 
The uncertainties due to the large top mass limit approximation 
(beyond Higgs masses of $2 \times m_t$) are difficult to estimate but
expected to be relatively small. Differential results
at NNLO are also available~\cite{Anastasiou:2004xq}.  NLO
(two-loop) EW corrections are known for Higgs masses below
$2m_W$, \cite{Aglietti:2004nj,Degrassi:2004mx}, and range
between 5\% and 8\% of the lowest order term. These EW corrections,
however, are not included in Figs.~\ref{fig:tev}, \ref{fig:lhc}, and
they are also omitted in the MSSM evaluations below.
The same holds for the recent corrections obtained in
\citeres{Moch:2005ky,Laenen:2005uz}.

\item $qq \to qqh + X$: vector boson fusion\\

This process is known at NLO in
QCD~\cite{Han:1992hr,Berger:2004pc,Figy:2003nv}.  Results plotted
here have been obtained with MCFM\cite{Campbell:1999ah}. 
Leading EW corrections are taken into account by using $\alpha(M_Z)$ as
the (square of the) electromagnetic coupling.
The PDF
used is CTEQ6M~\cite{Pumplin:2002vw} and the renormalization and
factorization scales are 
set equal to the Higgs-boson mass. The theoretical uncertainty is
rather small, less than 10\%.

\item $q \bar q \to Vh + X$: $W,Z$ associated production\\

These processes are known at NNLO in the QCD
expansion~\cite{Brein:2003wg} and at NLO in the electroweak
expansion~\cite{Ciccolini:2003jy}.  The results plotted here have
been obtained by the LH2003 Higgs working group by combining NNLO QCD
and NLO EW corrections~\cite{Assamagan:2004mu}.  The PDF used is
MRST2001 and the renormalization and factorization scales are set
equal to the Higgs-vector-boson invariant mass.  The residual
theoretical uncertainty is rather small, less than 5\%.

\item $b\bar b \to h + X$: bottom fusion\\

This process is known at NNLO in QCD in the five-flavor
scheme~\cite{Harlander:2003ai}.  The cross section in the
four-flavor scheme is known at
NLO~\cite{Dittmaier:2003ej,Dawson:2003kb}.  Results obtained in the
two schemes have been shown to be
consistent~\cite{Assamagan:2004mu,Dawson:2004sh,Dawson:2005vi}. 
The results plotted here are from Ref.~\cite{Harlander:2003ai}.
MRST2002 at NNLO has been used, with the renormalization scale set
equal to $m_h$ and the factorization scale set equal to $m_h/4$.  For
results with one final-state $b$-quark at high-$p_T$ we refer to
\citere{Campbell:2002zm,Dawson:2004sh}.  For results with two
final-state $b$-quarks at high-$p_T$ we refer to
\citere{Dittmaier:2003ej,Dawson:2003kb}.

\item $q\bar q,gg \to t \bar t h + X$: $t \bar t$ associated
       production\\

This process is known at NLO in
QCD~\cite{Beenakker:2001rj,Reina:2001sf,Dawson:2002tg}.  The results
plotted here are from Ref.~\cite{Dawson:2002tg}.  The PDF used is
CTEQ6M and the renormalization and factorization scales are set equal
to $m_t+m_h/2$.

\item $q b \to qth$ : single-top associated production\\

This process is known at LO in QCD~\cite{Ballestrero:1992bk}.  The
results plotted here ($t$-channel production, LHC only) are from
Ref.~\cite{Maltoni:2001hu}. The PDF used is CTEQ5L and the
renormalization and factorization scales are set equal to the
Higgs-boson mass.

\end{itemize}


\subsection{MSSM Higgs production cross sections}
\label{sec:MSSM}

The MSSM requires two Higgs doublets, resulting in five physical Higgs
boson degrees of freedom.  These are the light and heavy $\cp$-even
Higgs bosons, $h$ and $H$, the $\cp$-odd Higgs boson, $A$, and the
charged Higgs boson, $H^\pm$.  The Higgs sector of the MSSM can be
specified at lowest order in terms of $\MZ$, $\MA$, and $\tanb \equiv
v_2/v_1$, the ratio of the two Higgs vacuum expectation values.  The
masses of the $\cp$-even neutral Higgs bosons and the charged Higgs
boson can be calculated, including higher-order corrections, in terms
of the other MSSM parameters.

After the termination of LEP in the year 2000 (the final LEP
results can be found in \citeres{LEPHiggsSM,LEPHiggsMSSM}), the Higgs
boson search has shifted to the Tevatron and will later be continued
at the LHC.  For these analyses and investigations a precise
prediction of the Higgs boson masses, branching ratios and production
cross sections in the various channels is necessary.

Due to the large number of free parameters, a complete scan of the
MSSM parameter space is too involved. Therefore the search results at
LEP~\cite{LEPHiggsMSSM} and the
Tevatron~\cite{D0bounds,CDFbounds,Tevcharged}, as well as studies for the
LHC~\cite{schumi} have been performed in several benchmark
scenarios~\cite{benchmark,benchmark2,benchmark3}.

The code {\tt FeynHiggs}~\cite{Heinemeyer:1998yj,mhiggslong,mhiggsAEC}
provides a precise calculation of the Higgs boson mass spectrum, couplings and
the decay widths%
\footnote{
The code can be obtained from {\tt www.feynhiggs.de} .
}%
. This has now been supplemented by the evaluation of
all relevant neutral Higgs boson production cross sections at the
Tevatron and the LHC (and the corresponding three SM cross sections
for both colliders with $M_H^{\rm SM} = \mh, m_H, m_A$). 
They are calculated by using the effective
coupling approach, rescaling the SM result%
\footnote{
The inclusion of the charged Higgs production cross sections is
planned for the near future.
}%
.

In this section we will briefly describe the benchmark scenarios with
their respective features. The effective coupling approach, used to
obtain the production cross sections within {\tt FeynHiggs}, is
discussed. Results for the neutral Higgs production cross sections at
the Tevatron and the LHC are presented within the benchmark scenarios
for two values of $\tanb$, $\tanb = 5, 40$.

\subsection{The benchmark scenarios}
\label{ssec:bench}

We start by recalling the four benchmark scenarios~\cite{benchmark2}
suitable for the MSSM Higgs boson search at hadron colliders%
\footnote{
In the course of this workshop they have been refined to cover wider
parts of the MSSM parameter space relevant especially for heavy MSSM
Higgs boson production~\cite{benchmark3}.
}%
. 
In these scenarios the values of the parameters of the $\Stop$~and
$\Sbot$~sector as well as the gaugino masses are fixed, while
$\tanb$ and $\MA$ are the parameters that are varied. Here we fix $\tanb$
to a low and a high value, $\tanb = 5, 40$, but vary $\MA$. This also
yields a variation of $M_h$ and $\mH$.

In order to fix our notations, we list the conventions for the inputs
from the scalar top and scalar bottom sector of the MSSM:
the mass matrices in the basis of the current eigenstates $\StopL, \StopR$ and
$\SbotL, \SbotR$ are given by
\BEA
\label{stopmassmatrix}
{\cal M}^2_{\Stop} &=&
  \ML \MstL^2 + \mt^2 + \CZb (\edz - \frac{2}{3} \sw^2) \MZ^2 &
      \mt \Xt \\
      \mt \Xt &
      \MstR^2 + \mt^2 + \frac{2}{3} \CZb \sw^2 \MZ^2 
  \MR, \\
&& \non \\
\label{sbotmassmatrix}
{\cal M}^2_{\Sbot} &=&
  \ML \MsbL^2 + \mb^2 + \CZb (-\edz + \frac{1}{3} \sw^2) \MZ^2 &
      \mb \Xb \\
      \mb \Xb &
      \MsbR^2 + \mb^2 - \frac{1}{3} \CZb \sw^2 \MZ^2 
  \MR,
\EEA
where 
\BE
\mt \Xt = \mt (\At - \mu \CTb) , \quad
\mb\, \Xb = \mb\, (\Ab - \mu \tanb) .
\label{eq:mtlr}
\EE
Here $\At$ denotes the trilinear Higgs--stop coupling, $\Ab$ denotes the
Higgs--sbottom coupling, and $\mu$ is the higgsino mass parameter.
SU(2) gauge invariance leads to the relation
\BE
\MstL = \MsbL .
\EE
For the numerical evaluation, a convenient choice is
\BE
\MstL = \MsbL = \MstR = \MsbR =: \msusy .
\label{eq:msusy}
\EE
The parameters in the $\Stop/\Sbot$ sector are defined here as
on-shell parameters, see \citere{bse} for a discussion and a
translation to \drbar parameters.
The top-quark mass is taken to be 
$\mt = \mt^{\rm exp} = 172.7\GeV$~\cite{mt1727}.

\begin{itemize}
\item{The $\mhmax$ scenario:}

This scenario had been designed to obtain conservative $\tanb$ exclusion
bounds~\cite{tbexcl}.
The parameters are chosen such that the maximum possible 
Higgs-boson mass as a function of $\tanb$ is obtained
(for fixed $\msusy$ and $\mt$, 
and $\MA$ set to its maximal value, $\MA = 1\TeV$).
The parameters are%
\footnote{
As mentioned above, no external constraints are taken into account.
In the minimal flavor violation scenario,  better agreement with
${\rm BR}(b \to s \gamma)$ constraints would be 
obtained for the other sign of $\Xt$ (called the
``constrained $\mhmax$'' scenario~\cite{benchmark2}). 
}%
:
\BEA
&& \msusy = 1\TeV, \; 
\mu = 200\GeV, \;
M_2 = 200\GeV, \non \\
&& \Xt = 2\, \msusy  \; 
\Ab = \At, 
\mgl = 0.8\,\msusy~.
\label{mhmax}
\EEA

\item{The no-mixing scenario:}

This benchmark scenario is associated with
vanishing mixing in the $\Stop$~sector and with a higher SUSY mass
scale as compared to the $\mhmax$~scenario to increase the parameter
space that avoids the LEP Higgs bounds:
\BEA
&& \msusy = 2\TeV, \; 
\mu = 200\GeV, \;
M_2 = 200\GeV, \non \\
&& \Xt = 2\, \msusy  \; 
\Ab = \At, 
\mgl = 0.8\,\msusy~.
\label{nomix}
\EEA

\item{The gluophobic Higgs scenario:}

In this scenario the main production cross section for the light Higgs
boson at the LHC, $gg \to h$, can strongly suppressed for a wide range
of the $\MA-\tanb$-plane. This happens due to a
cancellation between the top quark and the stop quark loops in the
production vertex (see \citere{ggsuppr}). This cancellation is more
effective for small $\Stop$~masses and for relatively large
values of the $\Stop$~mixing parameter, $\Xt$. The partial width of
the most relevant decay mode, $\Gamma(h \to \gamma\gamma)$, is affected much
less, since it is dominated by the $W$~boson loop.
The parameters are:
\BEA
&& \msusy = 350\GeV, \; 
\mu = 300\GeV, \;
M_2 = 300\GeV, \non \\
&& \Xt = -750\GeV \; 
\Ab = \At, 
\mgl = 500\GeV~.
\label{ggsup}
\EEA

\item{The small $\aeff$ scenario:}

Besides the channel $gg \to h \to \ga\ga$ at the LHC, the other
channels for light Higgs 
searches at the Tevatron and at the LHC mostly rely on the decays 
$h \to b \bar b$ and $h \to \tau^+\tau^-$. Including Higgs-propagator
corrections the couplings 
of the lightest Higgs boson to down-type fermions is $\sim \sin\aeff$,
where $\aeff$ is the loop corrected mixing angle in the 
neutral $\cp$-even Higgs sector.
Thus, if $\aeff$ is small, the two main decay channels can be heavily
suppressed in the MSSM compared to the SM case. 
Such a suppression occurs for large $\tanb$ and not too large $\MA$.
The parameters of this scenario are:
\BEA
&& \msusy = 800\GeV, \;
\mu = 2.5 \, \msusy, \;
M_2 = 500\GeV, \non \\
&& \Xt = -1100\GeV, \; 
\Ab = \At, 
\mgl = 500\GeV~.
\label{smallaeff}
\EEA

\end{itemize}


\subsection{The effective coupling approximation}
\label{ssec:eff}

We consider the following neutral Higgs production cross sections at
the Tevatron and the LHC ($\phi$ denotes all neutral MSSM Higgs
bosons, $\phi = h, H, A$):
\BEA
gg &\to& \phi + X~, \\
qq &\to& qq\phi + X~, \\
q \bar q &\to& W/Z\phi + X~, \\
b \bar b &\to& \phi + X~, \\
gg, qq &\to& t \bar t \phi~.
\EEA
The MSSM cross sections have been obtained by rescaling the corresponding 
SM cross sections of Section~\ref{sec:SM} either with ratio of the corresponding MSSM decay with
(of the inverse process) over the SM decay width, or with the square
of the ratio of the corresponding couplings.
More precisely, we apply the following factors:
\begin{itemize}

\item{$gg \to \phi + X$:}
\BE
\frac{\Ga(\phi \to gg)_{\MSSM}}{\Ga(\phi \to gg)_{\SM}}
\label{eq:phigg}
\EE
We include the full one-loop result with SM QCD corrections. MSSM
two-loop corrections~\cite{gghMSSM2L} have been neglected.

\item{$qq \to qq\phi + X$:}
\BE
\frac{|g_{\phi VV, \MSSM}|^2}{|g_{\phi VV, \SM}|^2} , \; V = W, Z~.
\EE
We include the full set of Higgs propagator corrections in the
effective couplings.

\item{$qq \to W/Z\phi + X$:}
\BE
\frac{|g_{ \phi VV, \MSSM}|^2}{|g_{\phi VV, \SM}|^2} , \; V = W, Z~.
\EE
We include the full set of Higgs propagator corrections in the
effective couplings.

\item{$b \bar b \to \phi + X$:}
\BE
\frac{\Ga(\phi \to b \bar b)_{\MSSM}}{\Ga(\phi \to b \bar b)_{\SM}}~.
\EE
We include here one-loop SM QCD and SUSY QCD corrections, as well as
the resummation of all terms of \order{(\al_s \tanb)^n}.

\item{$gg, qq \to t \bar t \phi$:}
\BE
\frac{|g_{\phi t \bar t, \MSSM}|^2}{|g_{\phi t \bar t, \SM}|^2}~,
\label{eq:phitt}
\EE
where $g_{\phi t \bar t, \MSSM}$ and $g_{\phi t \bar t, \SM}$ are
composed of a left- and a right-handed part.
We include the full set of Higgs propagator corrections in the
effective couplings.

\end{itemize}

In the effective couplings introduced in
eqs.~(\ref{eq:phigg})--(\ref{eq:phitt}) we have used the proper
normalization of the external (on-shell) Higgs bosons as discussed in
\citere{Hahn:2002gm}.

It should be noted that the effective coupling approximation as
described above does not take into account the MSSM-specific
dynamics of the production processes. The theoretical uncertainty in
the predictions for the cross sections will therefore in general be
somewhat larger than for the decay widths.

\subsection{Results}

Results for the neutral Higgs production cross sections at
the Tevatron and the LHC are presented within the four benchmark scenarios
for two values of $\tanb$, $\tanb = 5, 40$, giving a total of eight plots
for each collider. 
 
\begin{figure}[p]
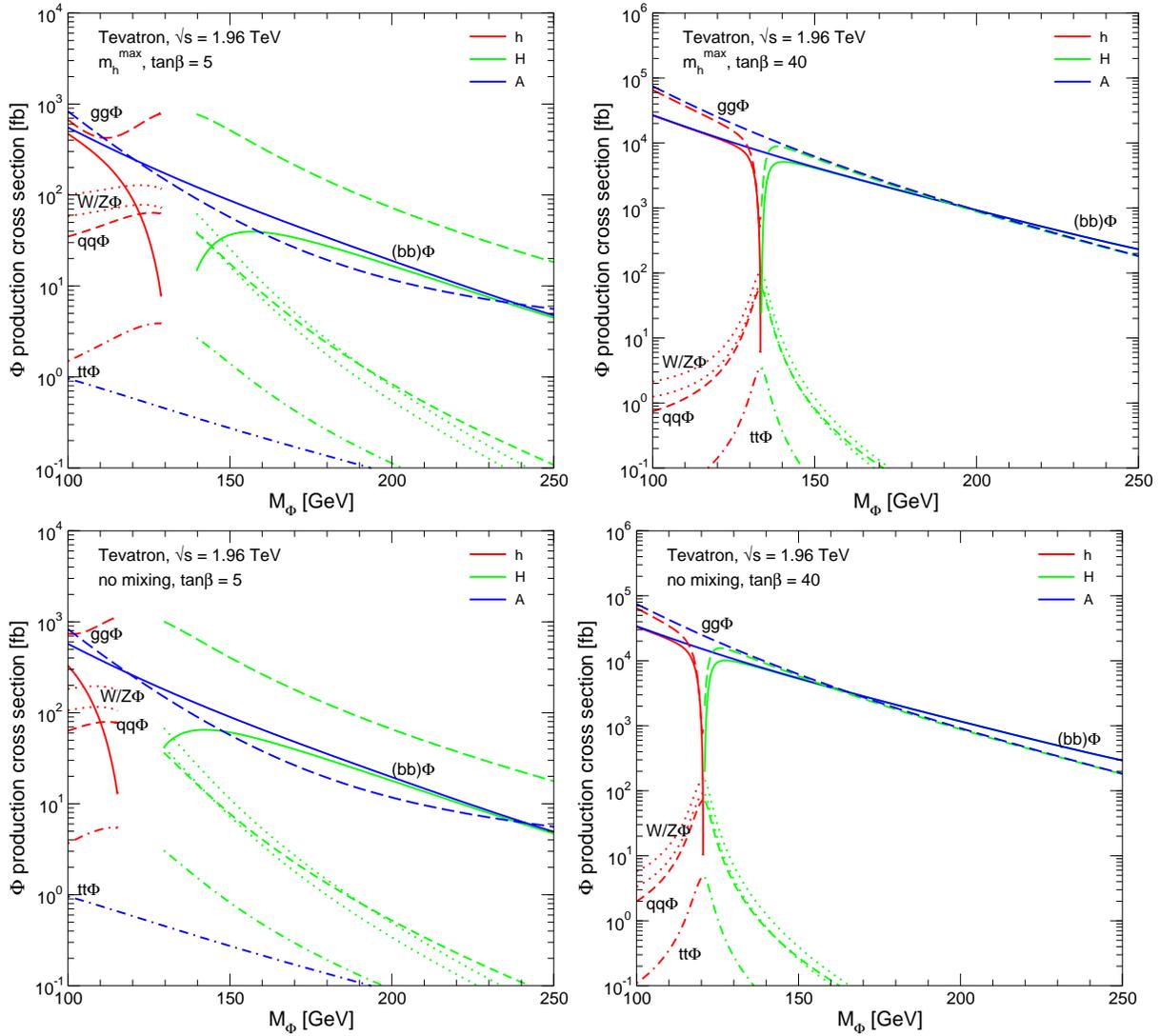

  \begin{minipage}[c]{0.49\textwidth}
    \includegraphics[width=\textwidth]{higgs/xsec/mhmax_tb05.cl.eps}
  \end{minipage}
  \begin{minipage}[c]{0.49\textwidth}
    \includegraphics[width=\textwidth]{higgs/xsec/mhmax_tb40.cl.eps}
  \end{minipage}
  \begin{minipage}[c]{0.49\textwidth}
    \includegraphics[width=\textwidth]{higgs/xsec/nomixing_tb05.cl.eps}
  \end{minipage}
  \begin{minipage}[c]{0.49\textwidth}
    \includegraphics[width=\textwidth]{higgs/xsec/nomixing_tb40.cl.eps}
  \end{minipage}
  \caption{Neutral Higgses production cross sections (fb) at the Tevatron, $\sqrt{s}=1.96$ TeV
    for the most relevant production mechanims as a function of the Higgs-boson mass. Results
    are based on the SM cross sections and evaluated through an effective coupling approximation
    in the $\mhmax$ and no-mixing scenarios, for $\tanb=5,40$.}
  \label{fig:mssm_tev_1}
\end{figure}

\begin{figure}[p]
  \begin{minipage}[c]{0.49\textwidth}
    \includegraphics[width=\textwidth]{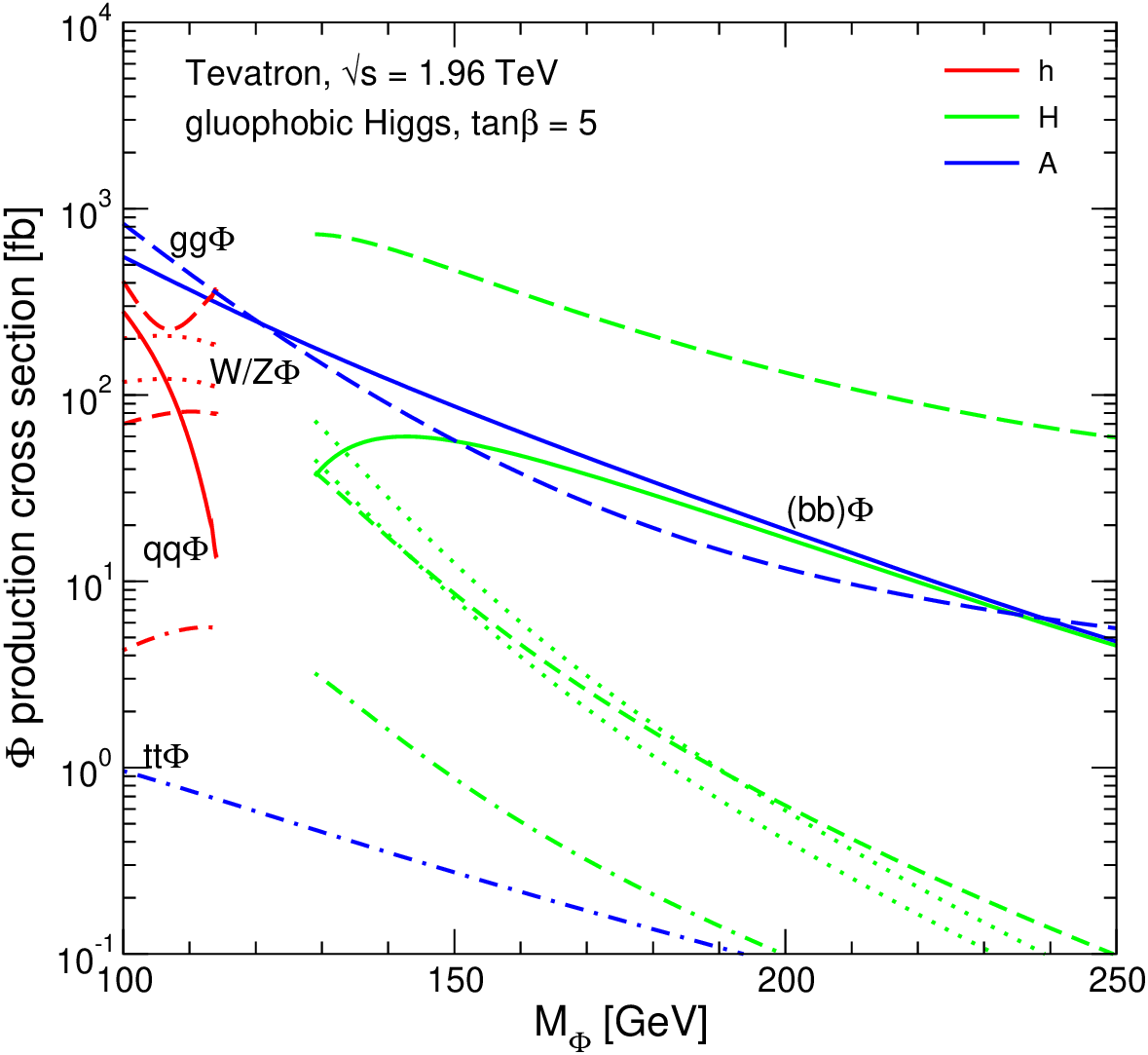}
  \end{minipage}
  \begin{minipage}[c]{0.49\textwidth}
    \includegraphics[width=\textwidth]{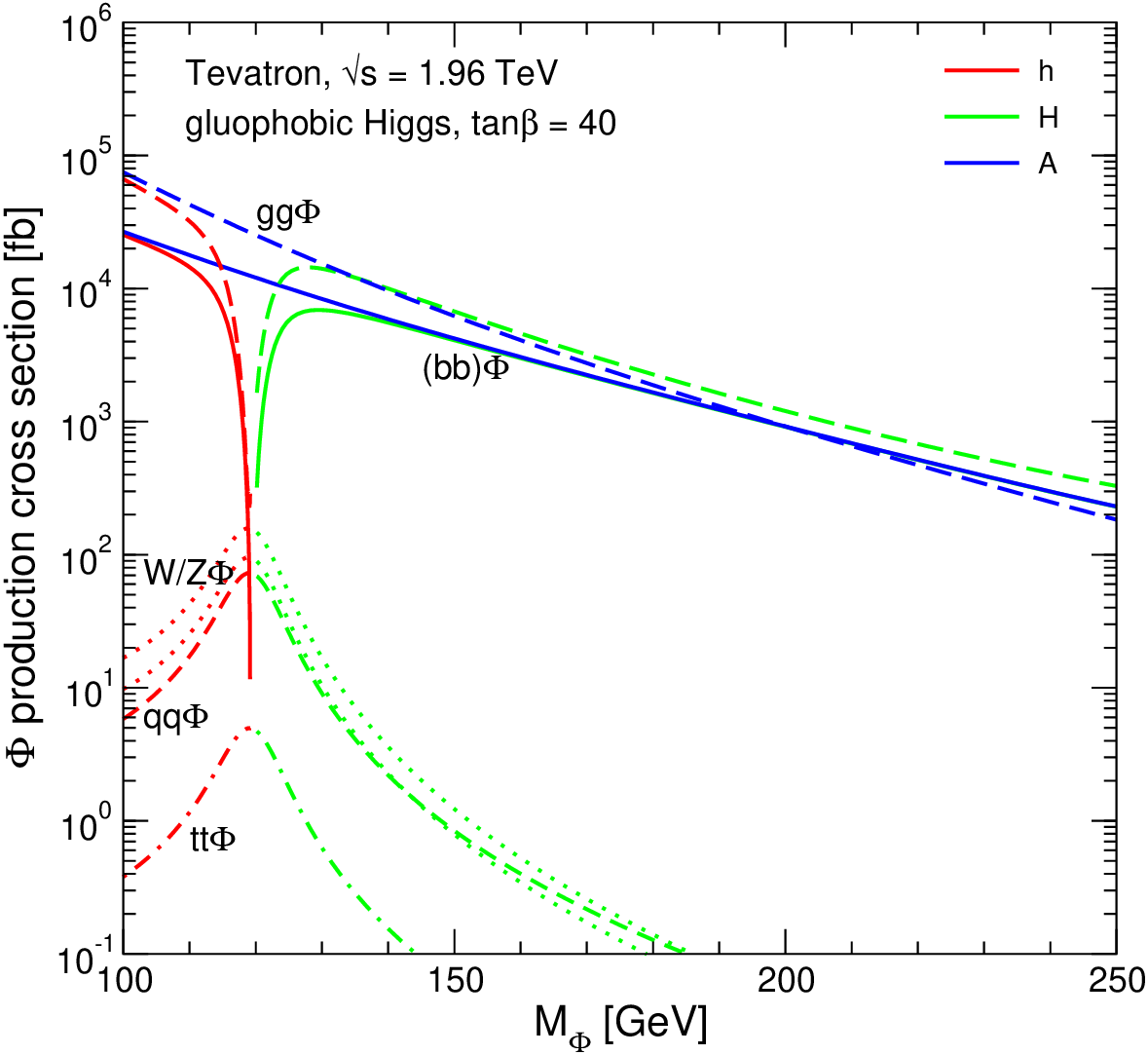}
  \end{minipage}
  \begin{minipage}[c]{0.49\textwidth}
    \includegraphics[width=\textwidth]{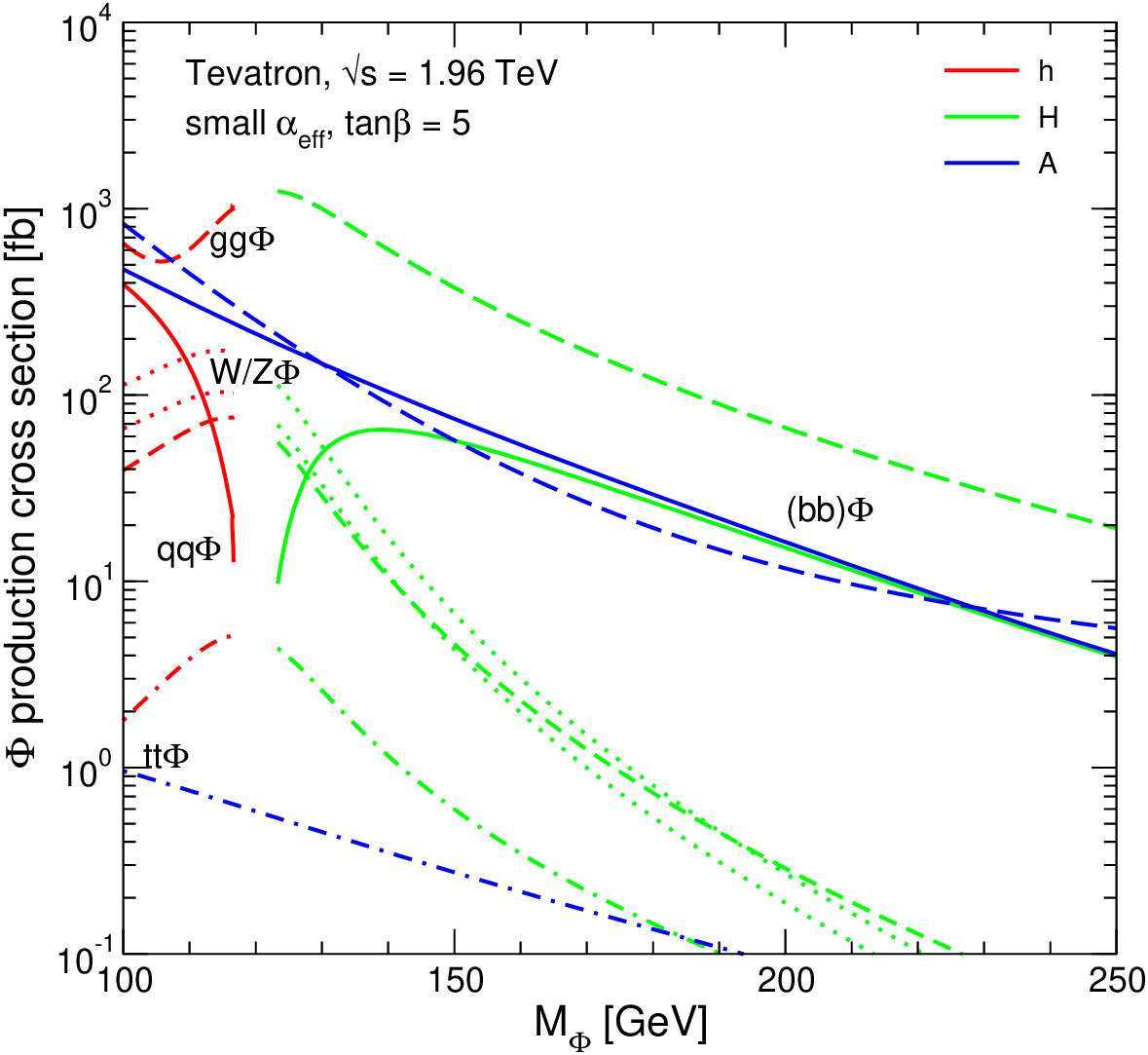}
  \end{minipage}
  \begin{minipage}[c]{0.49\textwidth}
    \includegraphics[width=\textwidth]{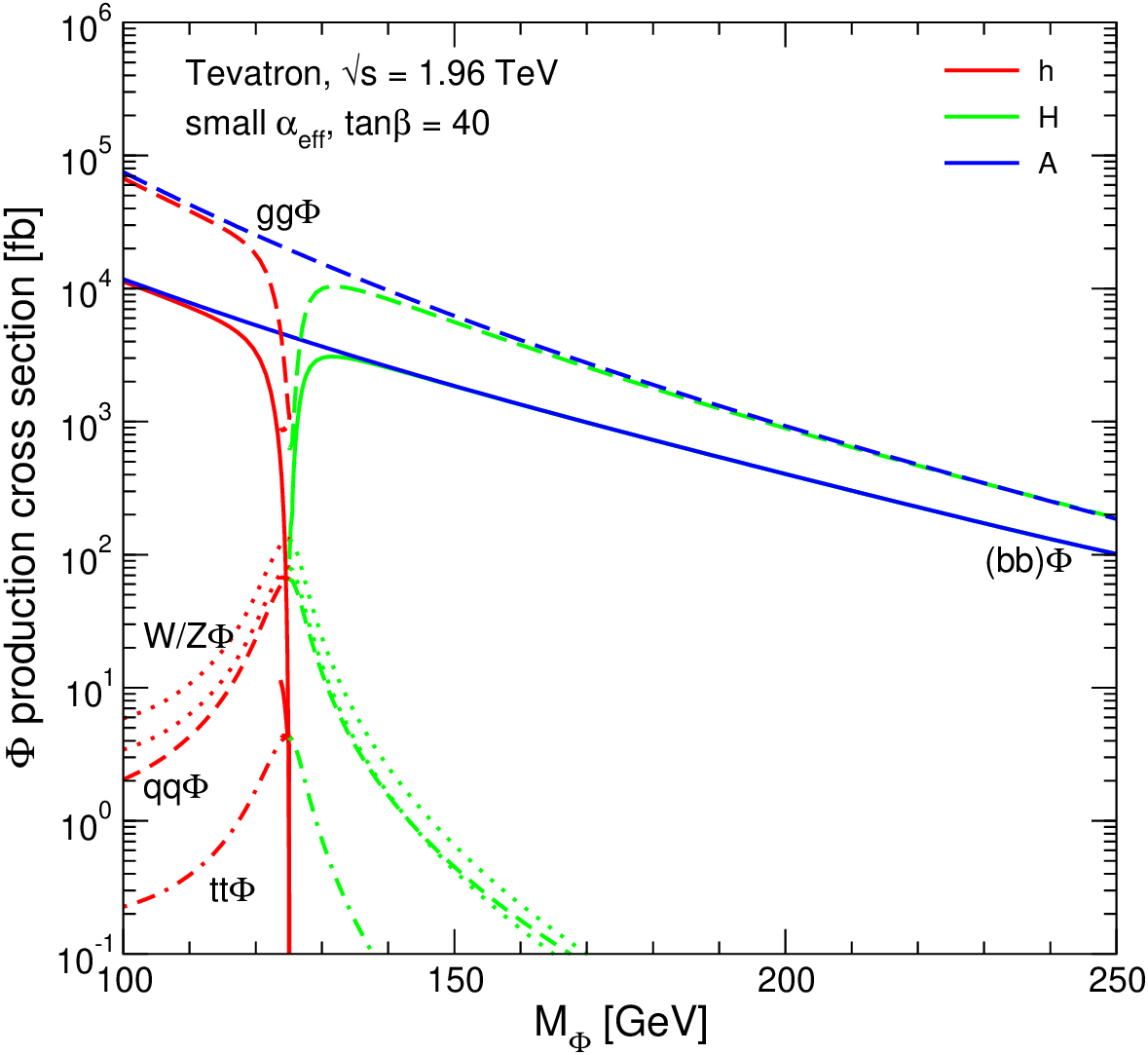}
  \end{minipage}
  \caption{Same as Fig.~\ref{fig:mssm_tev_1}, for the gluophobic Higgs and small $\aeff$ scenarios.}
  \label{fig:mssm_tev_2}
\end{figure}

\begin{figure}[p]
  \begin{minipage}[c]{0.49\textwidth}
    \includegraphics[width=\textwidth]{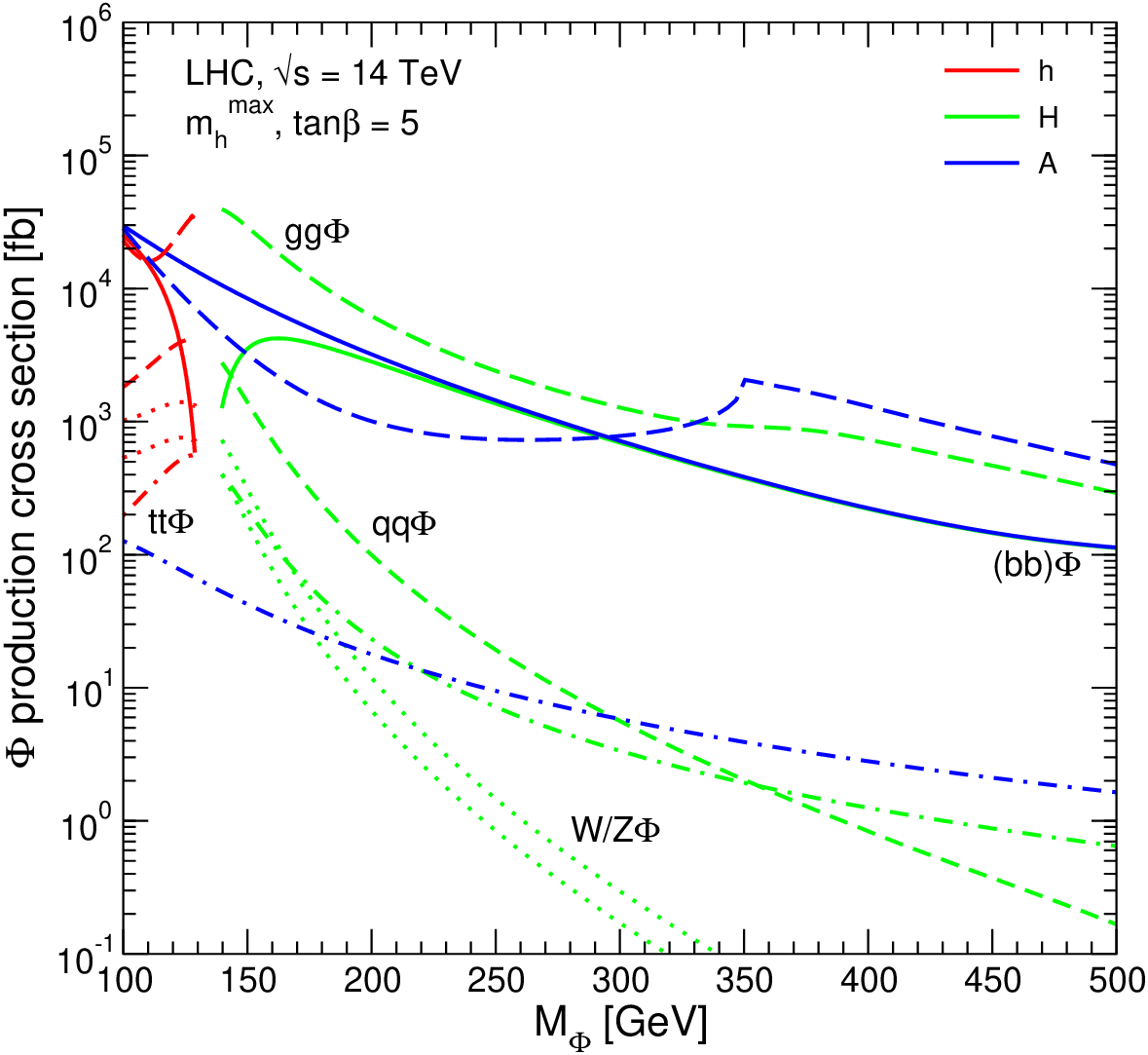}
  \end{minipage}
  \begin{minipage}[c]{0.49\textwidth}
    \includegraphics[width=\textwidth]{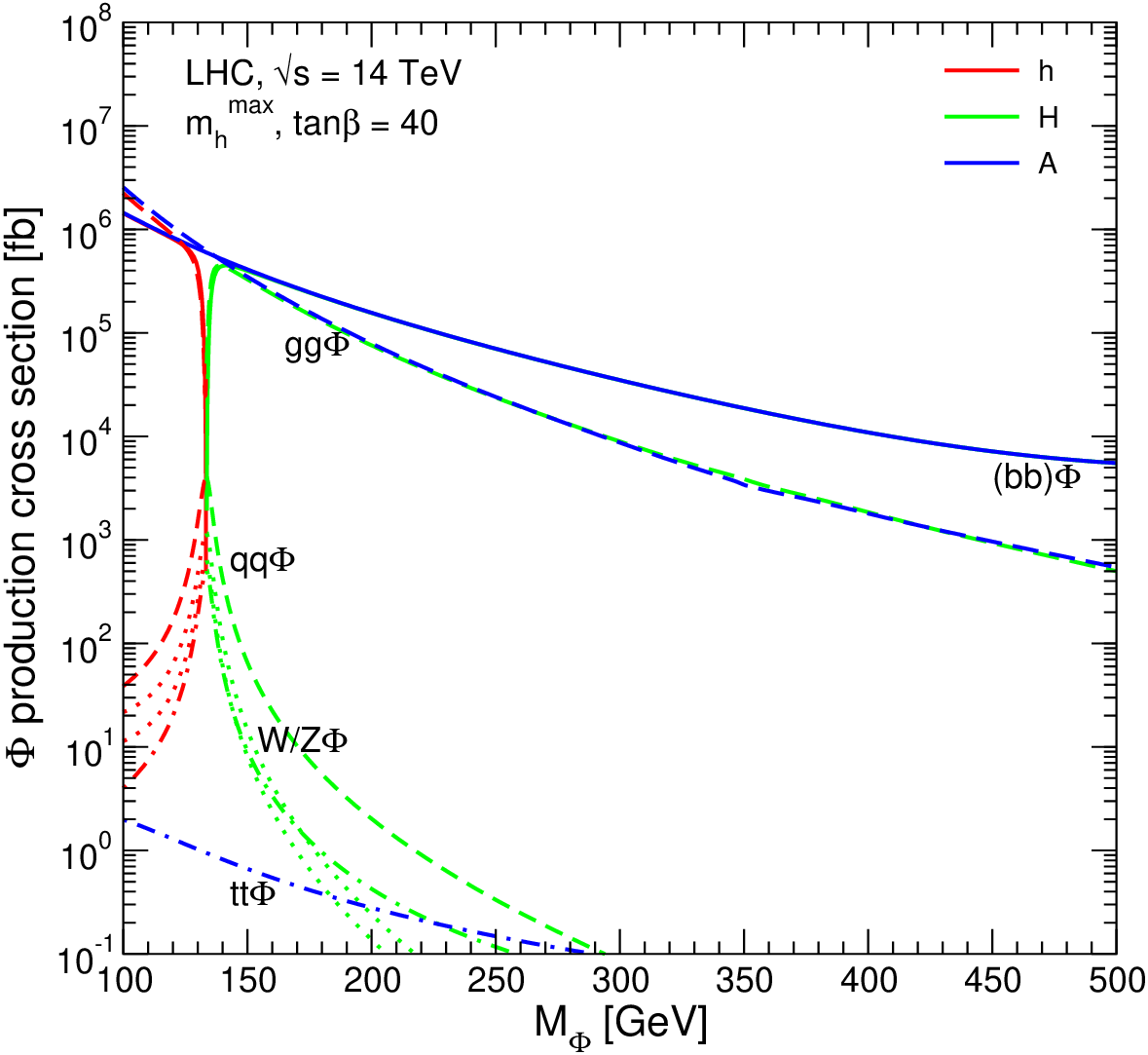}
  \end{minipage}
  \begin{minipage}[c]{0.49\textwidth}
    \includegraphics[width=\textwidth]{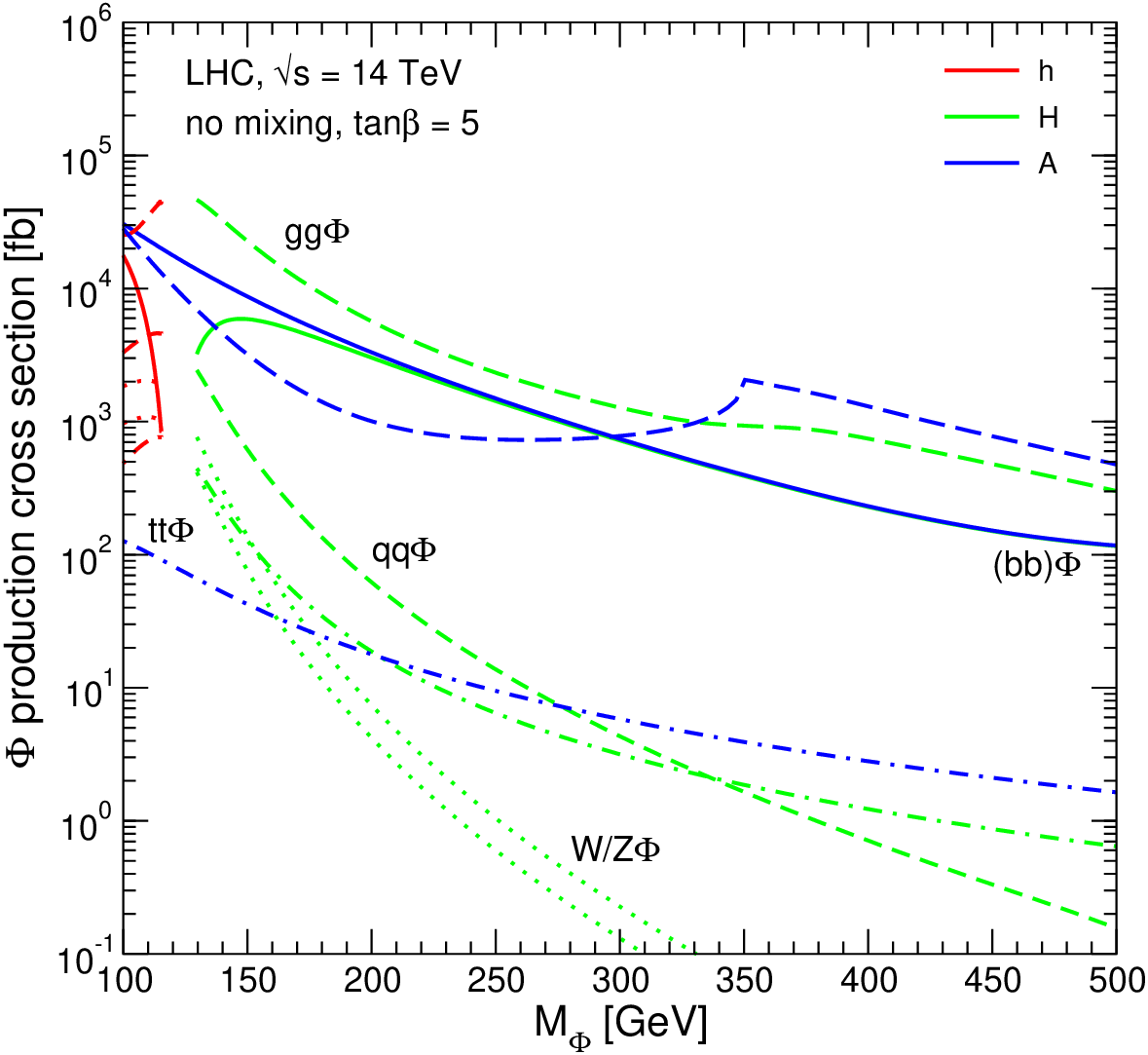}
  \end{minipage}
  \begin{minipage}[c]{0.49\textwidth}
    \includegraphics[width=\textwidth]{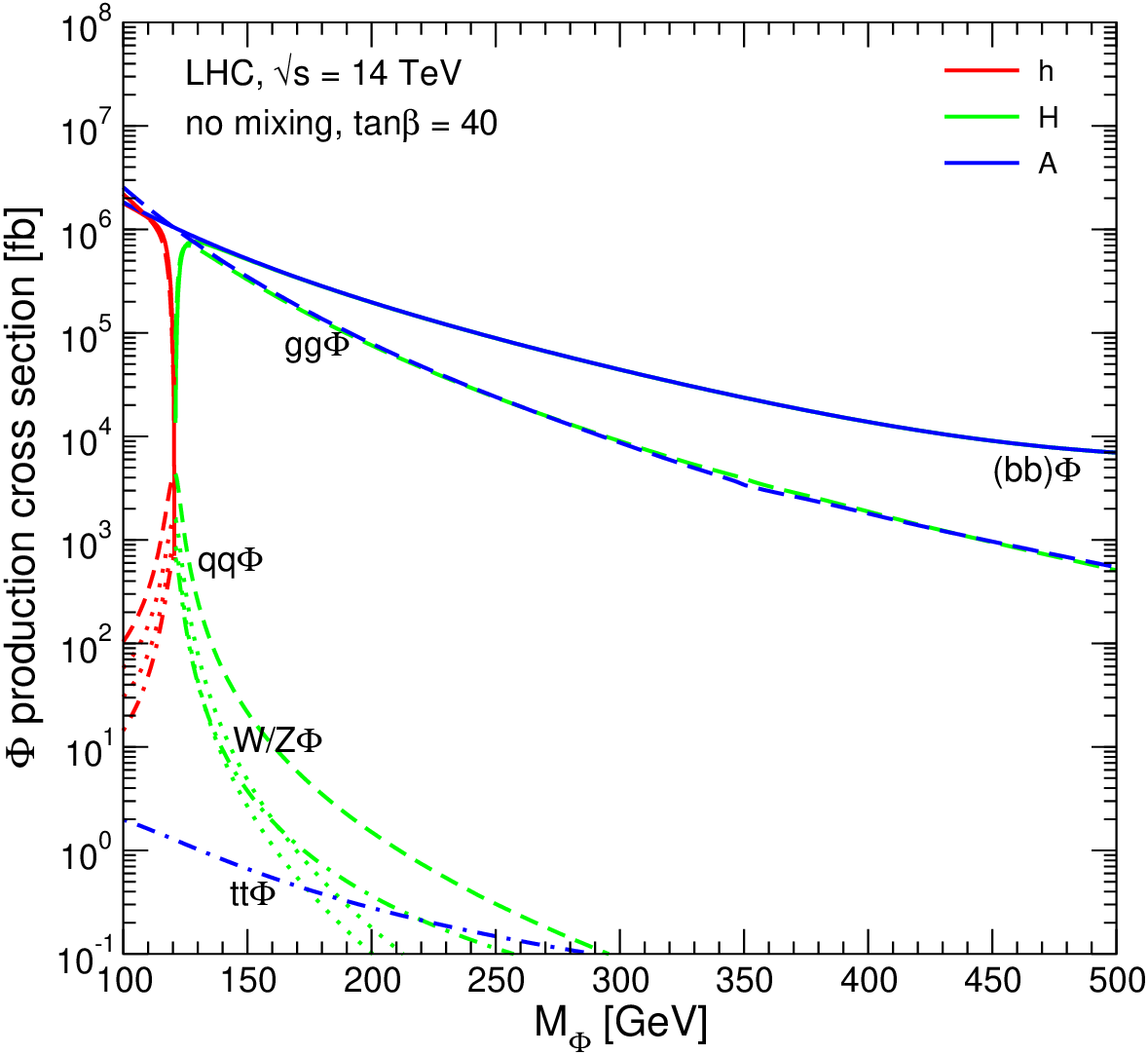}
  \end{minipage}
  \caption{Neutral Higgses production cross sections (fb) at the LHC, $\sqrt{s}=14$ TeV,
    for the most relevant production mechanims as a function of the Higgs-boson mass. Results
    are based on the SM cross sections and evaluated through an effective coupling approximation
    in the $\mhmax$ and no-mixing scenarios, for $\tanb=5,40$.}
  \label{fig:mssm_lhc_1}
\end{figure}

\begin{figure}[p]
  \begin{minipage}[c]{0.49\textwidth}
    \includegraphics[width=\textwidth]{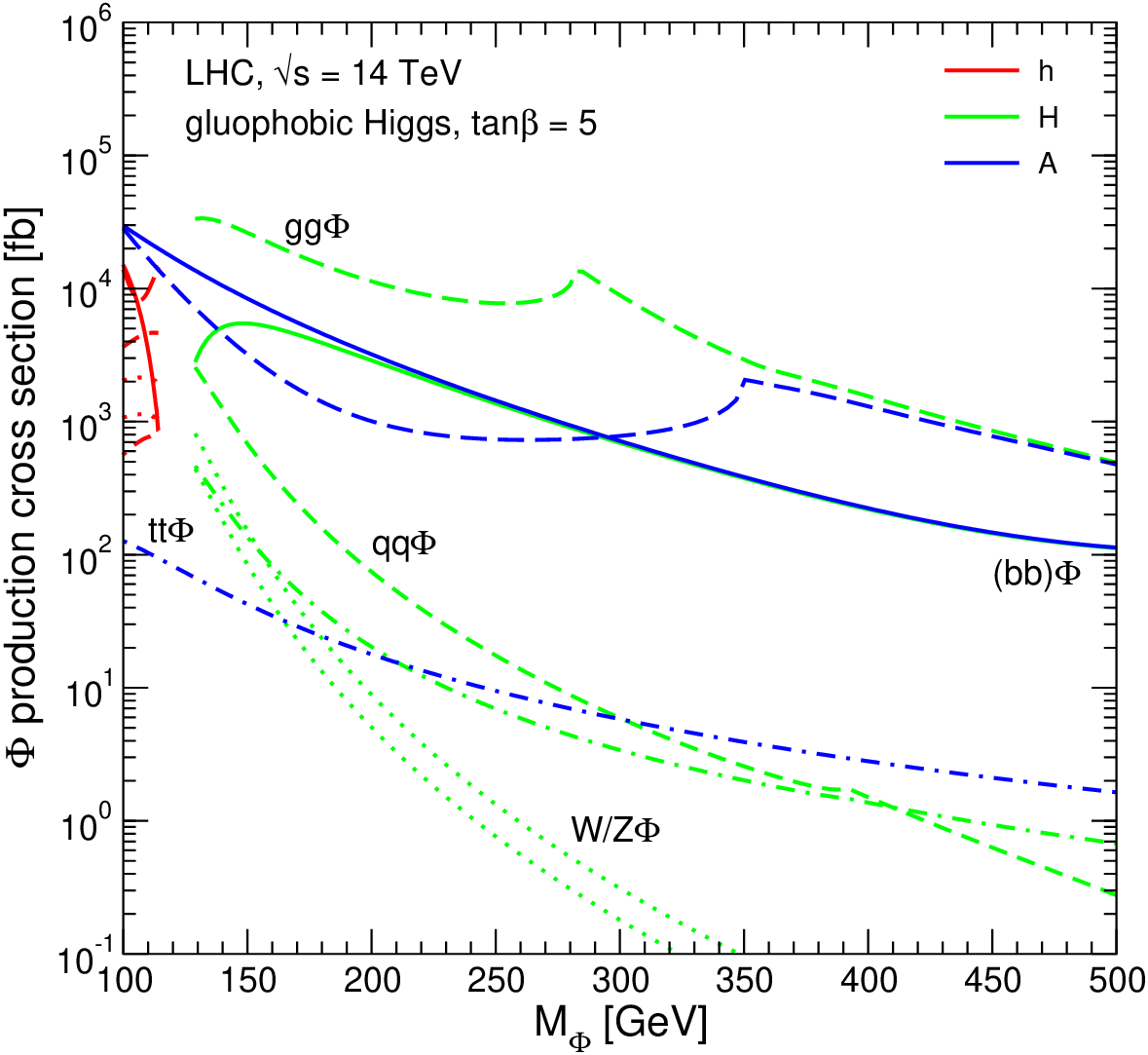}
  \end{minipage}
  \begin{minipage}[c]{0.49\textwidth}
    \includegraphics[width=\textwidth]{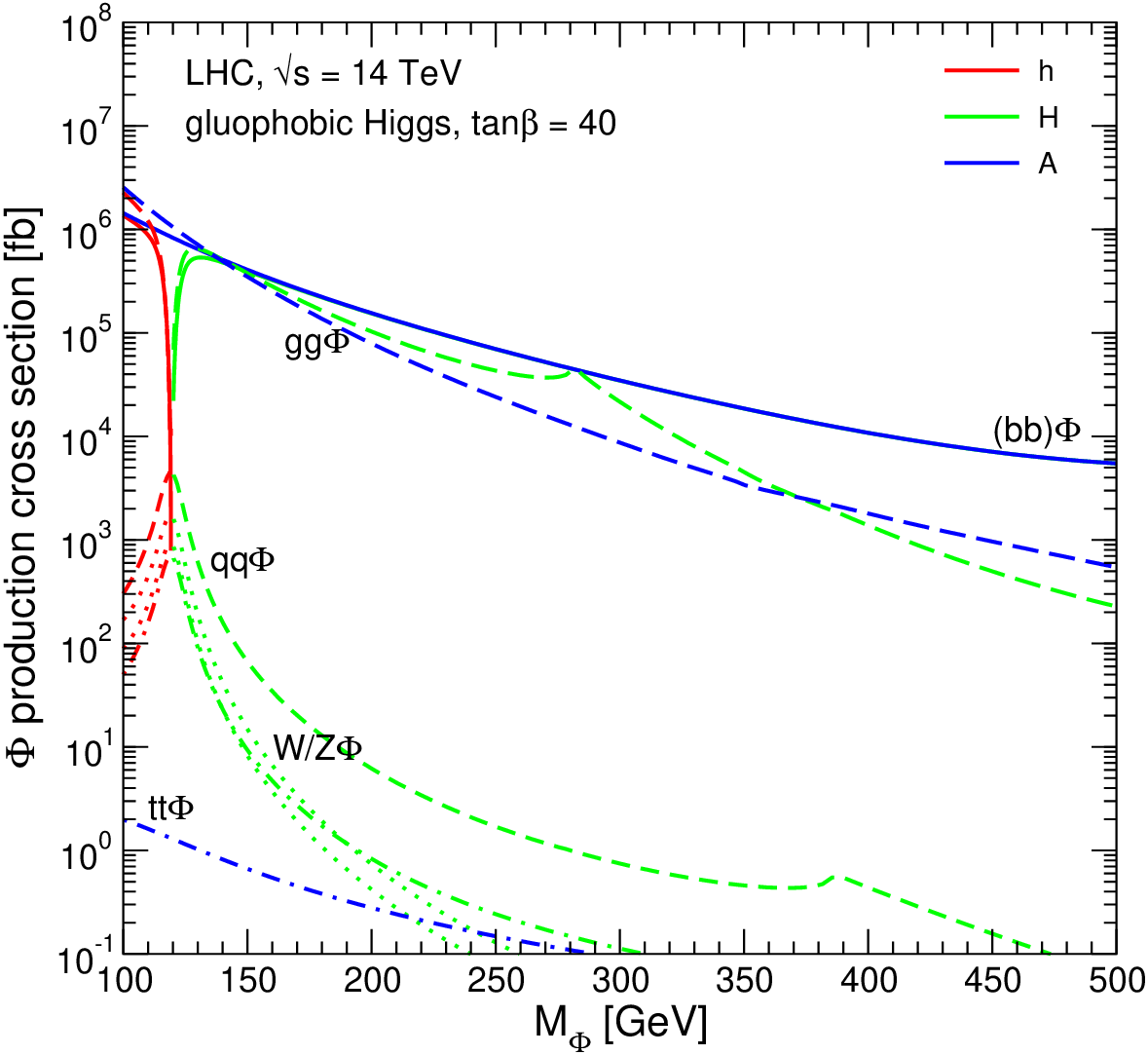}
  \end{minipage}
  \begin{minipage}[c]{0.49\textwidth}
    \includegraphics[width=\textwidth]{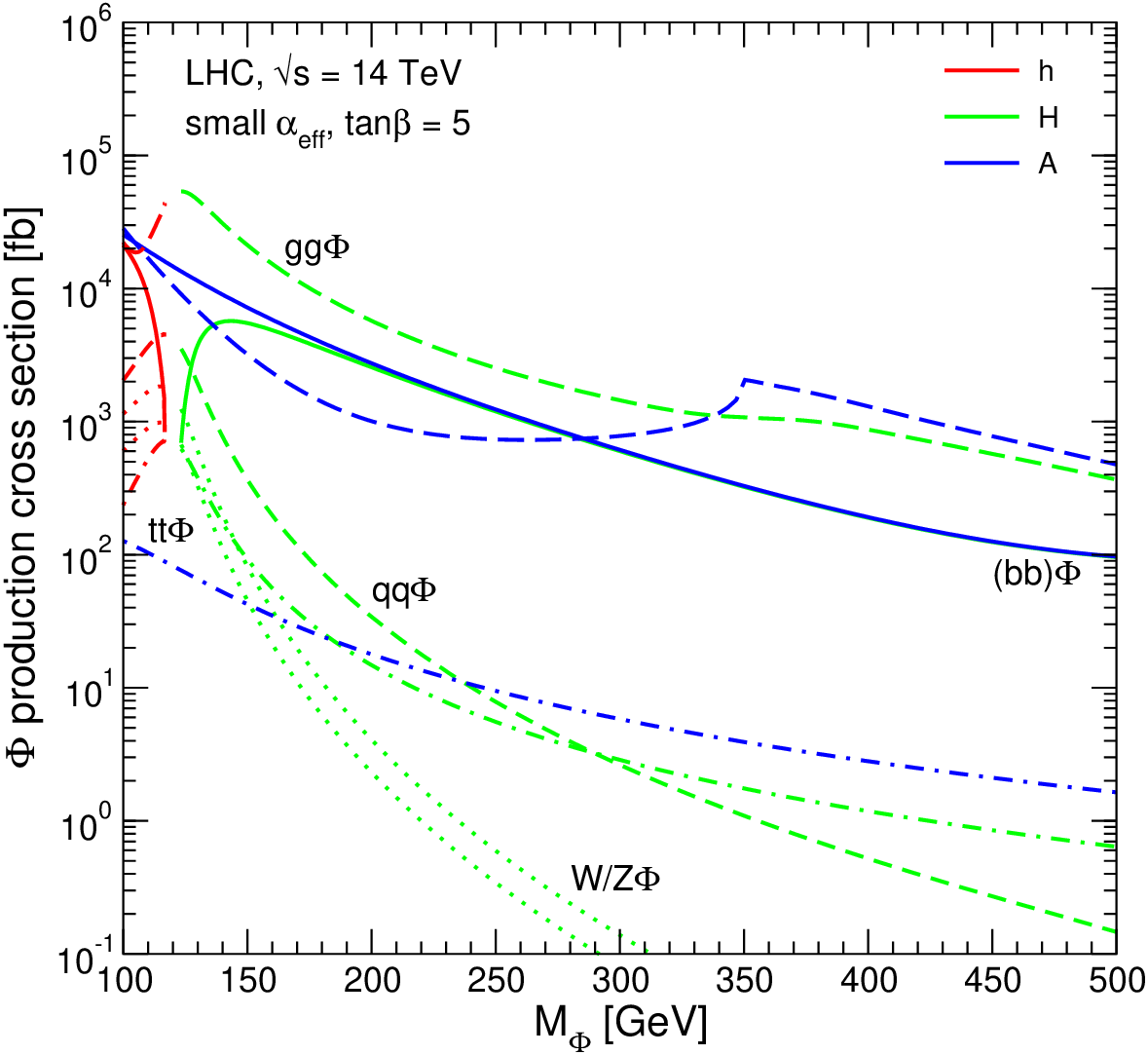}
  \end{minipage}
  \begin{minipage}[c]{0.49\textwidth}
    \includegraphics[width=\textwidth]{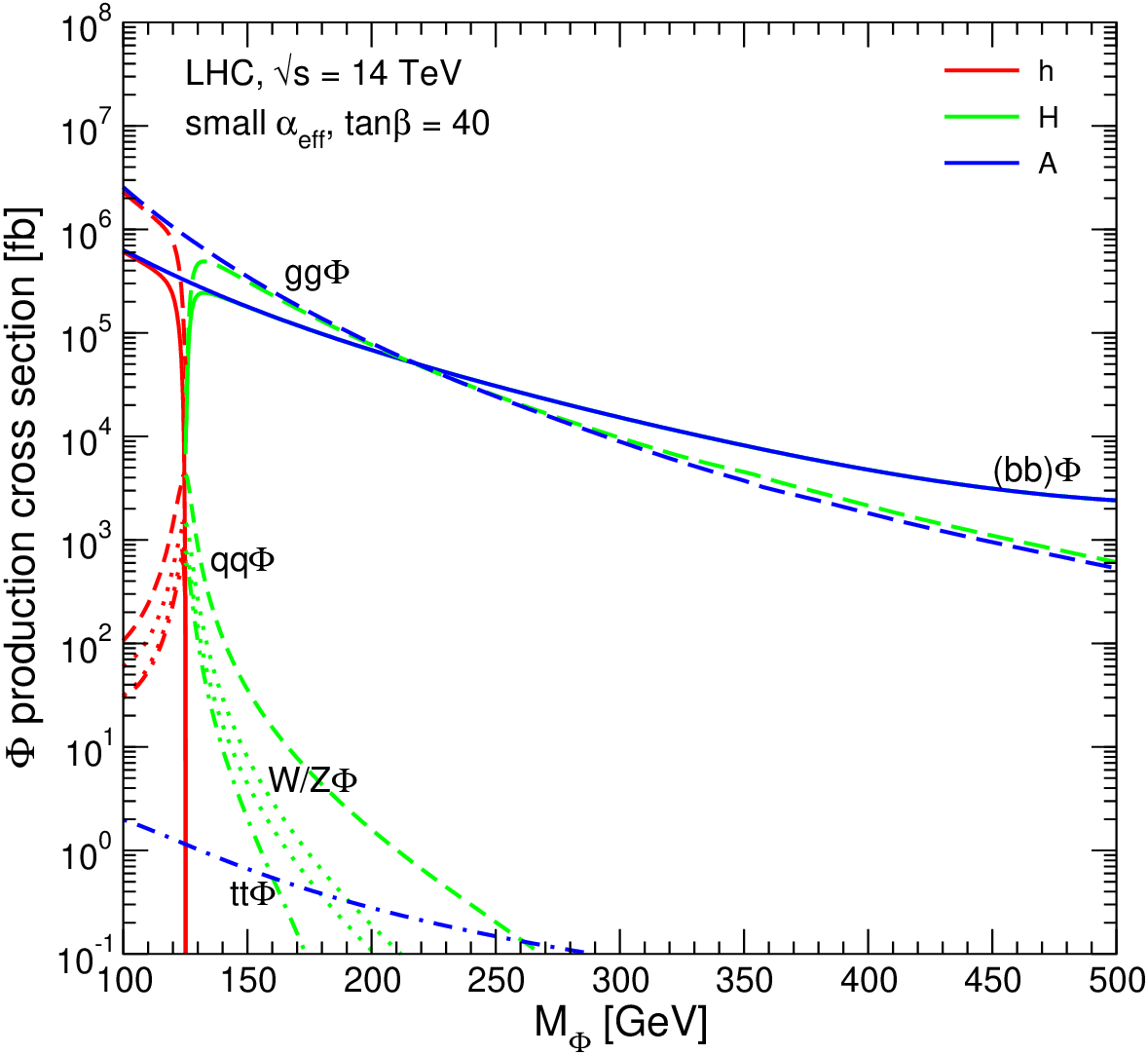}
  \end{minipage}
  \caption{Same as Fig.~\ref{fig:mssm_lhc_1}, for the gluophobic Higgs and small $\aeff$ scenarios.}
  \label{fig:mssm_lhc_2}
\end{figure}

Figs.~\ref{fig:mssm_tev_1} and~\ref{fig:mssm_tev_2}  
show the results for the Tevatron, while Figs.~\ref{fig:mssm_lhc_1} 
and~\ref{fig:mssm_lhc_2} show
the LHC results. In Fig.~\ref{fig:mssm_tev_1} (\ref{fig:mssm_lhc_1}) 
the Higgs production cross sections for the
neutral MSSM Higgs bosons at the Tevatron (LHC) in the $m_h^{\rm max}$
scenario (upper row) and the no-mixing scenario (lower row) can be found.
Fig.~\ref{fig:mssm_tev_2} (\ref{fig:mssm_lhc_2}) 
depicts the same for the gluophobic Higgs scenario (upper row)
and the small $\alpha_{\rm eff}$ scenario (lower row). 

For low $M_A$ values the production cross section of the $h$ and the $A$ 
are similar, while for large $M_A$ the cross sections of $H$ and $A$ 
are very close. This effect is even more pronounced for large $\tan\beta$.

The results presented in this paper have been obtained for the MSSM
with real parameters, i.e.\ the $\cp$-conserving case. They can can
easily be extended via the effective coupling approximation to the
case of non-vanishing complex phases (as implemented in {\tt FeynHiggs}).

\subsection*{Acknowledgements}
We are thankful to Mariano Ciccolini, Massimiliano Grazzini, 
Robert Harlander and Michael Kr\"amer for making some SM predictions
available to us.  
F.M.\ thanks Alessandro Vicini for useful discussions. 
S.H., F.M.\ and G.W.\ thank Michael Spira for lively
discussions.
S.H.\ is partially supported by CICYT (grant FPA2004-02948) and DGIID-DGA
(grant 2005-E24/2).

\clearpage
\section{Towards understanding the nature of Electroweak Symmetry
  Breaking at the Tevatron and LHC} 
\label{sec:belyaev}
\textbf{Contributed by: A.~Belyaev, A.~Blum, S.~Chivukula,
  E.~H.~Simmons}\\
PACS 14.80.Cp,11.30.Pb,11.15.Ex

\vspace{0.25in}

In this study 
we discuss how to extract  information about physics beyond the Standard Model (SM) from searches
for a light SM
Higgs at Tevatron Run II and CERN LHC.   
We demonstrate that new (pseudo)scalar states
predicted in both supersymmetric and dynamical models can have enhanced visibility in
standard Higgs search channels, making them potentially discoverable at Tevatron Run
II and CERN LHC.
We discuss the likely sizes of the enhancements in the various search channels for
each model and identify the model features having the largest influence on the degree
of enhancement.   We compare the
key signals for the non-standard scalars across models and also with expectations in
the SM, to show how one could start to identify which state has actually been found.
In particular, we suggest the likely mass reach of the Higgs search in $p\bar{p}/pp \to {\cal
H} \to \tau^+\tau^-$ for each kind of non-standard scalar state and we demonstrate that 
$p\bar{p}/pp \to {\cal H} 
\to \gamma\gamma$ may cleanly 
distinguish the scalars of supersymmetric models from those of dynamical models
and shed the light on the pattern of Electroweak Symmetry Breaking.



\subsection{Introduction}

The origin of electroweak symmetry breaking remains unknown.  While the
Standard Model (SM) of particle physics is consistent with existing data,
theoretical considerations suggest that this theory is only a low-energy effective
theory and must be supplanted by a more complete description of the underlying
physics at energies above those reached so far by experiment.

If the Tevatron or LHC  do find evidence for a new scalar state, it may not
necessarily be the Standard Higgs. Many alternative models of electroweak
symmetry breaking have spectra that include new scalar or pseudoscalar states
whose masses could easily lie in the range to which Run II is sensitive.  The
new scalars tend to have cross-sections and branching fractions that differ
from those of the SM Higgs.  The potential exists for one of these scalars to
be more visible in a standard search than the SM Higgs would be.

Here we discuss how to extract  information about non-Standard theories of electroweak symmetry breaking
from searches for a light SM Higgs at Tevatron Run II  and CERN LHC.
Ref.~\cite{Belyaev:2002zz} studied the potential of Tevatron Run II to augment its search
for the SM Higgs boson by considering the process $gg \to h_{SM} \to \tau^+\tau^-$.  
Authors determined what additional enhancement of scalar production and branching rate,
such as might be provided in a non-standard model like the MSSM, would enable a scalar to
become visible in the $\tau^+\tau^-$ channel alone at Tevatron Run II.   Similar work has
been done for  $gg\to h_{MSSM}\to \tau^+\tau^-$ at  the LHC ~\cite{Cavalli:2002vs} and
for $gg\to h_{SM} \to \gamma\gamma$ at the Tevatron \cite{Mrenna:2000qh} and LHC
\cite{Kinnunen:2005aq}. 

Our work builds on these results, considering an additional production
mechanism (b-quark annihilation), more decay channels ($b\bar{b}$, $W^+W^-$, $ZZ$,
and $\gamma\gamma$), and a wider range of non-standard physics (supersymmetry
and dynamical electroweak symmetry breaking) from which rate enhancement may
derive.  We discuss the possible sizes of the enhancements in the various search
channels for each model and pinpoint the model features having the largest
influence on the degree of enhancement.   We suggest the mass reach of the
standard Higgs searches for each kind of non-standard scalar state.   We also
compare the key signals for the non-standard scalars across models and also
with expectations in the SM, to show how one could identify which
state has actually been found.
Analytic formulas for the decay widths of the SM Higgs boson are taken from
\cite{Gunion:1989we}, \cite{Gunion:1992hs} and numerical values are calculated using the
HDECAY program \cite{Djouadi:1997yw}.

\subsection{Models of Electroweak Symmetry Breaking}

\subsubsection*{Supersymmetry}

One interesting possibility for addressing the hierarchy and triviality problems 
of the Standard Model is to
introduce supersymmetry. 

In order to provide masses to both 
up-type and down-type quarks, and to ensure anomaly cancellation,
the minimal supersymmetric Standard Model (MSSM) contains two
Higgs complex-doublet superfields:  $\Phi_d=(\Phi_d^0,\Phi_d^-)$ and 
$\Phi_u=(\Phi_u^+,\Phi_u^0)$ which aquire two vacuum expectation values $v_1$
and $v_2$ respectively.
Out of the original 8 degrees of freedom, 3 serve as Goldstone  bosons, absorbed
into longitudinal components of the $W^\pm$ and $Z$, making them massive.
The other 5 degrees of freedom remain in the spectrum as distinct scalar states, namely 
two neutral  CP-even states($h$, $H$), one neutral, CP-odd state ($A$)
and a charged pair ($H^\pm$).
It is conventional to choose $\tan\beta=v_1/v_2$ and $M_A=\sqrt{M_{H^\pm}^2-M_W^2}$
to define the SUSY Higgs sector. 
There are foloowing relations between Higgs masses
which will be useful for determining when Higgs 
boson interactions with fermions are enhanced: 
\begin{equation}
  M^2_{h,H} = {1\over 2}\left[ (M^2_A + M^2_Z) \mp 
    \sqrt{(M^2_A+M^2_Z)^2-4 M^2_A M^2_Z \cos^2 2\beta}\right];
  \cos^2(\beta-\alpha) = {M_h^2(M_Z^2-M_h^2)\over
    M_A^2(M_H^2-M_h^2)},
  \label{eq:higgs-mass}
\end{equation}
where $\alpha$ is the mixing angle of CP-even Higgs bosons.
The Yukawa interactions of the Higgs fields with the quarks and leptons
can be written as:
\footnote{Note that the interactions of the $A$ are pseudoscalar,
  {\it i.e.} it couples to $\bar{\psi} \gamma_5 \psi$.}
\BEA 
Y_{ht\bar{t}} /  Y^{SM}_{ht\bar{t}}   =  { \cos\alpha/\sin\beta} \ \ \ \ \ \ \ 
&
Y_{Ht\bar{t}} /  Y^{SM}_{ht\bar{t}}  =  { \sin\alpha/\sin\beta}  \ \ \ \ \ \ \ 
&
Y_{At\bar{t}}/   Y^{SM}_{ht\bar{t}}  =  { \cot\beta}  \nonumber\\
Y_{hb\bar{b}} / Y^{SM}_{hb\bar{b}}  = {-\sin\alpha/\cos\beta} 	\ \ \ \ \ \ \ 
& 
Y_{Hb\bar{b}} / Y^{SM}_{hb\bar{b}}  = {\  \cos\alpha/\cos\beta}	\ \ \ \ \ \ \  
&
Y_{Ab\bar{b}}/  Y^{SM}_{hb\bar{b}}  =  { \tan\beta}    
\label{eq-yukawas} 
\EEA
relative to the Yukawa couplings of the Standard Model ($Y^{SM}_{hf\bar{f}} = m_f/v)$.  Once again, the same pattern holds
for the tau lepton's  Yukawa couplings as for those of the $b$ quark.
There are several circumstances under which various Yukawa couplings are enhanced relative to 
Standard Model values.
For high $\tan\beta$ (small $\cos\beta$), eqns. (\ref{eq-yukawas}) show that the
interactions of all neutral Higgs bosons with the down-type fermions
are enhanced by a factor of $1/\cos\beta$.  In the decoupling limit, where $M_A\to\infty$, 
applying eq. (\ref{eq:higgs-mass})
to eqns. (\ref{eq-yukawas}) shows that the $H$ and $A$
Yukawa couplings to down-type fermions are enhanced by a factor of $\simeq\tan\beta$.
Conversely, for low $m_A\simeq m_h$, one can check that
$Y_{hb\bar{b}}  /Y^{SM}_{hb\bar{b}} =  
Y_{h\tau\bar{\tau}}/Y^{SM}_{h\tau\bar{\tau}}\simeq \tan\beta
$
that $h$ and $A$ Yukawas are enhanced instead.

\subsubsection*{Technicolor}

Another intriguing class of theories, dynamical electroweak symmetry
breaking (DEWSB), supposes that the scalar states involved in
electroweak symmetry breaking could be manifestly composite at scales
not much above the electroweak scale $v \sim 250$ GeV.  In these
theories, a new asymptotically free strong gauge interaction  (technicolor 
\cite{Susskind:1978ms,Weinberg:1975gm, Weinberg:1979bn}) breaks the chiral symmetries of massless fermions $f$ at
a scale $\Lambda \sim 1$ TeV.  If the fermions carry appropriate
electroweak quantum numbers (e.g. left-hand (LH) weak doublets and right-hand (RH) weak
singlets), the resulting condensate $\langle \bar f_L f_R \rangle \neq
0$ breaks the electroweak symmetry as desired.  Three of the Nambu-Goldstone Bosons
(technipions) of the chiral symmetry breaking become the
longitudinal modes of the $W$ and $Z$. The logarithmic running of the
strong gauge coupling renders the low value of the electroweak scale
natural.  The absence of fundamental
scalars obviates concerns about triviality.

Many models of DEWSB have additional light neutral pseudo Nambu-Goldstone bosons which could potentially be accessible to
a standard Higgs search; these are called ``technipions" in technicolor models.   Our analysis will assume, for
simplicity, that the lightest PNGB state is significantly lighter than other neutral (pseudo) scalar technipions, so as to
heighten the comparison to the SM Higgs boson.  

The specific models we examine are: 1) the traditional one-family model \cite{Farhi:1980xs} with a full family of
techniquarks and technileptons, 2) a variant on the one-family model \cite{Casalbuoni:1998fs} in which the lightest
technipion contains only down-type technifermions and is significantly lighter than the other pseudo Nambu-Goldstone
bosons,  3) a multiscale walking technicolor model \cite{Lane:1991qh} designed to reduce flavor-changing neutral currents,
and  4) a low-scale technciolor model (the Technicolor Straw Man model) \cite{Lane:1999uh} with many weak doublets of
technifermions, in which the second-lightest technipion $P'$ is the state relevant for our study (the lightest, being
composed of technileptons, lacks the anomalous coupling to gluons required for $gg \to P$ production).  For simplicity the
lightest relevant neutral technipion of each model will be generically denoted $P$; where a specific model is meant, a
superscript will be used.

One of the key differences among these models is the value of the technipion decay constant $F_P$, which is related to the
number $N_D$ of weak doublets of technifermions that contribute to electroweak symmetry breaking.  We refer reader
to~\cite{Belyaev:2005ct} for details.

\subsection{Results For Each Model}

\subsubsection*{Supersymmetry}

Let us consider how the signal of a light Higgs boson could be changed in the MSSM,
compared to expectations in the SM.   
There are several important sources of alterations in the predicted signal, some of which are
interconnected.

First, the MSSM includes three neutral Higgs bosons ${\cal H}=(h,H,A)$ states. The
apparent signal of a single light Higgs could be enhanced if two or three neutral
Higgs species are nearly degenerate, and  we  take advantage of this
near-degeneracy by combining the signals of the different neutral Higgs bosons when
their masses are closer than the experimental resolution.  

Second, the alterations of the couplings between Higgs bosons and ordinary fermions
in the MSSM can change the Higgs decay widths
and branching ratios relative to those in the SM. 
Radiative effects on the
masses and couplings can substantially alter decay branching fractions  in a non-universal way. 
For instance, $B(h\to\tautau)$ could be enhanced by up to an order of magnitude due
to the suppression of $B(h\to b\bar{b})$ in certain regions of parameter space
\cite{Carena:1998gk,Carena:1999bh}.  However, this gain in branching  fraction
would be offset
 to some degree by a reduction in Higgs production through channels
involving  $Y_{{\cal H}b\bar{b}}$~\cite{Belyaev:2002zz}.

Third, a large value of $\tan\beta$ enhances the
bottom-Higgs coupling (eqns. (\ref{eq-yukawas}) ), making gluon fusion through a 
$b$-quark loop significant, and possibly even dominant over the top-quark loop contribution.

Fourth, the presence of superpartners in the MSSM gives rise to new squark-loop
contributions  to Higgs boson production through gluon fusion. 
Light squarks with masses of order 100 GeV have been argued to lead
to a considerable universal enhancement (as much as a factor of five)
\cite{Dawson:1996xz,Harlander:2003bb,Harlander:2003kf,Harlander:2004tp}
for MSSM Higgs production compared to the SM.

Finally, enhancement of the $Y_{{\cal H}b\bar{b}}$ coupling at moderate to large $\tan\beta$
makes  $b\bar{b} \to {\cal H}$ a significant means of Higgs production in the
MSSM -- in contrast to the SM where it is negligible.   To include both production
channels when looking for a Higgs decaying as ${\cal H} \to xx$, we define
a combined enhancement factor 
\begin{equation}
\kappa_{total/xx}^{\cal H} 
=
{\sigma(gg\to{\cal H}\to xx)+\sigma(bb\to{\cal H} \to xx)\over
{\sigma(gg\to h_{SM} \to xx)+\sigma(bb\to h_{SM} \to xx)}}
\equiv
[\kappa_{gg/xx}^{\cal H}+\kappa_{bb/xx}^{\cal H} R_{bb:gg}]/
[{1+  R_{bb:gg}}]  .
\label{kappab}
\end{equation}
Here 
 $R_{bb:gg}$ is the ratio of $b\bar{b}$ and $gg$ initiated Higgs boson production in the 
 Standard Model, which can be calculated using HDECAY.  
\begin{figure}
  \begin{minipage}[c]{0.49\textwidth}
    \includegraphics[width=\textwidth]{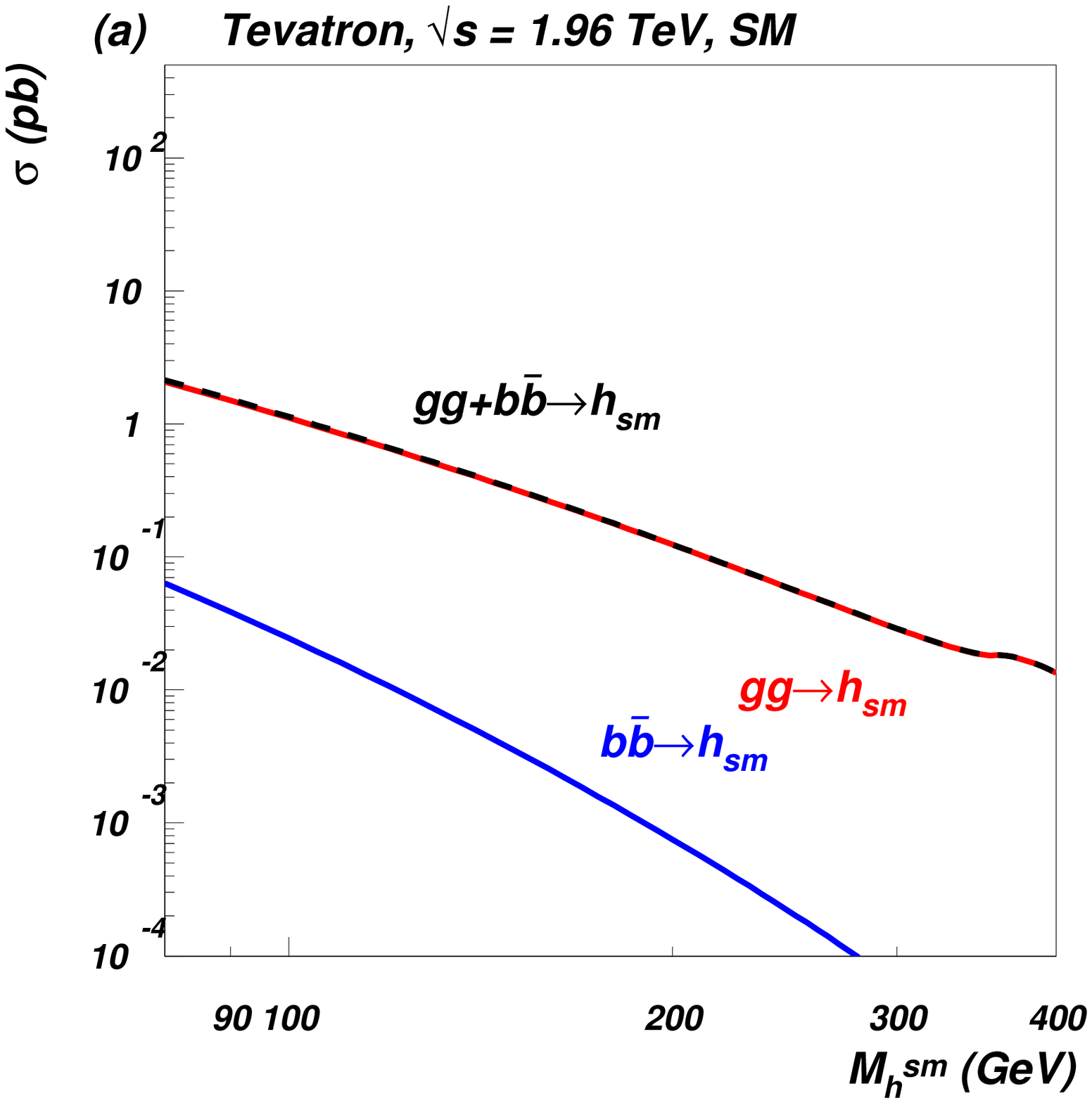}
  \end{minipage}
  \begin{minipage}[c]{0.49\textwidth}
    \includegraphics[width=\textwidth]{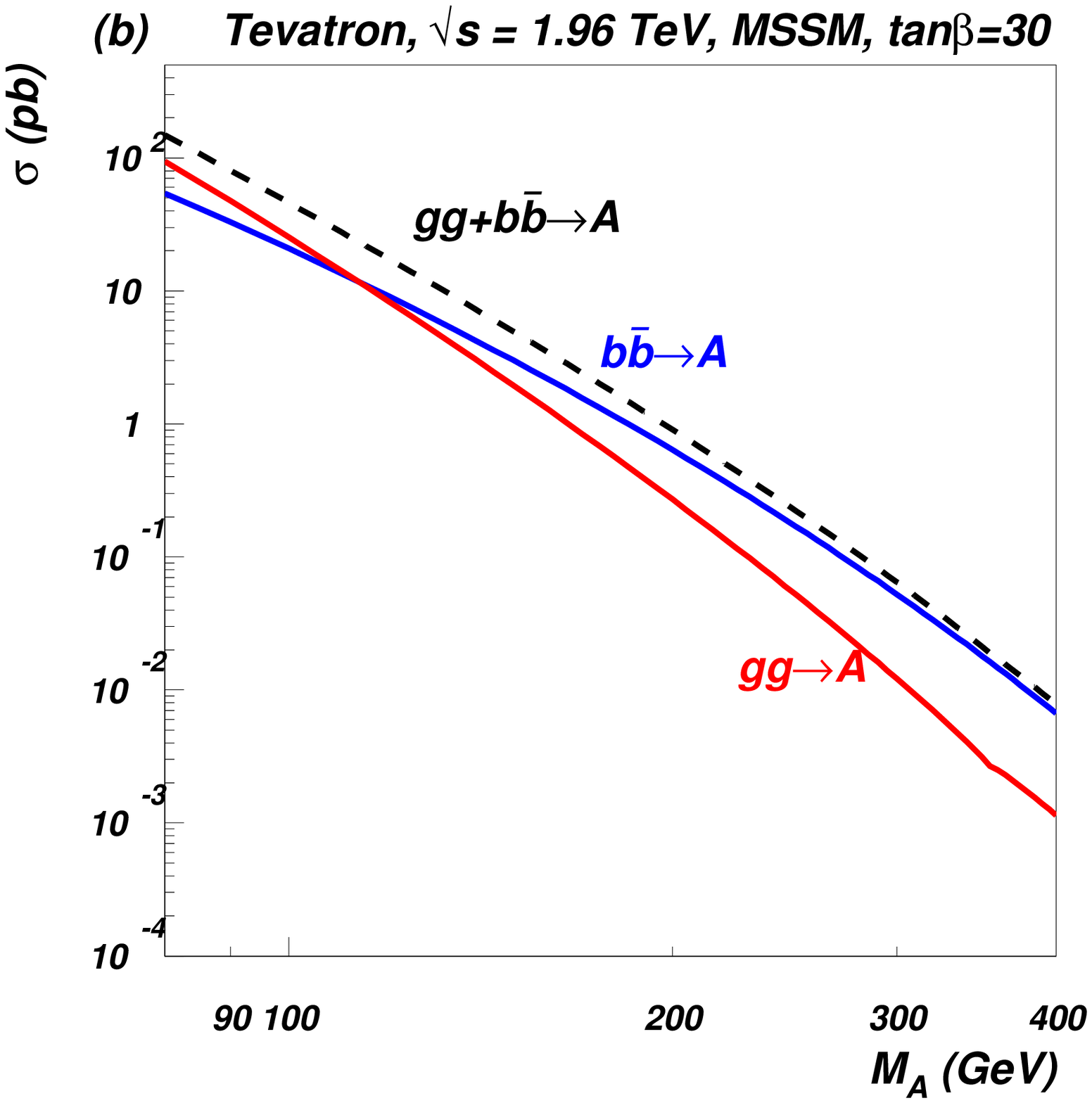}
  \end{minipage}
\caption[mssm-tot-prod]{
NLO cross sections for Higgs production via the $b\bar{b}\to {\cal H}$ and $gg\to {\cal H}$
processes (as well as their sum) at the Tevatron for  the SM Higgs (a)
the Supersymmetric  axial Higgs
boson with $\tan\beta=30$ (b).}
\label{fig:mssm-tot-prod}
\end{figure}

\begin{figure}
  \begin{minipage}[c]{0.49\textwidth}
    \includegraphics[width=\textwidth]{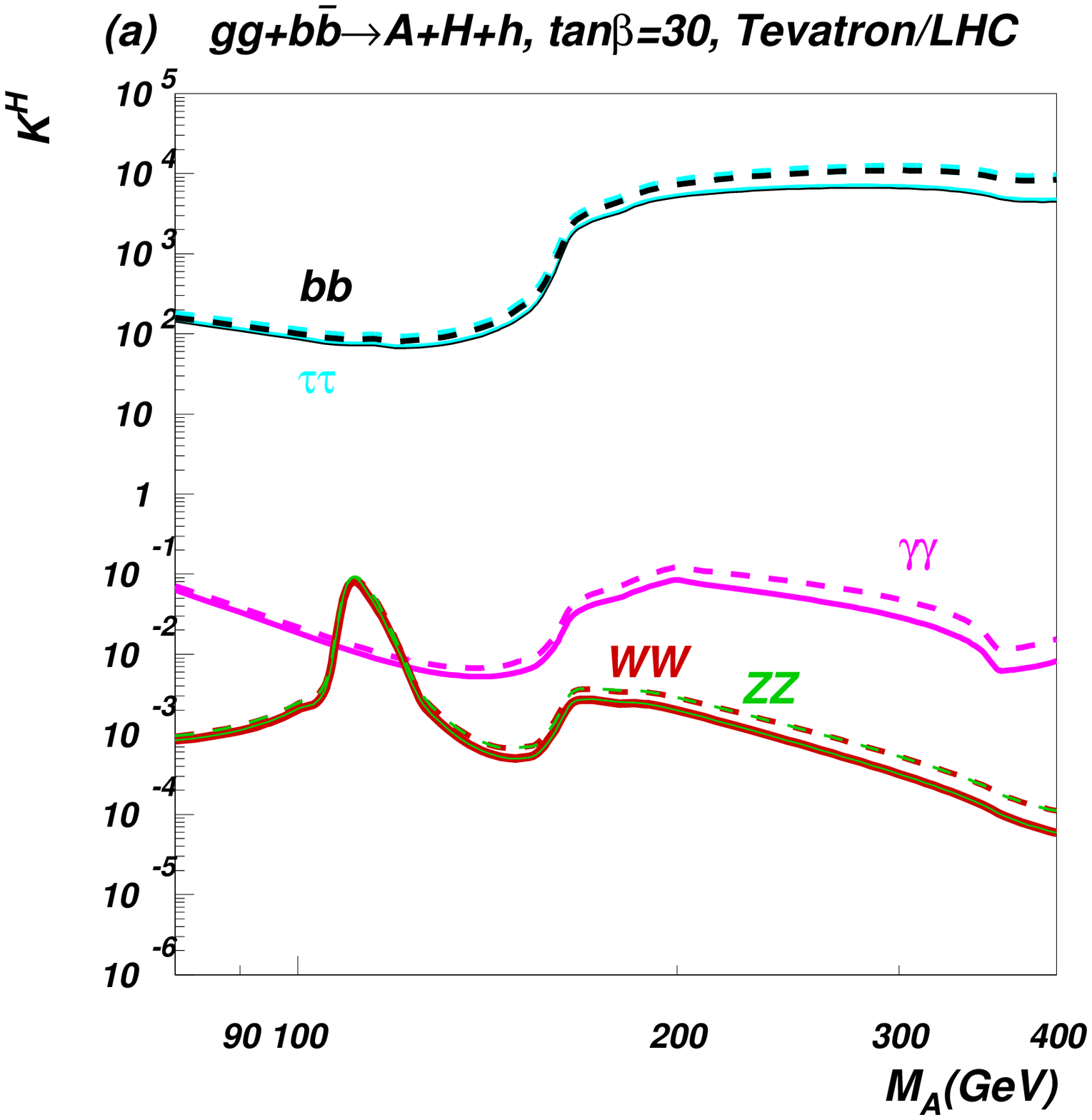}%
  \end{minipage}
  \begin{minipage}[c]{0.49\textwidth}
    \includegraphics[width=\textwidth]{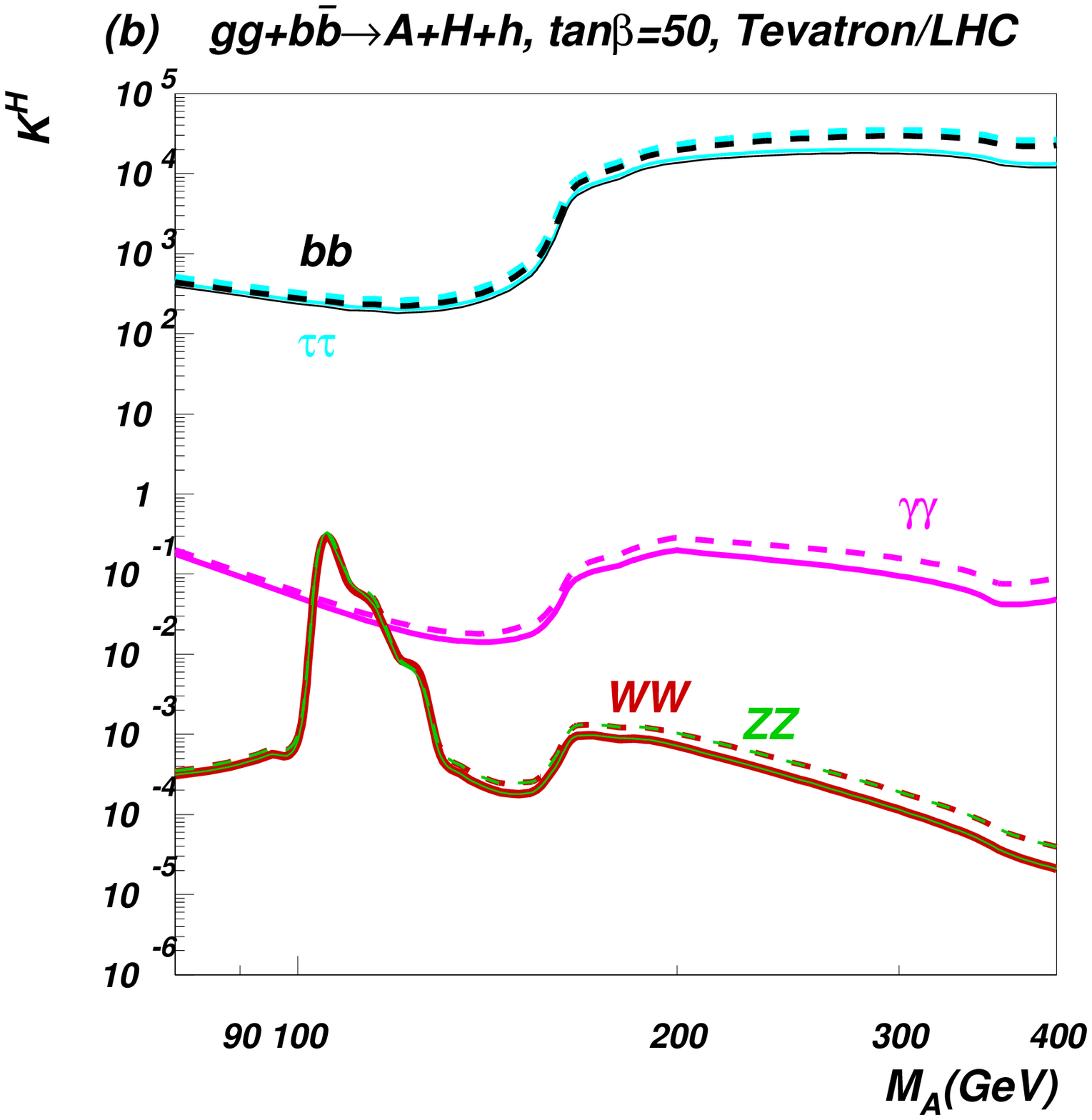}
  \end{minipage}
\caption[fig:kthtot]{
Enhancement factor $\kappa^{\cal H}_{tot/xx}$ for final states $xx = b\bar{b},\ \tau^+\tau^-,\ WW,\  ZZ,\
\gamma\gamma$ when both $gg\to {\cal H}$ and $b\bar{b}\to {\cal H}$ are included and the signals of all
three MSSM Higgs states are combined.  Frames (a) and  (b) correspond to $\tan\beta=30$ 
and 50, respectively,
at the Tevatron (solid lines) and at the LHC (dashed lines).
}
\label{fig:kthtot}
\end{figure}

Figure~\ref{fig:mssm-tot-prod} 
presents NLO cross sections at the Tevatron.
For $b\bar{b}\to {\cal H}$ we are using the 
code of Ref.~\cite{Balazs:1998sb}, \footnote{
Note that  $b\bar{b}\to {\cal H}$
has been recently calculated at NNLO
in \protect\cite{Harlander:2003ai}.}
while for $gg\to {\cal H}$ we use HIGLU \cite{Spira:1996if}
and HDECAY \cite{Djouadi:1997yw} .\footnote{
Specifically, we use the HIGLU package to calculate 
the $gg\to h_{sm}$ cross section. We then use
the ratio of the Higgs decay widths from HDECAY
(which includes a more complete set of one-loop MSSM corrections
 than HIGLU)
to get the MSSM $gg\to {\cal H}$ cross section:
$\sigma^{MSSM}= 
\sigma^{SM}\times \Gamma({\cal H}\to gg)/ \Gamma(h_{SM}\to gg)$.
} One can see
that in the MSSM  the contribution from  $b\bar{b}\to {\cal H}$ becomes  important even
for moderate values of $\tan\beta\sim 10$. For $M_{\cal H}<110-115$~GeV the contribution
from  $gg\to {\cal H}$ process is a bit bigger than that from $b\bar{b}\to {\cal H}$,  while for
$M_{\cal H}>115$~GeV $b$-quark-initiated production begins to outweigh gluon-initiated
production.  Results for LHC are qualitatively similar,
except the rate, which is about two orders of magnitude higher compared to that at the Tevatron.

Using the Higgs branching fractions  with these NLO cross sections for
$gg\to H$ and $b\bar{b}\to H$ allows us to derive $\kappa_{total/xx}^{\cal H}$, as
presented in Fig.~\ref{fig:kthtot} for the Tevatron and LHC. 
There are
several ``physical" kinks and peaks  in the enhancement factor for various Higgs
boson final states related to  $WW$, $ZZ$ and top-quark thresholds which can be
seen  for the respective  values of $M_A$. At very large  values of $\tan\beta$
the top-quark threshold effect for the $\gamma\gamma$ enhancement factor is almost gone
because the b-quark contribution dominates in the loop. 
One can see from Fig.~\ref{fig:kthtot} that the enhancement factors at the Tevatron
and LHC are very similar.  On the other hand, the values of the total rates at the LHC
are about two orders of magnitude higher than the corresponding rates at the Tevatron.
In contrast to strongly enhanced $b\bar{b}$ and $\tau\bar{\tau}$ signatures,
the  $\gamma\gamma$ signature is always strongly suppressed! This particular feature 
of SUSY models, as we will see below, may be important for distinguishing
supersymmetric models from models with dynamical symmetry breaking.

It is important to note that combining the signal from the neutral Higgs bosons $h, A, H$ in the MSSM
turns out to make our results more broadly applicable across SUSY parameter space.       
Combining the signals from  $A,h,H$ has the virtue of making  the enhancement factor independent of the
degree of top squark mixing (for fixed $M_A$, $\mu$ and $M_{S}$ and medium to high values of $\tan\beta$), 
which greatly reduces the
parameter-dependence of our results.

\subsubsection*{Technicolor}

Single production of a technipion can occur through the axial-vector anomaly which
couples the technipion to pairs of gauge bosons. For an $SU(N_{TC})$ technicolor group
with technipion decay constant $F_P$, the anomalous coupling between the technipion and
a pair of gauge bosons is given, in direct analogy with the coupling of a QCD pion to
photons,
 by \cite{Dimopoulos:1980yf,Ellis:1980hz,Holdom:1981bg}.
Comparing a PNGB to a SM Higgs boson of the same mass, 
we find the enhancement in the gluon fusion production  is 
\begin {equation}
\kappa_{gg\ prod} = \frac{ \Gamma (P \rightarrow gg)}{ \Gamma (h \rightarrow gg)} = \frac{9}{4} N_{TC}^2 {\cal A}_{gg}^2 \frac{v^2}{F_P^2}
\label{eq:kappagg}
\end {equation}

The main factors influencing $\kappa_{gg\ prod}$ for a fixed value of $N_{TC}$ are the anomalous coupling to gluons and
the technipion decay constant.   The value of $\kappa_{gg\ prod}$ for each model (taking $N_{TC} = 4$) is given in 
Table~\ref{tab:tc-pr}.

\begin {table} [tb]
  \label{tab:tc-pr}
  \begin{center}
    \caption{
      Calculated enhancement factors for production at the Tevatron and LHC of a 130 GeV technipion  via $gg$ alone, via
      $b\bar{b}$ alone, and combined. Note that the small enhancement in the $b\bar{b}$ process slightly reduces the
      total enhancement relative to that of $gg$ alone. In all cases, $N_{TC}=4$.}
    \begin{tabular}{|c||c||c||c||c|}\hline
      & 1) one family & 2) variant one-family & 3) multiscale & 4) low scale \\
      \hline
      $ \kappa^P_{gg\ prod} $ & 48 & 6 & 1200 & 120 \\
      \hline
      $ \kappa^P_{bb\ prod} $ & 4  & 0.67 & 16 & 10 \\
      \hline\hline
      $ \kappa^P_{prod} $ & 47 & 5.9 & 1100 & 120 \\
      \hline
    \end{tabular}
  \end{center}
\end{table}

The value of $\kappa_{bb\ prod}$ (shown in Table~\ref{tab:tc-pr}) is controlled by the size of the technipion decay
constant.   

We see from Table~\ref{tab:tc-pr} that $\kappa_{bb\ prod}$ is at least one order of magnitude smaller than $\kappa_{gg\
prod}$ in each model.  From the ${\kappa_{gg\ prod}}/{\kappa_{bb\ prod}}$ ratio 
which reads as
\begin{equation}
\frac{\kappa_{gg\ prod}}{\kappa_{bb\ prod}} = \frac{9}{4} N_{TC}^2 {\cal A}_{gg}^2 \lambda_b^{-2}
\left(1 - \frac{4m_b^2}{m_h^2}\right)^{\frac{3-s}{2}},
\end{equation}
we see that the larger size of $\kappa_{gg\ prod}$ is due to the factor of $N_{TC}^2$ coming from the fact
that gluons couple to a technipion via a techniquark loop.  The
extended technicolor (ETC) interactions coupling $b$-quarks to a
technipion have no such enhancement.
With a smaller SM cross-section and a smaller enhancement factor, it is
clear that technipion production via $b\bar{b}$ annihilation is essentially  negligible  at these hadron colliders.  

We now calculate the technipion branching ratios from the above information, taking $N_{TC} = 4$.  The
values are essentially independent of the size of $M_P$ within the range 120 GeV - 160 GeV; the branching
fractions for $M_P = 130$ GeV are shown in Table~\ref{tab:tc-br}. The branching ratios for the SM Higgs at NLO are given
for comparison; they were calculated using HDECAY \cite{Djouadi:1997yw}.  
\begin{table}[tb]
  \begin{center}
    \caption{\label{tab:tc-br}Branching ratios of Technipions/Higgs of mass 130 GeV}
    \begin{tabular}{|c||c||c||c||c|||c|}\hline
      Decay & 1) one family & 2) variant & 3) multiscale & 4) low scale & SM Higgs \\
      Channel & &one family & & & \\
      \hline
      $b \overline{b}$ & 0.60 & 0.53 & 0.23  & 0.60 & 0.53 \\
      $ \tau^+ \tau^-$ & 0.03  & 0.25  & 0.01 & 0.03 & 0.05  \\
      $ \gamma \gamma $ & $ 2.7 \times 10^{-4}  $ & $ 2.9 \times 10^{-3}  $ & $ 6.1 \times 10^{-4} $ & $ 6.4 \times 10^{-3}$  & $ 2.2 \times 10^{-3} $ 
      \\
      \hline
    \end{tabular}
  \end{center}
\end{table}
Comparing the technicolor and SM branching ratios in Table~\ref{tab:tc-br}, 
we see immediately that all decay enhancements.  
Model 2 is an exception; its unusual Yukawa couplings yield a decay enhancement in the
$\tau^+\tau^-$ channel of order the technipion's (low) production enhancement.  In the $\gamma\gamma$ channel,
the decay enhancement strongly depends on the group-theoretical structure of the model, through the anomaly
factor.   
\begin{table}[bt]
  \begin{center}
    \caption {\label{tab:tc-en}Enhancement Factors for 130 GeV technipions produced at the Tevatron and LHC,
      compared to production and decay of a SM Higgs Boson of the same mass. The
      slight suppression of \protect{$\kappa^P_{prod}$} due to the b-quark
      annihilation channel has been included. The rightmost column shows the
      cross-section (pb) for \protect{$p\bar{p}/pp \to P \to xx$} at Tevatron Run II/LHC. }
    \renewcommand\tabcolsep{9pt}
    \begin{tabular}{|c|c||c||c||c||c|}
      \hline
      Model & Decay mode &  $\kappa^P_{prod}$ & $\kappa^P_{dec}$ & $\kappa^P_{tot/xx}$ & $\sigma$(pb) Tevatron/LHC \\
      \hline
      &$b \overline{b}$ & 47 & 1.1 & 52 & 14 / 890 \\
      1) one family&$ \tau^+ \tau^-$ & 47 & 0.6 & 28 & 0.77 / 48 \\
      &$ \gamma \gamma $ & 47 & 0.12 & 5.6 & $ 6.4 \times 10^{-3}$ / 0.4  \\
      \hline
      \hline
      &$b \overline{b}$ & 5.9 & 1 & 5.9 & 1.8 / 100\\
      2) variant&$ \tau^+ \tau^-$ & 5.9 & 5 & 30 & 0.84 / 52\\
      one family&$ \gamma \gamma $ & 5.9 & 1.3 & 7.7 & $ 8.7 \times 10^{-3}$ / 0.55 \\
      \hline
      \hline
      &$b \overline{b}$ & 1100 & 0.43 & 470 & 130 / 8000\\
      3) multiscale &$ \tau^+ \tau^-$ & 1100 & 0.2 & 220 & 6.1 / 380\\
      &$ \gamma \gamma $ & 1100 & 0.27 & 300 & 0.34 /22 \\
      \hline
    \end{tabular}
  \end{center}
\end{table}
Our results for the Tevatron Run II and LHC production enhancements (including both $gg$
fusion and $b\bar{b}$ annihilation), decay enhancements, and overall enhancements
of each technicolor model relative to the SM are shown in Table ~\ref{tab:tc-en} for a technipion
or Higgs mass of 130 GeV.  Multiplying $\kappa^P_{tot/xx}$ by the cross-section for
SM Higgs production via gluon fusion \cite{Spira:1996if}
yields an approximate technipion production cross-section, as shown in the right-most column of Table~\ref{tab:tc-en}.   

In each technicolor model, the main enhancement of the possible technipion signal
relative to that of an SM Higgs arises at production, making the size of the
technipion decay constant the most critical factor in determining the degree of
enhancement for fixed $N_{TC}$.

\subsection{Interpretation}

We are ready to put our results in context.  The large QCD background for $q\bar{q}$
states of any flavor makes the tau-lepton-pair and di-photon final states the most
promising for exclusion or discovery of the Higgs-like states of the MSSM or
technicolor.  We now illustrate how the size of the enhancement factors for these two
final states vary over the parameter spaces of these theories at the Tevatron and LHC. 
We use this information to display the likely reach of each experiment in each of these
standard Higgs search channels.   Then, we compare the signatures of the MSSM Higgs
bosons and the various technipions to see how one might tell these states apart from one
another.

%
\begin{figure}
  \begin{minipage}[c]{0.49\textwidth}
    \includegraphics[width=\textwidth]{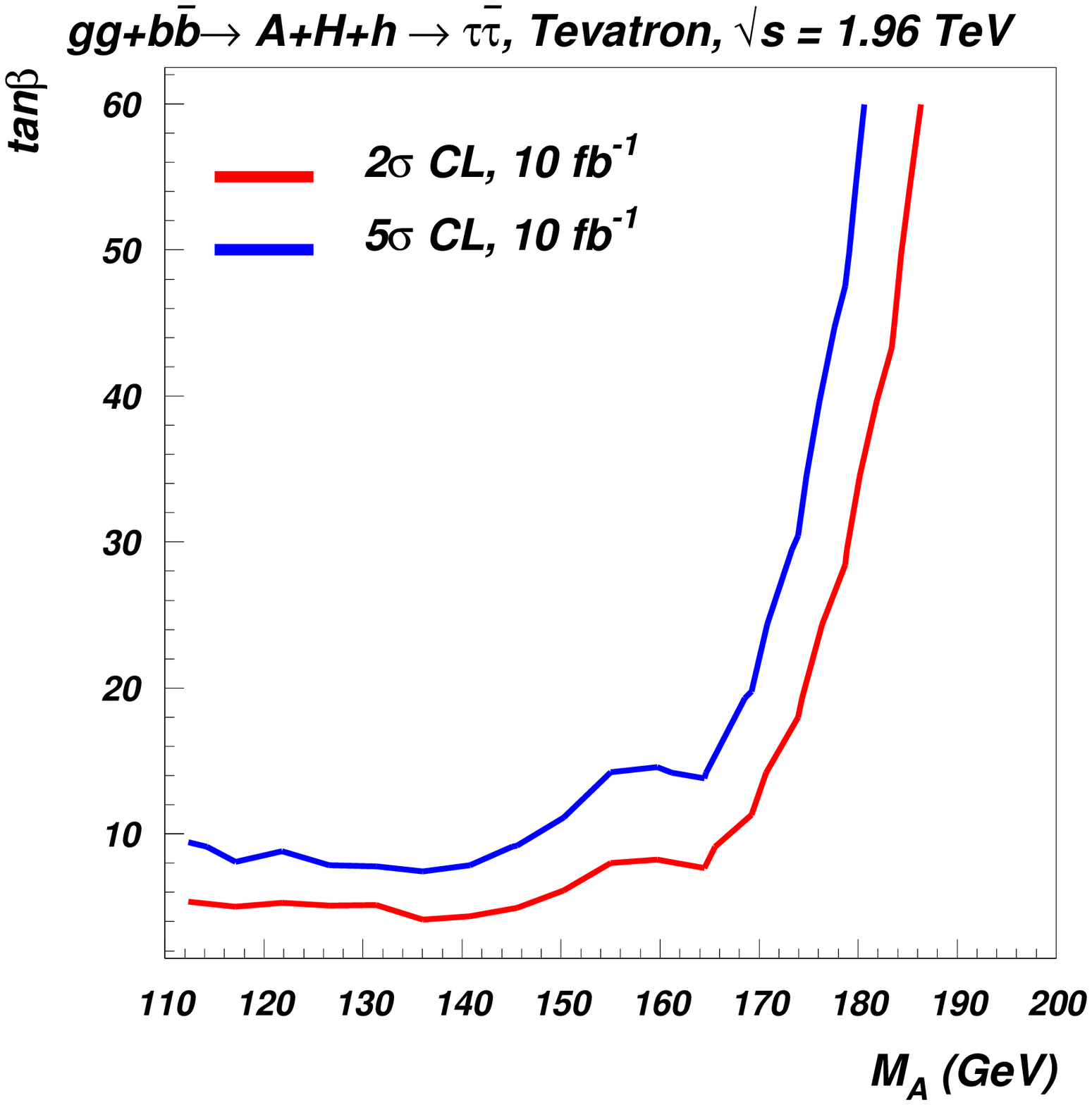}
  \end{minipage}
  \begin{minipage}[c]{0.49\textwidth}
    \includegraphics[width=\textwidth]{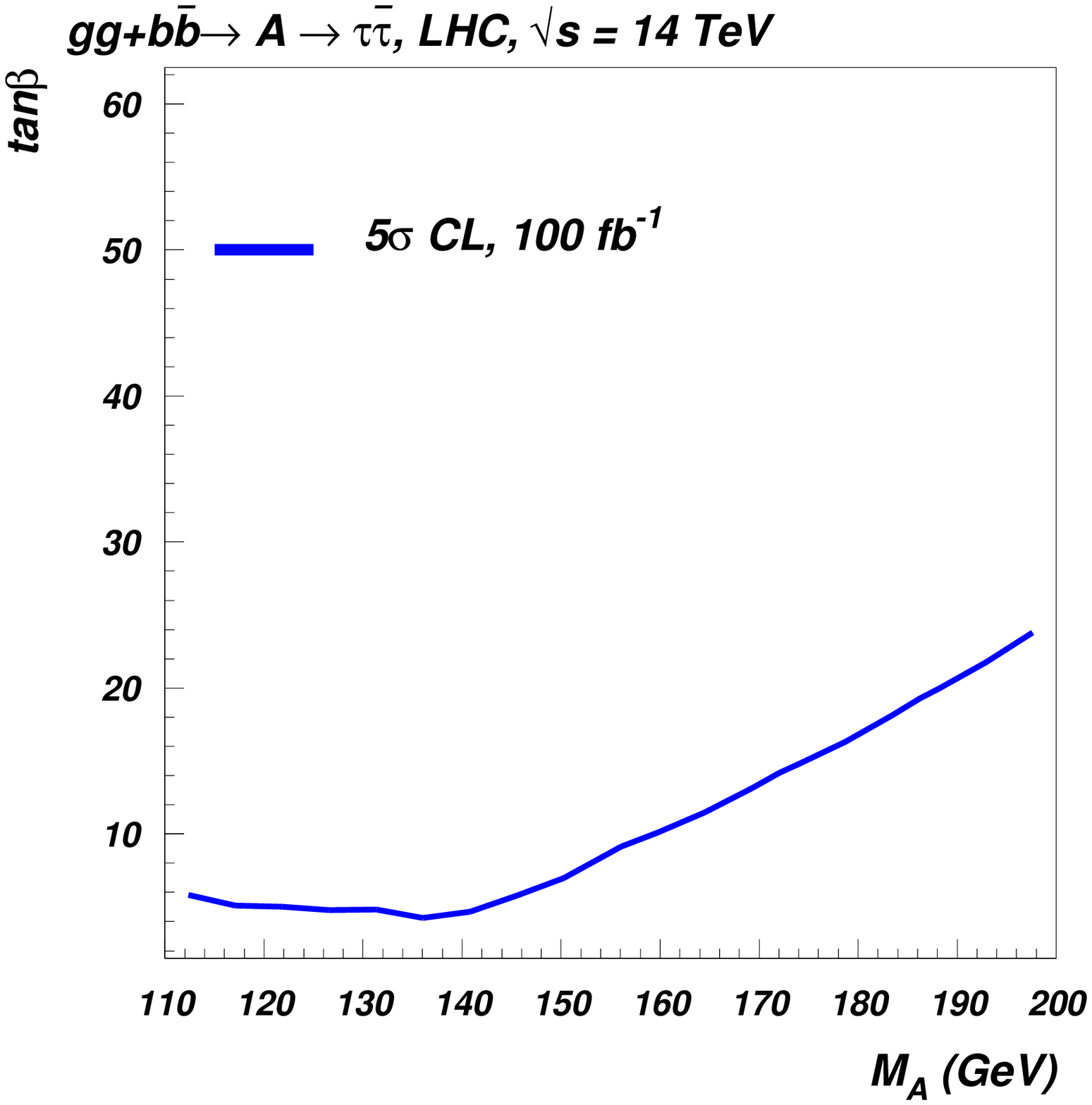}
  \end{minipage}
\caption[mssm-plots]{
Results for  \protect{$gg + b\bar{b} \to h + H + A \to \tau^+\tau^-$} at LHC.  
Left frame: Selected contours of given enhancement factor values
\protect{$\kappa^{\cal H}_{total/\tau\tau}$} in the MSSM.  Right frame: 
Predicted LHC reach, based on the $h_{SM} \to \tau^+\tau^-$ studies of  \cite{Cavalli:2002vs},
 in the MSSM parameter space.}
\label{fig:mssm-tev-lhc}
\end{figure}

In of Figure~\ref{fig:mssm-tev-lhc} we summarize the  ability of Tevatron (left) and LHC
(right)  to explore the MSSM  parameter space (in terms of both a $2\sigma$ exclusion
curve and a $5\sigma$ discovery curve) using the process $gg + b\bar{b} \to h + A + H \to
\tau^+ \tau^-$.  Translating the enhancement factors  into this reach plot draws on the
results of \cite{Belyaev:2002zz}.    As the $M_A$ mass increases up to about 140 GeV, the
opening of the $W^+W^-$ decay channel drives the $\tau^+\tau^-$ branching fraction down,
and increases the $\tan\beta$  value required to make  Higgses  visible in the
$\tau^+\tau^-$ channel.   At still larger $M_A$, a very steep drop in the gluon
luminosity (and the related $b$-quark luminosity) at large $x$ reduces the phase space
for ${\cal H}$ production.  Therefore for $M_A>$170 GeV,  Higgs bosons would only be
visible at very high values of $\tan\beta$.  The pictures for tevaron and LHC are
qualitatively similar, the main differences compared to the Tevatron are that the 
required value of  $\tan\beta$ at the LHC  is lower for a given $M_A$ and it does not
climb steeply for  $M_A>$170 GeV because there is much less phase space suppression.

It is important to notice that both, Tevatron and LHC,  could observe MSSM Higgs bosons
in the $\tau^+\tau^-$  channel even for moderate values of $\tan\beta$ for $M_A\lesssim
200$~GeV, because of significant enhancement of this channel. However the   
$\gamma\gamma$ channel is so suppressed that even the LHC will not be able to observe it
in any point of the $M_A < 200$ GeV parameter space studied in this paper! {\footnote{ In
the decoupling limit with large values of $M_A$ and low values of $\tan\beta$, the
lightest MSSM Higgs could be dicovered in the $\gamma\gamma$ mode just like the  SM model
Higgs boson}

The  Figure~\ref{fig:techni-limits} presents the Tevatron and LHC potentials to observe
technipions. For the Tevatron, the observability is presented in terms of enhancement
factor, while for the LHC we present signal rate in term of $\sigma\times Br(P\to
\tau\tau/\gamma\gamma)$. At the Tevatron, the available enhancement is well above what is
required to render the $P$ of any of these models visible in the $\tau^+\tau^-$
channel.   Likewise, the right frame of that figure shows that in the $\gamma\gamma$
channel at the Tevatron the technipions of models 3 and 4 will be observable at the
$5\sigma$ level while model 2 is subject to exclusion at the $2\sigma$ level.  The
situation at the LHC is even more promising: all four models could be observable at the
$5\sigma$ level in both the $\tau^+\tau^-$ (left frame) and $\gamma\gamma$ (right frame) 
channels.

\begin{figure}
  \begin{minipage}[c]{0.49\textwidth}
    \includegraphics[width=\textwidth]{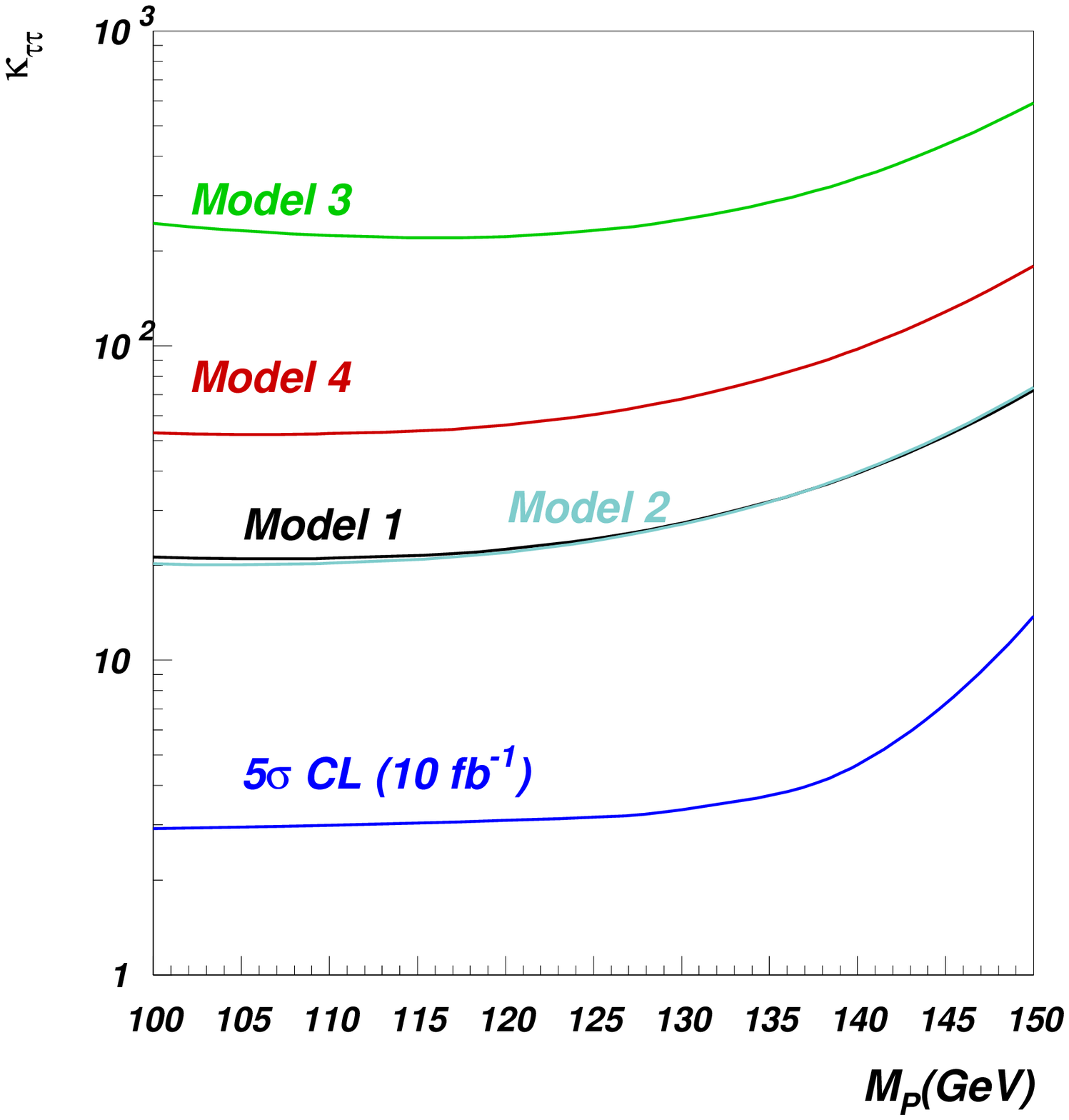}
  \end{minipage}
  \begin{minipage}[c]{0.49\textwidth}
    \includegraphics[width=\textwidth]{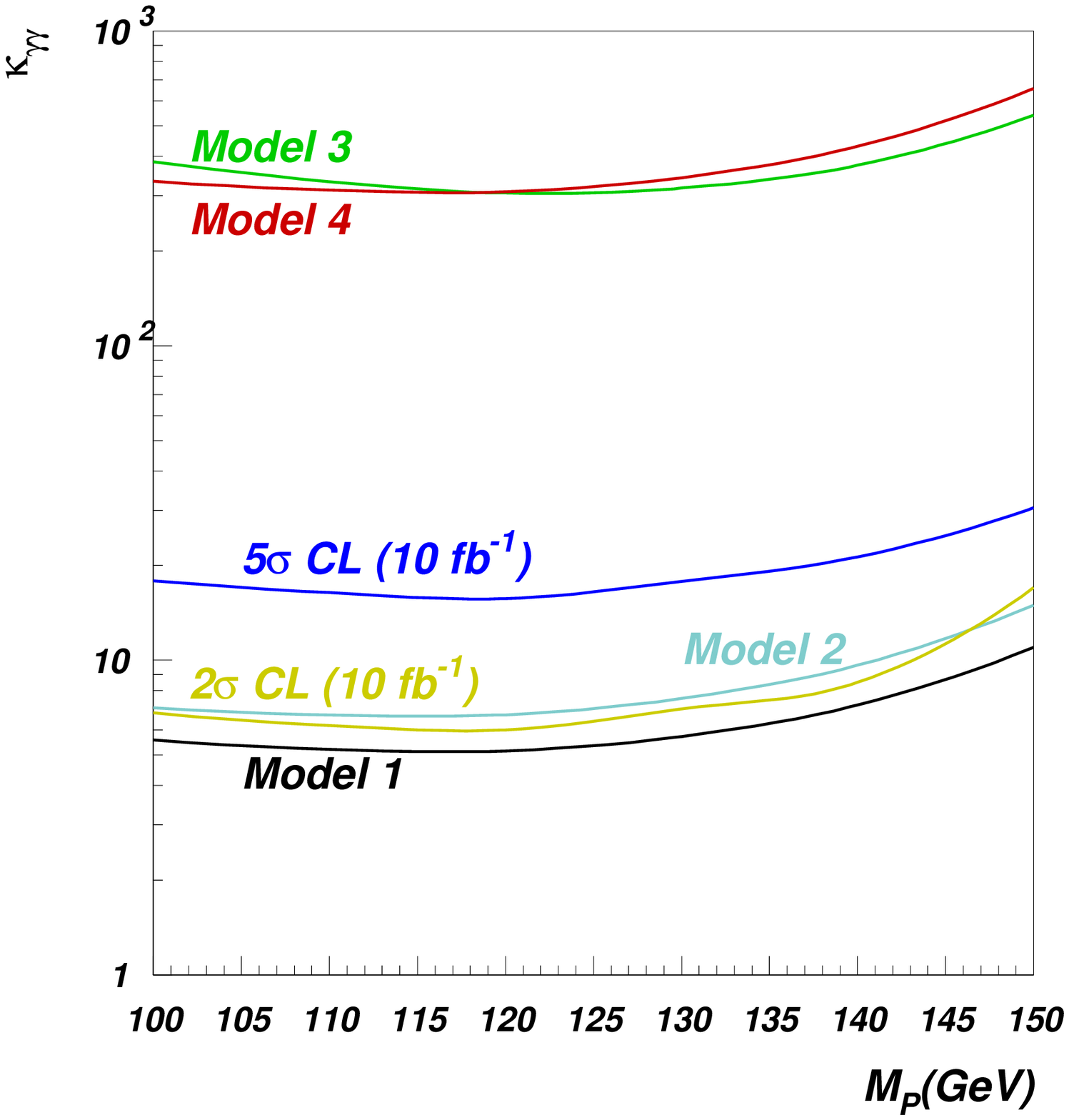}
  \end{minipage}
  \begin{minipage}[c]{0.49\textwidth}
    \includegraphics[width=\textwidth]{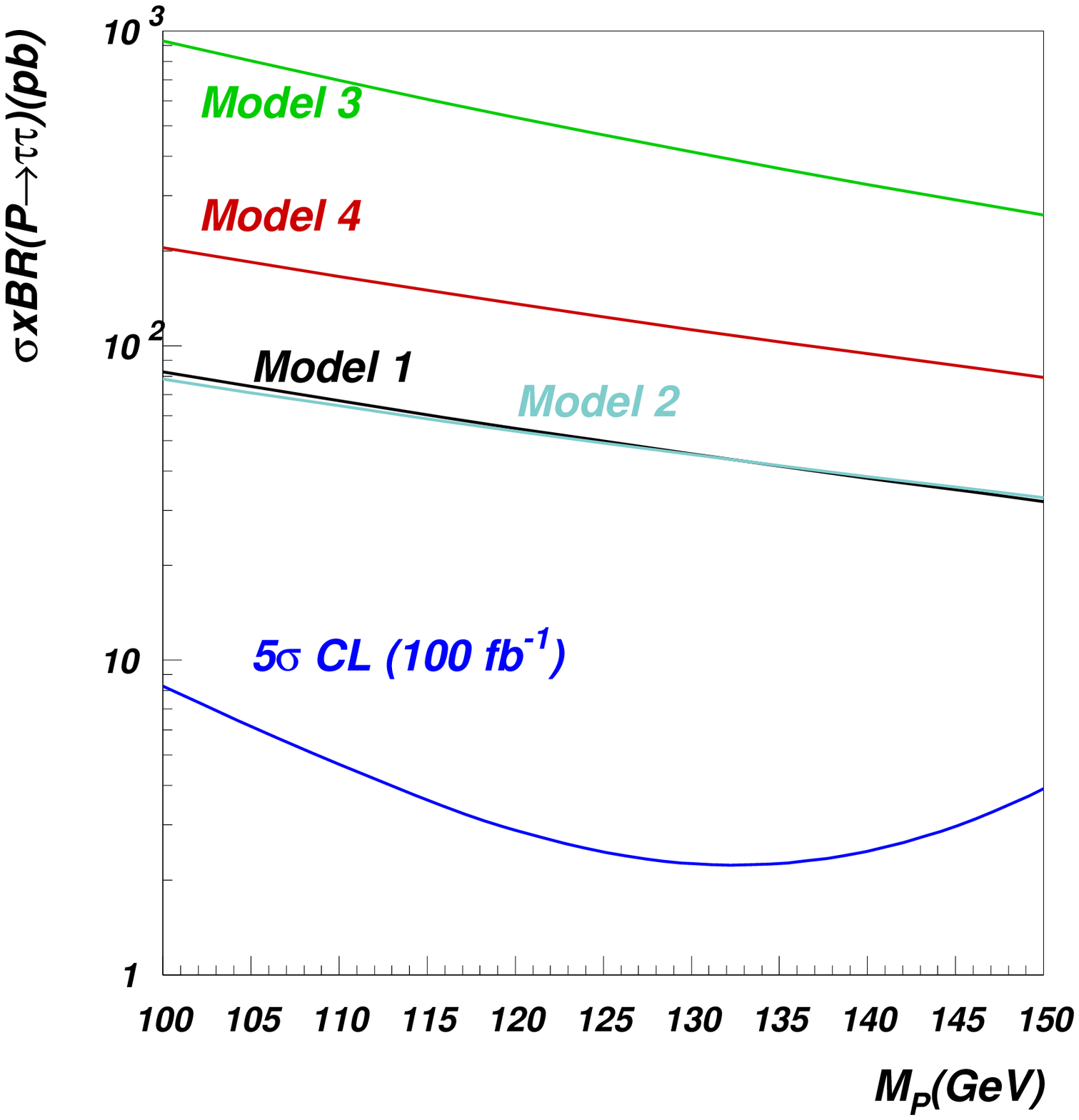}
  \end{minipage}
  \begin{minipage}[c]{0.49\textwidth}
    \includegraphics[width=\textwidth]{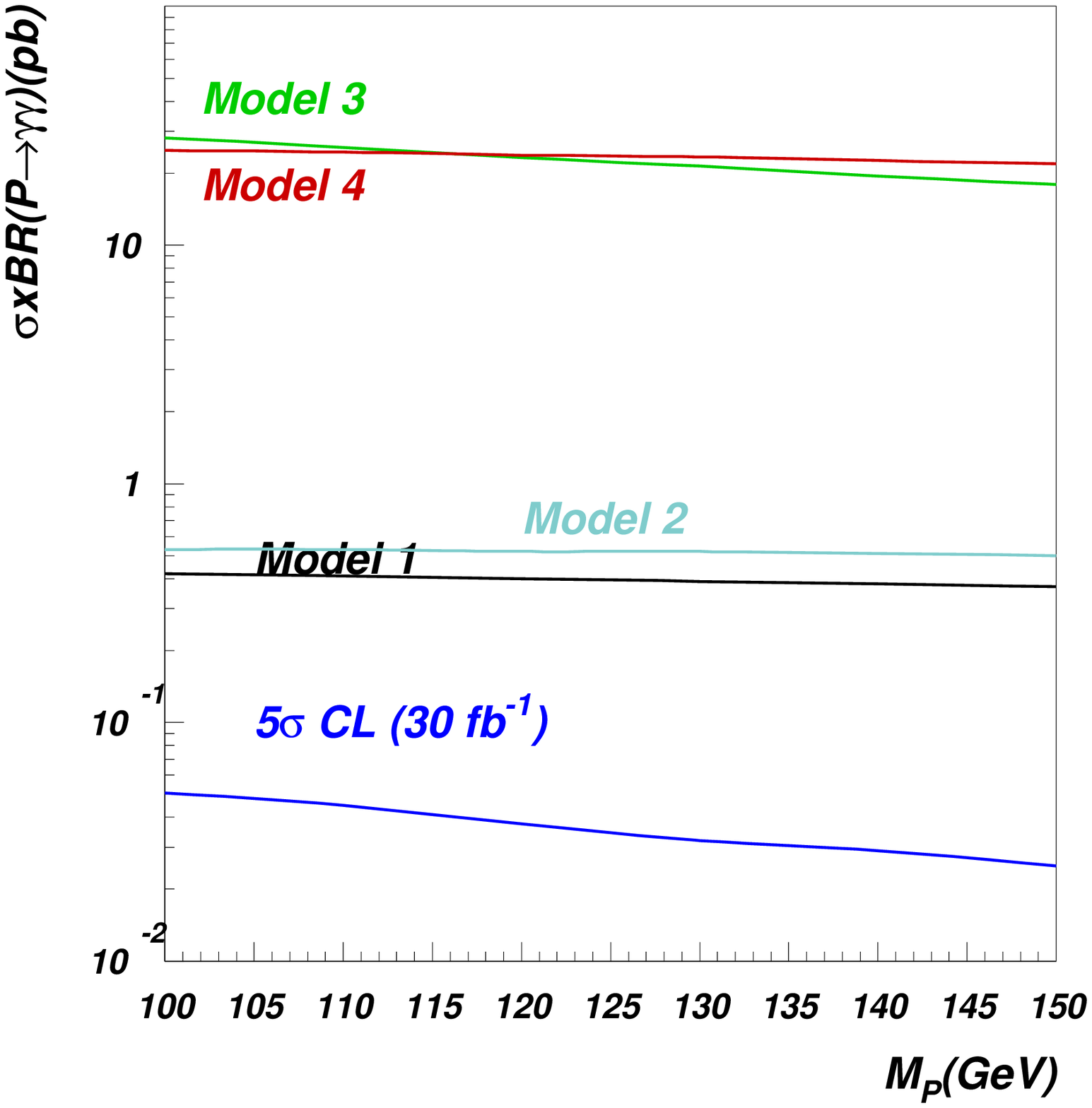}
  \end{minipage}
\caption[sm-h-br]{
Observability of technipions as a function of
technipion mass and assuming the final state is a tau pair (left frame) or photon pair (right frame)
at the Tveatron   and LHC.  
Top raw: the $5\sigma$ discovery and $2\sigma$ exclusion curves indicate the required enhancement
factor for a Higgs-like particle at Tevatron Run II when the final state is $\tau^+ \tau^-$
 \protect{\cite{Belyaev:2002zz}} (left frame) or $\gamma \gamma$
 \protect{\cite{Mrenna:2000qh}}(right frame).
Bottom raw: the lowest curve is the $\sigma\times Br$ required to make a
Higgs-like particle visible (\protect{$5\sigma$} discovery) in $\tau^+\tau^-$ \protect{\cite{Cavalli:2002vs}}
or in $\gamma \gamma$  \protect{\cite{Kinnunen:2005aq}}  at LHC.
\label{fig:techni-limits}}
\end{figure}

Once a supposed light ``Higgs boson'' is observed in a collider experiment, 
 an immediate  important task will be  to identify the new state 
 more precisely, i.e. to discern 
``the meaning of Higgs'' in this context.  Comparison of the enhancement
factors for different channels will aid in this task.  Our study has shown that comparison
of the  $\tau^+\tau^-$ and $\gamma\gamma$ channels can be particularly informative in
distinguishing supersymmetric from dynamical models.  In the case of supersymmetry, when the
  $\tau^+\tau^-$ channel is enhanced, the  $\gamma\gamma$ channel is suppressed,
   and this suppression is strong enough that even the LHC would not observe the 
   $\gamma\gamma$ signature.  In contrast, for  the dynamical symmetry breaking models studied
   we expect {\it simultaneous} enhancement of both
   the $\tau^+\tau^-$  and  $\gamma\gamma$ channels.
   The enhancement of  the $\gamma\gamma$ channel
   is so significant, that even at the Tevatron
   we may observe technipions via  this signature at the  $5\sigma$
   level for Models 3 and 4, while 
   Model 2 could be excluded at 95\% CL at the Tevatron.
   The LHC collider, which will have better sensitivity
   to the signatures under study,
   will be able to observe all four models of  dynamical symmetry breaking
   studied here  in the $\gamma\gamma$ channel, and can therefore distinguish 
   more conclusively between the supersymmetric and dynamical models.

\subsection{Conclusions}

In this paper we have shown that searches for a light Standard Model Higgs boson at
Tevatron Run II and CERN LHC have the power to provide significant information about
important classes of physics beyond the Standard Model.   
We demonstrated that the new
scalar and pseudo-scalar states predicted in both supersymmetric and dynamical models
can have enhanced visibility in standard $\tau^+\tau^-$ and $\gamma\gamma$ search
channels,  making them potentially discoverable at both the Tevatron Run II and the CERN
LHC. 
In comparing the key signals for the non-standard
scalars across models
 we investigated the
likely mass reach of the Higgs search in $pp/p\bar{p} \to {\cal H} \to \tau^+\tau^-$ for
each kind of non-standard scalar state, and we demonstrated that  $pp\, \ \, p\bar{p}
\to {\cal H} \to \gamma\gamma$ may cleanly distinguish the scalars of supersymmetric
models from those of dynamical models.

\subsection*{Acknowledgments}
This work was supported in part by the U.S. National Science Foundation under awards
PHY-0354838 (A.~Belyaev) and PHY-0354226 (R.~S.~Chivukula and E.~H.~Simmons).
A.B. thanks organizers of Tev4LHC workshop for 
the creative atmosphere and hospitality.

\clearpage
\section{MSSM Higgs Boson Searches at the Tevatron and the LHC: Impact
  of Different Benchmark Scenarios}
\textbf{Contributed by: M.~Carena, S.~Heinemeyer, C.E.M.~Wagner, G.~Weiglein}

\vspace{0.25in}


The MSSM requires two Higgs doublets, resulting in five
physical Higgs boson degrees of freedom. 
These are the light and heavy $\cp$-even Higgs bosons, $h$ and $H$, the
$\cp$-odd Higgs boson, $A$, and the charged Higgs boson, $H^\pm$.
The Higgs sector of the MSSM can be specified at lowest
order in terms of $\MZ$, $\MA$, and $\tanb \equiv v_2/v_1$, the ratio of the
two Higgs vacuum expectation values. 
The masses of the $\cp$-even neutral Higgs bosons and the
charged Higgs boson can be calculated,
including higher-order corrections, 
in terms of the other MSSM parameters. 

After the termination of LEP in the year 2000 (the close-to-final LEP
results can 
be found in \citeres{LEPHiggsSM,LEPHiggsMSSM}), the Higgs boson search
has shifted to the Tevatron and will later be continued at the LHC. 
Due to the large number of 
free parameters, a complete scan of the MSSM parameter space is too
involved. Therefore  the search results at LEP have been
interpreted~\cite{LEPHiggsMSSM} in 
several benchmark scenarios~\cite{benchmark,benchmark2}. Current
analyses at the Tevatron and investigations of the LHC~\cite{schumi}
potential also have been performed in the scenarios proposed in
\citeres{benchmark,benchmark2}. 
The $\mhmax$~scenario
has been used to obtain conservative bounds on
$\tanb$ for fixed values of the top-quark mass and the scale of the
supersymmetric particles~\cite{tbexcl}. 
These scenarios are conceived to study particular cases of 
challenging and interesting phenomenology
in the searches for the SM-like Higgs boson, i.e.\ 
mostly the light $\cp$-even Higgs boson. 

The current searches at the Tevatron are not yet sensitive to a 
SM-like Higgs in the mass region allowed by the LEP
exclusion bounds~\cite{LEPHiggsSM,LEPHiggsMSSM}. On the other hand,
scenarios with enhanced Higgs boson production cross sections can be
probed already with the currently accumulated luminosity. Enhanced
production cross sections can occur in 
particular for low $\MA$ in combination with large $\tanb$ due to the
enhanced couplings of the Higgs bosons to down-type fermions. 
The corresponding limits on the Higgs production cross section times
branching ratio of the Higgs decay into down-type fermions can be
interpreted in MSSM benchmark scenarios. Limits
from Run~II of the Tevatron have recently been published for the
following channels~\cite{Abazov:2005yr,Abulencia:2005kq,Tevcharged} (here and in the
following $\phi$ denotes all three neutral MSSM Higgs bosons, 
$\phi = h, H, A$):
\begin{eqnarray}
\label{scen1}
    && b \bar b \phi,  \phi \to b \bar b 
       ~(\mbox{with one additional tagged } b \mbox{ jet}) , \\[.3em] 
\label{scen2}
       && p \bar p \to \phi \to \tau^+\tau^-  
        ~(\mbox{inclusive}) , \\[.3em]
\label{scen3}
     &&p \bar p \to t \bar t \to H^\pm W^\mp \, b \bar b,
     H^{\pm} \to \tau \nu_{\tau} ~.
\end{eqnarray}
The obtained cross section limits have been interpreted in the
$\mhmax$ and the no-mixing scenario with a value for the higgsino mass
parameter of $\mu = - 200 \GeV$~\cite{Abazov:2005yr} and  
$\mu = \pm 200 \GeV$~\cite{Abulencia:2005kq}.
In these scenarios for $\MA \approx 100 \GeV$ the limits on $\tanb$ are
$\tanb \lsim 50$.

Here we investigate the dependence of the CDF and D0
exclusion bounds in the $\MA$--$\tanb$ plane on the parameters entering
through the most relevant supersymmetric radiative corrections
in the theoretical predictions for Higgs boson
production and decay processes.
We will show that the bounds obtained from the
$b \bar b \phi, \phi \to b \bar b$ channel depend
very sensitively on the radiative corrections affecting the relation
between the bottom quark mass and the bottom Yukawa coupling.
In the channels with $\tau^+ \tau^-$ final states, on the other hand,
compensations between large corrections in the Higgs production and
the Higgs decay occur. 
In this context we investigate the impact of a
large radiative correction in the $gg \to \phi$ 
production process that had previously been omitted.
 
In order to reflect the impact of the corrections to the bottom Yukawa
coupling on the exclusion bounds we suggest to supplement the existing
$\mhmax$~ and no-mixing scenarios, mostly designed to
search for the light $\cp$-even MSSM Higgs boson, $h$, with additional
values for the higgsino mass parameter $\mu$. In fact, varying the
value and sign of $\mu$, while keeping fixed the values of the gluino
mass and the common third generation squark mass parameter $\msusy$,
demonstrates 
the effect of the radiative corrections on the production and
decay processes.  
The scenarios discussed here are designed specifically to
study the MSSM Higgs sector without assuming any particular soft
supersymmetry-breaking scenario and taking into account constraints 
only from the Higgs boson sector itself. In particular, constraints
from requiring the correct cold dark matter density, 
$\br(b \to s \ga)$ or $(g - 2)_\mu$, 
which depend on other parameters of the theory, are not crucial in defining
the Higgs
boson sector, and may be avoided. 

We also study the non-standard MSSM Higgs boson search sensitivity 
at the LHC, focusing on the processes 
$pp \to H/A +X, \, H/A \to \tau^+ \tau^-$ and 
$pp \to t H^{\pm} +X, \, H^{\pm} \to \tau \nu_{\tau}$,
and stress the relevance of the proper inclusion of supersymmetric
radiative corrections to the production cross sections and decay widths.
We show the impact of these corrections by investigating the variation
of the Higgs boson discovery reach in the benchmark scenarios for
different values of $\mu$. 
In particular, we discuss the resulting
modification of the parameter region in which only the
light $\cp$-even  MSSM Higgs boson can be detected at the LHC.


\subsection{Predictions for Higgs boson production and decay processes}

\subsubsection*{Notation and renormalization}

The tree-level values for the $\cp$-even Higgs bosons of the
MSSM, $\mh$ and $\mH$, are determined by $\tanb$, 
the $\cp$-odd Higgs-boson mass $\MA$, and the $Z$ boson mass $\MZ$. 
The mass of the charged Higgs boson, $\mHplusminus$, is given in terms of $\MA$
and the $W$ boson mass, $\MW$.
Beyond the tree-level, the main correction to the Higgs boson masses
stems from the 
$t/\Stop$ sector, and for large values of $\tanb$ also from the 
$b/\Sbot$ sector.

In order to fix our notations, we list the conventions for the inputs
from the scalar top and scalar bottom sector of the MSSM:
the mass matrices in the basis of the current eigenstates $\StopL, \StopR$ and
$\SbotL, \SbotR$ are given by (modulo numerically small $D$-term
contributions) 
\begin{equation}
\label{stopmassmatrix}
{\cal M}^2_{\Stop} =
  \ML \MstL^2 + \mt^2 & 
      \mt \Xt \\
      \mt \Xt &
      \MstR^2 + \mt^2 
  \MR, \quad
{\cal M}^2_{\Sbot} =
  \ML \MsbL^2 + \mb^2 &
      \mb \Xb \\
      \mb \Xb &
      \MsbR^2 + \mb^2
  \MR,
\end{equation}
where 
\begin{equation}
\mt \Xt = \mt (\At - \mu \CTb) , \quad
\mb\, \Xb = \mb\, (\Ab - \mu \tanb) .
\label{eq:mtlr}
\end{equation}
Here $\At$ denotes the trilinear Higgs--stop coupling, $\Ab$ denotes the
Higgs--sbottom coupling, and $\mu$ is the higgsino mass parameter.
SU(2) gauge invariance leads to the relation
$
\MstL = \MsbL$ .
For the numerical evaluation, a convenient choice is
\begin{equation}
\MstL = \MsbL = \MstR = \MsbR =: \msusy .
\label{eq:msusy}
\end{equation}
The Higgs sector observables
furthermore depend on the SU(2) gaugino mass
parameter, $M_2$. The other gaugino mass parameter, $M_1$, is usually
fixed via the GUT relation 
$ M_1 = \frac{5}{3} \frac{\sw^2}{\cw^2} M_2$. 
At the two-loop level also the gluino mass, $\mgl$, enters the
predictions for the Higgs-boson masses.

\smallskip
Corrections to the MSSM Higgs boson sector have been evaluated in
several approaches. 
The status of the available corrections to the masses and mixing
angles in the MSSM Higgs sector (with real parameters) 
can be summarized as follows. For the
one-loop part, the complete result within the MSSM is 
known~\cite{mhiggsf1lB1,mhiggsf1lB2}. The by far dominant
one-loop contribution is the \order{\alpha_t} term due to top and stop 
loops ($\alpha_t \equiv h_t^2 / (4 \pi)$, $h_t$ being the 
top-quark Yukawa coupling).
Concerning the two-loop
effects, their computation is quite advanced and has now reached a
stage such that all the presumably dominant contributions are 
known~\cite{mhiggsEP1b,mhiggsRG11,mhiggsRG12,mhiggsRG13,mhiggsRG14,mhiggslong,Haber:1996fp,Zhang:1998bm,Espinosa:1999zm,Espinosa:2000df,Degrassi:2001yf,Brignole:2001jy,Brignole:2002bz,Heinemeyer:2004xw,Hempfling:1993kv,Hall:1993gn,Carena:1994bv,Carena:1999py,Eberl:1999he,Dedes:2003km}.
The remaining theoretical uncertainty on the light $\cp$-even Higgs
boson mass has been estimated to be below 
$\sim 3 \GeV$~\cite{mhiggsAEC,Heinemeyer:2004gx}. 
The above calculations have been implemented into public 
codes. The program 
{\tt FeynHiggs}~\cite{Heinemeyer:1998yj,Hahn:2004td} 
is based on the results obtained in the Feynman-diagrammatic (FD)
approach~\cite{mhiggslong,mhiggsAEC,Dedes:2003km}. It
includes all the above corrections. The code 
{\tt CPsuperH}~\cite{Lee:2003nt} is based on the renormalization group (RG)
improved effective potential
approach~\cite{mhiggsRG11,mhiggsRG12,mhiggsRG13,mhiggsRG14,Carena:2000dp}. 
For the MSSM with real parameters the two codes can differ by up to
$\sim 4 \GeV$ for the light $\cp$-even Higgs boson mass, mostly due to
formally subleading two-loop corrections that are included only in 
{\tt FeynHiggs}. 

\medskip
It should be noted in this context that the FD result has been obtained
in the on-shell (OS) renormalization scheme, whereas the RG result has been 
calculated using the \msbar\ scheme; see \citeres{Carena:2000dp,Heinemeyer:1999be} for a
detailed comparison. Owing to the different
schemes used in the FD and the RG approach for the
renormalization in the scalar top sector, the
parameters $\Xt$ and $\msusy$ are also scheme-dependent
in the two approaches. 


\subsubsection*{Leading effects from the bottom/sbottom sector}

The relation between the bottom-quark mass and the Yukawa coupling
$h_b$, which controls also the interaction between the Higgs fields and
the sbottom quarks, is affected at one-loop order by large radiative
corrections \cite{Hempfling:1993kv,Hall:1993gn,Carena:1994bv,Carena:1999py,Eberl:1999he}.
The leading effects are included in the effective Lagrangian
formalism developed in \citere{Carena:1999py}.
Numerically this is by far the dominant part of the
contributions from the sbottom sector (see also
\citeres{Brignole:2002bz,Dedes:2003km,Heinemeyer:2004xw}). The effective Lagrangian is
given by 
\begin{eqnarray}
\cal L = \frac{g}{2\MW} \frac{\mbms}{1 + \Delta_b} \Bigg[ 
&& \tanb\; A \, i \, \bar b \ga_5 b 
   + \sqrt{2} \, V_{tb} \, \tanb \; H^+ \bar{t}_L b_R \\
&+& \left( \frac{\sin\alpha}{\cos\beta} - \Delta_b \frac{\cos\alpha}{\sin\beta} \right) h \bar{b}_L b_R 
- \left( \frac{\cos\alpha}{\cos\beta} + \Delta_b \frac{\sin\alpha}{\sin\beta} \right) H \bar{b}_L b_R
    \Bigg] + {\rm h.c.}~. \nonumber
\label{effL}
\end{eqnarray}
Here $\mbms$ denotes the running bottom quark mass including SM QCD
corrections. In the numerical evaluations obtained with 
{\tt FeynHiggs} below we choose  
$\mbms = \mbms(\mt) \approx 2.97 \GeV$. 
The prefactor $1/(1 + \Delta_b)$ in Equation~\ref{effL} arises from the
resummation of the leading corrections to all orders. 
The additional terms $\sim \Delta_b$ in the $h\bar b b$ and $H\bar b b$
couplings arise from the mixing and coupling of the ``other'' Higgs
boson, $H$ and $h$, respectively, to the $b$~quarks.

As explained above, 
the function $\Delta_b$ consists of two main contributions, 
an \order{\alpha_s} correction from a
sbottom--gluino loop and an \order{\alpha_t} correction
from a stop--higgsino loop. The explicit
form of $\Delta_b$ in the limit of $M_S \gg \mt$ and $\tanb \gg 1$
reads~\cite{Hempfling:1993kv,Hall:1993gn,Carena:1994bv}
\begin{equation}
\Delta_b = \frac{2\alpha_s}{3\,\pi} \, \mgl \, \mu \, \tanb \,
                    \times \, I(m_{\tilde{b}_1}, m_{\tilde{b}_2}, \mgl) +
      \frac{\alpha_t}{4\,\pi} \, \At \, \mu \, \tanb \,
                    \times \, I(m_{\tilde{t}_1}, m_{\tilde{t}_2}, \mu) ~.
\label{def:dmb}
\end{equation}
The function $I$ is given by
\begin{eqnarray}
I(a, b, c) &=& \frac{1}{(a^2 - b^2)(b^2 - c^2)(a^2 - c^2)} \,
               \left( a^2 b^2 \log\frac{a^2}{b^2} +
                   b^2 c^2 \log\frac{b^2}{c^2} +
                   c^2 a^2 \log\frac{c^2}{a^2} \right) \\
 &\sim& \frac{1}{\mbox{max}(a^2, b^2, c^2)} ~. \nonumber
\end{eqnarray}
The large $\Sbot-\tilde{g}$~loops are resummed to all orders of
$(\alpha_s\tanb)^n$ via the inclusion of $\Delta_b$~\cite{Hempfling:1993kv,Hall:1993gn,Carena:1994bv,Carena:1999py,Eberl:1999he}.
The leading electroweak contributions are taken into account via the
second term in Equation~\ref{def:dmb}.

For large values of $\tanb$ and the ratios of $\mu \mgl/ \msusy^2$ and 
$\mu \At/ \msusy^2$,  the $\Delta_b$ correction can
become very important. Considering positve values of $\At$ and $\mgl$, the 
sign of the $\Delta_b$ term is governed by the sign of $\mu$. 
Cancellations can occur if $\At$ and $\mgl$ have opposite signs.
For $\mu, \mgl, \At > 0$ the $\Delta_b$ correction is positive, leading 
to a suppression of the bottom Yukawa coupling. On the other hand,
for negative values of 
$\Delta_b$, the bottom Yukawa coupling
may be strongly enhanced and can even acquire 
non-perturbative values when $\Delta_b \to -1$.


\subsubsection*{Impact on Higgs production and decay at large $\tanb$}

\label{sec:impact}

Higgs-boson production and decay processes at the Tevatron and the LHC
can be affected by different kinds of large radiative corrections. 
For large $\tanb$ the supersymmetric radiative corrections to the
bottom Yukawa coupling described above become particularly 
important~\cite{Carena:1999bh,Carena:1998gk}. 
Their main effect on the Higgs-boson production and decay processes  
can be understood from the way the 
leading contribution $\Delta_b$ enters. In the following we present simple 
analytic approximation formulae for the most relevant Higgs-boson 
production and decay processes. They are meant for illustration only so
that the impact of the $\Delta_b$ corrections can easily be traced. In our 
numerical analysis below, we use the full result from {\tt FeynHiggs} rather 
than the simple formulae presented in this section. 
No relevant modification
to these results would be obtained using {\tt CPsuperH}.

We begin with a simple approximate formula that represents well the MSSM
parametric variation of the decay rate of the $\cp$-odd Higgs boson
in the large $\tanb$ regime. One should recall, for that purpose,
that in this regime the $\cp$-odd Higgs boson decays mainly into 
$\tau$-leptons and bottom-quarks, and that the partial decay widths are
proportional to the square of the Yukawa couplings evaluated at
an energy scale of about the Higgs boson mass. Moreover, for Higgs
boson masses of the order of 100~GeV, the approximate relations
$m_b(\MA)^2 \simeq 9$~GeV$^2$, and $m_\tau(\MA)^2 \simeq 3$~GeV$^2$
hold.  Hence, since the number of colors is $N_c = 3$, 
for heavy supersymmetric particles, with masses far above
the Higgs boson mass scale, one has 
\begin{equation}
\label{eq:BRAbb}
{\rm BR}(A \to b \bar{b})  \simeq   
\frac{ 9}{\left(1 + \Delta_b \right)^2 + 9} ~ , \quad
{\rm BR}(A \to \tau^+\tau^-) \simeq 
\frac{\left(1 + \Delta_b\right)^2}{\left(1 + \Delta_b \right)^2 + 9} ~ .
\end{equation}

On the other hand, the production cross section for a $\cp$-odd Higgs boson
produced in association with a pair of bottom quarks is proportional
to the square of the bottom Yukawa coupling and therefore is proportional
to $\tan^2\beta/(1 + \Delta_b)^2$. Also in the gluon fusion channel,
the dominant contribution in the large $\tanb$ regime
is governed by the bottom quark loops, and therefore is also proportional
to the square of the bottom Yukawa coupling. Hence, the total 
production rate of bottom quarks and $\tau$ pairs
mediated by the production of a $\cp$-odd Higgs boson in the large $\tanb$
regime is approximately given by
\begin{eqnarray}
\label{eq:bbA}
\sigma(b\bar{b} A) \times {\rm BR}(A \to b \bar{b}) &\simeq&
\sigma(b\bar{b} A)_{\rm SM} \;
\frac{\tan^2\beta}{\left(1 + \Delta_b \right)^2} \times
\frac{ 9}{
\left(1 + \Delta_b \right)^2 + 9} ~, \\
\label{eq:Atautau}
\sigma(gg, b\bar{b} \to A) \times {\rm BR}(A \to \tau^+ \tau^-) &\simeq&
\sigma(gg, b\bar{b} \to A)_{\rm SM} \;
\frac{\tan^2\beta}{
\left(1 + \Delta_b \right)^2 + 9} ~,
\end{eqnarray} 
where $\sigma(b\bar{b}A)_{\rm SM}$ and $\sigma(gg, b\bar{b} \to A)_{\rm SM}$ 
denote the values of the corresponding SM Higgs boson production cross
sections for a Higgs boson mass equal to $\MA$.

As a consequence, the $b\bar{b}$ production rate depends sensitively on
$\Delta_b$ because of the factor $1/(1 + \Delta_b)^2$, while this leading
dependence on $\Delta_b$ cancels out in the $\tau^+\tau^-$ production rate.
There is still a subdominant parametric dependence in the $\tau^+\tau^-$ 
production rate on $\Delta_b$ that may lead
to variations of a few tens of percent of the $\tau$-pair production
rate (compared to variations of the rate by up to 
factors of a few in the case of bottom-quark pair
production).

The formulae above apply, within a good approximation, also to the
non-standard $\cp$-even 
Higgs boson in the large $\tanb$ regime. Depending on $\MA$ this can be
either the $h$ (for $\MA \lsim 120 \GeV$) or the $H$ 
(for $\MA \gsim 120 \GeV$). 
This non-standard Higgs
boson becomes degenerate in mass with the $\cp$-odd Higgs scalar.
Therefore, the production and decay
rates of $H$ ($h$) are governed by similar formulae as the ones presented
above, leading to an approximate enhancement of a factor 2 of the production
rates with respect to the ones that would be obtained in the case of the
single production of the $\cp$-odd Higgs boson as given in
Equations~\ref{eq:bbA}, (\ref{eq:Atautau}). 

\smallskip
We now turn to the production and decay processes of the charged Higgs
boson. In the MSSM, the masses and couplings of the 
charged Higgs boson in the large $\tanb$ regime
are closely related to the ones of the $\cp$-odd Higgs
boson. 
The tree-level relation
$\mHplusminus^2 = \MA^2 + \MW^2$
receives sizable corrections  
for large values of $\tanb$, $\mu$, $\At$ and $\Ab$. These corrections
depend on the ratios $\mu^2/\msusy^2$, 
$(\mu^2 - \Ab\At)^2/\msusy^4$, $(\At + \Ab)^2/\msusy^2$~\cite{mhiggsRG11,mhiggsRG12,mhiggsRG13,mhiggsRG14}.
The coupling of the charged Higgs boson to a top and a bottom quark
at large values of $\tanb$ is governed by the bottom Yukawa coupling and is
therefore affected by the same $\Delta_b$ corrections that
appear in the couplings of
the non-standard neutral MSSM Higgs bosons~\cite{Carena:1999py}.

The relevant channels for charged Higgs boson searches
depend on its mass.
For values of $\mHplusminus$ smaller than the top-quark mass,
searches at hadron colliders
concentrate on the possible emission of the charged
Higgs boson from top-quark decays. In this case, for
large values of $\tanb$, the charged Higgs decays
predominantly into a $\tau$ lepton and a neutrino, i.e.\ one has to a
good approximation $\br(H^\pm \to \tau \nu_\tau) \approx 1$. 
The partial decay width of the top quark into a charged Higgs
and a bottom quark is proportional to the square of
the bottom Yukawa coupling and therefore scales with
$\tan^2\beta/(1 + \Delta_b)^2$, see e.g.~\citere{Carena:1999py}. 

For values of the charged Higgs mass larger than $\mt$,
instead, the most efficient production channel is the
one of a charged Higgs associated with a top quark 
(mediated, for instance, by gluon-bottom fusion)~\cite{Berger:2003sm}. 
In this case, the production cross section is proportional
to the square of the bottom-quark Yukawa coupling. The
branching ratio of the charged Higgs decay into a 
$\tau$ lepton and a neutrino is, apart from threshold corrections, 
governed by a similar formula
as the branching ratio of the decay of the $\cp$-odd Higgs
boson into $\tau$-pairs, namely
\begin{equation}
\br(H^{\pm} \to \tau \nu_\tau) \simeq \frac{(1+\Delta_b)^2}{(1+\Delta_b)^2 + 9
\; (1 - r_t)^2}~,
\label{def:brcharged}
\end{equation}
where the factor $(1 - r_t)^2$ is associated with threshold corrections,
and $r_t = \mt^2/\mHplusminus^2$.

\bigskip
As mentioned above, our
numerical analysis will be based on the complete expressions for the
Higgs couplings rather than on the simple approximation formulae given
in this section.


\subsection{Interpretation of cross section limits in MSSM\\ scenarios}
\label{sec:interpretation}

\subsubsection*{Limits at the Tevatron}

The D0 and CDF Collaborations have recently published cross section
limits from the Higgs search at the Tevatron in the channel where at
least three bottom quarks are identified in the final state 
($b \bar b \phi, \phi \to b \bar b$)~\cite{Abazov:2005yr}
and in the inclusive channel with $\tau^+\tau^-$ final
states ($p \bar p \to \phi \to \tau^+\tau^-$)~\cite{Abulencia:2005kq}.
The CDF Collaboration has also done analyses searching for a charged Higgs
boson in top-quark decays~\cite{Tevcharged}. 
While the cross section for a SM Higgs boson is significantly below the
above limits, a large enhancement of these cross sections is possible in
the MSSM.
It is therefore of interest to interpret the cross section limits within
the MSSM parameter space. 
One usually displays the limits in the $\MA$--$\tanb$
plane.
As the whole structure of the MSSM enters via
radiative corrections, the limits in the $\MA$--$\tanb$ plane depend on the
other parameters of the model. One usually chooses certain benchmark
scenarios to fix the other MSSM parameters~\cite{benchmark,benchmark2}.
In order to understand the physical meaning of the exclusion bounds in
the $\MA$--$\tanb$ plane it is important to investigate how sensitively they
depend on the values of the other MSSM parameters, i.e.\ on the choice
of the benchmark scenarios.


\subsubsection*{Limits from the process $b \bar b \phi, \phi \to b \bar b$}

The D0 Collaboration has presented the limits in the $\MA$--$\tanb$ plane
obtained from the $b \bar b \phi, \phi \to b \bar b$ channel for the
$\mhmax$ and no-mixing scenarios as defined in \citere{benchmark}.
The $\mhmax$ scenario according to the definition of \citere{benchmark} reads
\begin{eqnarray}
&& \mt = 174.3 \GeV, \;
\msusy = 1000 \GeV, \;
\mu = -200 \GeV, \;
M_2 = 200 \GeV, \nonumber \\
&& \Xt^{\mathrm{OS}} = 2\, \msusy  \; \mbox{(FD calculation)}, \;
\Xt^{\msbar} = \sqrt{6}\, \msusy \; \mbox{(RG calculation)}, \nonumber \\
&& \Ab = \At, \;
\mgl = 0.8\,\msusy~.
\label{oldmhmax}
\end{eqnarray}
The no-mixing scenario defined in \citere{benchmark} differs from the
$\mhmax$ scenario only in
\begin{equation}
\Xt = 0 \; \mbox{(FD/RG calculation)}~.
\label{oldnomix}
\end{equation}
The condition $\Ab = \At$ implies that the different mixing in the stop
sector gives rise to a difference between the two scenarios also in the
sbottom sector. 
The definition of the $\mhmax$ and no-mixing scenarios given in
\citere{benchmark} was later updated 
in \citere{benchmark2}, see the discussion below.

For their analysis, the D0 Collaboration has used the
following approximate formula~\cite{Abazov:2005yr},
\begin{equation}
\sigma(b \bar b \phi) \times {\rm BR}(\phi \to b \bar b) =
2 \; \sigma(b \bar b \phi)_{\rm SM} \;
\frac{\tan^2\beta}{\left(1 + \Delta_b \right)^2} \times
\frac{ 9}{
\left(1 + \Delta_b \right)^2 + 9}~,
\label{eq:d0formula}
\end{equation}
which follows from Equation~\ref{eq:bbA} and the discussion in
Section~\ref{sec:impact}. The cross section 
$\sigma(b \bar b \phi)_{\rm SM}$ has been evaluated with the code
of~\citere{Campbell:2002zm}, while $\Delta_b$ has been calculated using 
{\tt CPsuperH}~\cite{Lee:2003nt}.
{}From the discussion in Section~\ref{sec:impact} it follows  that the choice
of negative values of $\mu$ 
leads to an enhancement of the bottom
Yukawa coupling and therefore to an enhancement of the signal cross
section in Equation~\ref{eq:d0formula}.
For $\tanb = 50$ the quantity $\Delta_b$ takes on the following values in
the $\mhmax$ and no-mixing scenarios as defined in
Equations~\ref{oldmhmax}, (\ref{oldnomix}),
\begin{eqnarray}
\mbox{$\mhmax$ scenario, $\mu = -200 \GeV$, $\tanb = 50$} &:&
\Delta_b = -0.21~, \label{eq:db1} \\
\mbox{no-mixing scen., $\mu = -200 \GeV$, $\msusy = 1000 \GeV$,
                                                 $\tanb = 50$} &:&
\Delta_b = -0.10~. \label{eq:db2}
\end{eqnarray}
While the \order{\alpha_s} contribution to $\Delta_b$, see Equation~\ref{def:dmb},
is practically the same in the two scenarios, the \order{\alpha_t}
contribution to $\Delta_b$ in the $\mhmax$ scenario differs significantly
from the one in the no-mixing scenario. In the $\mhmax$ scenario the 
\order{\alpha_t} contribution to $\Delta_b$ is about as large as the \order{\alpha_s}
contribution. In the no-mixing scenario, on the other hand, the 
\order{\alpha_t} contribution to $\Delta_b$ is very small, because $\At$ is close
to zero in this case.
Reversing the sign of $\mu$ in Equations~\ref{eq:db1}, (\ref{eq:db2}) 
reverses the sign of $\Delta_b$, leading
therefore to a significant suppression of the signal cross section in
Equation~\ref{eq:d0formula} for the same values of the other MSSM parameters.

The predictions for $b \bar b \phi$, $\phi \to b \bar b$ evaluated with
{\tt FeynHiggs} have been compared with the exclusion bound for
$\sigma \times \br$ as given in \citere{Abazov:2005yr}. 
As mentioned above, in our analysis we use the full Higgs couplings
obtained with  
{\tt FeynHiggs} rather than the approximate formula given in 
Equation~\ref{eq:d0formula}. Similar results would be
obtained with {\tt CPsuperH}.

The impact on the limits in the $\MA$--$\tanb$ plane from varying $\mu$ 
while keeping all other parameters fixed can
easily be read off from Equation~\ref{eq:d0formula}. 
For a given value of the $\cp$-odd mass and
$\tanb$, the bound on $\sigma(b\bar{b}\phi) \times 
\br (\phi \to b\bar{b})$ provides an upper bound on the 
bottom-quark Yukawa coupling. The main effect therefore is that
as $\mu$ varies, the bound on $\tanb$ also changes in
such a way that the value of the bottom Yukawa coupling
at the boundary line in the
$\MA$--$\tanb$ plane remains the same. 

The dependence of the limits in the $\MA$--$\tanb$ plane obtained
from the process $b \bar b \phi, \phi \to b \bar b$ on the parameter $\mu$
is shown in Figure~\ref{fig:d0change}. 
The limits for $\mu = -200\GeV$ in the 
$\mhmax$ and no-mixing scenarios, corresponding to the limits presented
by the D0 Collaboration in \citere{Abazov:2005yr}, are compared with the limits
arising for different $\mu$ values,
$\mu = +200, \pm 500, \pm 1000 \GeV$.
Figure~\ref{fig:d0change} illustrates that the effect
of changing the sign of $\mu$ on the limits in the $\MA$--$\tanb$ plane
obtained from the process $b \bar b \phi, \phi \to b \bar b$ is quite
dramatic. In the $\mhmax$ scenario the exclusion bound degrades from 
about $\tanb = 50$ for $\MA = 90\GeV$ in the case of $\mu = -200\GeV$ to
about $\tanb = 90$ for $\MA = 90\GeV$ in the case of $\mu = +200\GeV$.
We extend our plots to values of $\tanb$ much larger than 50 mainly for
illustration purposes; the region $\tanb \gg 50$ in the MSSM is theoretically
disfavoured, if one demands that the values of the bottom and $\tau$
Yukawa couplings remain in the perturbative regime up to energies of
the order of the unification scale. The situation for the
bottom-Yukawa coupling can be ameliorated for large positive values of
$\mu$ due to the $\Delta_b$ corrections.
The curves for $\mu = +500, +1000\GeV$ 
do not appear in the plot for the $\mhmax$ scenario, since for
these $\mu$ values there is no $\tanb$ exclusion below $\tanb = 130$
for any value of $\MA$. On the other hand, the
large negative values of $\mu$ shown in Figure~\ref{fig:d0change}, 
$\mu = -500, -1000\GeV$,
lead to an even stronger
enhancement of the signal cross section than for $\mu = -200 \GeV$ and,
accordingly, to an improved reach in $\tanb$. It should be noted that for 
$\mu = -500, -1000\GeV$ the bottom Yukawa coupling becomes so large for
$\tanb \gg 50$ that a perturbative treatment would no longer be reliable
in this region.


\begin{figure}
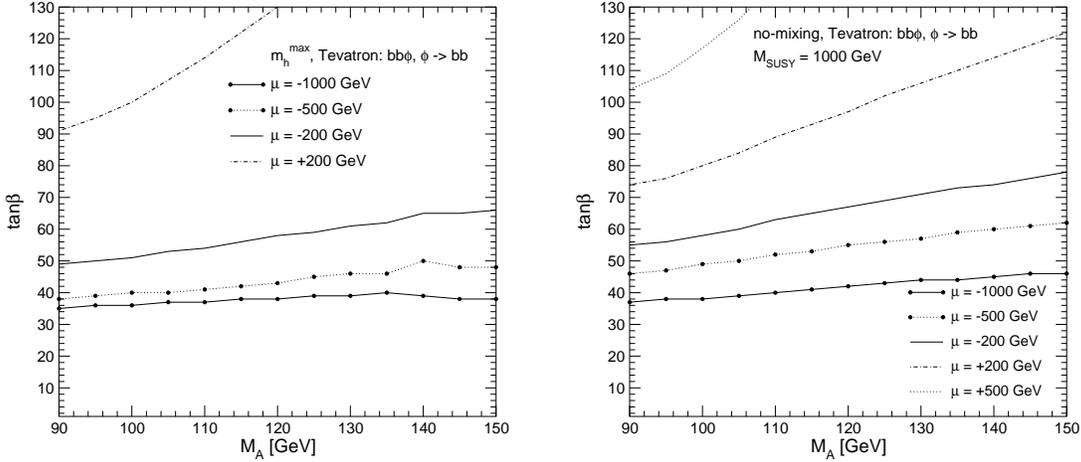

\begin{center}
\includegraphics[width=.42\textwidth]{higgs/benchmark3/t4l-benchP02c.bw.eps}\hspace{2em}
\includegraphics[width=.42\textwidth]{higgs/benchmark3/t4l-benchP03c.bw.eps}
\end{center}
\vspace{-1.5em}
\caption{Change in the limits obtained from the $b \bar b \phi$, 
$\phi \to b \bar b$ channel in the $\mhmax$ (left) and no-mixing (right)
benchmark scenarios for different values of $\mu$. The value 
$\mu = -200 \GeV$ was chosen by the D0 Collaboration in
\citere{Abazov:2005yr}. 
The other curves indicate the corresponding limits
for $\mu = +200, \pm 500, \pm 1000 \GeV$. The curves for 
$\mu = +500, +1000\GeV$ ($\mu = +1000 \GeV$)
do not appear in the left (right) plot for the $\mhmax$ (no-mixing) 
scenario, since for
these $\mu$ values there is no $\tanb$ exclusion below $\tanb = 130$
for any value of $\MA$.
}
\label{fig:d0change}
\vspace{-0.6em}
\end{figure}

In \citere{benchmark2} the definition of the $\mhmax$ and no-mixing
scenarios given in \citere{benchmark} has been updated.
The sign of $\mu$ in
the $\mhmax$ and no-mixing scenarios has been reversed to
$\mu = +200\GeV$ in 
\citere{benchmark2}. 
This leads typically to a better agreement with the constraints from 
$(g - 2)_{\mu}$.
Furthermore, the value of $\msusy$ 
in the no-mixing scenario was increased from $1000 \GeV$~\cite{benchmark}
to $2000 \GeV$ in order to ensure that most of the parameter space of this
scenario is in accordance with the LEP exclusion
bounds~\cite{LEPHiggsSM,LEPHiggsMSSM}. 

Another scenario defined in \citere{benchmark2} is the 
``constrained-$\mhmax$'' scenario. It differs from the $\mhmax$~scenario
as specified in \citere{benchmark2} by the reversed sign of $\Xt$,
\begin{equation}
\Xt^{\mathrm{OS}} = -2\, \msusy  \; \mbox{(FD calc.)}, \;
\Xt^{\msbar} = -\sqrt{6}\, \msusy \; \mbox{(RG calc.)}, \;
\mu = +200 \GeV ~.
\label{constrainedmhmax}
\end{equation}
For small $\MA$ and minimal flavor violation this results in better
agreement with the constraints from  
$\br(b \to s \ga)$. For large $\tanb$ one has $\At \approx \Xt$, thus
$\At$ and $\mgl$ have opposite signs. This can lead to cancellations in
the two contributions entering $\Delta_b$, see Equation~\ref{def:dmb}. 
In contrast to the $\mhmax$ scenario, where the two contributions
entering $\Delta_b$ add up, see Equation~\ref{eq:db1}, the constrained-$\mhmax$
scenario typically 
yields relatively small values of $\Delta_b$ and therefore a correspondingly
smaller effect
on the relation between the bottom-quark mass
and the bottom Yukawa coupling, e.g.\
\begin{eqnarray}
\mbox{constrained-$\mhmax$ scenario, $\mu = +200 \GeV$, $\tanb = 50$} &:&
\Delta_b = -0.001~. \label{eq:db3} 
\end{eqnarray}
For large values of $|\mu|$ the compensations between the two terms
entering $\Delta_b$ are less efficient, since the function $I$ in the second
term of  Equation~\ref{eq:db1} scales like $1/\mu^2$ for large $|\mu|$.

The impact of the benchmark definitions of
\citere{benchmark2} on the limits in the $\MA$--$\tanb$ plane arising from
the $b \bar b \phi$, $\phi \to b \bar b$ channel has been analyzed in
\citere{benchmark3}. The effect of
changing $\msusy = 1000 \GeV$ to $\msusy = 2000 \GeV$ in the no-mixing
scenario for $\mu = \pm 200 \GeV$ results in substantially weaker
(stronger) limits for $\mu = +(-) 200 \GeV$
Also the constrained-$\mhmax$ scenario has been analyzed in
\citere{benchmark3}. As expected the variation of the exclusion bounds
with a variation of $\mu$ is much weaker than in the other scenarios.


\subsubsection*{Limits from the process $p \bar p \to \phi \to \tau^+\tau^-$}

\label{sec:CDFtautau}

The limits obtained from the $p \bar p \to \phi \to \tau^+\tau^-$
channel by the CDF Collaboration were presented in the $\MA$--$\tanb$
plane for the $\mhmax$ and no-mixing scenarios as defined in
\citere{benchmark2} and employing two values of the $\mu$ parameter,
$\mu = \pm 200 \GeV$. According to the discussion in Section~\ref{sec:impact},
the limits obtained from the $p \bar p \to \phi \to \tau^+\tau^-$
channel are expected to show a weaker dependence on the sign and
absolute value of $\mu$ than the limits arising from the 
$b \bar b \phi, \phi \to b \bar b$ channel. 
On the other hand, for large values of $\tanb$ and
negative values of $\mu$, 
the large corrections to the bottom Yukawa coupling discussed above can
invalidate a perturbative treatment for this channel.

The MSSM prediction for 
$\sigma(p \bar p \to \phi) \times {\rm BR} (\phi \to \tau^+\tau^-)$ as 
a function of $\tanb$ has been evaluated by the CDF collaboration using
the {\tt HIGLU} program~\cite{Spira:1996if} for the gluon fusion channel.
The prediction for the $b \bar b \to \phi + X$ channel was obtained from the
NNLO result in the SM from \citere{Harlander:2003ai}, and 
$\left[\sigma \times {\rm BR}\right]_{\rm MSSM} /
\left[\sigma \times {\rm BR}\right]_{\rm SM}$ was calculated with the 
{\tt FeynHiggs} 
program~\cite{Heinemeyer:1998yj,Hahn:2004td}. While
the full $\Delta_b$ correction to the bottom Yukawa
correction was taken into 
account in the $b \bar b \to \phi + X$ production channel and the 
$\phi \to \tau^+\tau^-$ branching ratios, the public version of the 
{\tt HIGLU} program~\cite{Spira:1996if} does not include the $\Delta_b$ correction for
the bottom Yukawa coupling entering the bottom loop contribution to the 
$gg \to \phi$ production process. In order to 
treat the two contributing production processes in a uniform way, 
the $\Delta_b$ correction should be included (taking into account the
\order{\alpha_s} and the \order{\alpha_t} parts, see Equation~\ref{def:dmb}) in the 
$gg \to \phi$ production process 
calculation. For the large value of $\msusy$ chosen 
in the $\mhmax$ and no-mixing benchmark scenarios other higher-order
contributions involving sbottoms and stops can be 
neglected (these effects are small provided $\msusy \gsim 500$~GeV).

In \citere{benchmark3} a comparison of the ``partial $\Delta_b$'' and the
``full $\Delta_b$'' results has been performed. It was shown that the
inclusion of the $\Delta_b$ corrections everywhere can lead to a variation
of $\Delta\tanb \sim 10$ in the $\mhmax$ scenario, but has a much smaller
effect in the no-mixing scenario.
Following our analysis, the CDF Collaboration has adopted the
prescription outlined above for incorporating the $\Delta_b$ correction into
the $gg \to \phi$ production process. The limits given in
\citere{Abulencia:2005kq} are based on the MSSM prediction where the $\Delta_b$
correction is included everywhere in the production and decay processes
(see e.g.\ \citere{cdfold} for a previous analysis).

We next turn to the discussion of the sensitivity of the limits obtained 
from the $p \bar p \to \phi \to \tau^+\tau^-$ channel (including the $\Delta_b$
correction in all production and decay processes) on the sign and
absolute value of $\mu$. As discussed above, similar variations in the 
exclusion limits will occur if the absolute values of $\mu$, $\mgl$, $\At$ 
and $\msusy$ are varied, while keeping the ratios appearing in 
$\Delta_b$ constant.  
The results are given in
Figure~\ref{fig:cdfchange} for the $\mhmax$ scenario (left) and the
no-mixing scenario (right). In the $\mhmax$ scenario we find a
sizable dependence of the $\tanb$ bounds on the sign and absolute value of
$\mu$.%
\footnote{
For $\mu = -300 \GeV$ the curve stops at around $\tanb = 75$ because 
the bottom Yukawa coupling becomes very large, leading to 
instabilities in the calculation of the Higgs properties.
For the same reason, even more negative values of $\mu$ are not
considered here.
}
The effect grows with $\MA$ and, 
for the
range of parameters explored in Figure~\ref{fig:cdfchange},
leads to a variation
of the $\tanb$ bound larger than  $\Delta\tanb \sim 30$. 
In the
no-mixing scenario the effect is again smaller, but it can still 
lead to a variation of the $\tanb$ bounds by as much as
$\Delta\tanb \sim 10$.  

\begin{figure}[htb!]
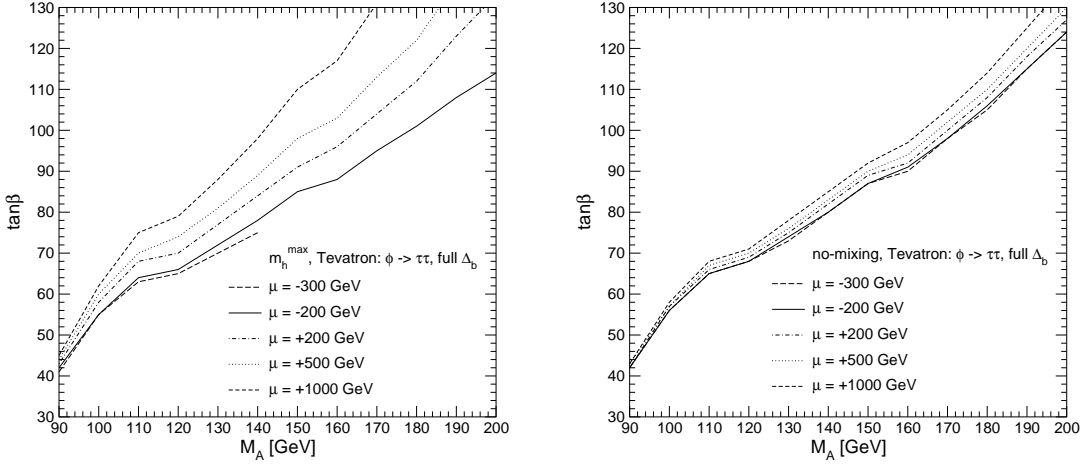

\begin{center}
\includegraphics[width=.42\textwidth]{higgs/benchmark3/t4l-benchP16.bw.eps}\hspace{2em}
\includegraphics[width=.42\textwidth]{higgs/benchmark3/t4l-benchP17.bw.eps}
\end{center}
\vspace{-1.5em}
\caption{Variation of the limits obtained from the 
$p \bar p \to \phi \to \tau^+\tau^-$ channel at the Tevatron 
in the $\mhmax$ (left) and 
no-mixing (right) benchmark scenarios for different values of $\mu$.}
\label{fig:cdfchange}
\vspace{-0.6em}
\end{figure}

The results obtained in the constrained-$\mhmax$ scenario are again 
very robust with respect to varying $\mu$.
All values of $\mu$ result
practically in the same $\tanb$ exclusion bounds~\cite{benchmark3}.



\subsubsection*{Prospects for Higgs sensitivities at the LHC}

The most sensitive channels for detecting heavy MSSM Higgs bosons at the
LHC are the channel $pp \to H/A +X, \, H/A \to \tau^+\tau^-$ (making use of
different decay modes of the two $\tau$ leptons) and the channel 
$t H^{\pm}, H^{\pm} \to \tau \nu_{\tau}$
(for $\mHplusminus \geq \mt$)~\cite{TDR,Abdullin:2005yn}.
We consider here the parameter region $\MA \gg \MZ$, for which the heavy
states $H$, $A$ are widely separated in mass from the 
light $\cp$-even Higgs boson $h$.
Here and in the following we do not discuss search channels where the
heavy Higgs bosons decay into
supersymmetric particles, which depend very sensitively on the model
parameters~\cite{Datta:2003iz,Weiglein:2004hn,Abdullin:2005yn}, but we will comment below on how these
decays can affect the searches with bottom-quarks and $\tau$-leptons
in the final state.


\subsubsection*{Discovery region for the process 
$pp \to H/A +X, \, H/A \to \tau^+\tau^-$}

To be specific, we concentrate in this section on the analysis carried
out by the CMS Collaboration~\cite{KinNik,Abdullin:2005yn}. Similar results for 
this channel have also
been obtained by the ATLAS Collaboration~\cite{TDR,atlasnote0303,atlasnote0309}.
In order to rescale the SM cross sections and branching ratios, 
the CMS Collaboration has used for the branching ratios the {\tt HDECAY} 
program~\cite{Djouadi:1997yw} and for the production cross sections 
the {\tt HIGLU} program~\cite{Spira:1996if} ($gg \to H/A$)
and the {\tt HQQ} program~\cite{hqq} ($gg \to b \bar bH$). In the
{\tt HDECAY} program the $\Delta_b$ corrections are partially included 
for the decays of the neutral Higgs bosons (only the \order{\alpha_s}
contribution to $\Delta_b$ is included, see Equation~\ref{def:dmb}).
The {\tt HIGLU} program (see also the discussion in
Section~\ref{sec:CDFtautau}) and {\tt HQQ}, on the other hand,
do not take into account the
corrections to the bottom Yukawa coupling.%
\footnote{
Since {\tt HQQ} is a leading-order program, non-negligible changes can
also be expected from SM-QCD type higher-order corrections.
}%
~The prospective $5 \sigma$
discovery contours for CMS 
(corresponding to the upper bound of the LHC ``wedge''
region, where only the light $\cp$-even Higgs boson may be observed
at the LHC) have been presented 
in \citeres{KinNik,Abdullin:2005yn}
in the 
$\MA$--$\tanb$ plane, for an integrated luminosity of 
30~fb$^{-1}$ and 60~fb$^{-1}$. 
The results
were presented in the $\mhmax$ scenario and for different $\mu$ values, 
$\mu = -200, +300, +500 \GeV$. It should be noted that decays of
heavy Higgs bosons into charginos and neutralinos open up for small
enough values of the soft supersymmetry-breaking parameters $M_2$ and $\mu$.
Indeed, the results presented in \citeres{KinNik,Abdullin:2005yn} show a 
degradation of the discovery reach in the $\MA$--$\tanb$ plane for
smaller absolute values of $\mu$, which is due to
an enhanced branching ratio of $H$,
$A$ into supersymmetric particles, and accordingly a reduced branching
ratio into $\tau$ pairs.

We shall  now study the impact of including the $\Delta_b$ corrections into the
production cross sections and branching ratios for different values of
$\mu$. The inclusion of the $\Delta_b$ corrections leads to a modification of 
the dependence of the production cross section on $\tanb$, as well
as of the branching ratios of the Higgs boson decays into $\tau^+ \tau^-$.
For a fixed value of $\MA$, the results obtained by the CMS
Collaboration for the discovery region in $\tanb$
can be interpreted in terms of a cross section limit 
using the approximation of rescaling the SM rate for the 
$pp \to H +X, \, H \to \tau^+ \tau^-$ process by the factor
\begin{equation}
\tan^2\beta_{\rm CMS} \times \frac{\br(\Htott)_{\rm CMS} +
\br(\Atott)_{\rm CMS}}
{\br(\Htott)_{\rm SM}}~. 
\label{HAtautauold}
\end{equation}
In the above, $\tanb_{\rm CMS}$ refers to the value of $\tanb$ on the
discovery contour (for a given
value of $\MA$) that was obtained in the analysis of the CMS
Collaboration with
30~fb$^{-1}$~\cite{Abdullin:2005yn}.
These $\tanb$ values as a function of $\MA$ correspond to
the edge of the area in the $\MA$--$\tanb$ plane
in which the signal $pp \to H/A +X, \, H/A \to \tau^+\tau^-$ is visible (i.e.\
the upper bound of the LHC wedge region). The branching ratios
$\br(\Htott)_{\rm CMS}$ and $\br(\Atott)_{\rm CMS}$ 
in the CMS analysis have been evaluated with {\tt HDECAY}, 
incorporating
therefore only the gluino-sbottom contribution to $\Delta_b$. 

After including all $\Delta_b$ corrections, we evaluate the
$pp \to H / A +X, \, H / A \to \tau^+ \tau^-$ process by rescaling the SM
rate with the new factor, 
\begin{equation}
\frac{\tan^2\beta}{(1 + \Delta_b)^2} \times 
\frac{\br(\Htott) + \br(\Atott)}{\br(\Htott)_{\rm SM}}~,
\label{HAtautaunew}
\end{equation}
where $\Delta_b$ depends on \tanb. The quantities have been
evaluted with {\tt FeynHiggs}, allowing also decays into
supersymmetric particles.  
The resulting shift in the discovery reach for the
$pp \to H/A +X, \, H/A \to \tau^+ \tau^-$ channel can be obtained 
by demanding that Equation~\ref{HAtautauold} and Equation~\ref{HAtautaunew} should
give the same numerical result for a given value of $\MA$.

This procedure has been carried out in two benchmark
scenarios for various values of $\mu$. The results are shown in
Figure~\ref{fig:LHCtautau} for the $\mhmax$~scenario (left) and for the
no-mixing scenario (right). The comparison of these results 
with the ones obtained by the CMS Collaboration~\cite{KinNik,Abdullin:2005yn} 
shows that
for positive values of $\mu$ the inclusion of the supersymmetric 
radiative corrections leads to a slight shift of the discovery region
towards higher values of $\tanb$, i.e.\ to a small increase of the LHC wedge
region. For $\mu = -200 \GeV$ the result remains approximately the
same as the one obtained by the CMS Collaboration. 
Due to the smaller considered $\tanb$ values compared to the analysis of
the Tevatron 
limits in Section~\ref{sec:CDFtautau}, the corrections to the
bottom Yukawa coupling from $\Delta_b$ are smaller, leading to a better
perturbative behavior. As a consequence, also the curves for 
$\mu = -500, -1000 \GeV$ are shown in
Figure~\ref{fig:LHCtautau}.

\begin{figure}[htb!]
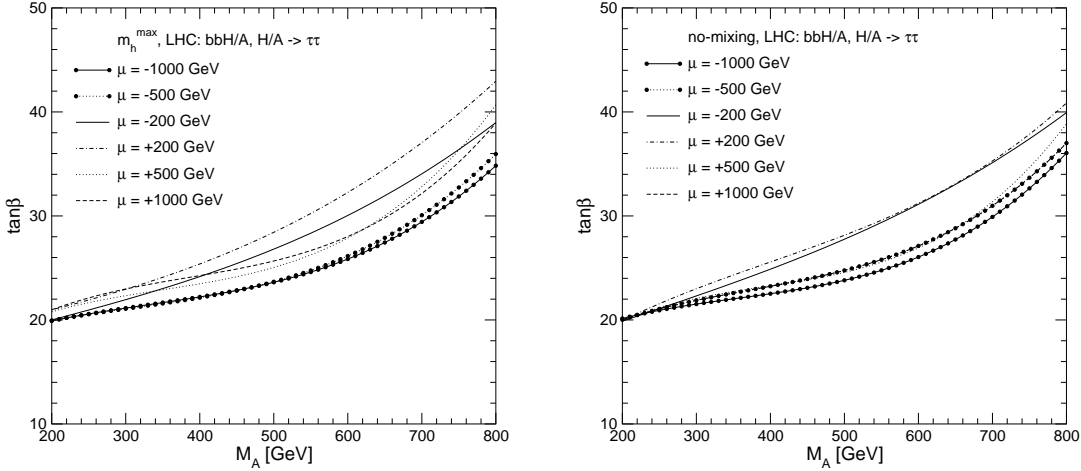

\begin{center}
\includegraphics[width=.42\textwidth]{higgs/benchmark3/t4l-benchP22c.bw.eps}\hspace{2em}
\includegraphics[width=.42\textwidth]{higgs/benchmark3/t4l-benchP23c.bw.eps}
\end{center}
\vspace{-1.5em}
\caption{Variation of the $5 \sigma$ discovery potential for the 
$pp \to H/A +X, \, H/A \to \tau^+\tau^-$ process at the LHC
in the $\mhmax$ (left) and 
no-mixing (right) benchmark scenarios for different values of $\mu$.}
\label{fig:LHCtautau}
\vspace{-0.6em}
\end{figure}

The change in the
upper limit of the LHC wedge region due to the variation of $\mu$ does
not exceed $\Delta\tanb \sim 8$. As explained above, this is a consequence
of cancellations of the leading $\Delta_b$ effects
in the Higgs production and the Higgs decay.
Besides the residual $\Delta_b$ corrections, a further variation of 
the bounds is caused by the
decays of the heavy Higgs bosons into supersymmetric particles. 
For a given value of $\mu$, the rates of these
decay modes are strongly dependent on the particular values of the
weak gaugino mass parameters $M_2$ and $M_1$. In our analysis,
we have taken $M_2 = 200$~GeV, as established by the benchmark scenarios
defined in \citere{benchmark2}, while $M_1 \simeq 100$~GeV.
In general,
the effects of the decays $H/A \to \tilde\chi^0_i \tilde\chi^0_j, 
                                   \tilde\chi^\pm_k \tilde\chi^\mp_l$
only play a role for $\MA \gsim |\mu| + M_1$. 
Outside this range the cancellations of the
$\Delta_b$ effects result in a very weak dependence of the rates on $\mu$.
The combination of the effects from supersymmetric radiative corrections
and decay modes into supersymmetric particles gives rise to a rather
complicated dependence of the discovery contour on $\mu$, see
\citere{benchmark3} for more details.


\subsubsection*{Discovery region for the process 
$t H^{\pm}, H^{\pm} \to \tau \nu_{\tau}$}

For this process we also refer to the analysis carried out by the CMS
Collaboration~\cite{Abdullin:2005yn,Kin2}. The corresponding analyses of the ATLAS
Collaboration can be found in \citeres{TDR,Assamagan:2002ne,Assamagan:2004tt}. The results of
the CMS Collaboration were given for an integrated luminosity of
30~fb$^{-1}$ in the $\MA$--$\tanb$ plane using the $\mhmax$ scenario with 
$\mu = -200 \GeV$. No $\Delta_b$ corrections were included in the 
$gb \to t H^{\pm}$ production process~\cite{tilman} and the 
$H^{\pm} \to \tau \nu_{\tau}$ decay~\cite{Djouadi:1997yw}.

\begin{figure}[htb!]
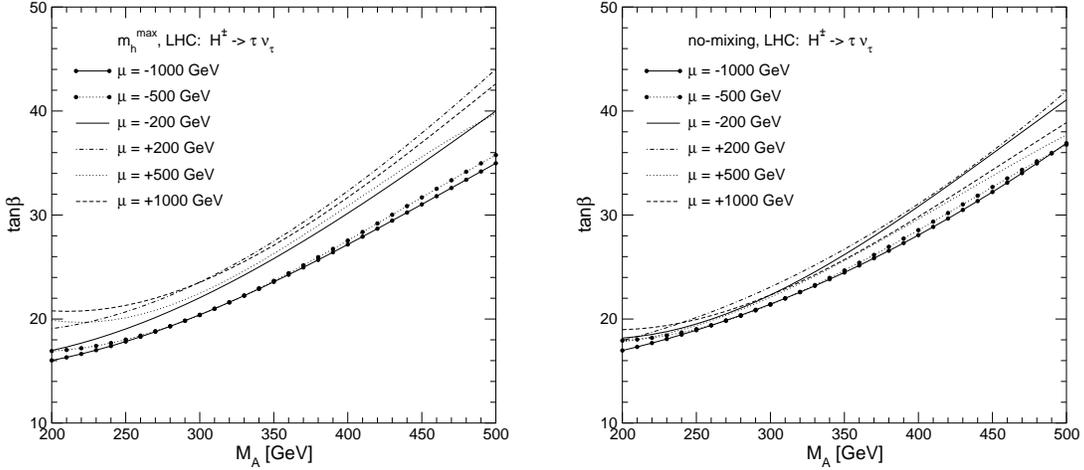

\begin{center}
\includegraphics[width=.42\textwidth]{higgs/benchmark3/t4l-benchP32c.bw.eps}\hspace{2em}
\includegraphics[width=.42\textwidth]{higgs/benchmark3/t4l-benchP33c.bw.eps}
\end{center}
\vspace{-1.5em}
\caption{Variation of the $5 \sigma$ discovery contours obtained 
from the $t H^{\pm}, H^{\pm} \to \tau \nu_{\tau}$
channel in the $\mhmax$ (left) and 
no-mixing (right) benchmark scenarios for different values of~$\mu$.}
\label{fig:LHCcharged}
\vspace{-0.6em}
\end{figure}

In Figure~\ref{fig:LHCcharged} we investigate the impact of including the
$\Delta_b$ corrections into the production and decay processes and of varying
$\mu$.
In order to rescale the original result for the discovery reach in $\tanb$ 
we have first
evaluated the $\tanb$ dependence of the production and decay processes.
If no supersymmetric 
radiative corrections are included, for a fixed $\MA$ value,
the discovery potential can be inferred 
by using a rate approximately proportional to
\begin{equation}
\tan^2\beta_{\rm CMS} \times \br(H^{\pm} \to \tau \nu_{\tau})_{\rm CMS}~. 
\label{Hptaunuold}
\end{equation}
Here 
$\tanb_{\rm CMS}$ is given by the edge of the area in the $\MA$--$\tanb$ plane
in which the signal $H^{\pm} \to \tau \nu_\tau$ 
is visible, as obtained in the CMS analysis. 
The $\br(H^{\pm} \to \tau \nu_{\tau})_{\rm CMS}$ has been
evaluated with {\tt HDECAY}.

The rescaled result for the discovery contour,
including all relevant $\Delta_b$ corrections, is obtained by demanding 
that the contribution
\begin{equation}
\frac{\tan^2\beta}{(1 + \Delta_b)^2} \times \br(H^{\pm} \to \tau \nu_{\tau})~,
\label{Hptaununew}
\end{equation}
where $\Delta_b$ depends on $\tanb$, is numerically equal to the one of
Equation~\ref{Hptaunuold}.  The quantities in Equation~\ref{Hptaununew} have been
evaluated with {\tt FeynHiggs}.

This procedure has been carried out in two benchmark
scenarios for various values of $\mu$. The results are shown in
Figure~\ref{fig:LHCcharged} for the $\mhmax$~scenario (left) and for the
no-mixing scenario (right). As a consequence of the 
cancellations of the leading $\Delta_b$ effects
in the Higgs production and the Higgs decay 
the change in the discovery contour
due to the variation of
$\mu$ does not exceed $\Delta\tanb \sim 10 (6)$ in the $\mhmax$ (no-mixing)
scenario. Also in this case there is a variation of the contour caused
by decays
into supersymmetric particles that, as in the neutral Higgs boson
case, are only relevant for small values of $|\mu|$.


\subsection{Benchmark Scenarios}

The benchmark scenarios defined in \citere{benchmark2}, which were
mainly designed for the search for the light $\cp$-even Higgs boson $h$
in the $\cp$-conserving case,
are also useful in the search for the heavy MSSM Higgs bosons $H$, $A$
and $H^{\pm}$. 
In order to take into account the dependence on $\mu$, which as
explained above is particularly
pronounced for the $b \bar b \phi, \phi \to b \bar b$ 
channel, we suggest to extend the definition of the $\mhmax$ and
no-mixing scenarios given in \citere{benchmark2} by several discrete
values of $\mu$. The scenarios defined in \citere{benchmark2} read\\[1em]
\underline{$\mhmax:$}\\[-3.0em]
\begin{eqnarray} 
&& \mt = 174.3 \GeV, \;
\msusy = 1000 \GeV, \;
\mu = 200 \GeV, \;
M_2 = 200 \GeV, \nonumber \\
&& \Xt^{\mathrm{OS}} = 2\, \msusy  \; \mbox{(FD calculation)}, \;
\Xt^{\msbar} = \sqrt{6}\, \msusy \; \mbox{(RG calculation)} \nonumber \\
&& \Ab = \At, \;
\mgl = 0.8\,\msusy~.
\label{eq:mhmax}
\end{eqnarray}
\underline{\rm no-mixing:}\\[-3.0em]
\begin{eqnarray}
&& \mt = 174.3 \GeV, \;
\msusy = 2000 \GeV, \;
\mu = 200 \GeV, \;
M_2 = 200 \GeV, \nonumber \\
&& \Xt = 0 \; \mbox{(FD/RG calculation)} \;
\Ab = \At, \;
\mgl = 0.8\,\msusy~.
\end{eqnarray}
\underline{\rm constrained $\mhmax$:}\\[-2.0em]
\begin{eqnarray}
&& \mt = 174.3 \GeV, \;
\msusy = 1000 \GeV, \;
\mu = 200 \GeV, \;
M_2 = 200 \GeV, \nonumber \\
&& \Xt^{\mathrm{OS}} = -2\, \msusy  \; \mbox{(FD calculation)}, \;
\Xt^{\msbar} = -\sqrt{6}\, \msusy \; \mbox{(RG calculation)}, \nonumber \\
&& \Ab = \At, \;
\mgl = 0.8\,\msusy~.
\label{constmhmax}
\end{eqnarray}
The constrained-$\mhmax$ scenario differs from Equation~\ref{eq:mhmax} only by the
reversed sign of $\Xt$. 
While the positive sign of the product $(\mu\,M_2)$ results in
general in better agreement with the $(g-2)_\mu$ experimental results,
the negative sign of the product ($\mu\,\At$) yields in general 
(assuming minimal flavor violation) better agreement with the 
$\br(b \to s \ga)$ measurements.

Motivated by the analysis in Section~\ref{sec:interpretation}
we suggest to investigate the following values of $\mu$
\begin{equation}
\mu = \pm 200, \pm 500, \pm 1000 \GeV ~,
\label{eq:musuggest}
\end{equation}
allowing both an enhancement and a suppression of the bottom Yukawa
coupling and taking into account the limits from direct searches for
charginos at LEP~\cite{Abbiendi:2003sc}.
As discussed above, the results in the constrained-$\mhmax$ scenario
are expected to yield more robust bounds against the variation of
$\mu$ than in the other scenarios.
It should be noted that the values $\mu = -500, -1000 \GeV$ can lead to
such a large enhancement of the bottom Yukawa coupling that a
perturbative treatment is no longer possible in the region of very large 
values of $\tanb$. Some care is therefore necessary to assess up to which
values of $\mu$ reliable results can be obtained, see e.g.\ the
discussion of Figure~\ref{fig:cdfchange}.

The value of the top-quark mass in \citere{benchmark2} was chosen
according to the experimental central value at that time. We propose 
to substitute this value with the most up-to-date experimental
central value for $\mt$.


\subsection{Conclusions}

In this paper we have analyzed the impact of supersymmetric radiative
corrections on the current MSSM Higgs boson exclusion limits at the 
Tevatron and the prospective discovery reach at the LHC.
In particular, we have
studied the variation of the exclusion and discovery contours obtained 
in different MSSM benchmark scenarios under 
changes of the higgsino mass parameter
$\mu$ and the supersymmetry breaking parameters associated with the
third generation squarks. These parameters determine the most important
supersymmetric radiative corrections in the large $\tanb$ region that are
associated with a change of the effective Yukawa
couplings of the bottom quarks to the Higgs fields (since the squarks
are relatively heavy in the considered benchmark scenarios, other
squark-loop effects are sub-dominant). These corrections had been
ignored or only partially considered
in some of the previous analyses of Higgs searches at hadron
colliders. We have shown that their inclusion leads to a significant
modification of the discovery and exclusion regions. 

We have investigated the exclusion bounds obtained from the Tevatron searches 
for non SM-like Higgs bosons in different channels. For the 
$b \bar b \phi, \phi \to b \bar b$ channel ($\phi = h, H, A$)
we find that the effects of
the supersymmetric radiative corrections on the exclusion bounds in the
$\MA$--$\tanb$ plane are quite dramatic. While in the $\mhmax$ scenario
the current data allow to rule out values of $\tanb \gsim 50$~(35) for 
$\MA \approx 100 \GeV$ if the higgsino mass parameter is chosen as 
$\mu = -200 \GeV$~($-1000 \GeV$), hardly any bound 
on $\tanb$ can be set if positive
values of $\mu$ are chosen. The shifts are smaller, but still 
important, for the no-mixing benchmark scenario.
We have shown that the 
constrained-$\mhmax$ scenario yields results that are much more stable
against variations of $\mu$ than the other benchmark scenarios.

For the inclusive channel with $\tau^+\tau^-$ final states, 
$p \bar p \to \phi \to \tau^+\tau^-$, compensations occur between large
corrections to Higgs production and decay, so that the limits in the 
$\MA$--$\tanb$ plane obtained from this channel turn out to be less
affected by varying $\mu$ than the ones from the
associated production with bottom
quarks. Nevertheless we have found that the exclusion limit is shifted
by up to $\Delta\tanb = 30$ as a consequence of choosing different input
values for $\mu$. We have investigated the impact of including the
dominant supersymmetric radiative corrections to the gluon fusion 
production process, which had previously been omitted. The inclusion of
these corrections leads to a shift of up to $\Delta\tanb = 10$ in the 
exclusion limit. 
Following our analysis, the CDF Collaboration has adopted the
prescription outlined in this paper for incorporating the correction into
the $gg \to \phi$ production process.
The Tevatron experiments are expected to collect further data at
higher luminosities, 
up to 4--8~fb$^{-1}$, in the next few years. This will extend the Tevatron
MSSM Higgs boson discovery and exclusion reach  in the $M_A$--$\tanb$ 
plane to lower values of $\tanb$,  decreasing the sensitivity of the
obtained bounds to variations of the low energy supersymmetry mass parameters.

For the LHC we have analyzed the channels 
$pp \to H/A +X, \, H/A \to \tau^+\tau^-$ and 
$t H^{\pm}, H^{\pm} \to \tau \nu_{\tau}$, which yield the best
sensitivities in the search for heavy MSSM Higgs bosons. Accordingly,
the discovery contours for these channels in the $\MA$--$\tanb$ plane
determine the boundary of the region where only the (SM-like) light 
$\cp$-even Higgs boson can be detected at the LHC. Since the discovery
contours for the LHC are at smaller values of $\tanb$ compared to those
accessible via the
current exclusion bounds at the Tevatron, the impact of the
$\tanb$-enhanced supersymmetric corrections is less pronounced
in this case. We have
studied the effect of including the dominant supersymmetric
corrections, which had been omitted in the analyses of the 
production processes at the LHC, and their variation with the relevant
parameters. Possible decays of the heavy MSSM Higgs bosons into
charginos and neutralinos have been taken into account.
We have found that the prospective discovery contours at the LHC 
are shifted by up to $\Delta\tanb \lsim 10$.

Based on our analysis of the sensitivities of the searches for MSSM
Higgs bosons at the Tevatron and the LHC we have defined benchmark
scenarios for the analysis of MSSM Higgs-boson searches at hadron
colliders. They are based on a generalization of similar benchmark
scenarios proposed for the searches for SM-like MSSM Higgs bosons at
the Tevatron and the LHC.


\subsection*{Acknowledgements}
We thank
A.~Anastassov,
J.~Conway,
A.~Goussiou, 
A.~Haas,
B.~Heinemann, 
A.~Kharchilava,
R.~Kinnunen,
A.~Lath,
A.~Nikitenko,
T.~Plehn, 
M.~Schumacher 
and
M.~Spira
for helpful discussions.

\clearpage
\section{Sensitivity of CDF's Higgs Boson Searches}
\label{sec:xsec}
%
\textbf{Contributed by: T. Junk for the CDF Higgs Group}

\vspace{0.25in}

The search for the Standard Model (SM) Higgs boson is one of the central
pieces of the current High Energy Physics program.  The $SU(2)\times U(1)$
gauge model of electroweak interactions makes a number of predictions
which have been experimentally verified to high precision, but its validity
depends on the breaking of this symmetry to the $U(1)_{\mathrm{EM}}$ symmetry group
at low energies.  Many differing proposals of the details of this symmetry
breaking have been advanced, most of which predict one or more observable
scalar bosons.  If the minimal SM Higgs mechanism describes nature, then
precision electroweak data~\cite{Group:2005di} provide evidence that the scalar
Higgs boson should be lighter than about 200~GeV, with a preferred
value at around 115~GeV.  Direct searches at LEP have
excluded~\cite{Barate:2003sz} a SM Higgs boson with a mass below 114.4~GeV.
If there is a SM Higgs boson with a mass between $\sim 115$~GeV and $\sim 200$~GeV
it is produced in $p{\bar{p}}$ collisions at
the Tevatron, and, with enough data, it should be possible to exclude or discover
such a particle.

Data are being accumulated by the Tevatron experiments CDF and D\O, whose runs
are expected extend until 2009.   Currently, more than 1~fb$^{-1}$ of data have been
recorded by each experiment, although the Higgs boson searches reported here
are based on approximately 300~pb$^{-1}$ of data.  The exact luminosities
used in the channels is listed in Table~\ref{tab:lumi}.

With 300~pb$^{-1}$ of data
and the expected signal-to-background ratios in the channels, the SM Higgs boson
hypothesis cannot be tested for any value of $m_H$.  Nonetheless, with additional
data, and improvements to the analyses, sensitivity at the 95\% CL level
may be obtained for $m_H$ up to 180~GeV, assuming the design integrated
luminosity of 8~fb$^{-1}$ is collected with good quality by both
detectors, according to a 1999 study~\cite{Carena:2000yx}.
An updated study~\cite{Babukhadia:2003zu} was conducted in 2003 to check the earlier projections
with more realistic simulations and preliminary data samples which could be used
to calibrate some backgrounds.  The later study did not consider searches for
Higgs bosons with $m_H$ greater than 130~GeV, and also did not include the
effects of systematic uncertainties on the amount of luminosity required to
test for Higgs bosons.    Each report includes
calculations of the estimated amounts of luminosity required for a combination of all of CDF's
channels and D\O's channels to exclude at the 95\% CL, assuming a Higgs boson is not present,
as well as the luminosity requirements for a combined $3\sigma$ evidence and
$5\sigma$ discovery.  The luminosity thresholds are shown in Figure~\ref{fig:shw}
for the 1999 study and in Figure~\ref{fig:hswg} for the 2003 study.

\begin{figure}
\begin{center}
\includegraphics[width=0.65\textwidth]{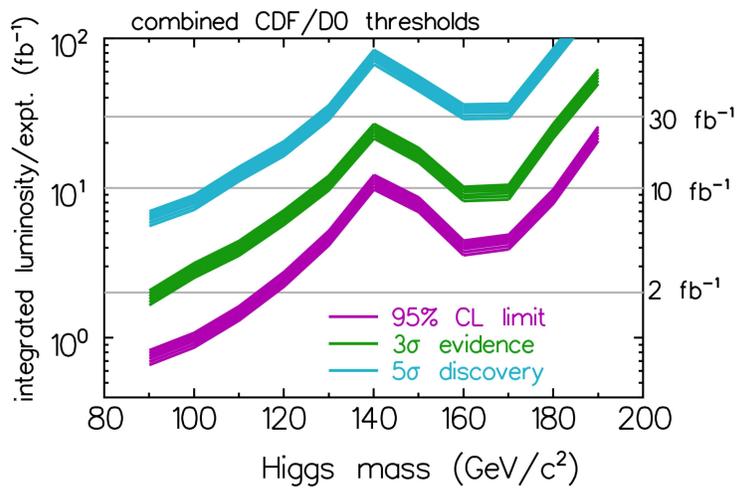}
\end{center}
\caption{SUSY/Higgs Working Group estimations of the luminosity required for
95\% exclusion, $3\sigma$ evidence, and $5\sigma$ discovery for the combined
CDF+D\O\ search channels. (2000).}
\label{fig:shw}
\end{figure}

\begin{figure}
\begin{center}
\includegraphics[width=0.65\textwidth]{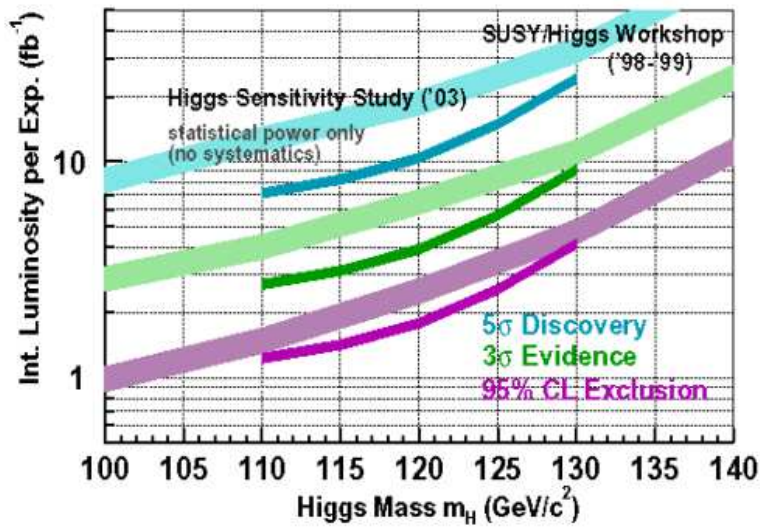}
\end{center}
\caption{Higgs Sensitivity Working Group estimations of the luminosity required for
95\% exclusion, $3\sigma$ evidence, and $5\sigma$ discovery for the combined
CDF+D\O\ search channels, compared against the earlier SUSY/Higgs Working Group's
calculation.}
\label{fig:hswg}
\end{figure}

The CDF channels as they stand as of the Summer 2005 conferences are not as
powerful as those assumed in the two sensitivity studies.  The following sections
provide a snapshot of the sensitivity of the CDF channels separately
and combined, as of the October 2005 TeV4LHC workshop, with plans for improvement.

\begin{table}
\caption{Integrated luminosities by channel.}
\label{tab:lumi}
\begin{center}
\begin{tabular}{|l|c|c|}\hline
Channel & $\int{\cal{L}}dt$ (pb$^{-1}$) & Reference \\ \hline
$W^\pm H\rightarrow\ell^\pm\nu b{\bar{b}}$ & 319 & \cite{Ishizawa:2005} \\
$ZH\rightarrow\nu{\bar{\nu}}b{\bar{b}}$ & 289 & \cite{Veszpremi:2005} \\
$gg\rightarrow H\rightarrow W^+W^-$ & 360 & \cite{Chuang:2005} \\
$W^\pm H\rightarrow W^\pm W^+W^-$ & 194 & \cite{Hirokazu:2004} \\ \hline
\end{tabular}
\end{center}
\end{table}

\subsection{Sensitivity by Channel}

The expected signal and background rates and shape distributions were collected from
each of the channel analysis teams and combined using the CL$_\mathrm{s}$ technique~\cite{Read:2002hq,Junk:1999kv}
to find the expected limits on the cross-section multiplied by the branching fractions.
Candidate information was not included in the combination, so the observed limit
of the combination is not computed.  All of the observed limits in the channels
are close to expectations, the observed limit
of the combination is expected to be close to the expected combined limit.

\begin{boldmath}
\subsubsection*{The $W^\pm H\rightarrow\ell^\pm\nu b{\bar{b}}$ Channel}
\end{boldmath}
\label{lvbb}

The results of the $W^\pm H\rightarrow\ell^\pm\nu b{\bar{b}}$ search are described in~\cite{Ishizawa:2005}.
The reconstructed mass distribution in the single-tagged analysis is used in computing the
expected limits, with each bin counted as an independent counting experiment.  Systematic
errors are taken on the background and signal rates, but the shapes are not varied.  Each
bin is assumed to have fully correlated systematic uncertainties with all other bins of
the mass distribution.  The systematic uncertainties are detailed in Table~\ref{tab:syst}.
Acceptances and signal distributions are linearly interpolated~\cite{Read:1999kh}
between the supplied test points at which Monte Carlo samples are available.
The observed and expected cross-section times branching ratio limits are shown at the 95\% CL
 in  Figure~\ref{fig:lvbblimit} as a function
of $m_H$.

\begin{table}
\caption{Relative systematic uncertainties by channel.  Errors from the same source
are considered correlated, across channels, and between signal and background.  The
``uncorrelated'' errors are uncorrelated across channels and between signal and background.}
\label{tab:syst}
\begin{scriptsize}
\begin{center}
\begin{tabular}{|l|cc|cc|cc|cc|cc|}\cline{2-11}
\multicolumn{1}{c|}{ } & \multicolumn{10}{|c|}{Channel} \\\cline{2-11}
\multicolumn{1}{c|}{ } 
 & \multicolumn{2}{|c}{$W^\pm H\rightarrow\ell^\pm\nu b{\bar{b}}$}
 & \multicolumn{2}{|c}{$ZH\rightarrow\nu{\bar{\nu}}b{\bar{b}}$}
 & \multicolumn{2}{|c}{$ZH\rightarrow\ell^+\ell^-b{\bar{b}}$}
 & \multicolumn{2}{|c}{$gg\rightarrow H\rightarrow W^+W^-$}
 & \multicolumn{2}{|c|}{$W^\pm H\rightarrow W^\pm W^+W^-$} \\\cline{1-1}
Source & s [\% ] & b [\% ] & s [\% ] & b [\% ] & s [\% ] & b [\% ] & 
         s [\% ] & b [\% ] & s [\% ] & b [\% ] \\\hline
lumi & 6 & 6 &         6 & 6      & 6      &   6 & 6 & 6 & 6 & 6\\
b-tag s.f. & 5 & &     6.4 & 1.9  & 15 & 15      &   &   & 0.37 & \\
lepton ID  & 5 & &     &          & 7  & 7       &    & & & \\
lepton trig & 0.6 & &  &          & 1  & 1       &    & & 2.4 & \\
PDF & 1 & &            &         &         &    &    & & 1.5 & \\
ISR & 3 & &            &         &         &    &    & & 3.0 & \\
FSR & 7 & &            & &       7         &    &    & & 3.2 & \\
JES & 3 & &            7.8 & 3.5 &         &    &    & & & \\
Jet model & 1.4 & &    & &                 &    &    & & & \\
$\nu\nu b{\bar{b}}$ trig & & &  3 & 1.5 & & & & & & \\
$\nu\nu b{\bar{b}}$ veto & & &  2 & 2 & & & & & & \\
uncorrelated  & & 15 & 2 & 22.1  & & 9      &    6 & 7 & 3.7 & 66 \\\hline
\end{tabular}
\end{center}
\end{scriptsize}
\end{table}

\begin{figure}
\begin{center}
\includegraphics[width=0.65\textwidth]{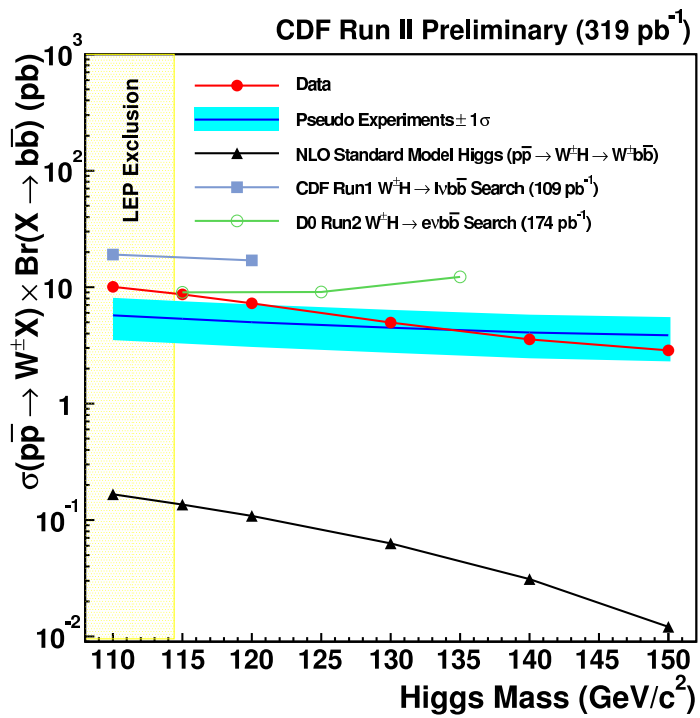}
\end{center}
\vspace{-0.5cm}
\caption{The observed and expected 95\% CL limits on the production cross-section times the
Higgs decay branching ratio as a function of $m_H$ for the
$W^\pm H\rightarrow\ell^\pm\nu b{\bar{b}}$ channel.  The limits are compared with
the SM prediction.}
\label{fig:lvbblimit}
\end{figure}

\begin{boldmath}
\subsubsection*{The $ZH\rightarrow\nu{\bar{\nu}}b{\bar{b}}$ Channel}
\end{boldmath}
\label{vvbb}

The results of the $ZH\rightarrow\nu{\bar{\nu}}b{\bar{b}}$ search are described in~\cite{Veszpremi:2005}.
The reconstructed mass distribution was not provided for combination, but the numbers of
events for the expected signal and background after a mass window cut which moves with
the Higgs boson mass under test are used.  They are linearly interpolated between
the model points listed in~\cite{Veszpremi:2005}.  The systematic uncertainties on the signal
and background are detailed in Table~\ref{tab:syst}.
The observed and expected cross-section times branching ratio limit is shown at the 95\% CL in 
Figure~\ref{fig:vvbblimit} as a function
of $m_H$, and compared to the
SM expectation. 

\begin{figure}
\begin{center}
\includegraphics[width=0.65\textwidth]{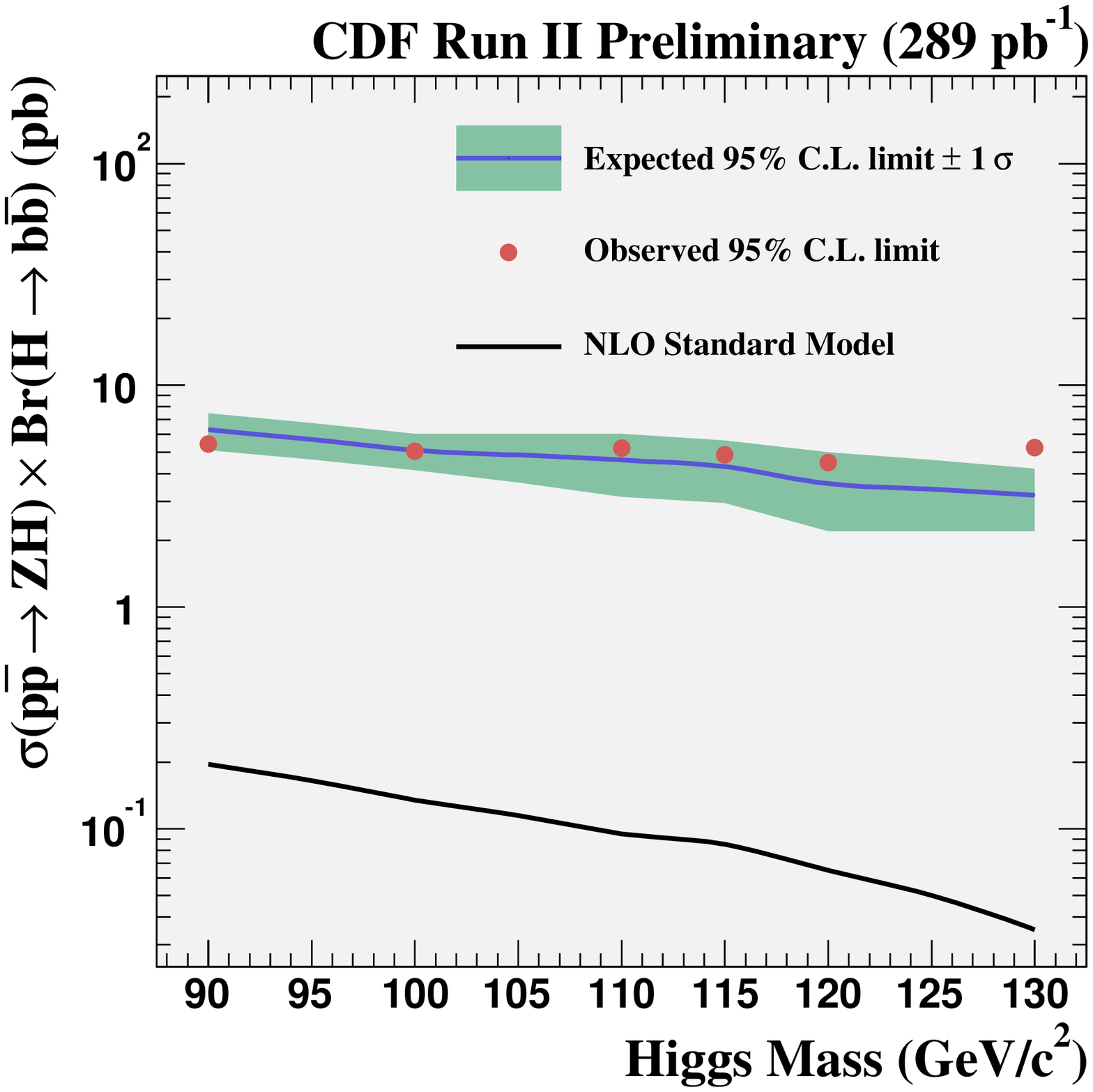}
\end{center}
\vspace{-0.5cm}
\caption{The observed and expected 95\% CL limits on the production cross-section times the
Higgs decay branching ratio as a function of $m_H$ for the
$ZH\rightarrow\nu{\bar{\nu}}b{\bar{b}}$ channel. The liimts are compared with
the SM prediction.}
\label{fig:vvbblimit}
\end{figure}

\begin{boldmath}
\subsubsection*{The $ZH\rightarrow\ell^+\ell^-b{\bar{b}}$ Channel}
\end{boldmath}
\label{llbb}

The  $ZH\rightarrow\ell^+\ell^-b{\bar{b}}$ channel is still in development
and the analysis is still in its ``blind'' stage.  Hence, data candidate
information is not yet available.  The current status is described
in~\cite{Efron:2005}.  The selection starts with a very clean sample of
$Z\rightarrow\ell^+\ell^-$ decays, identifying isolated leptons with $m_{\ell\ell}$
close to $m_Z$, and two or three jets, at least one of which must be b-tagged.
The systematic uncertainties on the signal
and background are detailed in Table~\ref{tab:syst}.  The neural net has seventeen
input variables described in~\cite{Efron:2005}.  The most powerful ones are the
invariant mass of the two leading jets taken together, the event $H_T$ (which is the
scalar sum of all the $P_T$'s of the observed particles), and the $E_T$ of the
leading jet.  The median expected limit on the cross-section times the branching
ratio for this process is approximately 2.2~pb for 300~pb$^{-1}$ of data.  This expected
limit is lower than that for other channels mainly due to the very small background
prediction.  It must be compared, however, against a much smaller SM signal expectation.

\begin{figure}
\begin{center}
\includegraphics[width=0.65\textwidth]{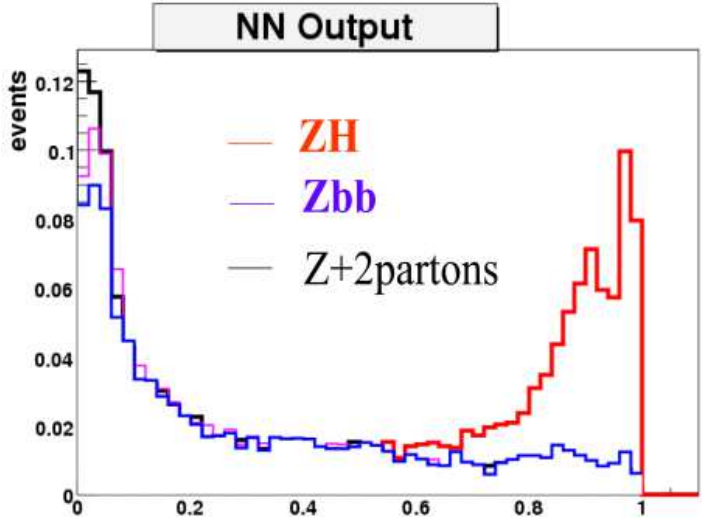}
\end{center}
\vspace{-0.5cm}
\caption{The distribution of the neural net discriminant function for the
$ZH\rightarrow\ell^+\ell^-b{\bar{b}}$ channel, shown separately for the
signal and for the major backgrounds, $Zb{\bar{b}}$ and $Z+2$~partons.  The data
in this channel are still blind.  }
\label{fig:llbbnn}
\end{figure}

\begin{boldmath}
\subsubsection*{The $gg\rightarrow H\rightarrow W^+W^-$ Channel}
\end{boldmath}
\label{ww}

The results of the $gg\rightarrow H\rightarrow W^+W^-$ search are described in~\cite{Chuang:2005}.
The histograms of $\Delta\phi_{\ell\ell}$ are used as the discriminant variable input
to the limit calculation -- each bin is a separate counting experiment.  The
shapes are interpolated~\cite{Read:1999kh} between $m_H$ points, as are the signal
rates and background rates.  The analysis uses $m_H$-dependent cuts, and so the background
rates depend on the $m_H$ under test.  The systematic uncertainties on the signal
and background are detailed in Table~\ref{tab:syst}.

The median expected 95\% CL cross-section times branching ratio limit is shown in 
Figure~\ref{fig:wwlimit} as a function
of $m_H$ compared to the SM expectation and to the computation of~\cite{Chuang:2005}.

\begin{figure}
\begin{center}
\includegraphics[width=0.65\textwidth]{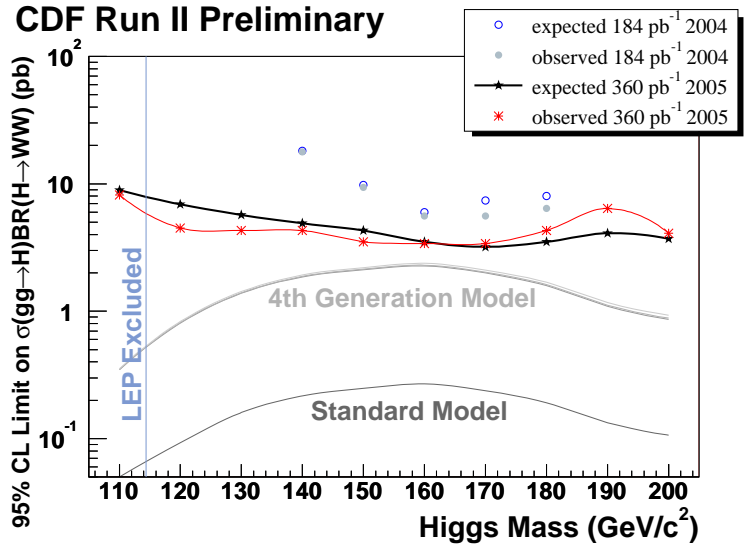}
\end{center}
\vspace{-0.5cm}
\caption{The observed and expected 95\% CL limits on the production cross-section times the
Higgs decay branching ratio as a function of $m_H$ for the
$gg\rightarrow H\rightarrow W^+W^-$ channel.  The limits are compared with
the SM prediction, and also the prediction of a model with a heavy fourth generation
of SM-like fermions.}
\label{fig:wwlimit}
\end{figure}

\begin{boldmath}
\subsubsection*{The $W^\pm H\rightarrow W^\pm W^+W^-$ Channel}
\end{boldmath}
\label{www}

The results of the $W^\pm H\rightarrow W^\pm W^+W^-$ search are described in~\cite{Hirokazu:2004}.
It is a single counting experiment -- there are no discriminant variables whose histograms
have different $s/b$ ratios to use.  The acceptance is interpolated between the
$m_H$ points listed in~\cite{Hirokazu:2004}.  The systematic uncertainties on the signal
and background are detailed in Table~\ref{tab:syst}.  For this calculation, the
data statistical uncertainty on the residual conversion background is treated as independent
of the other errors on the background and the errors add in quadrature instead of linearly
as they do in~\cite{Hirokazu:2004}.  Furthermore, the FSR systematic uncertainty is almost
certainly truly uncorrelated with other channels' FSR uncertainty, but it has been treated
as correlated.  As is seen below, the entire systematic error treatment in this channel
matters little to the sensitivity.

The observed cross-section times branching ratio limit is shown at the 95\% CL in 
Figure~\ref{fig:wwwlimit} as a function
of $m_H$ compared to the SM expectation and to the computation of~\cite{Hirokazu:2004}.

\begin{figure}
\begin{center}
\includegraphics[width=0.65\textwidth]{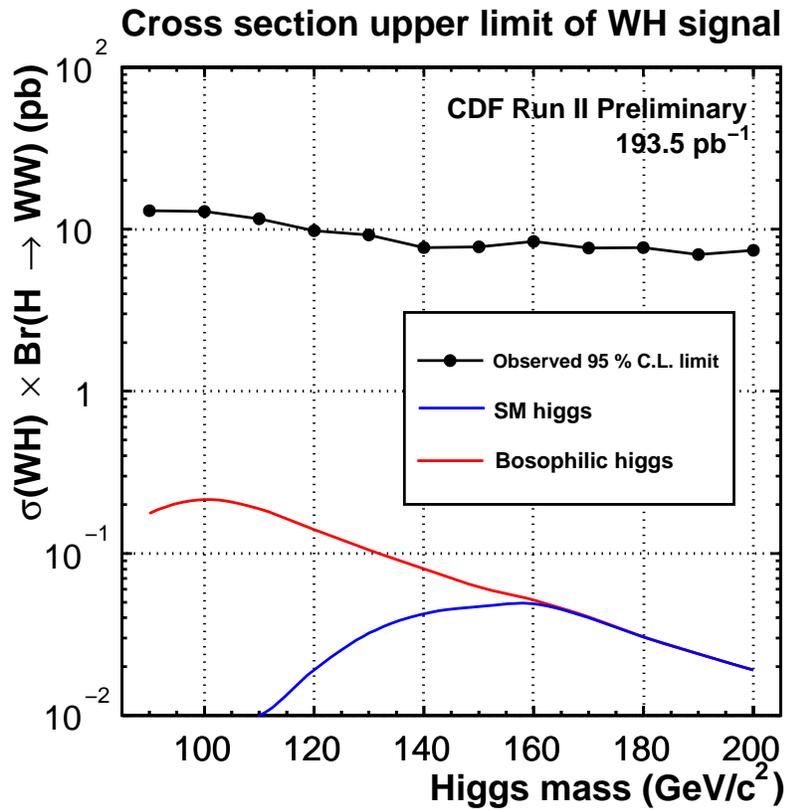}
\end{center}
\caption{The observed 95\% CL limit on the production cross-section times the
Higgs decay branching ratio as a function of $m_H$ for the
$W^\pm H\rightarrow W^\pm W^+W^-$ channel.  The limits are compared with
the SM prediction, and also the prediction of a ``bosophilic'' (also known
as ``fermiophobic'') model.}
\label{fig:wwwlimit}
\end{figure}

\subsection{Sensitivity of the SM Channels when Combined}

The observed 95\% CL limits in all of CDF's SM Higgs channels are shown, compared with SM predictions,
and also compared with observed limits from D\O's channels, in Figure~\ref{fig:smcompare}.
The different searches search for different processes which have different rates, and
thus contribute differently to the combined sensitivity.  It is somewhat easier to
compare the channels' sensitivity to a SM Higgs when the ratio of the limit in each
channel to the SM prediction is formed.  This ratio is shown for the same collection
of CDF and D\O\ channels in Figure~\ref{fig:smratio}.

\begin{figure}
\begin{center}
\includegraphics[width=0.9\textwidth]{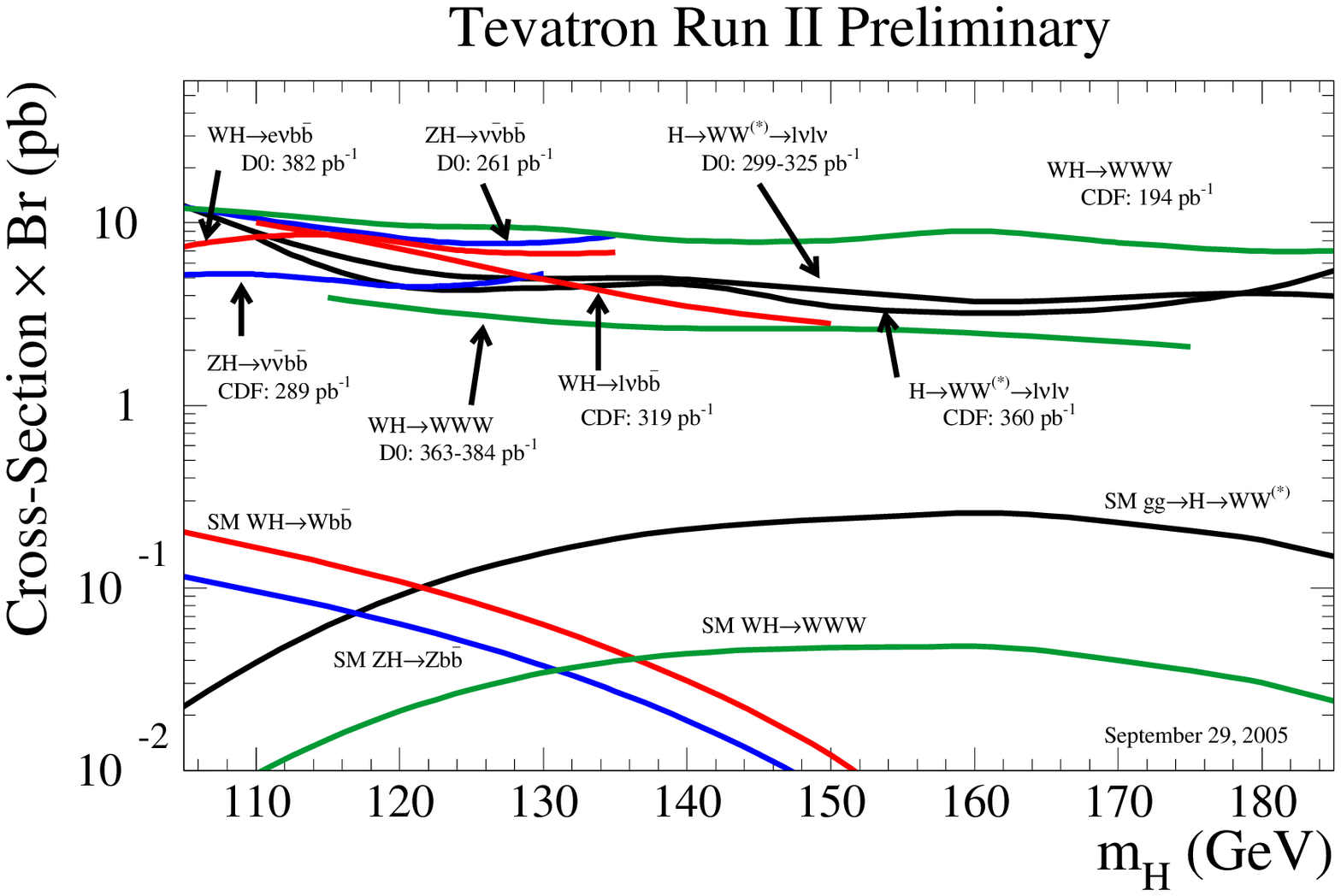}
\end{center}
\caption{The observed 95\% CL limits on the production cross section
times the Higgs decay branching ratio for each of the five
search channels, compared with D\O's limits, and also
compared with SM expectations.}
\label{fig:smcompare}
\end{figure}

\begin{figure}
\begin{center}
\includegraphics[width=0.9\textwidth]{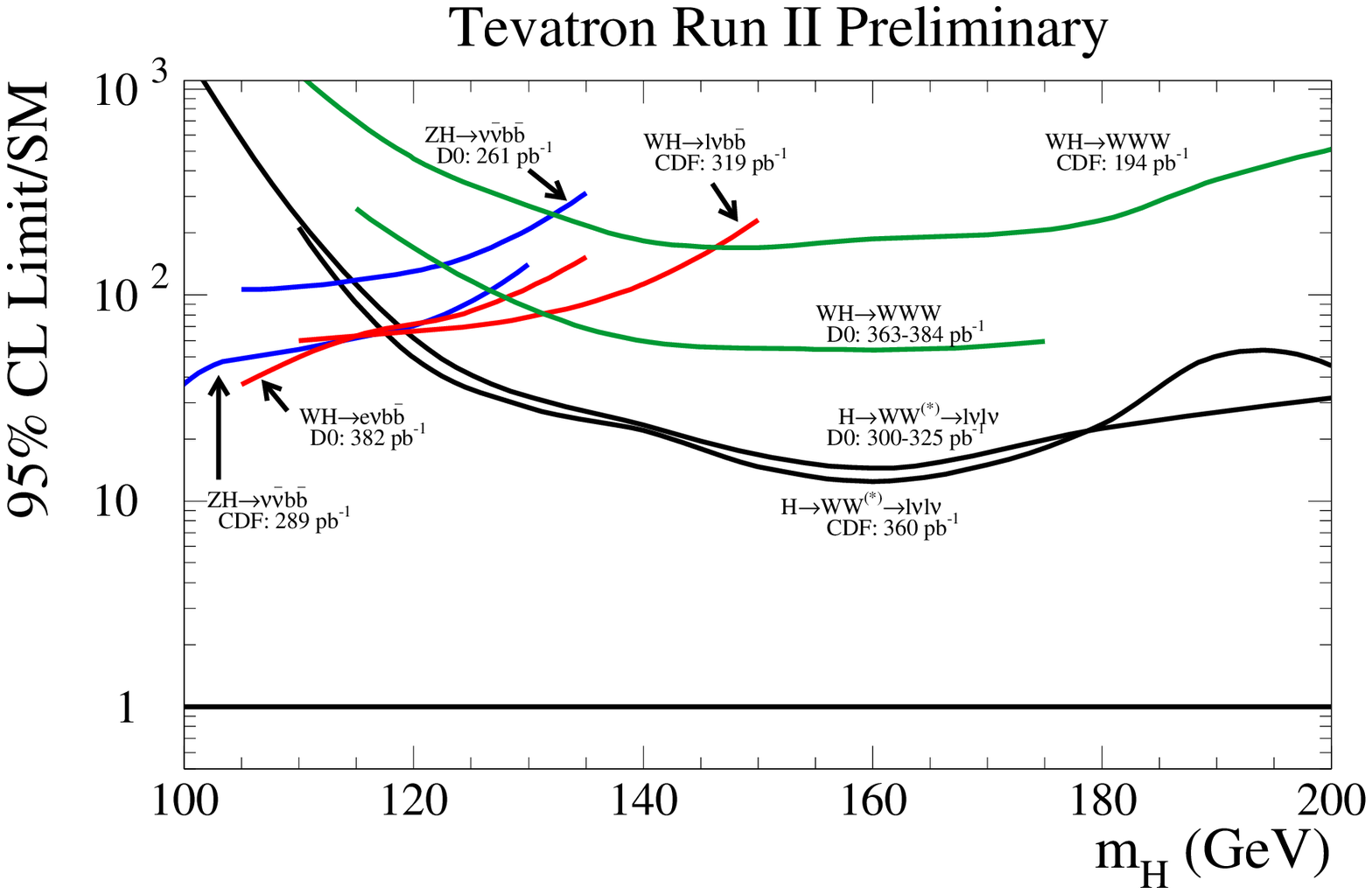}
\end{center}
\caption{CDF and D\O's observed 95\% CL limits on the production cross section
times the Higgs decay branching ratio, divided by the corresponding SM predictions,
for each of the five search channels.}
\label{fig:smratio}
\end{figure}

  The CL$_{\mathrm{s}}$ method is used on the collection of CDF's
five SM Higgs boson search channels
to compute the multiplicative scale factor $s_{95}$ on the total signal which can just
barely be expected to be excluded in a median experimental outcome.  This procedure
doesn't make much physical sense for scale factors exceeding unity,
as there isn't a well-motivated physical model which
scales all of the production mechanisms for SM Higgs bosons in the same way, but it
provides a technical benchmark of how far we are from the SM in our sensitivity.
The results of this combination are shown in Figure~\ref{fig:combineds95}.  It must
be shown as a multiplicative factor of the SM prediction because of the different SM
predictions used for each search channel.

\begin{figure}
\begin{center}
\includegraphics[width=0.9\textwidth]{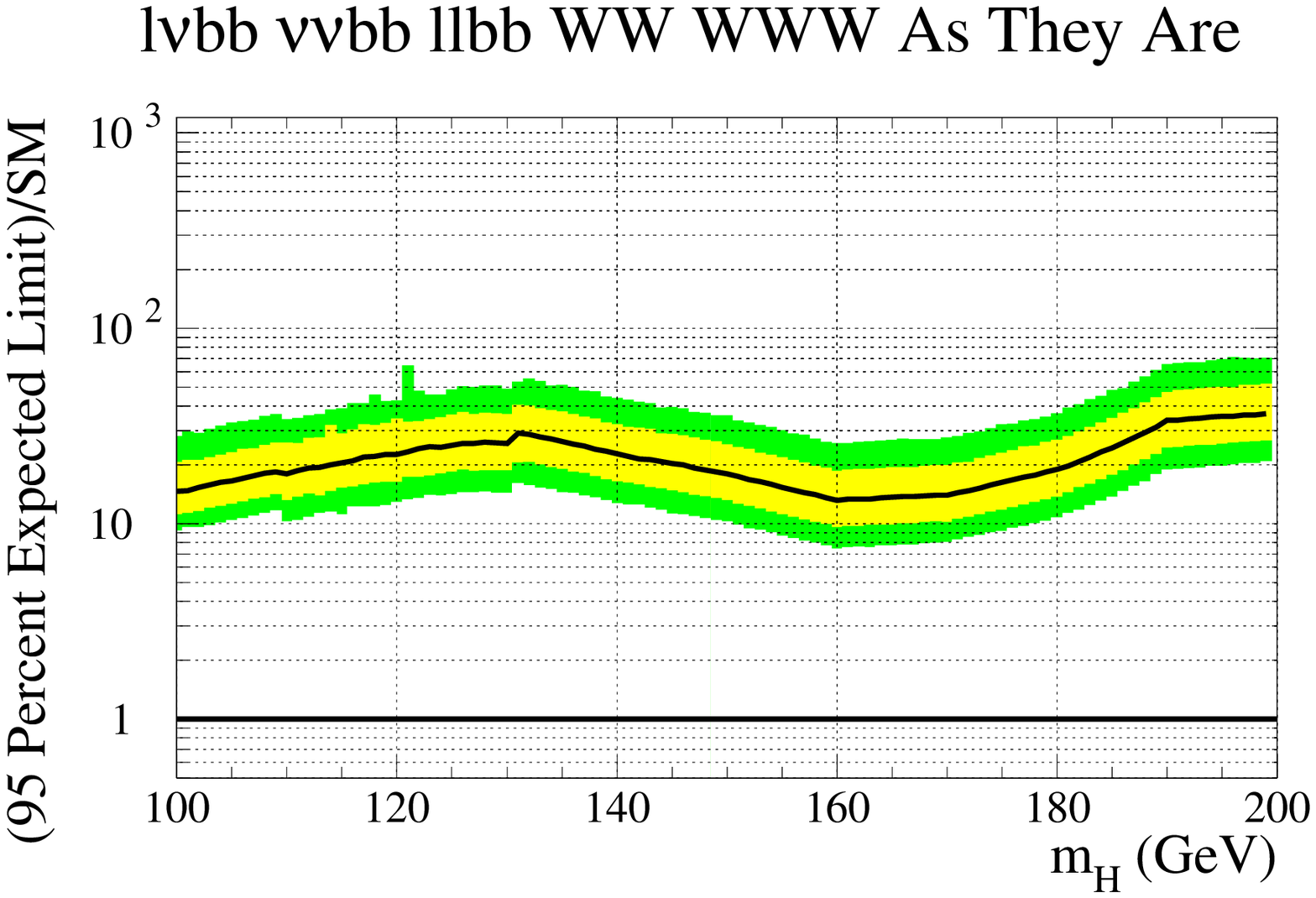}
\end{center}
\vspace{-0.5cm}
\caption{The expected 95\% CL limit on the multiplicative scale factor
of SM Higgs boson production for CDF's five SM
Higgs boson search channels combined,
as a function of $m_H$, assuming the absence of a Higgs boson.
The yellow and green bands show the
$\pm 1\sigma$ and $\pm 2\sigma$ expectations, which fluctuate
depending on the possible data which may be observed.}
\label{fig:combineds95}
\end{figure}

\subsection{Necessary SM Channel Improvements}

The current channels as we have them are insufficient to test for the presence or absence
of the Standard Model Higgs boson, even if the projected 8~fb$^{-1}$ of data
are collected.  Improvements must be made to increase the acceptance,
reduce the background, and to separate the selected events into disjoint subsets with
different $s/b$ ratios, and to combine them together.  Furthermore, the results must
be combined with D\O.

The Higgs Sensitivity Working Group report~\cite{Babukhadia:2003zu} lists changes which can be made to
the analyses which can get us to the desired level of sensitivity.  Much of this work
has already been done to improve our resolutions, to increase our lepton acceptance
to the forward region, and to develop neural nets.  But the work has been done by a variety
of different people separated in space, time, and institution.  The work of many groups
must be collected together in the analysis channels in order to achieve the sensitivity
reported in~\cite{Carena:2000yx,Babukhadia:2003zu}.

The factors on the expected amount of luminosity needed to get exclusion at the 95\%~CL,
$3\sigma$ evidence and $5\sigma$ discovery can be computed for most of the improvements
rather easily.  For acceptance increases, the background ought to increase as the signal
acceptance increases.  In fact, it should increase faster, because as we expand our
acceptance to forward regions of the detector or to include leptons of lower quality,
a larger fraction of background is expected to creep in.  For this estimation, the
estimations are taken from the HSWG report's Sections~2.3 and~4.2 (for the Neural Net factor).
A listing of improvements and their factors in luminosity is given in Table~\ref{tab:improvements}.
It is
assumed in the luminosity projections that the systematic uncertainties will scale inversely
with the square root of the integrate luminosity.  Furthermore, accounting of the shape
uncertainties may make the systematic errors larger.
 
\begin{table}
\caption{Luminosity factors expected from analysis improvements, separated by
channel.}
\label{tab:improvements}
\begin{center}
\begin{tabular}{|l|c|c|c|}\hline
Improvement & $W^\pm H\rightarrow\ell^\pm\nu b{\bar{b}}$ & 
              $ZH\rightarrow\nu{\bar{\nu}}b{\bar{b}}$ & 
              $ZH\rightarrow\ell^+\ell^-b{\bar{b}}$ \\ \hline
$m_H$ Resolution & 1.7 & 1.7 & 1.7 \\
Continuous b-tags & 1.5 & 1.5 & 1.5 \\\hline
Forward B-tags & 1.1 & 1.1 & 1.1 \\
Forward Leptons & 1.3 & 1.0 & 1.6 \\
Neural Nets & 1.75 & 1.75 & 1.0\\
Track-Only Leptons & 1.4 & 1.0 & 1.6 \\
WH signal in ZH & 1.0 & 2.7 & 1.0 \\ 
Product of above & 8.9 & 13.3 & 7.2 \\ \hline
CDF+D\O\ Combination & 2.0 & 2.0 & 2.0\\
All Combined & 17.8 & 26.6 & 14.4 \\ \hline
\end{tabular}
\end{center}
\end{table}

The neural net factor of 1.75 is not uniformly applicable to all channels, as the
$ZH\rightarrow\ell^+\ell^-b{\bar{b}}$ channel estimations already take advantage of
a neural net.  The forward lepton acceptance improvement cannot strictly be multiplied
by the track-only lepton factor since the forward tracking is not sufficient.  Nonetheless,
a naive product of the factors from the analysis improvements is approximatley 20.
The analysis improvements will not be made all at once -- work is ongoing to develop
and characterize the techniques.

\subsection{SM Sensitivity Projections}

Assuming that the acceptances of the channels are increased and neural nets or other
advanced techniques are used to reduce the backgrounds, the projected reach of of the
Tevatron SM Higgs search program is estimated.  It is assumed that the systematic
uncertainties scale inversely with the square root of the integrated luminosity,
and that D\O\ contributes channels with the same sensitivity as CDF's and that they
are combined together.  Figure~\ref{fig:cdfd0foreseen} shows how the significance of
an excess of events is expected to develop, as a function of the integrated luminosity
collected per experiment, assuming a SM Higgs boson is present with a mass $m_H=115$~GeV.
The actual evolution of such an excess, if a signal is actually present, will be more
of a random walk as data are collected, so the figure also includes the width of
the expected distribution.  Figure~\ref{fig:probscdfd0foreseen} shows the
evolution of the probability of seeing a $2\sigma$, a $3\sigma$, or a $5\sigma$ excess
in the combined data when searching for a SM Higgs boson of mass $m_H=115$~GeV, if
it is truly present, as a function of the luminosity collected by each experiment.
After collecting 8~fb$^{-1}$ per experiment, it is 10\% likely that a $5\sigma$ excess
will be observed if $m_H$ is truly 115~GeV.

\begin{figure}
\begin{center}
\includegraphics[width=0.65\textwidth]{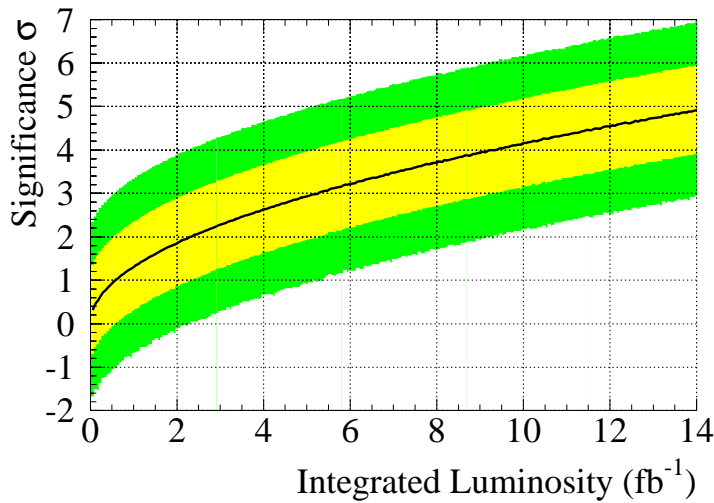}
\end{center}
\vspace{-0.5cm}
\caption{The evolution of the expected significance of an excess in the
data if a Standard Model Higgs boson is present with a mass of 115~GeV.
The yellow (light) interior band shows the $\pm 1\sigma$ distribution
of the expected significance, and the green (darker) exterior band shows
the $\pm 2\sigma$ range around the expectation.  CDF and D\O\ are combined,
and the foreseen sensitivity improvements have been assumed.  The integrated
luminosity is per experiment.}
\label{fig:cdfd0foreseen}
\end{figure}

\begin{figure}
\begin{center}
\includegraphics[width=0.65\textwidth]{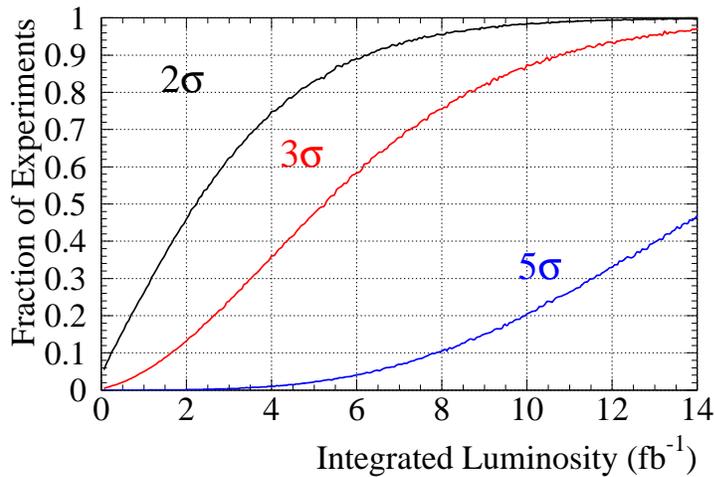}
\end{center}
\vspace{-0.5cm}
\caption{The fraction of experiments expected to make an observation of
a 115~GeV SM Higgs boson if it is truly there, as a function of the
integrated luminosity.  CDF and D\O\ are combined,
and the foreseen sensitivity improvements have been assumed.  Separate
curves are shown for the fraction of experiments observing a $\ge 2\sigma$
excess in the data, a $\ge 3\sigma$ excess, or a $\ge 5\sigma$ excess.}
\label{fig:probscdfd0foreseen}
\end{figure}

\begin{boldmath}
\subsection{The MSSM $H/h/A\rightarrow\tau^+\tau^-$ Sensitivity}
\end{boldmath}

CDF has published its search for $H/h/A\rightarrow \tau^+\tau^-$ search, using 310~pb$^{-1}$
of Run~2 collision data~\cite{Abulencia:2005kq}.   Tau pairs are selected in which one tau decays leptonically,
and the other decays semi-hadronically.  Kinematic selection requirements were designed
to separate tau pairs from $W$+jets and QCD backgrounds, in which jets are misidentified
as taus.  The dominant remaining background is $Z\rightarrow\tau^+\tau^-$ production.
In order to separate $H/h/A\rightarrow \tau^+\tau^-$ from this and other backgrounds,
the invariant mass of the visible tau decay products is formed, shown in Figure~\ref{fig:taumvis}.
The reconstructed invariant mass of Higgs boson signal events peaks near the signal mass,
with a width which grows rapidly with increasing Higgs boson signal mass.  This is
offset by the fact that the background is very small for large reconstructed masses.
The observed and expected limits on the production cross section times the decay
branching ratio to tau pairs is shown in Figure~\ref{fig:tauxsbr}.

\begin{figure}
\begin{center}
\includegraphics[width=0.65\textwidth]{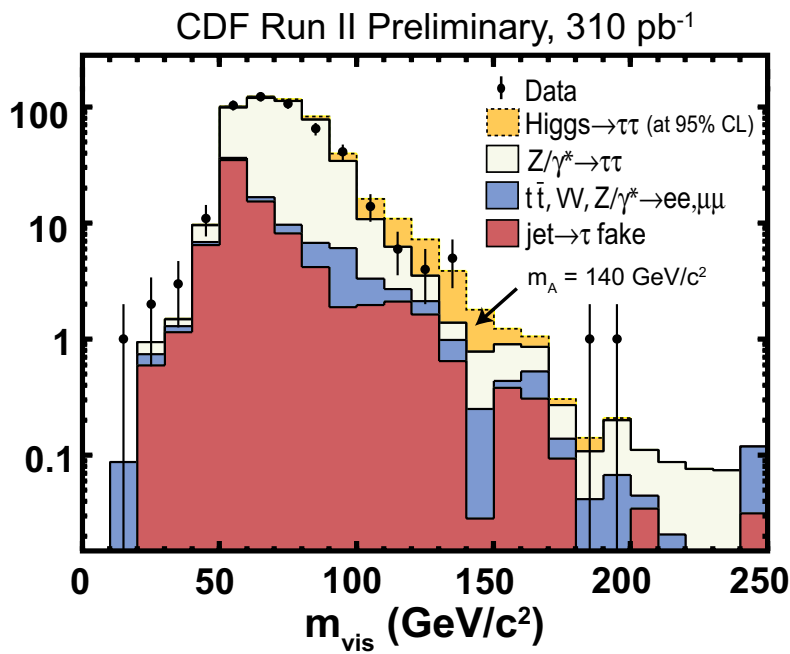}
\end{center}
\vspace{-0.5cm}
\caption{The invariant mass of the reconstructed tau decay products
in the MSSM $H\rightarrow\tau^+\tau^-$ search.  The data (points) are
compared to a sum of background predictions.  A Higgs boson signal
of mass $m_A=140$~GeV, with a production cross section just at the
exclusion threshold, is shown.}
\label{fig:taumvis}
\end{figure}

\begin{figure}
\begin{center}
\includegraphics[width=0.65\textwidth]{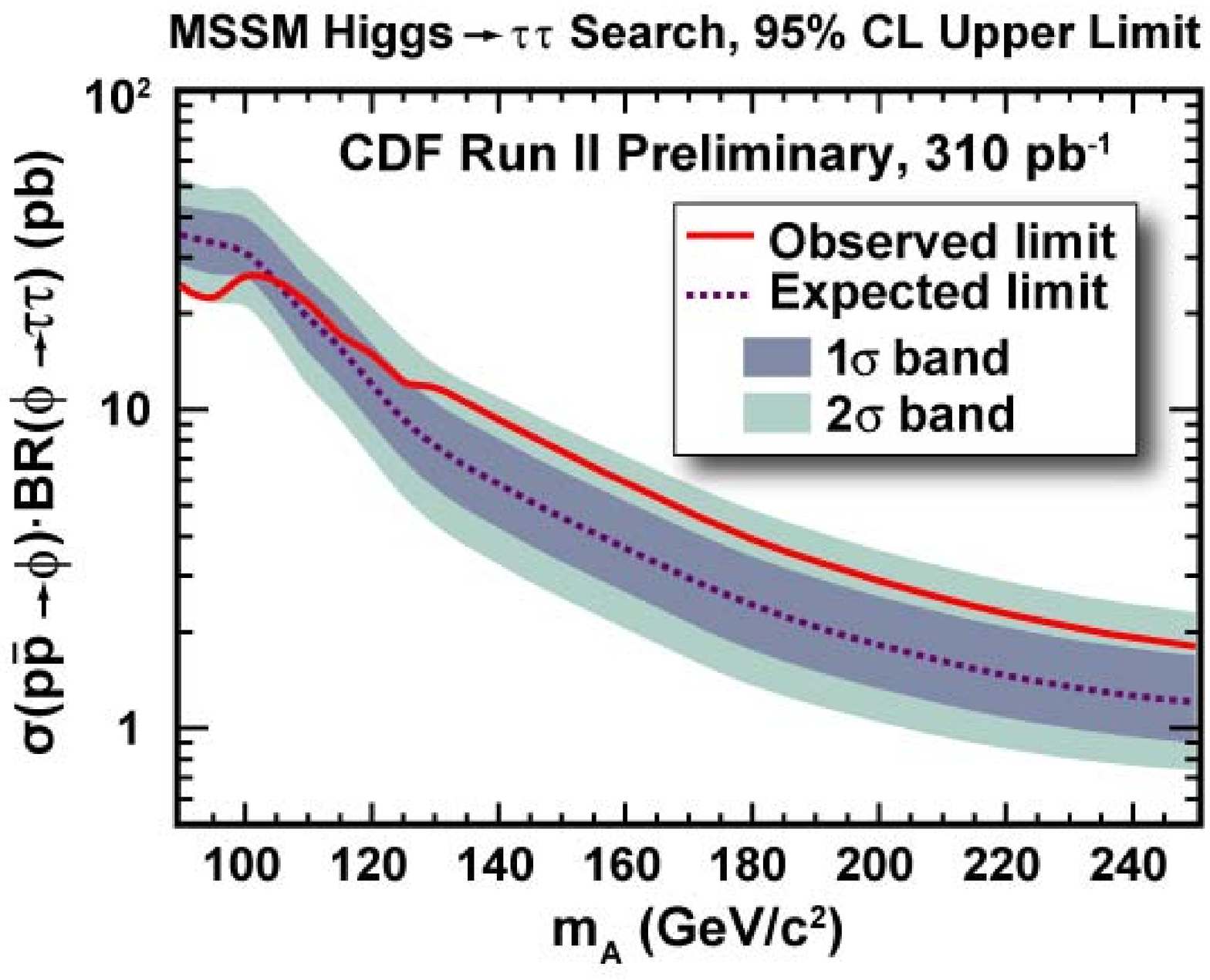}
\end{center}
\vspace{-0.5cm}
\caption{The 95\% CL limit on the production cross section times the decay branching
ratio for Higgs bosons decaying to tau pairs, using 310~pb$^{-1}$ of CDF data,
as a function of the Higgs boson mass.}
\label{fig:tauxsbr}
\end{figure}

This cross-section limit can be interpreted in the MSSM; we choose to represent
it as an exclusion in the $(m_A,\tan\beta)$ plane in the $mh-max$ and $no-mixing$ MSSM benchmark
scenarios~\cite{Carena:1999xa}.  This interpretation benefits from the fact that
for larte $\tan\beta$, two Higgs bosons (either $h$ and $A$, or $H$ and $A$),
are nearly degenerate in mass and contribute
rougly equally to the expected signal.  CDF's observed 95\% CL
limits are shown in Figure~\ref{fig:tausensitivity}, along with projected 
CDF+D\O\ combined sensitivity
contours for 2, 4, and 8~fb$^{-1}$ of data collected by both CDF and D\O.  The large
improvement in sensitivity at larger Higgs boson masses comes from the fact that the
background rate is very low for large invariant-mass tau pairs.  For a search with
a large background rate, the expected signal limit is roughly
inversely proportional to the square root of the integrated luminosity,
while for searches with very small backgrounds, the expected limit
is roughly inversely proportional to the integrated luminosity.

\begin{figure}
\begin{center}
\includegraphics[width=0.65\textwidth]{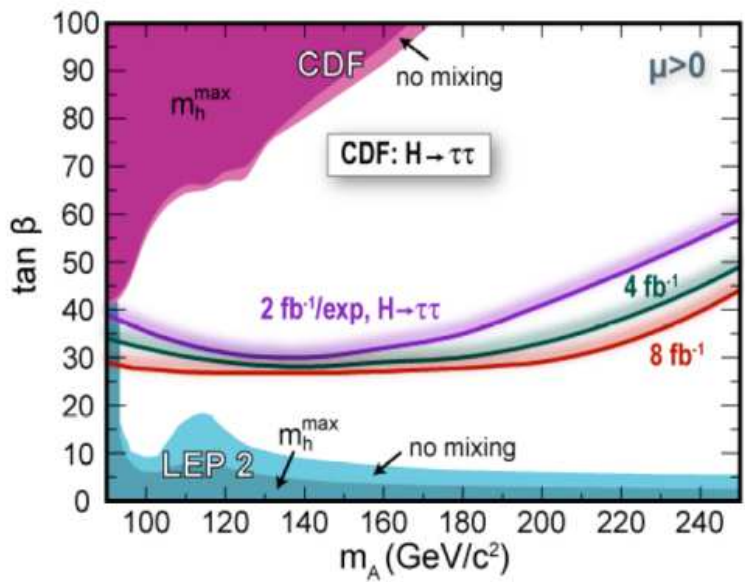}
\end{center}
\vspace{-0.5cm}
\caption{The observed 95\% CL limits in the tau channel in the $(m_A,\tan\beta)$
plane, for the $m_H$-max MSSM benchmark scenerio and also the no-mixing
benchmark scenario, using 310~pb$^{-1}$ of CDF data.  Projections are shown 
for the expected combined CDF+D\O\ 
exclusion reach for 2, 4, and 8~fb$^{-1}$ per experiment.}
\label{fig:tausensitivity}
\end{figure}


\clearpage
\section{Two-Loop EW Corrections to Higgs Production}
\textbf{Contributed by: U.~Aglietti, R.~Bonciani,  G.~Degrassi, A.~Vicini}
\vspace{0.25in}

We study the impact of the two-loop electroweak corrections
on the production of a Higgs boson via gluon-fusion in proton-proton
collisions at LHC energies.
We discuss the prescritpion to include the corrections to the hard
scattering matrix element in the calculation of the hadronic
cross-section $\sigma (p+p\to H+X)$.
Under the hypothesis of factorization of the electroweak corrections
with respect to the dominant soft and collinear QCD radiation,
we observe an increase of the total cross-section from 4 to 8\%, for
$\mh\leq160~{\rm GeV}$. This increase is comparable with the present 
QCD uncertainties originating from hard scattering matrix elements.

\subsection{Introduction}
We study the impact of the two-loop electroweak corrections
on the production of a Higgs boson via gluon-fusion in proton-proton
collisions at LHC energies.
We discuss the prescritpion to include the corrections to the hard
scattering matrix element in the calculation of the hadronic
cross-section $\sigma (p+p\to H+X)$.
Under the hypothesis of factorization of the electroweak corrections
with respect to the dominant soft and collinear QCD radiation,
we observe an increase of the total cross-section from 4 to 8\%, for
$\mh\leq160~{\rm GeV}$. This increase is comparable with the present 
QCD uncertainties originating from hard scattering matrix elements.


The Higgs boson is one of the missing ingredients of the Standard
Model and its discovery represents one of the most important physics 
goals of the LHC.
This goal will be achieved only if we can predict with high accuracy
all the production cross sections of this particle and if we understand 
in detail the different decay channels and the relative backgrounds.

At the LHC, the gluon-fusion is the dominant production mode over the entire 
range of interesting values of the mass of the Higgs particle ($100 \GeV 
\lesssim \mH \lesssim 1\TeV$). In particular, in the range $100 \GeV \lesssim \mH 
\lesssim 2 m_t$ this production mode is larger by almost one order of magnitude 
with respect to the next important mechanism, the vector boson fusion. It is,
therefore, very important to have a precise prediction of its cross section 
and a reliable estimate on the remaining theoretical accuracy.

The total cross section for the Higgs boson production by gluon fusion in the 
LO approximation was calculated in the late seventies \cite{H2gQCD0}. It is 
an ${\mathcal O}( \alpha_S^2 \, \alpha)$ 
calculation, since the Higgs couples to the 
gluons only via an heavy-quark loop (the most important contribution is the
one due to the loop of top). For what concerns the higher orders, the 
calculation of the NLO QCD corrections have 
been done in the infinite $m_t$ (mass of the top) limit in \cite{H2gQCD1},
and, with the full quark mass dependence, in \cite{QCDg2}.
Besides of the fact that the infinite $m_t$ approximation should be valid 
in the Higgs mass range $\mH \lesssim 300$ GeV, it has been noticed 
\cite{Kramer:1996iq} that this approximation works also for values of $\mH$ 
beyond the top threshold, and up to masses of ${\mathcal O}(1\TeV)$. 
The total effect of the NLO QCD corrections is the increase of the LO 
cross section by a factor 1.5--1.7, giving a residual 
renormalization/factorization scale dependence of about 30\%. The unexpected 
size of the NLO QCD radiative corrections made in such a way that the 
electroweak corrections, evaluated in the infinite $m_t$ limit 
in \cite{Djouadi:1994ge,hep-ph/9712330} and turning out to amount of less that 1\%, were totally 
neglected. The attention was driven by the evaluation of the NNLO QCD 
corrections, carried out in the infinite $m_t$ limit by several groups 
\cite{Harlander:2000mg,hep-ph/0102227,hep-ph/0102241,hep-ph/0201206,hep-ph/0207004,hep-ph/0302135}. The calculation shows a good convergence of the 
perturbative series: while the NNLO corrections are sizable, they are, 
nevertheless, smaller that the NLO ones. Moreover, the NNLO corrections 
improve the stability agaist renormalization/factorization scale 
variations. The effect due to the resummation of soft-gluon radiation has been 
included in \cite{hep-ph/0306211}, and the remaining theoretical uncertainty, due 
to higher-order QCD corrections, has been estimated to be smaller than 10\%.
Finally, several efforts were also devoted to the calculation of QCD radiative 
corrections to less inclusive quantities, such as the rapidity distribution,
recently evaluated at the NNLO \cite{hep-ph/0211141,hep-ph/0409088}, or the transverse momentum ($q_T$) 
distribution \cite{Nucl.Phys.B297.221,Phys.Lett.B211.335,Phys.Rev.D38.3475,Nucl.Phys.B339.38,Phys.Rev.D44.1415,Phys.Rev.D45.1512,Phys.Lett.B283.395,hep-ph/9902483,hep-ph/0001103,hep-ph/0002032,hep-ph/0008152,hep-ph/0108273,hep-ph/0201114,hep-ph/0209248,hep-ph/0210135,hep-ph/0304267,hep-ph/0307208,hep-ph/0309264,hep-ph/0309303,hep-ph/0309096,hep-ph/0501098,hep-ph/0501130,hep-ph/0501172}, which, in \cite{hep-ph/0302104,hep-ph/0508068}, is evaluated using the 
fixed-order perturbative results up to NLO in QCD and the 
resummation up to the NNLL.

Motivated by this accurate scenario, the NLO electroweak corrections to
the gluon fusion were again taken into account recently. In \cite{hep-ph/0404071,hep-ph/0407162} the  
contribution to the partonic cross section due to the light fermions were 
calculated. It turned out that they are sizeable. In particular, in the 
intermediate Higgs mass range, from 114 GeV up the the $2 \mW$ threshold, 
these corrections increase the LO partonic cross section by an amount of 
4--9\%. For larger values of the mass of the Higgs, $\mH > 2 \mW$,
they change sign and reduce the LO cross section; however, in this region 
the light-fermion corrections are quite small, reaching at most a -2\%. 
In \cite{hep-ph/0407249}, also the remaining electroweak corrections due to the top quark 
were calculated as a Taylor expansion in $\mH^2/(4\mW^2)$. They are valid 
for $\mH \lesssim 2 \mW$, range in which they have opposite sign with respect 
to the light-fermion corrections. However, the corrections due to the top quark
are smaller in size, reaching at most a 15\% of the light-quark ones.

The impact of the NLO electroweak corrections on the hadronic cross 
section has not been discussed yet. We present here the effect of their
inclusion in the calculation at the hadronic level.

\subsection{Inclusion of the Two-Loop Electroweak Corrections}

The partonic gluon fusion process occurs, in lowest order, via one-loop
diagrams, as the one depicted in Fig.~\ref{diagabc} (a);
\begin{figure}
\begin{center}
\vskip-10.5cm
\hspace*{-2.5cm}
\includegraphics[height=280mm,angle=0]{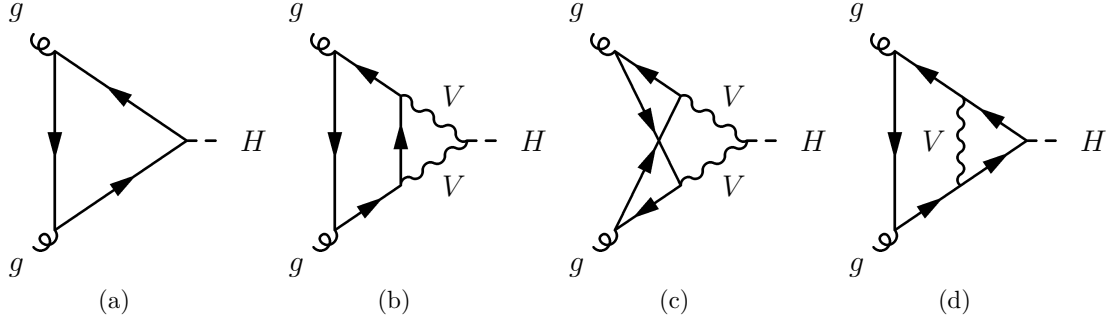}
\vskip-12cm
\caption{Lowest order (a) and generic NLO-EW (b), (c), (d) Feynman
  diagrams. The solid lines are fermions. The wavy lines are gauge
  bosons $(V=W,Z)$.}
\label{diagabc}
\end{center}
\end{figure}
in the loop run only the top and the bottom quarks, because of the
Yukawa suppression of the lighter quarks.
The NLO-EW corrections are schematically represented by the diagrams
in Figs.~\ref{diagabc} (b), (c) and (d).
In particular, in Figs.~\ref{diagabc} (b) and (c) the WWH/ZZH couplings 
avoid the Yukawa suppression, and, therefore, in these diagrams the fermionic
line represents all the possible flavours: light flavours, evaluated in 
\cite{hep-ph/0404071,hep-ph/0407162}, and top quark, evaluated in \cite{hep-ph/0407249}.
In Fig.~\ref{diagabc} (d), instead, the fermionic line can represent only the
top quark \cite{hep-ph/0407249}. 

At the hadronic level, we consider the Higgs boson production at the LHC,
and therefore in proton-proton collisions. The hadronic cross section can be 
written as:
\begin{eqnarray}
&&\sigma(p+p\to H+X) =
\sum_{a,b}\int_0^1 dx_1 dx_2 \,\,f_{a,p}(x_1,M^2)\,
f_{b,p}(x_2,M^2)\times\nonumber\\
&&~~~~~~~~~~~~~~~~~~~~~~~~~~~\times
\int_0^1 dz~ \delta\left(z-\frac{\tau_H}{x_1 x_2} \right)
~\Big(1+\delta_{EW}(\mh)\Big)\hat\sigma_{ab}(z)\nonumber\\[2mm]
&&\hat\sigma_{ab}(z)=
\hat\sigma_0\,
\left(1+K^{QCD~only}_{ab}(\alpha_s(\mu^2),\mu^2,M^2)  \right)
\label{sigmafull}
\end{eqnarray}
where the partonic processes initiated by partons $(a,b)$ are
convoluted with the corresponding parton densities
$f_{i,p}(x,M^2),~~(i=a,b)$, evaluated at a scale $M$.
The effect of the higher order QCD and EW corrections is described by
the two functions $K^{QCD-only}$ and $\delta_{EW}$, obtained by factoring 
the lowest order cross section $\hat\sigma_0$.

In the partonic cross section, QCD and EW corrections have been factorized.
This ansatz is valid up to subleading higher order corrections which
start at the 3-loop level 
(i.e. ${\cal O}(\alpha\alpha_s)$ with respect to the lowest order).
The factorization of the QCD initial state collinear divergences
holds for the hard process described by the electroweak NLO
corrections, following from general arguments of the factorization
theorems and from the universal nature of the initial state
collinear radiation.
In fact, the whole set of EW corrections
is characterized by a scale $\mw$, much harder than the one typical
of the leading collinear emission.
In addition, in the limit of light Higgs, the EW corrections can be
expanded as a Taylor series in powers of $\mh/\mw$ and the EW
corrections vertex becomes effectively pointlike.
In this regime the factorization of the QCD collinear divergences
becomes rigorous. 
For heavier Higgs masses, 
the factorization should still be valid only in leading order,
due to the modifications induced by the EW form factor.

\subsection{Numerical Results}

The hadronic proton-proton cross section has been calculated at LHC
energy, in NNLO-QCD accuracy, i.e. setting $\delta_{EW}=0$,
using the MRST2002 NNLO parton distribution functions \cite{Martin:2002dr}.
The theoretical uncertainty due to the choice of the renormalization
scale $\mu$ and of the factorization scale $M$ has been canonically estimated
by setting $M=\mu$ equal to $\mh/2,\mh,2\mh$ respectively.
The predictions, shown in Fig.~\ref{csLHC} (dotted lines),
vary by approximately $\pm 8$\% with respect to the central value.
This uncertainty is further reduced when including the effect of the
resummation of all the initial state soft gluon radiation \cite{hep-ph/0306211}.

\begin{figure}
\begin{center}
\begin{picture}(0,0)%
\includegraphics{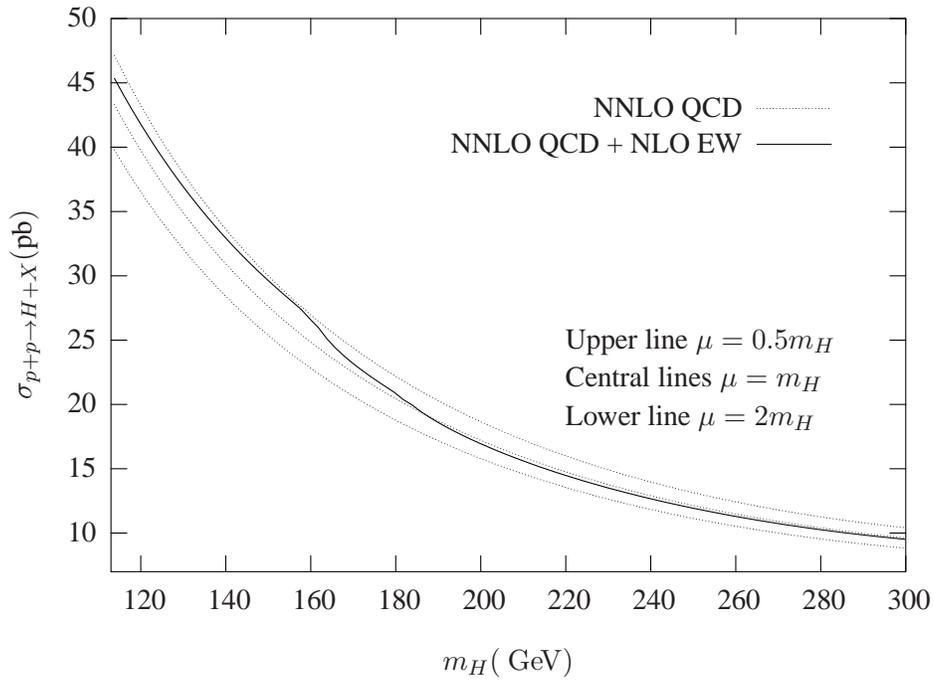}%
\end{picture}%
\setlength{\unitlength}{0.0200bp}%
\begin{picture}(18000,12960)(0,0)%
\put(1925,2707){\makebox(0,0)[r]{\strut{} 10}}%
\put(1925,3920){\makebox(0,0)[r]{\strut{} 15}}%
\put(1925,5133){\makebox(0,0)[r]{\strut{} 20}}%
\put(1925,6345){\makebox(0,0)[r]{\strut{} 25}}%
\put(1925,7558){\makebox(0,0)[r]{\strut{} 30}}%
\put(1925,8771){\makebox(0,0)[r]{\strut{} 35}}%
\put(1925,9984){\makebox(0,0)[r]{\strut{} 40}}%
\put(1925,11197){\makebox(0,0)[r]{\strut{} 45}}%
\put(1925,12410){\makebox(0,0)[r]{\strut{} 50}}%
\put(2761,1429){\makebox(0,0){\strut{} 120}}%
\put(4362,1429){\makebox(0,0){\strut{} 140}}%
\put(5964,1429){\makebox(0,0){\strut{} 160}}%
\put(7565,1429){\makebox(0,0){\strut{} 180}}%
\put(9167,1429){\makebox(0,0){\strut{} 200}}%
\put(10769,1429){\makebox(0,0){\strut{} 220}}%
\put(12370,1429){\makebox(0,0){\strut{} 240}}%
\put(13972,1429){\makebox(0,0){\strut{} 260}}%
\put(15573,1429){\makebox(0,0){\strut{} 280}}%
\put(17175,1429){\makebox(0,0){\strut{} 300}}%
\put(550,7029){\rotatebox{90}{\makebox(0,0){\strut{} $\sigma_{p+p \to H+X} (\mbox{pb})$ }}}%
\put(9687,275){\makebox(0,0){\strut{} $\mH (\GeV)$ }}%
\put(10769,6345){\makebox(0,0)[l]{\strut{}Upper line $\mu=0.5 \mH$}}%
\put(10769,5618){\makebox(0,0)[l]{\strut{}Central lines $\mu=\mH$}}%
\put(10769,4890){\makebox(0,0)[l]{\strut{}Lower line $\mu=2 \mH$}}%
\put(14097,10712){\makebox(0,0)[r]{\strut{}NNLO QCD}}%
\put(14097,10037){\makebox(0,0)[r]{\strut{}NNLO QCD + NLO EW}}%
\end{picture}
\caption{The cross section $\sigma_{p+p \to H+X}$, in pb, is plotted as a 
function of the mass of the Higgs boson, 
between 114 GeV and 300 GeV. 
The dotted lines describe the band of NNLO-QCD uncertainty, for three values
of the QCD factorization/renormalization scale $\mu=\mh/2,\mH,2\mh$.
The solid line is the NNLO-QCD $(\mu=\mh)$ with the two-loop EW corrections,
according to Eq.~(\ref{sigmafull}).
The two-loop EW corrections include also the top-quark effect, for
$\mh\leq 155$GeV, but only the light quarks contribution for larger
values of $\mh$. }
\label{csLHC}
\end{center}
\end{figure}

The two-loop electroweak corrections have been added according to 
Eq.~(\ref{sigmafull}) and setting $M=\mu=\mh$. The light fermion corrections 
can be evaluated for any choice of $\mh$, whereas the top quark contribution 
has been computed by means of a Taylor expansion and is limited to the region 
$\mh\leq 160$GeV. The hadronic cross section increases from 4 to 8\%, for 
$\mh\leq 160$ GeV. As we can observe in Fig.~\ref{csLHC}, the effect of the 
electroweak corrections is an increase of the cross section by an amount which 
is of the same order of magnitude of the NNLO-QCD theoretical uncertainty, and 
possibly larger than the uncertainty estimated after the resummation of soft 
gluon radiation. The main source of uncertainty on the hadronic cross section
remains in the accurate determination of the parton distribution functions of 
the proton.

The effect of the NLO-EW corrections is of great interest, 
because it enhances the most important Higgs production mechanism
and, in turn, affects the absolute number of events of all the Higgs
decay modes.

Following Eq.~(\ref{sigmafull}),
the NLO-EW corrections can be implemented as a simple rescaling
of the QCD hadronic cross section.
This multiplicative factor is presented in Table \ref{tabrescale} as a
function of $\mh$ and can be fitted, in the range $114\GeV
\lesssim \mH \lesssim 155\GeV$, by the following simple formula:
\begin{equation}
\delta_{EW}(\mH) = 1.00961 + 6.9904 \cdot 10^{-5} \, \mh 
+ 2.31508 \cdot 10^{-6} \, \mh^2 \, .
\end{equation}
\begin{table}[t]
\begin{center}
\begin{tabular}{|c|c||c|c||c|c||c|c|}
\hline
$\mh$ (GeV) & $\delta_{EW}$ & $\mh$ (GeV) & $\delta_{EW}$ 
& $\mh$ (GeV) & $\delta_{EW}$ & $\mh$ (GeV) & $\delta_{EW}$ \\
\hline
114 &  1.048 & 136 &  1.062 & 158 &  1.077 & 180 &  1.020\\
\hline
116 &  1.049 & 138 &  1.063 & 160 &  1.069 & 182 &  1.010\\
\hline
118 &  1.050 & 140 &  1.065 & 162 &  1.063 & 184 &  1.010\\
\hline
120 &  1.051 & 142 &  1.066 & 164 &  1.049 & 186 &  1.002\\
\hline
122 &  1.053 & 144 &  1.068 & 166 &  1.041 & 188 &  0.997\\
\hline
124 &  1.054 & 146 &  1.069 & 168 &  1.035 & 190 &  0.994\\
\hline
126 &  1.055 & 148 &  1.071 & 170 &  1.031 & 192 &  0.991\\
\hline
128 &  1.056 & 150 &  1.073 & 172 &  1.028 & 194 &  0.989\\
\hline
130 &  1.058 & 152 &  1.074 & 174 &  1.026 & 196 &  0.987\\
\hline
132 &  1.059 & 154 &  1.076 & 176 &  1.024 & 198 &  0.986\\
\hline
134 &  1.060 & 156 &  1.077 & 178 &  1.022 & 200 &  0.985 \\
\hline
\end{tabular}
\caption{Rescaling factor $\delta_{EW}$ as a function of the Higgs
  boson mass.}
\end{center}
\label{tabrescale}
\end{table}
The computation of the NLO-EW corrections to the gluon fusion process
has been described in detail in \cite{hep-ph/0404071,hep-ph/0407162,hep-ph/0407249}.
The analytical expression of the probability amplitude has been expressed in
terms of Generalized Harmonic PolyLogarithms (GHPL) \cite{hep-ph/0304028} and
has been implemented in a FORTRAN routine\footnote{available upon request 
from the authors}.
The GHPL can be evaluated numerically in several different ways: by
direct numerical integration of the basic functions, by power
expansions or by solving the associated differential equations.
We have checked that these fully independent approaches agree.

\subsection{Conclusions}
In conclusion, the calculation of the QCD corrections to the production of a 
Higgs boson via gluon-fusion has reached a very high level of accuracy; 
it is now mandatory the inclusion of the two-loop electroweak corrections, 
whose typical size for $\mh\leq 160$ GeV is larger than 5\%,
comparable or larger than the QCD uncertainty.
The main source of uncertainty on the hadronic cross section
remains in the accurate determination
of the parton distribution functions of the proton.

\subsection*{Acknowledgments}
The authors wish to thank S. Catani, D. de Florian and M. Grazzini
for allowing the use of the numerical program 
of JHEP {\bf 0105} (2001) 025 [arXiv:hep-ph/0102227], and for useful 
discussions.

This work was partly supported by the European Union under the 
contract HPRN-CT2002-00311 (EURIDICE) and by MCYT (Spain) under 
Grant FPA2004-00996, by Generalitat Valenciana (Grants GRUPOS03/013 
and GV05/015).

\clearpage
\section{Higgs Resummation}
\label{sec:bfield:resummation}
%
\textbf{Contributed by: B.J. Field}
\vspace{0.25in}

Resummation when combined with fixed-order calculations yield the most
accurate theoretical definition for differential cross-sections for
use by the experimental community to correctly determine and
unambiguously reconstruct the mass of any Higgs bosons that may exist
in nature. There has been excellent progress in recent years at
obtaining higher-order fixed-order results for all the relevant Higgs
production processes. It is therefore necessary to follow these
calculations with resummed calculations to complete our theoretical
understanding ahead of future collider data.

Resummation of processes involving both scalar and pseudoscalar Higgs
bosons have been studied extensively\cite{Catani:1989ne,
Kauffman:1991jt, Yuan:1991we, Kauffman:1991cx, Catani:1996yz,
Kramer:1996iq, Balazs:2000wv, deFlorian:2000pr, deFlorian:2001zd,
Glosser:2002gm, Berger:2002ut, Berger:2003pd, Bozzi:2003jy,
Catani:2003zt, Kulesza:2003wn, Field:2004tt, Field:2004nc,
Belyaev:2005bs, Bozzi:2005wk}. In order to understand the
small-$p_{T}$ behavior of a differential cross-section, we need to
employ the resummation formalism\cite{Collins:1981uk, Collins:1981va,
Collins:1984kg}. Resummation will smooth out any numeric instabilities
and divergencies that occur in fixed order calculations.

There are several ways to approach resummation of both total
cross-sections and differential quantities. Here we will discuss only
the differential quantities as they are more relevant for experimental
concerns. The formalism of resummation is different depending on what
kind of observable one is interested in studying. Although much of the
work of resummation has focused on inclusive production, it is
possible to study exclusive processes\cite{Kidonakis:2003tx}.

For inclusive processes where typically only one particle is produced
at lowest order, the standard formalism allows a resummed differential
cross-section to be written as an inverse Bessel transformation in
terms of an expansion in $\alpha_s$ of both universal and
process-dependent terms. For instance,
\begin{eqnarray*}
\lefteqn{\frac{d\sigma^{\textrm{resum}}}{dp_t^2 \, dy \, d\phi} = 
\sum_{a,b} \int_{x_{1,\textrm{min}}}^1 \!\!\! dx_1
\int_{x_{2,\textrm{min}}}^1 \!\!\! dx_2
\int_0^\infty db \, \frac{b}{2} J_0(bp_t)} \\
& \times f_{a/h_1}\!(x_1,b_0/b) \, f_{b/h_2}\!(x_2,b_0/b) \,
\frac{S}{Q^2} W_{ab}(x_1 x_2 S;Q,b,\phi),
\end{eqnarray*}
where we define,
\begin{eqnarray*}
\lefteqn{W_{ab}(s;Q,b,\phi) = \sum_c \int_0^1 dz_1 \int_0^1 dz_2 \,
  \bar{C}_{ca}(\alpha_s(b_0/b),z_1)} \\
& \times \bar{C}_{\bar{c}b}(\alpha_s(b_0/b),z_2) \,
\delta(Q^2-z_1 z_2 s) \, \frac{d\sigma_{\bar{c}c}^{LO}}{d\phi}
\, S_c(Q,b),
\end{eqnarray*}
and the Higgs mass is $M_\Phi^2 = Q^2$, $d\phi$ is any unintegrated
phase space of the system under consideration, and
$\hat{\sigma}^{(LO)}_{c\bar{c}}$ is the lowest order cross-section
with a $c\bar{c}$ initial state which is therefore defined at
$p_t=0$. Since this is an inclusive process, the transverse momentum
and rapidity in the differential cross-section $(p_t,y)$ are that of
the Higgs boson produced. The integration variable $b$ is the impact
parameter, $S$ is the center-of-momentum energy of the hadronic
system, and the $f_{i/A}$ are the parton distribution functions for a
parton $i$ in hadron $A$. The constant $b_0$ is written in terms of
the Euler-Mascheroni constant $\gamma_E=0.57721\ldots$ as
$b_0=2e^{-\gamma_E}$ to simplify the coming coefficient
expressions. The coefficients $C_{ab}$ are process dependent and can
be written as power series to be described below. $J_0(bp_t)$ is the
first order Bessel function. The Sudakov form factor $S_c$, which
makes the integration over the Bessel function convergent, can be
written as,
\begin{equation}
S_c(Q,b) = \exp \biggl\{
- \int^{Q^2}_{b_0^2/b^2} \frac{dq^2}{q^2} \biggl[
A_c(\alpha_s(q)) \ln\frac{Q^2}{q^2} + B_c(\alpha_s(q))
\biggr] \biggr\}.
\end{equation}
The coefficient functions $A_c$, $B_c$, and $C_{ab}$ can be written as
power series in $\alpha_s$ as
\begin{equation}
A_c(\alpha_s) = \sum_{n=1}^{\infty}
                \biggl( \frac{\alpha_s}{\pi} \biggr)^n A_c^{(n)}, \quad
B_c(\alpha_s) = \sum_{n=1}^{\infty}
                \biggl( \frac{\alpha_s}{\pi} \biggr)^n B_c^{(n)},
\end{equation}
\begin{equation}
\bar{C}_{ab}(\alpha_s,z) = \delta_{ab}\delta(1-z) +
                      \sum_{n=1}^{\infty} \biggl( \frac{\alpha_s}{\pi}
                      \biggr)^n \bar{C}_{ab}^{(n)}(z).
\end{equation}
The $A^{(1)}_c$, $A^{(2)}_c$, and $B^{(1)}_c$ coefficients have been
shown to be universal.

Although this particular method of handling resummed differential
cross-sections is useful, it ignores several of the more interesting
channels where a Higgs boson would be produced in association with
another particle, and therefore has nothing to say about the other
particles in the process. There has been interest in a Higgs bosons
produced in association with weak vector bosons, light and heavy
quarks, and combinations of these particles as can easily be seen in
the cross-sections that have been compiled for this workshop. In
particular, there is continuing theoretical\cite{Campbell:2002zm,
Maltoni:2003pn, Dawson:2003kb, Dittmaier:2003ej, Campbell:2004pu,
Dawson:2004sh, Maltoni:2005wd, Dawson:2005vi} and
experimental\cite{Abazov:2005yr} interest in a Higgs produced in
association with bottom quarks, particularly in supersymmetric models
where bottom quarks can play a role equal to or greater than top
quarks in Higgs processes.

To understand resummed processes for a more general particle
configuration can be calculated with a different
mechanism\cite{Kidonakis:2003tx} which can be used for inclusive,
exclusive, and the resummation of pair-invariant-mass quantities. This
formalism is setup as a unified approach that allows one to perform
the resummation based on the color flow of the chosen process as well
as several previously calculated quantities.

If we define a generic plus-distribution related object in terms of
the variable that becomes soft $(s_4)$ or zero $(1-z)$ on threshold as
\begin{equation}
\mathcal{D}_l(s_4) \equiv 
\biggl[ \frac{\ln^l(s_4/M^2)}{s_4}
\biggr]_+, \qquad \textrm{or}, \qquad
\mathcal{D}_l(z) \equiv 
\biggl[ \frac{\ln^l(1-z)}{1-z}
\biggr]_+,
\end{equation}
then at next-to-leading-log (NLL) accuracy we can define a total or
differential cross-section, where we generically call the threshold
variable $x_{\textrm{th}}$, as
\begin{equation}
d\hat{\sigma} = d\sigma^{\textrm{B}} \frac{\alpha_s}{\pi}
\biggl\{ c_3 \mathcal{D}_1(x_\textrm{th}) 
       + c_2 \mathcal{D}_0(x_\textrm{th})
       + c_1        \delta(x_\textrm{th})
\biggr\},
\end{equation}
where $d\hat{\sigma}$ can be either a total or differential quantity
of interest and $d\sigma^{\textrm{B}}$ is the total or differential
Born cross-section. There exists a similar equation of the NNLL
corrections that will not be reproduced here but can be found in the
literature\cite{Kidonakis:2003tx}. The coefficients $c_i$ can be
calculated in terms of the color Casimir invariants of the partons
involved in the process, the soft anomalous dimension matrix, and the
kinematic invariants of the process. At higher orders, the expressions
become more complicated, but straightforward to calculate in a unified
manner for several different processes.

The results of the Higgs calculations in the literature tells us
several qualitative facts about resummed Higgs processes. First, the
resummed quantities are finite at small-$p_T$ and removed the
fixed-order divergencies. We also find smaller scale uncertainties at
higher-orders as expected. 

The two primary inclusive Higgs processes that have been studied are
$gg \rightarrow \Phi$ and $b\bar{b} \rightarrow \Phi$, where $\Phi$ is
generically any Higgs boson of interest. Some sample resummation
calculation for inclusive Higgs from Refs~\cite{Field:2004tt,
Field:2004nc} are shown in Figure~\ref{fig:bfield:resum}.

Aside from the usual observations for these differential
cross-sections, we can see that the total uncertainty in the magnitude
of the cross-section at the peak, from parton distribution functions
and a scale variation by a factor of two, is approximately
$35$\%. This level of theoretical uncertainty is on par with similar
fixed order calculations.

\begin{figure}[!ht]
  \begin{center}
  \begin{tabular}{cc}
    \includegraphics[width=0.5\textwidth]{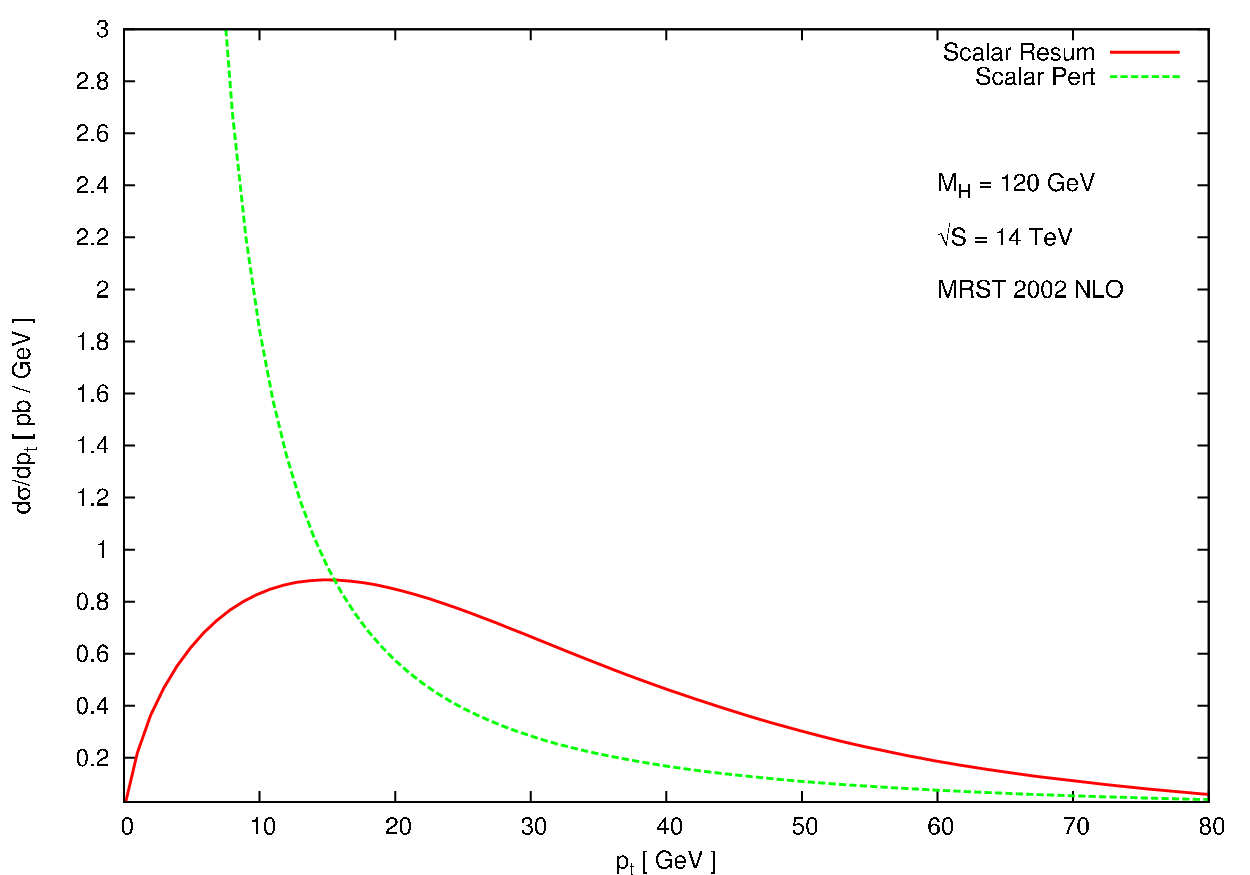} &
    \includegraphics[width=0.5\textwidth]{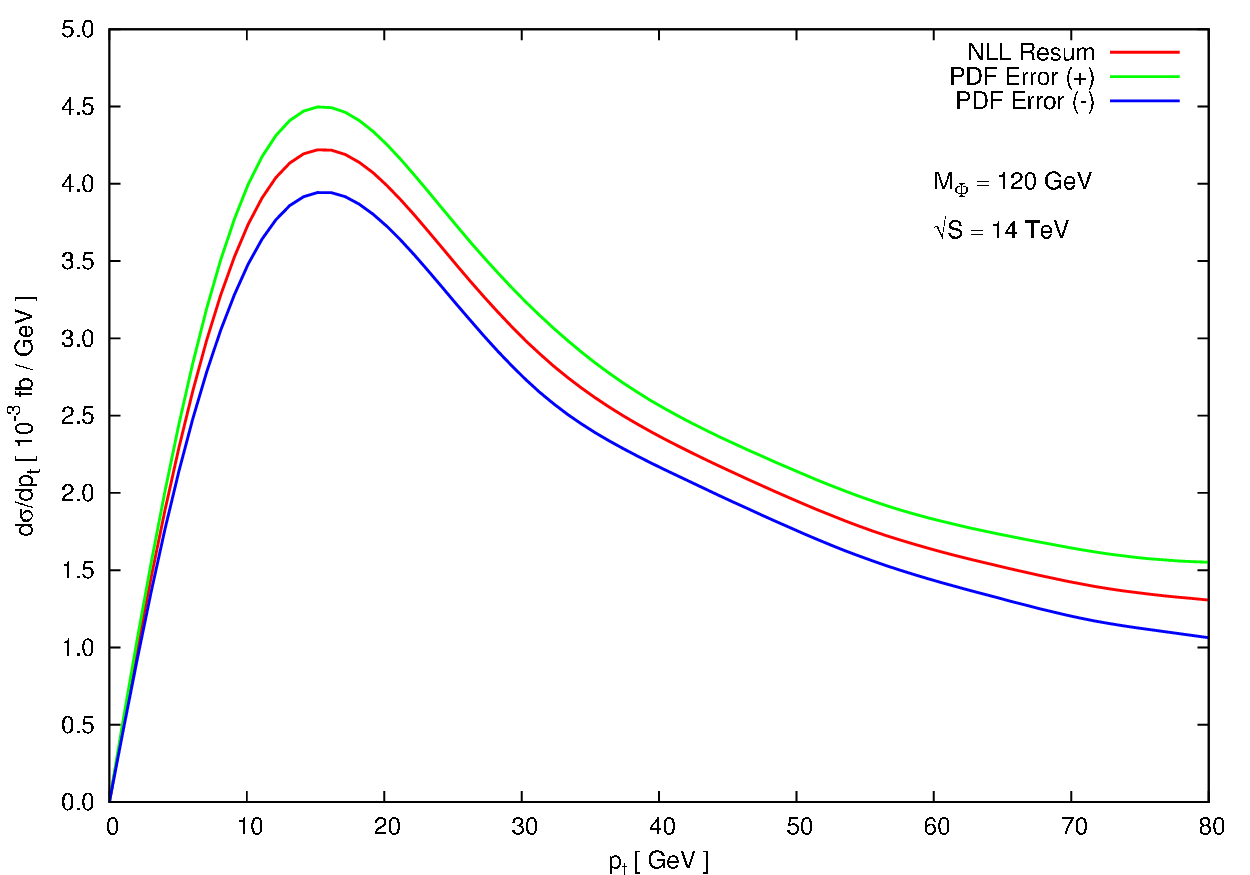} \\
    (a) & (b)
  \end{tabular}
  \end{center}
  \caption{Figure~\ref{fig:bfield:resum}a shows the transverse
momentum spectrum for a scalar Higgs boson produced via gluon-fusion
at the LHC for $|y| \leq 2.5$. The resummed curve is the NLL
resummation, and the perturbative curve is the NLO fixed order
calculation. The NLO fixed order calculation diverges in the negative
direction at small $p_t$. This piece of the differential cross-section
is not shown for clarity. These two curves cross at approximately
$p_t=100$~GeV/c and stay very close
thereafter. Figure~\ref{fig:bfield:resum}b shows the errors associated
with the CTEQ6.1M PDF set for the inclusive resummation $b\bar{b}
\rightarrow H$. The variation is approximately $8 - 12$\%.}

  \label{fig:bfield:resum}
\end{figure}

Beyond the inclusive processes, there has been an excellent
experimental use of tagged bottom jets to constrain the
$(M_A,\tan\beta)$ plane in supersymmetric models\cite{Abazov:2005yr},
where $M_A$ is the pseudoscalar mass and $\tan\beta$ is the ratio of
the vacuum expectation values of the up- and down-quark
sectors. However, the fixed-order results for the Higgs plus bottom
quark process show a numeric instability in the region of the cut on
the bottom quark\cite{Dawson:2005vi} when calculated in the five
flavor number scheme (5FNS).

Using the unified approach for the resummation of exclusive
quantities, the instabilities that occurred in the fixed-order
calculation can be smoothed out in the region around the cut in the
bottom quark transverse momentum in a fixed-order calculation become
well behaved. However, it is important to understand that the
exclusive resummation formalism is built on $2 \rightarrow 2$
kinematics and there is no way to add a cut in the transverse momentum
of one of the outgoing particles without setting the differential
cross-section below that point to zero. However, one could use the
smooth resummed calculation to further constrain the $(M_A,\tan\beta)$
limits as the transverse momentum cut on the bottom jet is further
pushed down into the region where the fixed order calculation becomes
increasingly unreliable. This investigation is currently underway.


\clearpage
\section{Hadronic Higgs Production with Heavy Quarks at the Tevatron and 
the LHC}
\label{sec:djrw}
\textbf{Contributed by: S.~Dawson, C.B.~Jackson, L.~Reina, D.~Wackeroth}

\vspace{0.25in}

A light Higgs boson is preferred by precision fits of the Standard
Model (SM) and also theoretically required by the Minimal
Supersymmetric extension of the Standard Model (MSSM).  The production
of a Higgs boson in association with a heavy quark and antiquark pair,
both $t\bar{t}$ and $b\bar{b}$, at the Tevatron and the Large Hadron
Collider (LHC) will be sensitive to the Higgs-fermion couplings and
can help discriminate between models.

The associated production of a Higgs boson with a pair of $t\bar{t}$
quarks has a distinctive signature and can give a direct measurement
of the top quark Yukawa coupling.  This process is probably not
observable at the Tevatron, but will be a discovery channel at the LHC
for $M_h < 130$~GeV.  The associated production of a Higgs boson with
a pair of $b\bar{b}$ quarks has a small cross section in the Standard
Model, and can be used to test the hypothesis of enhanced bottom quark
Yukawa couplings in the MSSM with large values of $\tan\beta$. Both
the Tevatron and the LHC will be able to search for enhanced
$b\bar{b}h$ production, looking for a final state containing no bottom
quarks (inclusive production), one bottom quark (semi-inclusive
production) or two bottom quarks (exclusive production).

The rates for $t {\overline t} h$ production at the Tevatron and the
LHC have been calculated at NLO QCD several years
ago\cite{Reina:2001sf,Beenakker:2001rj,Reina:2001bc,Beenakker:2002nc,
Dawson:2002tg,Dawson:2003zu}.  The theoretical predictions for
$b\bar{b}h$ production at hadron colliders involve several subtle
issues, and depend on the number of bottom quarks identified in the
final state. In the case of no or only one tagged bottom quark there
are two approaches available for calculating the cross sections for
$b\bar{b}h$ production, called the four flavor number schemes
(4FNS)\cite{Dittmaier:2003ej,Dawson:2004sh} and five flavor number
scheme (5FNS)\cite{Maltoni:2003pn}.  The main difference between these
two approaches is that the 4FNS is a fixed-order calculation of QCD
corrections to the $gg$ and $q\bar{q}$-induced $b\bar{b}h$ production
processes, while in the 5FNS the leading processes arise from $bg$
($\bar{b}g$) and $b\bar{b}$ initial states and large collinear
logarithms are resummed using a pertubatively defined bottom quark
Parton Distribution Function (PDF). Very good agreement is found for
the NLO QCD corrected cross sections for $b\bar b$ Higgs associated
production when the two schemes are
compared\cite{Dawson:2005vi,Campbell:2004pu}.

In the following sections, we present numerical results at NLO QCD for
$t {\overline t} h$ and $b {\overline b} h$ production at the Tevatron
and the LHC.  If not stated otherwise, numerical results have been
obtained in the 4FNS. We emphasize theoretical uncertainties from
scale and PDF uncertainties and also present differential cross
sections at NLO for $b {\overline b} h$ production in the case when
two $b$ quarks are tagged.

\subsection{Results for $t\bar{t}h$ Production}
\label{sec:djrw_tth}
\unboldmath 
The observation of a $t\bar{t}h$ final state will allow for the
measurement of the $t\bar{t}h$ Yukawa coupling.  If
$M_{h}\!\le\!130$~GeV, $pp\to t\bar{t}h$ is an important discovery
channel for a SM-like Higgs boson at the LHC
($\sqrt{s}\!=\!14$~TeV)~\cite{Beneke:2000hk,Drollinger:2001yy}.  Given
the statistics expected at the LHC, $pp\to t\bar{t}h$, with $h\to
b\bar{b},\tau^+\tau^-,W^+W^-,\gamma\gamma$ will be instrumental for
the determination of the couplings of the Higgs boson.  Precisions of
the order of 10-15\% on the measurement of the top quark Yukawa
coupling can be obtained with integrated luminosities of 100~fb$^{-1}$
per detector\cite{Zeppenfeld:2000td,Belyaev:2002ua,Maltoni:2002jr,Duhrssen:2004uu}.

\begin{figure}[t]
\begin{center}
\includegraphics[bb=150 500 430 700,scale=0.65]{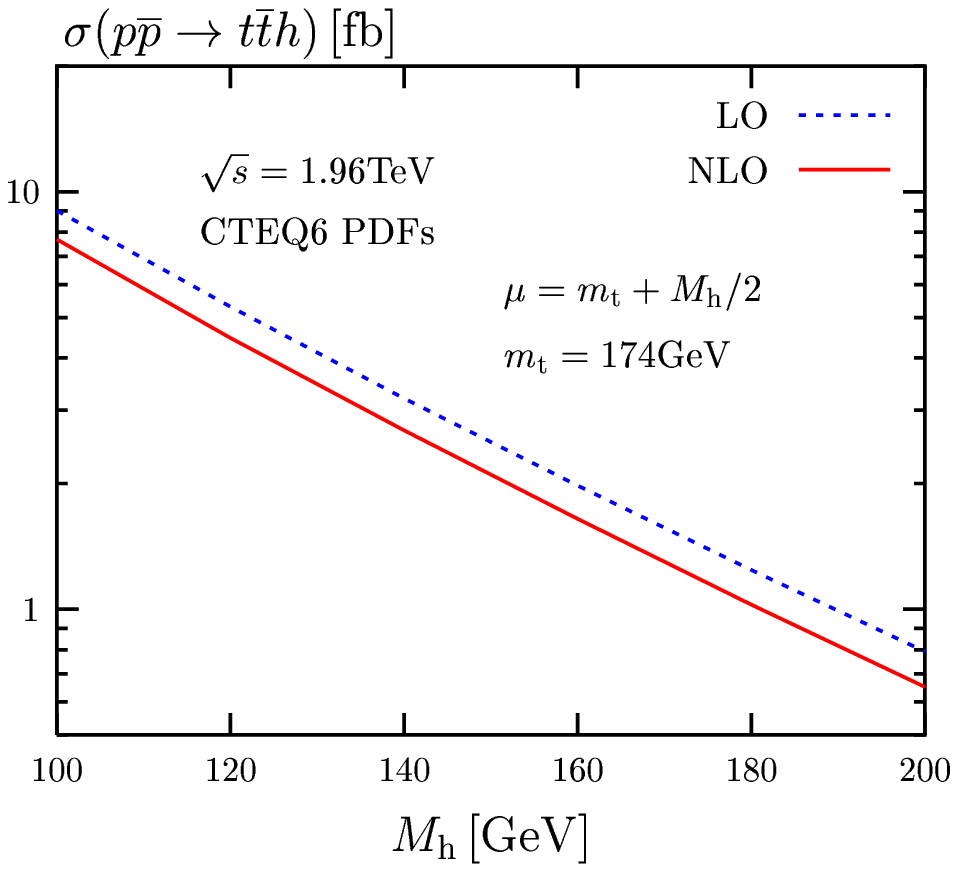} 
\includegraphics[bb=150 500 430 700,scale=0.65]{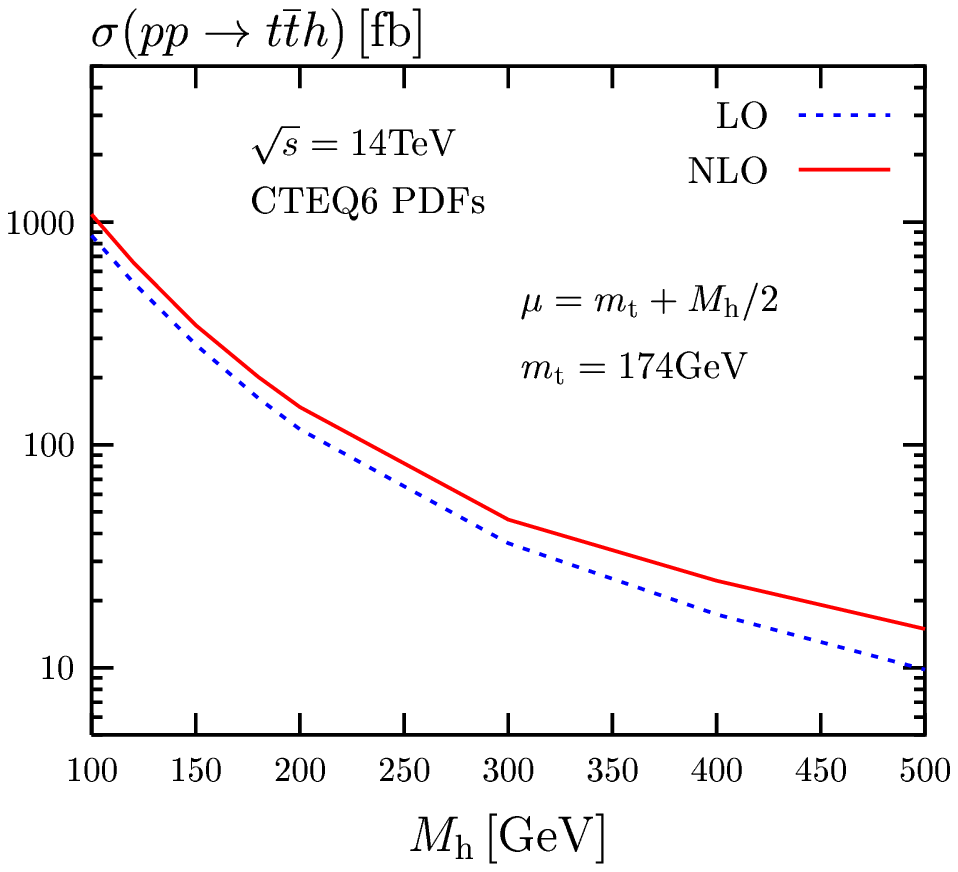} 
\vspace{0.5truecm}  
\caption[]{Total LO and NLO cross sections for $pp,p\overline{p}\to t\bar{t}h$
as functions of $M_h$, at $\sqrt{s} \!=\!1.96$~TeV and $\sqrt{s}
\!=\!14$~TeV, for $\mu\!=m_t+M_h/2$.}
\label{fg:tth_mhdep}
\end{center}
\end{figure}
\begin{figure}[t]
\begin{center}
\begin{tabular}{rl}
\includegraphics[scale=0.4]{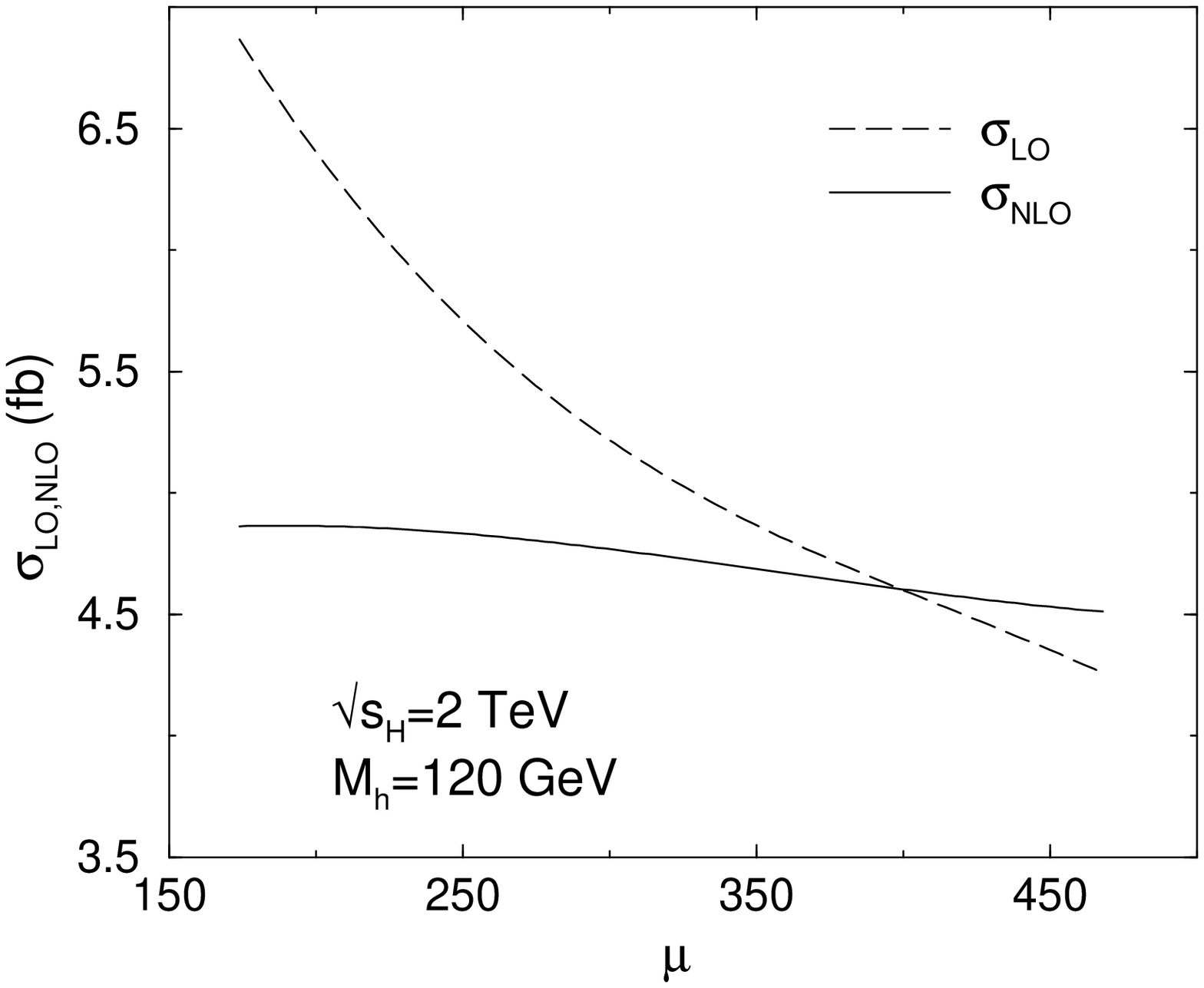} &
\includegraphics[scale=0.4]{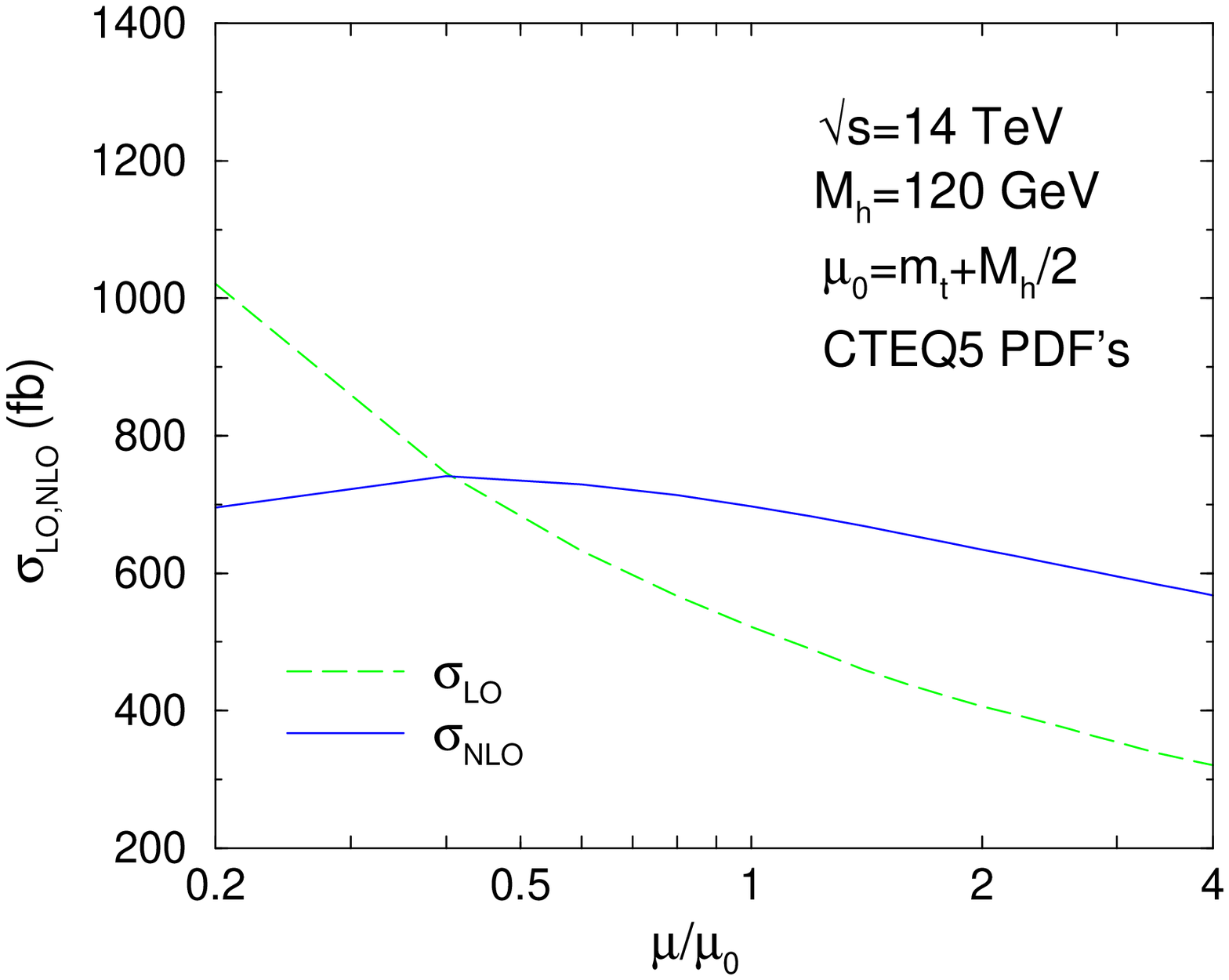}
\end{tabular} 
\caption[]{Dependence of $\sigma_{\scriptscriptstyle LO,NLO}(pp,p\bar p\to t\bar{t}h)$ on 
  the renormalization/factorization scale $\mu$, at $\sqrt{s}\!=\!2$~TeV (l.h.s.)and
$\sqrt{s}\!=\!14$~TeV (r.h.s.), for $M_h\!=\!120$ GeV.}
\label{fg:tth_mudep}
\end{center}
\end{figure}

The impact of NLO QCD corrections on the total cross section for
$p\overline{p},pp\to t\bar{t}h$ production in the Standard Model is
illustrated in
Fig.~\ref{fg:tth_mhdep}\cite{Dawson:2003zu,Dawson:2002tg,Reina:2001sf,Reina:2001bc}
and Fig.~\ref{fg:tth_mudep}\cite{Dawson:2003zu,Reina:2001bc}. The
dependence of the total cross sections on the renormalization and
factorization scales is strongly reduced at NLO as shown in
Fig.~\ref{fg:tth_mudep}.  The numerical results at NLO are obtained
using CTEQ4M (Fig.~\ref{fg:tth_mudep} (l.h.s.)), CTEQ5M
(Fig.~\ref{fg:tth_mudep} (r.h.s.)), and CTEQ6M
(Fig.~\ref{fg:tth_mhdep}) parton distribution functions.  The NLO
cross section is evaluated using the 2-loop evolution of
$\alpha_s(\mu)$ with $\alpha_s^{NLO}(M_Z)=0.116$
(Fig.~\ref{fg:tth_mudep} (l.h.s.)) and $\alpha_s^{NLO}(M_Z)=0.118$
(Fig.~\ref{fg:tth_mudep} (r.h.s.)) and Fig.~\ref{fg:tth_mhdep}), and
$m_t=174$~GeV.  The renormalization/factorization scale dependence,
uncertainty on the PDFs, and the error on the top quark pole mass,
$m_t$, are estimated to give a 15-20$\%$ uncertainty.

\subsection{Results for $b\bar{b}h$ Production}
\label{sec:djrw_bbh}

The $b\bar bh$ production processes are only relevant discovery modes
in the MSSM with large $\tan\beta$. To a good approximation, the
predictions for the MSSM rates can easily be derived from the Standard
Model results by rescaling the Yukawa couplings\cite{Dawson:2005vi}.
The dominant MSSM radiative correction to $b\bar bh$ production can be
taken into account by including the MSSM corrections to the $b\bar bh$
vertex only, i.e. by replacing the tree level Yukawa couplings by the
radiative corrected ones. We follow the treatment of the program {\sc
FeynHiggs}~\cite{Hahn:2005cu,Hahn:2004td} and take into account the
leading, $\tan\beta$ enhanced, radiative corrections that are
generated by gluino-sbottom and chargino-stop loops.  For large
$\tan\beta$, the bottom quark Yukawa coupling is enhanced and the top
quark Yukawa coupling coupling is strongly suppressed, resulting in a
MSSM $b\bar bh$ cross section that is about three orders of magnitude
larger than the Standard Model cross section.  For the Tevatron, we
calculate the production rates for the lightest MSSM Higgs boson,
$h^0$, while for the LHC we consider the rate for the heavier neutral
Higgs boson, $H^0$.\footnote{We assume $M_{SUSY}=1$~TeV,
$M_{\tilde{g}} =1$~TeV, $A_b=A_t=2$~TeV ($h^0$), $A_b=A_t=25$~GeV
($H^0$), $\mu=M_2=200$~GeV ($h^0$), and $\mu=M_2=1$~TeV ($H^0$).  For
$M_{h^0}=120$~GeV, the $bbh^0$ coupling is enhanced by a factor of
$33$ relative to the SM coupling, while for $M_{H^0}$ between $200$
and $800$~GeV, the $bbH^0$ coupling is enhanced by a factor of $27$
relative to the SM coupling.}

%
In the numerical evaluation of cross sections for the exclusive and
semi-inclusive channels ($b{\overline b}h$ and $bh+\bar{b}h$
production), it is required that the final state bottom quarks have
$p_T\!>\!20~$GeV and pseudorapidity $\mid\!\eta\!\mid<\!2.0$ for the
Tevatron and $\mid\!\eta\!\mid<\!2.5$ for the LHC. In the NLO real
gluon emission contributions, the final state gluon and bottom quarks
are considered as separate particles only if their separation in the
pseudorapidity-azimuthal angle plane, $\Delta
R\!=\!\sqrt{(\Delta\eta)^2+(\Delta\phi)^2}$, is larger than $0.4$. For
smaller values of $\Delta R$, the four momentum vectors of the two
particles are combined into an effective bottom/anti-bottom quark
momentum four-vector.

If not stated otherwise, the numerical results at NLO are obtained
using CTEQ6M PDFs, the 2-loop evolution of $\alpha_s(\mu)$ with
$\alpha_s^{NLO}(M_Z)=0.118$, and the $\overline{MS}$ renormalization
scheme for the bottom quark mass and Yukawa coupling with 2-loop
renormalization group improved $\overline{MS}$ masses. The bottom quark
pole mass is chosen to be $m_b=4.62$~GeV. 

\subsection{Total Cross Sections for  $b\bar bh$ Production}
\label{subsec:djrw_0btag}
We present total cross section results at NLO in the 4FNS in
Fig.~\ref{fg:tot} for associated $b\bar b$ Higgs production in the
MSSM with $\tan\beta=40$.  The bands represent the theoretical
uncertainty due to the residual scale dependence. They have been
obtained by varying the renormalization ($\mu_r$) and factorization
($\mu_f$) scales independently from $\mu_0/4$ to $\mu_0$, where
$\mu_0\!=\!m_b+M_h/2$.

If the outgoing bottom quarks cannot be observed then the dominant
MSSM Higgs production process at large $\tan\beta$ is $gg\rightarrow
(b\bar{b})h$ (the curve labelled '0 b').  The inclusive cross section
is experimentally relevant only if the Higgs boson can be detected
above the background without tagging bottom quarks.  At the LHC, this
process can be identified at large $\tan\beta$ by the decays to
$\mu^+\mu^-$ and $\tau^+\tau^-$ for the heavy Higgs bosons, $H^0$ and
$A^0$, of the MSSM.  At the Tevatron this process, with
$h^0\rightarrow \tau^+\tau^-$, has been used to search for the neutral
MSSM Higgs boson.
If a single bottom quark is tagged then the final state is $bh$ or
$\bar{b}h$ (the curve labelled '1 b'). Although requiring a $b$ quark in
the final state significantly reduces the rate, it also reduces the
background.  A recent Tevatron study~\cite{Abazov:2005yr} used the search
for neutral MSSM Higgs bosons in events with three bottom quarks in
the final state ($bh^0+\bar{b}h^0$ production with $h^0\rightarrow
b\bar{b}$) to impose limits on the $\tan\beta$ and $M_{A^0}$ parameter
space.
 
Finally, we show the fully exclusive cross sections for $b\bar{b}h$
production, where both the outgoing $b$ and $\bar{b}$ quarks are
identified (the curve labelled '2 b').  The exclusive measurement
corresponds to the smallest cross section, but it also has a
significantly reduced background. Moreover, both the exclusive and
semi-inclusive $b\bar bh$ production modes are the only ones that can
unambiguously measure the bottom quark Yukawa coupling.

\begin{figure}[btp]
\begin{center}
\begin{tabular}{rl}
\hskip 2.2in
\includegraphics[bb=90 700 524 13,scale=0.25,angle=-90]{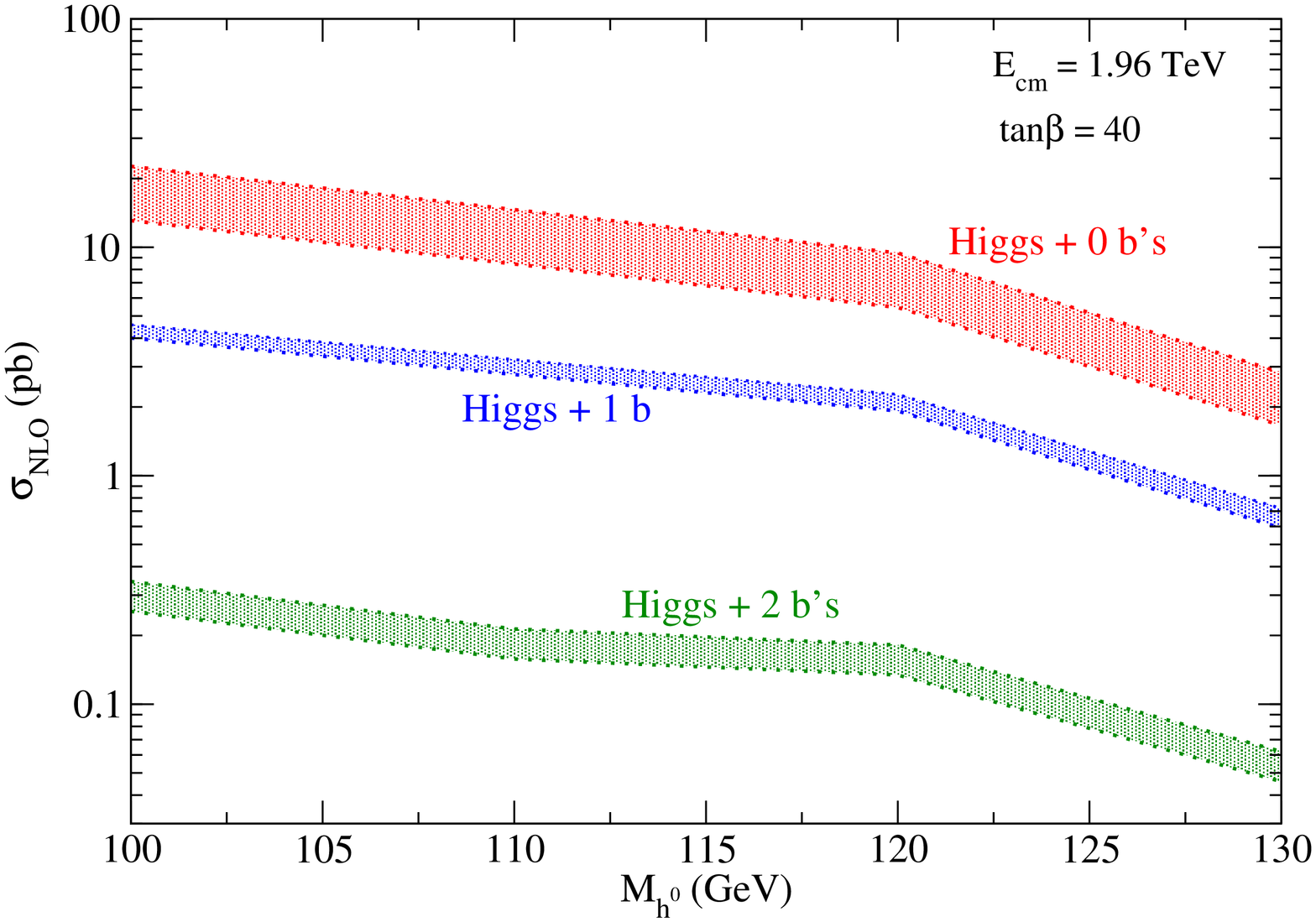} & 
\hskip 3in
\includegraphics[bb=90 700 524 12 ,scale=0.25,angle=-90]{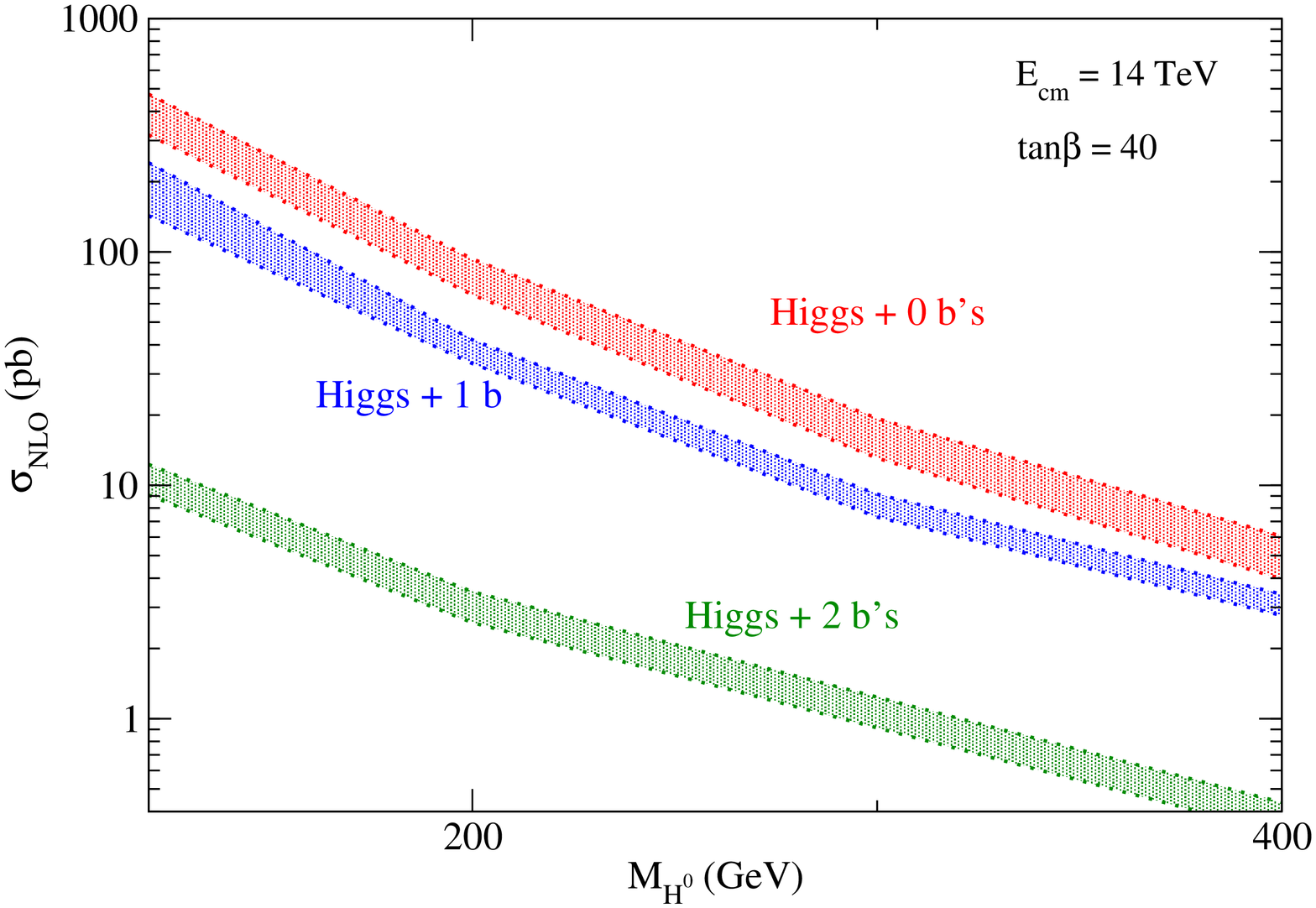}
\end{tabular}
\vspace*{8pt}
\caption[]{Total cross sections for $pp,p\bar p\rightarrow b\bar{b}h$
  in the MSSM in the 4FNS at NLO for the Tevatron and the LHC in 
the MSSM with $\tan\beta=40$ and with 0,1 or 2 $b$
quarks identified.  The Tevatron (LHC)  plot is for the 
lightest (heaviest) neutral Higgs boson, $h^0$ ($H^0$).
  The error bands have been obtained by varying
  the renormalization and factorization scales as
  described in the text.}
\label{fg:tot}
\end{center}
\end{figure}

\subsection{Differential Cross Sections for $b\bar bh $ Production}

In assessing the impact of the NLO corrections it is particularly
interesting to study the kinematic distributions.  In
Figs.~\ref{fg:bbh_ptrel} and \ref{fg:bbh_etarel} we illustrate the
impact of NLO QCD corrections on the transverse momentum and
pseudorapidity distribution of the SM Higgs boson and the bottom quark
by showing the relative correction, $d\sigma_{NLO}/d\sigma_{LO}-1$ (in
percent) for the exclusive case ($b {\overline b} h$ where both $b$
quarks are observed).  For the renormalization/factorization scale we
choose $\mu=2 \mu_0$ at the Tevatron and $\mu=4 \mu_0$ at the LHC, with
$\mu_0=m_b+M_h/2$, and use the CTEQ5 set of PDFs.  As
can be seen, the NLO QCD corrections can considerably affect the shape
of kinematic distributions, and their effect cannot be obtained from
simply rescaling the LO distributions with a K-factor of $\sigma_{\scriptscriptstyle
NLO}/\sigma_{\scriptscriptstyle LO}\!=\!1.38\pm 0.02$ (Tevatron, $\mu\!=\!2\mu_0$)
and $\sigma_{\scriptscriptstyle NLO}/\sigma_{\scriptscriptstyle LO}\!=\!1.11\pm 0.03$ (LHC,
$\mu\!=\!4\mu_0$).\footnote{The kinematic distributions have been
calculated within the Standard Model and using the on-shell scheme for
the definition of the $b$ quark mass, but we see a similar behavior
when using the $\overline{MS}$ bottom quark Yukawa coupling.}

\begin{figure}[t]
\begin{center}
\includegraphics[bb=150 500 430 700,scale=0.5]{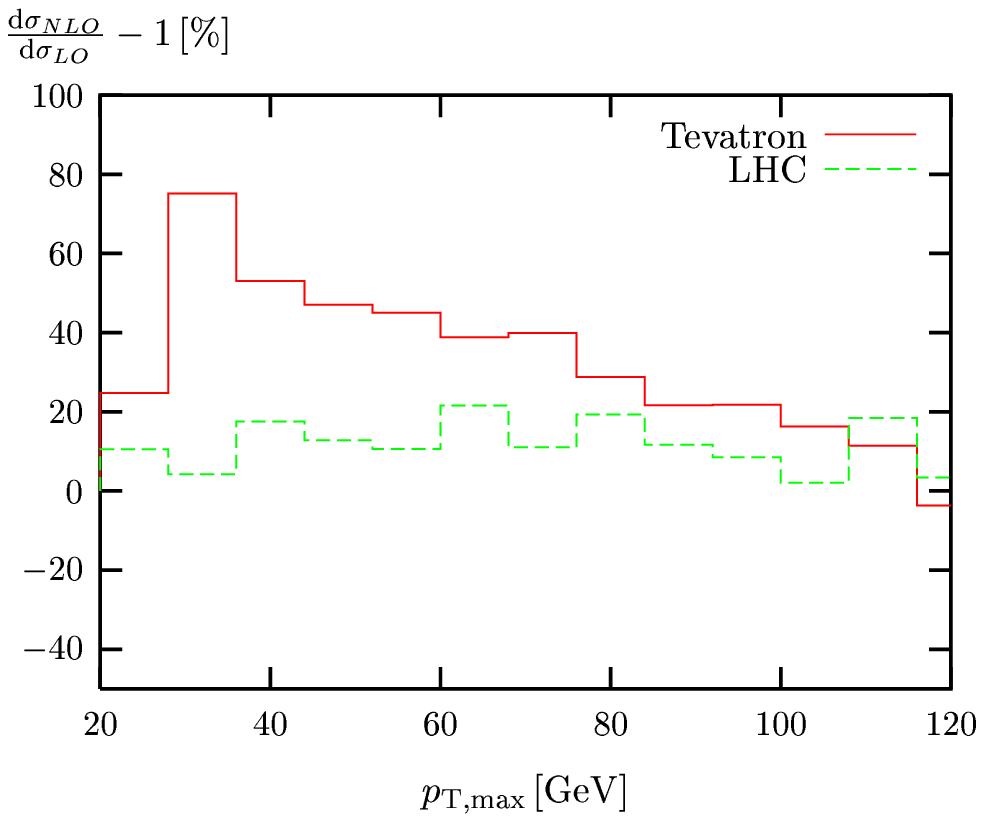} 
\includegraphics[bb=150 500 430 700,scale=0.5]{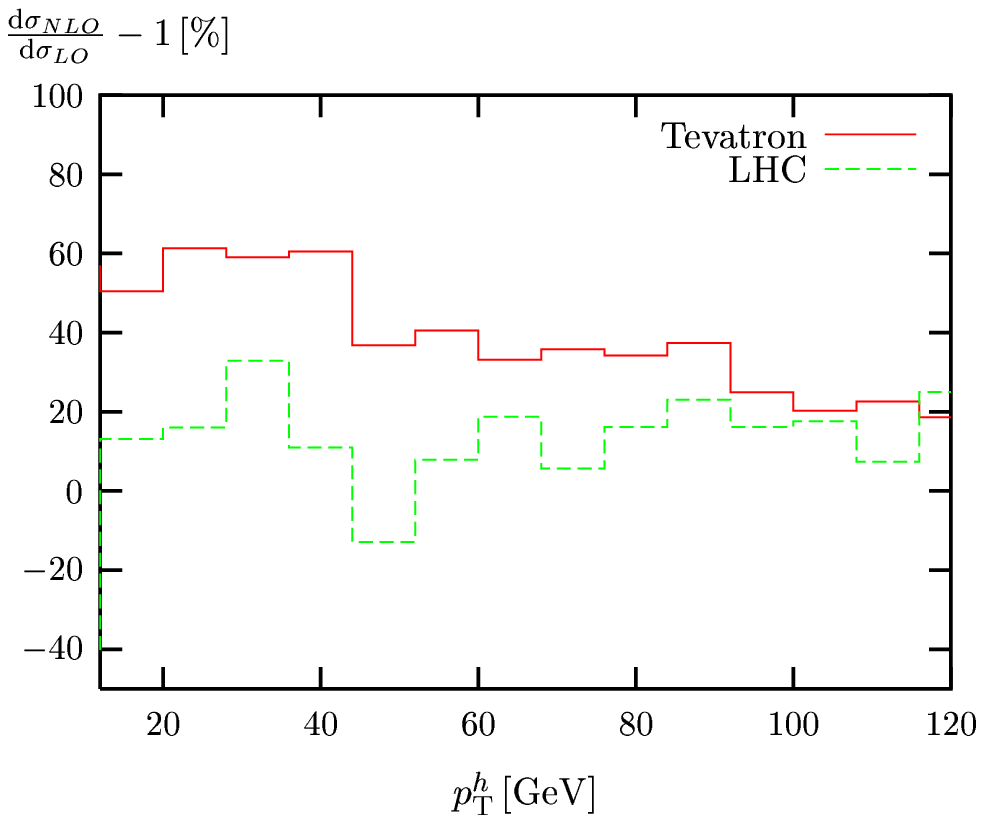} 
\caption[]{The relative corrections $d\sigma_{\scriptscriptstyle NLO}/d \sigma_{\scriptscriptstyle
LO}-1$ for the $p_T$ distribution of the bottom or anti-bottom quark
with the largest $p_T$ ($p_{T,max}$) (left) and of the SM Higgs boson
($p_{T}^h$) (right) to $b\bar b h$ production in the SM at the
Tevatron (with $\sqrt{s}\!=\!2$~TeV and $\mu\!=\!2 \mu_0$) and the LHC
(with $\sqrt{s}\!=\!14$~TeV and $\mu\!=\!4 \mu_0$).}
\label{fg:bbh_ptrel}
\end{center}
\end{figure}

\begin{figure}[t]
\begin{center}
\includegraphics[bb=150 500 430 700,scale=0.5]{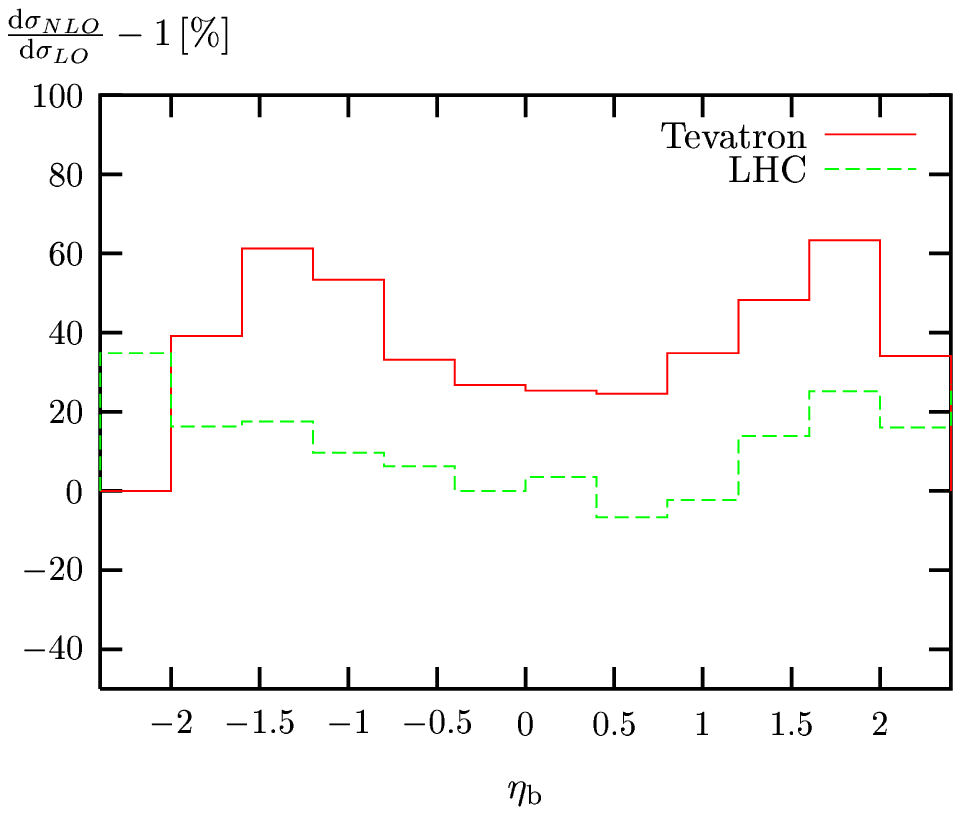} 
\includegraphics[bb=150 500 430 700,scale=0.5]{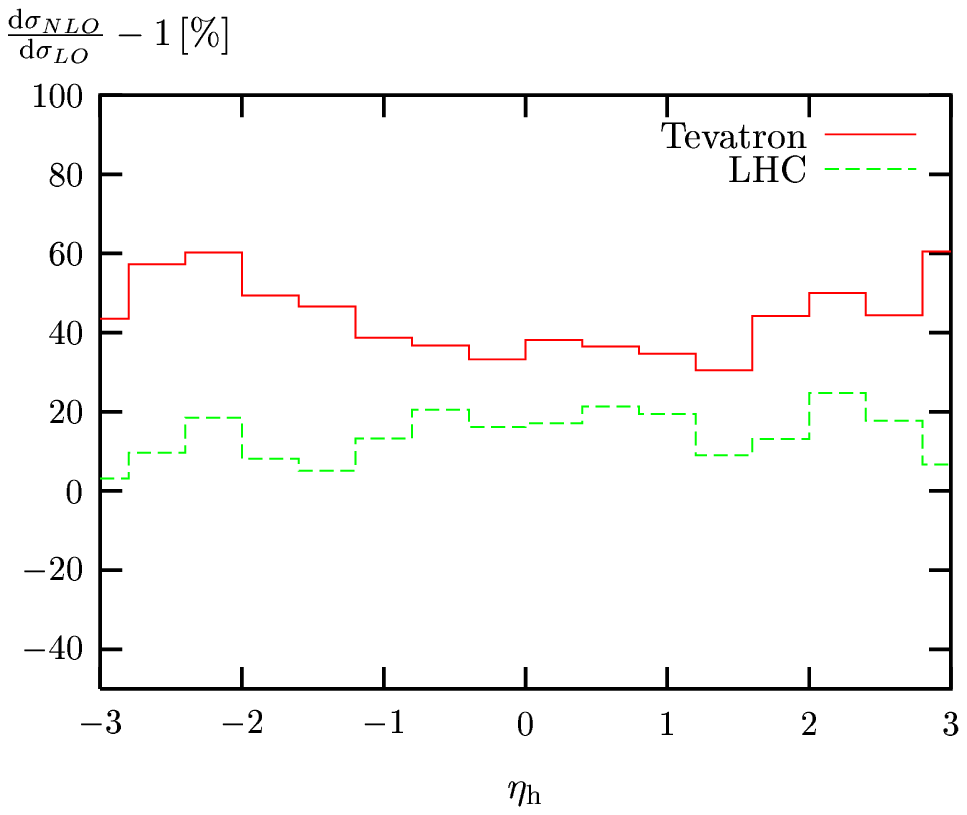} 
\caption[]{The relative corrections $d\sigma_{\scriptscriptstyle NLO}/d \sigma_{\scriptscriptstyle
LO}-1$ for the $\eta$ distribution of the bottom quark $\eta_b$ (left)
and of the SM Higgs boson ($\eta_{h}$) (right) to $b\bar b h$
production in the SM at the Tevatron (with $\sqrt{s}\!=\!2$~TeV and
$\mu\!=\!2 \mu_0$) and the LHC (with $\sqrt{s}\!=\!14$~TeV and
$\mu\!=\!4 \mu_0$).}
\label{fg:bbh_etarel}
\end{center}
\end{figure}

\subsection{PDF and Renormalization/Factorization Scale Uncertainties}
\label{sec:djrw_pdfs}
A major source of theoretical uncertainty for cross section
predictions comes from the PDFs.  We study the uncertainties of
semi-inclusive $bh$ production rates from the uncertainties in
the PDFs using the CTEQ PDF sets\cite{Pumplin:2002vw}.  First, the
central value cross section $\sigma_0$ is calculated using the global
minimum PDF (i.e. CTEQ6M).  The calculation of the cross section is
then performed with the additional 40 sets of PDFs to produce 40
different predictions, $\sigma_i$.  For each of these, the deviation
from the central value is calculated to be $\Delta\sigma_i^{\pm} =
|\sigma_i-\sigma_0|$ when ${\sigma_i}_{<}^{>} \sigma_0$.  Finally, to
obtain the uncertainties due to the PDFs the deviations are summed
quadratically as $\Delta\sigma^{\pm} = \sqrt{ \sum_i
{\Delta\sigma^{\pm}_{i}}^{2}}$ and the cross section including the
theoretical uncertainties arising from the PDFs is quoted as
$\sigma_0|^{+\Delta\sigma^{+}}_{-\Delta\sigma^{-}}$.

In Fig.~\ref{fg:ggvbg}, we plot the normalized total SM NLO cross
sections for semi-inclusive $bh$ production, calculated in the
5FNS ($bg\rightarrow bh$) as implemented in MCFM~\cite{Campbell:2003dd} and
in the 4FNS ($gg\rightarrow b(\bar{b})h$), and compare their
respective uncertainties due to the PDFs.  We see that, at both the
Tevatron and the LHC, the PDF uncertainties are almost identical for
both the $gg$ and $bg$ initial states.

\begin{figure}[t]
\begin{center}
\includegraphics[scale=0.22,angle=-90]{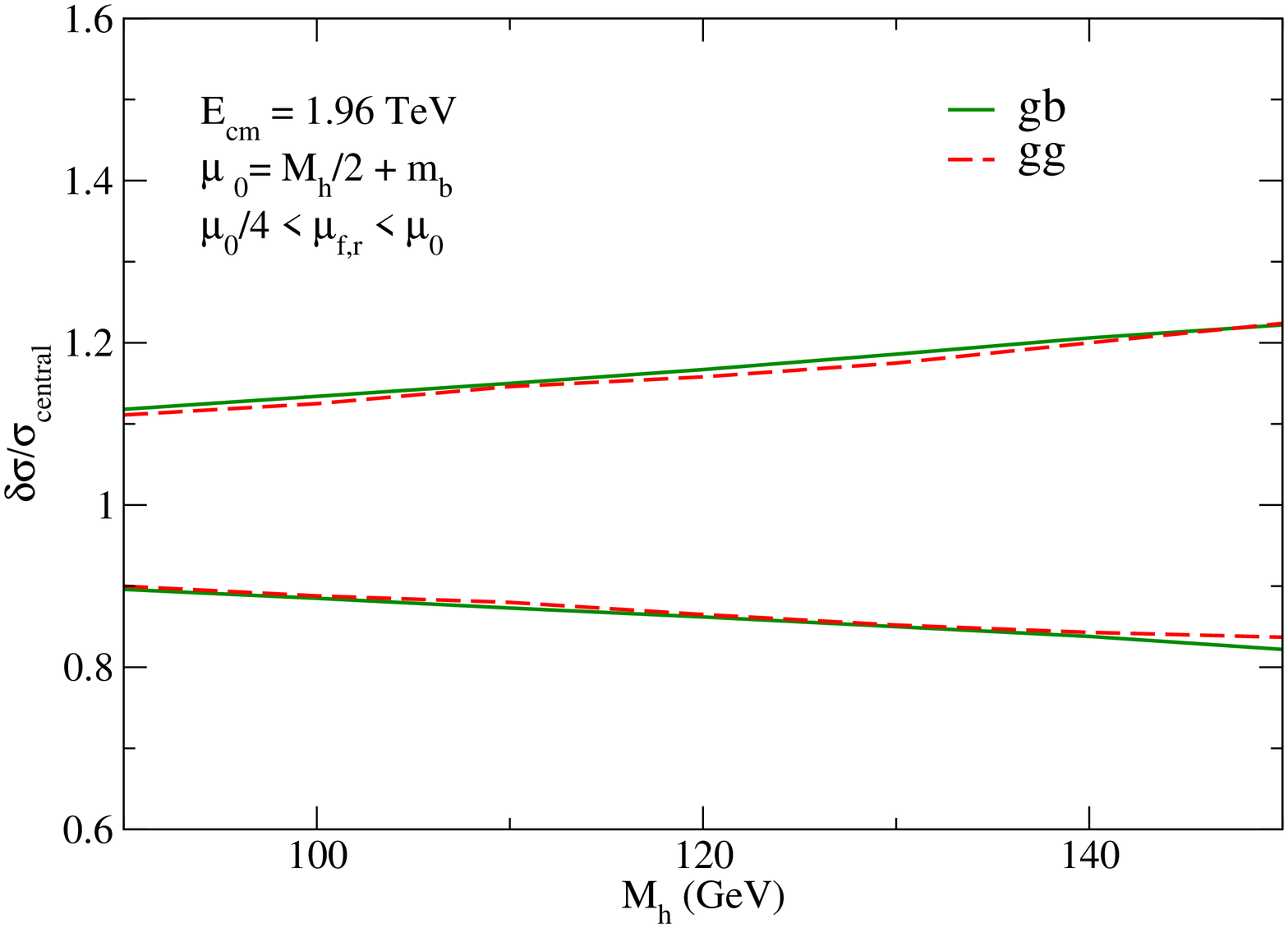}
\includegraphics[scale=0.22,angle=-90]{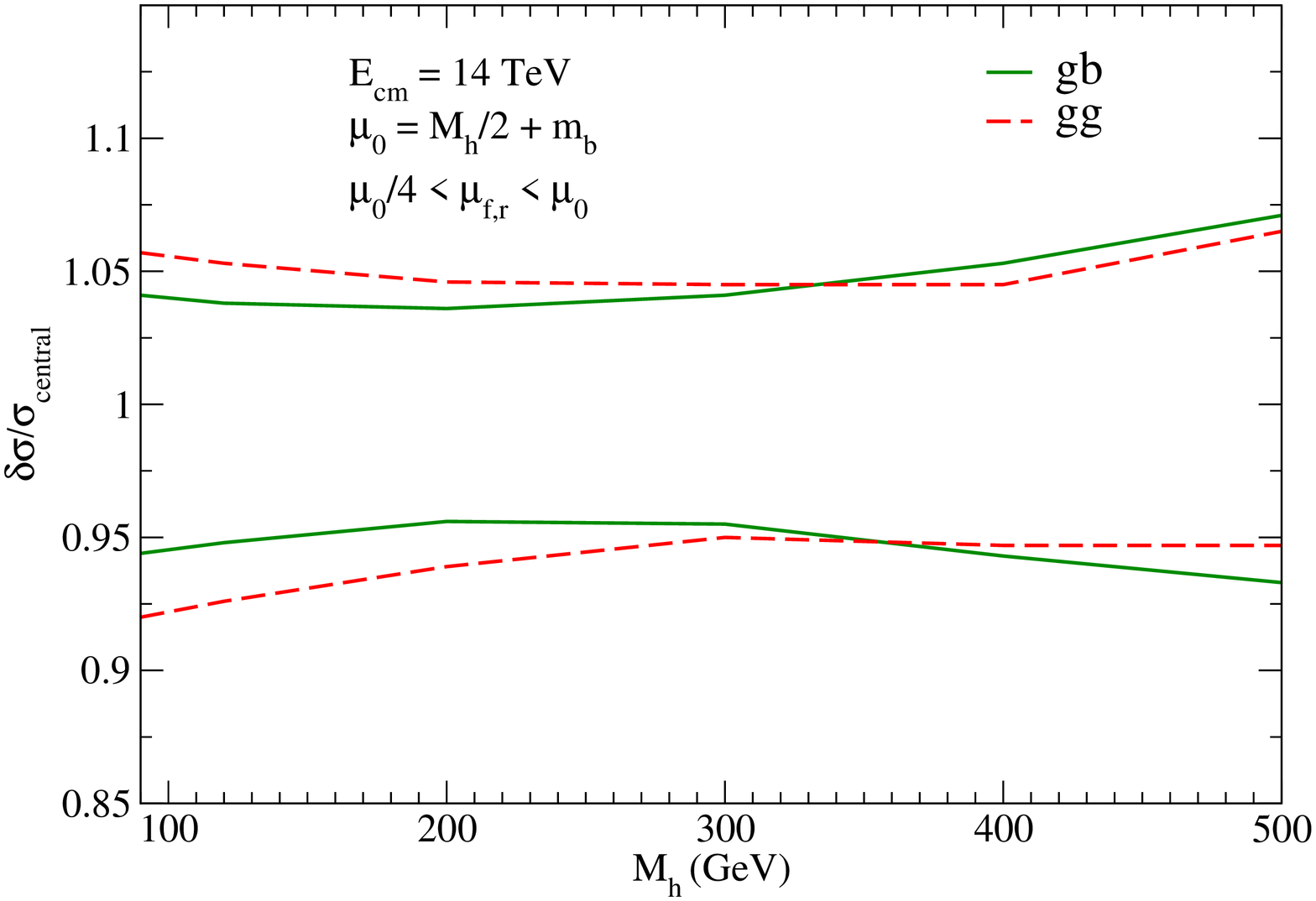} 
\caption[]{Normalized cross sections for Higgs production with one $b$
jet at the Tevatron (l.h.s) and the LHC (r.h.s) showing the uncertainty
from PDFs for both the $gg$ (4FNS) and $bg$ (5FNS) initial states.}
\label{fg:ggvbg}
\end{center}
\end{figure}

\begin{figure}[t]
\begin{center}
\begin{tabular}{rl}
\includegraphics[scale=0.25,angle=-90]{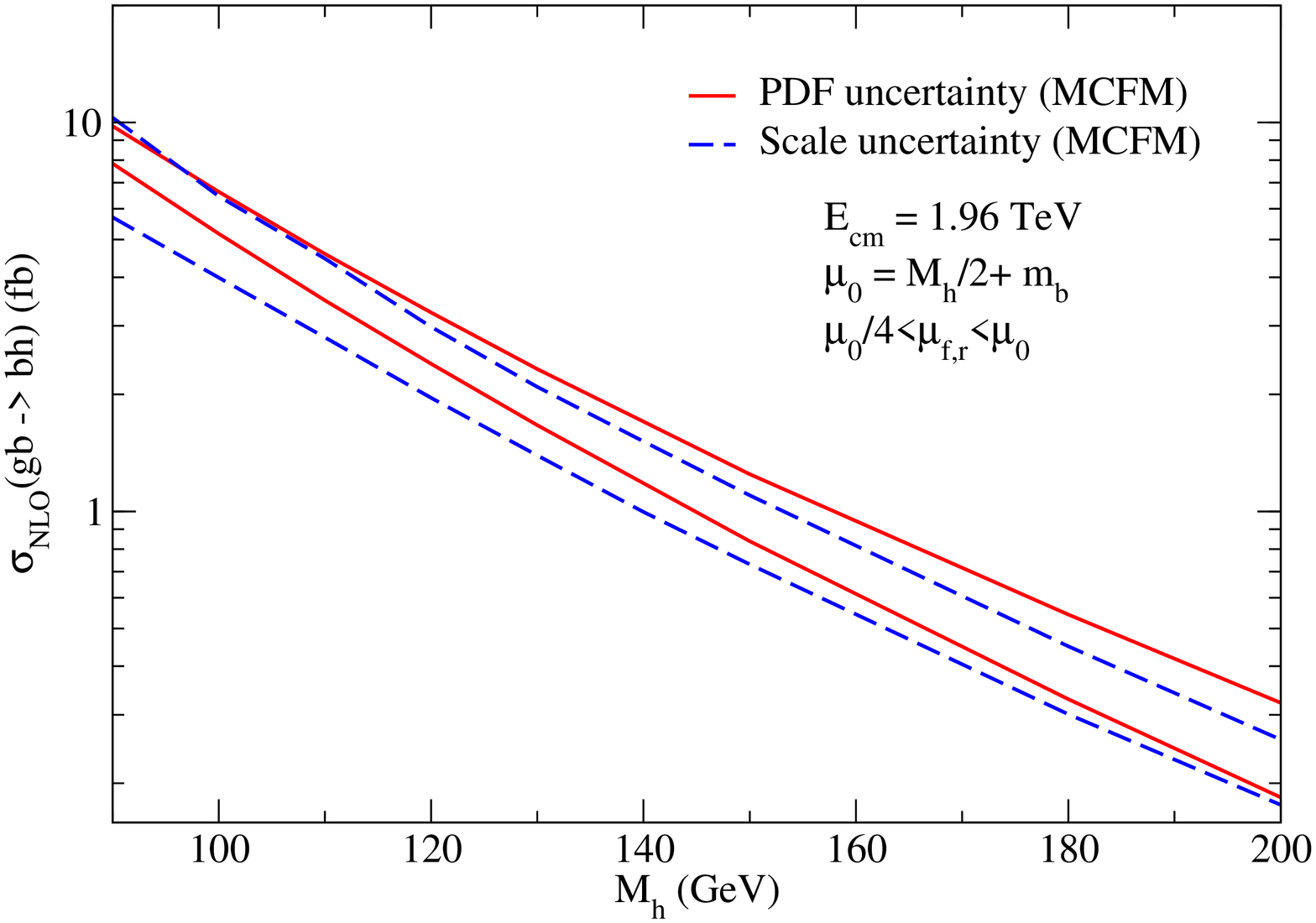} &
\includegraphics[scale=0.25,angle=-90]{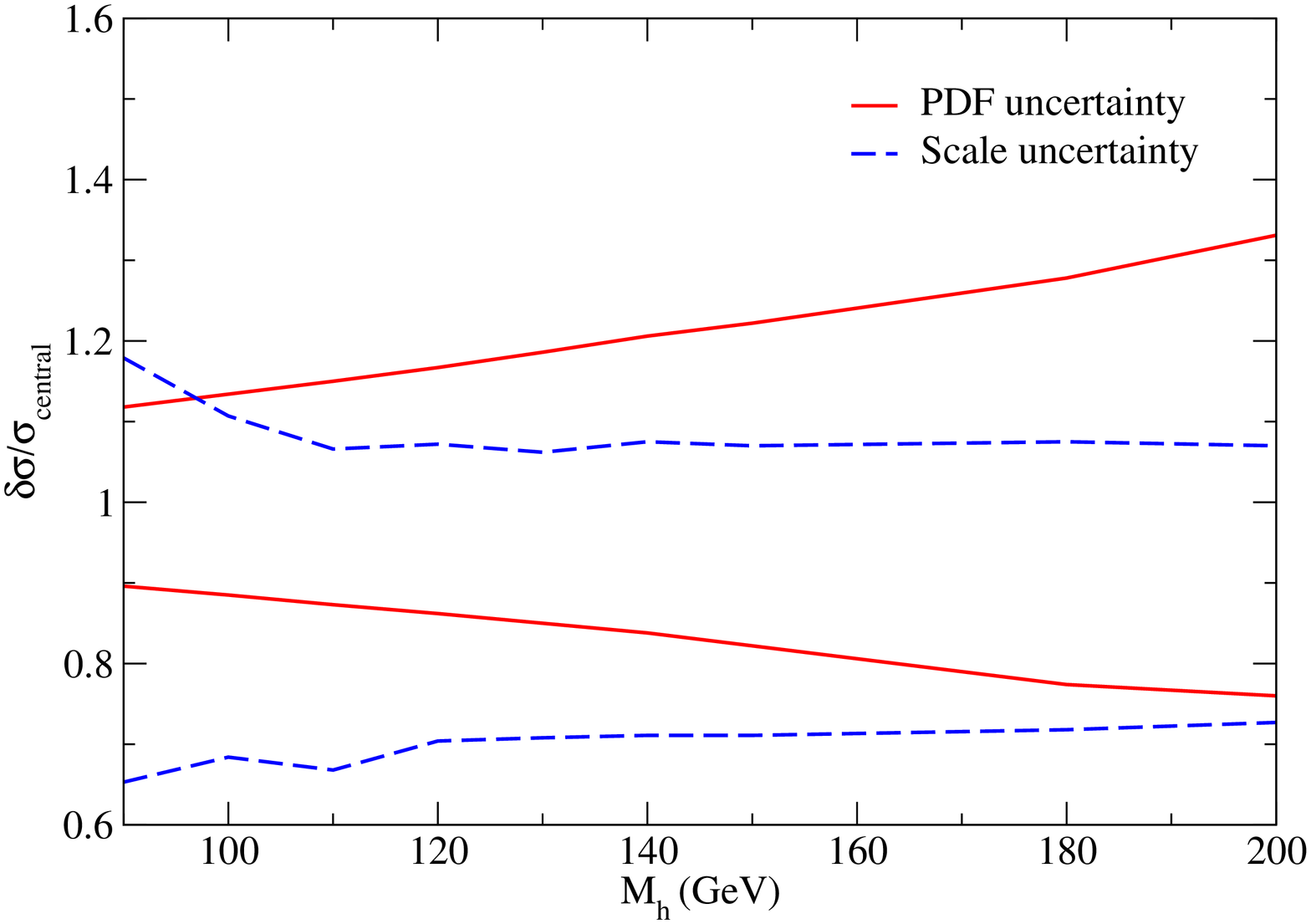}
\end{tabular} 
\caption[]{Comparison between theoretical uncertainties due to scale
dependence and uncertainties arising from the PDFs at the Tevatron
for
semi-inclusive $bh$ production in the Standard Model.
In the right-hand plot, both uncertainty bands have been normalized to the 
central value of the total cross section $\sigma_0$.}
\label{fg:tevPDF}
\end{center}
\end{figure}

\begin{figure}[t]
\begin{center}
\begin{tabular}{rl}
\includegraphics[scale=0.25,angle=-90]{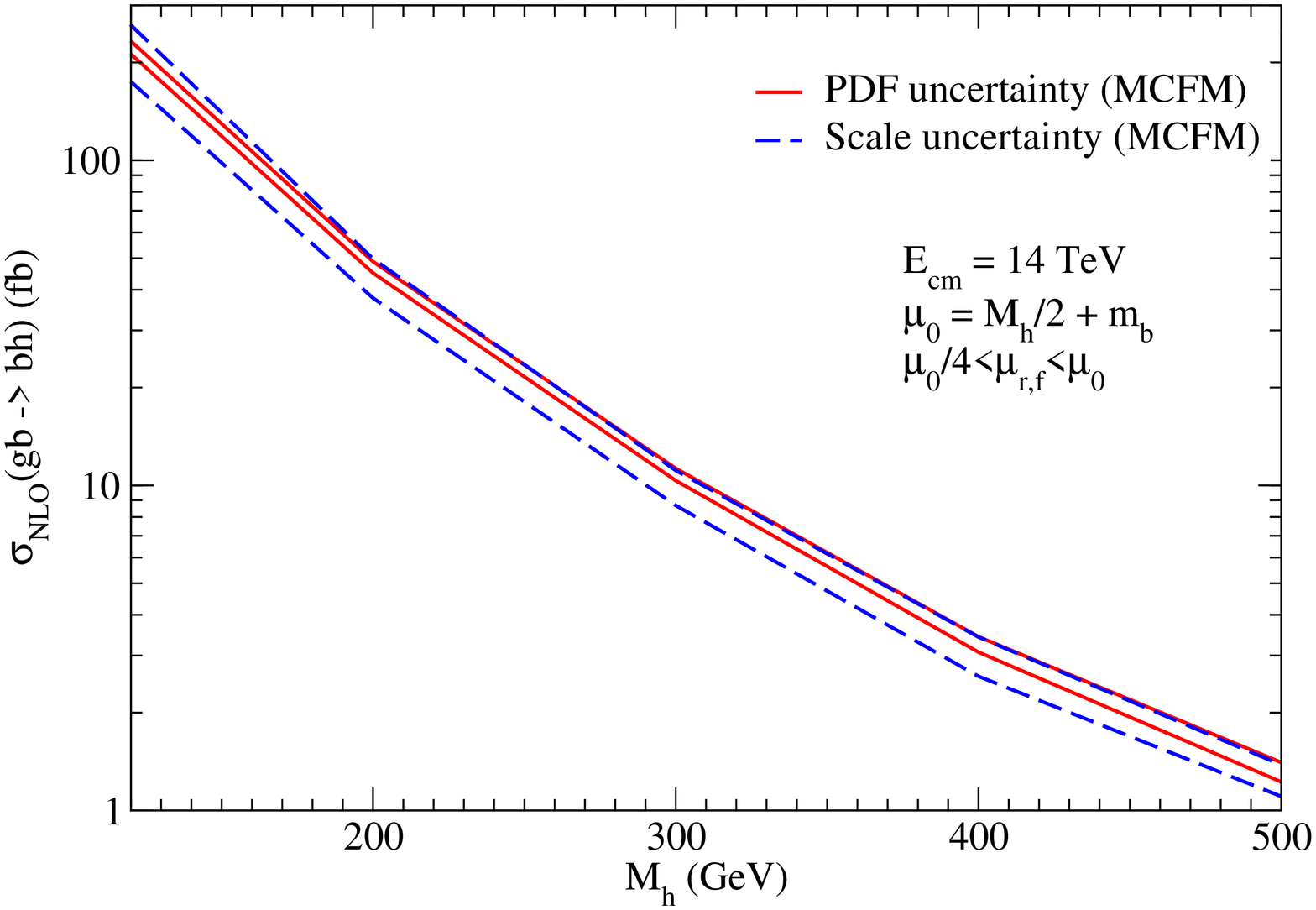} &
\includegraphics[scale=0.25,angle=-90]{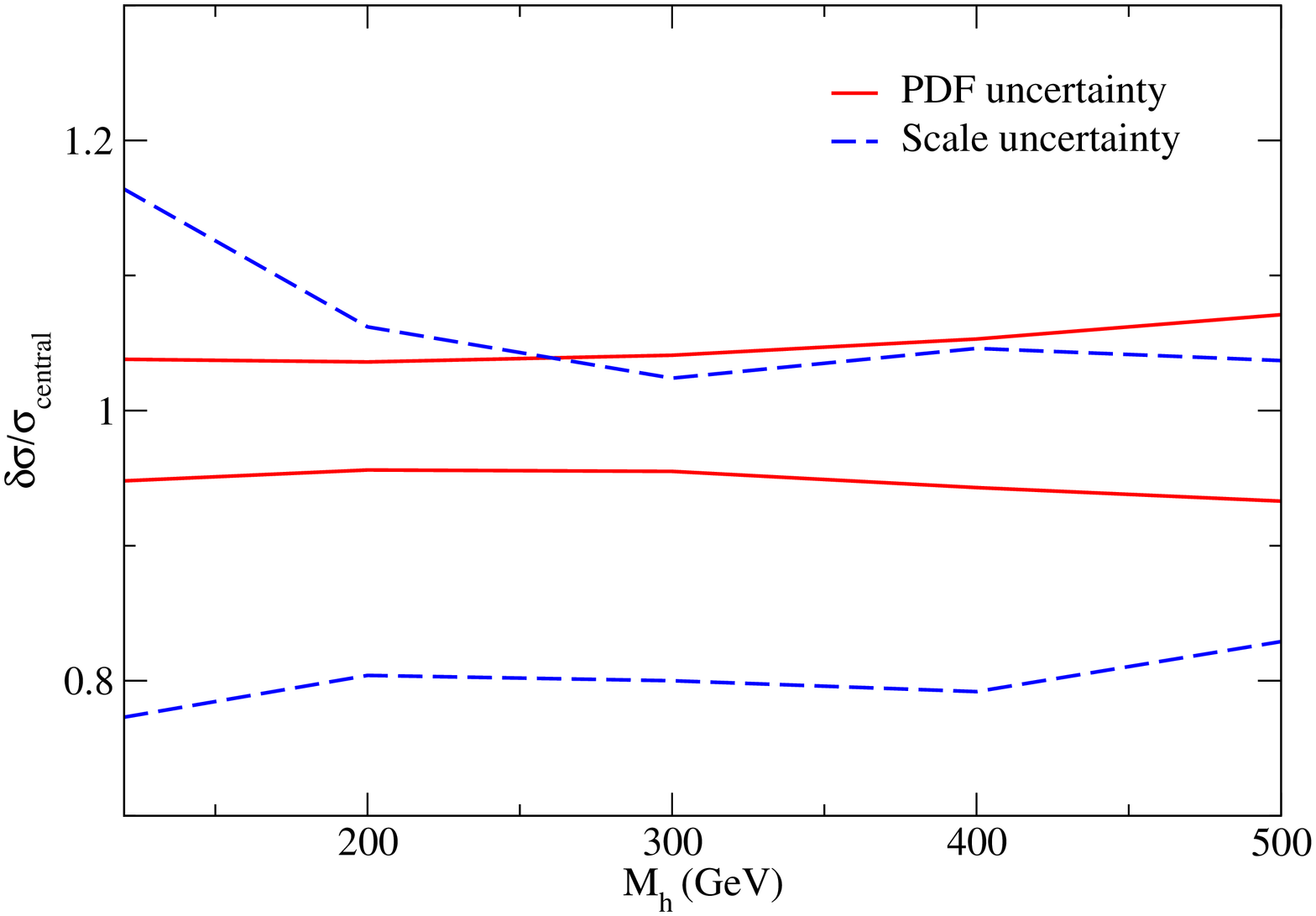}
\end{tabular} 
\caption[]{Comparison between theoretical uncertainties due to scale
dependence and uncertainties arising from the PDFs at the LHC for
semi-inclusive $bh$ production in the Standard Model.
In the bottom plot, both uncertainty bands have been normalized to the 
central value of the total cross section $\sigma_0$.}
\label{fg:lhcPDF}
\end{center}
\end{figure}
In Figs.~\ref{fg:tevPDF} and~\ref{fg:lhcPDF} we 
compare the uncertainties from residual scale dependence and the PDFs
on the example of $bg\rightarrow bh$ (5FNS)   
at the Tevatron and LHC
respectively\cite{Dawson:2005vi}.  Here, we perform the comparison 
for both the total cross
section (left) and the total cross section normalized to the central
value calculated with CTEQ6M (right). Similar results are obtained in the 4FNS.

From Fig.~\ref{fg:lhcPDF} one can see that, at the LHC, the
theoretical uncertainty is dominated by the residual scale dependence.
Due to the large center of mass (c.o.m.) energy of the LHC, the gluons
and bottom quarks in the initial state have small momentum fraction
($x$) values and, hence, small PDF uncertainties typically in the
5-10\% range.

In contrast, due to the smaller c.o.m. energy, the PDF uncertainties
at the Tevatron (Fig.~\ref{fg:tevPDF}) are comparable and even larger
than the uncertainties due to residual scale dependence over the full
Higgs mass range.  The smaller c.o.m. energy results in higher-$x$
gluons and bottom quarks in the initial state which corresponds to
large PDF uncertainties in the 10-30\% range.

\subsection{Conclusion}

The NLO cross sections for $t{\overline t}h$ and $b{\overline b}h$
have been presented for the Tevatron and the LHC with emphasis on the
renormalization/factorization scale and PDF dependences.
\subsection*{Acknowledgments}
The work of S.D. and C.J. ( L.R) is supported in part by the U.S.
Department of Energy under grant DE-AC02-98CH10886
(DE-FG-02-91ER40685). The work of D.W. is supported in part by the
National Science Foundation under grant No.~PHY-0244875.
%



\clearpage
\section{Heavy-flavor effects in supersymmetric Higgs boson production
  at hadron colliders}
\label{sec:bbh}
\textbf{Contributed by: A. Belyaev, S. Berge, P.M. Nadolsky,
  F.I. Olness, C.P. Yuan}
\vspace{0.25in}

The Higgs sector may be represented by one complex scalar doublet,
as it is economically realized in the Standard Model (SM), or by two
or more doublets, as it takes place in the Minimal Supersymmetric
Standard Model (MSSM) and its extensions. An important feature of
MSSM is that, for large values of $\tan\beta$, the Yukawa couplings
of the $b$-quarks to the neutral Higgs bosons $\mH$ (where $\mH=h$,
$H$, or $A$) are strongly enhanced compared to the SM $b\bar{b}H_{SM}$
Yukawa coupling. Consequently, production of supersymmetric Higgs
bosons in $b\bar{b}$ fusion can have a large cross section in supersymmetric
extensions of the Standard Model \cite{Carena:2002es,Spira:1997dg,Assamagan:2004mu,Balazs:1998nt,Diaz-Cruz:1998qc}.

The partonic processes contributing to the inclusive Higgs boson production
with enhanced $b\bar{b}\mH$ coupling are represented by (a) $b\bar{b}\rightarrow\mH$;
(b) $gb\rightarrow\mH b$; and (c) $gg\rightarrow b\bar{b}\mH$ scattering.
The three processes (a,b,c) all give rise to the same hadronic final
states, with two $B$-mesons appearing in different, but overlapping,
regions of phase space. The distinction between the three processes
depends very much on the factorization scheme adopted for the QCD
calculation, as has been recently reviewed in Ref.~\cite{Belyaev:2005nu}.

As shown in Refs.~\cite{Belyaev:2002zz,Belyaev:2005ct}, the correct
model for the transverse momentum distribution of the Higgs boson
is crucial for unambiguous reconstruction of the Higgs boson mass
in the $\mH\rightarrow\tau\tau$ decay channel. It is also important
for discriminating the signal events from the backgrounds by examining
the $q_{T}$ distribution of the Higgs boson in ${\mH}b\bar{b}$ associated
production, followed by ${\mH}\rightarrow b\bar{b}$ decay~\cite{Carena:2000yx}.
The transverse momentum ($q_{T}$) distributions of Higgs bosons may
be sensitive to the mass $m_{b}$ of the bottom quark when $q_{T}$
is comparable to $m_{b}$. In Refs.~\cite{Berge:2005rv,Belyaev:2005bs}
, we study the effect of the initial-state multiple parton radiation
and heavy-quark masses on the transverse momentum distribution in
the $b\bar{b}\rightarrow\mH$ process. Here we summarize the results
of those two papers.

\subsection{\label{sec:Formalism}Transverse Momentum Resummation for Massive
Quarks}

The resummed differential cross section for inclusive production of
Higgs bosons in scattering of initial-state hadrons $A$ and $B$
takes the form~\cite{Collins:1984kg}\begin{equation}
\frac{d\sigma}{dQ^{2}dydq_{T}^{2}}=\int_{0}^{\infty}\frac{{\rm b}d{\rm b}}{2\pi}\, J_{0}(q_{T}{\rm b})\, W({\rm b},Q,x_{A},x_{B},m_{b})\,\,+\,\, Y(q_{T},Q,y,m_{b}),\label{WYDY}\end{equation}
 where $y$ is the rapidity of the Higgs boson, $x_{{A,B}}\equiv Qe^{\pm y}/\sqrt{S}$
are the Born-level partonic momentum fractions, $S$ is the square
of the center-of-mass energy of the collider, and $J_{0}(q_{T}{\rm b})$
is the Bessel function. The resummed form factor $W$ is given in
impact parameter ($\mathrm{b}$) space and factorizes as\begin{equation}
W({\rm b},Q,x_{{A}},x_{{B}},m_{b})=\frac{\pi}{S}\,\sum_{j,k}\sigma_{jk}^{(0)}\, e^{-\mathcal{S}({\rm b},Q,m_{b})}\,\,{\mathcal{\overline{P}}}_{j/A}(x_{A},{\rm b},m_{b})\,\,{\mathcal{\overline{P}}}_{k/B}(x_{{B}},{\rm b},m_{b}),\label{WCSS}\end{equation}
 where the summation is performed over the relevant parton flavors
$j$ and $k$. Here, $\sigma_{jk}^{(0)}$ is a product of the Born-level
prefactors, $e^{-{\mathcal{S}}({\rm b},Q,m_{b})}$ is an exponential
of the Sudakov integral \begin{eqnarray}
{\mathcal{S}}({\rm b},Q,m_{b})\equiv\int_{b_{0}^{2}/{\rm b}^{2}}^{Q^{2}}\frac{d\bar{\mu}^{2}}{\bar{\mu}^{2}}\biggl[{\mathcal{A}}(\alpha_{s}(\bar{\mu}),m_{b})\,\mathrm{ln}\biggl(\frac{Q^{2}}{\bar{\mu}^{2}}\biggr)+{\mathcal{B}}(\alpha_{s}(\bar{\mu}),m_{b})\biggr],\label{Sudakov}\end{eqnarray}
 with $b_{0}\equiv2e^{-\gamma_{E}}\approx1.123$, and ${\mathcal{\overline{P}}}_{j/A}(x,{\rm b},m_{b})$
are the ${\rm b}$-dependent parton distributions for finding a parton
of type $j$ in the hadron $A$. In the perturbative region (${\rm b}^{2}\ll\nolinebreak\Lambda_{QCD}^{-2}$),
the distributions ${\mathcal{\overline{P}}}_{j/A}(x,\mathrm{b},m_{b})$
factorize as\begin{eqnarray}
\left.\overline{{\mathcal{P}}}_{j/A}(x,\mathrm{b},m_{b})\right|_{\mathrm{b}^{2}\ll\Lambda_{QCD}^{-2}} & = & \sum_{a=g,u,d,...}\,\int_{x}^{1}\,\frac{d\xi}{\xi}\,{\mathcal{C}}_{j/a}(x/\xi,\mathrm{b},m_{b},\mu_{F})\, f_{a/A}(\xi,\mu_{F})\label{CxF}\end{eqnarray}
 into a convolutions of the Wilson coefficient functions ${\mathcal{C}}_{j/a}(x,\mathrm{b},m_{b},\mu_{F})$
and $k_{T}$-integrated parton distributions $f_{a/A}(\xi,\mu_{F})$.
The Sudakov exponential and ${\rm b}$-dependent parton densities
resum contributions from soft and collinear multi-parton radiation,
respectively. $Y\equiv\mbox{PERT}-\mbox{ASY}$ is the difference between
the finite-order cross section (PERT) and its asymptotic expansion
in the small-$q_{T}$ limit (ASY).

The Higgs cross sections depend on the mass $m_{b}$ of the bottom
quark. The distributions $\overline{{\mathcal{P}}}_{j/A}(x,{\rm b},m_{b})$
for the heavy quarks ($j=c,b$) cannot be reliably evaluated at all
impact parameters if a conventional factorization scheme, such as
the zero-mass variable-flavor number (ZM-VFN, or massless) scheme,
is used. The reason is that $m_b$ acts as an additional large momentum
scale, which, depending on the value of ${\rm b}$, introduces large
logarithms $\ln^{n}(m_b{\rm b})$ or non-negligible terms $\propto(m_b{\rm b})^{n}$.
The situation encountered here is reminiscent of the heavy-quark contributions
to the DIS structure functions $F_{i}(x,Q^{2}),$ which are not adequately
described by the conventional factorization schemes at either small
or large momentum transfers $Q^{2}$ (see, for instance, \cite{Tung:2001mv}).
To work around this complication, Ref.~\cite{Nadolsky:2002jr} proposed
to formulate the CSS formalism in a general-mass variable flavor number
(GM-VFN) scheme~\cite{Collins:1998rz}, which correctly evaluates
the heavy-quark mass effects at all momentum scales. Among all GM-VFN
factorization schemes, the S-ACOT scheme \cite{Collins:1998rz,Kramer:2000hn}
was found to be well-suited for the efficient calculation of the CSS
resummed cross sections. In particular, in this heavy-quark CSS (CSS-HQ)
formalism \cite{Nadolsky:2002jr} the dependence on $m_{b}$ is dropped
in all ${\cal O}(\alpha_{s})$ terms in Eq.~(\ref{WYDY}) except
for $\overline{{\mathcal{P}}}_{b/A}(x,{\rm b},m_{b})$. 

\begin{figure}[tb]
\begin{center}\includegraphics[%
  clip,
  width=0.49\columnwidth]{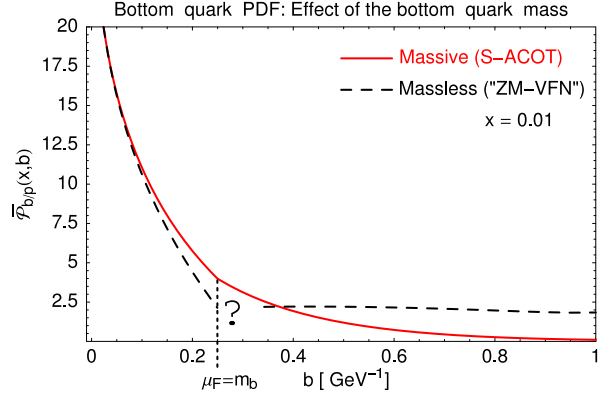}\end{center}

\caption{The bottom-quark distributions $\overline{{\mathcal{P}}}_{b/p}(x,{\rm b},m_b)$
in the proton vs. the impact parameter~${\rm b}$. The solid and
dashed curves correspond to the S-ACOT and massless ({}``ZM-VFN'')
factorization schemes, respectively.\label{fig:PbA}}
\end{figure}

The dependence of the bottom-quark parton density $\overline{{\mathcal{P}}}_{b/p}(x,{\rm b},m_b)$
on the impact parameter is shown in Fig.~\ref{fig:PbA}. The ZM-VFN
parton density $\overline{{\mathcal{P}}}_{b/p}(x,{\rm b},m_b)$
is not properly defined below the threshold $\mu_{F}=m_b$ (or above
${\rm b}=b_{0}/m_b$). It was continued to large ${\rm b}$ in the
previous calculations using an effective {}``ZM-VFN'' approximation
described in Ref.~\cite{Berge:2005rv}. The S-ACOT parton density
$\overline{{\mathcal{P}}}_{b/p}(x,{\rm b},m_b)$ is well-defined
at all ${\rm b}.$ It reduces to the ZM-VFN result at ${\rm b}\ll b_{0}/m_b$
and is strongly suppressed at ${\rm b}\gg b_{0}/m_b$. The suppression
is caused by the decoupling of the heavy quarks in the parton densities
at $\mu_{F}$ much smaller than $m_b$ (${\rm b}$ \nolinebreak
much larger than $b_{0}/m_b$). Consequently the impact of the non-perturbative
contributions from ${\rm b}\gtrsim1\mbox{\, GeV}^{-1}$ is reduced
in the heavy-quark channels compared to the light-quark channels. 

\begin{figure}
\includegraphics[%
  scale=0.4]{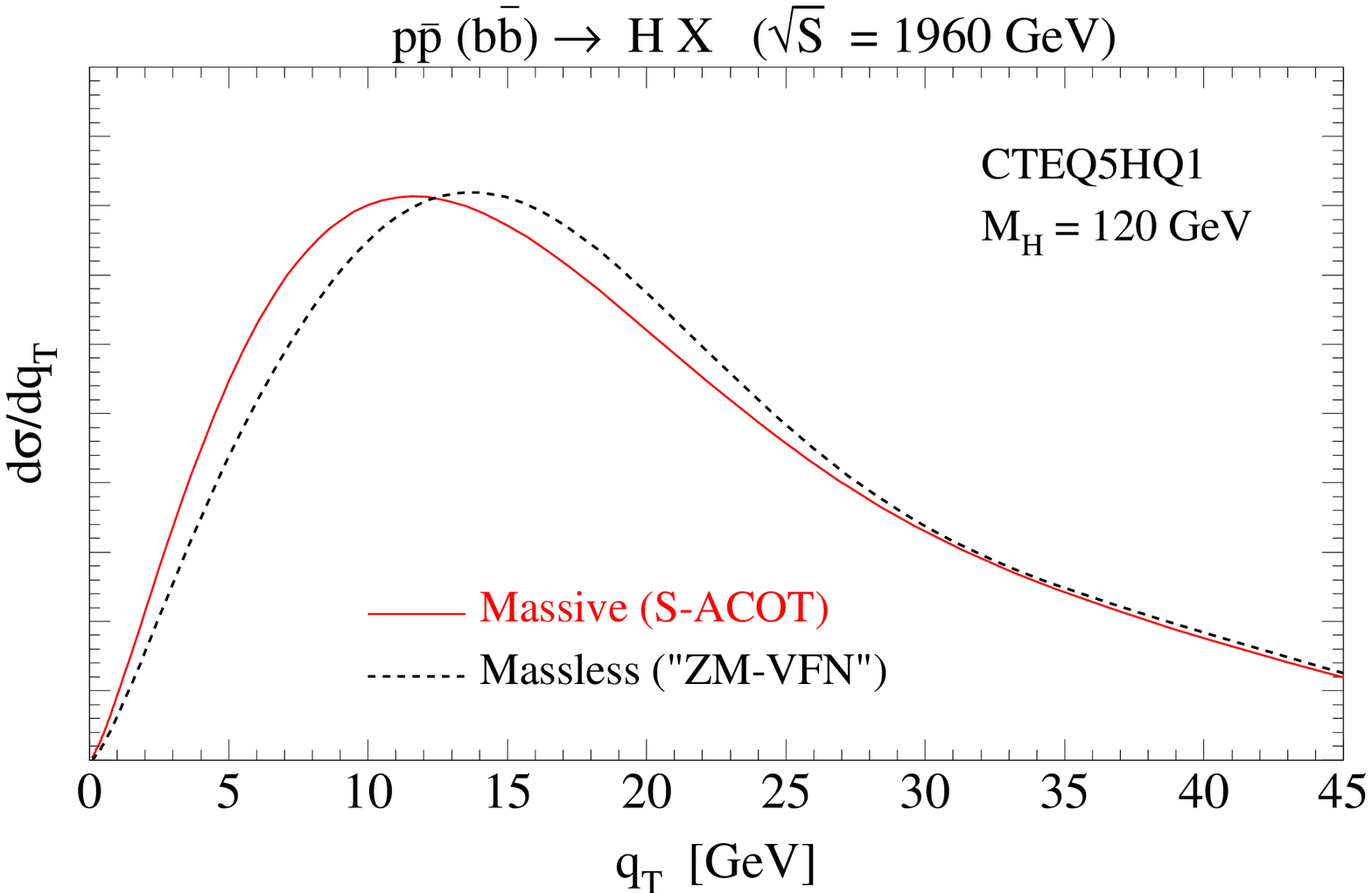}\includegraphics[%
  scale=0.4]{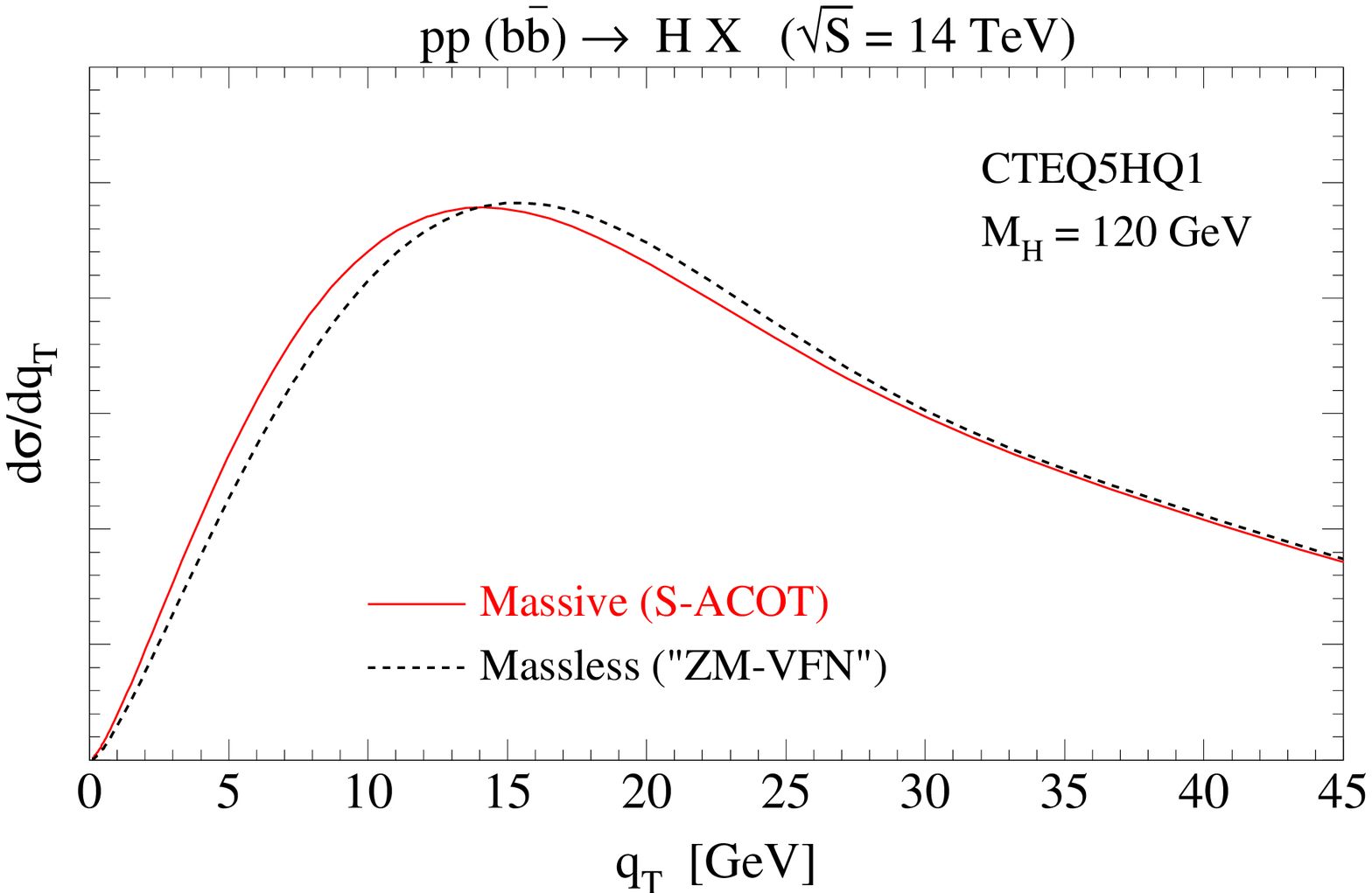}\vspace*{-17pt}

\begin{center}\hspace*{1.cm}(a)\hspace*{7.5cm}(b)\vspace*{-22pt}\end{center}

\caption{Transverse momentum distribution of on-shell Higgs bosons in the
$b\bar{b}\to\mH$ channel at (a)~the Tevatron and (b)~LHC. The solid
(red) lines show the $q_{T}$ distribution in the massive (S-ACOT)
scheme. The dashed (black) lines show the distribution in the massless
({}``ZM-VFN'') scheme. The numerical calculation was performed using
the programs Legacy and ResBos~\cite{Balazs:1997xd,Landry:2002ix}
with the CTEQ5HQ1 parton distribution functions~\cite{Lai:1999wy}.
The bottom quark mass is taken to be $m_{b}=4.5$~GeV.\label{fig:bbh_lhc}}
\end{figure}

The massless ({}``ZM-VFN'') calculation therefore underestimates
the true behavior at ${\rm b}>\nolinebreak0.1\mbox{ GeV}^{-1}$ and
small $q_{T}$. This effect can be seen in Fig.~\ref{fig:bbh_lhc},
which displays $d\sigma/dq_{T}$ for $b\bar{b}\rightarrow\mH$~boson
production at (a)~the Tevatron and (b)~LHC.%
\footnote{Fig.~\ref{fig:bbh_lhc} does not specify the overall normalization
of $q_{T}$ distributions. It is valid for both Standard Model and
supersymmetric Higgs bosons, since at leading order the supersymmetric
result can be obtained by rescaling the Standard Model $b\bar{b}H_{SM}$
coupling: $g_{b\bar{b}\{ h,H,A\}}^{MSSM}=\{-\sin\alpha,\cos\alpha,\sin\beta\,\gamma_{5}\} g_{bbH}^{SM}/\!\cos\beta$.
The net effect of $m_{b}$ on $q_{T}$ distributions will be the same
for the SM and MSSM Higgs bosons, up to an overall normalization constant. %
} At the Tevatron, the $q_{T}$ maximum shifts in the {}``ZM-VFN''
approximation to larger $q_{T}$ by about $2$~GeV out of $11.7$~GeV
(about $17\,$\%). For a Higgs mass $M_{H}=200$~GeV, the maximum
of $d\sigma/dq_{T}$ shifts by about 1.9 GeV out of $12.7$~GeV.
At the LHC, the difference between the {}``ZM-VFN'' and S-ACOT calculations
is smaller compared to the Tevatron, because the influence of the
${\rm b}>0.1\,\mbox{GeV}^{-1}$ region is reduced at smaller momentum
fractions~$x$ probed at the LHC~\cite{Qiu:2000hf}. The maximum
of the $q_{T}$ distribution shifts in the {}``ZM-VFN'' approximation
by about $1.3$~GeV (9\% out of $14.1$~GeV) to larger $q_{T}$.
The results for other Higgs masses can be found in Ref.~\cite{Berge:2005rv}.

\subsection{Numerical Comparison with PYTHIA\label{sec:Numerical-results}}

\begin{figure}
\begin{center}\includegraphics[%
  width=0.50\textwidth]{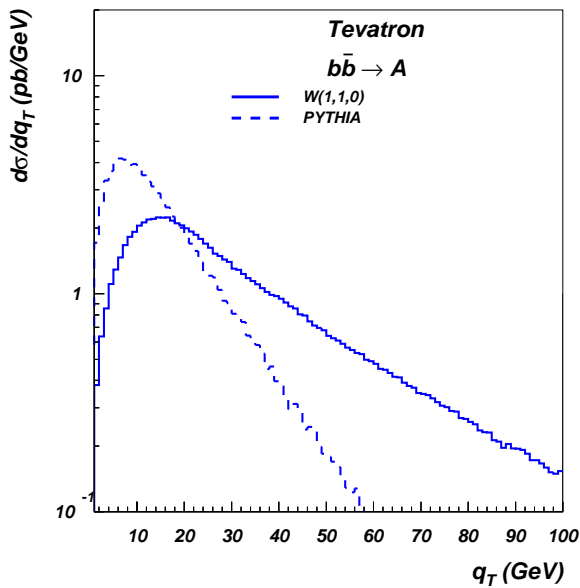}\end{center}

\caption{$q_{T}$ distributions for production of 100 GeV CP-odd Higgs bosons
$A$ via $b\bar{b}$ fusion in the Tevatron Run-2. The solid and dashed
curves correspond to the lowest-order $W$-term $W(1,1,0)$ (with
functions ${\cal A}(\alpha_{s}(\bar{\mu}))$ and ${\cal B}(\alpha_{s}(\bar{\mu}))$
evaluated at ${\cal O}(\alpha_{s})$) and PYTHIA. \label{Fig:one}}
\end{figure}

The full $q_{T}$ dependence of the $b\bar{b}\rightarrow\mH$ process
is affected by constraints on phase space available for QCD radiation
(less relevant at small $q_{T}$). We illustrate the interplay of
various effects by comparing the CSS-HQ resummation to the PYTHIA
Monte Carlo program~\cite{Sjostrand:2001yu}. We focus on production
of the CP-odd Higgs particle $A$ for $\tan\beta=50$ (predictions
for the other Higgs bosons can be obtained by rescaling the $b\bar{b}A$
coupling).

As compared to the CSS-HQ formalism, the PYTHIA calculation does not
include contributions generated from the ${\cal C}$-functions and
$Y$-term, and it evaluates the soft parton contributions at ${\cal O}(\alpha_{s}).$
Therefore, we start by comparing the PYTHIA $q_{T}$ distribution
to the resummed $W$-term $W(1,1,0)$ in Eq.~(\ref{WYDY}), with
the functions ${\cal A},$ ${\cal B}$, and ${\cal C}$ in Eqs.~(\ref{Sudakov}),
(\ref{CxF}) being evaluated at orders $\alpha_{s},$ $\alpha_{s}$,
and $\alpha_{s}^{0}$, respectively. The orders of $\alpha_{s}$ in
${\cal A},$ ${\cal B}$, and ${\cal C}$ are shown as the arguments
of $W(1,1,0)$. 

It is evident from Fig.~\ref{Fig:one} that the shapes of $W(1,1,0)$
and PYTHIA $q_{T}$ distribution are very different, though the integrated
rates (\emph{i.e.,} the areas under the two curves) are about the
same. The $q_{T}$ distribution from PYTHIA is narrower and peaks
at lower $q_{T}$ than $W(1,1,0)$. The large discrepancy between
the two curves is in contrast to the case of $W$ and $Z$ production
via light-quark scattering, where the above two calculations predict
similar, though not identical, $q_{T}$ distributions~\cite{Balazs:1997xd}. 

A closer examination reveals that additional features must be implemented
in the resummed cross section in order to reliably describe the $q_{T}$
distributions of Higgs bosons produced via $b\bar{b}$ fusion. 

\begin{itemize}
\item The kinematical effects account for a large part of the disparity
between $W(1,1,0)$ and PYTHIA. The bottom-quark PDF is a rapidly
decreasing function of $x$ in the probed range of $x$. Consequently,
approximations for the true partonic kinematics (especially those
made for the light-cone momentum fractions $x$) may have a strong
impact on the rate of $b\bar{b}$ scattering. This feature should
be contrasted to the behavior of the light-quark PDF's in $W$ and
$Z$ production, which include a substantial valence component and
vary slower with $x$. As a result, the kinematical approximations
are less consequential in the $W$ and $Z$ case.\\
{\small ~}\\
When PYTHIA generates QCD radiation, the kinematical distributions
of the final-state particles, including the quarks and gluons from
the QCD showering, are modified to satisfy energy-momentum conservation
at each stage of the showering. In the resummation calculation, information
about the exact parton kinematics is included in the finite-order
term (PERT). The resummed cross section is therefore expected to be
closer to PYTHIA once the ${\cal O}(\alpha_{s})$ finite term, PERT(1)-ASY(1),
is included. In the $W(1,1,0)$ calculation, the emitted gluons are
assumed not to carry any momentum at all in the soft limit. To compensate
for small, but nonzero energy of the soft gluon emissions, we introduce
a {}``kinematical correction'' (KC) in the W and ASY terms. This
correction modifies the minimal values of partonic momentum fractions
$x_{A}$ and $x_{B}$ in order to account for reduction of phase space
available for collinear QCD radiation at large $q_{T}$. 
\item The lowest-order cross section $W(1,1,0)$ does not evaluate effects
of the bottom-quark mass, which is first included in the ${\cal C}$-function
of order $\alpha_{s}$. Also, additional, though not complete, $O(\alpha_{s}^{2})$
contributions arise in the Sudakov form factors inside PYTHIA when
the next-to-leading order PDF's are used. To account for both features,
we evaluate the $W$ term at one higher order (2,2,1) and include
the $m_{b}$ dependence using the CSS-HQ scheme. 
\end{itemize}
\begin{figure}
\begin{center}\includegraphics[%
  width=0.50\textwidth]{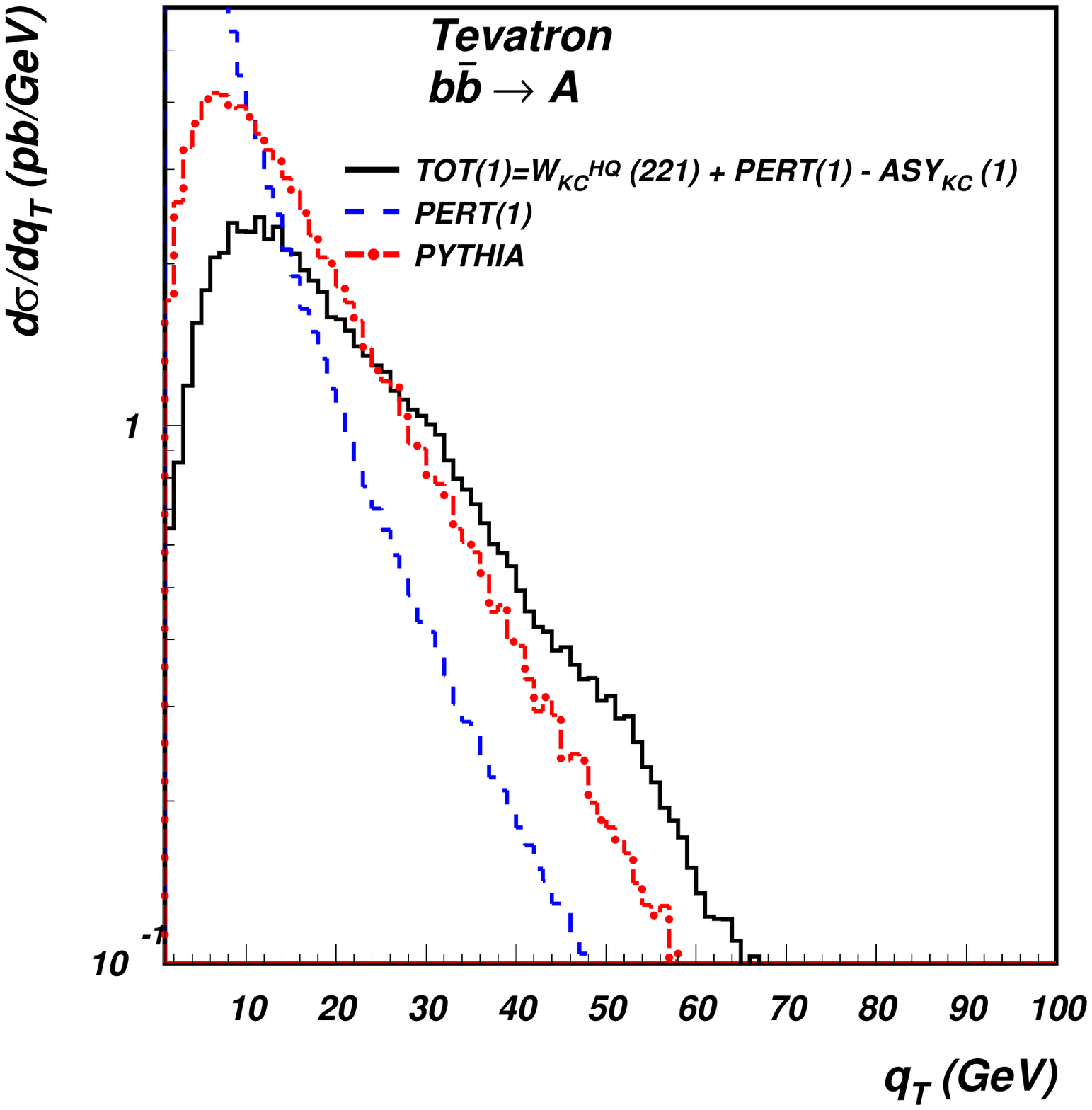}\includegraphics[%
  width=0.50\textwidth]{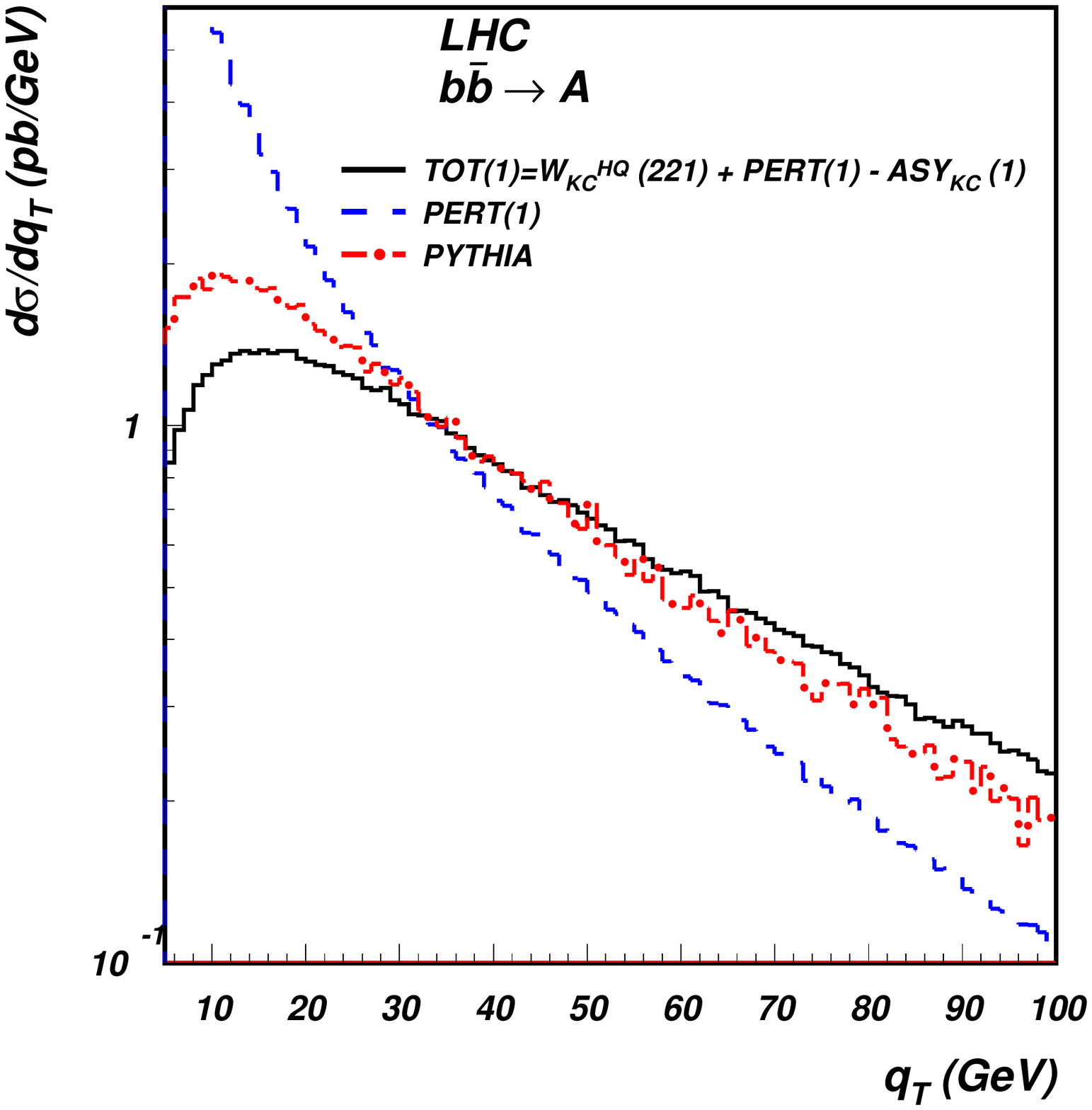}\end{center}

\begin{center}(a)\hspace{3in}(b)\end{center}

\caption{Comparison of $q_{T}$ distributions predicted by TOT(1), PERT(1)
and PYTHIA, for Higgs boson produced via $b\bar{b}$ fusion at (a)
the Tevatron Run-2 and (b) LHC, for $M_{A}=$100 and 300 GeV respectively.
\label{fig:three}}
\end{figure}

Thus, our full prediction TOT(1) is obtained by adding ${\textrm{W}}_{\textrm{KC}}^{\textrm{CSS-HQ}}$(2,2,1)
(evaluated in the CSS-HQ formalism with the kinematical correction)
and PERT(1), and subtracting ${\textrm{ASY}}_{\textrm{KC}}$(1). It
is shown for $M_{A}=100$~GeV at the Tevatron in Fig.~\ref{fig:three}(a)
and $M_{A}=300$~GeV at the LHC in Fig.~\ref{fig:three}(b). TOT(1)
(solid line) is compared to the fixed-order prediction PERT(1) (dashed)
and the PYTHIA prediction (dot-dashed). As one can see, the results
for Tevatron and LHC are qualitatively similar. TOT(1) is closer to
the PYTHIA prediction than $W(1,1,0)$, though the two distributions
are not identical. The PYTHIA $q_{T}$ distribution peaks at lower
$q_{T}$ than TOT(1). In the large $q_{T}$ region, the TOT(1) rate
is larger than the PYTHIA rate.

\begin{figure}
\includegraphics[%
  width=0.50\textwidth]{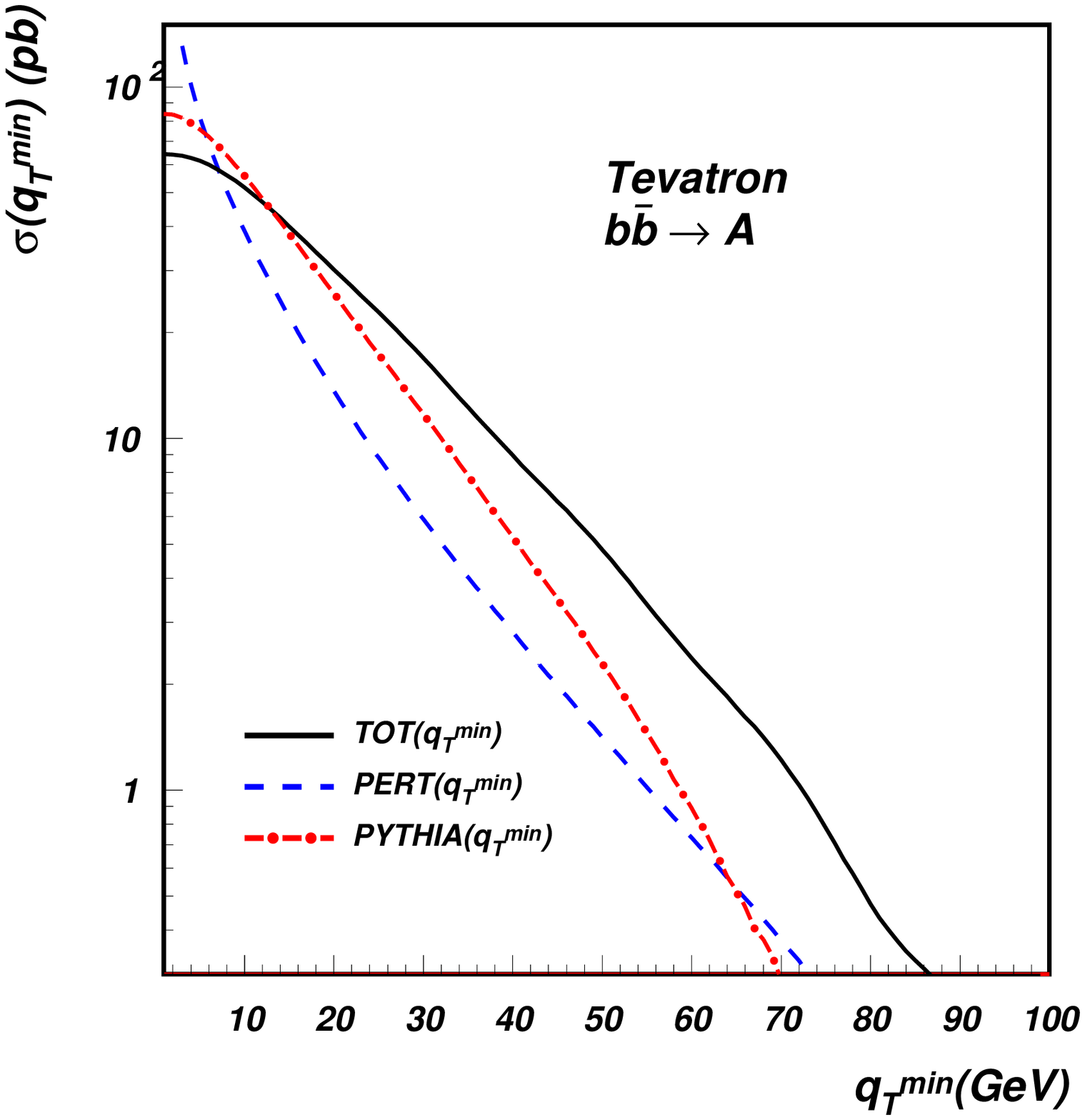}\includegraphics[%
  width=0.50\textwidth]{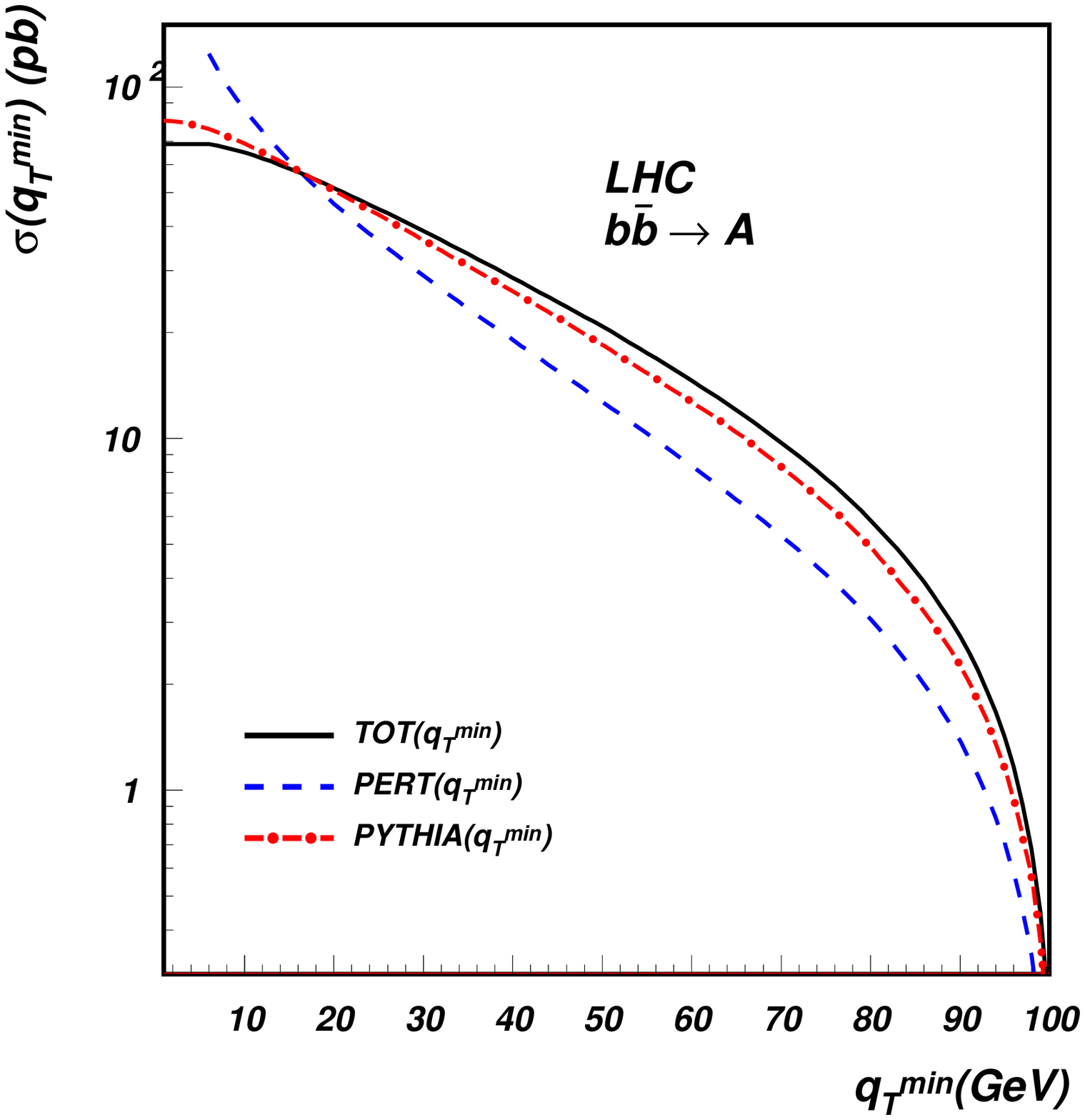}

\caption{Comparison of the integrated rates, deduced form Fig.~\ref{fig:three},
as a function of the minimal $q_{T}$ value taken in the integration
over $q_{T}$ at the Tevatron Run-2 (left) and LHC (right) for $M_{A}=$100
and 300 GeV, respectively. \label{fig:four} }
\end{figure}

Finally, Fig.~\ref{fig:four} shows the integrated cross section
as a function of the minimal $q_{T}$ in the calculation for the Tevatron
(left) and LHC (right). This is another way to illustrate the differences
in the shapes of $q_{T}$ distributions obtained in the resummation,
fixed-order, and PYTHIA calculations.

\subsection{Conclusion\label{sec:Conclusion}}

Multiple parton radiation in $b$-quark scattering is conspicuously
sensitive to effects of large bottom-quark mass $m_{b}$ and phase-space
constraints on collinear emissions. Both $m_{b}$ dependence and phase-space
dependence tangibly modify the shape of Higgs $q_{T}$ distributions
in the $b\bar{b}\rightarrow\mH$ processes. The two types of effects
were consistently implemented within the CSS resummation formalism
for heavy-quark scattering \cite{Nadolsky:2002jr,Berge:2005rv,Belyaev:2005bs},
realized in a massive (GM-VFN) factorization scheme. These corrections
act on different $q_{T}$ regions. When the dependence on $m_{b}$
is taken into account, the position of the peak in the $d\sigma/dq_{T}$
distribution shifts to a lower $q_{T}$ value, leaving the rate at
large $q_{T}$ essentially unchanged. The kinematical correction is
effective in the high-$q_{T}$ region, where it largely reduces the
Higgs production rate. As a result, we obtain an improved prediction
for the full $q_{T}$ spectrum of Higgs bosons, an important piece
of information needed for the future Higgs searches.

\subsection*{Acknowledgments}

This work was supported in part by the U.S. Department of Energy under
grant DE-FG03-95ER40908, contract W-31-109-ENG-38, and the Lightner-Sams
Foundation. We also acknowledge the support in part by the U.S. National
Science Foundation under awards PHY-0354838 and PHY-0244919.

\clearpage
\section{Higgs Signal for $h\rightarrow aa$ at the Tevatron}
\label{sec:h2aa}
%
%
%
%
%



%
%
%
%
\textbf{Contributed by: M. Carena, T. Han, G.-Y. Huang, C.E.M Wagner}

\vspace{0.25in}

The elucidation of the mechanism leading to the origin of mass
of all observed elementary particles is one of the main goals in
high energy physics. The simple Standard Model (SM) picture, based
on the spontaneous breakdown of the electroweak symmetry by the
vacuum expectation value of an elementary Higgs field, seems to
lead to a picture that is consistent with all experimental observables,
provided the Higss boson mass is smaller than about
250\GeV. Moreover, the best fit to the precision electroweak observables 
measured at the LEP, SLC and Tevatron experiments lead to 
values of the Higgs mass of the order of or
smaller than the present bound coming from direct searches at LEP,
$m_{H_{\rm SM}} \gtrsim 114\GeV$.

In spite of the extraordinary good agreement of the experimental
observations with the standard model predictions, there are many
theoretical motivations to go beyond the Standard Model description.
Several extensions of the Standard Model exist in the literature,
and in most of them the Higgs sector is extended to a more complicated
structure, often including at least two Higgs doublets. The requirement
of preserving the good agreement with experimental data can be
easily fulfilled in extensions, like supersymmetry, in which the
effect of the additional particles on the precision electroweak observables
rapidly vanish with increasing values of the new particle masses.
Independently of the particular extension, the direct and indirect
limits on the Higgs mass must be revised. In particular, the
direct search for Higgs bosons may be affected by additional decay
modes that are beyond the ones analysed by the LEP collaborations.

As an example, let us consider the minimal supersymmetric extension of
the Standard Model (MSSM). In the MSSM,
there is an additional Higgs doublet, 
leading, in the absence of CP-violation in the Higgs sector, 
to two CP-even and one CP-odd Higgs 
boson states.  At large values of $\tan\beta$, the ratio of the v.e.v.
of the two Higgs doublets,
one of the CP-even Higgs bosons acquires Standard Model properties, while the
second Higgs boson may only be produced in association with the CP-odd
Higgs boson state. In addition, the masses of the non-standard CP-even
Higgs and the CP-odd Higgs are close to each other. Under these conditions,
the mass bound on the SM-like CP-even Higgs is similar to the SM one, while
the CP-odd and the second CP-even Higgs boson mass bound reads
$m_h>90\GeV$~\cite{Heister:2001kr}. 

In this note, we will depart from these simple assumptions, by 
breaking the mass relations that appear in the simplest supersymmetric
models, and studying the consequences of such modifications of the
parameters of the theory. 
Indeed, while it has been a common belief that 
the Higgs boson 
will be eventually discovered at the upcoming LHC experiments, 
one would like to fully utilize the potential 
to search for the Higgs bosons at the Tevatron in these non-conventional
scenarios as well. Non-standard mass relations are already present 
in  extensions of the  MSSM  including 
an additional singlet (NMSSM)~\cite{Dermisek:2005ar},  
or when explicit CP-violations exist in the Higgs sector~\cite{Carena:2002bb}.
In these cases, the SM-like Higgs ($h$) may  dominantly 
decay into a pair of lighter Higgs ($a$), often the CP odd state. 
Therefore it is possible that the Higgs escaped detection at the 
LEP experiments
by avoiding the usual decay modes such as $h\to 2b,\ 2\tau,\ WW^*$ and $ZZ^*$, 
and the lower limit on Higgs mass should be 
re-evaluated~\cite{Abdallah:2004wy}. We are interested in analysing the
sensitivity of the Tevatron experiments in the search for  
a light, SM-like  Higgs boson with such an exotic 
decay mode.
In particular, we will consider the case when the Higgs boson 
decays into a pair of scalars $h\to aa$, which in turn 
cascade into a heavy fermion pair $a\to b\bar{b}$ and $a\to \tau\bar{\tau}$, 
respectively. 

The dominant production of the Higgs boson at hadron colliders
comes from the gluon fusion, but the above channel would encounter 
huge SM backgrounds. 
We therefore consider the Higgs signal produced in association 
with a $W$ or $Z$ boson, 
in the hope that the leptonic decays of the weak bosons will
provide a clean trigger, and will significantly reduce the background
as well. The events being searched are
\begin{eqnarray}
Wh\to & l\nu_l, aa &\to \left\{ \begin{array}{l}
l\nu_l, b\bar{b}, b\bar{b} \\
l\nu_l, b\bar{b}, \tau\bar{\tau} 
\end{array} \right. \\
Zh\to & l^+l^-, aa &\to \left\{ \begin{array}{l}
l^+l^-, b\bar{b}, b\bar{b} \\
l^+l^-, b\bar{b}, \tau\bar{\tau} ,
\end{array} \right. 
\end{eqnarray}
with $l=e,\mu$.

\subsection {Parameter Choices}

We would like to perform a relatively model-independent search for 
the typical signal.
The direct search
for a Higgs boson with SM-like couplings to the
gauge bosons, in a model and  
decay mode-independent way, leads to a lower bound 
on $m_h$ of about 82\GeV~\cite{Abbiendi:2002qp}. Moreover, the proposed search
is expected to become inefficient for $m_h>130\GeV$, since
the standard decays into the $WW^*$ and $ZZ^*$ channels
are still expected to be dominant.  Therefore, 
the optimal setting to detect the Higgs decaying into an $aa$ pair is to 
have the mass $m_h$ within the range of $90-130\GeV$. 
The choice for $m_a$ can be more flexible.  
As long as  $m_a>2m_b$ to kinematically allow the decay 
$a \to b\bar{b}$,   our analyses are rather insensitive to the mass choices
(see below for a more detailed analysis of this question).

In a generic model, the $Wh/Zh$ production rate differs from that in the SM.
The change can be characterized by a prefactor $\kappa^2_{hWW}$
($\kappa^2_{hZZ}$), where $\kappa_{hVV}$ is the coupling strength of Higgs to
vector boson $V$ relative to that in the SM.  The production cross section can
thus be written in terms of the SM result with an overall factor to account for
the modification of the coupling
\begin{eqnarray}
\sigma(Vh) = \kappa^2_{hVV} \sigma^{SM}(Vh) .
\end{eqnarray}
We are interested in the range of
$\kappa\sim 0.5-1.0$, so that this Higgs contributes to the
electroweak symmetry breaking and consequently the 
associated productions are still sizeable.

In order for the $h\to aa$ decay to be dominant and thus escape the LEP bound, 
$BR(h\to aa)$
is required to be close to unity. For instance, in the NMSSM, 
$BR(h\to aa)>0.9$ turns out to be very
general in terms of the naturalness of $c$
in the trilinear coupling term $(cv/2)haa$~\cite{Dobrescu:2000jt}.
Moreover, if the down quark and lepton coupling to the Higgs
is proportional to their masses, then 
$BR(a\to b\bar{b})$ and
$BR(a\to \tau\bar{\tau})$ are set to be 0.92 and 0.08, respectively.
In general, however, the relations between the coupling and the masses
may be modified by radiative corrections, which can lead to a large
increase of the $BR(h \to \tau\tau)$~\cite{Carena:1998gk}.
The representative values and the ranges of the parameters
are summarized in Table \ref{param}. 
\begin{table}[htb]
\begin{centering}
    \begin{tabular}[t]{ l || l ||  c|c}
       &   & representative & considered  \\
       & parameters  & value & range  \\
      \hline \hline 
            & $m_h$ & ~120~ & ~90$-$130~   \\
      \raisebox{1.5ex}[0pt]{masses}
            & $m_a$ & ~30~ & ~20$-$40~   \\
      \hline
	coupling
            & $\kappa_{hVV}$ &  ~0.7~  & ~0.5$-$1.0~  \\
      \hline
            & $BR(h\to aa)$ &  ~0.85~  & ~0.8$-$1.0~  \\
      \raisebox{1.5ex}[0pt]{branching}
            & $BR(a\to b\bar{b})$ &  ~0.92~ & ~ 0.95$-$0.70 ~ \\
      \raisebox{1.5ex}[0pt]{fractions}
            & $BR(a\to \tau\bar{\tau})$ &  ~0.08~ & ~ 0.05$-$0.30 ~ \\
      \hline
      \hline
	$2b2\tau$
            & $C^2$ &  ~0.061~  & ~0.019$-$0.42~  \\
    \end{tabular}
    \caption{Parameter choices for $h\to aa$ decays. 
    \label{param}}
\end{centering}
\end{table}

Including the decay branching fractions, for instance for
$a_1\to b\bar b,\ a_2\to \tau \bar\tau$, we obtain the  cross section as
\begin{eqnarray}
   \sigma(aa) =  \kappa^2_{hVV}\  \sigma^{SM}(Vh) \ BR(V)\  
	2BR(h\to aa) BR(a\to b\bar{b}) BR(a\to\tau\bar{\tau}) .
    \label{aadecayxsec}
\end{eqnarray}
where $BR(V)=0.213\ (0.067)$ is
the leptonic branching ratio of the decay of $W\ (Z)$. 
into $l=e,\mu$.

The overall factor modifying the SM result 
in Eq.~(\ref{aadecayxsec}), {\it i.e.}
\begin{equation}
C^2 \equiv 2\kappa^2_{hVV}BR(h\to aa) BR(a\to b\bar{b}) BR(a\to\tau\bar{\tau}),
\end{equation} 
corresponds to the process-dependent $C^2$ factor defined in 
the DELPHI search~\cite{Abdallah:2004wy}. 
Our parameter choice (range) is equivalent to 
a $C^2_{2b2\tau}$  of 0.061 (0.019$-$0.42), 
consistent with the bounds for a large range of our $m_h,m_a$ choices
set forth in Ref.~\cite{Abdallah:2004wy}~\footnote{Conversion 
between $C^2_{2b2\tau}$ 
and $C^2_{4b}$ involves a factor $BR(a\to bb)/2BR(a\to \tau\tau)$. }.
A value of 0.061 for $C^2$ is assumed 
for all numerics from here on, unless explicitly noted otherwise.

\subsection{Signal Event Rate}
The associated production of $p\bar{p}\to Wh$ 
usually features a larger cross section than that of $Zh$,
and the leptonic branching fraction of $W$ is 
about 3 times larger than $Z$'s.
For illustration purposes, we choose to concentrate on the $Wh$ channel henceforth.
\begin{figure}[tb]
    \includegraphics[scale=0.35]{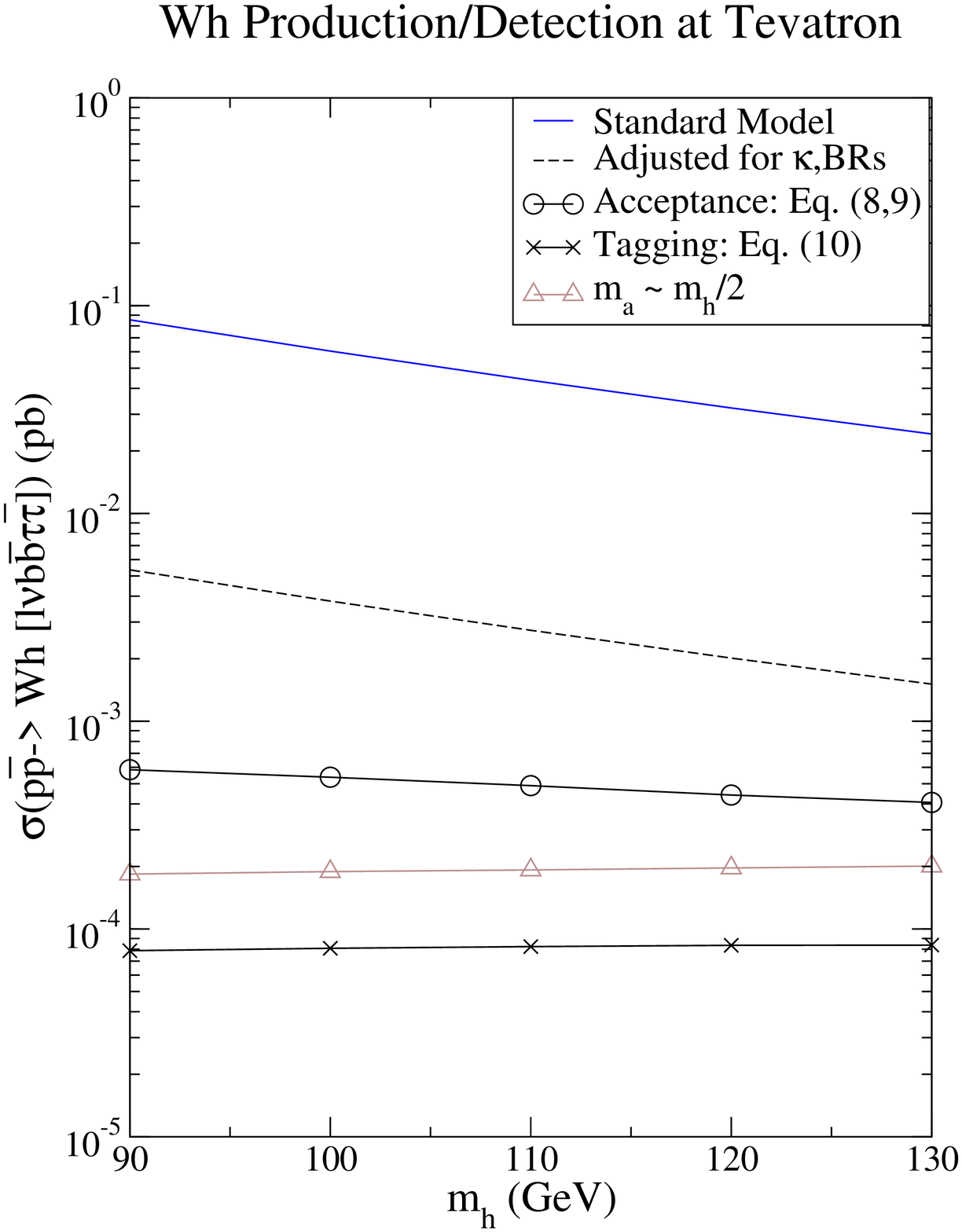}
    \includegraphics[scale=0.35]{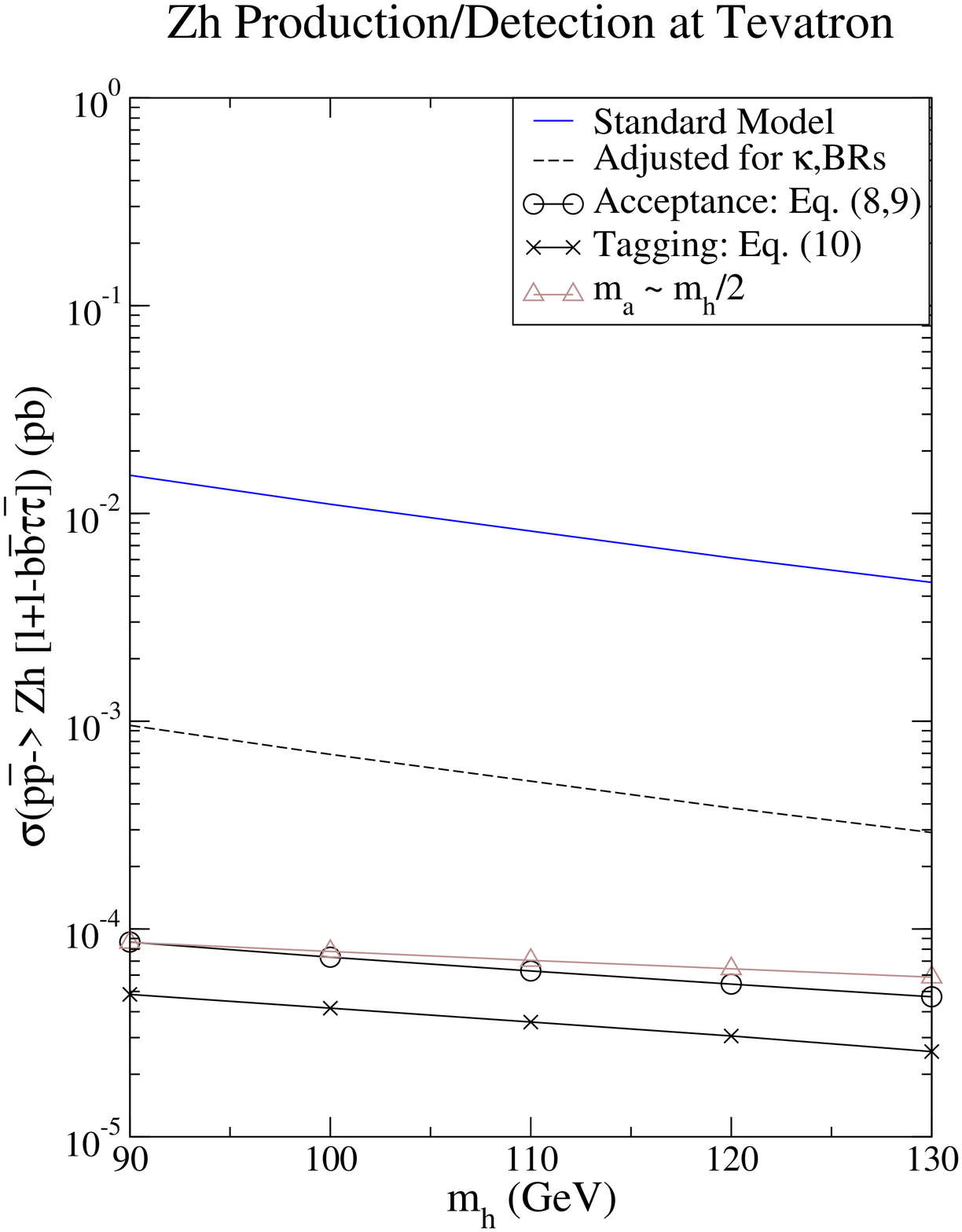}
    \caption {Cross sections of Higgs signal at the Tevatron
	in the $2b2\tau$ 
	channel produced by Higgs-strahlung with a leptonically 
	decaying $W$ (left) or $Z$ (right).
$m_a=30\GeV$ is assumed except for the two curves where 
$m_a=(m_h-10\GeV)/2 $.
$C^2=0.061$ is understood.}
    \label{xsec}
\end{figure}

The Standard Model rate of a Higgs produced in association with a leptonically 
decaying $W$ is
\begin{equation}
 \sigma^{SM}(Wh) \ BR(W\to l\nu_l) \sim 85\ (24) \ {\rm fb}
\end{equation}
 at $\sqrt s=1.96$ TeV for $m_h=90\ (130)\GeV$.

Including the branching fractions and couplings,
the cross section of the signal in Eq.~(\ref{aadecayxsec}) is
\begin{equation}
 \sigma(aa) \sim 5.3\ (1.5) \ {\rm fb \qquad for \quad} C^2 = 0.061
\end{equation}
as illustrated in Fig.~\ref{xsec}.
The solid curve on top represents the total cross section for $Vh$ production, 
with $V$ decaying leptonically, but without any cuts.
The dashed curve represents the cross section after 
adjusting for the couplings and branching fractions, 
as in Eq.~(\ref{aadecayxsec}).
Cross sections for $Zh$ are also plotted for completeness. 

The events have yet to pass the acceptance cuts, or to have the taus and
$b$'s tagged. Both bring significant reductions to the event rate.
Our challenges are to retain as many events as possible, and to control
the backgrounds from various sources.

\subsection{Background and Acceptance Cuts}

We look for events with 5 particles plus missing energy 
in the final states: 
$b\bar{b}\tau\bar{\tau}l\nu_l$. We wish to trigger the events
by the isolated lepton $l$, tag the $b$'s and $\tau$'s,
and demand significant missing transverse energy 
(\MET)
in the events.
With neutrinos in the decay products,
 tau momenta cannot be fully reconstructed.
Therefore we cannot reconstruct the invariant masses $m_{\tau\tau}$ or 
$m_h\sim m_{bb\tau\tau}$. 
Instead, the signal should appear as a peak
in the $m_{bb}$ plot, around the value of $m_a$.

\subsubsection*{Acceptance Cuts}

The following cuts are employed to mimic the detector acceptance:
\begin{eqnarray} \nonumber
p_T&>&10\GeV {\rm \qquad for } \ b,l^\pm \\
\nonumber
\MET&>&10\GeV {\rm \qquad for } \ Wh \ {\rm only} \\
|\eta|&<&3.0 \qquad\qquad {\rm for } \ b,l^\pm \\
\nonumber
\Delta R &>& 0.4 \qquad\qquad {\rm for } \ bb, bl^\pm \\
\nonumber
m_{inv} &>&20\GeV {\rm \qquad  for } \ bb,
\label{accept-cut}
\end{eqnarray}
and
\begin{eqnarray} \nonumber
p_T&>&10,8,5\GeV {\rm \qquad for }\ \tau_h, \tau_e, \tau_{\mu}  \\
|\eta|&<&1.5 {\rm \qquad\qquad\qquad  for }\ \tau\\  
\nonumber
\Delta R &>& 0.4 {\rm \qquad\qquad\qquad for } \ \tau\tau, \tau b, \tau l^\pm \\
\nonumber
m_{inv} &>&10\GeV {\rm \qquad\qquad for } \ \tau\tau,
\label{tau-cut}
\end{eqnarray}
where 
$\tau_e$,$\tau_\mu$ and $\tau_h$ stand for the decays of
$\tau\to e \nu_e \nu_\tau$,
$\tau\to \mu \nu_\mu \nu_\tau$, and
$\tau\to \rm{ hadrons }+ \nu_\tau$,
respectively.
Lower cuts on $bb$ and $\tau\tau$ invariant masses are to eliminate the
large number of background events from soft photons and gluons.
 
The
momentum of the tau-lepton cannot be fully reconstructed since all tau decays
involve at least one neutrino, therefore the cuts on tau are applied
to the visible decay products, and are decay-mode dependent.
After these acceptance cuts,
$10-25$\% of the signal events survive, and the cross section becomes 
$0.6 \ (0.4)\ $fb for $m_h=90\ (130)\GeV$ 
with the given set of input parameters, or $C^2\sim 0.06$. 
The cross sections passing acceptance
are plotted in Fig.~\ref{xsec} 
 versus the Higgs mass, represented by the circled curve. 
There would be 
a few events to several tens of events with a few fb$^{-1}$
integrated luminosity, for $C^2 \sim 0.019-0.42$.

\subsubsection*{Irreducible Background}

The dominant source of the irreducible background
the $b\bar{b}$ pair from a virtual gluon splitting, 
the $\tau\bar{\tau}$ pair from an intermediate $Z^*/ \gamma^*$ and
the charged lepton plus missing energy from a $W$ boson. 
Our Monte Carlo simulations with MadEvent~\cite{Maltoni:2002qb} 
show that the $Z^*$ is almost on-shell.
It can be readily removed
with a cut on $\tau\bar{\tau}$ invariant mass. 
However as we shall see below, 
due to the small size of the background,
we can afford not to do so. 
A $\tau\bar{\tau}$ pair from a virtual photon 
can be more easily confused with the signal, 
but such a background is further suppressed by the 
electromagnetic couplings.

After applying the acceptance cuts, 
the irreducible background is estimated 
to be around 0.01 fb,
which is very small compared to the signal size.
It is entirely absent given the luminosity at the Tevatron.

\subsubsection*{Reducible Background and Tagging}

Taus and $b$'s need to be identified. 
During the identification (tagging), signal events are lost due to
the tagging efficiency and to additional cuts. For example,
the efficiency of tagging a single bottom is around 50\% 
in the region $p_T>15\GeV$, 
and falls off rapidly as we approach lower $p_T$.
Tagging (hadronic) taus faces the same problem.
Therefore tightening the kinematical cuts on $b$'s and taus
are necessary to assure favorable tagging efficiencies.
We decide to tag one $b$ and one tau.

The $b$- and hadronic $\tau$-tagging efficiencies are taken 
to be~\cite{Acosta:2005ij,Jeans:2005ew}
\begin{eqnarray}
\nonumber
b-tagging: & 50\% & {\rm for}\ E_T^{jet}>15\GeV \ and \ |\eta_{jet}|<1.0\ , \\
\tau-tagging: & 40\% & {\rm for}\ E_{vis}>20\GeV \ and \ |\eta|<1.5\ . 
\label{tautagging}
\end{eqnarray}
Outside these kinematical regions, 
the tagging efficiencies drop sharply~\cite{Acosta:2005ij,Jeans:2005ew}.


Reducible background arise from jets mis-identified as $b$'s, or
hadronically decaying taus. 
The mistag rate per jet is taken to be 
around $0.5-1.0$\% ($0.5$\%) for 
tau ($b$)~\cite{Acosta:2005ij,Jeans:2005ew}. 
In addition, the experiments cannot distinguish 
directly produced electrons (muons) from
leptonically decaying taus.
\begin{itemize}
\item 
The background due to misidentified bottom comes from the process
$2\tau 2j l + \MET$, 
which has a cross section of 5 fb. Considering the mistag rate and 
the additional cuts, it contributes 0.01 fb to the background events.

\item 
The background due to misidentified $\tau$
differs from the decay modes of $\tau$'s: 
  \begin{itemize}
    \item
For
$\tau_l \tau_h 2b l \MET$ ($2l2b \tau_h \MET$),
it comes from $2\tau 2b j$ with
$\MET$ from the leptonic decays of both taus. 
The contribution is estimated at 
0.003 fb.

    \item
For
$\tau_h \tau_h 2b l \MET$, the background
comes from $2j 2b l \MET$ estimated at $50$ fb 
after events of the $bb$ and $jj$ resonances around the $Z$ mass
are rejected. 
It's further reduced by a factor of 
$0.01-0.02$ from the tau-mistag rate,
and a factor of 0.8 due to b-tagging.

This results in a background rate of $0.4-0.8$ fb.
In a continuum distribution of 
$m_{bb}$, it is at or below the level of the signal. 
We notice that the $b$ jets are harder in this 
background than in the signal (see Fig.~\ref{ptdist}). 
Imposing a upper $p_T$ cut of $50\GeV$ will reduce the
background by a factor of about 4, 
while the signal is minimally affected.
  \end{itemize}

\item
The backgrounds from both a mistagged tau and a mistagged $b$ 
mostly come from the $4j l \MET$ events, 
which has a cross section of about 16 pb.
After the cuts and folding in the mistag rates, 
this contributes $0.3-0.6$ fb of background events.
It can be further reduced by imposing upper $p_T$ cuts, 
similarly to the  $2j 2b l \MET$ background.
\end{itemize}

\begin{figure}[htb]
    \includegraphics[scale=0.35]{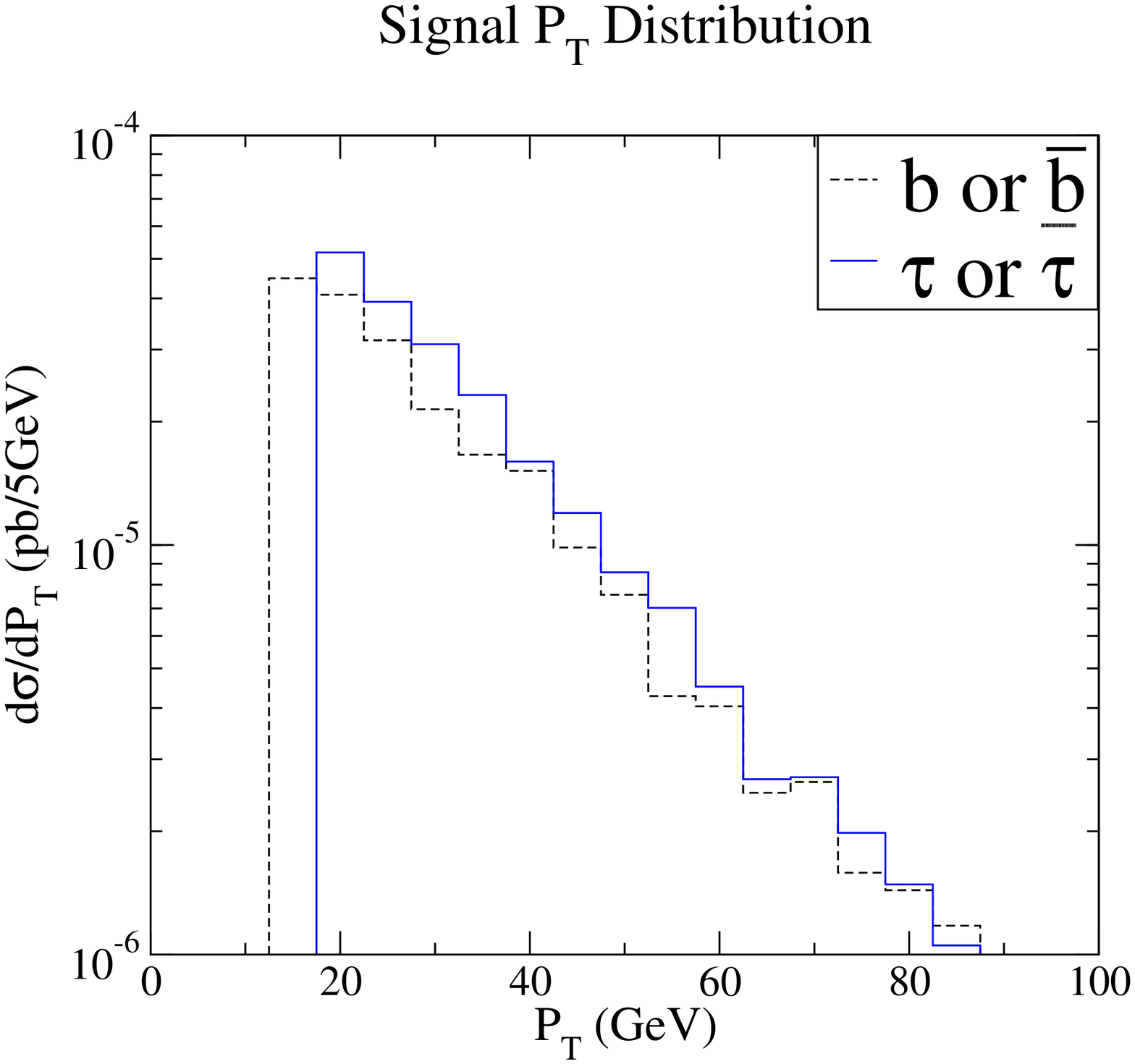}
    \includegraphics[scale=0.35]{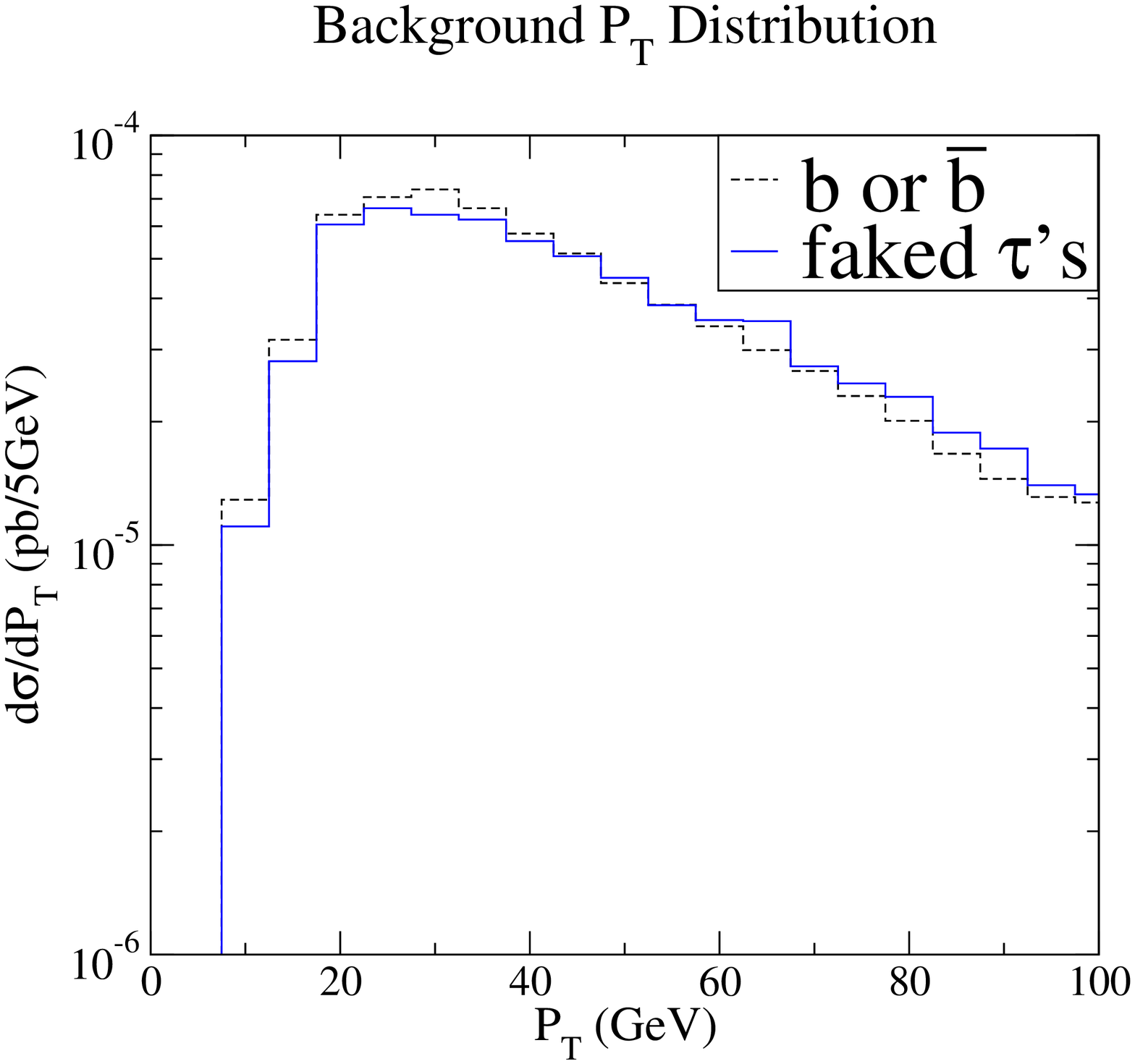}
    \caption {Transverse momenta distribution of the $b$, $\tau$ or jets in the
$Wh$ signal (left) and background (right) events.
}
    \label{ptdist}
\end{figure}

After carefully tightening the cuts, 
the reducible background can be a factor of a few to ten
smaller than the signal, 
but unfortunately, the cuts and the tagging efficiencies 
together reduce the signal greatly to about 0.08 fb for $Wh$ and
$0.03-0.05$~fb for $Zh$, with $C^2 \sim 0.06$.
With an optimistic $C^2 \sim 0.42$, 
the cross section is 0.55 fb,
we would expect to see about a couple of signal events
with an integrated luminosity of a few fb$^{-1}$. 

To illustrate a most optimistic situation, we explore the
mass relations of $m_a$ and $m_h$.
The signal loss is mainly due to the softness of the $b$ and $\tau$'s, 
therefore most events are rejected from the lower $p_T$ threshold. 
Increasing $m_a$ would stretch the $p_T$ distributions to the higher
$p_T$ end.
To achieve this without significantly affecting the decay phase space
of $h$, we set 
\begin{equation}
m_a=(m_h-10\GeV )/2,
\end{equation}
which resulted in almost doubling the signal rate, 
as seen from the curve with triangles in Fig.~\ref{xsec}.
The cross section ($\sim 0.2\ {\rm fb}$ for $C^2=0.06$, 
and $\sim 1.4\ {\rm fb}$ for $C^2=0.42$) 
is still challenging for observation 
with the Tevatron's projected luminosity.

\subsection{Summary}
The search for a Higgs boson with couplings to the gauge bosons of
the order of the SM-one, and decaying into two lighter CP-odd Higgs bosons 
states may  be performed at the Tevatron collider in the
$Wh\to l\nu aa (2b2\tau)$ channel. For $90 \leq m_h\leq 130\GeV$,  we 
found a sizable number of events, with negligible irreducible background. 
However, further cuts and tagging $b$ and $\tau$,  necessary to
remove the much larger reducible background 
worsen the signal event rate to the order 0.1 fb 
for a value of $C^2 \sim 0.06$, a factor determined by  the product of the 
relevant branching ratios times the ratio of the 
Higgs production cross section to the SM-one. Therefore, the
signal observation becomes statistically limited. For an optimal
choice of the value of the CP-odd Higgs mass $m_a$, the signal rate 
may be twice as large. With favorable couplings and branching fractions, 
the $C^2$ factor can be as large as 0.42, 
which would enhance the signal rate by a factor of 7 to around 1.4 fb.
It can be further improved by 
another $40-60\%$ by combining $Zh$ events with the $Wh$ events, leading
to a possible observation of a few events for a Tevatron luminosity of
the order of a few fb$^{-1}$.




\clearpage
\section {The $Z \to b\bar{b}$ decay as a b-jet energy calibration
  tool}
%
%
%
%
%
\textbf{Contributed by: T. Dorigo, J. Donini}

\vspace{0.25in}

We use a sample of $Z \to b\bar{b}$ decays free collected by the CDF
experiment to assist in both a precise measurement of the energy scale
of $b$-quark jets and a determination of the $b$-jet energy
resolution.  The more precise determination of $b$-jet energy scale
helps all precision measurements of the top quark mass and a
determination of the $b$-jet energy resolution is important for the
search of a low-mass Higgs boson.  This technique could also prove useful
at the LHC.
\subsection {Introduction}

\noindent
Since their discovery in 1983\cite{Phys.Lett.B122.103,Phys.Lett.B126.398,Phys.Lett.B122.476,Phys.Lett.B129.130},
$W$ and $Z$ bosons have been studied
at hadronic colliders only using their leptonic decays. As a matter of fact
the hadronic decays of these particles are generally so difficult 
to separate from the huge QCD background that, after 
the extraction of a nice mass bump in the dijet mass distribution 
by the UA2 collaboration in 1987\cite{Phys.Lett.B186.452,Z.Phys.C49.17}, little more has emerged. 

In fact, at the Tevatron things are more complicated for the 
direct observation of hadronic decays of vector bosons. With respect to
the $Sp\bar{p}S$, the Tevatron's higher center-of-mass 
energy is a disadvantage for once, because in the face of a four-fold
increase in signal cross section, the irreducible background from 
QCD processes yielding jet pairs increases by over an order of magnitude,
due to the steeply falling gluon PDF $g(x)$. 

Using Run I data, hadronic $W$ decays were successfully
used by the CDF and D0 experiments in the discovery and 
measurement of the top quark both in the single lepton and fully 
hadronic final state; the $W \to q \bar{q}'$
decay was used both in the event selection and as a constraint
in kinematical fits to extract the top quark mass. 
A handful of dijet masses peaking at 80 GeV were also directly 
observed in a subset of high-purity $t \bar{t}$ events\cite{Phys.Rev.Lett.80.5720}.
In Run II, with increased sample sizes, it has become possible to 
exploit the hadronic decay of $W$ bosons in top events even more, 
by using them for a direct 
calibration of the energy measurement of light-quark jets in the
reconstruction of the $t\bar{t}$ decay\cite{Abulencia:2005ak}.  
That technique has allowed a significant reduction of the systematic 
uncertainty arising from the knowledge of the jet energy scale, 
which is by now the largest contribution to the top mass measurement error.

For the $Z$ boson, which is not produced in top
decays and whose inclusive cross section in $p\bar{p}$ collisions
is three times smaller than that
of the $W$, the extraction of hadronic decays is even more complicated;
only the decay to $b$-quark pairs reaches the level of observability,
thanks to the significant reduction of QCD processes provided by the
distinct signature of $b$-quark jets. Indeed, a small signal of 
$Z \to b\bar{b}$ decays was extracted by CDF in Run I data
exploiting the semileptonic
decay of $b$ quarks with an inclusive muon trigger of low $P_T$\cite{HEP-EX/9806022}. 
The signal was too small to allow any study of $b$-jet energy and 
resolution, but its 
demonstrated observability in the Tevatron environment gave hope to the
searches for the analogous signature of a low-mass Higgs boson decay, 
and spurred the development of a dedicated trigger for Run II, capable of 
collecting a large $Z$ signal without the need to 
rely on the semileptonic decay of $b$ quarks.

A large-sized signal of $Z \to b\bar{b}$ 
decays free from selection biases allows both a precise
measurement of the energy scale of $b$-quark jets and a determination
of the $b$-jet energy resolution. The reduction of the uncertainty in 
the $b$-jet energy scale helps all precision measurements of the top 
quark mass, while a determination of the $b$-jet energy resolution is
important for the search of a low-mass Higgs boson. The signal, most
notably, opens the doors to a direct test of algorithms that attempt
to increase the resolution of the $b$-jet energy measurement. These
algorithms are a critical ingredient for the observability of the 
Higgs boson at the Tevatron if $M_H<135$ GeV.

\subsection {Triggering on $Z\to b\bar{b}$ decays}

\noindent
In Run II CDF benefits from a hardware tracker using silicon detector
hits at the second trigger level, the Silicon Vertex Tracker (SVT)\cite{Nucl.Instr.Meth.A518.532}.

The SVT works by comparing the pattern of hits in the five layers of 
silicon sensors of the Silicon Vertex Detector (SVX)\cite{Nucl.Instr.Meth.A447.1} 
to those expected by charged tracks of given transverse
momentum, azimuth, and impact parameter, which are stored in 256 
associative memory chips. Use of the 12-fold azimuthal symmetry of the
SVX reduces the number of needed patterns and allows a
parallelization of the task of finding track candidates and
performing linearized fits. On average as little as 15 $\mu s$ are 
needed to process
an event and determine the impact parameter of tracks with a resolution
of 35 $\mu m$. The efficiency to reconstruct fiducial tracks with 
$P_T>2$ GeV is close to $90\%$.

Using SVT information as well as calorimetric
input, the $Z \to b\bar{b}$ trigger selects events containing two 
back-to-back $E_T>10$ GeV jets and two $P_T>2$ GeV tracks whose 
impact parameter with respect to the beam line is larger than 160 $\mu m$; 
a veto on forward jets with $E_T>3$ GeV is also applied to reduce 
QCD backgrounds. These requirements have an efficiency of about $4\%$ on
$Z \to b\bar{b}$ decays, and they result in an effective cross section 
lower than 100 nb, which corresponds to a manageable rate
for machine luminosity up to $L = 10^{32} cm^{-2} s^{-1}$. 

As the luminosity grows, so does the average number of multiple 
interactions occurring in the same bunch crossing. Since the $Z$ signal 
can only be isolated in clean events with two back-to-back jets 
and little extra jet activity, it is reasonable to foresee a dynamic
prescaling of the trigger, which should anyway allow the collection
of at least 2 fb$^{-1}$ of data with the base data collection plan
of Tevatron's Run II.
A sample of 80,000 signal events is thus achievable.

\subsection {Preliminary Run II results}

\noindent
A signal of $Z$ decays to $b$-quark pairs has been observed in 
333 pb$^{-1}$ of CDF data collected by the trigger described above. 
After a reconstruction of jets with a $R=0.7$ cone algorithm\cite{Bhatti:2005ai}, 
events were selected by requiring two jets of raw transverse energy 
exceeding 20 GeV in the rapidity interval $|\eta|<1.5$, both of them 
containing a secondary vertex ($b$-tag) reconstructed by the SecVtX 
algorithm\cite{Phys.Rev.D71.052003}. 

After those requirements the $Z$ signal is still buried in a 
very large background consisting predominantly of QCD direct 
$b\bar{b}$ production, which needs to be reduced further. 

Most direct $b\bar{b}$ pairs are produced at the Tevatron 
by gluon fusion, whose high color charge in the initial 
state and color flow topology are distinctive characteristics. 
To exploit the smaller probability of QCD radiation from the 
initial state quarks in $Z$ boson production, the two leading jets 
are required to be back-to-back in azimuth within $\Delta \Phi_{jj}>3$, 
and events containing a third jet with raw $E_T^3>10$ GeV are discarded.

\begin{figure}[h!]
\includegraphics[width=16cm]{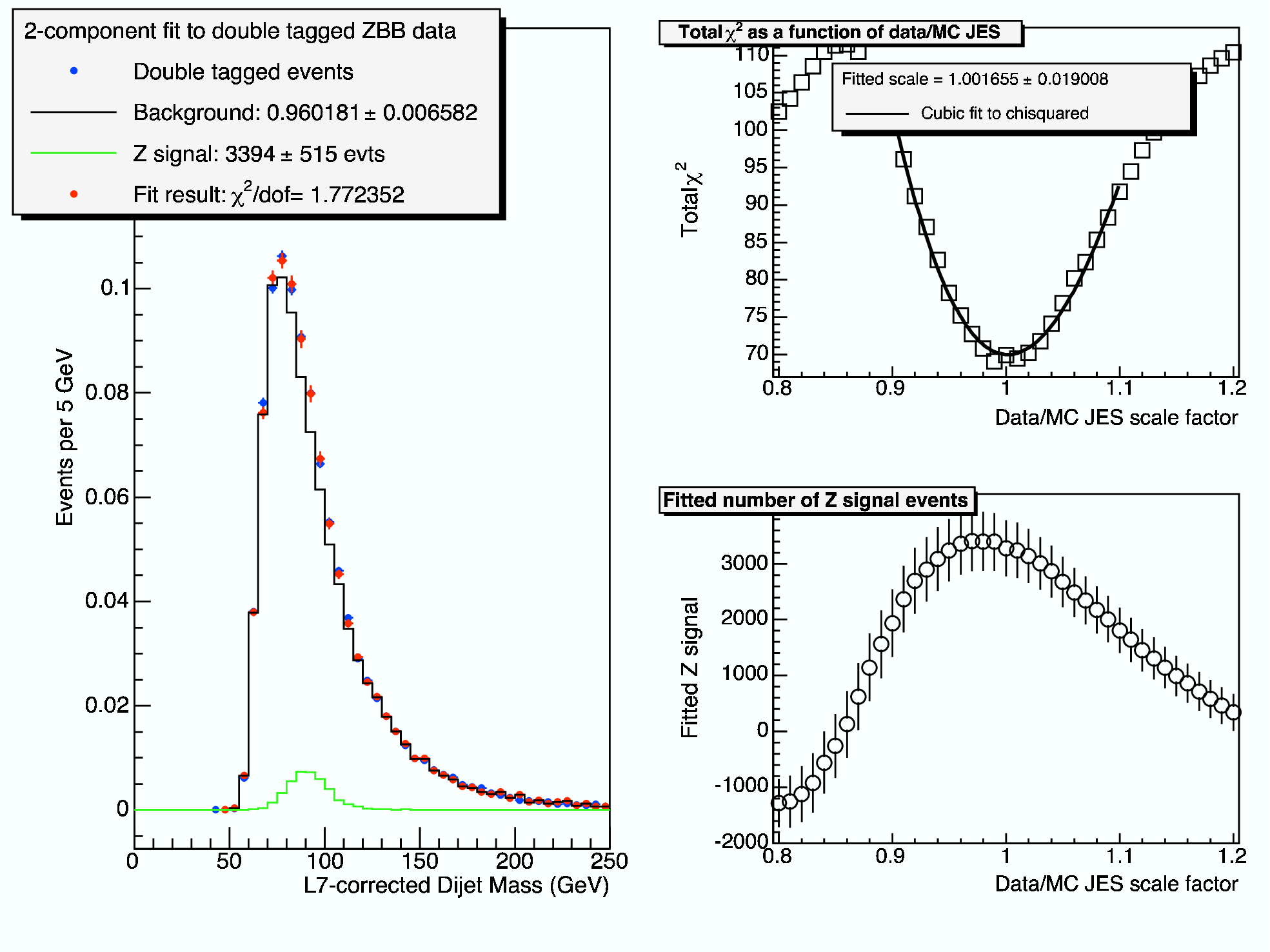}
\caption{The $Z\to b\bar{b}$ signal extracted by CDF with 333 pb$^{-1}$ of
Run II data. Left: the dijet mass of events with two $b$-tags is fit as the
sum of a background template (in grey) and a signal template (in green).
Top right: fit $\chi^2$ as a function of the $b$-JES. Bottom right:
number of signal events from the fit as a function of the $b$-JES.}
\label{f:zbbprplot}
\end{figure}

The resulting sample of 86,000 events contains roughly 3400 $Z$ boson
decays. Their reconstructed dijet invariant mass can be fit using as a 
background template the mass distribution of dijet events
which do not contain secondary vertices, by accounting for the
bias due to the non-flat $b$-tag probability versus dijet mass using
a correction function; the latter is obtained from events 
failing the kinematical requirements. 
For the $Z$ signal, 40 Monte Carlo-derived templates with a varying 
$b$-jet energy scale factor (JES) from 0.8 to 1.2 in steps of 0.01 
are used in turn. One thus obtains the dependence of fit $\chi^2$ 
on the $b$-JES, from which a measurement of the latter can be extracted.
Fig.~\ref{f:zbbprplot} shows the results for the best fit, which 
corresponds to a $b$-JES of unity, with a statistical uncertainty smaller
than $2\%$.

\subsection {Prospects for the B-jet energy scale extraction}

\noindent
The largest contribution to the total uncertainty in the top quark
mass determination at the Tevatron originates from the knowledge of the
jet energy scale, a factor which measures
the discrepancy between the effect of detector response and energy 
corrections in real and simulated hadronic jets.
The JES can be determined from studies of photon-jet balancing\cite{jetclu}, 
but modeling and selection biases limit the accuracy of the method;
a determination which is mostly statistics-limited comes 
instead from the measurement of $W \to q \bar{q}'$ decays in top events.
With these methods, the Tevatron experiments 
can reach a precision close to $1\%$ on the top quark mass in Run II 
by reducing the JES uncertainty to a similar level.
 
When dealing with $b$-jets, however, one has to cope with several
peculiarities of their fragmentation and decay properties, and with
the different color connection of $b$-jets and light quark jets in
top quark decay. All these effects have to be accurately modeled if one is to
use a generic JES factor extracted from jets not containing heavy flavors
to the two $b$-jets always present in a $t\bar{t}$ decay.
A recent study predicts that the uncertainties in the modeling of 
fragmentation, decay characteristics and color flow may affect the 
knowledge of the $b$-JES by as little as $0.6\%$\cite{arguin}, 
but a direct determination of that quantity is of course preferable.

Due to the small cross section of production processes yielding 
events with a high-energy photon recoiling against a $b$-quark jet, 
a measurement of the $b$-JES with balancing techniques is difficult,
although both D0 and CDF have recently started exploring that 
option.

The preliminary $Z \to b\bar{b}$ signal extracted by CDF appears to 
demonstrate that the data have a sufficient statistical power to allow
the determination of a precise $b$-JES factor: one expects that 2 fb$^{-1}$
of data will reduce the statistical uncertainty of template fits well 
below $1\%$.
However, systematics are a concern: given the smallness of the signal
to noise ratio of the selected sample, a meaningful determination 
of the $b$-JES from bump fitting requires that the background shape 
be modeled with the utmost accuracy, especially if its most probable value 
occurs close to that of the $Z$ signal, as in Fig.~\ref{f:zbbprplot}. 

Reducing the 20 GeV threshold on raw transverse energy of the jets, which
directly affects the peak position of the background distribution, is
however problematic, since low-$E_T$ jets suffer from subtle trigger
effects which are hard to model correctly; moreover, at very low $E_T$
it has been shown that $c\bar{c}$ production is a sizable contribution
to the SVT-triggered dataset, 
and its presence complicates the understanding of the 
$b$-tagging bias in the data.

Recent studies have shown that using large Monte Carlo samples of 
direct $b\bar{b}$ production processes and careful parametrizations of 
trigger biases it is possible to check the
background shape extracted from the data as described above, and
reduce the associated systematic uncertainty in the determination of
the $b$-JES. CDF plans to use that additional information to finally
achieve a significant measurement of the $b$-jet energy scale with
$Z \to b\bar{b}$ decays, thus justifying several years of investigation
of the extraction of a well-known signal.

\subsection {B-jet energy resolution studies}

The mass resolution of pairs of $b$-jets has been duly stressed 
as one of the critical factors in the search for a light Higgs
boson decay at the Tevatron. While the 1999 study
of the Tevatron Higgs Working Group\cite{Carena:2000yx} could only make the educated
guess that a $\sigma_M/M_{bb} \sim 10\%$ relative mass resolution 
was attainable with
a dedicated effort, the Higgs Sensitivity Working Group\cite{Babukhadia:2003zu}
went as far as producing some evidence that such precision was indeed
reachable, by a careful use of several
corrections in series, followed by the exploitation of the
correlations between kinematic variables measured in $WH \to l \nu b\bar{b}$ 
events and the induced biases in the dijet mass measurement 
(see Fig.~\ref{f:hswg}).

\begin{figure}[h!]
\includegraphics[width=16cm]{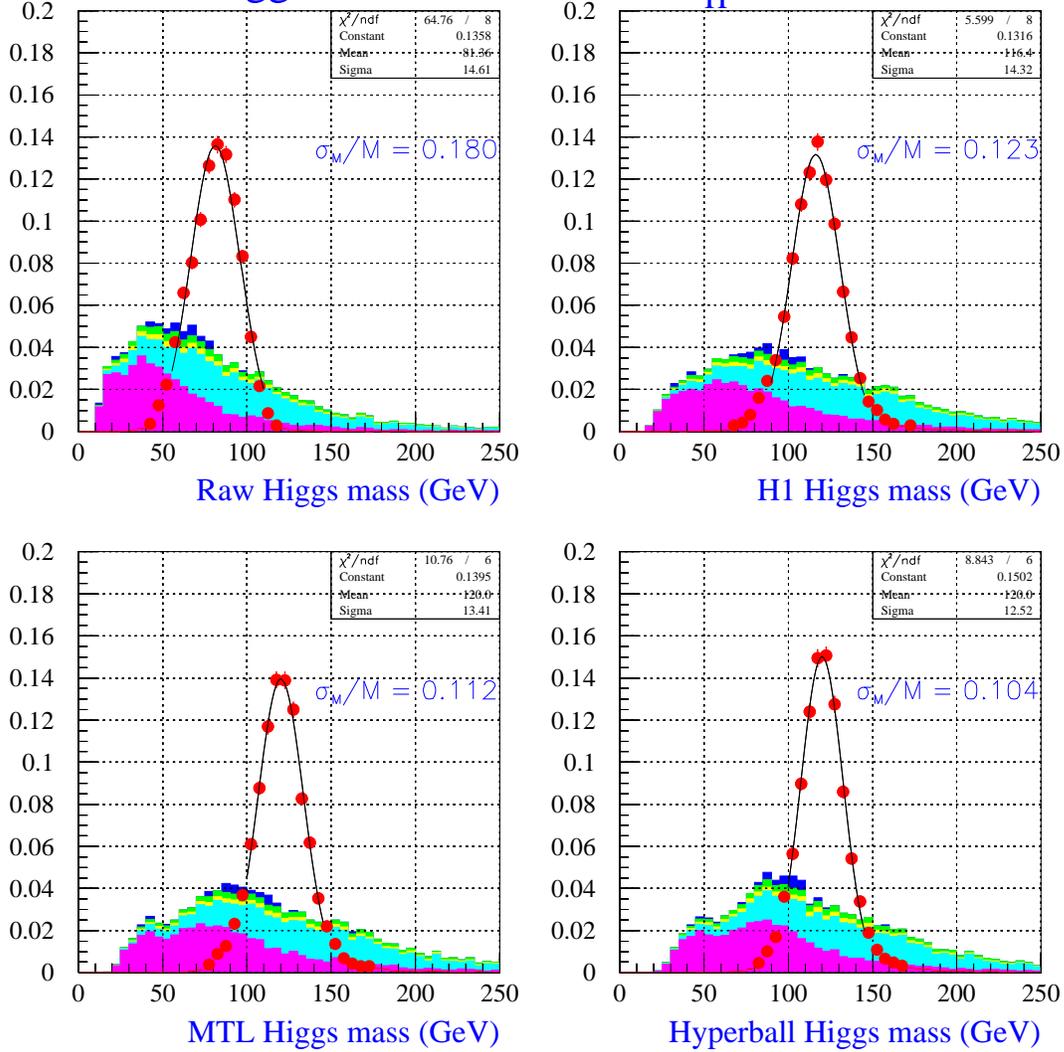}
\caption{Dijet mass distribution for pairs of $b$-jets after different
levels of jet energy corrections: 
raw jet energies (top left), energies corrected
with the H1 algorithm (top right), then after the subsequent application
of $b$-specific corrections (bottom left), and finally after the
use of the hyperball algorithm, a method that corrects the dijet mass
accounting for the correlation with event observables. The red points
describe the behavior of $WH \to l \nu b\bar{b}$ events; the stacked
histograms are Monte Carlo simulations of $W+$ jets (purple), $t\bar{t}$ production (cyan), single top production (green and yellow), and $WZ$ production
(blue).  } 
\label{f:hswg}
\end{figure}

Indeed, when compared to any selection applied on the data to increase
the signal to noise ratio, an algorithm that reduces the width of a
resonance sitting on top of a large background has the obvious advantage
of keeping intact the size of the signal. If signal significance is on
the yardstick, a $20\%$ decrease of $\sigma_M/M_{bb}$ can be shown to 
have the same effect of a $20\%$ increase in collected luminosity\cite{Babukhadia:2003zu}.

\begin{figure}[h!]
\includegraphics*[width=16cm]{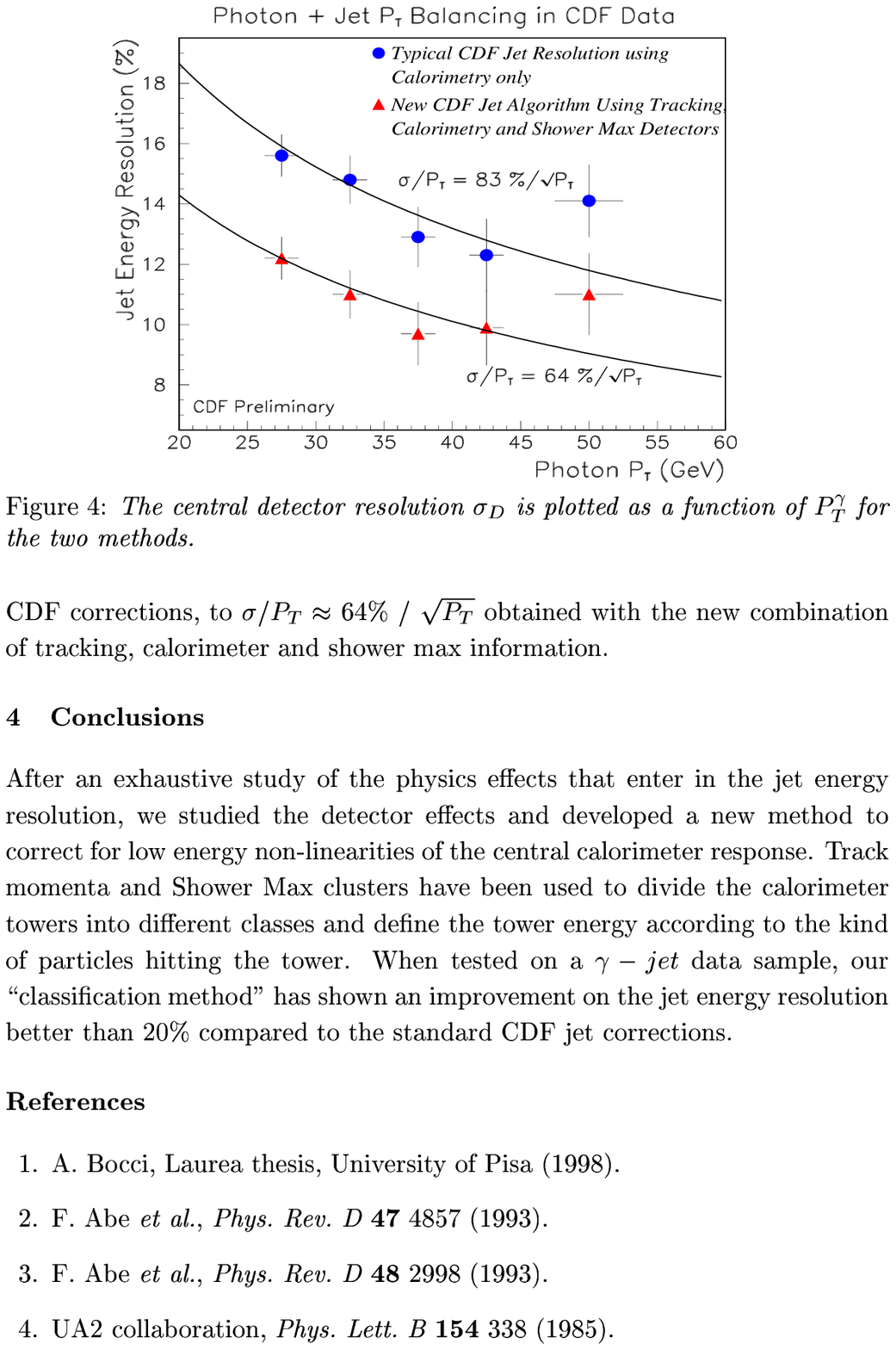}
\caption{ Jet energy resolution measured in photon+jet data as a function
of photon $P_T$. The blue points show the resolution of the standard
jet energy corrections, the red points show the results of a dedicated
algorithm exploiting information from many subdetectors.}
\label{f:trackcal}
\end{figure}

The resolution in the transverse energy of generic jets can be
measured with $\gamma-jet$ events. Those events have in fact constituted
the basis of CDF studies of an algorithm exploting both the tracker, the
shower max detector, and the calorimeter to increase the precision of 
the transverse energy measurement\cite{jcor2k}. As Fig.~\ref{f:trackcal}
shows, a $30\%$ improvement in the resolution of generic jets can be
achieved by a combined use of the information from different subdetectors. 
Unfortunately, the lack of sizable samples
of data containing a $b$-quark recoiling against an energetic photon
prevents a data-driven study of the $b$-jet energy resolution, and
a check of the effect of generic corrections applied to $b$-jets.

The development of a $b$-specific algorithm, aimed at increasing the 
$E_T$ resolution on jets containing $b$-tags, highly
profits from the availability of a statistically significant sample
of $Z \to b\bar{b}$ decays: one can then both check the effectiveness of
any recipe and measure the resulting mass resolution.

As was noted above, $b$-jets are different from generic jets originated
from light quarks or gluons in several aspects. $23\%$ of $b$-quarks
decay semileptonically, and more soft leptons are yielded by the
following charm quark decay; the large mass of the $b$-quark produces
tracks with significant transverse momentum with respect to the jet axis;
and finally,  $b$-quarks have a hard fragmentation function, 
which may translate in an average detector response different
from that of generic jets. The total effect of these peculiarities
is a worse $E_T$ resolution for $b$-jets and a significant
negative bias, mainly due to the neutrinos from semileptonic decay.

$B$-jets which are tagged by a vertex-finding algorithm are also 
different from an experimental point of view, since the detection of
a displaced vertex allows the measurement of several ancillary characteristics:
the distance between primary and secondary vertex, the total charge of
tracks forming the secondary vertex, the total transverse momentum and
combined mass of the charged decay tracks. 

All these observable quantities can be exploited by algorithms 
detecting the correlation between their values and the average bias
on the jet $E_T$ measurement. For instance, the presence of a muon in
a jet is strongly correlated with the resulting calorimeter response,
such that the muon $P_T$ can be used with success to increase the $E_T$
resolution. The best results are obtained when all correlations are
exploited together, by finding the most probable bias in the $E_T$
measurement as a function of the value of all observed jet variables.

\begin{figure}[h!]
\includegraphics[width=16cm]{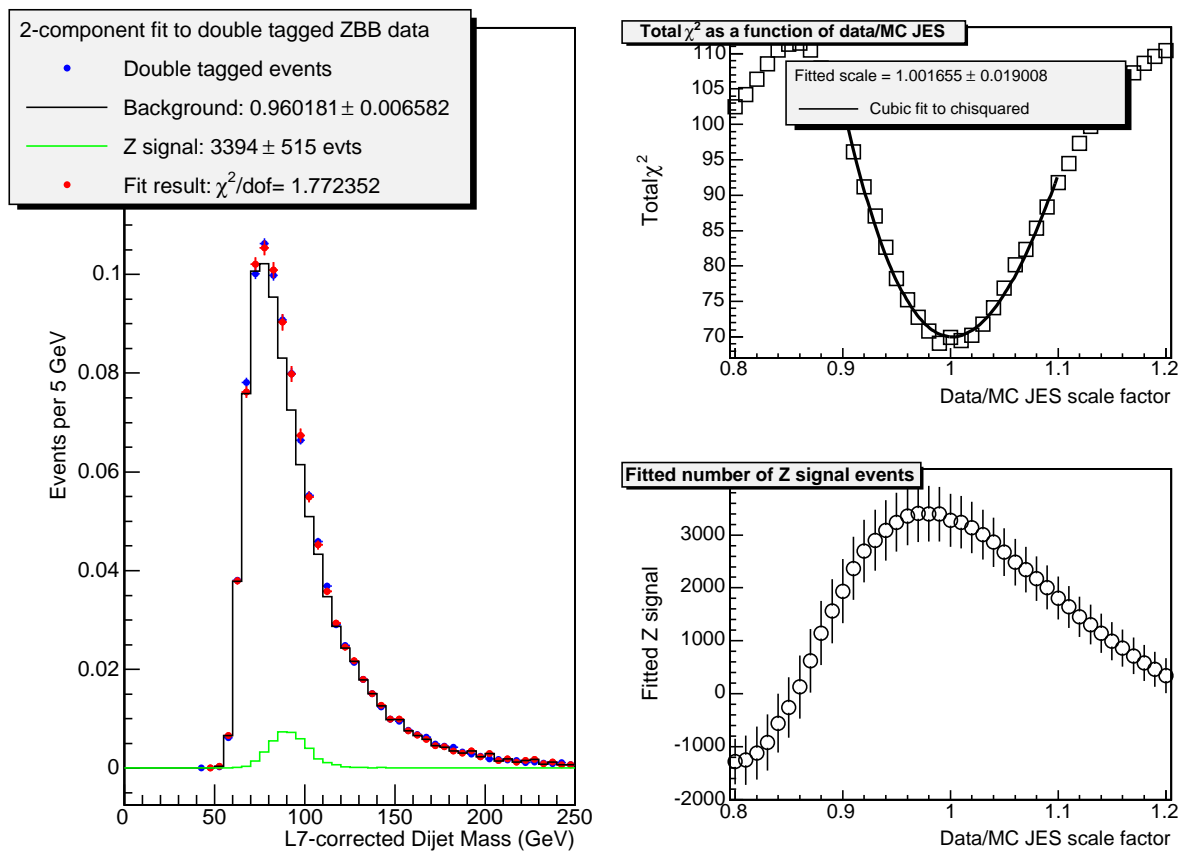}
\caption{Relative $E_T$ resolution on $b$-tagged jets from a QCD $b\bar{b}$
Monte Carlo simulation, after generic jet corrections (top left, red points)
and after $b$-specific jet corrections using all jet observables 
(bottom left, blue points). Right: a comparison of the $E_T$ resolution 
obtained with the two corrections.}
\label{f:hbjc}
\end{figure}

Preliminary results by the CDF collaboration have determined that the
$E_T$ resolution for $b$-jets can be improved by as much as $30\%$ from
the baseline resolution yielded by the application of standard, 
non-$b$-specific energy corrections (see Fig.~\ref{f:hbjc}). 
Two algorithms are being developed
for that purpose, and the study on the $Z \to b\bar{b}$ signal will
prove their effectiveness in the near future.

%
%
%
%
%
%
%
%
%
%
%
%
%
%
%
%
%
%
%


\clearpage
\section{Selected Topics in Standard Model Higgs searches using $H\rightarrow W^+W^-$ and
$H\rightarrow\tau^+\tau^-$ Decays at the LHC}
\label{sec:atlas}
\textbf{Contributed by: B. Mellado, W.~Quayle, S.~L.~Wu}

\vspace{0.25in}

We study control samples for an in-situ determination of the major
backgrounds to $H\rightarrow W^+W^-\rightarrow l^+l^-\nu\nu$, where
a full jet veto is applied. We find that the theoretical uncertainty
on the extrapolation of the QCD $W^+W^-$ background from the control
sample to the signal-like region is $5\,\%$ and that the impact of
the singly-resonant top background on the extrapolation from a
b-tagged control sample is about 10\,\%. We therefore conclude that
it is possible to perform a data-driven estimate of the background
in the signal-like region with sufficient accuracy to achieve a
$5\,\sigma$ discovery of a $160\,\gev$ Higgs with less than
2~fb$^{-1}$ of integrated luminosity. We evaluated the
Next-to-Leading Order corrections to Higgs production in the
analysis $H\rightarrow\tau^+\tau^-$ in association with one High
$P_T$ jet. The ratio of NLO to LO cross-sections after the
application of analysis cuts is in the range $1.5\div 1.6$ for Higgs
production via gluon-gluon fusion.

\subsection{Introduction}

We study control samples for an in-situ determination of the major
backgrounds to $H\rightarrow W^+W^-\rightarrow l^+l^-\nu\nu$, where
a full jet veto is applied. We find that the theoretical uncertainty
on the extrapolation of the QCD $W^+W^-$ background from the control
sample to the signal-like region is $5\,\%$ and that the impact of
the singly-resonant top background on the extrapolation from a
b-tagged control sample is about 10\,\%. We therefore conclude that
it is possible to perform a data-driven estimate of the background
in the signal-like region with sufficient accuracy to achieve a
$5\,\sigma$ discovery of a $160\,\gev$ Higgs with less than
2~fb$^{-1}$ of integrated luminosity. We evaluated the
Next-to-Leading Order corrections to Higgs production in the
analysis $H\rightarrow\tau^+\tau^-$ in association with one High
$P_T$ jet. The ratio of NLO to LO cross-sections after the
application of analysis cuts is in the range $1.5\div 1.6$ for Higgs
production via gluon-gluon fusion.

The search for the Higgs boson called for by the Standard Model is
arguably one of the most important topics in high-energy particle
physics today. For the early observation of a Higgs boson of mass
$115<M_H<135\,\gev$ the most relevant final states involve
$H\rightarrow\gamma\gamma$ and
$H\rightarrow\tau^+\tau^-$~\cite{Asai:2004ws}.
For the range of masses $135<M_H<190\,\gev$ the most promising decay
mode of the Standard Model Higgs boson is $H\rightarrow
W^+W^-$~\cite{pr_55_167}.

 In this work, we discuss selected topics related to
the search for the Standard Model Higgs boson at LHC using
$H\rightarrow W^+W^-$ and $H\rightarrow\tau^+\tau^-$. In
Section~\ref{sec:hww} we describe our Monte Carlo samples, event
selection, methods for in-situ background determination for the
channel $H\rightarrow W^+W^-\rightarrow l^+l^-\nu\nu$ with a full
jet veto. Section~\ref{sec:htautau} reports on studies of QCD higher
order corrections to Higgs signal production in the analysis of
$H\rightarrow\tau^+\tau^-$ in association with one high $P_T$
jet~\cite{pl_b611_60}.

\subsection{Selected Topics in \boldmath{$H\rightarrow W^+W^-\rightarrow l^+l^-\nu\nu$} Analysis}
\label{sec:hww}

This Section is subdivided into four sub-sections.
Sub-section~\ref{Monte Carlo and Analysis Method} describes the
Monte Carlo samples and the analysis method used in the analysis for
the search of the Higgs boson with $H\rightarrow W^+W^-\rightarrow
l^+l^-\nu\nu$ when applying a veto on events with high $P_T$ jets.
We also discuss data-driven methods for the extraction of the
backgrounds. In sub-sections~\ref{Theoretical Uncertainties in the
WW Background} and~\ref{Theoretical Uncertainties in the Top
Background} we discuss the theoretical uncertainties in the
background extraction procedures. In sub-section~\ref{Comparison of
MC@NLO, Alpgen, and Sherpa}, we perform a brief comparison of three
generators for the $W^+W^-$ background for validation purposes.

\subsection{Monte Carlo and Analysis Method}\label{Monte Carlo and
Analysis Method} We consider the following signal and background
processes:
\begin{itemize}
\item Higgs production.
We model the gluon-initiated process with the generator provided in
MC@NLO and normalize the cross-section for the signal to the values
obtained used HIGLU~\cite{Spira:1996if}.  The small contribution from Weak
Boson Fusion (VBF) is modelled with
Pythia~\cite{Sjostrand:1993yb,Sjostrand:2000wi}.
\item QCD $W^+W^-$ production is modelled with the generator provided in MC@NLO version
3.1~\cite{JHEP_0206_029,JHEP_0308_007}. A non-negligible number of
$W^+W^-$ events come from $gg\rightarrow W^+W^-$ diagrams that are
not included in MC@NLO; we model this contribution using the
generator documented in~\cite{JHEP_0503_010}.
\item $t\overline{t}$ production. The (dominant) doubly-resonant contribution is modelled with MC@NLO.
To estimate the impact of the singly-resonant and non-resonant
$W^+W^-bb$ contributions to the background, we perform a comparison
between leading-order calculations of $pp\rightarrow W^+W^-bb$ and
$pp\rightarrow t\overline{t}\rightarrow W^+W^-bb$ using
MadEvent~\cite{Stelzer:1994ta,hep-ph_0208156}.
\item QCD $Z/\gamma$ production, with $Z\rightarrow ee/\mu\mu/\tau\tau$.  We model this background with MC@NLO.
\end{itemize}
Although we do not expect detector effects to be important in this
calculation, it is convenient to simulate a detector using the last
fortran-based release of ATLFAST, and we apply the jet energy
corrections in ATLFAST-B~\cite{ATL-PHYS-98-131}.\footnote{We also apply a
small correction to the energy of jets for which HERWIG was used for
the parton showering and hadronization; the correction is given by
$(1-5\times10^{-5} P_{T}^{jet}+0.042)$ where the jet $P_{T}$ is
measured in GeV.}

\begin{table}[t]
\begin{center}
\begin{tabular}{ c | c c | c c c c c}
\hline \hline
Cut             & $gg\rightarrow H$     & VBF   & $t\overline{t}$       & EW $WW$         & $gg\rightarrow WW$    & $qq\rightarrow WW$        & $Z/\gamma^{*}$\\
\hline \hline
Trigger and $Z$ rej. & 185    & 25.1   & 7586   & 11.4 & 48.5   & 792    & 151   \\
Hard Jet Veto        & 90.0   & 1.48   & 51.6   & 0.16 & 21.2   & 451    & 31.4  \\
B Veto          & 89.6   & 1.46   & 37.6   & 0.16 & 21.1   & 449    & 30.8  \\
$P_{T}^{Higgs}$ & 53.2   & 1.23   & 33.0   & 0.09 & 13.1   & 177    & 23.6  \\
$M_{ll}$        & 42.9   & 1.10   & 7.85   & 0.02 & 6.31   & 65.2   & 22.0  \\
$\Delta\phi_{ll}$ & 33.1 & 0.93   & 5.23   & 0.02 & 5.14   & 42.8   & 0.07  \\
\hline \hline
\end{tabular}
\caption{Cut flows (in fb) for $M_{H}=160$ GeV in the $e\mu$
channel. } \label{cutflow}
\end{center}
\end{table}
\begin{table}[tbp]
\begin{center}
\begin{tabular}{ c | c c | c c c c c c }
\hline \hline
Sample  & $gg\rightarrow H$     & VBF   & $t\overline{t}$       & EW $WW$         & $gg\rightarrow WW$ & $qq\rightarrow WW$       & $Z\rightarrow \tau\tau$\\
\hline \hline
Primary          & 1.86   & 0.03        & 33.4   & 0.08 & 6.19   & 121.0  & 7.96\\
b-tagged         & 0.18 & 0.007        & 17.02  & 0.0001      & 0.08 & 1.51  & 1.29\\
\hline \hline
\end{tabular}
\caption{Cross-sections (in fb) in the two control samples discussed
in Section~\ref{Monte Carlo and Analysis Method} for $M_{H}=160$
GeV, summed over lepton flavor. } \label{controlSamples}
\end{center}
\end{table}

Our event selection consists of the following cuts:
\begin{itemize}
\item Trigger and Topology cuts.  We require that the event has exactly two leptons with transverse
momentum greater than 15 GeV in the region with $|\eta|<2.5$, and we
apply a lepton identification efficiency of 90\% for each lepton.
The dilepton invariant mass is required to be less than 300~GeV.
\item $Z$ rejection.  The event is rejected if the leptons have an invariant mass between 82 and 98 GeV.
We require a large missing transverse momentum $P_T>30$GeV,
which is raised to 40 GeV if the two leptons have the same flavor.
To reduce the nontrivial background from the decay
$Z\rightarrow\tau\tau\rightarrow ll +P_T$, we calculate,
using the collinear approximation, $x_{\tau}^{1}$ and
$x_{\tau}^{2}$, the energy fractions carried by the visible decay
products of the $\tau$ leptons, and $M_{\tau\tau}$, the invariant
mass of the two $\tau$ leptons.  We reject the event if
$x_{\tau}^{1}>0$, $x_{\tau}^{2}>0$, and
$|M_{\tau\tau}-M_{Z}|<25$GeV.
\item Jet Veto.  We reject the event if there are any jets with $P_{T}>30$GeV anywhere in the detector, or
if it contains any b-tagged jets with $P_{T}>20$GeV and
$|\eta|<2.5$.  We assume a b-tagging efficiency of 60\% with
rejections of 10 and 100 against jets from $c$ quarks and light
jets, respectively.
\item Transverse momentum of the Higgs candidate. We require that $P_{T}^{Higgs}>11.1$~GeV.
\end{itemize}
In the signal-like region, we apply three more cuts:  we require
that the dilepton mass have $6.3<M_{ll}<64.1$~GeV, that the
azimuthal opening angle between the leptons satisfy
$\Delta\phi_{ll}<1.5$ radians, and that the transverse mass obey
$50<M_{T}<M_{H}+10$~GeV.  The cross-sections after successive cuts
for a representative Higgs mass of 160~GeV in the $e\mu$ channel are
shown in Table~\ref{cutflow}.
\begin{figure}[tb]
\begin{center}
\includegraphics[width=12cm]{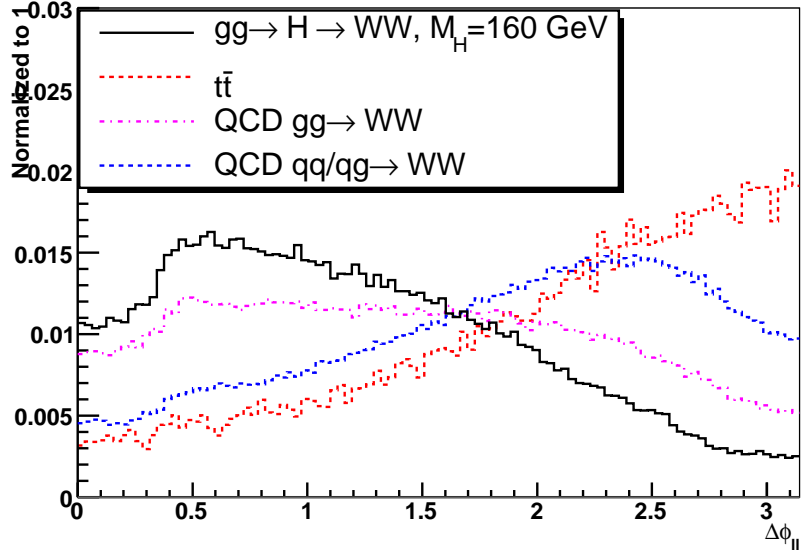}
\caption{ The
distribution of the azimuthal angle (in the transverse plane)
between the leptons after cuts. } \label{dPhill_control}
\end{center}
\end{figure}
We also consider two control samples:
\begin{itemize}
\item The primary control sample is defined the same way as the signal-like region, but with
different cuts on the dilepton opening angle in the transverse plane
and the dilepton invariant mass. We require $\Delta\phi_{ll}>1.5$
radians and $80<M_{ll}<300$~GeV; we remove the cut on the transverse
mass.
\item The b-tagged control sample cuts are the same as in the primary
control sample, except that instead of applying a b-jet veto, we
require that there be a b-tagged jet with $P_{T}$ between 20 GeV and
30~GeV; we also remove the lower bound on the dilepton invariant
mass.
\end{itemize}
Table~\ref{controlSamples} shows the cross-sections in these two
control samples. In order to make meaningful estimates of systematic
errors, it is helpful to define the following three quantities:
\begin{itemize}
\item $\alpha_{WW}$:  The ratio of the QCD $W^+W^-$ cross-section in the signal-like region over the QCD $W^+W^-$ cross-section
in the primary  control sample.
\item $\alpha_{tt}$:  The ratio of the $t\overline{t}$ cross-section in the signal-like region over the $t\overline{t}$
cross-section in the b-tagged control sample.
\item $\alpha_{tt}^{WW}$:  The ratio of the $t\overline{t}$ cross-section in the primary control sample over the
$t\overline{t}$ cross-section in the b-tagged control sample.
\end{itemize}
With these ratios taken from Monte Carlo, we estimate the number of
$t\overline{t}$ events in the signal-like region as
$N_{tt}^{signal-like}=\alpha_{tt} N_{b-tagged}$ and the number of
$W^+W^-$ background events in the signal-like region as
\begin{eqnarray}
N_{WW}^{signal-like}=\alpha_{WW} N_{WW}^{primary}=\alpha_{WW}
(N_{total}^{primary}-\alpha_{tt}^{WW}
N_{b-tagged}-small~backgrounds)\nonumber
\end{eqnarray}
where $N_{tt}^{b-tagged}$ is the number of events in the b-tagged
control sample, $N_{total}^{primary}$ is the total number of events
in the primary control sample, and the ``$small~backgrounds$''
consist mostly of Drell-Yan events.


\subsection{Theoretical Uncertainties in the $W^+W^-$ Background}\label{Theoretical Uncertainties in the WW Background}
We begin with the theoretical uncertainties in the extrapolation
coefficient $\alpha_{WW}$. Here, the theoretical error is dominated
by the uncertainty in the normalization of
 the $gg\rightarrow W^+W^-$ contribution;  recent studies
have shown that this contribution can be in excess of 30\% for the
cuts used in those studies~\cite{JHEP_0503_010,JHEP_0505_064}.

We compute the the theoretical error as the sum in quadrature of the
uncertainty due to the fit error in the parton density function
parameterization and the uncertainty due to the choice of $Q^{2}$
scale.  To estimate the parton density function (PDF) uncertainty,
we have used the CTEQ6 PDF set and its error sets; using equation
(3) in~\cite{JHEP_0207_012}, we find that the uncertainty in
$\alpha_{WW}$ is 2.8\%. To assess the uncertainty due to the choice
of $Q^{2}$ scale, we have varied the renormalization and
factorization scales by factors of 8.\footnote{This is an unusually
large scale variation to choose; typically, a scale uncertainty will
be quoted based on a scale variation of 2 or at most 4.  Our
motivation for this choice is the fact that we expect the K-factor
for $gg\rightarrow W^+W^-$ to be large, since the K-factor for
$gg\rightarrow\gamma\gamma$ has been calculated and it is slightly
less than 2~\cite{pr_66_074018}.} We examine four choices of scale
variations: Scale 1 has $Q_{ren}\rightarrow 8 Q_{ren}$,
$Q_{fac}\rightarrow Q_{fac}/8$; Scale 2 has $Q_{ren}\rightarrow
Q_{ren}/8$, $Q_{fac}\rightarrow 8 Q_{fac}$; Scale 3 has
$Q_{ren}\rightarrow 8 Q_{ren}$, $Q_{fac}\rightarrow 8 Q_{fac}$; and
Scale 4 has $Q_{ren}\rightarrow Q_{ren}/8$, $Q_{fac}\rightarrow
Q_{fac}/8$. Table~\ref{scale_uncert_24Sep2005} shows the
cross-sections before and after cuts in the signal-like region and
primary control sample for the $gg\rightarrow W^+W^-$ and
$qq\rightarrow W^+W^-$ contributions, with the central-value $Q^{2}$
scales and the four modified scale choices. The largest variation in
$\alpha_{WW}$ we observe is 4.1\%, and we take this to be the
theoretical error due to the choice of $Q^{2}$ scale. The total
theoretical uncertainty we calculate on the prediction of
$\alpha_{WW}$ is therefore 5\%.
\begin{table}[tbp]
\begin{center}
\begin{tabular}{ c | c c | c c | c c | c}
\hline \hline
              & No cuts & & Sig. Reg.  & & Cont. Samp. & & \\
\hline
Scale Choice  & $gg\rightarrow WW$    & $qq\rightarrow WW$ &  $gg$    & $qq$    & $gg$            & $qq$            & $\alpha_{WW}$\\
\hline
Central       & 487.77 & 11302.44 & 6.45                  & 63.20                 & 6.38               & 130.10              & 0.5103\\
scale1        & 239.93 & 12862.82 & 2.92                  & 69.25                 & 3.33               & 143.83              & 0.4904\\
scale2        & 1058.97 & 9076.86 & 14.5                  & 49.03                 & 13.46              & 107.44              & 0.5255\\
scale3        & 278.17 & 11189.52 & 3.81                  & 65.02                 & 3.54               & 131.92              & 0.5081\\
scale4        & 913.38 & 11702.80 & 11.1                  & 61.81                 & 12.66              & 133.51              & 0.4988\\
\hline \hline
\end{tabular}
\caption{ Cross-sections before and after cuts for the signal-like
region and the Primary control sample, with the corresponding
extrapolation coefficients, using the nominal assumptions and the 4
altered scale choices. } \label{scale_uncert_24Sep2005}
\end{center}
\end{table}

\subsection{Theoretical Uncertainties in the Top Background}\label{Theoretical Uncertainties in the Top Background}
We now turn our attention to the uncertainties in $\alpha_{tt}$ and
$\alpha_{tt}^{WW}$. Here, the most important question to ask is how
to handle single top production. A procedure for generating both
$pp\rightarrow t\overline{t}$ and $pp\rightarrow Wt$ without
double-counting at leading order was presented
in~\cite{pr_63_034012}, and a calculation including off-shell
effects and spin correlations in the $W^+W^-bb$ system at tree level
was presented in~\cite{pr_62_014021}. Unfortunately, we know of no
event generator available at the time of this writing which also
takes into account the one-loop radiative corrections to $W^+W^-bb$
production, so we will perform our uncertainty estimate at
tree-level.

In addition to the $t\overline{t}$ Monte Carlo sample (from MC@NLO)
that we have used in the other sections of this note, we have
generated two separate $W^+W^-bb$ Monte Carlo samples using
MadGraph. One includes only doubly-resonant top quark pair
production, and the other includes the full $W^+W^-bb$ final state.
For this generation, we have allowed the b-quarks to be generated
with $P_{T}$ as low as 1 GeV, and with pseudorapidity as high as
100. One would expect a disproportionately large contribution from
the region where one b-quark is soft or forward, and we therefore
feel it is likely that the single-top contribution is overestimated
in our nonresonant $W^+W^-bb$ Monte Carlo. This is exactly what we
want if we are to prove that our analysis is robust. We have applied
the cuts for the signal-like region and both of the control regions
to these two Monte Carlo samples to assess the importance of
single-top production in this analysis.
\begin{table}[tbp]
\begin{center}
\begin{tabular}{ c | c c c | c c }
\hline \hline
Process               & Signal-like   & Cont. Samp.  & b-tagged     & $\alpha_{tt}$ & $\alpha_{tt}^{WW}$\\
\hline
$W^+W^-bb$                & 13.34         & 109.41        & 47.13         & 0.2829        & 2.3211\\
$tt\rightarrow W^+W^-bb$  & 9.80          & 80.77         & 37.72         & 0.2599        & 2.1413\\
\hline \hline
\end{tabular}
\caption{ Cross-sections (in fb) and extrapolation coefficients for
the $t\overline{t}$ background for various masses, using MadGraph to
model the $W^+W^-bb$ background. }
\label{singletop_uncert_30Aug2005}
\end{center}
\end{table}

Table~\ref{singletop_uncert_30Aug2005} shows the $W^+W^-bb$
background cross-sections in the signal-like region, the primary
control sample, and the b-tagged control sample obtained with the
leading-order doubly-resonant $t\overline{t}$ and inclusive
$W^+W^-bb$ samples. We note that although the difference in the
absolute cross-section given by the two samples is approximately
30\%, the corresponding differences in the predictions of
$\alpha_{tt}$ and $\alpha_{tt}^{WW}$ are only about 9\%. It is worth
noting that this figure is only a general guideline, since the exact
values of $\alpha_{tt}$ and $\alpha_{tt}^{WW}$ are strongly
dependent on the particulars of the b-tagging algorithm used.  Our
intent in this section is only to give a rough idea of what the
theoretical uncertainty on the extrapolation from a b-tagged sample
to a b-vetoed sample should be. In practice, this uncertainty should
be addressed in detail using full detector simulation by any
experimenter performing a $H\rightarrow W^+W^-$ search like the one
outlined here.

\subsection{Comparison of MC@NLO, Alpgen, and
Sherpa}\label{Comparison of MC@NLO, Alpgen, and Sherpa} In this
subsection, we check the agreement among the predictions of the QCD
$W^+W^-$ background given by MC@NLO and by tree-level jet-parton
matching algorithms like the ones in Sherpa and Alpgen. We begin
with a few general comments about the generators under study. In
this sub-section we ignore the process $gg\rightarrow W^+W^-$.
\begin{itemize}
\item  All three generators ignore the contribution from gluon-initiated diagrams that contain a quark box.
This contribution is not negligible; in practice, we treat the
gluon-initiated contribution as a separate process modelled with a
separate generator.
\item The matrix element calculations in MC@NLO and Alpgen were programmed by hand by their respective authors, while Sherpa
uses an automated matrix element generator to write code to compute
the (tree-level) matrix elements relevant to a particular process.
There are therefore some differences regarding which Feynman
diagrams are included in the two calculations.  In the case of this
analysis, where we are concerned with the production of W pairs
which decay leptonically, Sherpa includes the contribution from
diagrams where two $Z$ bosons are produced, with one $Z$ decaying to
leptons and the other to neutrinos.  This leads to a spike in the
dilepton invariant mass distribution in events with same-flavor
leptons; this feature does not appear to be present in MC@NLO and
Alpgen.  For this reason, we will consider only events with one
electron and one muon in this section.
\item MC@NLO includes the contribution from loop diagrams in its calculation; Sherpa and Alpgen rely instead
on jet-parton matching schemes like the one discussed
in~\cite{JHEP_0111_063}.
\end{itemize}

\begin{figure}[tbp]
\begin{center}
\includegraphics[width=12cm]{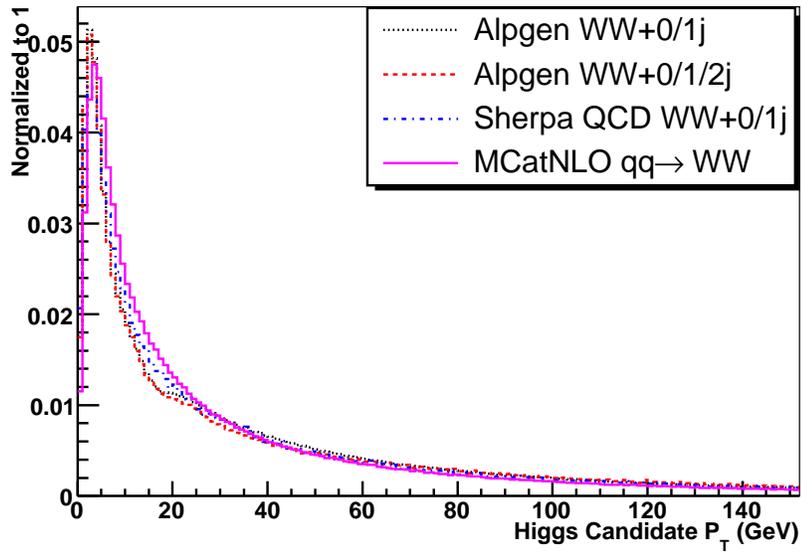}
\caption{ The
transverse momentum of the Higgs candidate in QCD $W$ pair
production as given by Alpgen, Sherpa, and MC@NLO. There is a small
shift in the location of the peak, but the difference is not
dramatic at all. } \label{pthiggs_comparison}
\end{center}
\end{figure}

It is worthwhile to point out that although the treatment of soft
hadronic physics in Alpgen, Sherpa, and MC@NLO are all quite
different, the result is nevertheless similar for the three
generators. Figure~\ref{pthiggs_comparison} shows the distribution
of the transverse momentum of the Higgs candidate (in the $e - \mu$
channel) given by Alpgen, Sherpa, and MC@NLO for the QCD $W^+W^-$
background.  Although a detailed study of the errors on these
distributions is beyond the scope of this work, we feel that the
similarity among all three generators is encouraging.  We note that
the Alpgen and Sherpa samples predict a slightly lower cross-section
for events with Higgs candidate $P_{T}$ between roughly 10 and 25
GeV.  This is no doubt an artifact of the jet-parton matching
method, and we expect that the behavior of this region could be
tuned by tuning the matching parameters in the respective generators
(although such a tuning is not necessary for our analysis).

\begin{figure}[tbp]
\begin{center}
\includegraphics[width=12cm]{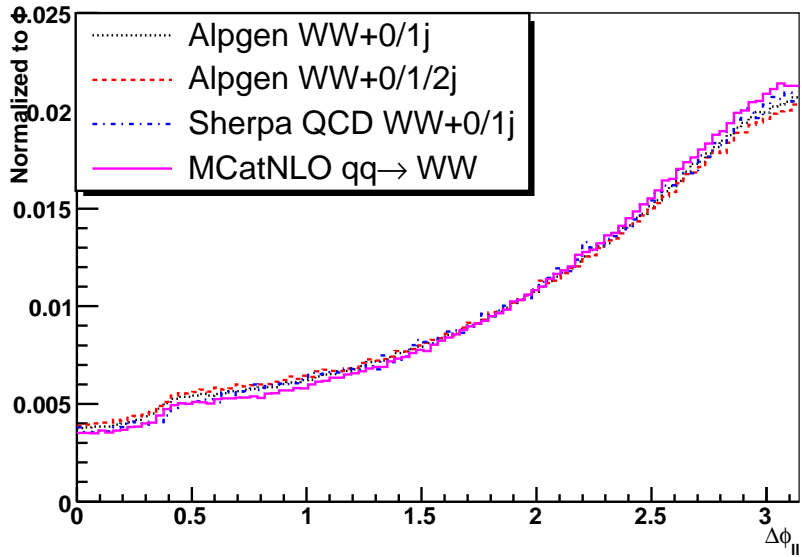}
\caption{ The
distribution of the azimuthal angle between the leptons (in the
$e-\mu$ channel) in QCD $W$ pair production as given by Alpgen,
Sherpa and MC@NLO. } \label{dphill_comparison}
\end{center}
\end{figure}

\begin{figure}[tbp]
\begin{center}
\includegraphics[width=12cm]{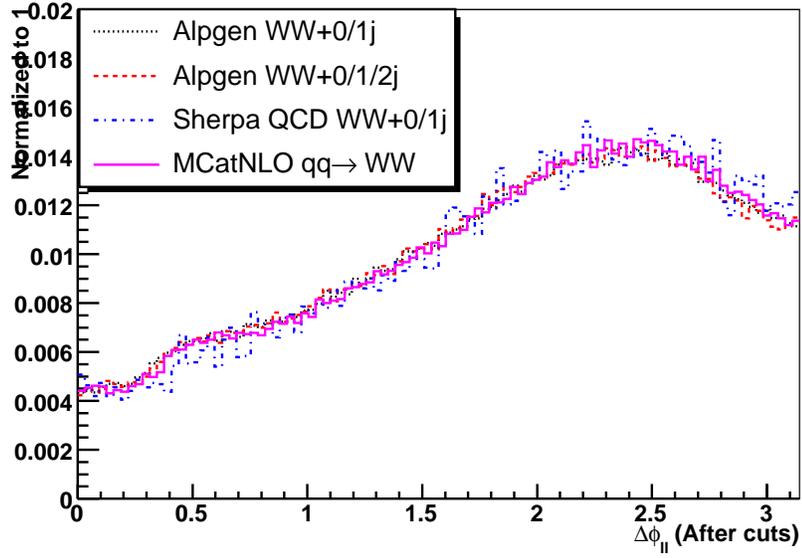}
\caption{
The distribution of the azimuthal angle between the leptons (in the
$e-\mu$ channel) in QCD $W$ pair production as given by Alpgen,
Sherpa and MC@NLO.  This figure plots the distributions after all
cuts except the cuts on $M_{ll}$, $\Delta\phi_{ll}$, and $M_{T}$. }
\label{dphill_comparison_aftercuts}
\end{center}
\end{figure}

\begin{figure}[tbp]
\begin{center}
\includegraphics[width=12cm]{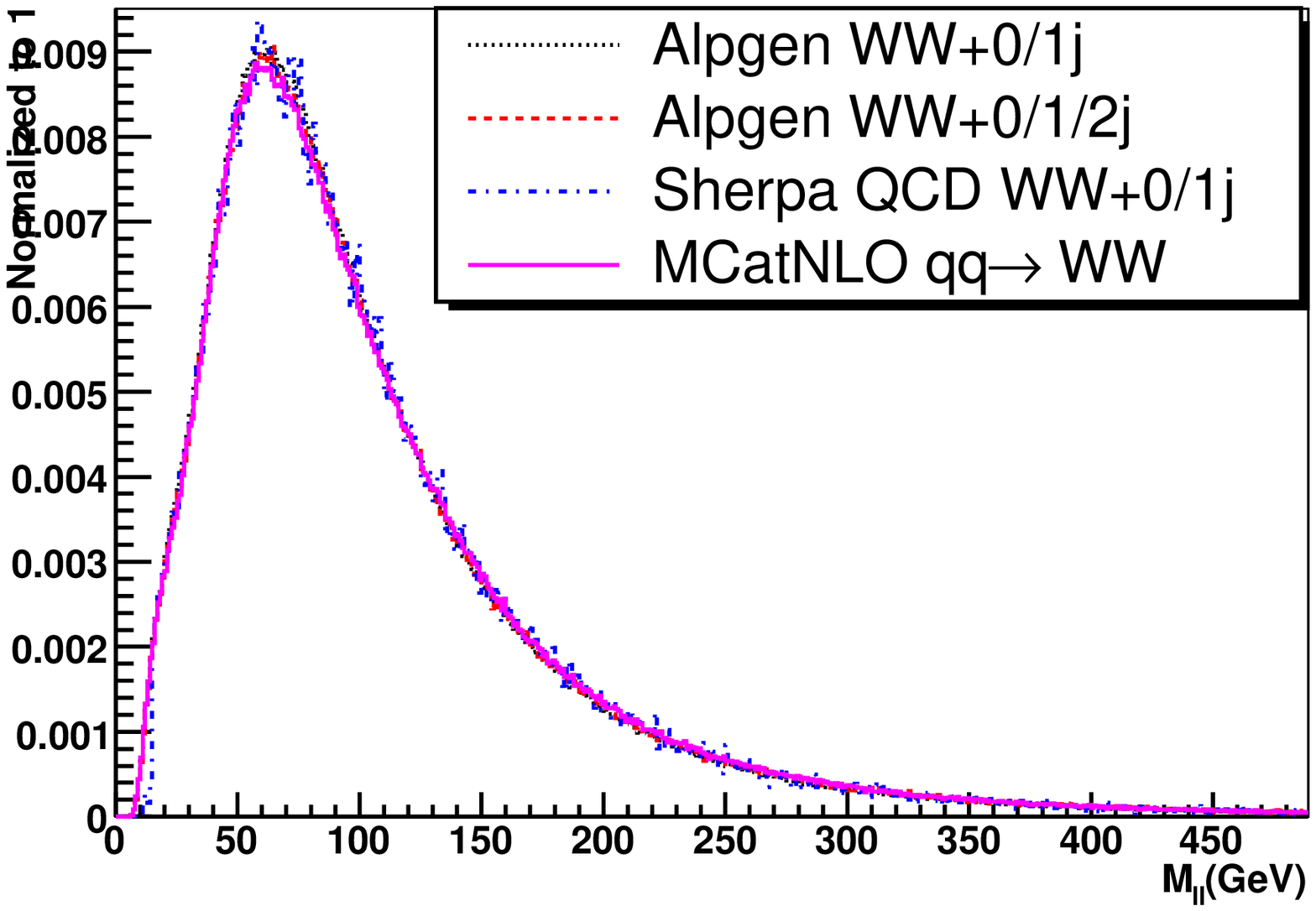}
\caption{ The
distribution of the dilepton invariant mass (in the $e-\mu$ channel)
in QCD $W$ pair production as given by Alpgen, Sherpa and MC@NLO. }
\label{mll_comparison}
\end{center}
\end{figure}

It is also worthwhile to compare the predictions of variables
related to spin correlations in the $W^+W^-$ system, as these
correlations are crucially important for the analysis.
Figure~\ref{dphill_comparison} shows the distribution of the
azimuthal angle between the leptons (in the $e - \mu$ channel)
before the cuts in the next section;  there is a slight difference
in the shape of these inclusive distributions.  The discrepancy is
not serious at all; we believe it is a kinematic effect caused by
the depletion in events with Higgs $P_{T}$ between 10 and 25 GeV
mentioned in the previous paragraph.
Figure~\ref{dphill_comparison_aftercuts} shows the distribution of
the dilepton opening angle in the transverse plane after all but the
last three cuts of Section~\ref{Monte Carlo and Analysis Method}
have been applied; there is good agreement for this distribution
among the various generators. Figure~\ref{mll_comparison} shows the
dilepton invariant mass for events with one electron and one muon
(before the cuts of the analysis are applied); it is clear from the
figure that the distribution of this variable is also very similar
in all three generators.

\subsection{QCD NLO Corrections for Higgs Production in
\boldmath{$H\rightarrow\tau^+\tau^-$} in Association with One High
\boldmath{$P_T$ Jet}} \label{sec:htautau}

In a recent publication it was demonstrated that the search for the
Higgs boson using the $H\rightarrow\tau^+\tau^-$ decay in
association with one high $P_T$ jet is a promising discovery channel
at the LHC~\cite{pl_b611_60}. Next-to-Leading order corrections (NLO)
were not evaluated for the final state considered
in~\cite{pl_b611_60}.

There are two main requirements in the analysis, which are
instrumental in achieving a good signal-to-background ratio: the
application of a large cut on the $P_T$ of the Higgs candidate
($P_{TH}>100\,\gev$) and the requirement that the invariant mass of
the Higgs candidate and the leading jet in the event be very large
($M_{HJ}>700\,\gev$). In addition to the two cuts just mentioned it
was necessary to require that there be no additional hard jets
(hadron level $P_T>20\,\gev$) in the central region of the detector
($\left|\eta\right|<2)$. The latter is introduced to suppress the
$t\overline{t}$ background.

It is meaningful to evaluate QCD higher order corrections to the
signal process after the application of the cuts mentioned above.
Apart from a chance in the overall normalization, the impact of
extra jets in the final state on the analysis is not expected to be
trivial.

In order to evaluate QCD Next-to-Leading corrections, we use the
MCFM program~\cite{pr_65_113007}. This package enables the user to
apply cuts at the parton level. Next-to-Leading Order matrix
elements to Higgs production in association with one jet are
available. In this calculation the infinite top mass approximation
is used. In addition, NLO matrix elements for Higgs production via
weak boson fusion are also available within MCFM.

Figure~\ref{fig:htautau_before} shows the Higgs $P_T$ (plots on the
left) and  the invariant mass of the Higgs and the leading jet
(plots on the right) for Higgs produced via gluon-gluon fusion. The
upper plots in Figure~\ref{fig:htautau_before} show the
distributions to Leading Order (LO, solid lines) and to NLO (dashed
lines). The lower plots in Figure~\ref{fig:htautau_before} show the
ratio of the NLO to the LO cross-sections.

Figure~\ref{fig:htautau_before} illustrates that the QCD NLO
corrections to the signal produced via gluon-gluon fusion in the
region of the phase space where the Higgs boson will be searched for
are large. The size of the NLO correction is larger than the
correction before the application of cuts on the Higgs $P_T$ and
$M_{HJ}$. The perturbative analysis shows that the NLO correction
grows with $M_{HJ}$. This can be understood qualitatively: for large
values of $M_{HJ}$ a large $P_T$ extra parton is likely to be
present in the final state, providing extra transverse momentum to
the system made by the Higgs and the leading jet and indirectly
enhancing its invariant mass.

\begin{figure}[tbp]
  \begin{minipage}[c]{0.49\textwidth}
    \includegraphics[width=\textwidth]{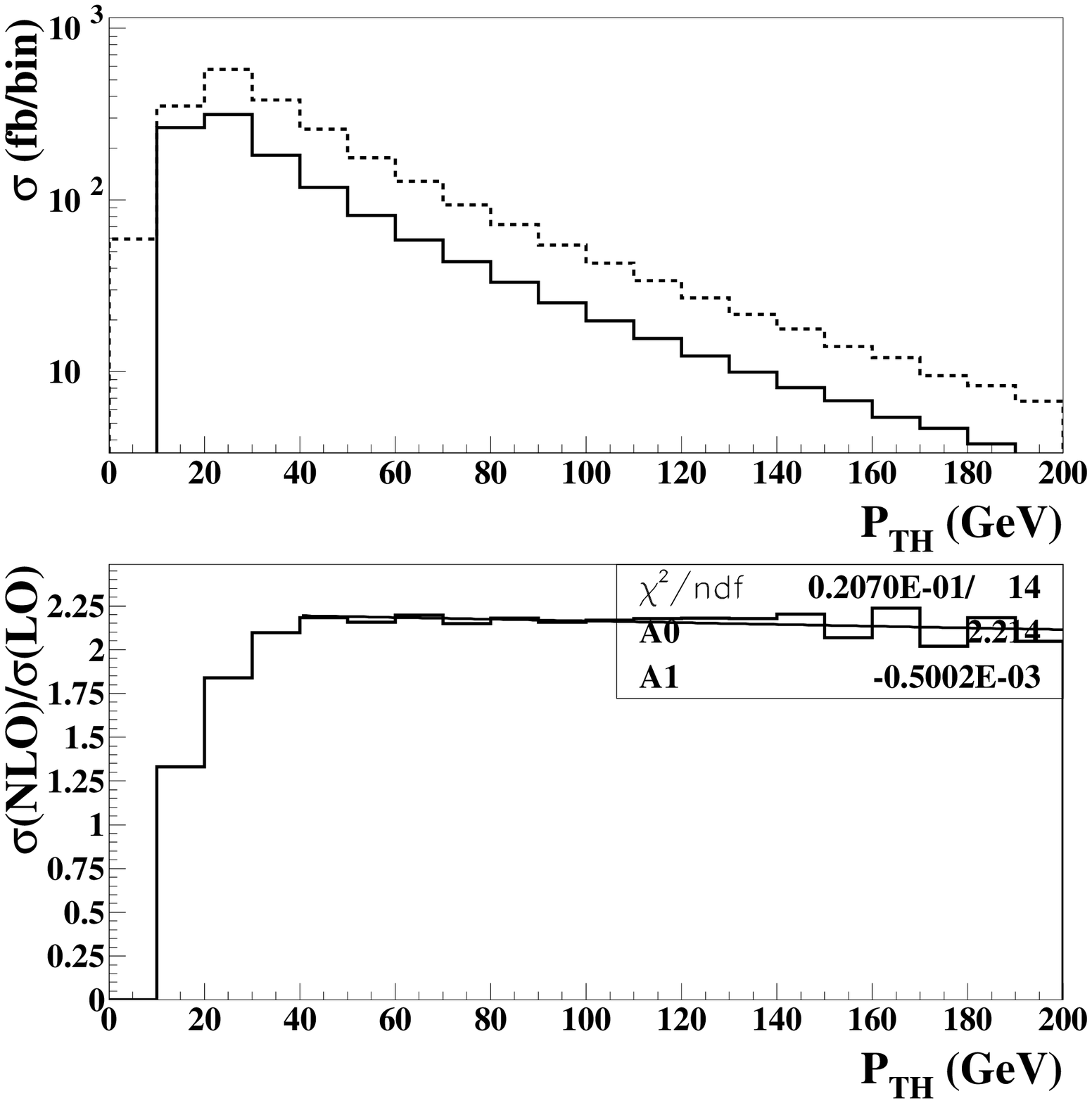}
  \end{minipage}
  \begin{minipage}[c]{0.49\textwidth}
    \includegraphics[width=\textwidth]{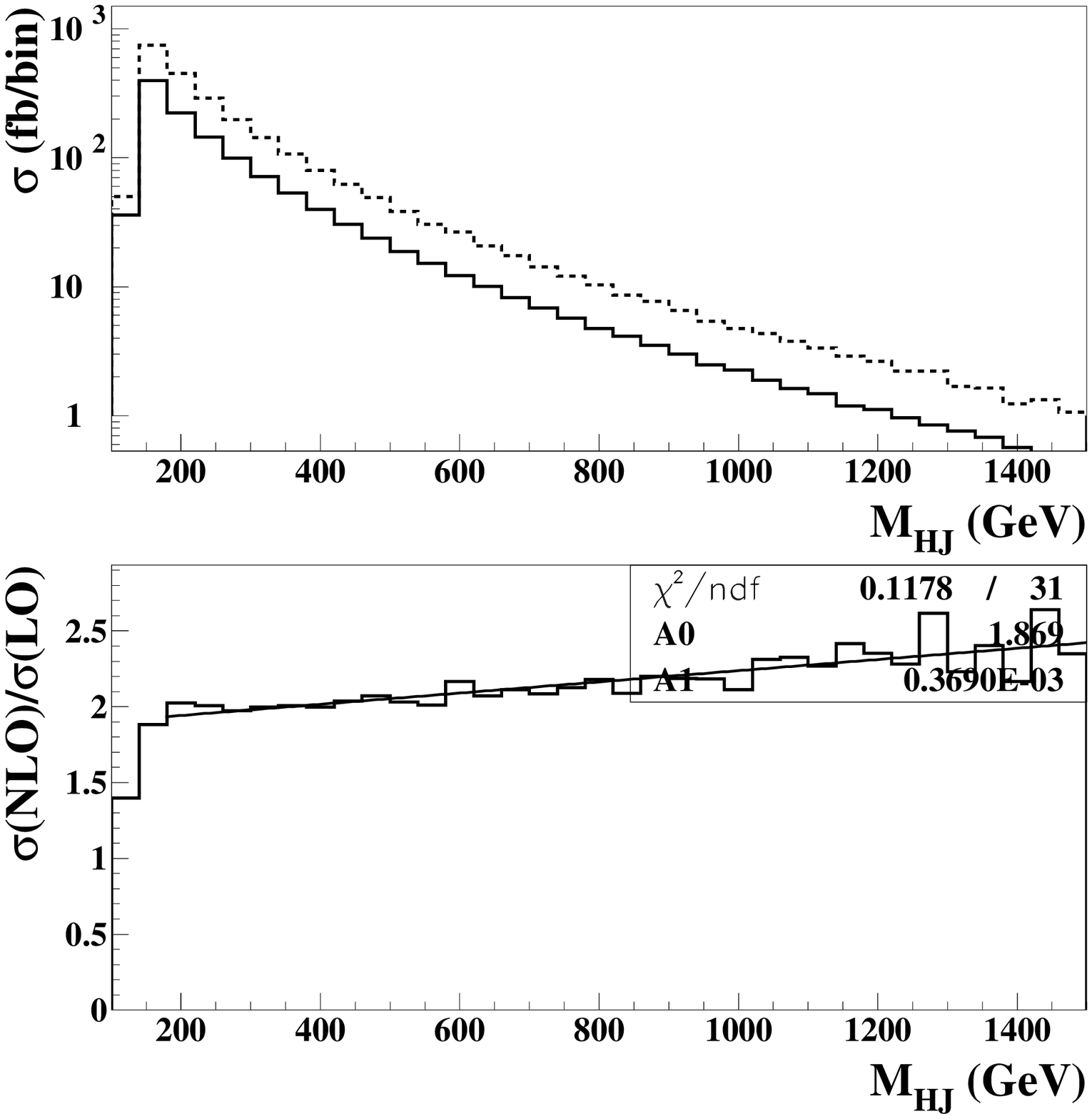}
  \end{minipage}
  \caption{Higgs $P_T$ (plots on the left) and  the invariant mass of
    the Higgs and the leading jet (plots on the right) for Higgs
    produced via gluon-gluon fusion. The upper plots show the
    distributions to Leading Order (solid lines) and to Next-to-Leading
    Order (dashed lines). The lower plots show the ratio of the
    Next-to-Leading Order to the Leading Order cross-sections. The
    package MCFM was used to evaluate the cross-sections. Cross-sections
    are given in fb per bin.} \label{fig:htautau_before}
\end{figure}

As pointed out above in this Section, the application of a veto on
extra hadronic activity is important for the suppression of the
$t\overline{t}$ production. It is necessary to evaluate the QCD NLO
corrections after the addition of this jet veto. Plots in
Figure~\ref{fig:htautau_after} show the same distributions as in
Figure~\ref{fig:htautau_before} after the application of a veto on
events with an extra parton with $P_T>30\,\gev$ and
$\left|\eta\right|<2$. The plots on the left illustrate that after
the application of a veto on extra high $P_T$ partons changes size
of the NLO corrections takes place. The ratio of NLO to LO
cross-sections decreases with the Higgs $P_T$ for $P_T>50\,\gev$.
After the application of the veto the ratio of the NLO to LO
cross-sections becomes flat as a function of $M_{HJ}$ instead of
increasing, as illustrated in Figure~\ref{fig:htautau_before}.

\begin{figure}[tbp]
  \begin{minipage}[c]{0.49\textwidth}
    \includegraphics[width=\textwidth]{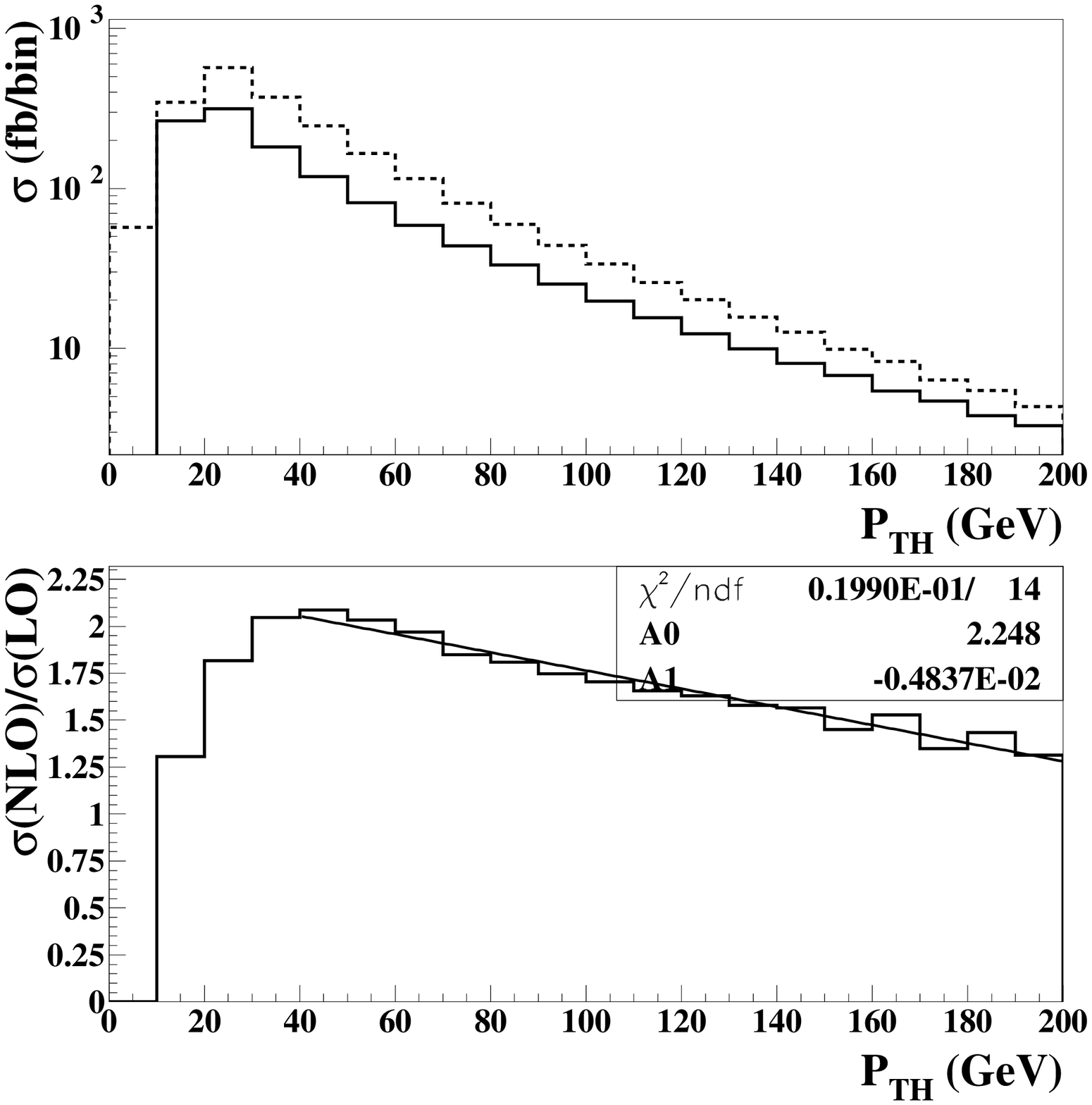}
  \end{minipage}
  \begin{minipage}[c]{0.49\textwidth}
    \includegraphics[width=\textwidth]{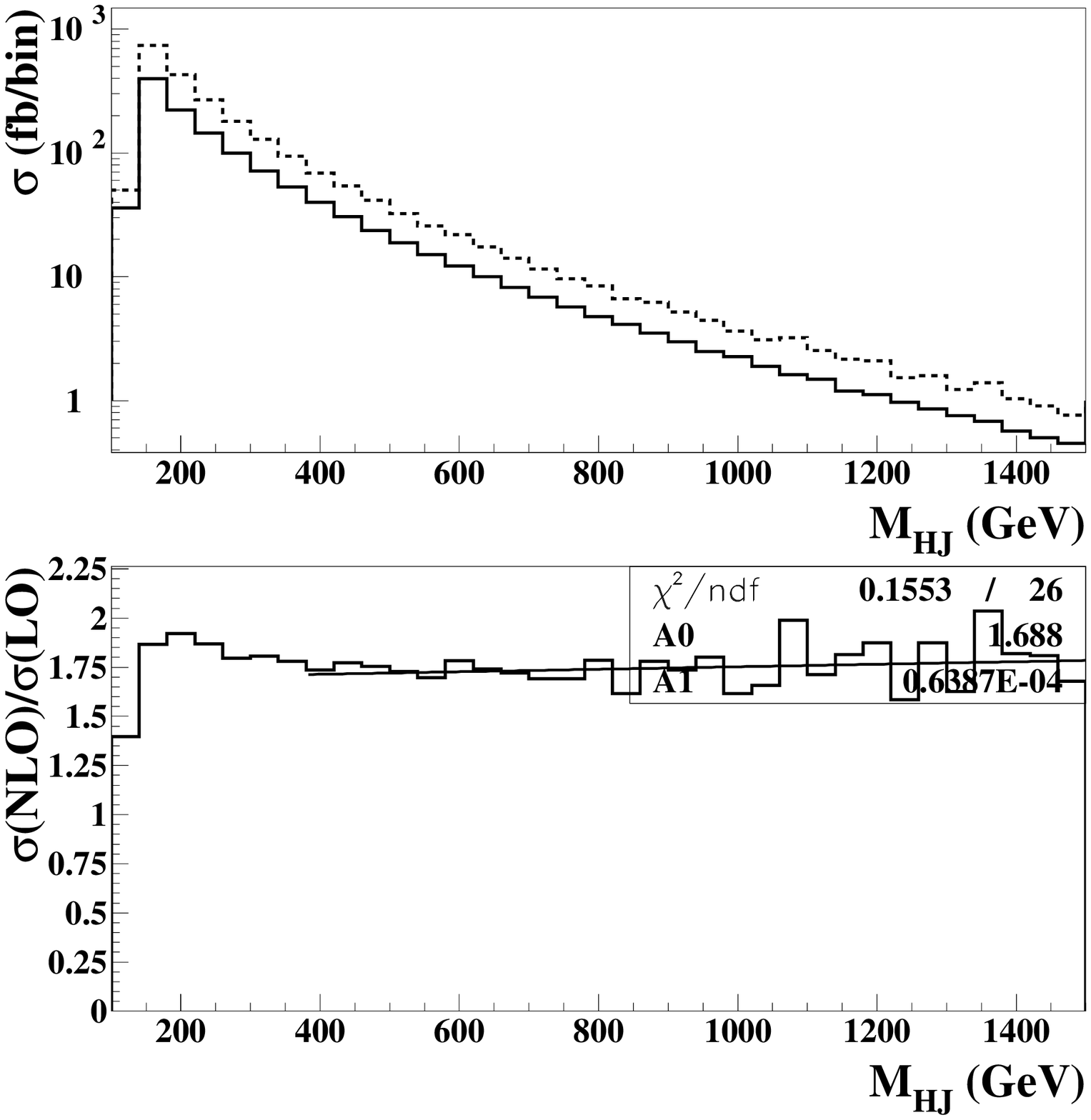}
  \end{minipage}

  \caption{Same as Figure~\ref{fig:htautau_before} after requiring a
    veto on an additional parton with $P_T>30\,\gev$ and
    $\left|\eta\right|<2$.} \label{fig:htautau_after}
\end{figure}

After the application of the cuts described above in this section
the ratio of the NLO to the LO cross-sections remains large, in the
range $1.5\div 1.6$.

A similar analysis was performed with the Higgs production via weak
boson fusion. Before the application of a jet veto the impact of QCD
NLO corrections on the Higgs $P_T$ and $M_{HJ}$ distributions is
mild and have little impact on the sensitivity of the channel.

\subsection{Summary}\label{Summary}

We have proposed a method to estimate the normalization of the
dominant backgrounds in the $H\rightarrow W^+W^-\rightarrow l^+
l^-\nu\nu$ channel using two control samples in the data, one
b-tagged, and the other b-vetoed; in our approach, the systematic
errors must be given in terms of the ratios $\alpha_{WW}$,
$\alpha_{tt}$, and $\alpha_{tt}^{WW}$. We have computed the
theoretical uncertainty on $\alpha_{WW}$; the result is 5\%. We have
shown that, for a b-tagging algorithm operating only on jets with
$P_{T}>20$~GeV and $|\eta|<2.5$, such that $\epsilon_{b}=60$\% and
the rejections against light quarks and c-quarks are 100 and 10
respectively, the effect of singly-resonant and non-resonant
$W^+W^-bb$ diagrams is less than 10\% on $\alpha_{tt}$ and
$\alpha_{tt}^{WW}$. A study using these uncertainties and this
background extraction technique is in progress; the preliminary
result is that a Higgs discovery at $M_{H}=160$~GeV would require
less than 2 fb$^{-1}$ of integrated
luminosity~\cite{ATL-PHYS-PUB-2005-016}. However, final calculations
of the uncertainties on these last two extrapolation coefficients,
as well as final results on the overall sensitivity of the search we
have presented here, must be computed within the context of the LHC
experiments.

We have evaluated the QCD NLO corrections for signal in the Higgs
search with $H\rightarrow\tau^+\tau^-$ in association with one high
$P_T$ jet. The ratio of the NLO to LO cross-sections for Higgs
production via gluon-gluon fusion is well above $2$ for Higgs
$P_T>50\,\gev$ and increases with $M_{HJ}$. The ratio drops
substantially with the application of a veto on events with an extra
parton with $P_T>30\,\gev$ and $\left|\eta\right|<2$. After the
application of analysis cuts the ratio of the NLO to LO
cross-sections for Higgs signal produced via gluon-gluon fusion is
in the range $1.5\div 1.6$.

\subsection{Acknowledgement}\label{Acknowledgement}
The authors are grateful to J.~Campbell, N.~Kauer and S.~Frixione.
We also would like to thank the organizers and the convenors of the
Higgs working group of the very successful TEV4LHC workshop. This
work was supported in part by the United States Department of Energy
through Grant No. DE-FG0295-ER40896.




\clearpage
\section{An invisibly decaying Higgs at Tevatron and LHC}
\label{sec:invisihiggs}
%
%
%
%
%
%
%
%
%
\textbf{Contributed by: H. Davoudiasl, T. Han, H.E. Logan}
\vspace{0.25in}

The Higgs particle is the only missing part of the highly successful Standard 
Model (SM) of particle physics.  The current experimental data 
from direct searches \cite{Barate:2003sz} 
and electroweak precision measurements
\cite{Grunewald:2003ij,Roth:Moriond,lepewwg} 
point to a Higgs mass in the range 
$114\GeV < m_h \lesssim 250\GeV$.   Thus, if the Higgs exists the Tevatron might 
detect it in the next several years and the LHC is expected to discover it.  

Most analyses assume that the Higgs will predominantly decay into detectable 
SM fields.  However, this may not be a good assumption if there are new weakly 
interacting particles with mass less than half the Higgs mass that couple to 
the Higgs with ${\cal O}(1)$ strength.  In this case, if 
$m_h < 160\GeV$ $ \simeq 2 \, m_W$ so that the Higgs partial width
into SM particles is very small, the Higgs will decay predominantly into
the new weakly interacting particles.  In particular, if these new
weakly interacting particles are neutral and stable, the Higgs will decay 
{\it invisibly}.  There are many models in which this situation is 
realized, such as the Minimal Supersymmetric Standard Model 
(MSSM, with Higgs decays to lightest neutralinos), 
models with extra dimensions (with Higgs decays to Kaluza-Klein 
neutrinos \cite{Arkani-Hamed:1998vp}), and Majoron 
models \cite{Joshipura:1992ua,Joshipura:1992hp}.  
An invisible Higgs is also quite 
generic in minimal models of dark matter containing a stable singlet 
scalar \cite{McDonald:1993ex,Burgess:2000yq,Davoudiasl:2004be}.  
The combined LEP experimental bound on the mass of an 
invisibly-decaying Higgs boson is 114.4\GeV at 95$\%$ confidence 
level \cite{Josa:2001fn}.   

In this work, we study the discovery potential for the invisible Higgs 
$h_{inv}$ at the LHC and the Tevatron in the channels $Z + \hin$
and $\hin + j j $ in Weak Boson Fusion (WBF).  
There have been a number of similar studies in 
the past \cite{Gunion:1993jf,Kersevan:2002zj,Choudhury:1993hv,
Frederiksen:1994me,
Martin:1999qf,Eboli:2000ze,Godbole:2003it,Battaglia:2004js,Belotsky:2004ex}.
We also examine the prospects for determining the mass of the invisible Higgs
from production cross sections at the LHC.  We show that the $Z + \hin$ 
channel gives a surprisingly good handle on the Higgs mass given 100 fb$^{-1}$
of integrated luminosity.  We also show how the $Z + \hin$ and WBF 
channels can be combined at the LHC to remove model assumptions from the
Higgs mass extraction.
A more detailed account of this study can be found in 
Ref.~\cite{Davoudiasl:2004aj}.

\subsection{Production of \boldmath $\hin$ via WBF at the Tevatron}
\label{sec:WBF}

WBF production of the invisible Higgs was studied for the LHC in 
Ref.~\cite{Eboli:2000ze}, which showed that WBF can provide significant
signals for invisible Higgs discovery, even at low luminosity.  
Here, we will use their approach to show that WBF contributes significantly
to the observation of $\hin$ at the Tevatron.  Even though a 
3$\sigma$ observation of a 
120\GeV $\hin$ in any single channel at the Tevatron  
is not possible with less than 12 fb$^{-1}$ per detector,
one can enhance the significance of the 
signal by combining data from various channels.  At the Tevatron, an 
important production mode is $Z+\hin$
\cite{Martin:1999qf} and yields a somewhat larger significance 
than the WBF 
channel that we study.  Combining these two channels and data from two 
Tevatron detectors, we show that a 3$\sigma$ observation of $\hin$ with 
$m_h = 120\GeV$ can be obtained with 7 fb$^{-1}$ of integrated luminosity 
per detector. 

At the LHC, the kinematic requirements for suppressing the backgrounds 
rely on the large energy 
and rapidity of the forward tagging jets characteristic of WBF at the LHC, 
together with the large rapidity coverage
of the LHC detectors.  Despite the more limited kinematic range
and rapidity coverage at the Tevatron, we show that the WBF 
production mode will indeed have a significant impact on the prospects 
for the observation of $\hin$ at the Tevatron, before data from the LHC 
becomes available.  

The signal here is $\PT + 2 j$.  
A large background comes from 
$Z (\to \nu {\bar \nu}) + 2 j$ with the jets produced via QCD.
A smaller, but less reducible, background comes from 
$Z (\to \nu {\bar \nu}) + 2 j$ in which the $Z$ is produced by WBF
and the jets have kinematics similar to that of the signal.
In addition, there are backgrounds from $W(\to \ell \nu) + 2j$, in which
the lepton from the $W$ decay is missed, and
QCD backgrounds with fake $\PT$ from missed jets in 
multi-jet events and jet energy mismeasurements in di-jet events.  

We generate the signal, $\hin + 2 j$, the QCD and electroweak backgrounds
with $Z(\to \nu \bar \nu) + 2 j$, and the QCD background with 
$W(\to \ell \nu) + 2 j$ for the Tevatron using 
Madgraph~\cite{Stelzer:1994ta,Maltoni:2002qb}.  
We start with the following ``minimal cuts'':
\begin{equation}
	p_T(j) > 10\GeV, \qquad \qquad
	|\eta(j)| < 3.0, \qquad \qquad 
	\Delta R(jj) > 0.4, \qquad \qquad 
	\PT > 90\GeV.
\label{eq:WBFbasiccuts}
\end{equation}
The $\PT > 90\GeV$ requirement provides a trigger.  We take the calorimeter
pseudo-rapidity coverage from, {\it e.g.}, Ref.~\cite{DZeroeta}.

We impose ``WBF cuts'': we 
require that the two jets reconstruct to a large invariant mass,
\begin{equation}
	m_{jj} > 320, \ 340, \ 360, \ 400\GeV,
	\label{eq:mjjcuts}
\end{equation}
and are separated by a large rapidity gap,
\begin{equation}
	\Delta \eta_{jj} > 2.8.
\label{eq:WBFdetacut}
\end{equation}
These two cuts eliminate most of the QCD $Z+2j$ and $W+2j$ backgrounds, 
in which the jets
tend to be softer and have a smaller rapidity gap, while preserving a 
significant fraction of the WBF signal. 

To reduce the $W+2j$ background further, we apply a lepton veto.  We
veto events that contain an isolated electron with~\cite{Acosta:2004wq}
\begin{equation}
	p_T(\ell) > 8\GeV, \qquad \qquad |\eta(\ell)| < 3.0.
\end{equation}
For simplicity, we apply the same veto to $W$ decays to muons or taus.
Loosening the veto requirements to 
$p_T(\ell) > 10\GeV, |\eta(\ell)| < 2.0$ increases
the $W+2j$ background by about a factor of two.

Background can also come from QCD multi-jet events with fake $\PT$ 
due to mismeasurement of jets and jet activity escaping down the beampipe.
We follow the techniques of a CDF study of $\PT + 2 j$ \cite{Acosta:2004zb} 
to deal with this
background.  Please see Ref.\cite{Davoudiasl:2004aj} for more details.

In Table~\ref{tab:WBF1} we show results
for signal and background cross sections for the $m_{jj}$ cuts given
in Eq.~(\ref{eq:mjjcuts}).
In Table~\ref{tab:WBF2} we show the resulting signal-to-background ratio and
significance for 10 fb$^{-1}$.

\begin{table}
\begin{tabular}{ccccc}
\hline \hline
$m_{jj}$ cut & S($\hin+2j$) & B($Z+2j$,QCD) & B($Z+2j$,EW) & B($W+2j$,QCD) \\
\hline
320\GeV & 4.1 fb & 55 fb & 1.7 fb & 7 fb \\
340\GeV & 3.6 fb & 43 fb & 1.6 fb & 5 fb \\
360\GeV & 3.2 fb & 34 fb & 1.4 fb & 5 fb \\
400\GeV & 2.4 fb & 21 fb & 1.2 fb & 2 fb \\
\hline \hline
\end{tabular}
\caption{Signal and background cross sections for $\hin+2j$ at Tevatron Run 2,
for $m_h = 120\GeV$.
The statistical uncertainty on B($Z+2j$,QCD) after cuts is roughly 
10\% due to our limited Monte Carlo sample.
There is an additional background from QCD with fake $\PT$ which is
taken from Ref.~\cite{Acosta:2004zb} to be 5 fb; this
represents a conservative overestimate of the fake $\PT$ background.}
\label{tab:WBF1}
\end{table}

\begin{table}
\begin{tabular}{cccc}
\hline \hline
$m_{jj}$ cut & S (10 fb$^{-1}$) & S/B & S/$\sqrt{\rm B}$ (10 fb$^{-1}$) \\
\hline
320\GeV & 41 evts & 0.060 & 1.6 \\
340\GeV & 36 evts & 0.066 & 1.5 \\
360\GeV & 32 evts & 0.070 & 1.5 \\
400\GeV & 24 evts & 0.082 & 1.4 \\
\hline \hline
\end{tabular}
\caption{Number of signal events, signal-to-background ratio, and
significance for $\hin+2j$ at Tevatron Run 2, for $m_h = 120\GeV$.
We include the background from QCD with fake $\PT$ of 5 fb \cite{Acosta:2004zb}
in S/B and S/$\sqrt{\rm B}$.
}
\label{tab:WBF2}
\end{table}
 
We find a signal significance of about 1.6$\sigma$ with 10 fb$^{-1}$ 
of luminosity at one Tevatron detector.  This significance is not much less 
than that found in Ref.~\cite{Martin:1999qf}
for $Z+\hin$ at the Tevatron, namely $1.9 \sigma$ with 10~fb$^{-1}$ for
$m_h = 120\GeV$.  Combining data from both Tevatron detectors, a $3 \sigma$ 
observation
would require at least 12~fb$^{-1}$ in the $Z+\hin$ channel,
or 18~fb$^{-1}$ in the WBF channel.
However, by combining these two channels, we find that a 3$\sigma$ 
observation of $\hin$ is possible with 7 fb$^{-1}$ per detector,
if the background can be determined to better than 10\%.
Thus, WBF provides an important second channel that brings an observation 
of $\hin$ into the realm of possibility at the Tevatron before the results 
of the LHC become available.  Here, we note that there may be other 
production channels, such as $g g \to h_{inv} jj$, that could contribute 
to the signal, even after the WBF cuts we have outlined.  However, this 
could only enhance $h_{inv}$ production, making our results for the WBF 
channel a lower bound on the number of signal events. 

In Refs.~\cite{Eboli:2000ze,Rainwater:1999gg}, it
is claimed that vetoing additional soft jets in the central
region improves the
signal-to-background ratio by a factor of three at the LHC.
If a similar background reduction could be achieved at the Tevatron,
the prospects for $\hin$ observation in the WBF channel would improve
considerably: a $3\sigma$ observation in the WBF channel alone would then
be possible with 6 fb$^{-1}$ per detector, with a signal-to-background
ratio close to 1/5.  Further discussion of background 
reduction is presented in Ref.~\cite{Davoudiasl:2004aj}.
We emphasize that we have \emph{not} applied a central jet veto to 
obtain the results in Tables~\ref{tab:WBF1} and \ref{tab:WBF2}.

\subsection{Associated \boldmath $Z + \hin$ Production at the LHC}

Discovery of the Higgs in the $Z + \hin$ channel was studied for the LHC in 
Refs.~\cite{Frederiksen:1994me,Godbole:2003it}.
This channel was also analyzed 
for the Tevatron in Ref.~\cite{Martin:1999qf}.  In 
Ref.~\cite{Frederiksen:1994me}, the $Z +$jet background at the LHC was found 
to diminish the significance of the signal considerably, and the 
electroweak backgrounds coming from $W W$ and $ZW$ final states were ignored.  
We update and refine the analysis of Ref.~\cite{Frederiksen:1994me} by
taking into account sources of background not included in 
that study and considering a wider acceptance range for the leptons.
We show that, with the kinematic acceptance and the cuts we adopt,
the prospects for the discovery of the invisible Higgs 
in $Z+\hin$ at the LHC are brighter than presented in
Ref.~\cite{Frederiksen:1994me}, even with the $WW$ and $ZW$ backgrounds
included.  Our results are consistent with those of Ref.~\cite{Godbole:2003it}.

We consider the production process
\begin{equation}
  p \, p \to Z(\to \ell^+\ell^-) + \hin \, \, ; \qquad \ell = e, \mu, 
  \label{pptozh}
\end{equation}
at the LHC.  We assume that the Higgs decays 100\% of the time to invisible
final states, and that the production cross section is the same
as in the SM.  Our results can be easily scaled for other invisible branching
fractions or non-SM production cross sections.
The signal rate is simply
scaled by the production rate and invisible branching fraction:
\begin{equation}
  S = S_0 \frac{\sigma}{\sigma_{SM}} \frac{{\rm BR}_{inv}}{1},
\end{equation}
where $S_0$ is the signal rate from our studies, $\sigma/\sigma_{SM}$ 
is the ratio of the nonstandard production cross section to that of the SM
Higgs, and BR$_{inv}$ is the invisible branching fraction.
Assuming that the SM is the only source of background, the luminosity 
required for a given signal significance then scales like
\begin{equation}
  \mathcal{L} = \mathcal{L}_0 \left[ \frac{\sigma}{\sigma_{SM}}
    \frac{{\rm BR}_{inv}}{1} \right]^{-2},
\end{equation}
where $\mathcal{L}_0$ is the luminosity required for a given significance
found in our studies.

\subsubsection*{Signal for $\hin$}

As the signal is $\ell^+ \ell^- \PT$, the most significant sources of 
background are 
\begin{equation}
  Z (\to \ell^+ \ell^-) Z (\to \nu {\bar \nu}), \qquad  
  W^+ (\to \ell^+ \nu) W^- (\to \ell^- {\bar\nu}), \qquad
  Z (\to \ell^+ \ell^-) W (\to \ell \nu),
  \label{zzwz}
\end{equation}
(with the lepton from the $W$ decay in $ZW$ missed)
and $Z + {\rm jets}$ final states with fake 
$\PT$~\cite{Frederiksen:1994me,Martin:1999qf}.  We simulate the signal and the 
first three backgrounds for the LHC using 
Madgraph~\cite{Stelzer:1994ta,Maltoni:2002qb}.  

We start with the following ``minimal cuts'':
\begin{equation}
	p_T(\ell^{\pm}) > 10\GeV, \qquad \qquad
	|\eta(\ell^\pm)| < 2.5, \qquad \qquad
	\Delta R(\ell^+\ell^-) > 0.4,
\label{mincuts}
\end{equation}
where $\eta$ denotes pseudo-rapidity and $\Delta R$ is the separation 
between the two particles in the detector, 
$\Delta R \equiv \sqrt{(\Delta \eta)^2 + (\Delta \phi)^2}$; 
$\phi$ is the azimuthal angle.  
The electromagnetic calorimeter at both ATLAS~\cite{:1999fq} and 
CMS~\cite{unknown:1997pw} covers the range 
$|\eta| < 3$; however, the electron trigger covers only $|\eta| < 2.5$ (2.6) 
at ATLAS (CMS).  The pseudo-rapidity acceptance for dielectrons could be 
expanded by requiring only one electron within $|\eta| < 2.5$ and the other
within $|\eta| < 3$.  Meanwhile, the muon trigger covers 
$|\eta| < 2.2$ (2.1) at
ATLAS (CMS), with muon identification and momentum measurement 
out to $|\eta| < 2.4$.  We require 
$|\eta(\ell^\pm)| < 2.5$ for both leptons, so that the larger acceptance
for dielectron events compensates the smaller acceptance for dimuon events.

Because we will cut on the invariant 
mass of the dilepton pair to keep only
events in which the dileptons reconstruct to the $Z$ mass,
we imitate the effects of LHC detector resolution by smearing the electron
momenta according to
\begin{equation}
	\Delta E/E = {0.1\over \sqrt{E (\GeV)} } \oplus 0.5\%,
\end{equation}
with the two contributions added in quadrature.  
This smearing has a negligible effect on our results. 
We have thus applied the same smearing to the final state with muons.

The $W W$ background can be largely eliminated by requiring 
that the $\ell^+ \ell^-$ invariant mass $m_{\ell^+ \ell^-}$ is close to $m_Z$:
\begin{equation}
	|m_{\ell^+ \ell^-}-m_Z|<10\GeV.
	\label{mllcut}
\end{equation}  
Also, the $\ell^+$ and $\ell^-$ from two different parent $W$ bosons tend 
to be more back-to-back than the leptons in the signal.  We therefore impose 
an azimuthal angle cut on the lepton pair, 
\begin{equation}
	\Delta\phi_{\ell^+ \ell^-}<2.5\ \ {\rm or}\ \ 143^\circ.
	\label{dphicut}
\end{equation}
This cut also eliminates Drell-Yan backgrounds with fake $\PT$ caused by 
mismeasurement of the lepton energies.

Our third cut is on $\PT$.  The number of $\ell^+ \ell^- \PT$ signal 
events typically falls more slowly with $\PT$ than those of the $ZZ$ 
or $WW$ backgrounds, as shown in Fig.~\ref{fig:ZHptmiss}. 
\begin{figure}
\resizebox{1.0\textwidth}{!}{\includegraphics{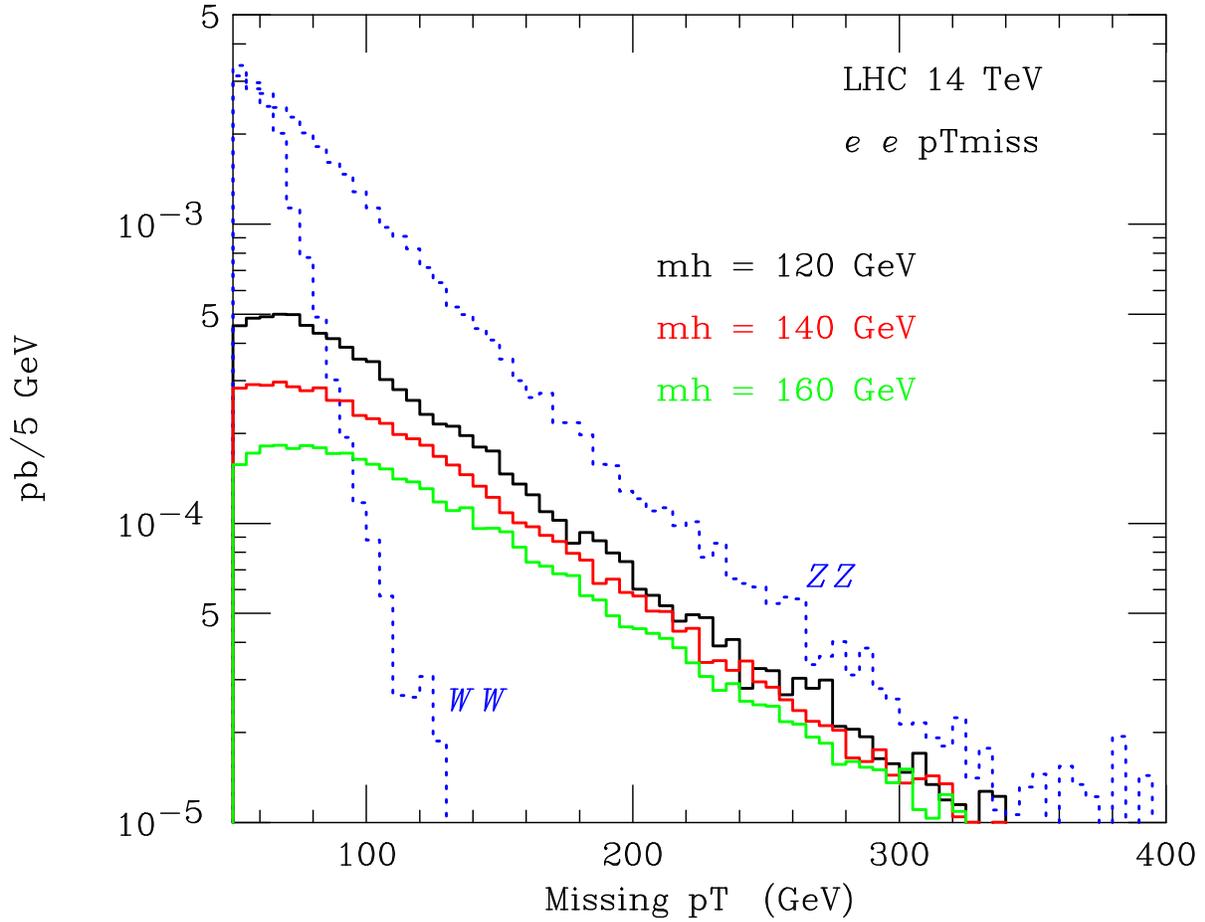}}
\caption{Missing $p_T$ distribution for $Z(\to e^+e^-)+\hin$ signal 
(solid lines, with $m_h = 120$, 140 and 160\GeV top to bottom) 
and backgrounds from $WW$ and $ZZ$ (dotted lines) at the LHC, after
applying the cuts in Eqs.~(\ref{mincuts}), (\ref{mllcut}) and (\ref{dphicut}).
}
\label{fig:ZHptmiss}
\end{figure}

The final state $Z(\to \ell^+\ell^-) W(\to \ell \nu)$, where the 
lepton from the $W$ decay is missed, can be a potential background.  
However, the probability of
missing the lepton from the $W$ decay is small given the kinematic 
coverage at the LHC.
To reduce this background, we veto events with a third isolated electron
with
\begin{equation}
	p_T > 10\GeV, \qquad \qquad |\eta| < 3.0.
\end{equation}
For simplicity, we apply the same veto to $W$ decays to muons or taus.
This veto reduces the $Z+W$ background to the level of 5--10 fb, so that
it has little effect on the significance of the signal.  

We also include the background from $Z+{\rm jets}$ with fake $\PT$.
As shown in Ref.~\cite{Frederiksen:1994me}, events of the type 
$Z +$jets can constitute a significant background due to jet energy
mismeasurements resulting in fake $\PT$, or when
one or more jets are emitted outside the fiducial region of the detector
and are therefore missed.  The majority
of those events can be eliminated by applying a jet veto, but those in 
which the jet(s) are soft and/or escape down the beampipe  
can fake $Z+\PT$ events.  A simulation of the latter
requires simulating the detector effects, which is beyond the scope of our 
analysis.  Instead, as explained in Ref.~\cite{Davoudiasl:2004aj}, 
we adopt the results for this background from 
Ref.~\cite{Frederiksen:1994me}.  

At this point, we note that there are other potentially large sources 
of background that need to be addressed \cite{Martin:1999qf}.  
The background events from $Z^* \to \tau^+ \tau^- \to \ell^+ \ell^- \PT$ 
are efficiently suppressed by our $Z$-mass cut on $m_{\ell^+ \ell^-}$, 
the $\PT$ cut, and the cut on $\Delta \phi_{\ell^+\ell^-}$ that requires 
that the leptons are not back-to-back.  This can be seen from Table 2 in 
Ref.~\cite{Godbole:2003it}, where it is shown that, after cuts similar to those we use, the resulting background from a single $Z$ is basically absent for the $ZH$ production channel.  The same conlcusion is reached for the $W + {\rm jet}$ background in the $ZH$ channel, in Table 2 of Ref.~\cite{Godbole:2003it}.  Hence,   
fake events from $W(\to \ell \nu) +$jet, where 
the jet is misidentified as a lepton of the appropriate charge and flavor, 
are also ignored in our analysis.

Our results for the background and signal cross sections are tabulated
in Table~\ref{table1}.  The corresponding signal to background ratio, $S/B$, 
and significance, $S/\sqrt B$,
are tabulated in Table~\ref{table2}.  We see from 
Table~\ref{table2} that a $>5\sigma$ discovery can be obtained for 
$m_h = 120\GeV$ with 10 fb$^{-1}$ of integrated luminosity, even with 
our conservative estimate for the $Z+$jets background for $\PT > 75\GeV$.  
With 30 fb$^{-1}$, discovery can be pushed out to $m_h = 160\GeV$.  

\begin{table}[htb]
\begin{tabular}{l|cccc|ccc}
\hline \hline
          &         &         &         &            &
	\multicolumn{3}{c}{S($Z+\hin$)} \\
$\PT$ cut & B($ZZ$) & B($WW$) & B($ZW$) & B($Z+j)^*$ & 
	$m_h = 120$~~~ & 140~~~  & 160\GeV \\
\hline
65\GeV & 48.0 fb & 10.6 fb & 10.2 fb & 22 fb &
	14.8 fb & 10.8 fb & 7.9 fb \\
75\GeV & 38.5 fb & 4.3 fb & 7.4 fb & 9 fb &
	12.8 fb & 9.4 fb & 7.0 fb \\
85\GeV & 30.9 fb & 1.8 fb & 5.5 fb & &
	11.1 fb & 8.3 fb & 6.3 fb \\
100\GeV & 22.1 fb & 0.6 fb & 3.6 fb & &
	8.7 fb & 6.8 fb & 5.3 fb \\
\hline \hline
\end{tabular}
\caption{
Background and signal cross sections for 
associated $Z(\to \ell^+\ell^-)+\hin$ production at the 
LHC, combining the $ee$ and $\mu\mu$ channels.
$^*$Estimated from Ref.~\cite{Frederiksen:1994me}
(see text for details).
}
\label{table1}
\end{table}

\begin{table}[htb]
\begin{tabular}{l|ccc|c|c}
\hline \hline
          & \multicolumn{3}{c|}{$m_h = 120\GeV$} 
		& $m_h = 140\GeV$ & $m_h = 160\GeV$ \\
$\PT$ cut & S/B & S/$\sqrt{\rm B}$ (10 fb$^{-1}$) 
	& S/$\sqrt{\rm B}$ (30 fb$^{-1}$)
		& S/$\sqrt{\rm B}$ (30 fb$^{-1}$) 
			& S/$\sqrt{\rm B}$ (30 fb$^{-1}$) \\
\hline
65\GeV & 0.22 (0.16) & 5.6 (4.9) & 9.8 (8.5) & 7.1 (6.2) & 5.2 (4.5) \\
75\GeV & 0.25 (0.22) & 5.7 (5.3) & 9.9 (9.1) & 7.3 (6.7) & 5.4 (5.0) \\
85\GeV & 0.29        & 5.7       & 9.8       & 7.4       & 5.6       \\
100\GeV & 0.33       & 5.4       & 9.3       & 7.3       & 5.7       \\
\hline \hline
\end{tabular}
\caption{
Signal significance for
associated $Z(\to \ell^+\ell^-)+\hin$ production at the LHC,
combining the $ee$ and 
$\mu\mu$ channels.  The numbers in the parentheses include the 
estimated $Z + $jets background discussed in the text.  
}
\label{table2}
\end{table}

The $Z+\hin$ channel can thus be used at the LHC for $m_h \lesssim 160\GeV$
to supplement the WBF channel \cite{Eboli:2000ze}, which has higher 
significance.  WBF production of $\hin$ at the LHC was studied in
Ref.~\cite{Eboli:2000ze}, which concluded that with only 10 fb$^{-1}$
of integrated luminosity, $\hin$ can be detected at the $\geq 5 \sigma$
level up to $m_h \simeq 480\GeV$.  They also showed that the invisible 
branching fraction of a 120\GeV Higgs can be constrained at the 95\% 
confidence level to be 
less than $13\%$ if no signal is seen in the WBF$\to \hin$ channel, 
again with 10 fb$^{-1}$.
However, we would like to emphasize that the $\PT$ measurements in the
process $\ell^+\ell^- \PT$ that we studied here 
are largely determined by $p_T(\ell\ell)$,
and the distribution will suffer much less from systematic uncertainties
compared to the WBF where $\PT$ is determined mainly from the forward jets.

\subsubsection*{Higgs boson mass}

The $Z+\hin$ channel may also provide an interesting handle on
the Higgs boson mass, as follows.
The mass of an invisibly-decaying Higgs boson obviously cannot be 
reconstructed from the Higgs decay products.  Unless the Higgs is also
observed in a visible channel, our only chance of determining the Higgs 
mass comes from the $m_h$ dependence of the production process.
Extracting $m_h$ from the production cross section requires the assumption
that the production couplings are the same as in the SM.  (Non-observation
of the Higgs in any visible final state implies that the invisible branching
fraction is close to 100\%.)

The Higgs mass extraction from measurements of the production cross
sections in $Z+\hin$ and WBF are shown in Tables~\ref{Tab:MH:Z+h} and
\ref{Tab:MH:WBF}, respectively.  There are two sources of uncertainty in the 
signal: statistical and from background normalization.  
The statistical uncertainty is 
$\Delta \sigma_S/\sigma_S = \sqrt{\rm S+B}/{\rm S}$.
We estimate the total background normalization 
uncertainty for 
$Z + \hin$ to be the same size as that of the dominant process involving 
$Z \to \nu \nu$: $\Delta {\rm B}/{\rm B} = \Delta {\rm B}(ZZ)/{\rm B}(ZZ)$.  
We assume that this background can be measured via
the corresponding channels in which $Z \to \ell^+ \ell^-$
and take the uncertainty to be the statistical uncertainty 
on the $Z \to \ell^+ \ell^-$ rate:  $\Delta {\rm B}(ZZ)/{\rm B}(ZZ) \simeq 
7.1\% \; (2.2\%)$, 
for an integrated luminosity of 10 (100) fb$^{-1}$.  
In Tables~\ref{Tab:MH:Z+h} and \ref{Tab:MH:WBF} we quote the resulting
uncertainty on the signal cross section, given by  
$\Delta \sigma_S/\sigma_S = ({\rm B}/{\rm S})\times \Delta {\rm B}/{\rm B}$.  The total uncertainty $[\Delta \sigma_S/\sigma_S]_{tot}$, presented in 
Tables~\ref{Tab:MH:Z+h} and \ref{Tab:MH:WBF}, is then the sum, in quadrature, 
of the statistical and background uncertainties, as well as the other 
uncertainties given in the table captions.  We then have 
$\Delta m_h$ = $(1/\rho)[\Delta \sigma_S/\sigma_S]_{tot}$.

\begin{table}
\begin{tabular}{lccc}
\hline \hline
$m_h$ (\GeV)
 &   120
 &   140
 &   160 \\
\hline
$\rho = (d\sigma_S/dm_h)/\sigma_S$ (1/\GeV)
 &  $-0.013$
 &  $-0.015$
 &  $-0.017$ \\
Statistical uncert.
 &   21\% (6.6\%)
 &   28\% (8.8\%)
 &   37\% (12\%) \\
Background normalization uncert.
 &   33\% (10\%)
 &   45\% (14\%)
 &   60\% (19\%) \\
Total uncert.
 &   40\% (16\%)
 &   53\% (19\%)
 &   71\% (24\%) \\
\hline
$\Delta m_h$ (\GeV)
 &   30 (12)
 &   35 (12)
 &   41 (14) \\
\hline \hline
\end{tabular}
\caption{Higgs mass determination from $Z + \hin$ with 10 (100) fb$^{-1}$, 
assuming Standard Model production cross section and 100\% invisible decays.  
The signal and background cross sections were taken from Table~\ref{table1} 
for $\PT > 75\GeV$.  The total uncertainty includes a
theoretical uncertainty on the signal cross section from QCD and PDF 
uncertainties of 7\% \cite{Brein:2004ue} and an estimated
lepton reconstruction efficiency uncertainty of 4\% (2\% per lepton)
and luminosity normalization uncertainty of 5\% \cite{Duhrssen}.
}
\label{Tab:MH:Z+h}
\end{table}

\begin{table}
\begin{tabular}{lccccccc}
\hline\hline
$m_h$ (\GeV)
 &   120
 &   130
 &   150
 &   200 \\
\hline
$\rho = (d\sigma_S/dm_h)/\sigma_S$ (1/\GeV)
 &  $-0.0026$
 &  $-0.0026$
 &  $-0.0028$
 &  $-0.0029$ \\
Statistical uncert.
 &   5.3\% (1.7\%)
 &   5.4\% (1.7\%)
 &   5.7\% (1.8\%)
 &   6.4\% (2.0\%) \\
Background norm.\ uncert.
 &   5.2\% (2.1\%)
 &   5.3\% (2.1\%)
 &   5.6\% (2.2\%)
 &   6.5\% (2.6\%) \\
Total uncert.
 &   11\% (8.6\%)
 &   11\% (8.6\%)
 &   11\% (8.6\%)
 &   12\% (8.8\%) \\
\hline
$\Delta m_h$ (\GeV)
 &   42 (32)
 &   42 (33)
 &   41 (31)
 &   42 (30) \\
\hline \hline
\end{tabular}
\caption{Higgs mass determination from ${\rm WBF} \to \hin$ 
with 10 (100) fb$^{-1}$, assuming Standard Model
production cross section and 100\% invisible decays.  The background 
and signal cross sections were taken from Tables II and III, respectively, of 
Ref.~\cite{Eboli:2000ze}, and include a central jet veto.  
The total uncertainty includes a 
theoretical uncertainty from QCD and PDF uncertainties of 
4\%~\cite{Figy:2003nv,Berger:2004pc},
and an estimated uncertainty on the efficiency of the WBF jet tag and 
central jet veto of 5\% and luminosity normalization uncertainty of 
5\%~\cite{Duhrssen}.
}
\label{Tab:MH:WBF}
\end{table}

The cross section for $Z+\hin$ production falls quickly with increasing $m_h$ 
due to the $s$-channel propagator suppression.  This is in contrast to the
WBF production, which provides a $>5\sigma$ signal up to 
$m_h \simeq 480\GeV$ with 10 fb$^{-1}$ if the 
Higgs decays completely invisibly \cite{Eboli:2000ze}.  
Thus, while the statistics are much better for the WBF measurement than for
$Z + \hin$, the systematic uncertainties hurt WBF more
because $(d\sigma_S/dm_h)/\sigma_S$ is much smaller for WBF than for
$Z + \hin$.  The $Z+\hin$ cross section
is therefore more sensitive to the Higgs mass than the WBF cross section.

More importantly, however, taking the ratio of the $Z+\hin$ and
WBF cross sections allows for a more model-independent determination
of the Higgs mass.  This is due to the fact that the production 
couplings in $Z+\hin$ ($hZZ$) and in WBF (contributions from $hWW$ and 
$hZZ$) are related by custodial SU(2) symmetry 
in any model containing only Higgs doublets and/or singlets.  The production
couplings thus drop out of the ratio of rates in this wide class 
of models (which includes the MSSM, multi-Higgs-doublet models, and
models of singlet scalar dark matter), 
leaving dependence only on the Higgs mass.  The resulting Higgs mass
extraction is illustrated in Table~\ref{Tab:MH:Ratio}.
Assuming SM event rates for the statistical uncertainties, 
we find that the Higgs mass can be extracted with an uncertainty of
35--50\GeV (15--20\GeV) with 10 (100) fb$^{-1}$ of integrated luminosity.
The ratio method also allows a test of the SM cross section assumption by
checking the consistency of the $m_h$ determinations from the 
$Z+\hin$ and WBF cross sections alone with the $m_h$ value
extracted from the ratio method.
Furthermore, observation of the invisibly-decaying Higgs in WBF but not in 
$Z+\hin$ allows one to set a lower limit on $m_h$ in this class of models.

\begin{table}
\begin{tabular}{lccc}
\hline \hline
$m_h$ (\GeV)
 &   120
 &   140
 &   160 \\
\hline
$r = \sigma_S(Zh)/\sigma_S({\rm WBF})$
 &   0.132
 &   0.102
 &   0.0807 \\
$(dr/dm_h)/r$ (1/\GeV)
 &  $-0.011$
 &  $-0.013$
 &  $-0.013$ \\
Total uncert., $\Delta r/r$
 &   41\% (16\%)
 &   54\% (20\%)
 &   72\% (25\%) \\
\hline
$\Delta m_h$ (\GeV)
 &   36 (14)
 &   43 (16)
 &   53 (18) \\
\hline \hline
\end{tabular}
\caption{Higgs mass determination from the ratio method discussed in the
text, with 10 (100) fb$^{-1}$.  
The event rates for WBF were interpolated linearly for Higgs 
masses of 140 and 160\GeV, which were not given explicitly in 
Ref.~\cite{Eboli:2000ze}.  
Statistical uncertainties were obtained assuming SM signal rates.
The total uncertainty includes
theoretical uncertainties from QCD and PDF uncertainties of 7\% for 
$Z + \hin$ \cite{Brein:2004ue} and 4\% for 
WBF~\cite{Figy:2003nv,Berger:2004pc},
and estimated uncertainties on the lepton reconstruction efficiency in 
$Z+\hin$ of 4\% (2\% per lepton) and on the efficiency of the WBF jet 
tag and central jet veto of 5\%~\cite{Duhrssen}.  
The luminosity normalization uncertainty cancels out in the
ratio of cross sections and is therefore not included.
}
\label{Tab:MH:Ratio}
\end{table}

We note that the $\PT$ distribution is also sensitive
to $m_h$: larger $m_h$ results in a larger average $\PT$ in $Z+\hin$
events.  At the LHC, the production cross section and $\PT$ distribution 
may be the only experimental handles on the mass of a Higgs boson with 
no visible decays.



\clearpage
\section{Studies of Spin Effects in Charged Higgs Boson Production 
 with an Iterative Discriminant Analysis}


\noindent\textbf{Contributed by: S. Hesselbach, S. Moretti, J. Rathsman, A. Sopczak}

\vspace*{0.25in}

We report on detailed Monte Carlo comparisons of selection variables to
separate $ tbH^\pm$ signal events from the Standard Model $t\bar t$ background
using an Iterative Discriminant Analysis (IDA) method.
While kinematic differences exist between the two processes whenever 
$m_{H^\pm}\ne m_{W^\pm}$, the exploration of the spin difference between the
charged Higgs and the $W^\pm$ gauge boson becomes crucial in the particularly
challenging case of near degeneracy of the charged Higgs boson mass with the
$W^\pm$ mass.
The TAUOLA package is used to decay the tau leptons emerging from the charged Higgs 
and $W^\pm$ boson decays taking the spin difference properly into account. 
We demonstrate that, even if the individual selection variables have limited discriminant 
power, the IDA method achieves a significant separation between the expected signal and background.
For both Tevatron and LHC energies, the impact of the spin effects and $ H^\pm$ mass 
on the separation of signal and background has been studied quantitatively. The
effect of a hard transverse momentum cut to remove QCD background has been
studied and it is found that the spin effects remain important.
The separation is expressed in purity versus efficiency curves. The study is performed 
for charged Higgs boson masses between the $W^\pm$ mass and near the top mass.

\subsection{Introduction}
The importance of charged Higgs boson searches 
has in the recent years been emphasized~\cite{lep94,lc95,Carena:2000yx,reviewLHC} 
for LEP, a future International Linear Collider (ILC), the Tevatron and  the
Large Hadron Collider (LHC),
as the detection of a charged Higgs boson would be a definite 
signal for the existence of New Physics beyond the Standard Model (SM). 
Charged Higgs bosons naturally arise in non-minimal Higgs scenarios, 
such as Two-Higgs Doublet Models (2HDMs). A Supersymmetric version of the
latter is the Minimal Supersymmetric Standard Model (MSSM). It is 
a Type II 2HDM with specific relations among neutral
and charged Higgs boson masses and couplings, dictated by
Supersymmetry (SUSY) \cite{Gunion:1989we}. 

The Tevatron collider at Fermilab is currently in its second stage
of operation, so-called
Run 2, with a center-of-mass (CM) energy of $\sqrt s=1.96$
TeV. This machine will be the first one to 
directly probe charged Higgs boson masses in the 
mass range up to $m_{H^\pm}\sim m_t$.
Starting from 2008, the LHC at CERN will be in a position
to confirm or rule out the existence of such a particle over a very
large portion of both the 2HDM and MSSM parameter space, 
$m_{H^\pm}\Ord 400$ GeV, depending on $\tan\beta$, 
the ratio of the vacuum expectation values of the two Higgs doublets 
(see the reviews~\cite{Roy:2004az,Roy:2004mm,Roy:2005yu} and a recent study~\cite{Mohn:2007fd}).

At present, a lower bound
on the charged Higgs boson mass exists from LEP \cite{leplimit},
$m_{H^\pm}\OOrd m_{W^\pm}$,
independently of the charged Higgs boson decay Branching Ratios (BRs).
This limit is valid within any Type II 2HDM whereas, in 
the low $\tan\beta$ region (below about 3), an indirect lower 
limit on $m_{H^\pm}$ can be derived  in the MSSM 
from the one on $m_{A}$ (the mass
of the pseudoscalar Higgs state of the model):
$m_{H^\pm}^2\approx m_{W^\pm}^2+m_{A}^2\OOrd (130~\mathrm{GeV})^2$.

If the charged Higgs boson mass $m_{H^\pm}$ satisfies 
$m_{H^\pm} < m_{t} - m_{b}$, where $ m_{t}$ is the top quark mass and 
$ m_{b}$ the bottom quark mass,
$H^\pm$ bosons could be produced in the decay of on-shell (i.e., $\Gamma_t\to0$)
top (anti-)quarks $ t \rightarrow bH^+$,
the latter being in turn produced in pairs via $gg$ fusion  and $q\bar q$ annihilation.
This approximation is the one customarily used in event generators
when $m_{H^\pm} \Ord m_{t}$. 
Throughout this study we adopt the same notation as in Ref.~\cite{Alwall:2003tc}: 
charged Higgs production is denoted by
$q\bar q$, $ gg \rightarrow t\bar t \rightarrow tbH^\pm$ if due to (anti-)top decays 
 and by $ q\bar q$, $ gg \rightarrow tbH^\pm$ if further production diagrams are included.
In fact, owing to the large top decay width ($ \Gamma_{t} \simeq 1.5$~GeV) and 
due to  the additional diagrams which do not proceed via direct $ t\bar t$ 
production~\cite{Borzumati:1999th,Miller:1999bm,Moretti:1999bw},
charged Higgs bosons could 
also be produced at and beyond the kinematic top decay threshold. 
The importance of these effects in the so-called `threshold' or 
`transition'  region ($m_{H^\pm}\approx m_t$) was emphasized in 
Les Houches proceedings~\cite{Cavalli:2002vs,Assamagan:2004mu}
as well as in Refs.~\cite{Alwall:2003tc,Guchait:2001pi,Moretti:2002ht,Assamagan:2004gv}, 
so that the calculations of Refs.~\cite{Borzumati:1999th,Miller:1999bm} 
(based on the appropriate $q\bar q,gg\to tb H^\pm$ description) 
are now implemented in 
HERWIG\,\cite{herwig,Corcella:2000bw,Corcella:2002jc,Moretti:2002eu}\,and 
PYTHIA\,\cite{Sjostrand:2003wg,Alwall:2004xw}. A comparison between
the two generators was carried out in Ref.~\cite{Alwall:2003tc}.
For any realistic simulation of $H^\pm$ production with  
$m_{H^\pm}\OOrd m_t$
the use of  these implementations is important.
In addition, 
in the mass region near the top quark mass, a matching of the calculations for the
$ q\bar q,~gg \rightarrow tbH^\pm$ and 
$ gb \rightarrow tH^\pm$ processes might be required
\cite{Alwall:2004xw}.

A charged Higgs boson with $m_{H^\pm}\Ord m_{t}$
decays predominantly into a $\tau$ lepton and
a neutrino. 
For large values of $\tan \beta$ ($\OOrd$ 5) the corresponding BR
is near 100\%. 
For $m_{H^\pm}\OOrd m_{t}$, $H^\pm\to \tau\nu_\tau$ is overtaken
by $H^\pm\to tb$, but the latter is much harder to disentangle from background
than the former.  
The associated top quark decays predominantly into a $W^\pm$ boson, or at times 
a second charged Higgs boson,
and a $b$ quark. 
The reaction
\begin{equation}
q\bar q, gg \to  tbH^\pm~~~(t\to bW)~~~(H^\pm \to \tau ^\pm \nu_{\tau})
\label{channel}
\end{equation}
is then a promising channel to search for a charged Higgs boson at both the Tevatron
(where the dominant production mode is $q\bar q$)
and the LHC (where $gg$ is the leading subprocess). If
the  $H^\pm\to\tau\nu_\tau$ decay channel is used to search for Higgs bosons, then
a key ingredient in the signal selection process should be the exploitation of
decay distributions that are sensitive to the spin nature of the particle
yielding the $\tau$ lepton ($H^\pm$ in the signal or $W^\pm$ in the
background), as advocated in 
Refs.~\cite{Roy:1991sf,Raychaudhuri:1995kv,Raychaudhuri:1995cc,Roy:1999xw}
(see also \cite{Assamagan:2002in,Assamagan:2002ne}).
The $\tau$ spin information affects both the energy and
the angular distribution of the $\tau$ decay products.

In the search for a charged Higgs boson signal
containing a $\tau$ lepton,
not only the magnitude of the production  
cross section is important, but also the efficiency of identifying the
$\tau$ lepton in the hadronic environment plays a crucial role.  
Since $\tau$ leptons have a very short life-time ($\sim10^{-6}$~s),
they decay within the detectors and can only be identified through
their decay products. In about 35\% of the cases they decay
leptonically and about 65\% of the times they do so hadronically. 
Both of these decay modes are usually addressed in charged Higgs boson
searches by employing dedicated $\tau$ lepton triggers. 
The identification of taus in hadronic $p\bar p$ collisions has recently
been studied, e.g. $Z\to\tau^+\tau^-$ events~\cite{Abazov:2004vd} and 
further details are given in~\cite{leshouches05}.

It is the purpose of this note to outline the possible improvements 
that can be achieved at the Tevatron and LHC in the search for charged 
Higgs bosons, with mass below the top mass and including the appropriate 
description of the spin effects in the $H^\pm\to\tau\nu_\tau$ decay.
In order to quantify the spin effect an Iterative Discriminant Analysis
(IDA) method has been applied, which is a powerful tool to separate signal 
and background, even in cases such as the one presently under study when 
several selection variables with limited discriminant power are present.

\subsection{Tevatron energy}

We start by studying charged Higgs production $q\bar q, gg \to  tbH^\pm$
with subsequent decays $t \to b W$, $H^\pm \to \tau \nu_\tau$
at the FNAL Tevatron with $\sqrt{s} = 1.96$~TeV.
In the following we analyze hadronic decays of
the $W^\pm$ boson and $\tau$ lepton
($W^\pm \to q\bar{q}'$, $\tau \to \mathrm{hadrons} + \nu_\tau$),
which results in the signature
$2 b + 2 j + \tau_\mathrm{jet} + p_t^{\rm miss}$
(2 $b$ jets, 2 light jets, 1 $\tau$ jet and missing transverse momentum).
The most important irreducible 
background process is $q\bar q, gg \to  t\bar{t}$
with the subsequent decays $t \to b W^+$ and $\bar{t} \to \bar{b} W^-$,
one $W^\pm$ boson decaying hadronically ($W^\pm \to q\bar{q}'$)
and one leptonically ($W^\mp \to \tau \nu_\tau$), which results in
the same final state particles as for the expected signal. 

\subsubsection{Simulation and detector response}
The signal process $q\bar q, gg \to  tbH^\pm$ is simulated with PYTHIA 
\cite{Sjostrand:2003wg}.
The subsequent decays $t \to b W^\pm$ (or its charge conjugate),
$W^\pm \to q\bar{q}'$ and
$H^\mp \to \tau \nu_\tau$ are also carried out within PYTHIA,
whereas the $\tau$ leptons are decayed externally with the program TAUOLA
\cite{Jadach:1990mz, Golonka:2003xt},
which includes the complete spin structure of the $\tau$ decay.
The background process $q\bar q, gg \to  t\bar{t}$ is also simulated
with PYTHIA with the built-in subroutines for $t\bar{t}$ production.
The decays of the top quarks and $W^\pm$ bosons are performed within PYTHIA
and that of the $\tau$ lepton within TAUOLA.

The momenta of the final $b$ and light quarks from the PYTHIA event
record are taken as the momenta of the corresponding jet, whereas for the
$\tau$ jet the sum of all non-leptonic final state particles as given by
TAUOLA is used.
The energy resolution of the detector and parton shower and hadronization effects
are emulated through a Gaussian
smearing $(\Delta(p_t)/p_t)^2 = (0.80/\sqrt{p_t})^2$
of the transverse momentum $p_t$
for all jets in the final state, including the $\tau$ jet~\cite{Carena:2000yx}.
As typical for fast simulation studies, no effects of underlying events, are simulated.
Events are removed which contain jets with less than 20 GeV transverse 
momentum\footnote{In order to be largely independent of the specific 
detector performance, no requirement on the jet resolution is applied.}, 
corresponding to about $|\eta|>3$.
The transverse momentum of the leading charged pion in the $\tau$ jet
is assumed to be measured in the tracker independently of the
transverse momentum of the $\tau$ jet. The identification and momentum 
measurement of the pion is important to fully exploit the $\tau$ spin information.
In order to take into account the tracker performance
we apply  Gaussian smearing on $1/p_t^\pi$ with
$ \sigma(1/p_t^\pi)[\mathrm{TeV}^{-1}] =
 \sqrt{0.52^2 + 22^2/(p_t^\pi[\mathrm{GeV}])^2 \sin\theta_\pi}$,
where $\theta_\pi$ is the polar angle of the $\pi$.
The missing transverse momentum \mbox{$p_t^{\rm miss}$} is constructed from
the transverse momenta of all visible jets 
(including the visible $\tau$ decay products)
after taking the modelling of the detector into account.
The generic detector description is a good approximation for both Tevatron experiments,
CDF and D0.

\subsubsection{Expected rates}
For completeness we present a brief discussion of the expected cross section 
of the charged Higgs boson signature under investigation.
The signal cross section has been calculated for
$\tan\beta = 30$
and $m_{H^\pm} = 80, 100, 130$ and $150$~GeV with PYTHIA, version 6.325,
using the implementation described in \cite{Alwall:2004xw},
in order to take the effects in the transition region into account. Furthermore,
it has been shown in \cite{Alwall:2003tc} that the signal cross section
for $tbH^\pm$ agrees with the one from the top-decay approximation
$t\bar t \to tbH^\pm$ for charged Higgs boson masses up to about 160~GeV
if the same factorization and renormalization scales are used.
Thus, we have used everywhere in this study the factorization scale $(m_t + m_{H^\pm})/4$ and the renormalization scale $m_{H^\pm}$ for both signal and background (i.e., those recommended in \cite{Alwall:2004xw} as most appropriate for the $tbH^\pm$ 
signal)\footnote{Clearly, for a proper
experimental study, factorization and renormalization scales for our background process 
$q\bar q$, $ gg \rightarrow t\bar t\to tbW^\pm$
ought to be chosen appropriately, i.e., unrelated to the charged Higgs boson mass.},
since the primary purpose of our study is to  single out variables that show a 
difference between our $W^\pm$ and $H^\pm$ data samples and that this
can unambiguously be ascribed to the different nature of the two kinds of bosons
(chiefly, their different mass and spin state).
In addition, the running $b$ quark mass entering in the Yukawa
coupling of the signal has been evaluated at $m_{H^\pm}$.
This procedure eventually results in a dependence of our background calculations on $\tan\beta$
and, especially, $m_{H^\pm}$ that is more marked than the one that would more naturally arise 
as only due to indirect effects through the top decay width.
Hence, the cross sections have been rescaled with a common factor
such that the total $t \bar t$ cross section is
$\sigma^{\rm prod}_{t\bar{t}} = 5.2$~pb \cite{ttbarxsec}.
To be more specific, we have first calculated the total cross section 
$\sigma^{\rm prod,PYTHIA}_{t\bar{t}}(m_{H^\pm})$ with the 
built-in routine for $t \bar t$ production
in PYTHIA for all $m_{H^\pm} = 80, 100, 130$ and $150$~GeV and then
calculated from this the respective rescaling factors
$c(m_{H^\pm}) = 
 5.2~\mathrm{pb}/\sigma^{\rm prod,PYTHIA}_{t\bar{t}}(m_{H^\pm})$
for each $m_{H^\pm}$.
Then we have calculated the background cross section 
for $m_{H^\pm}=80$~GeV into the final state with the signature
$2 b + 2 j + \tau_\mathrm{jet} + p_t^{\rm miss}$
by enforcing the respective decay channels in PYTHIA using the 
built-in routine for $t \bar t$ production and multiplied 
it with $c(80~\mathrm{GeV})$.
In the same manner we have calculated the signal cross sections with the
PYTHIA routines for $tbH^\pm$ production by enforcing the respective decay
channels in PYTHIA and multiplying with the rescaling factors
$c(m_{H^\pm})$ for $m_{H^\pm} = 80, 100, 130, 150$~GeV.
The resulting cross sections 
are given in Table~\ref{tab:Hpm:crosssec} before 
($\sigma^{\rm th}$) and after ($\sigma$) applying the basic cuts 
$p_t^{\rm jets} > 20$~GeV and the hard cut $p_t^{\rm miss} > 100$~GeV.
For the four signal masses, the $tbH^\pm$ and $ t\bar t \to tbH^\pm$
cross section calculations agree numerically.

\begin{table}[htbp]
\centering
\begin{tabular}{c|c|c|c|c|c}
 & $q\bar q, gg \to  t\bar{t}$ &
  \multicolumn{4}{c}{$q\bar q, gg \to  tbH^\pm$} \\
$m_{H^\pm}$ (GeV) & 80 & 80 & 100 & 130 & 150\\
  \hline
$\sigma^{\rm th}$ (fb) & 350 &  535    &  415    & 213 & 85  \\
$\sigma$ (fb) for $p_t^\mathrm{jets} > 20$ GeV 
         & 125   & 244     &   202   &  105 & 32 \\
$\sigma$ (fb) for $(p_t^\mathrm{jets},p_t^{\rm miss}) > (20,100)$ GeV 
         & 21   &   30   &  25    &  18 & 7 \\
\end{tabular}
\caption{\label{tab:Hpm:crosssec}
Tevatron cross sections of background $q\bar q, gg \to  t\bar{t}$
and signal $q\bar q, gg \to  tbH^\pm$
for $\tan\beta = 30$ and $m_{H^\pm} = 80, 100, 130$ and $150$~GeV
into the final state
$2 b + 2 j + \tau_\mathrm{jet} + p_t^{\rm miss}$
before ($\sigma^{\rm th}$) and after ($\sigma$)
the basic cuts ($p_t > 20$~GeV for all jets)
and the hard cut ($p_t^{\rm miss} > 100$ GeV).
}
\vspace*{-4mm}
\end{table}

\subsubsection{Event preselection and discussion of discriminant variables}
The expected cross sections of the 
$2 b + 2 j + \tau_\mathrm{jet} + p_t^{\rm miss}$ signature 
are of the same order of magnitude for the signal and background 
reactions, as shown in Table~\ref{tab:Hpm:crosssec}.
%
Thus, the same number of signal and background events is assumed for the 
analysis of different kinematic selection variables.
For the signal $5\cdot 10^5$ events have been simulated with PYTHIA for each 
charged Higgs mass at the Tevatron energy of 1.96~TeV using the built-in 
$t\bar t$ routine in the $t\bar t \rightarrow tbH^\pm$ approximation, 
while for the $t\bar t$ background also $5\cdot 10^5$ events have been 
simulated using the built-in $t\bar t$ routine.
Then the basic cuts $p_t^{\rm jets} > 20$~GeV are applied.
An additional hard cut on the missing transverse momentum
$p_t^{\rm miss} > 100$ GeV is used to suppress the QCD background, as 
for example demonstrated in Ref.~\cite{Assamagan:2002in}.
%
After the additional anti-QCD cut about 28000 to 42000 signal events,
depending on the simulated charged Higgs bosons mass, and about 30000 
$t\bar t$ background events remain.
Other background reactions, for example W+jet production, are expected to be 
negligible because they have either a much lower production cross section
or are strongly suppressed compared to $t\bar t$ background, 
as quantified for example in Ref.~\cite{Assamagan:2002in}.
In addition to the previous study 
(based on $5000\times \mathrm{BR}(\tau \to \mathrm{hadrons})$ events
each)~\cite{leshouches05},
the present one applies an IDA method~\cite{ida}
to explore efficiencies and purities. As already mentioned,
particular attention is devoted to the study of 
spin sensitive variables in the exploitation of polarization
effects for the separation of signal and background events.

Figures~\ref{fig:pttau}--\ref{fig:hjet} show examples of the signal and
background distributions of some of the kinematic 
variables used in the IDA method and the respective difference between
signal and background distributions, namely:
\begin{itemize}
\item the transverse momentum of the $\tau$ jet, $p_t^{\tau_{\rm jet}}$~(Fig.~\ref{fig:pttau}),
\item the transverse momentum of the leading $\pi^\pm$ in the $\tau$ jet, 
$p_t^{\pi^\pm}$ (Fig.~\ref{fig:ptpi})
\item the ratio $p_t^{\pi^\pm}/p_t^{\tau_{\rm jet}}$ (Fig.~\ref{fig:r1}),
\item the transverse momentum of the second (least energetic) $b$ quark jet, $p_t^{b_2}$ (Fig.~\ref{fig:ptb2}),
\item the transverse mass in the $\tau_{\rm jet} + p_t^{\rm miss}$ system,
      $m_t = \sqrt{2 p_t^{\tau_\mathrm{jet}} p_t^{\rm miss}
             [1-\cos(\Delta\phi)]}$, where $\Delta\phi$ is the azimuthal angle
      between $p_t^{\tau_\mathrm{jet}}$ and $p_t^{\rm miss}$ 
(Fig.~\ref{fig:mtransverse})\footnote{Strictly speaking this is not the transverse mass since there are two neutrinos in the decay chain of the charged Higgs boson we are considering, even so the characteristics of this mass are very similar to that of the true transverse mass.},
\item the invariant mass distribution of the two light quark jets and the
      second $b$ quark jet, $m_{jjb_2}$ (Fig.~\ref{fig:mjjb2}),
\item the spatial distance between the $\tau$ jet and the second $b$ quark jet, 
      $\Delta R(\tau,b_2) = \sqrt{(\Delta\phi)^2 + (\Delta\eta)^2}$,
      where $\Delta\phi$ is the azimuthal angle between the $\tau$ and $b$ jet (Fig.~\ref{fig:distance-tau-b})
      and 
\item the sum of the (scalar) transverse momenta of all the quark jets, 
      $H_{\rm jets} = p_t^{j_1} + p_t^{j_2} + p_t^{b_1} + p_t^{b_2}$ (Fig.~\ref{fig:hjet}).
\end{itemize}
The distributions of signal and background events are normalized 
to the same number of $10^4$ events, 
in order to make small differences better visible.

The signal and background distributions 
for the variables shown in Figs.~\ref{fig:ptb2}--\ref{fig:hjet} 
are as expected rather similar for $m_{H^\pm}=m_{W^\pm}$
and are hence mostly important to discriminate between signal and
background in the IDA for $m_{H^\pm} > m_{W^\pm}$.
Especially the transverse mass, Fig.~\ref{fig:mtransverse},
shows a large variation with the charged Higgs boson mass.
However, the different spin
of the charged Higgs boson and the $W^\pm$ boson has a large effect on
the $\tau$ jet variables $p_t^{\tau_{\rm jet}}$ and $p_t^{\pi^\pm}$
(Figs.~\ref{fig:pttau} and \ref{fig:ptpi}) resulting in
significantly different distributions of signal and background
even for $m_{H^\pm}=m_{W^\pm}$.
Moreover, the spin effects in the $p_t^{\tau_{\rm jet}}$ and $p_t^{\pi^\pm}$ 
distributions are correlated which can be seen in Fig.~\ref{fig:r1}
where the distributions of the ratio $p_t^{\pi^\pm}/p_t^{\tau_{\rm jet}}$
\cite{Roy:1991sf,Raychaudhuri:1995cc,Roy:1999xw}
show even larger differences.
This highlights the importance
of the additional variable $p_t^{\pi^\pm}$ (and hence
$p_t^{\pi^\pm}/p_t^{\tau_{\rm jet}}$),
compared to a previous study \cite{leshouches05}.
The large separation power of this variable
is indeed due to the different $\tau$ polarizations
in signal and background as can be inferred from the lower plots in
Figs.~\ref{fig:pttau}--\ref{fig:r1}.
There the signal and background distributions for $p_t^{\tau_{\rm jet}}$,
$p_t^{\pi^\pm}$ and $p_t^{\pi^\pm}/p_t^{\tau_{\rm jet}}$ are shown
for reference samples where the $\tau$ decay has been performed without the
inclusion of spin effects with the built-in routines of PYTHIA and
hence the differences between signal and background nearly vanish.

\subsubsection{Iterative discriminant analysis (IDA)}

The IDA method is a modified Fisher Discriminant Analysis~\cite{ida}
and is characterized by 
the use of a quadratic, instead of a linear, discriminant function and 
also involves iterations in order to 
enhance the separation between signal and background.

In order to analyze our events with the IDA method, signal
and background  have been split in two samples of equal size.
With the first set of samples the IDA training has been performed and then
the second set of samples has been analyzed.
We have used the following 20 variables in the IDA study:
the transverse momenta 
$p_t^{\tau_{\rm jet}}$, $p_t^{\pi^\pm}$,
$p_t^{\rm miss}$,
$p_t^{b_1}$, $p_t^{b_2}$, $p_t^{j_1}$, $p_t^{j_2}$, $p_t^{j j}$;
the transverse mass $m_t$;
the invariant masses
$m_{jj}$, $m_{jjb_1}$, $m_{jjb_2}$, $m_{bb}$ and
$\hat{s}=m_{jjbb\tau}$;
the spatial distances $\Delta R(\tau,b_1)$, $\Delta R(\tau,b_2)$,
$\Delta R(\tau,j_1)$, $\Delta R(\tau,j_2)$;
the total transverse momenta of all quark jets $H_{\rm jets}$ 
and of all jets $H_{\rm all} = H_{\rm jets} + p_t^{\tau_{\rm jet}}$.
In the analysis of real data, b-quark tagging probabilities and 
the reconstruction of $t$ and $W$ masses could be used to improve
the jet pairing, and replace the allocation of least and most energetic
$b$-jet by a probabilistic analysis.

%
The results of the IDA study are shown in Figs.~\ref{fig:ida1}
and \ref{fig:ida} for the event samples with spin effect in the
$\tau$ decays for $m_{H^\pm}=80, 100, 130, 150$~GeV and for the
reference samples without the spin effect for $m_{H^\pm}=80$~GeV
in order to illustrate the spin effect.
In all plots of the IDA output variable the number of 
background events has been normalized to the number of signal events.
Two IDA steps have been performed.
Figure~\ref{fig:ida1} shows the IDA output variable after the first
step, where 90\% of the signal is retained when a cut at zero is applied. 
The signal and background events after this cut are then passed to the 
second IDA step. 
Figure~\ref{fig:ida} shows the
IDA output variable distributions after the second step.
A cut on these distributions leads to the
efficiency and purity (defined as ratio of the number of signal events
divided by the sum of 
signal and background events) combinations as shown in the
lower right plot in Fig.~\ref{fig:ida}.
These combinations define the working point (number of expected
background events for a given 
signal efficiency) and the latter can be optimized to maximize the
discovery potential. 
The difference between the dashed (no spin effects in $\tau$ decay)
and solid (with spin effects in $\tau$ decay) lines for $m_{H^\pm}=80$~GeV
in the lower right plot in Fig.~\ref{fig:ida} stresses again the importance
of the spin effects to separate signal and background.

In order to illustrate the effect of the hard cut on the missing
transverse momentum ($p_t^{\rm miss}>100$~GeV), which is imposed to
suppress the QCD background, 
the final efficiency-purity plot of the IDA analysis is shown 
in Fig.~\ref{fig:ida_hard_cut} for $m_{H^\pm}=80$~GeV for two
reference samples (red, long dashed: with spin effects in the $\tau$
decay; red, dotted: without spin effects) without imposing the hard
cut. The black lines (dashed and solid) are for the samples with the
hard cut as also shown in the lower right plot in Fig.~\ref{fig:ida}.
As expected the achievable purity for a given efficiency decreases
with the hard cut,
therefore the spin effects become even more important to
separate signal and background.
In principle, by choosing the signal reduction rates in the previous IDA iterations, 
the signal and background rates in the final distributions can be varied appropriately.
However, we have checked that a different number of IDA iterations and/or different efficiencies 
for the first IDA iteration have only a minor effect on the final result.

\begin{figure}[htbp]
\epsfig{file=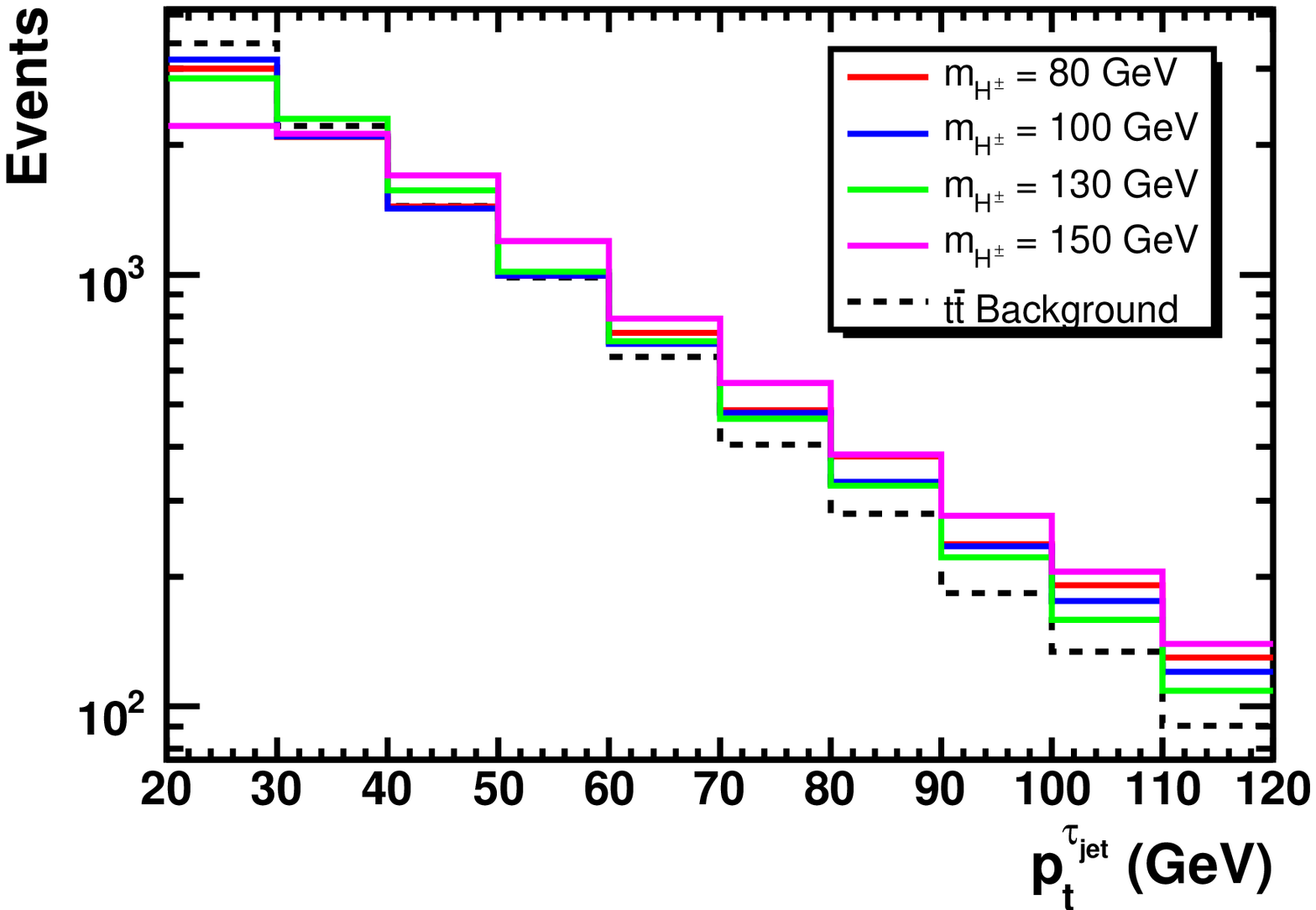, width=0.4\textwidth} \hfill
\epsfig{file=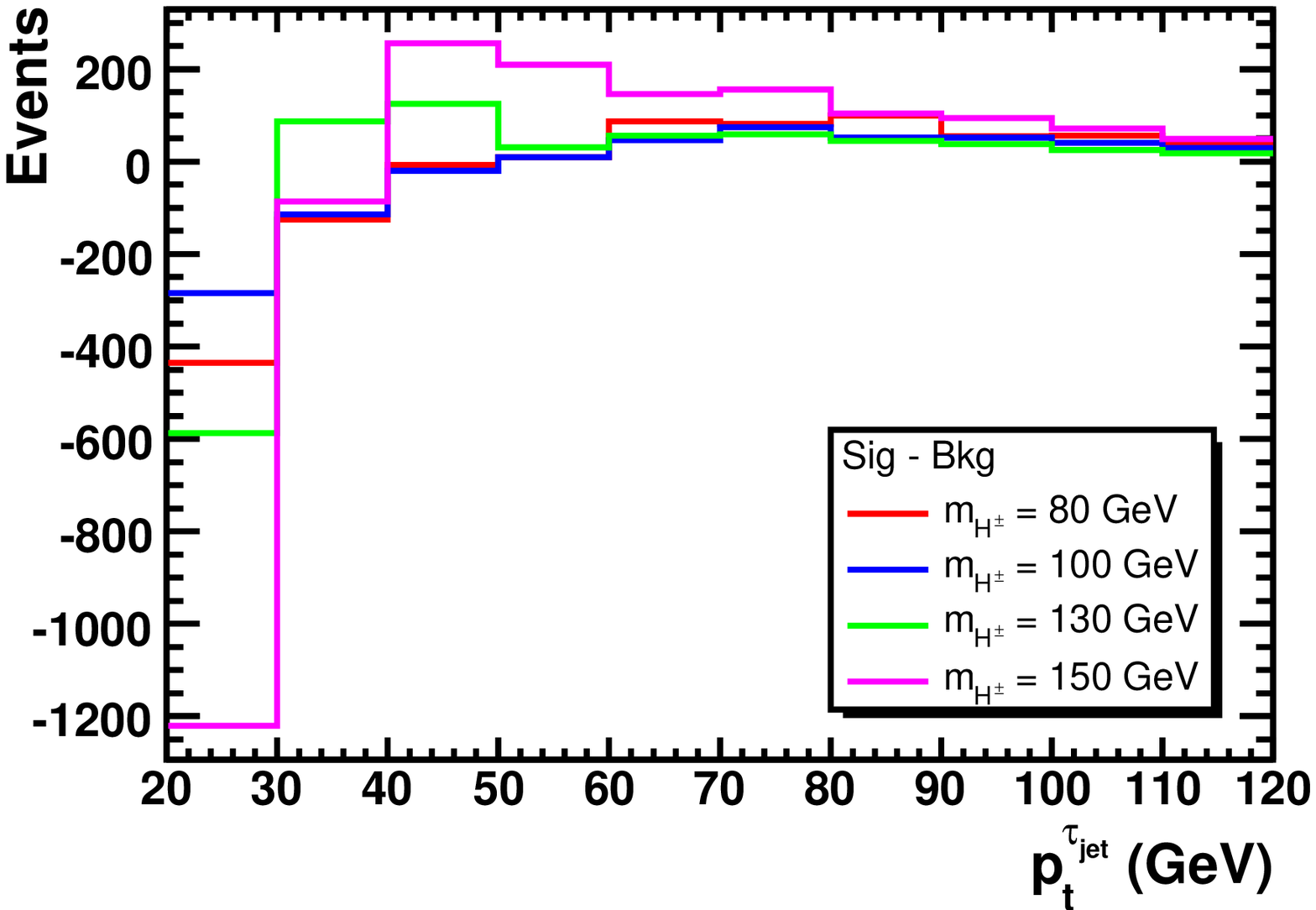, width=0.4\textwidth}
\epsfig{file=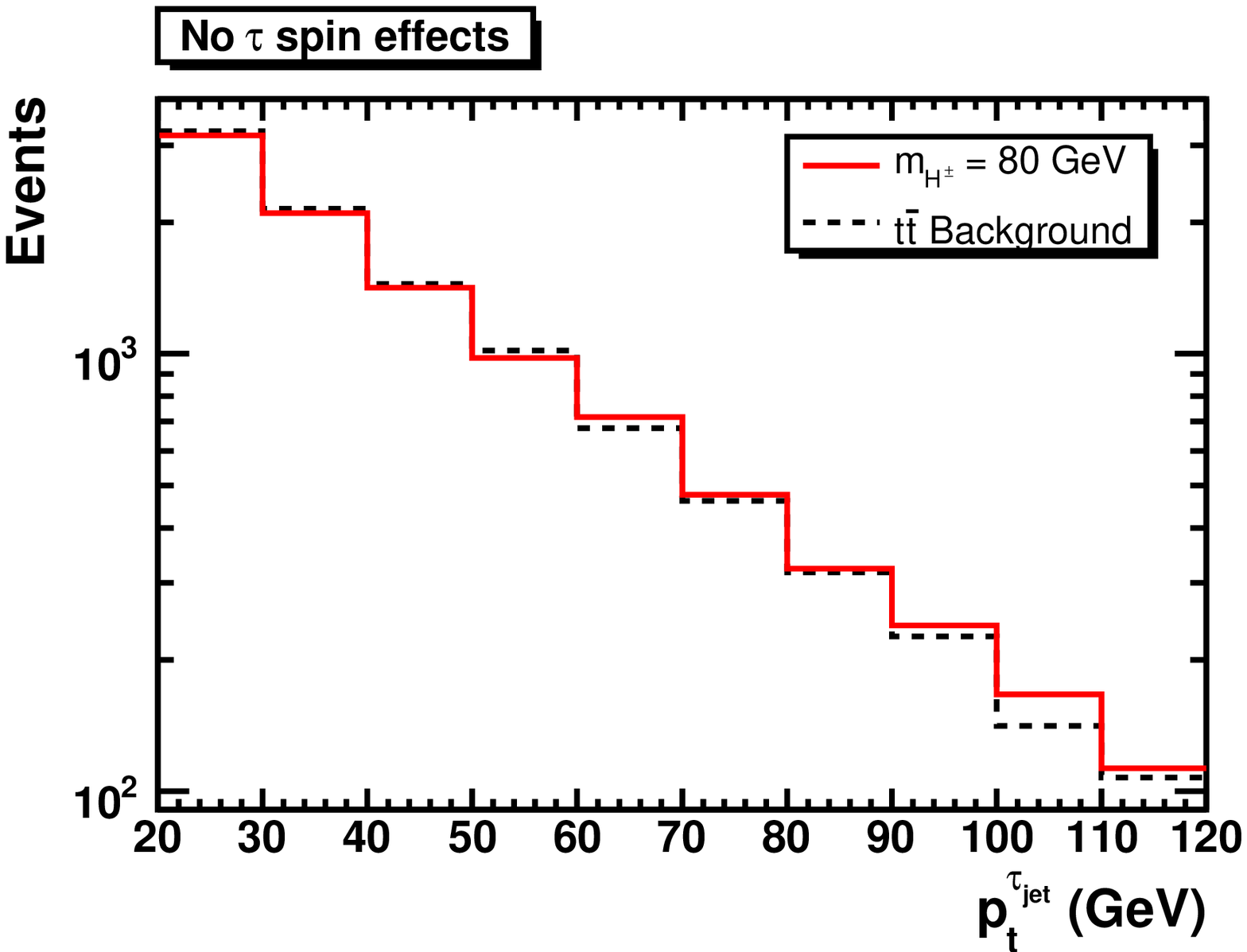, width=0.4\textwidth} \hfill
\epsfig{file=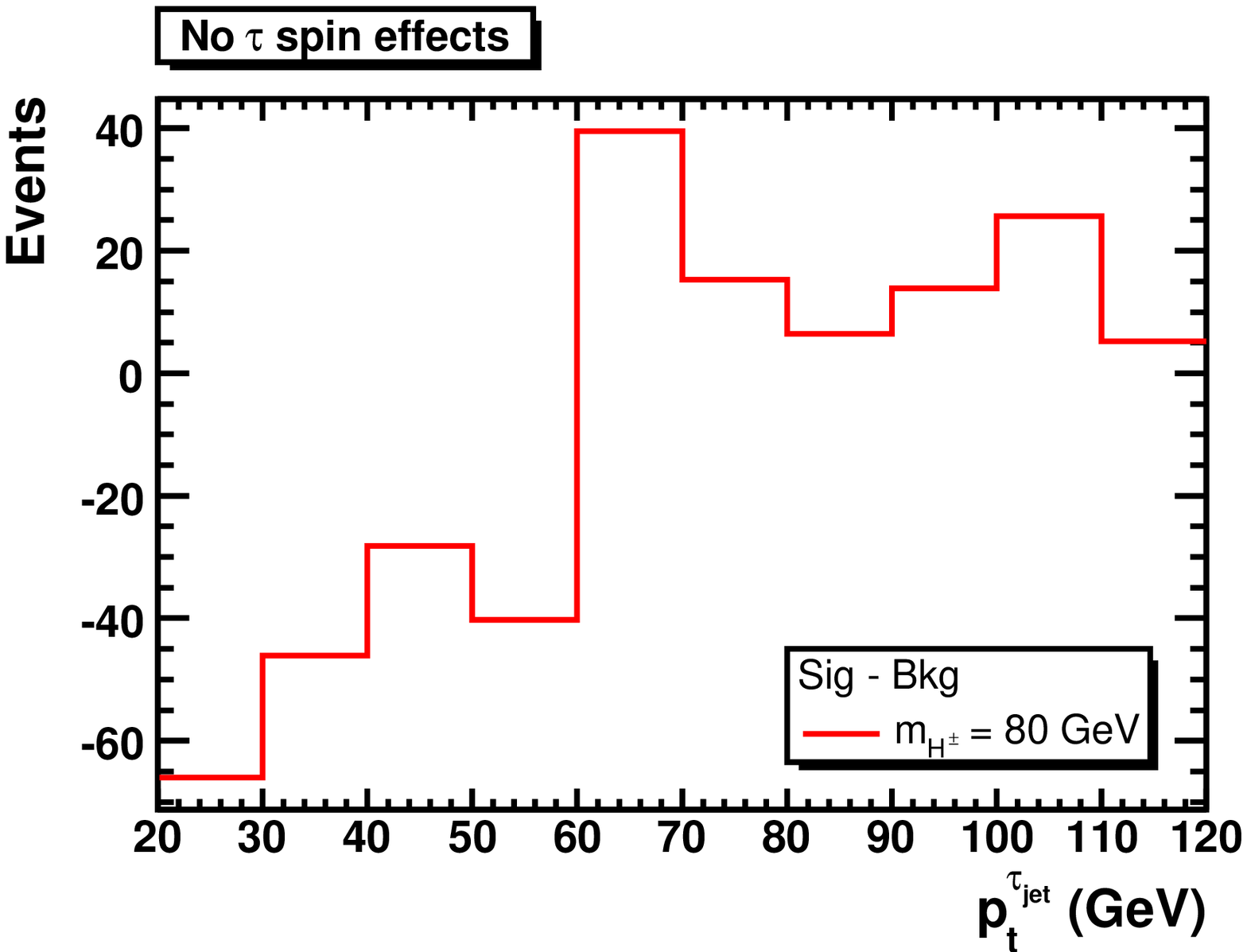, width=0.4\textwidth}
\vspace*{-0.2cm}
\caption{ 
$p_t$ distributions 
of the $\tau~{\mathrm{jet}}$ for the $tbH^\pm$ signal
and the $t\bar{t}$ background for $\sqrt{s}=1.96$~TeV (left)
and the respective differences between signal and background (right).
The lower plots show distributions without spin effects in the $\tau$
decays.
}
\label{fig:pttau}
\end{figure}

\begin{figure}[htbp]
\vspace*{-0.2cm}
\epsfig{file=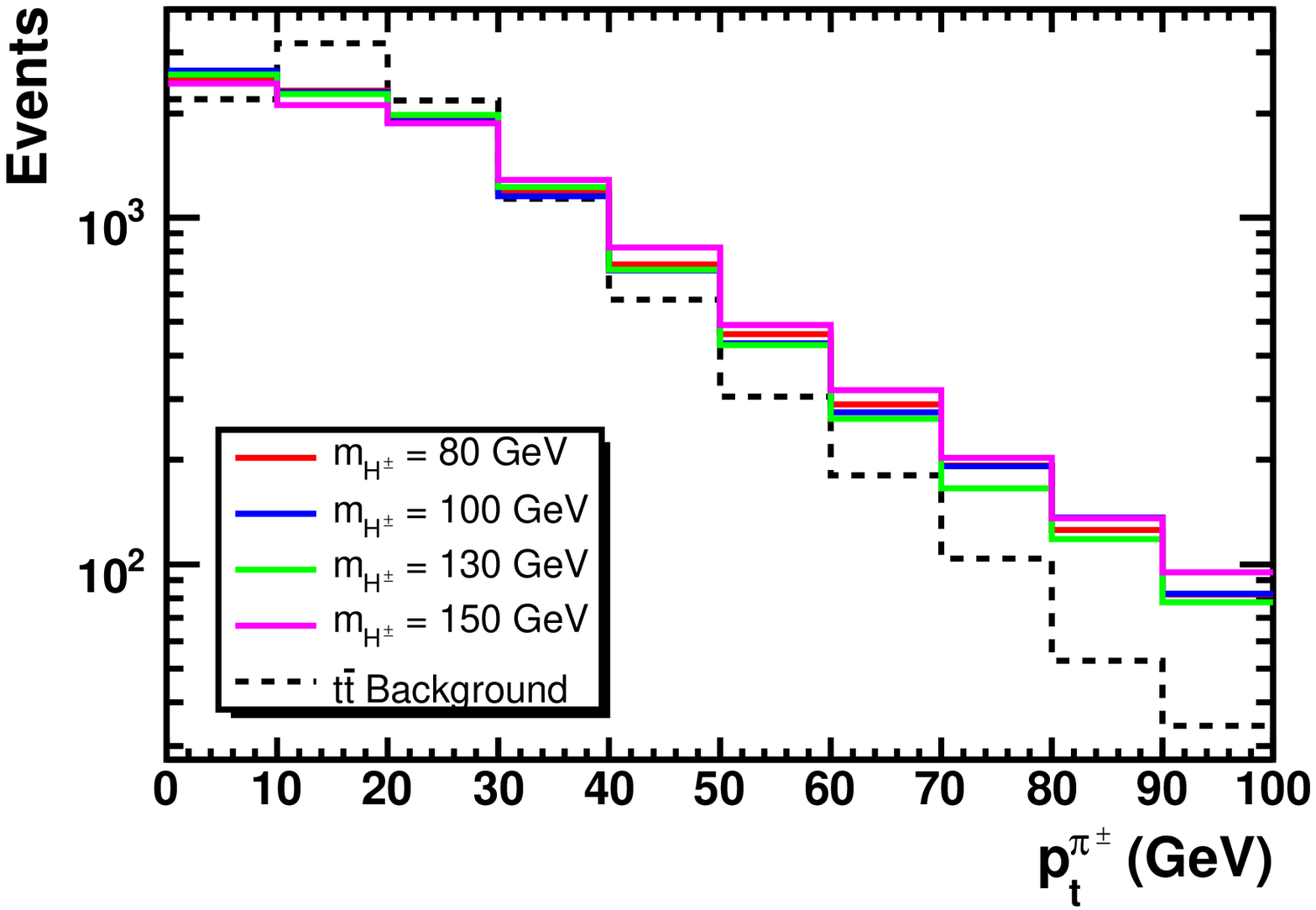, width=0.4\textwidth}  \hfill
\epsfig{file=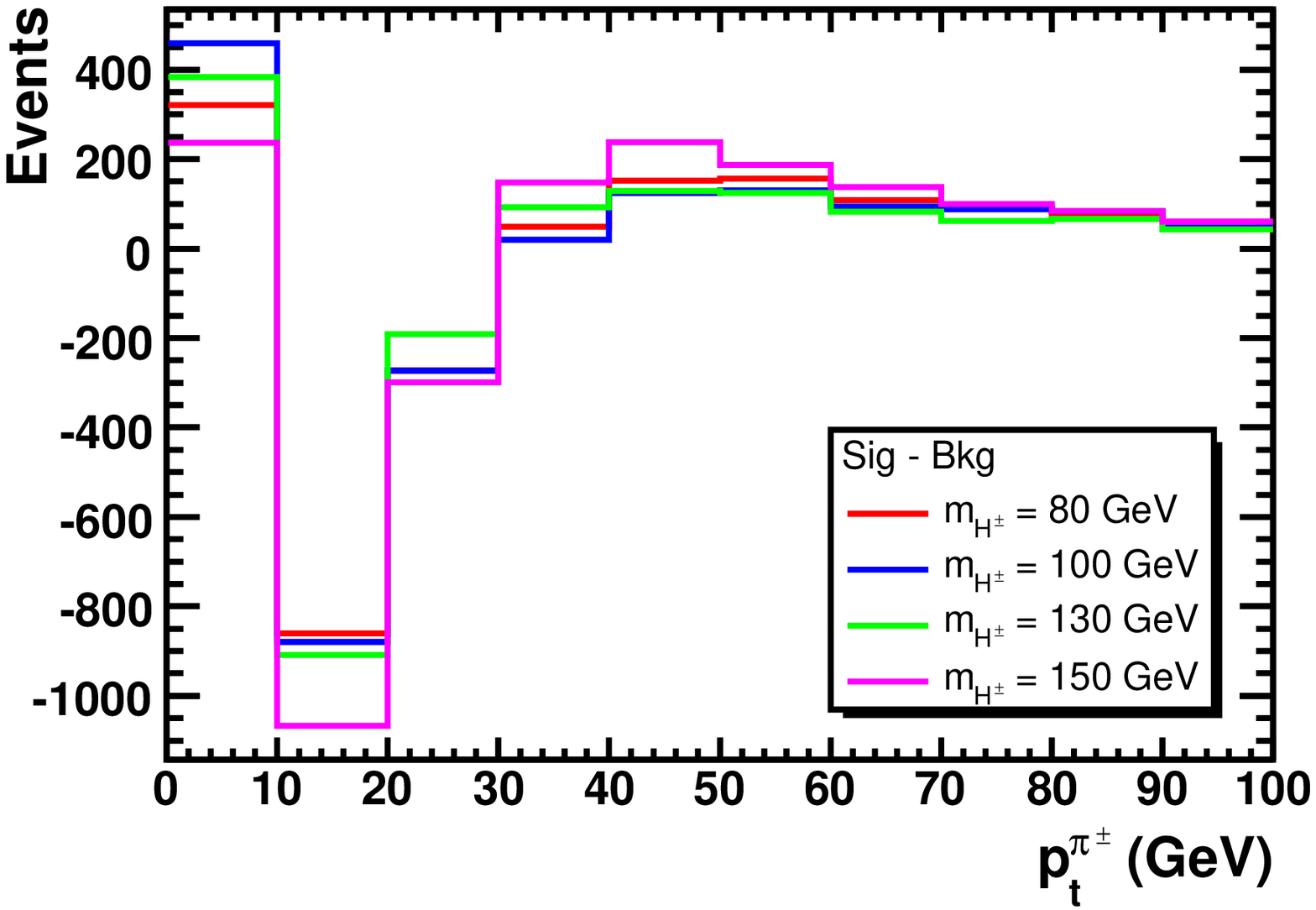, width=0.4\textwidth}
\epsfig{file=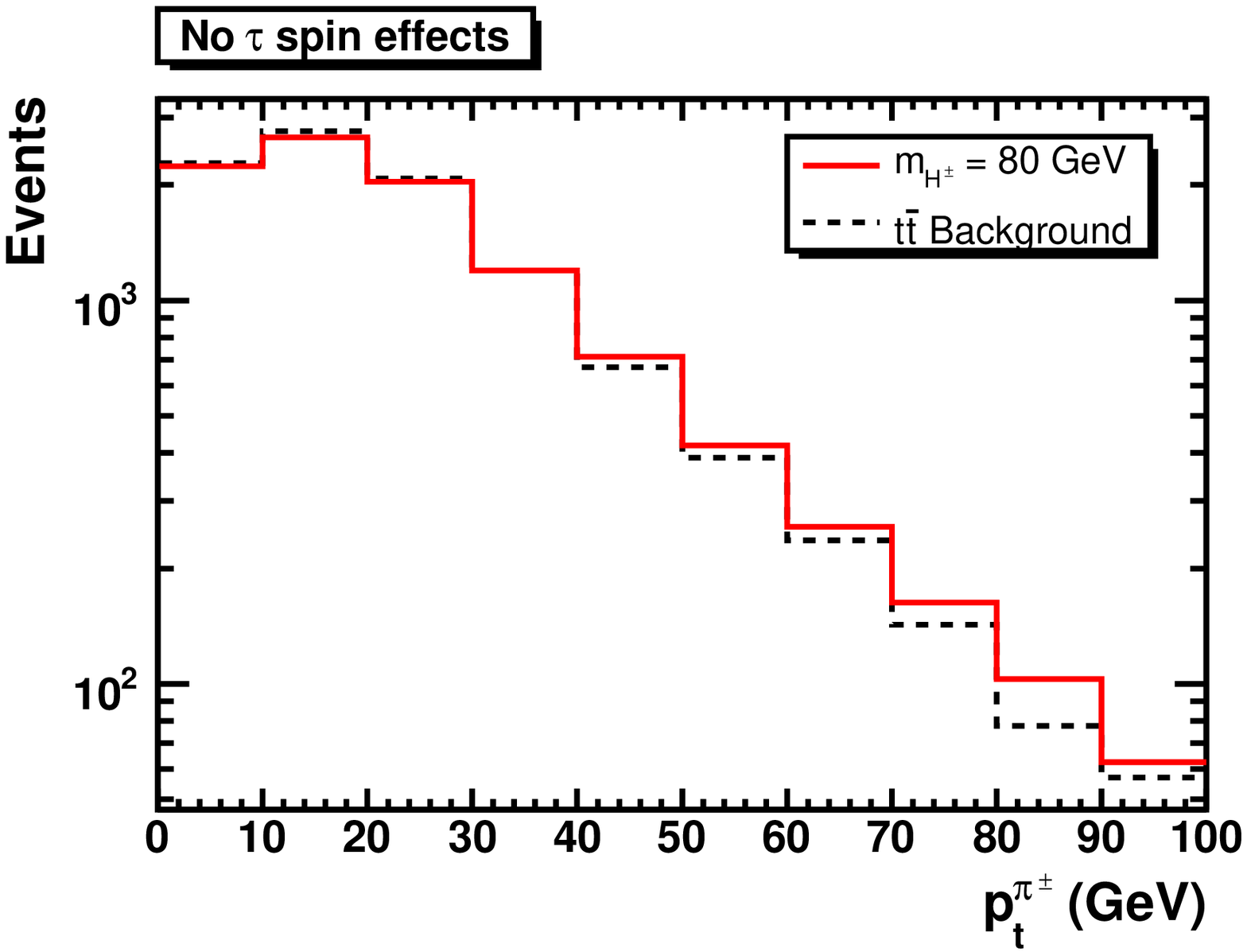, width=0.4\textwidth}  \hfill
\epsfig{file=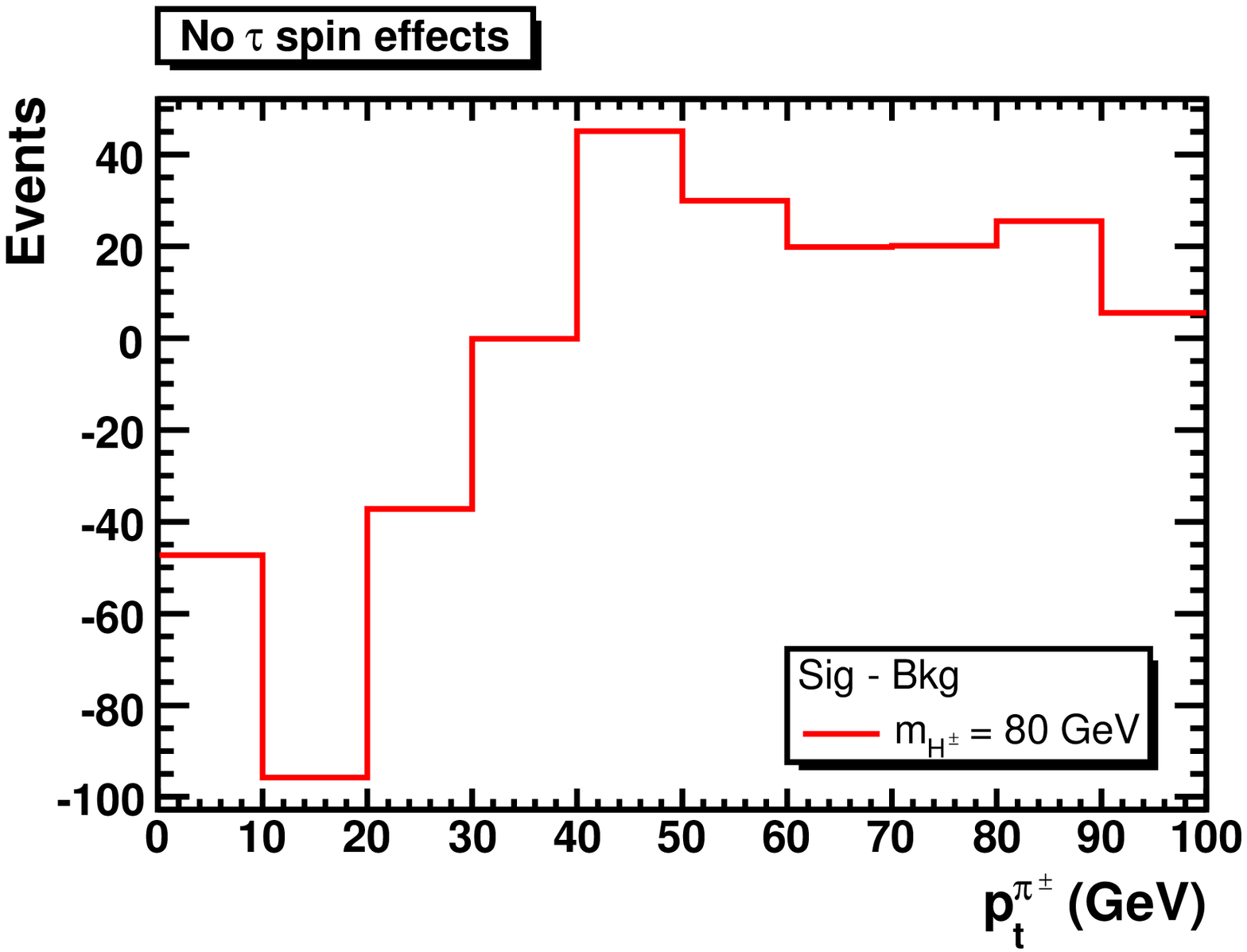, width=0.4\textwidth}
\vspace*{-0.2cm}
\caption{
$p_t$  distributions of the leading $\pi^\pm$ from the $\tau$ decay 
for the $tbH^\pm$ signal and the 
$t\bar{t}$ background for $\sqrt{s}=1.96$~TeV (left)
and the respective differences between signal and background (right).
The lower plots show distributions without spin effects in the $\tau$
decays.
}
\label{fig:ptpi}
\vspace*{-0.2cm}
\end{figure}

\begin{figure}[htbp]
\epsfig{file=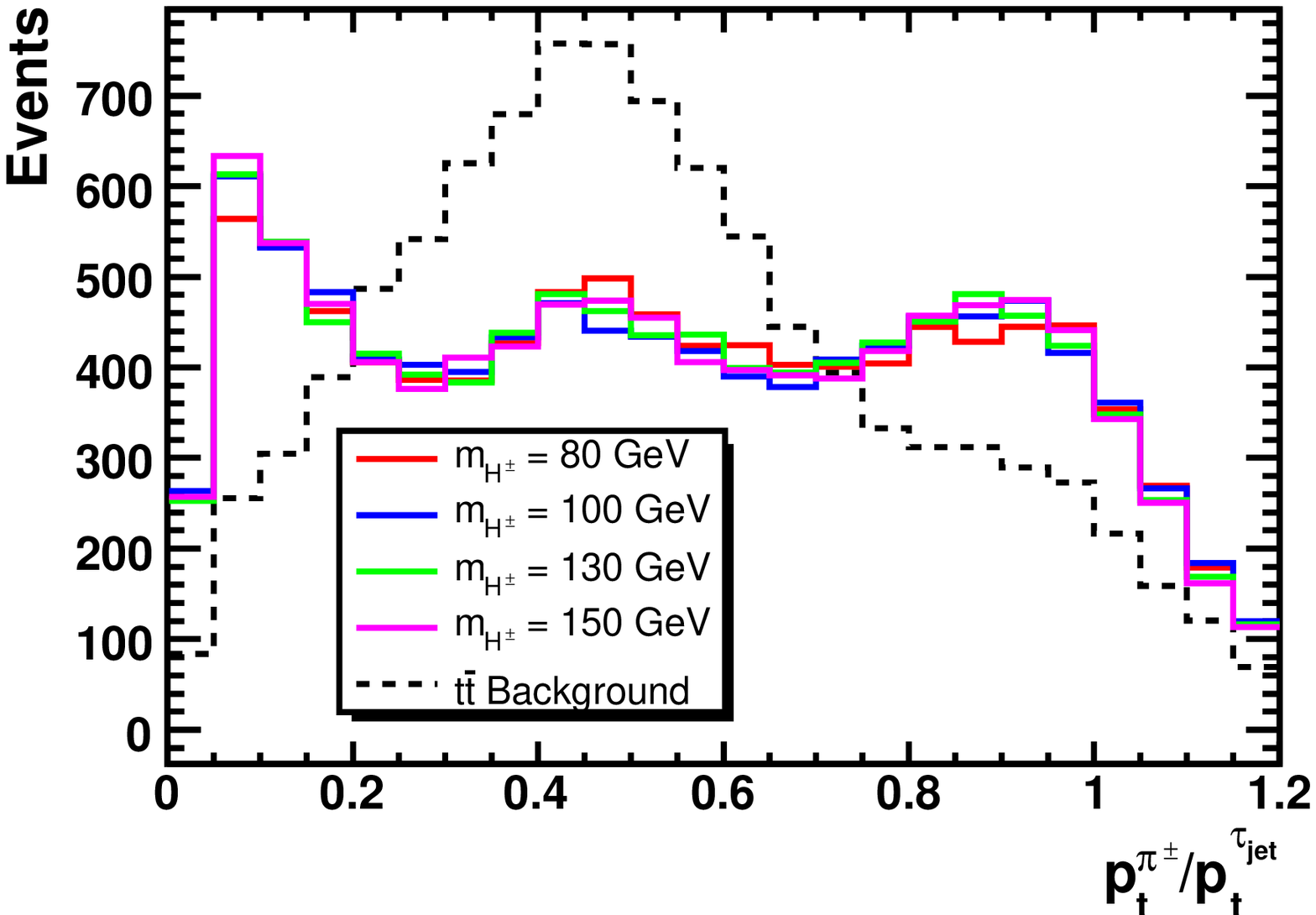, width=0.5\textwidth} \hfill
\epsfig{file=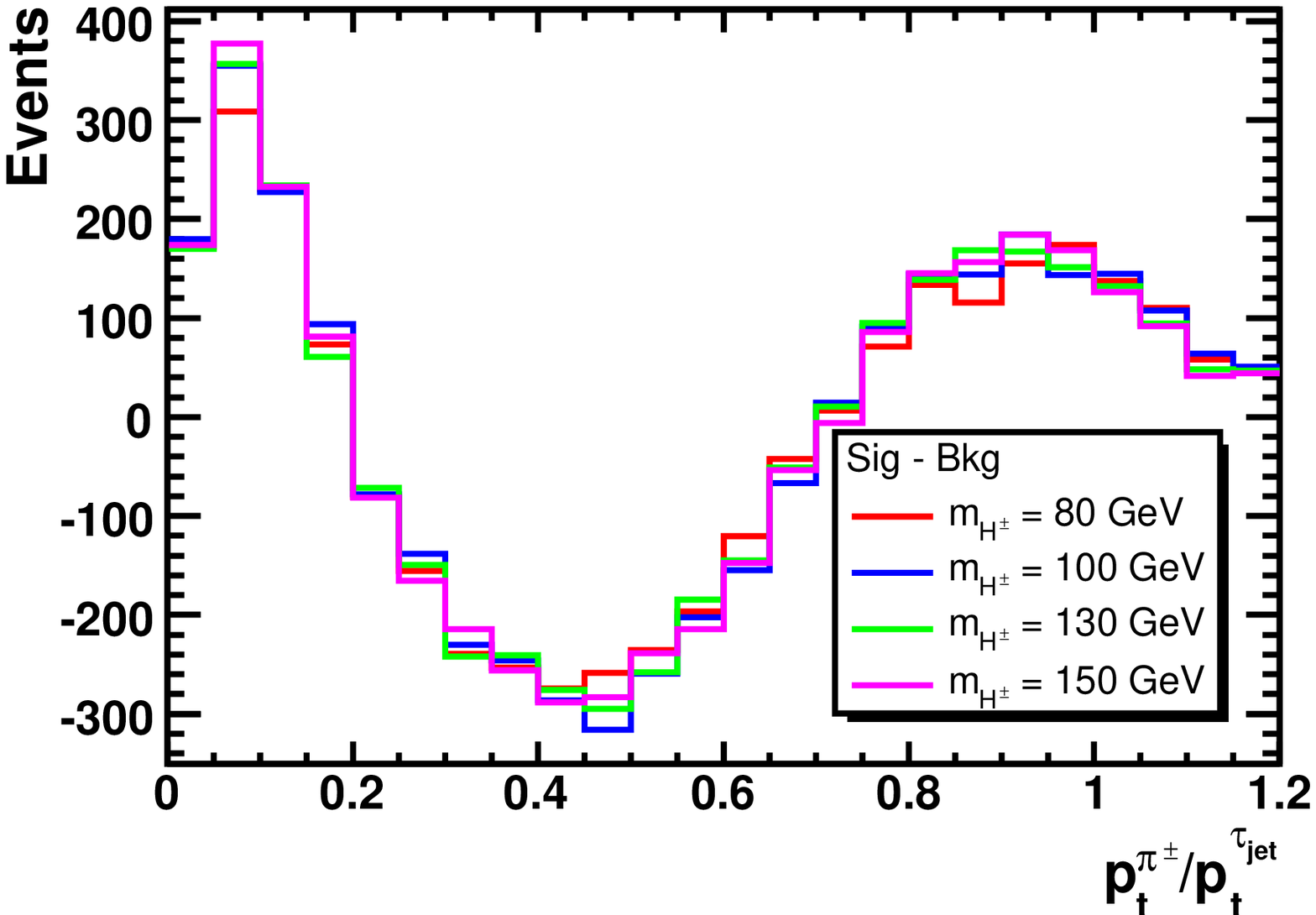, width=0.5\textwidth}
\epsfig{file=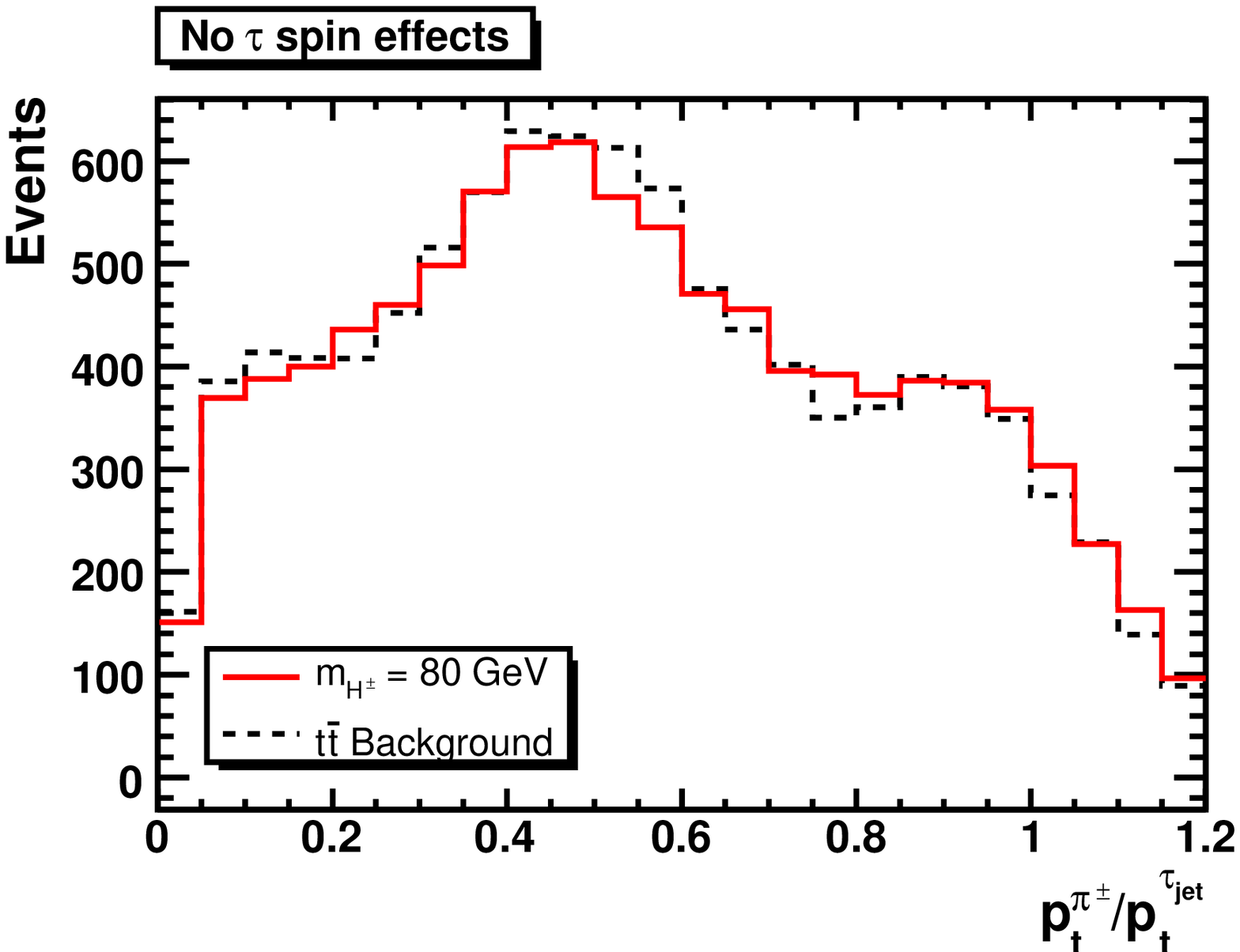, width=0.5\textwidth} \hfill
\epsfig{file=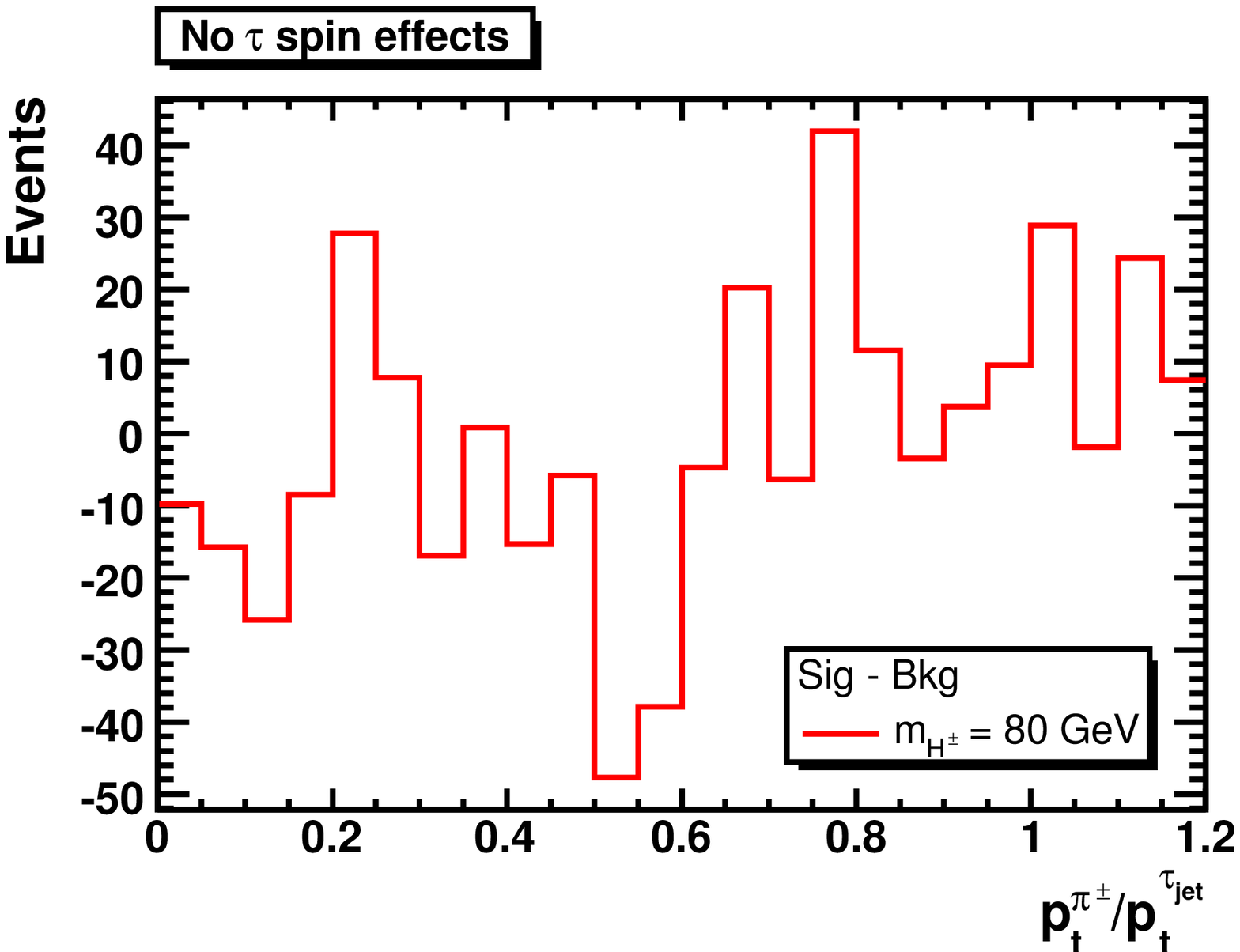, width=0.5\textwidth}
\caption{Distributions of the  
ratio $p_t^{\pi^\pm}/p_t^{\tau_\mathrm{jet}}$
for the $tbH^\pm$ signal
and the $t\bar{t}$ background for $\sqrt{s}=1.96$~TeV (left)
and the respective differences between signal and background (right).
The lower plots show distributions without spin effects in the $\tau$
decays.
}
\label{fig:r1}
\end{figure}

\begin{figure}[htbp]
\epsfig{file=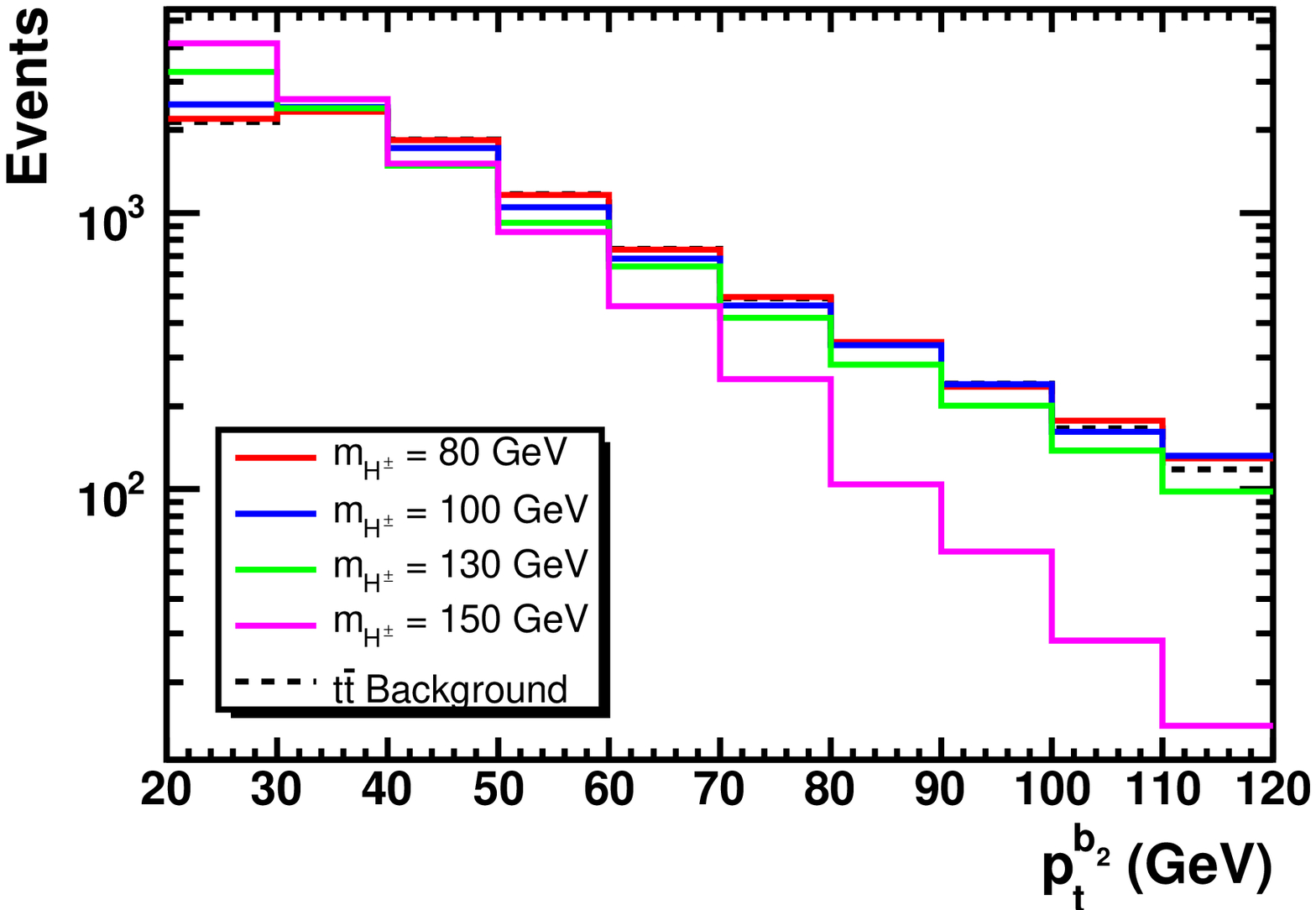, width=0.5\textwidth}  \hfill
\epsfig{file=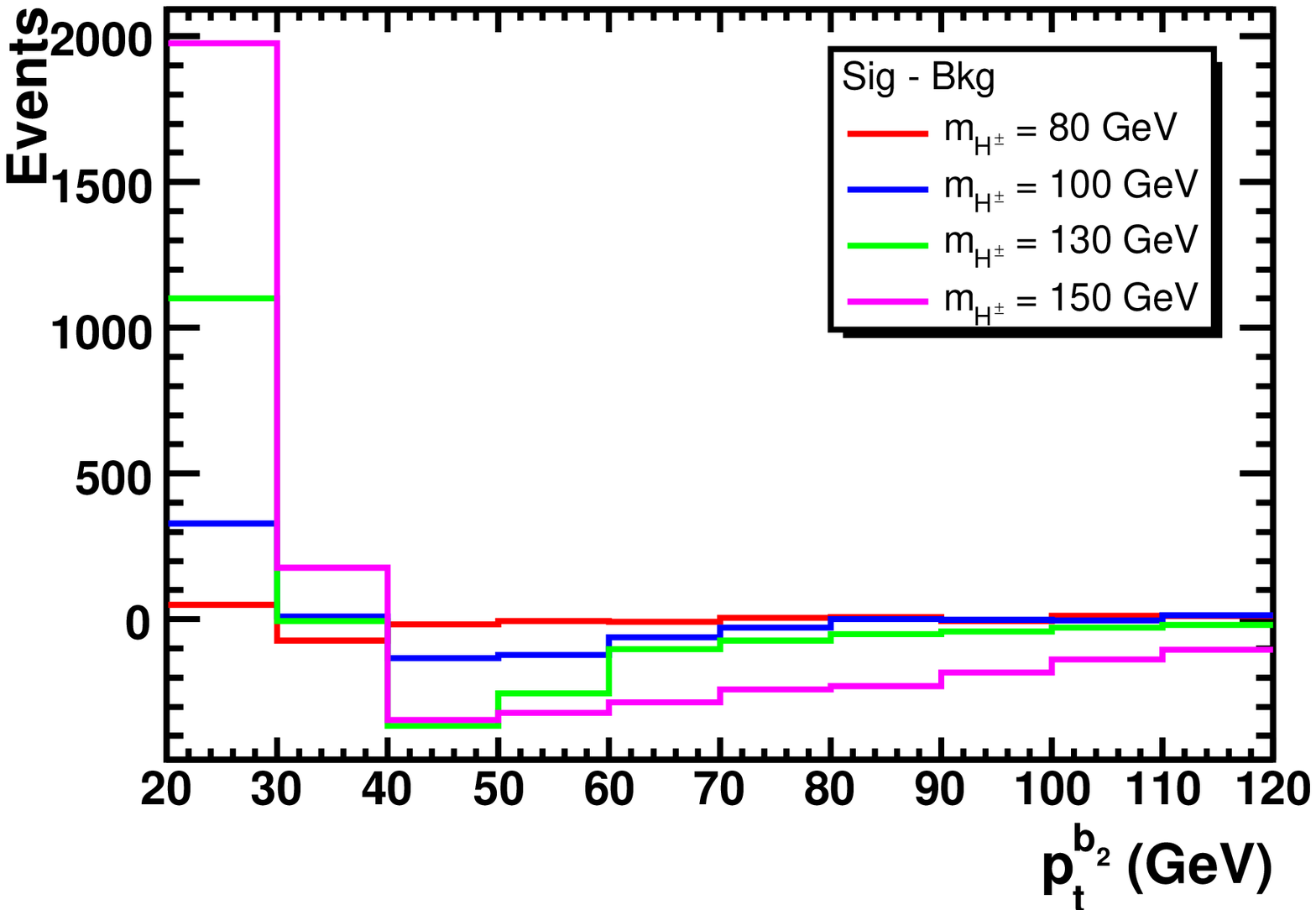, width=0.5\textwidth}
\caption{
$p_t$ distributions of the second (least energetic) $b$ quark jet
for the $tbH^\pm$ signal
and the $t\bar{t}$ background for $\sqrt{s}=1.96$~TeV (left)
and the respective differences between signal and background (right).
}
\label{fig:ptb2}
\end{figure}

\begin{figure}[htbp]
\vspace*{-2mm}
\epsfig{file=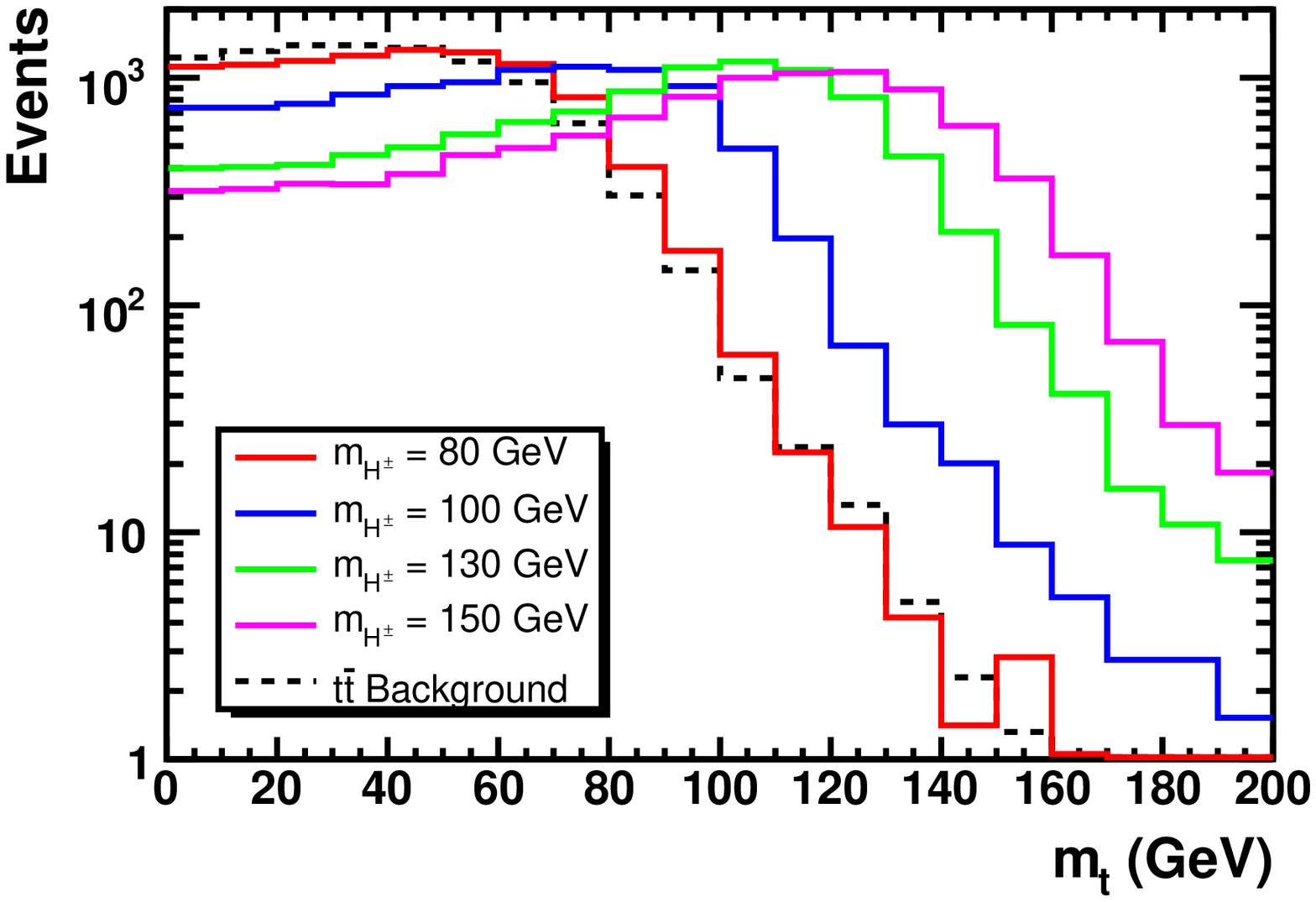, width=0.47\textwidth}  \hfill
\epsfig{file=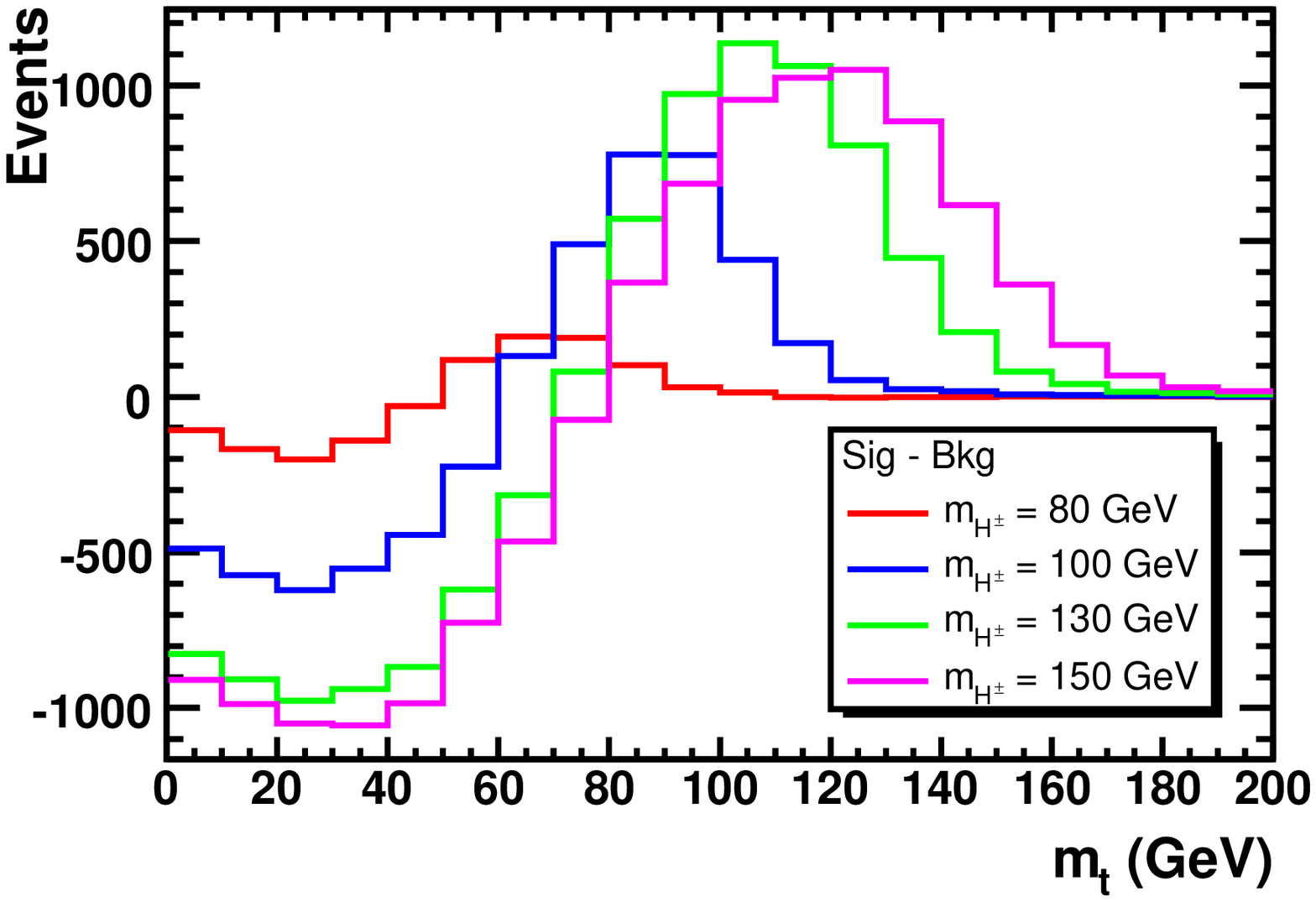, width=0.47\textwidth}
\vspace*{-4mm}
\caption{
Transverse mass
$m_t = \sqrt{2 p_t^{\tau_\mathrm{jet}} p_t^{\rm miss}
  [1-\cos(\Delta\phi)]}$
distributions of the
$\tau_{\rm jet} + p_t^{\rm miss}$ system
($\Delta\phi$ is the azimuthal angle
between $p_t^{\tau_\mathrm{jet}}$ and $p_t^{\rm miss}$)
for the $tbH^\pm$ signal
and the $t\bar{t}$ background for $\sqrt{s}=1.96$~TeV (left)
and the respective differences between signal and background (right).
}
\label{fig:mtransverse}
\vspace*{-2mm}
\end{figure}

\begin{figure}[htbp]
\epsfig{file=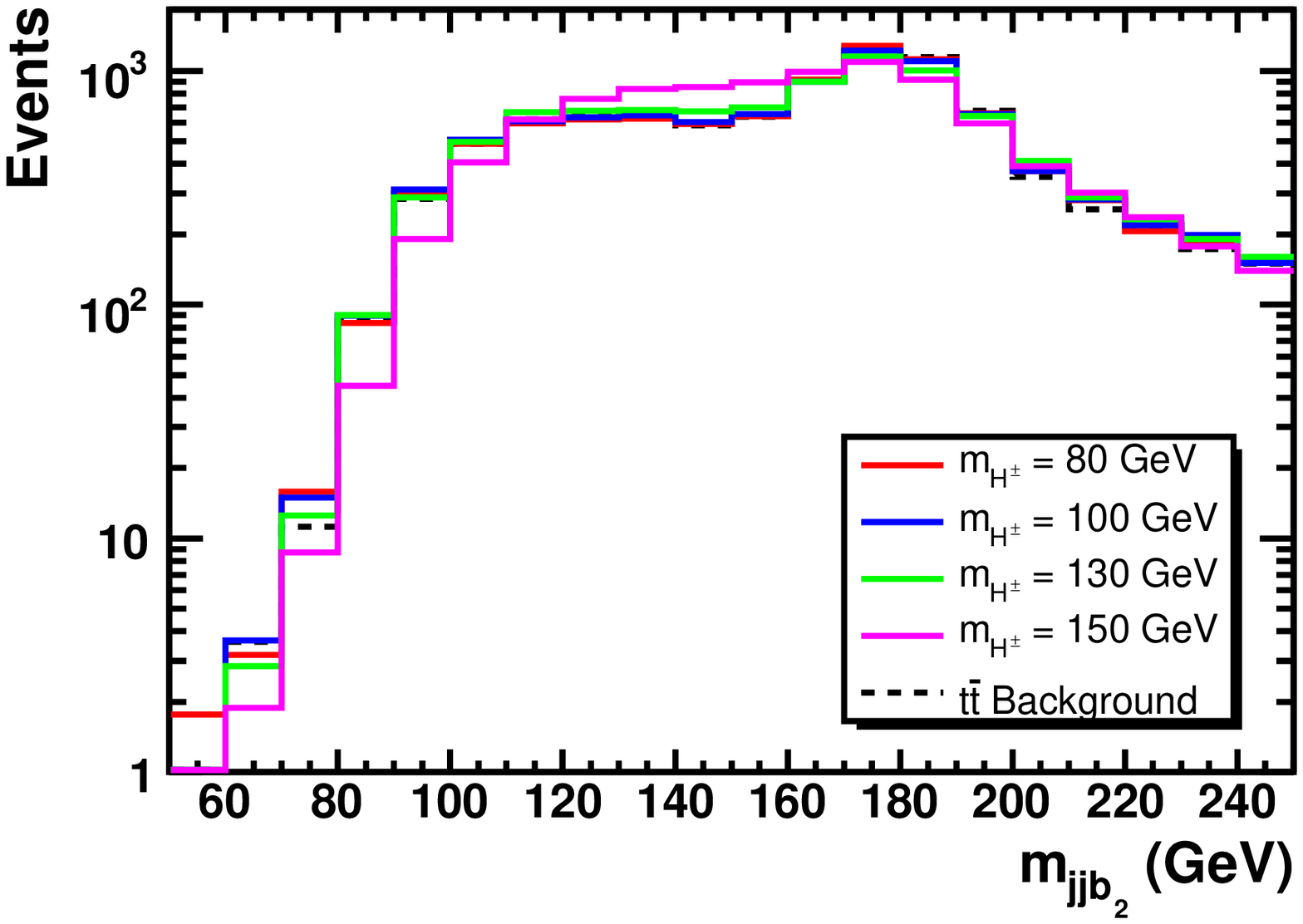, width=0.47\textwidth}  \hfill
\epsfig{file=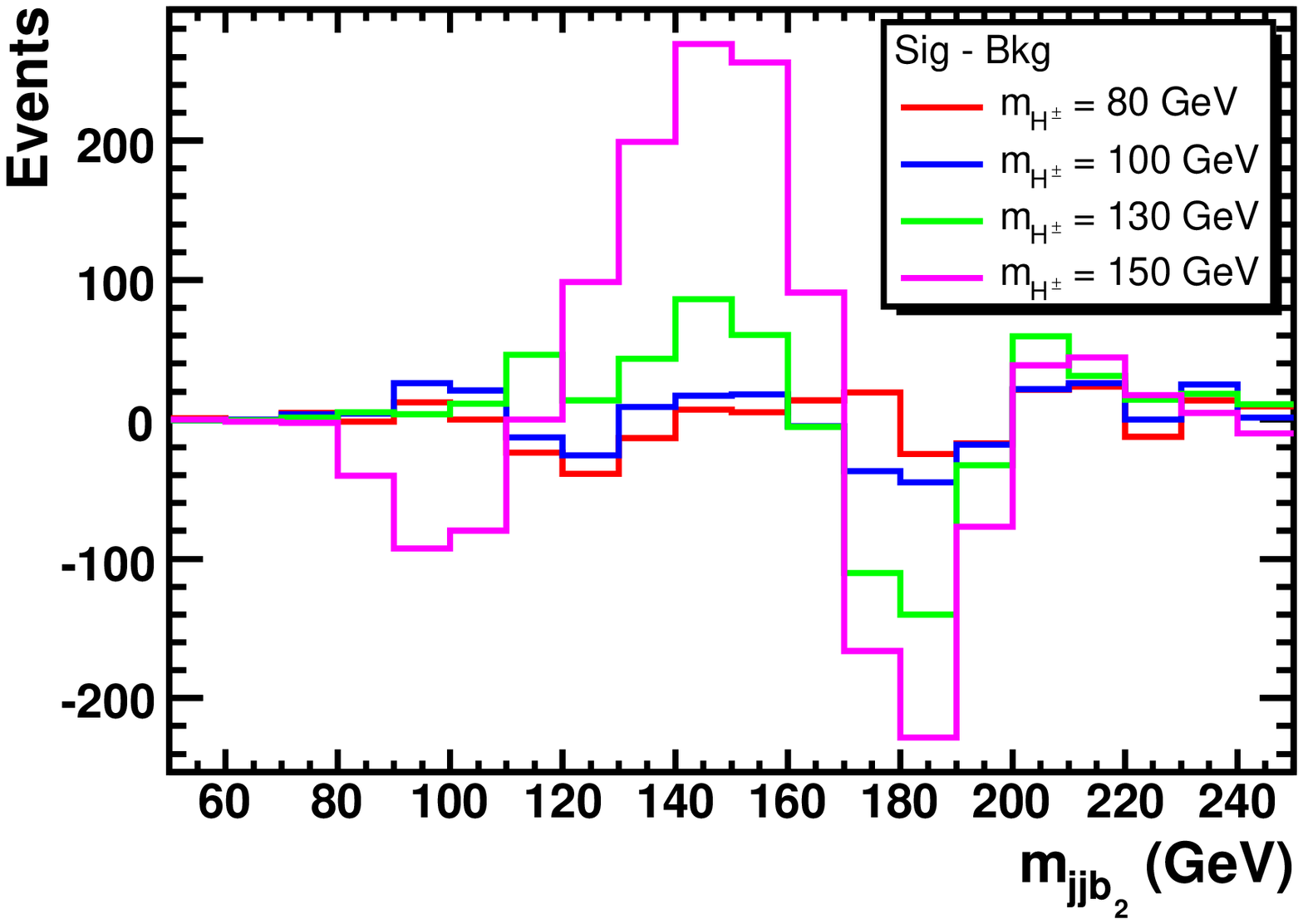, width=0.47\textwidth}
\vspace*{-4mm}
\caption{
Invariant mass distributions of the two light quark jets and the 
second (least energetic) $b$ quark jet
for the $tbH^\pm$ signal
and the $t\bar{t}$ background for $\sqrt{s}=1.96$~TeV (left)
and the respective differences between signal and background (right).
}
\label{fig:mjjb2}
\vspace*{-2mm}
\end{figure}

\begin{figure}[htbp]
\epsfig{file=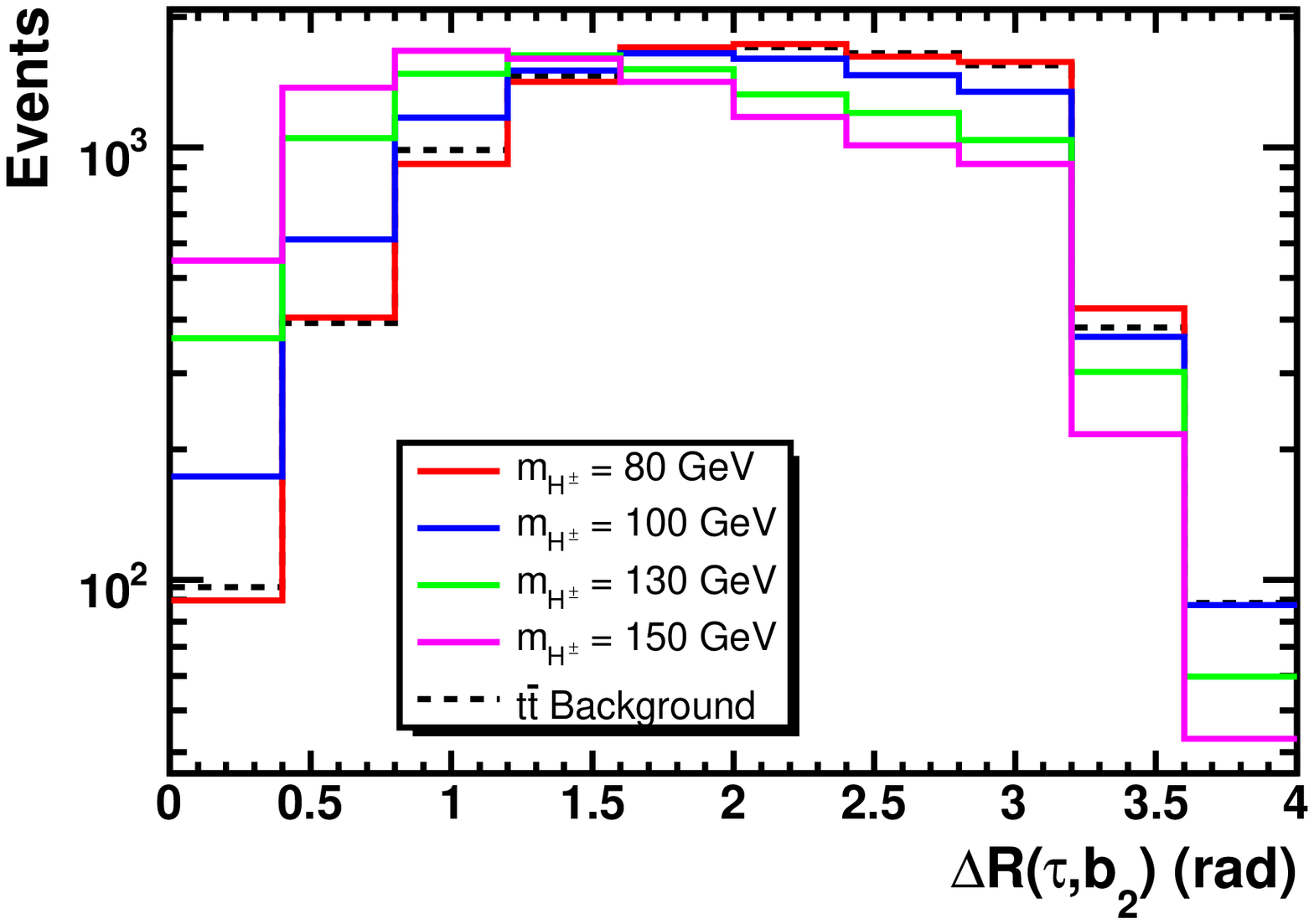, width=0.47\textwidth}  \hfill
\epsfig{file=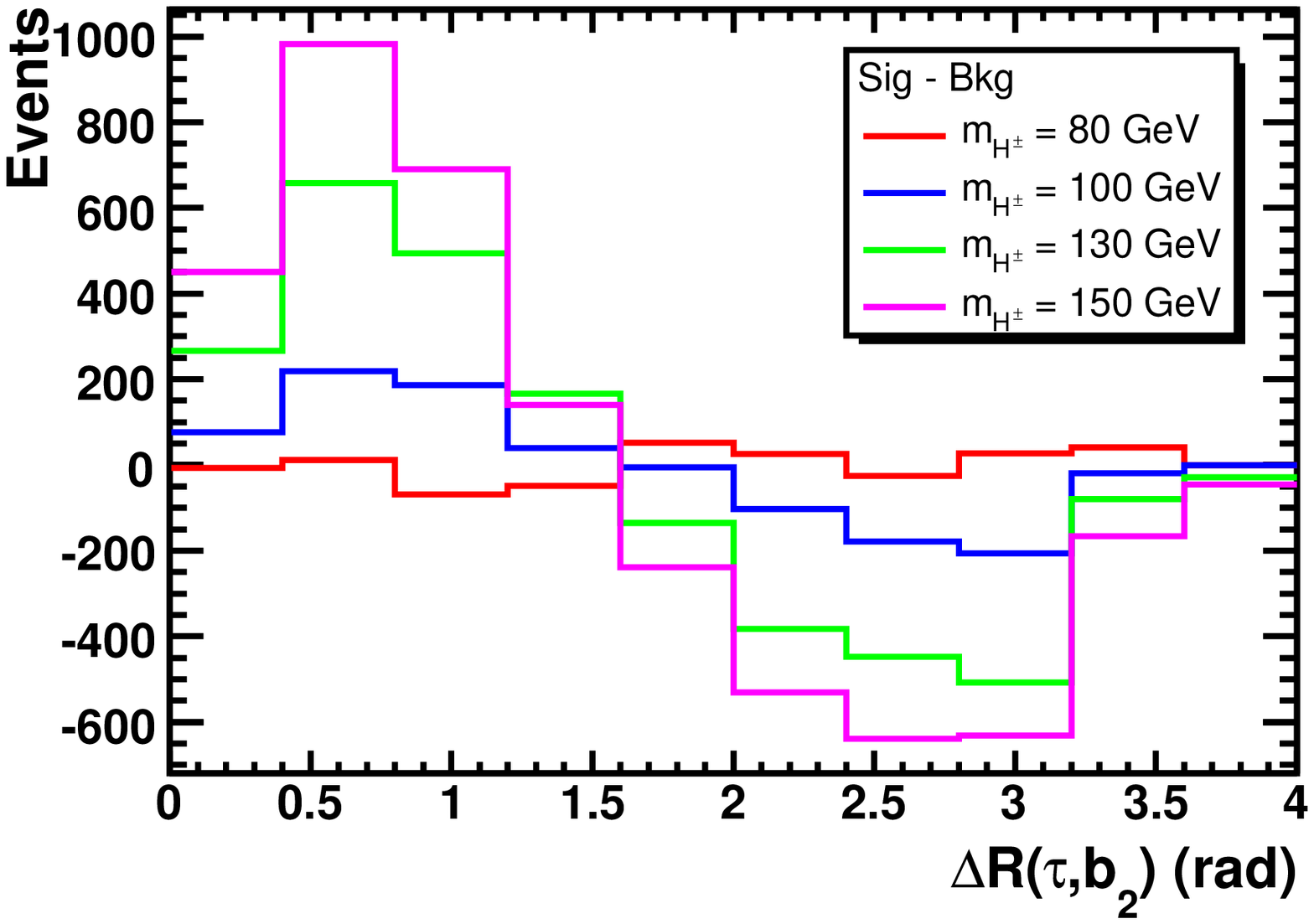, width=0.47\textwidth}
\vspace*{-4mm}
\caption{
Spatial distance
$\Delta R(\tau,b_2) = \sqrt{(\Delta\phi)^2 + (\Delta\eta)^2}$
distributions
(where $\Delta\phi$ is the azimuthal angle in rad between
the $\tau$ and $b$ jet)
for the $tbH^\pm$ signal
and the $t\bar{t}$ background for $\sqrt{s}=1.96$~TeV (left)
and the respective differences between signal and background (right).
}
\label{fig:distance-tau-b}
\vspace*{-2mm}
\end{figure}

\begin{figure}[htbp]
\epsfig{file=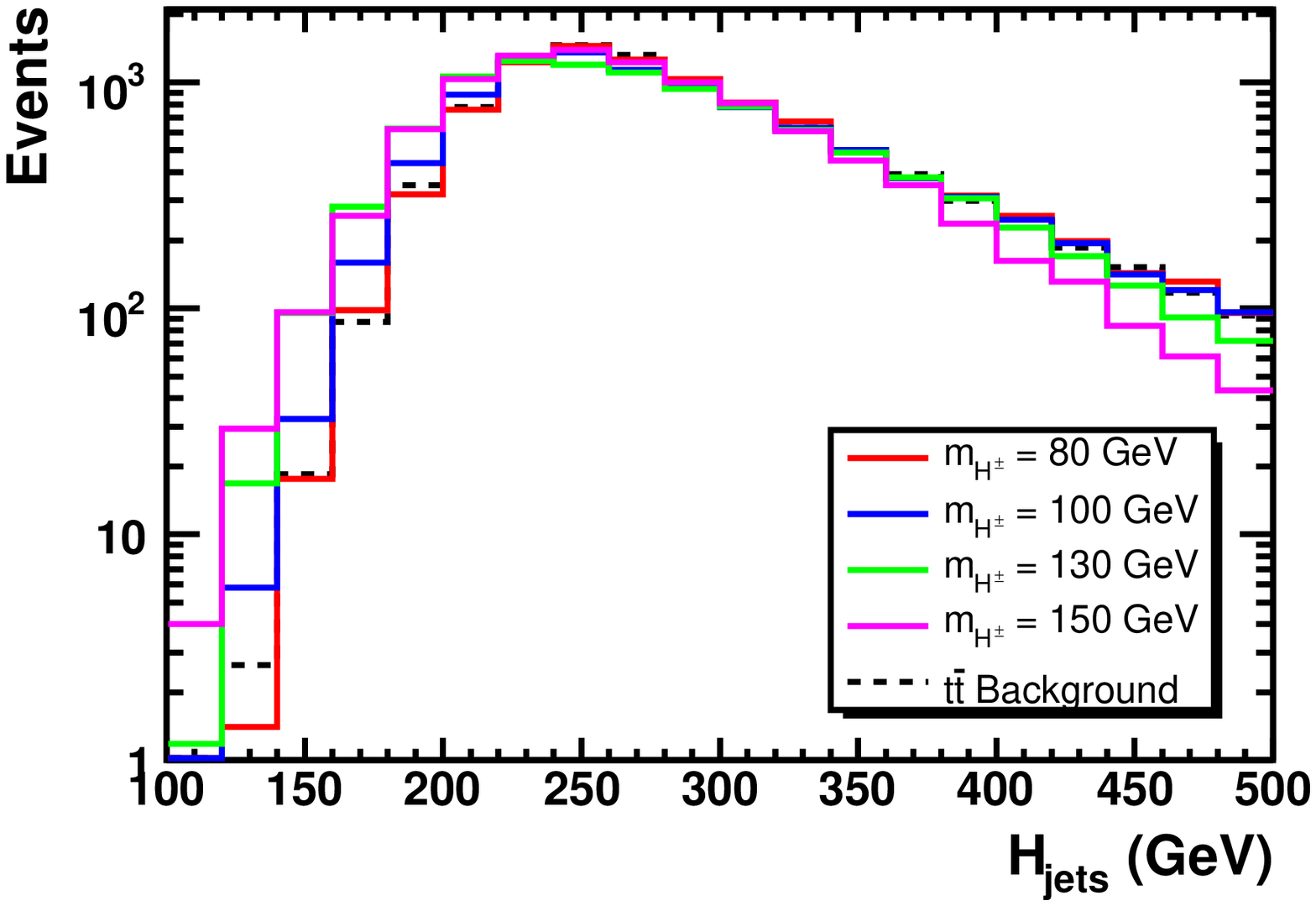, width=0.4\textwidth}  \hfill
\epsfig{file=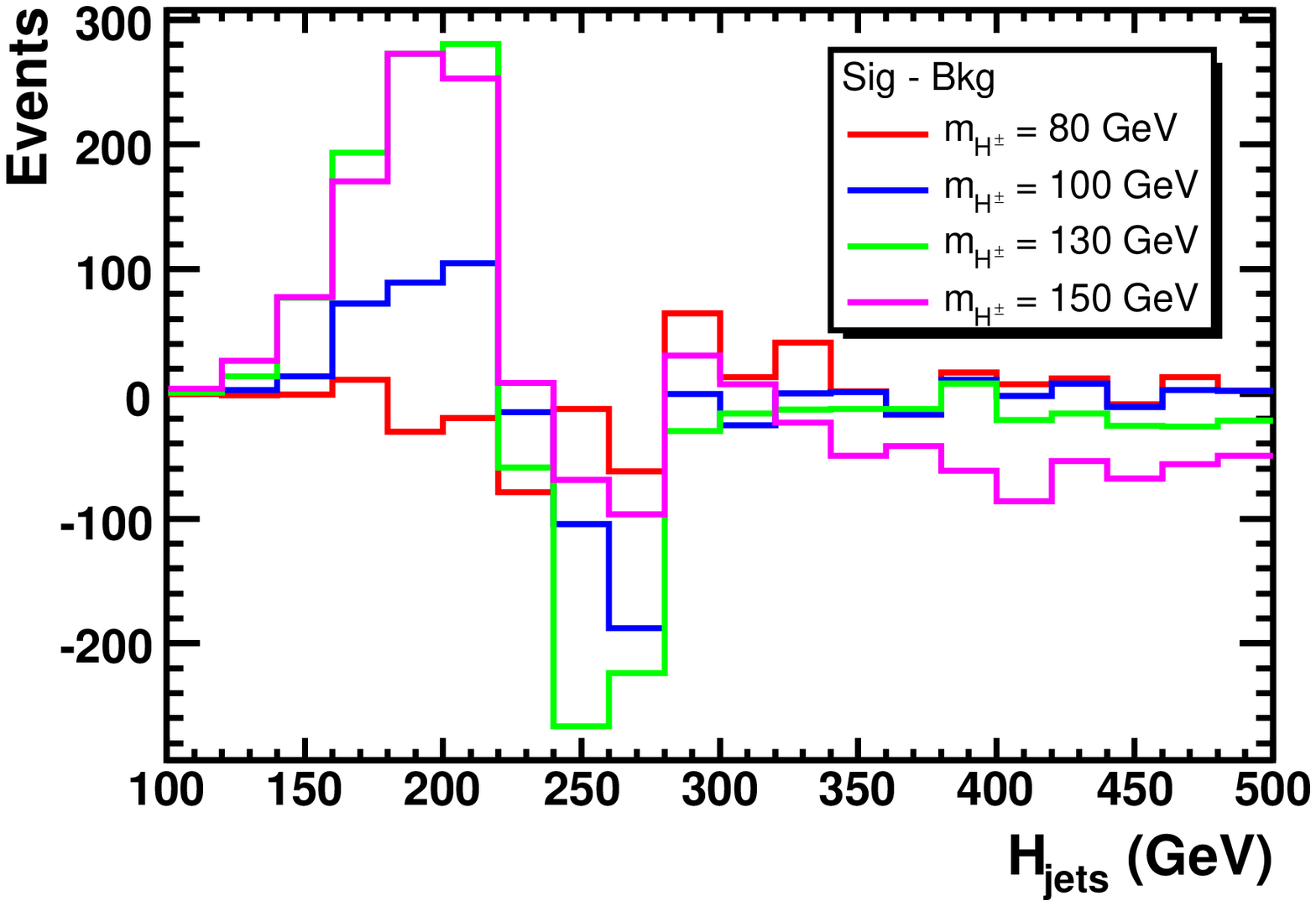, width=0.4\textwidth}
\vspace*{-4mm}
\caption{
Distributions of the
total transverse momentum of all quark jets,
$H_{\rm jets} = p_t^{j_1} + p_t^{j_2} + p_t^{b_1} + p_t^{b_2}$,
for the $tbH^\pm$ signal
and the $t\bar{t}$ background for $\sqrt{s}=1.96$~TeV (left)
and the respective differences between signal and background (right). 
}
\label{fig:hjet}
\vspace*{-4mm}
\end{figure}

\begin{figure}[htbp]
\epsfig{file=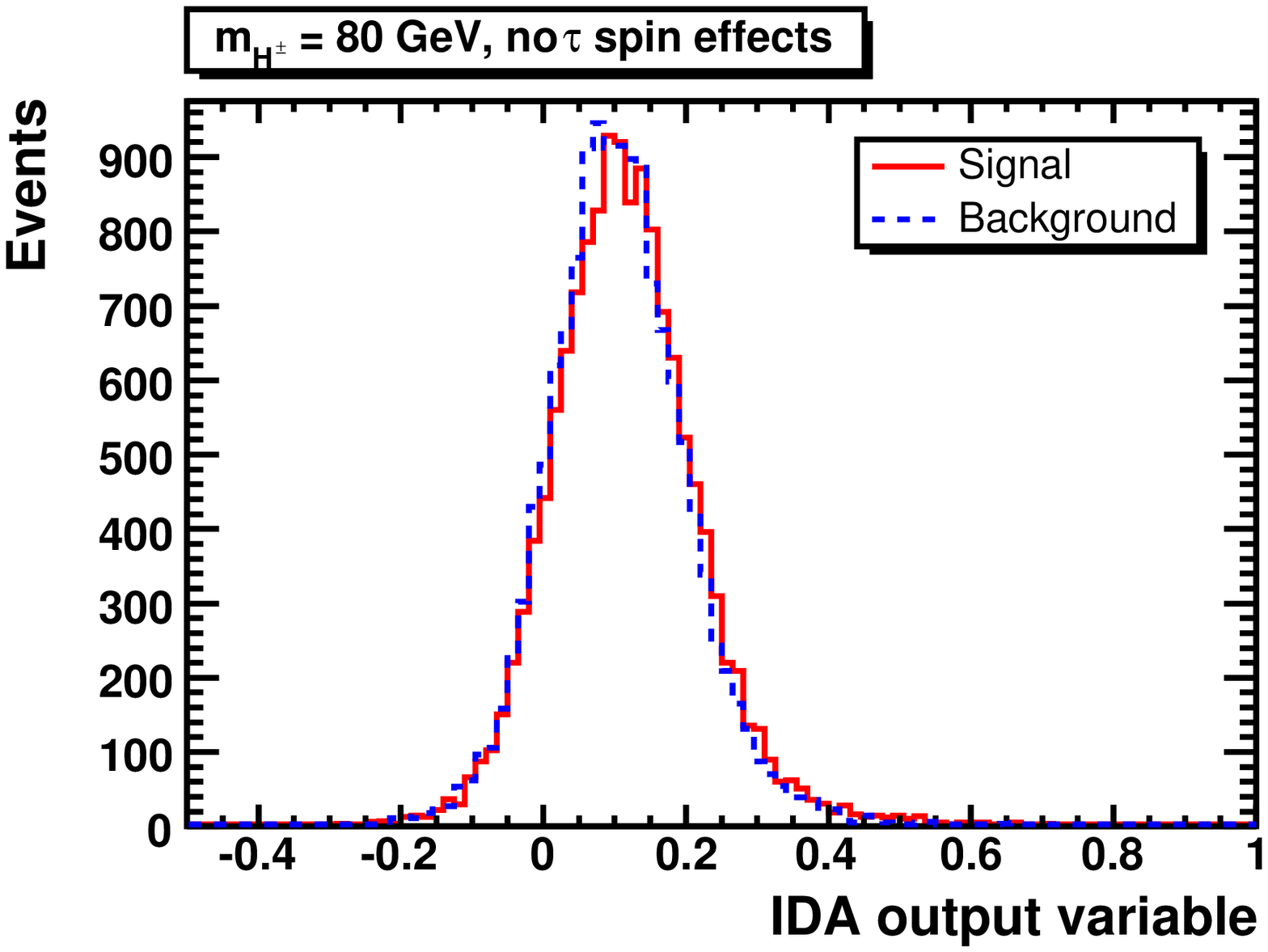, width=0.4\textwidth}  \hfill
\epsfig{file=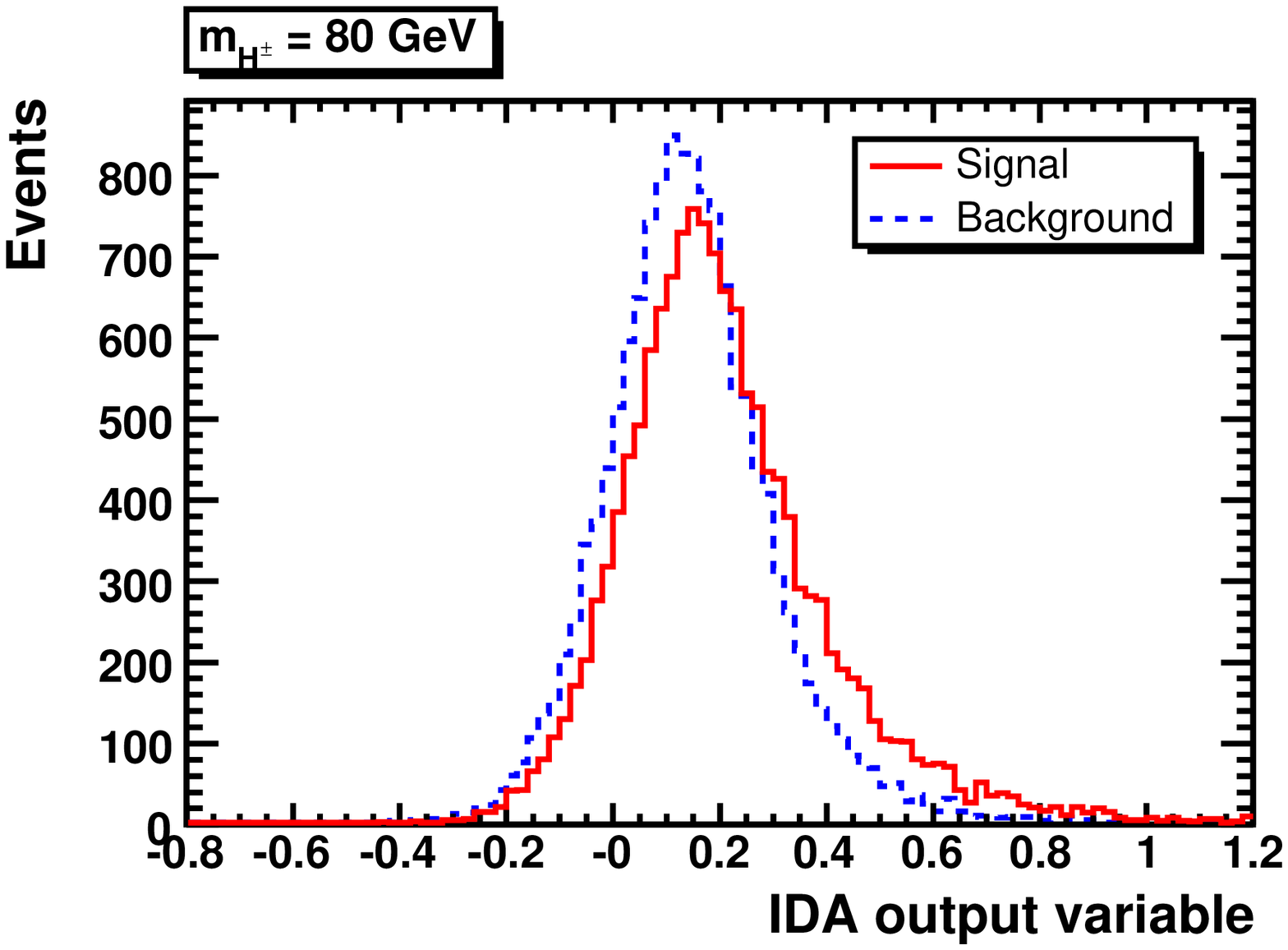, width=0.4\textwidth}
\epsfig{file=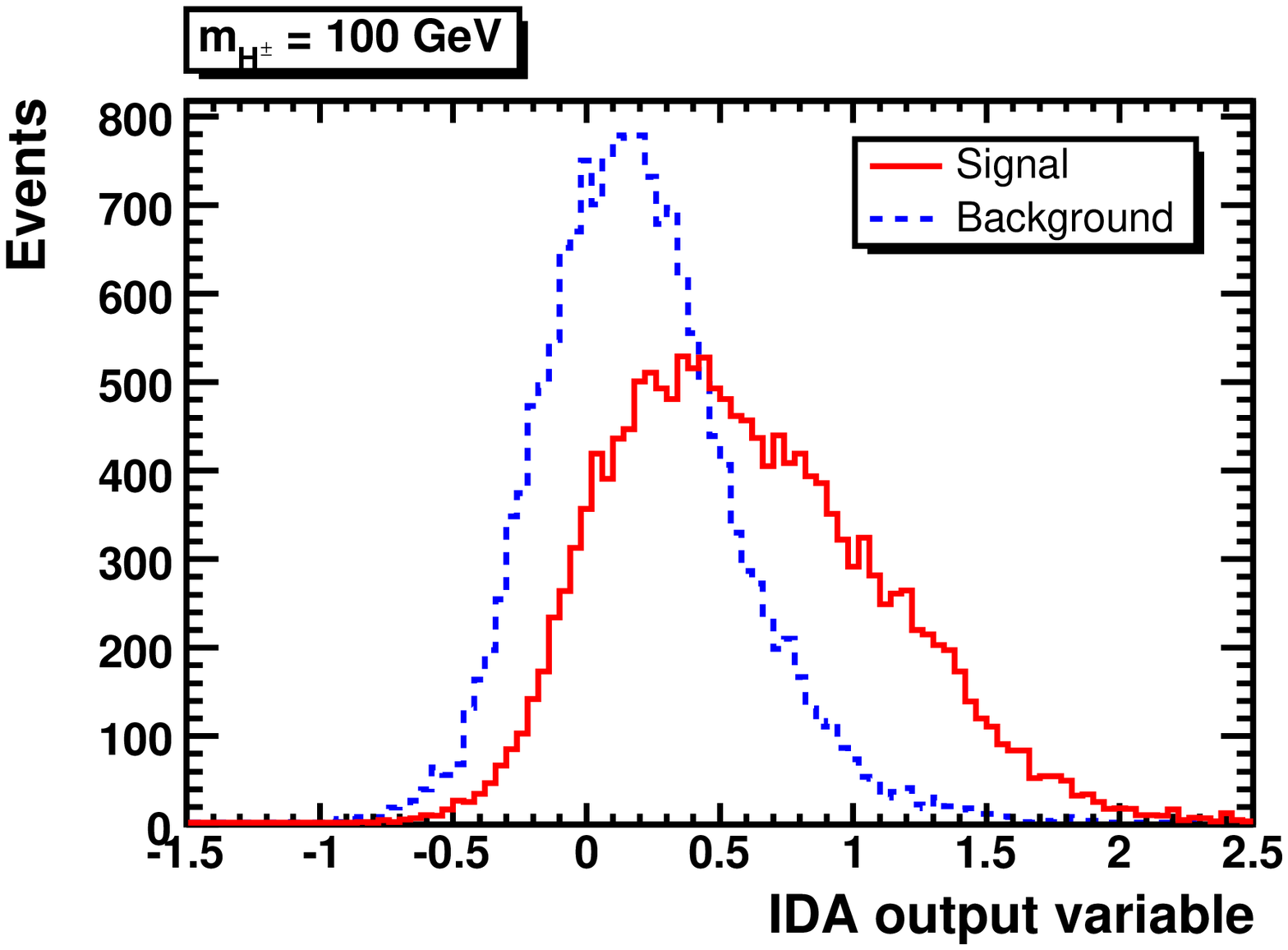, width=0.4\textwidth}  \hfill
\epsfig{file=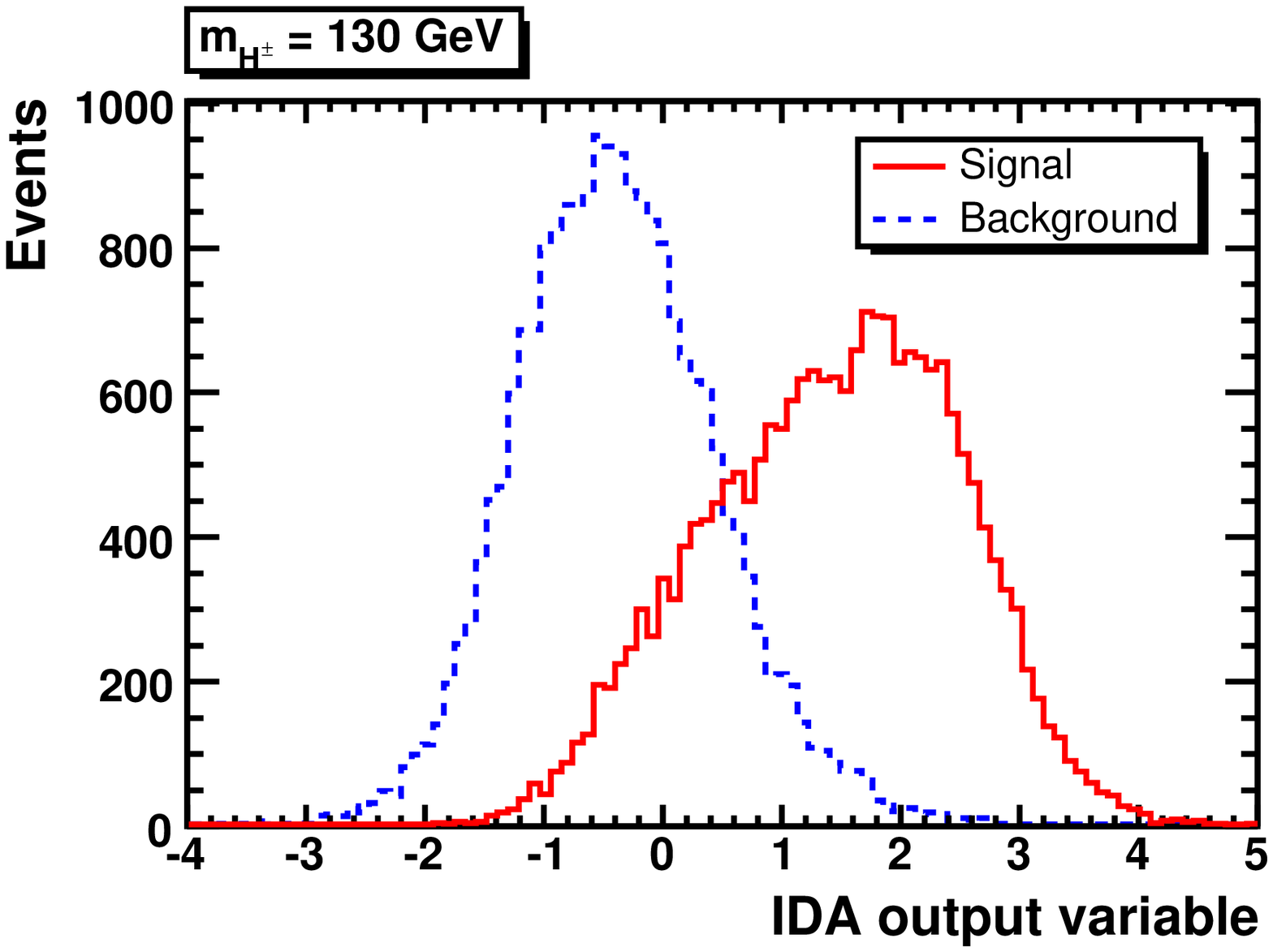, width=0.4\textwidth}
\epsfig{file=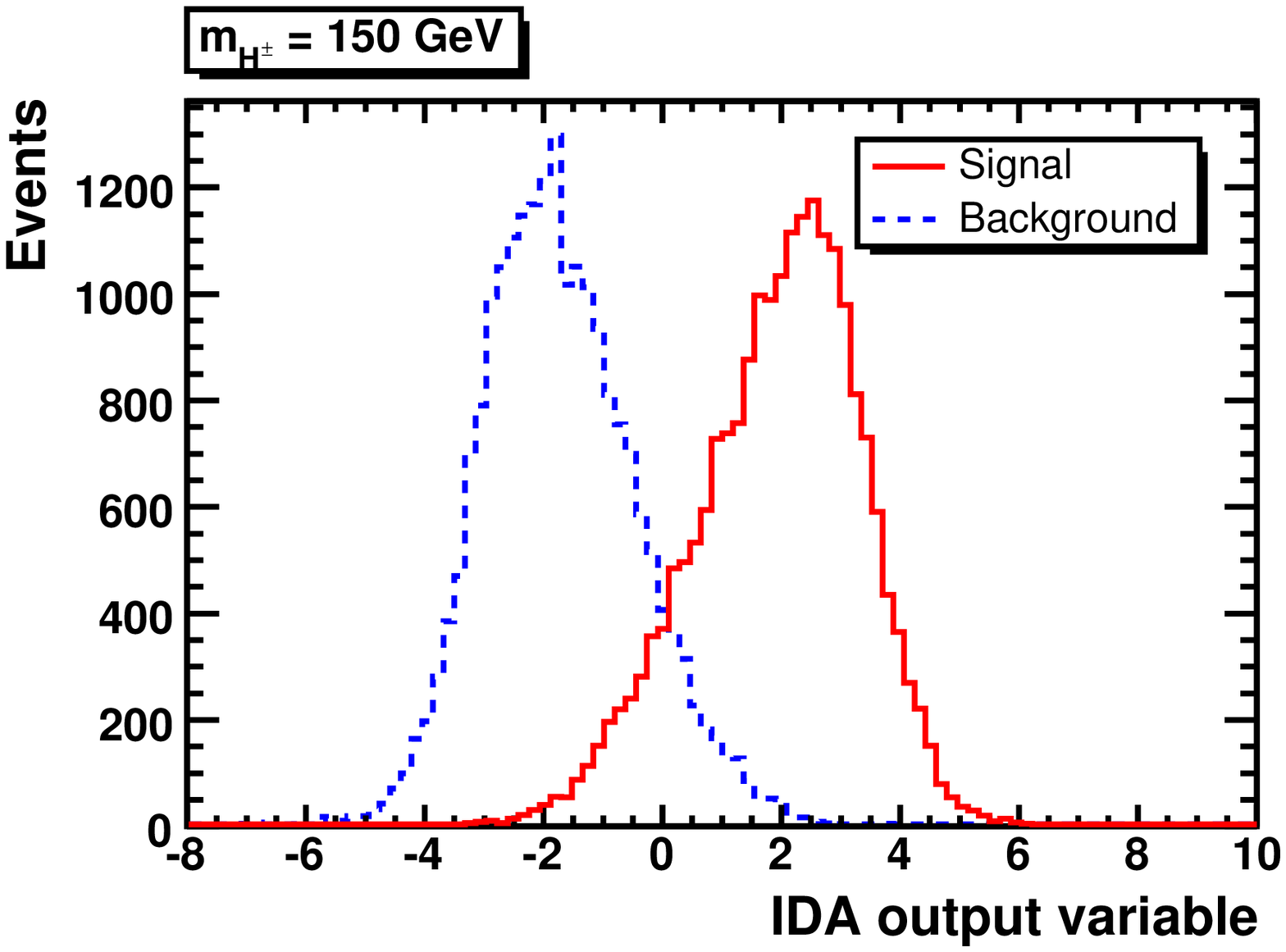, width=0.4\textwidth}  \hfill
\vspace*{-4mm}
\caption{
Distributions of the IDA output variable in the first IDA
step for the $tbH^\pm$ signal (solid, red)
and the $t\bar{t}$ background (dashed, blue) for $\sqrt{s}=1.96$~TeV.
}
\label{fig:ida1}
\end{figure}

\begin{figure}[htbp]
\epsfig{file=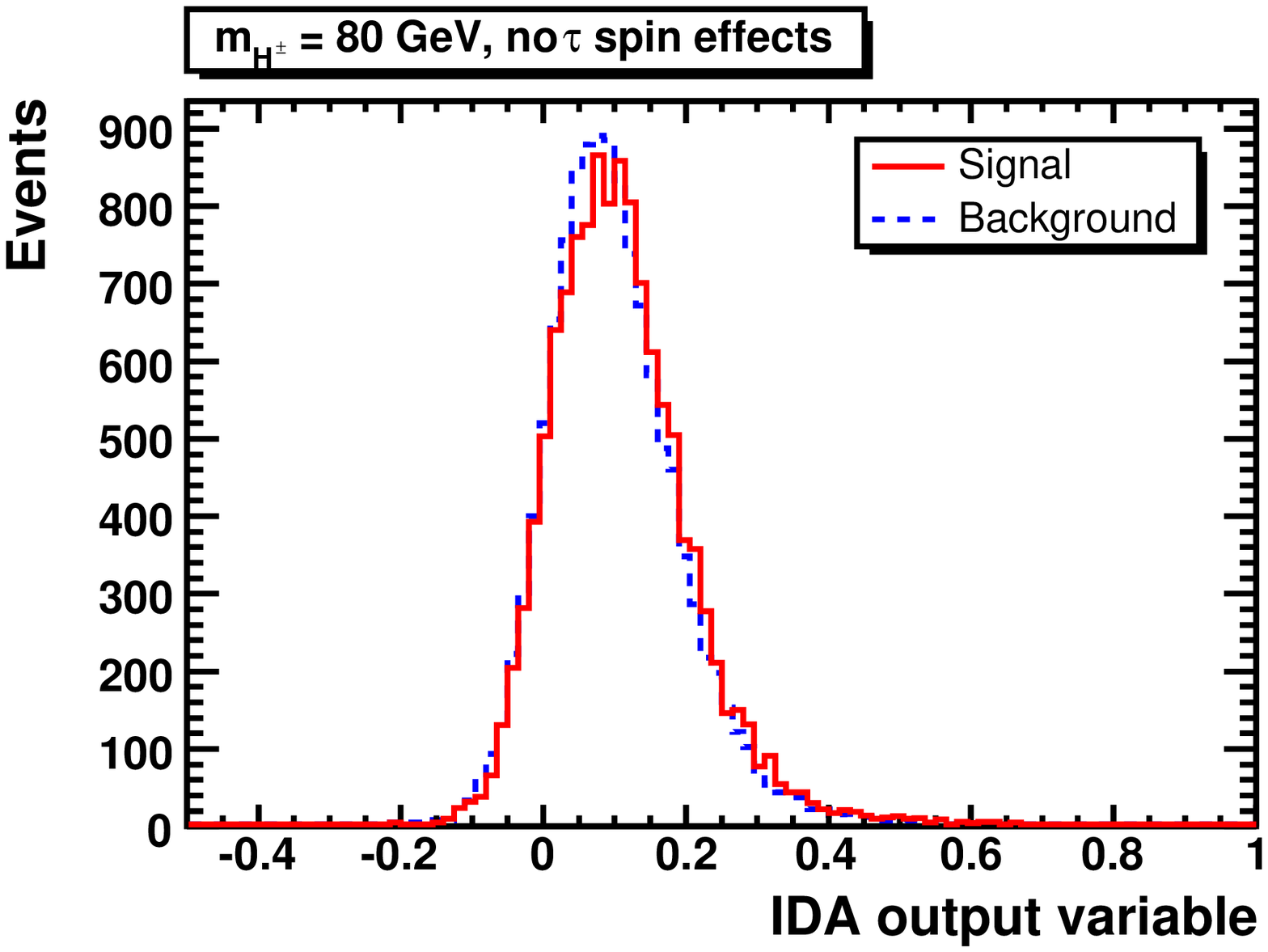, width=0.5\textwidth}  \hfill
\epsfig{file=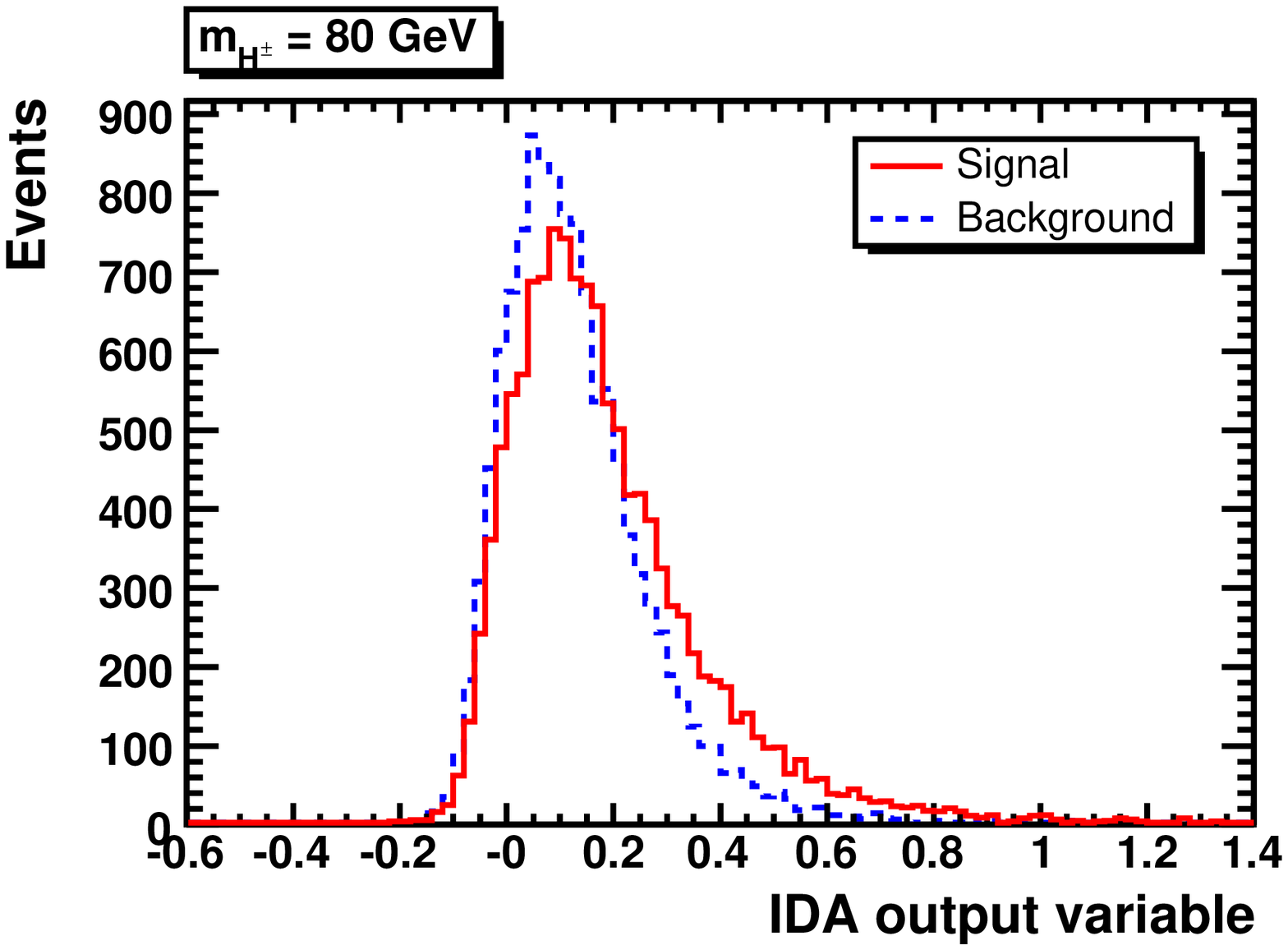, width=0.5\textwidth}
\epsfig{file=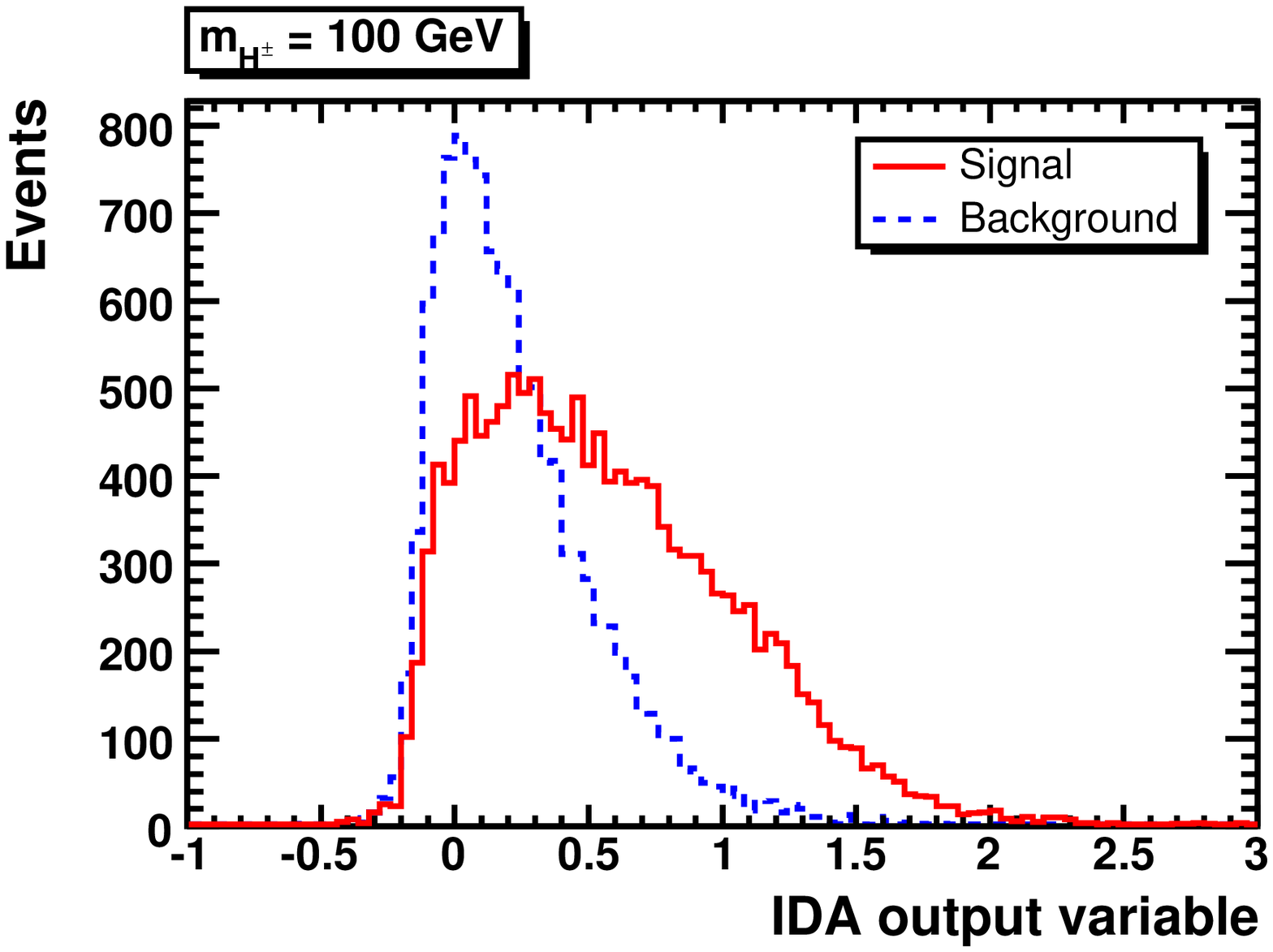, width=0.5\textwidth}  \hfill
\epsfig{file=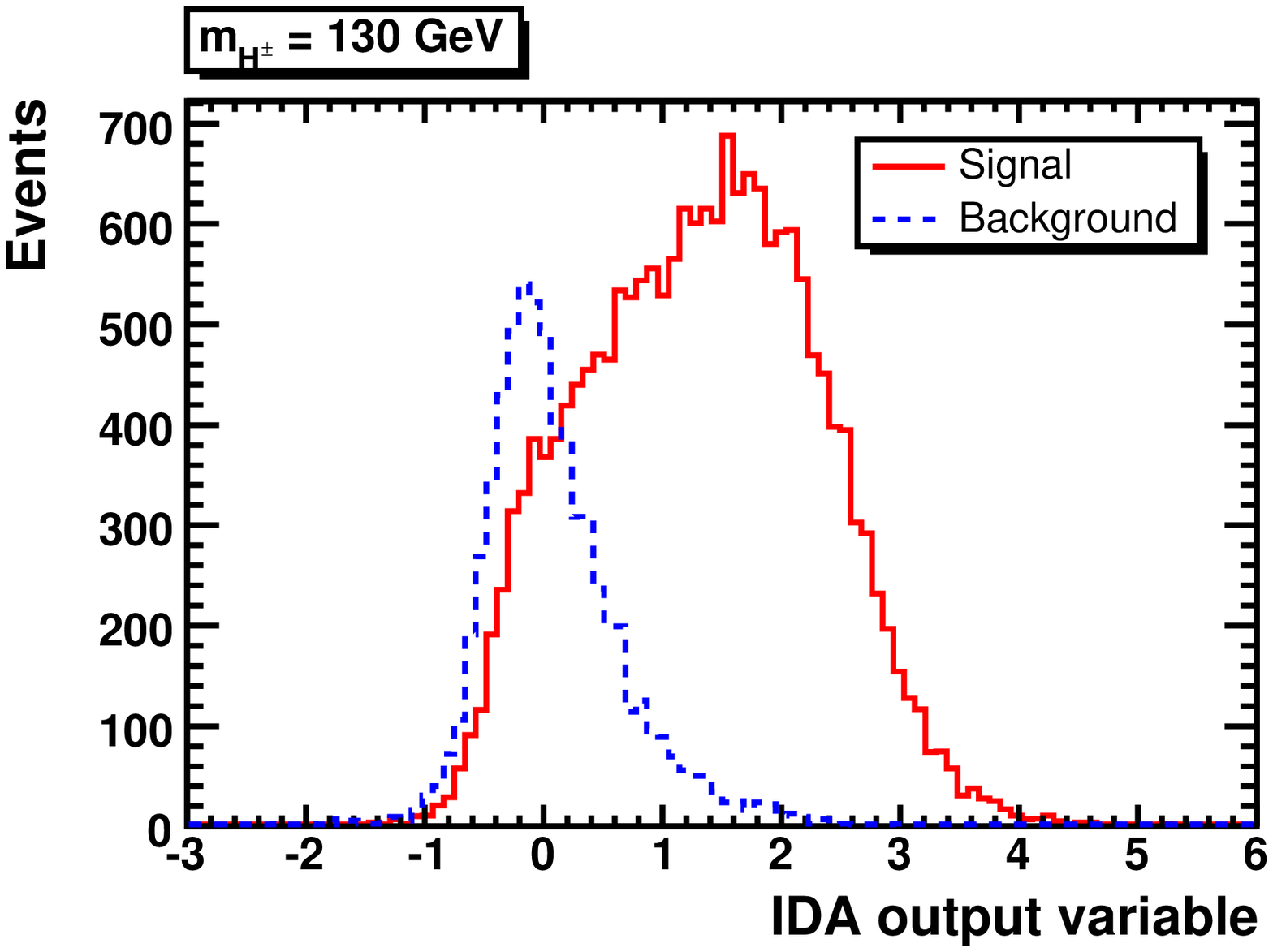, width=0.5\textwidth}
\epsfig{file=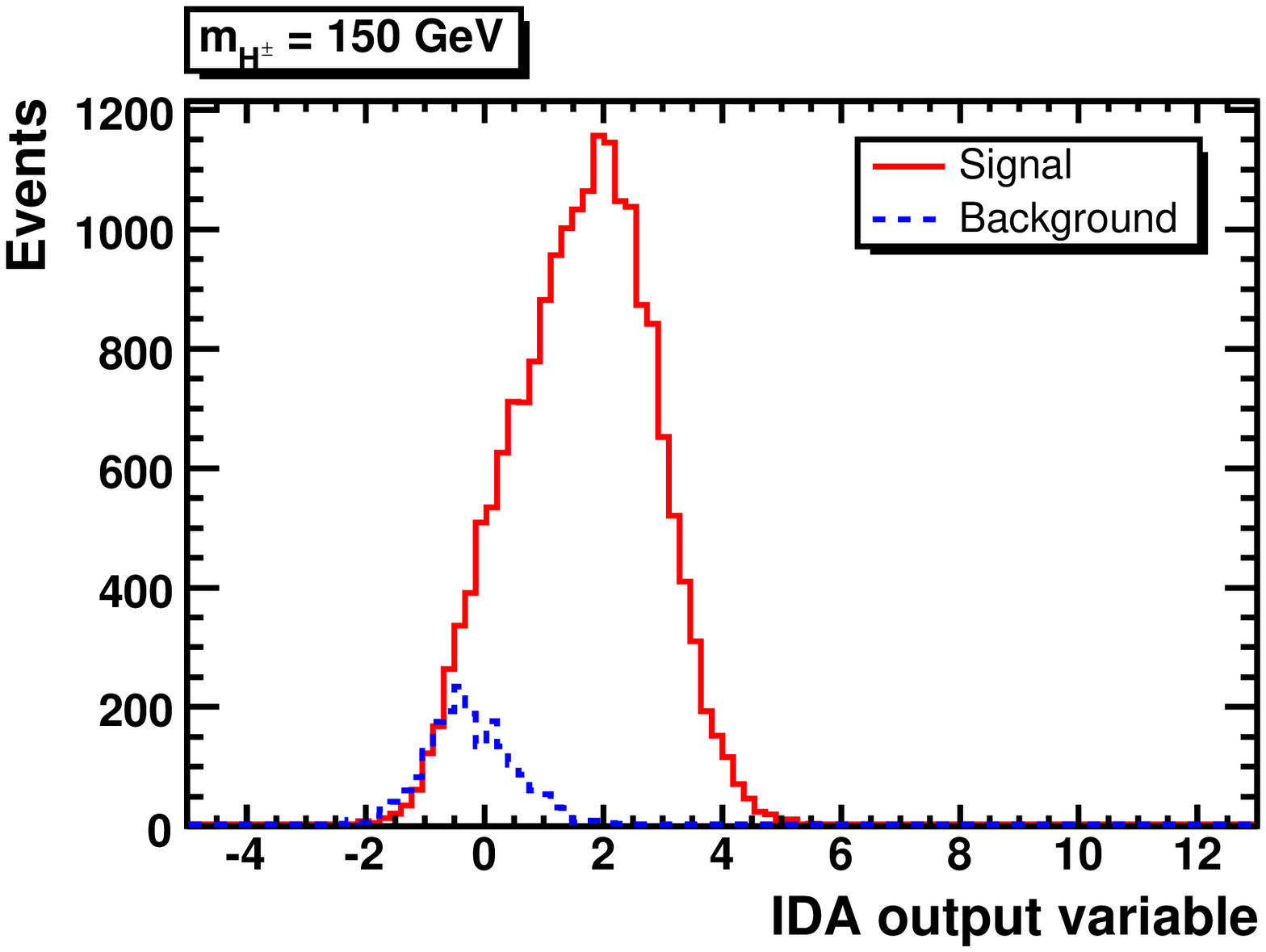, width=0.5\textwidth}  \hfill
\epsfig{file=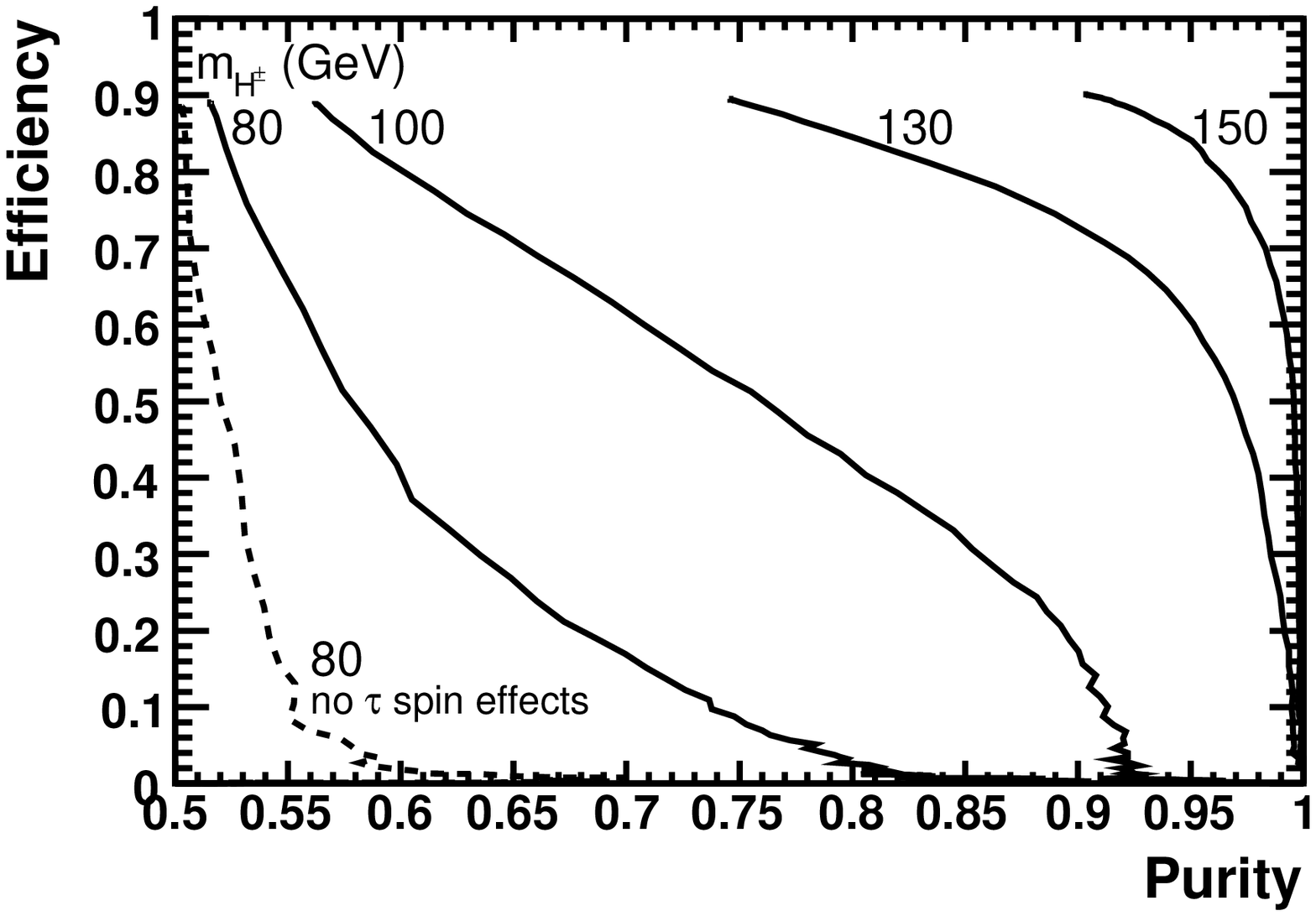, width=0.5\textwidth}
\caption{
Upper row, middle row and lower left figure:
distributions of the IDA output variable 
in the second IDA step
for 90\% efficiency in the first IDA
step (corresponding to a cut at 0 in Fig.~\ref{fig:ida1})
for the $tbH^\pm$ signal (solid, red)
and the $t\bar{t}$ background (dashed, blue).
Lower right figure: efficiency as a function of the purity
when not taking the spin effects in the $\tau$ decay into account for
$m_{H^\pm}=80$~GeV (dashed) and with 
spin effects in the $\tau$ decay for
$m_{H^\pm}=80,100,130,150$~GeV (solid, from left to right).
Results are for the Tevatron.
}
\label{fig:ida}
\end{figure}

\clearpage

\begin{figure}[htp]
\begin{minipage}{0.49\textwidth}
\begin{center}
\epsfig{file=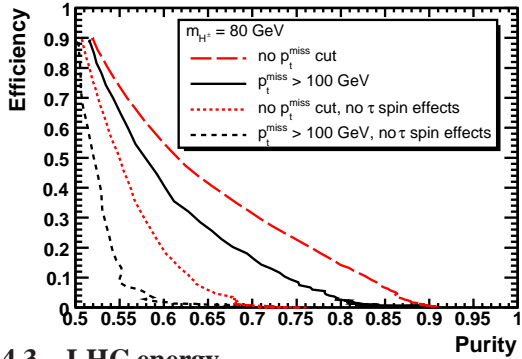, width=1\textwidth}
\end{center}
\end{minipage}
\begin{minipage}{0.49\textwidth}
\caption{
Efficiency as a function of purity for $m_{H^\pm}=80$~GeV and
$\sqrt{s}=1.96$ TeV.
The black lines are the results after applying the hard cut
\mbox{$p_t^{\rm miss} > 100$~GeV}
when not taking the spin effects in the $\tau$ decay into account
(dashed) and with spin effects in the $\tau$ decay (solid),
as also shown in Fig.\ref{fig:ida}.
The red lines are the results without applying the hard cut
on \mbox{$p_t^{\rm miss}$}
when not taking the spin effects in the $\tau$ decay into account
(dotted) and with spin effects in the $\tau$ decay (long dashed).
}
\label{fig:ida_hard_cut}
\end{minipage}
\vspace*{-15mm}
\end{figure}

\subsection{LHC energy}

The simulation procedure and the emulation of the detector response are the same as 
those outlined in Sect.~2.1 for the Tevatron, as well as, for the preselection and IDA method, 
as described in Sects. 2.3 and 2.4, respectively. Hence, only the expected LHC rates are 
discussed, followed by the description of changes in the distributions of the variables and 
the final IDA results.

Unlike the case of the Tevatron, where only charged Higgs masses smaller
than the top quark mass can be explored, and 2HDM/MSSM signatures
practically rely on $\tau\nu_\tau$ pairs only, at the LHC the phenomenology 
is more varied. Here, the  search strategies depend strongly 
on the charged Higgs boson mass.
If $m_{H^\pm} < m_{t} - m_{b}$ (later referred to as a light Higgs boson), 
the charged Higgs boson can be produced in top \mbox{(anti-)}\-quark decay. The main source of 
top (anti-)quarks at the LHC is again $t \bar{t}$ pair production ($\sigma_{t\bar{t}}=850$ pb at 
NLO)~\cite{beneke00}. 
For the whole ($\tan\beta, m_{H^\pm}$) parameter space there is a competition between the $ bW^\pm$ 
and $ bH^\pm$ channels in top decay keeping the sum 
$\mathrm{BR}(t \to b W^+) + \mathrm{BR}(t \to b H^+)$
at almost unity.
The top quark decay to $ bW^\pm$ is however the dominant mode for most of the parameter space. 
Thus, the best way to search for a (light) charged Higgs boson is by requiring that the top
quark produced in the $tbH^\pm$ process decays to a $W^\pm$. 
While in the case of $H^\pm$ decays $\tau$'s will be tagged via their hadronic decay producing low-multiplicity narrow jets 
in the detector, there are two different $W^\pm$ decays that can be explored. The leptonic signature
$ b \bar{b} H^\pm W^\mp \to b \bar{b} \tau \nu l \nu $ provides a clean selection 
of the signal via the identification of the lepton $l=e,\mu$.
In this case the charged Higgs transverse mass cannot be reconstructed because
of the presence of two neutrinos with different origin. In this channel charged Higgs 
discovery will be determined 
by the observation of an excess of such events over SM expectations through a simple counting experiment. In the case of hadronic decays 
$ b \bar{b}H^\pm W^\mp  \to  b \bar{b}\tau \nu jj$ the transverse mass can instead be 
reconstructed since all neutrinos are arising from the charged Higgs boson decay. 
This allows for an efficient separation of the signal and the main 
$t\bar{t} \to b \bar{b}W^\pm W^\mp  \to  b \bar{b}\tau \nu jj$ background
(assuming $m_{H^\pm}\OOrd m_{W^\pm}$). 
The absence of a lepton ($e$ or $\mu$) provides a less 
clean environment but the use of the transverse mass makes it possible to reach the same mass discovery region as 
in the previous case and also to extract the charged Higgs boson mass. Both these channels show that after an 
integrated luminosity of 30 fb$^{-1}$ the discovery could be possible up to a mass of 150 GeV 
for all tan$\beta$ values in both ATLAS and CMS~\cite{Mohn:2007fd,biscarat,abdullin}. 

If the charged Higgs is heavier than the top quark, the dominant decay
channels are 
$H^\pm \to \tau \nu$ and $H^\pm \to tb$ depending on $\tan\beta$. They
have both been studied by  
ATLAS and CMS~\cite{assamagan,kinnunen,salmi,lowette}.
The charged Higgs bosons are produced in the $pp \to tbH^\pm$ channel. For the 
$H^\pm \to tb$ decay, a charged Higgs boson can be discovered up to 
high masses ($m_{H^\pm} \sim 400$~GeV) in the case of very large $\tan\beta$ values and this reach
cannot be much improved because of the large multi-jet environment. For the 
$H^\pm \to \tau \nu$ decay mode this reach is larger due to a cleaner signal despite a 
lower BR. In this case the 5$\sigma$ reach ranges from $\tan\beta=20$ for 
$m_{H^\pm}=200$ GeV to $\tan\beta=30$ for $m_{H^\pm}=400$ GeV.

For the LHC, signal and background events have been simulated in the
same way as for the
Tevatron as described before, however, without implying any rescaling factor to
match a measured $t\bar{t}$ cross section.
Table~\ref{tab:Hpm:crosssecLHC} lists 
the resulting cross sections before 
($\sigma^{\rm th}$) and after ($\sigma$) applying the basic cuts 
$p_t^{\rm jets} > 20$~GeV and the hard cut $p_t^{\rm miss} > 100$~GeV.
The LHC rates allow for the discovery to be less challenging than at 
the Tevatron in the region $m_{H^\pm} \sim m_{W^\pm}$, 
yet the separation of  signal events from
background remains crucial for the measurement of the charged Higgs mass.

\begin{table}[htbp]
\vspace*{-4mm}
\centering
\begin{tabular}{c|c|c|c|c|c}
 & $q\bar q, gg \to  t\bar{t}$ &
  \multicolumn{4}{c}{$q\bar q, gg \to  tbH^\pm$} \\
$m_{H^\pm}$ (GeV) & 80 & 80 & 100 & 130 & 150 \\
  \hline
$\sigma^{\rm th}$ (pb) & 45.5 &  72.6    &  52.0    &  24.5 & 9.8  \\
$\sigma$ (pb) for $p_t^\mathrm{jets} > 20$ GeV 
   & 17.3 &  33.9   &  25.7    &  12.2  &  3.8 \\
$\sigma$ (pb) for $(p_t^\mathrm{jets},p_t^{\rm miss}) > (20,100)$ GeV
   & 4.6 &   6.0    &   4.8    &  2.9   &  1.2
\end{tabular}
\caption{\label{tab:Hpm:crosssecLHC}
LHC cross sections of background $q\bar q, gg \to  t\bar{t}$
and signal $q\bar q, gg \to  tbH^\pm$
for $\tan\beta = 30$ and $m_{H^\pm} = 80, 100, 130$ and $150$~GeV
into the final state
$2 b + 2 j + \tau_\mathrm{jet} + p_t^{\rm miss}$
before ($\sigma^{\rm th}$) and after ($\sigma$)
the basic cuts ($p_t > 20$~GeV for all jets)
and the hard cut ($p_t^{\rm miss} > 100$ GeV).
}
\end{table}

The kinematic distributions are shown in Figs.~\ref{fig:lhc_pttau} to
\ref{fig:lhc_hjet} 
for $\sqrt{s}=14$~TeV.
The choice of variables is identical to the one for the Tevatron and
allows for a one-to-one comparison, 
the differences being  due to a change in CM energy (and, to a
somewhat lesser extent, due to the leading partonic mode of the 
production process\footnote{As the latter
is dominated by $q\bar q$ annihilation at the Tevatron and $gg$ fusion
at the LHC.}). 
The main differences with respect to
Figs.~\ref{fig:pttau}--\ref{fig:hjet} are that 
the various transverse momenta and
invariant masses have longer high energy tails. 
In particular, it should be noted
that the effect of the spin differences between $W^\pm$ and $H^\pm$
events can be explored very effectively also at LHC energies,
e.g. the ratio $p_t^{\pi^\pm}/p_t^{\tau_\mathrm{jet}}$ is shown
in Fig.~\ref{fig:lhc_r1} which is very sensitive to the spin effects.
These observations lead to the conclusion
that the same method using spin differences can be used to separate
signal from background at both the Tevatron and the LHC.

The distributions of the IDA output variables for the study at
$\sqrt{s}=14$~TeV for two steps with 90\% efficiency in the first step
are shown in Figs.~\ref{fig:lhc_ida1} and \ref{fig:lhc_ida}.
These distributions are qualitatively similar to those for the  Tevatron 
(Figs.~\ref{fig:ida1} and \ref{fig:ida}) and the final achievable purity for a 
given efficiency is shown in Fig.~\ref{fig:lhc_ida}. As for the Tevatron energy 
a good separation of signal and background events can be achieved with the spin 
sensitive variables and the IDA method even in case $m_{H^\pm} \sim m_{W^\pm}$.
For heavier $H^\pm$ masses the separation of signal and background events increases due to 
the kinematic differences of the event topology.

\begin{figure}[htbp]
\vspace*{-4mm}
\epsfig{file=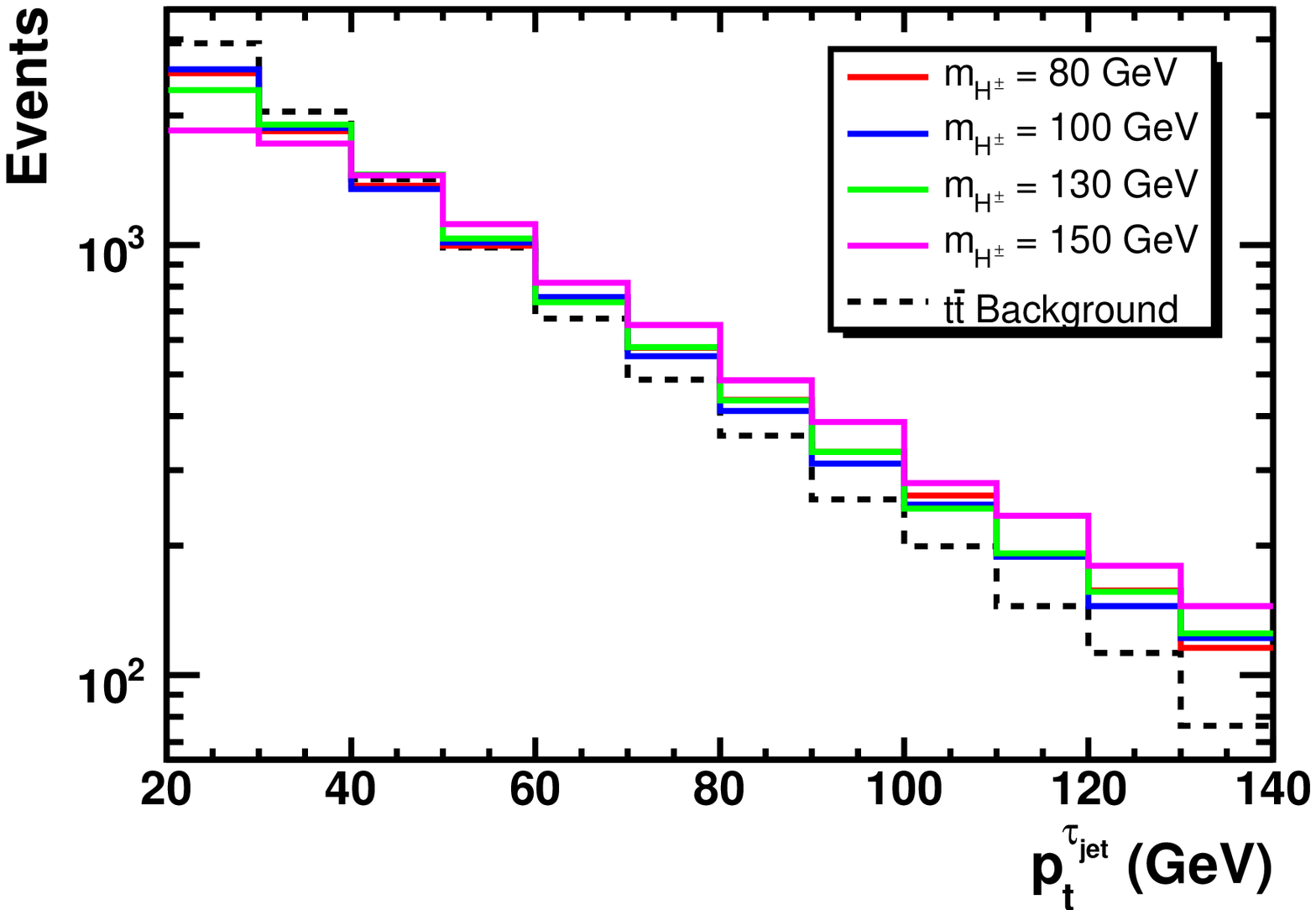, width=0.4\textwidth} \hfill
\epsfig{file=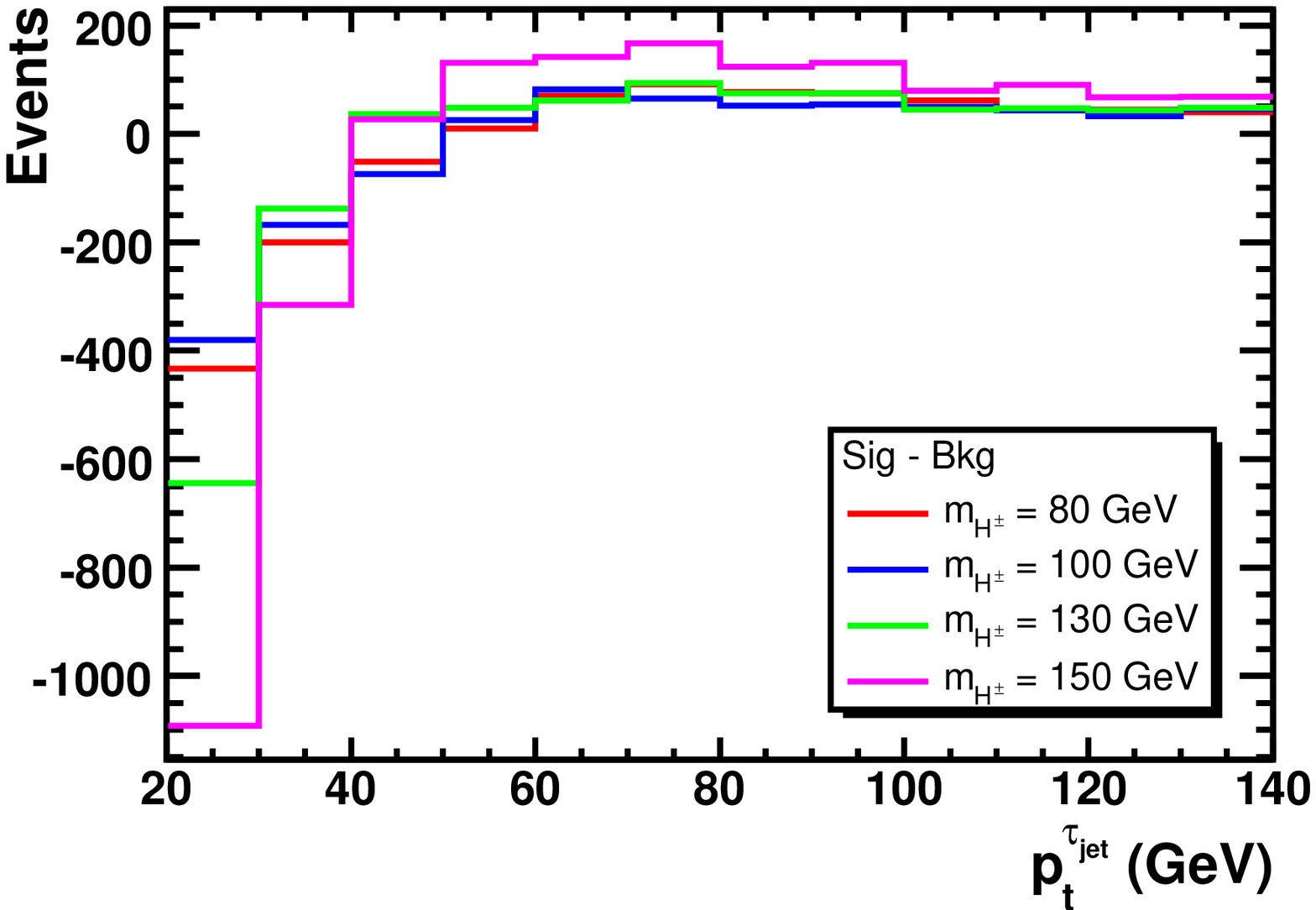, width=0.4\textwidth}
\epsfig{file=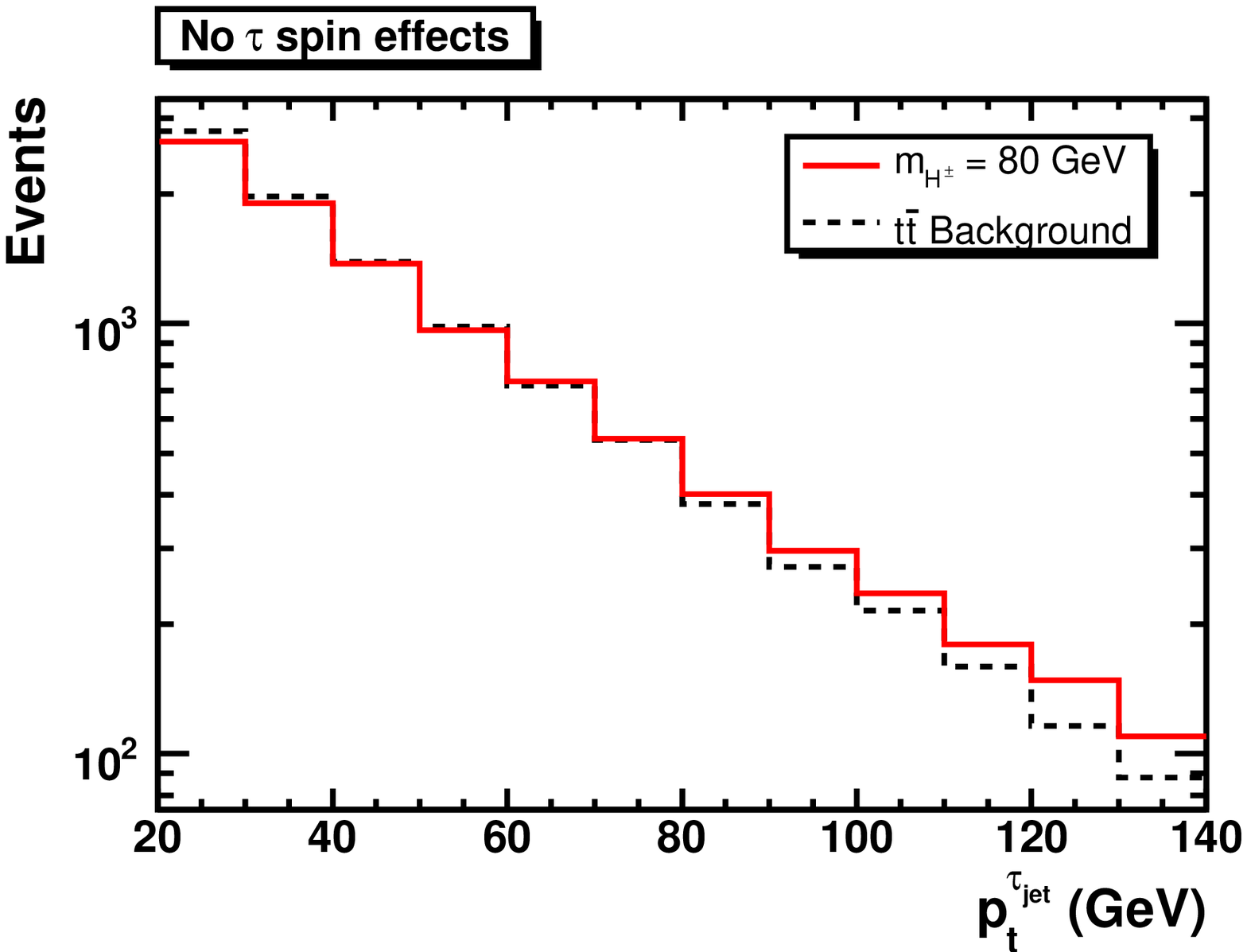, width=0.4\textwidth} \hfill
\epsfig{file=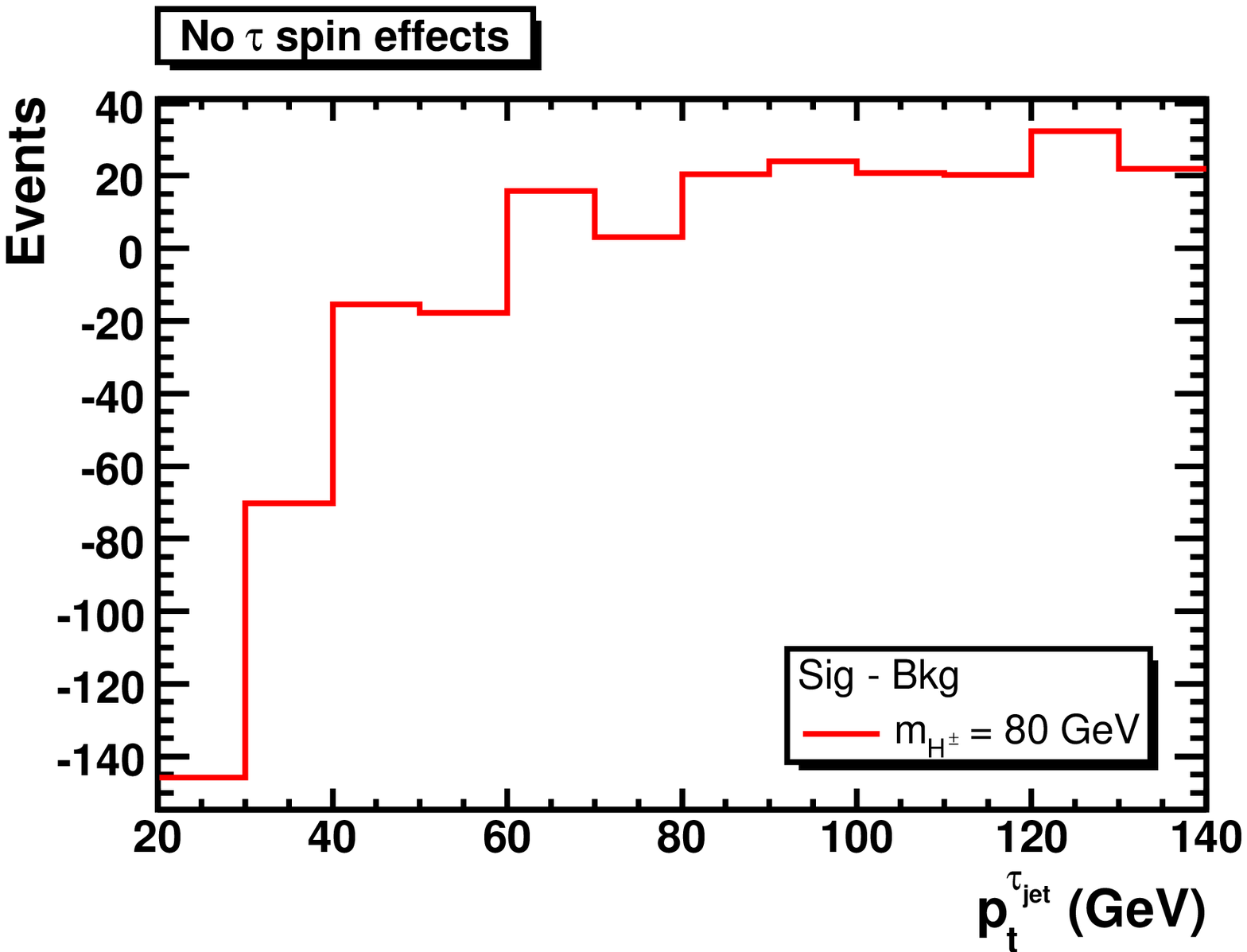, width=0.4\textwidth}
\vspace*{-4mm}
\caption{ 
$p_t$ distributions 
of the $\tau~{\mathrm{jet}}$ for the $tbH^\pm$ signal
and the $t\bar{t}$ background for $\sqrt{s}=14$~TeV (left)
and the respective differences between signal and background (right).
The lower plots show distributions without spin effects in the $\tau$
decays.
}
\label{fig:lhc_pttau}
\vspace*{-4mm}
\end{figure}

\begin{figure}[htbp]
\vspace*{-4mm}
\epsfig{file=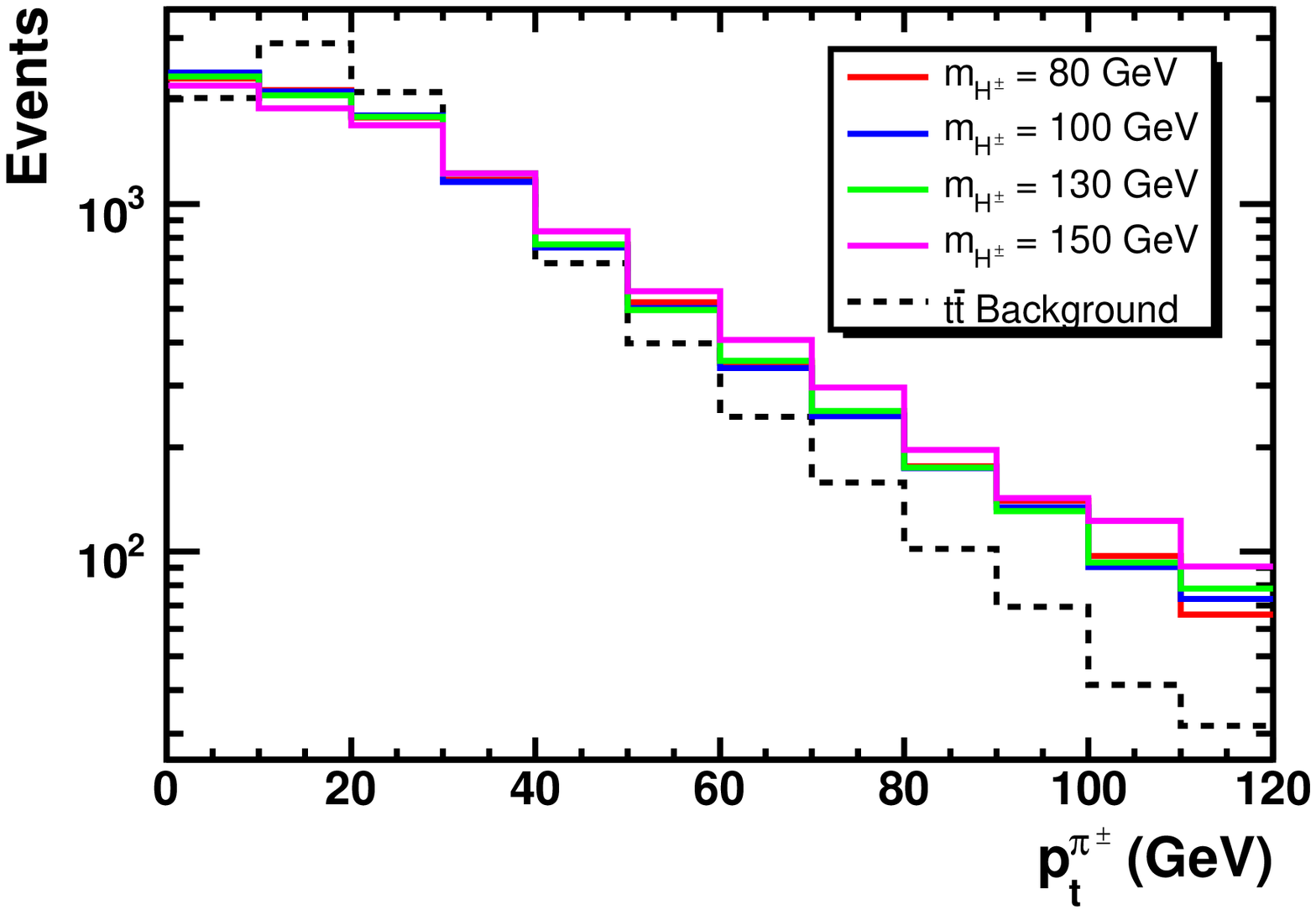, width=0.4\textwidth}  \hfill
\epsfig{file=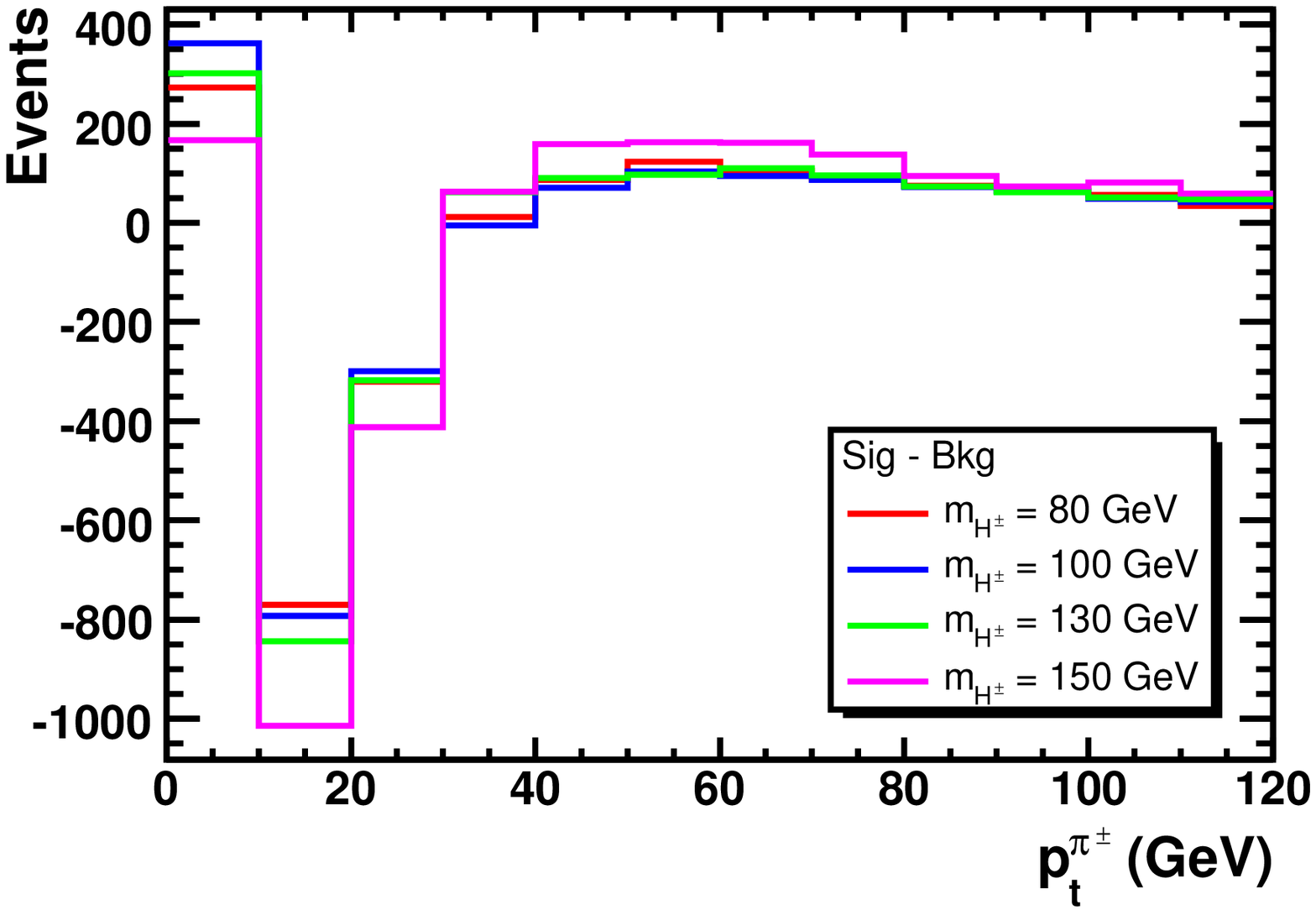, width=0.4\textwidth}
\epsfig{file=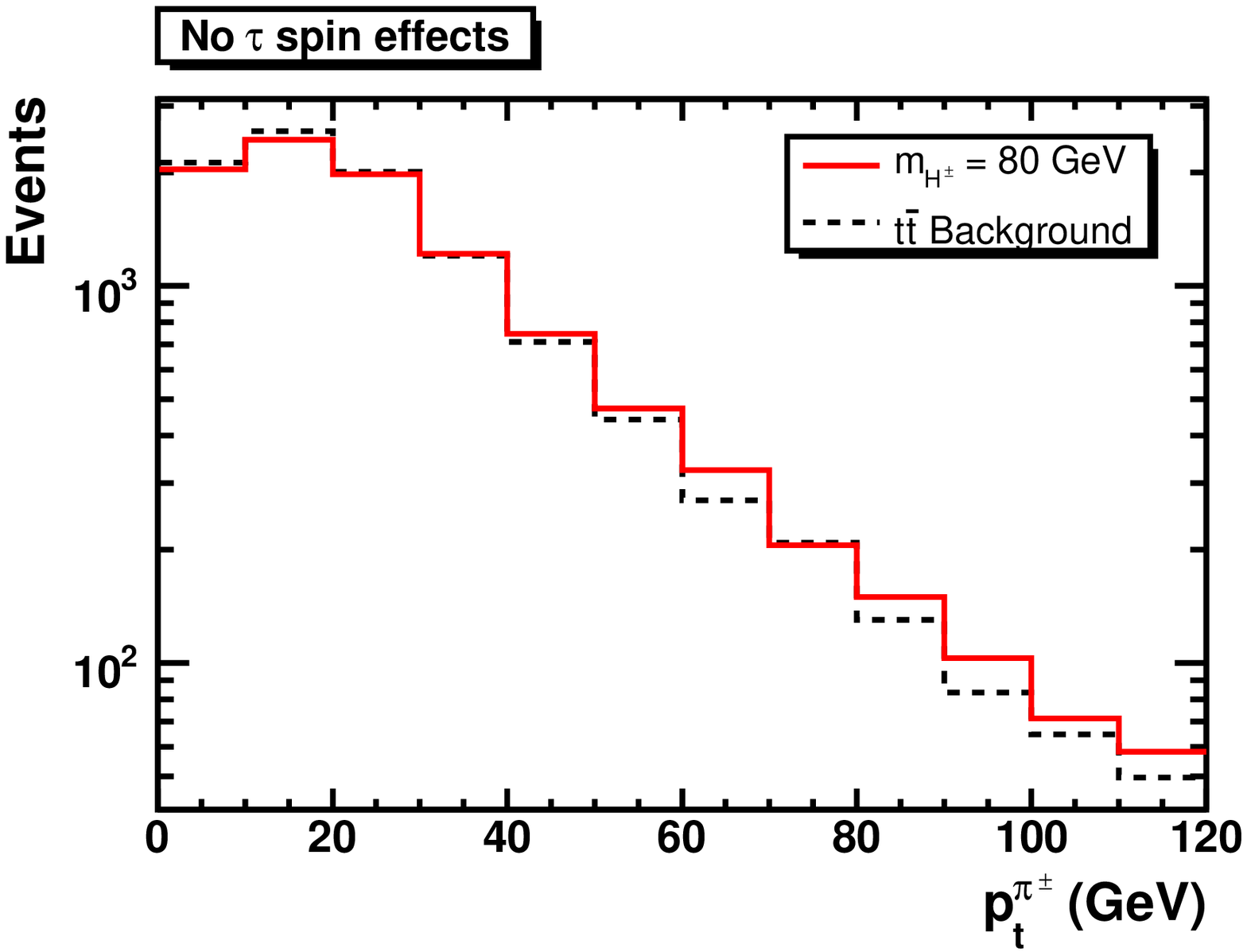, width=0.4\textwidth}  \hfill
\epsfig{file=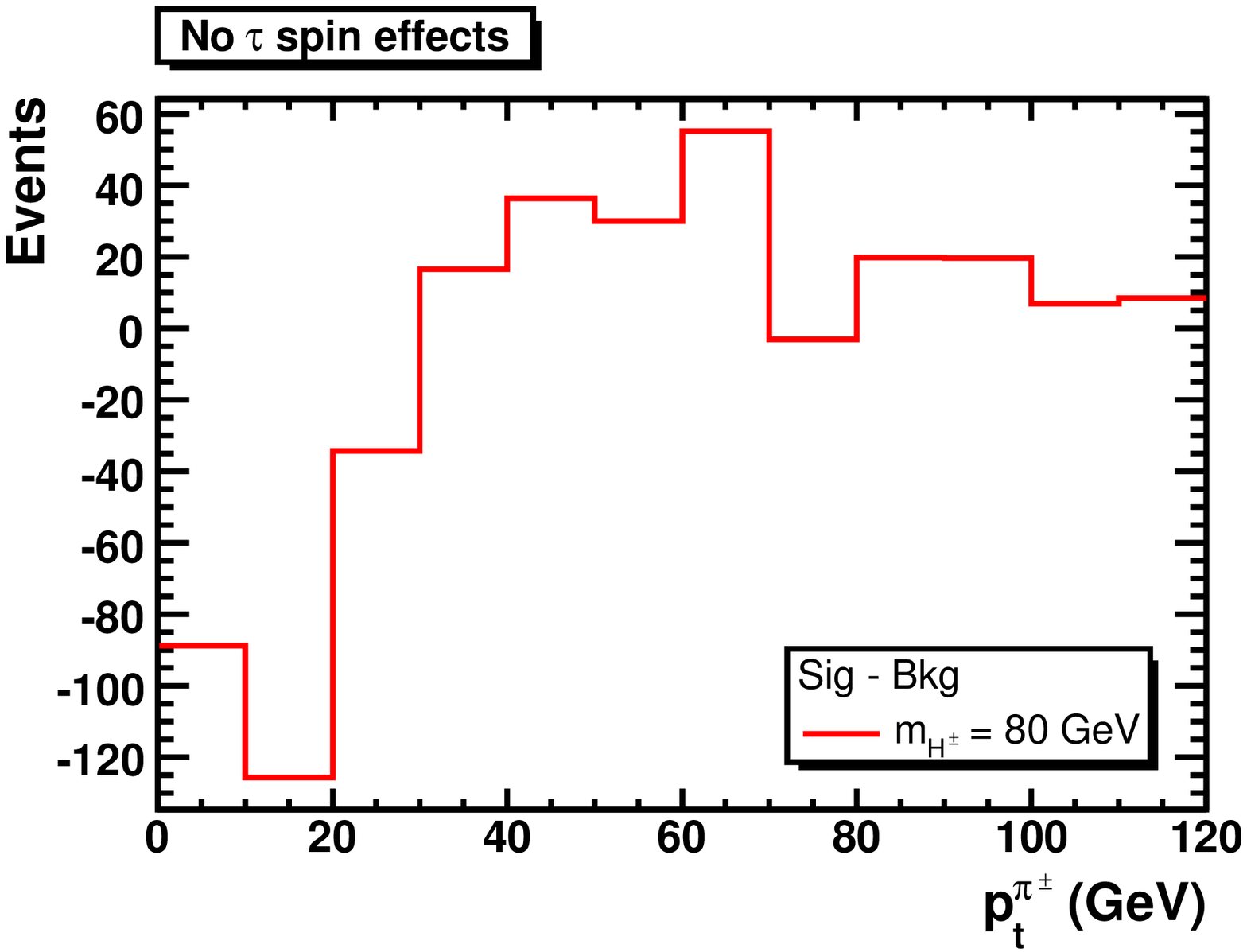, width=0.4\textwidth}
\vspace*{-4mm}
\caption{
$p_t$  distributions of the leading $\pi^\pm$ from the $\tau$ decay 
for the $tbH^\pm$ signal and the 
$t\bar{t}$ background for $\sqrt{s}=14$~TeV (left)
and the respective differences between signal and background (right).
The lower plots show distributions without spin effects in the $\tau$
decays.
\vspace*{-4mm}
}
\label{fig:lhc_ptpi}
\end{figure}

\begin{figure}[htbp]
\epsfig{file=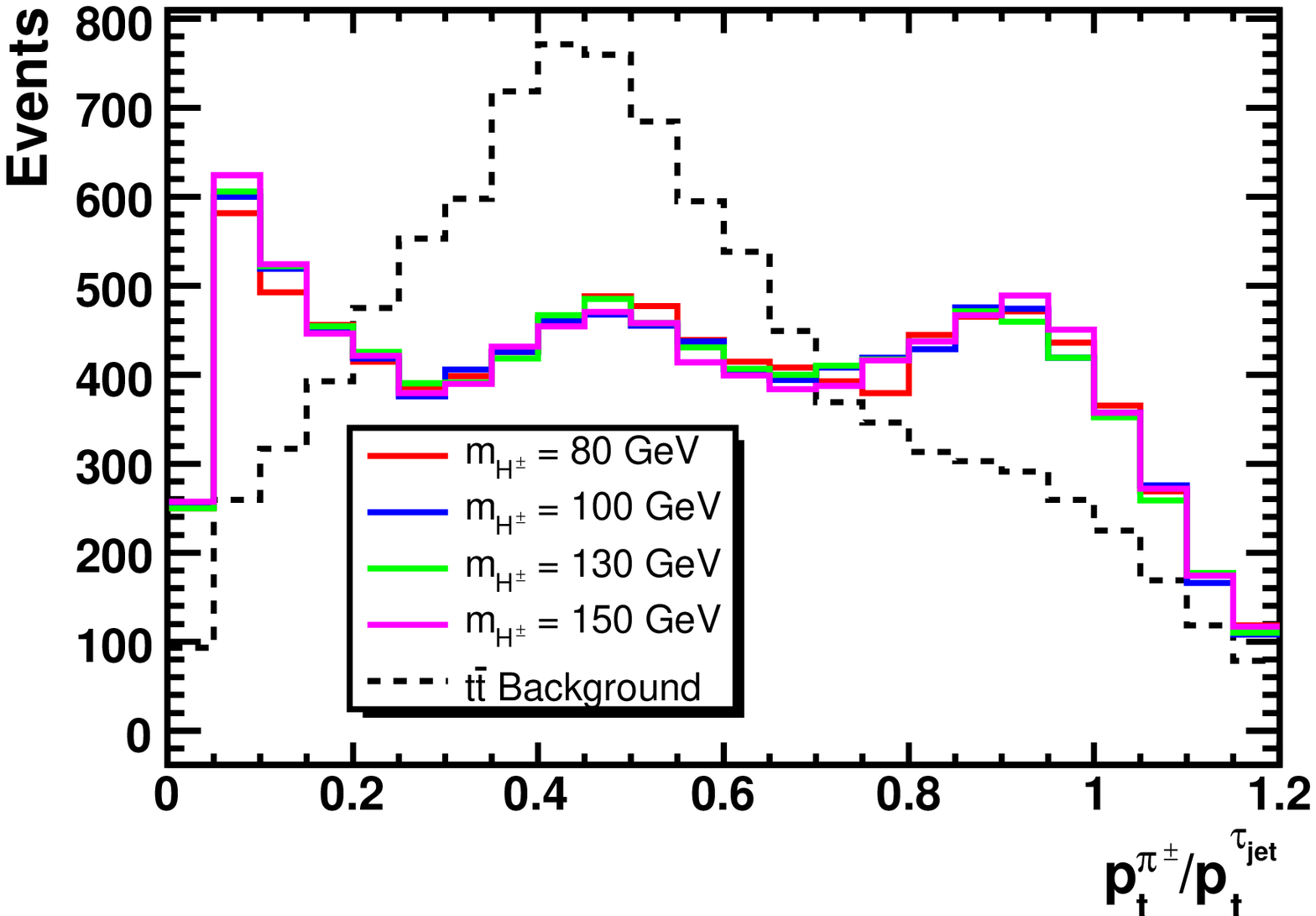, width=0.4\textwidth} \hfill
\epsfig{file=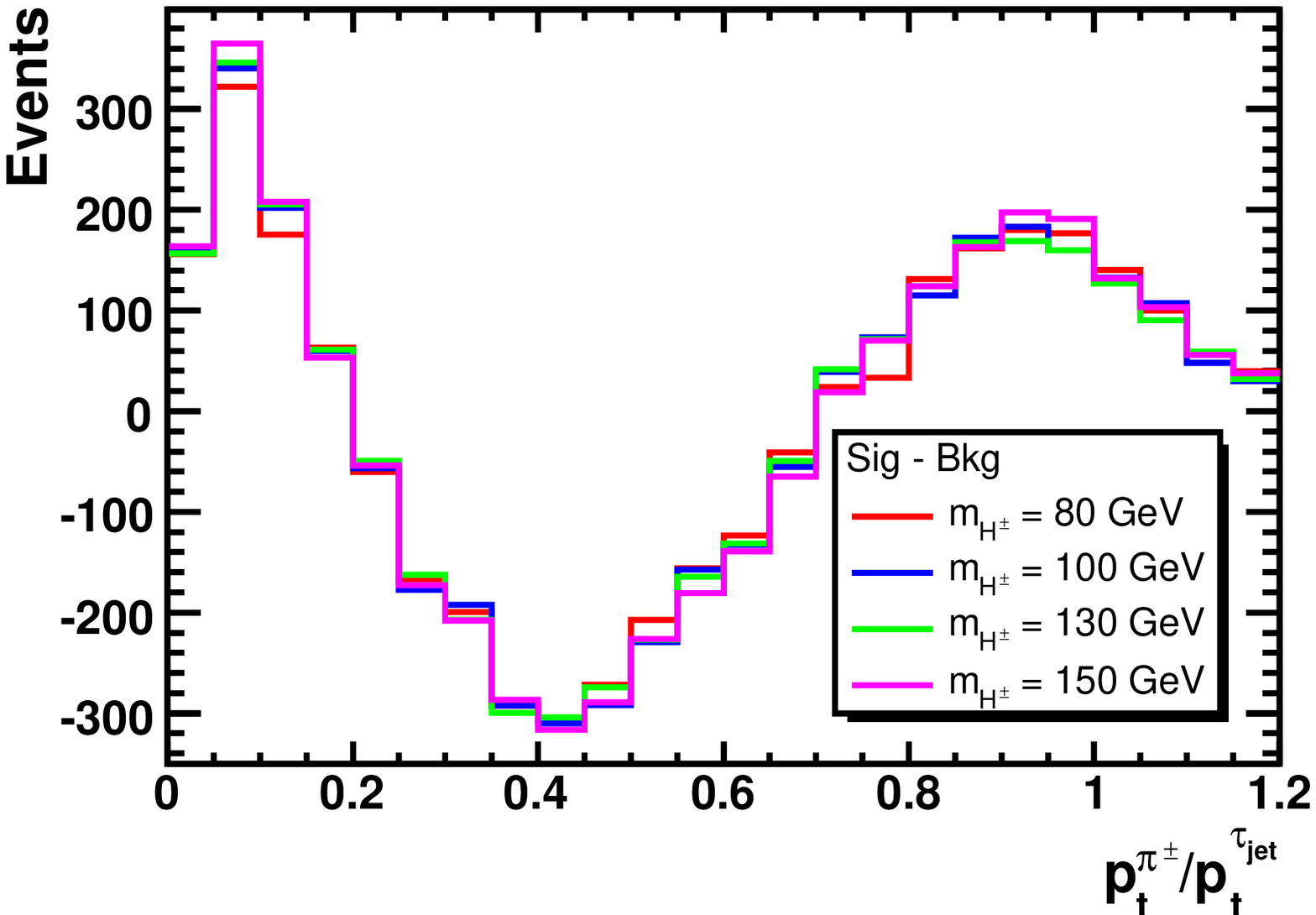, width=0.4\textwidth}
\epsfig{file=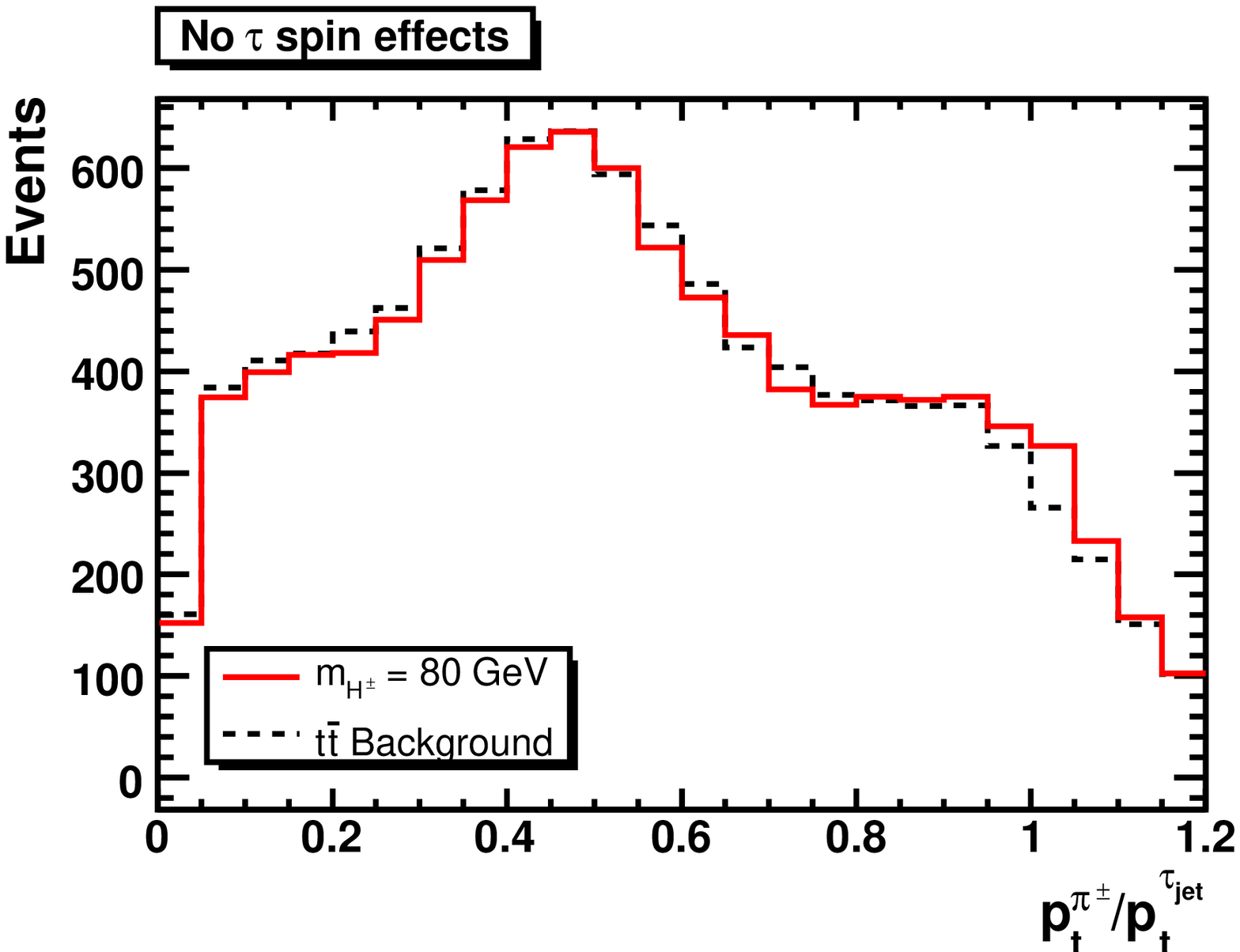, width=0.4\textwidth} \hfill
\epsfig{file=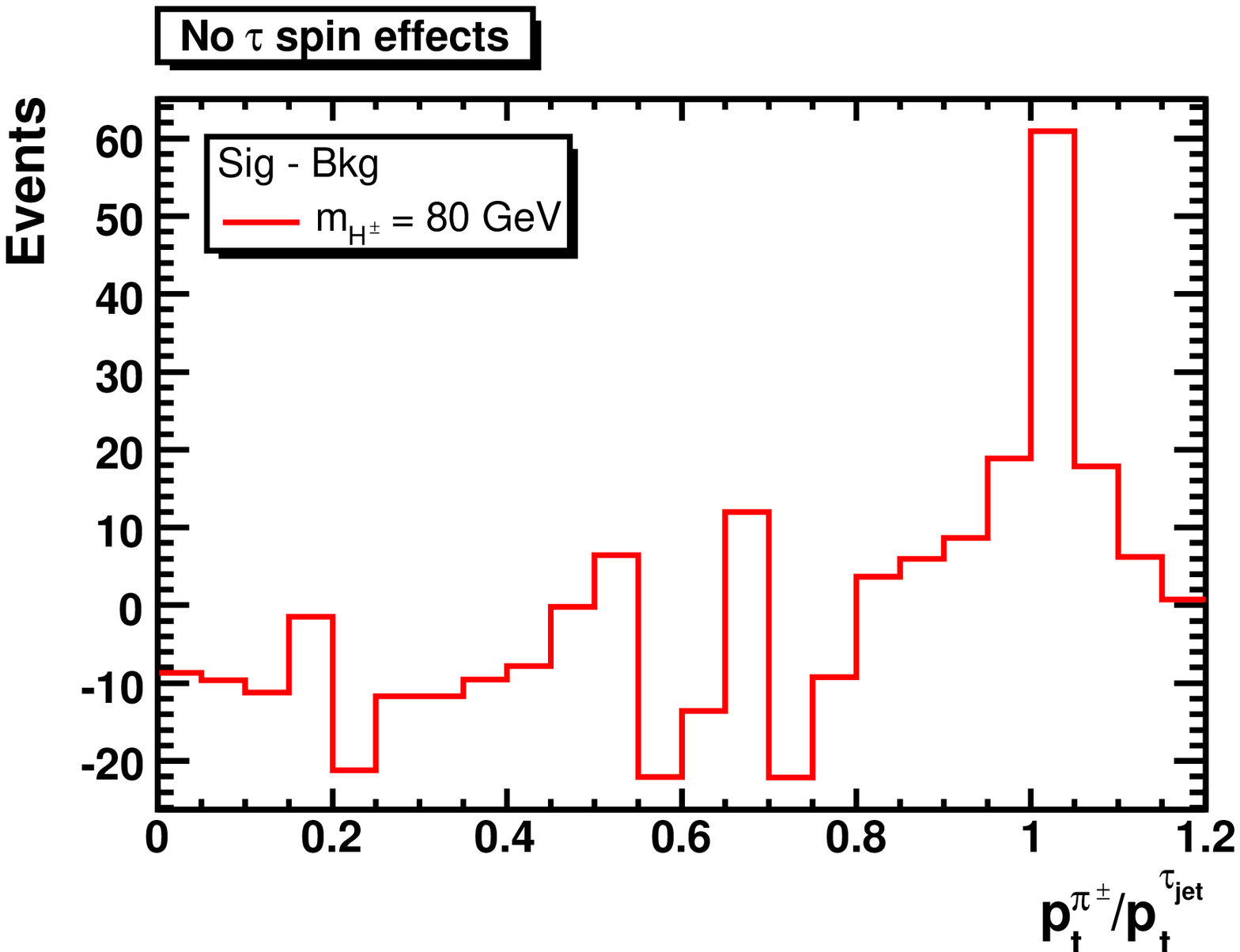, width=0.4\textwidth}
\vspace*{-4mm}
\caption{Distributions of the  
ratio $p_t^{\pi^\pm}/p_t^{\tau_\mathrm{jet}}$
for the $tbH^\pm$ signal
and the $t\bar{t}$ background for $\sqrt{s}=14$~TeV (left)
and the respective differences between signal and background (right).
The lower plots show distributions without spin effects in the $\tau$
decays.
}
\label{fig:lhc_r1}
\end{figure}

\begin{figure}[htbp]
\epsfig{file=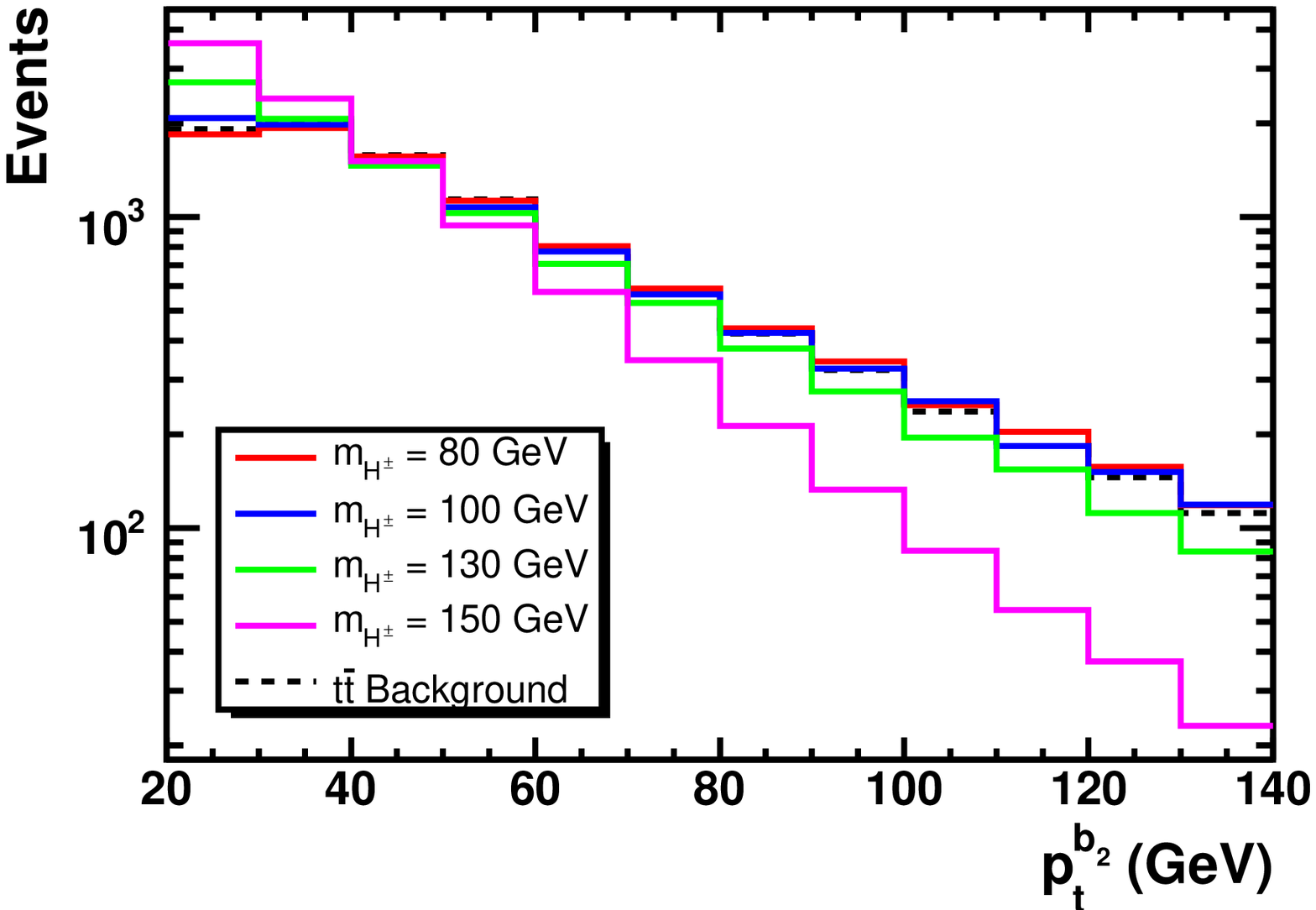, width=0.5\textwidth}  \hfill
\epsfig{file=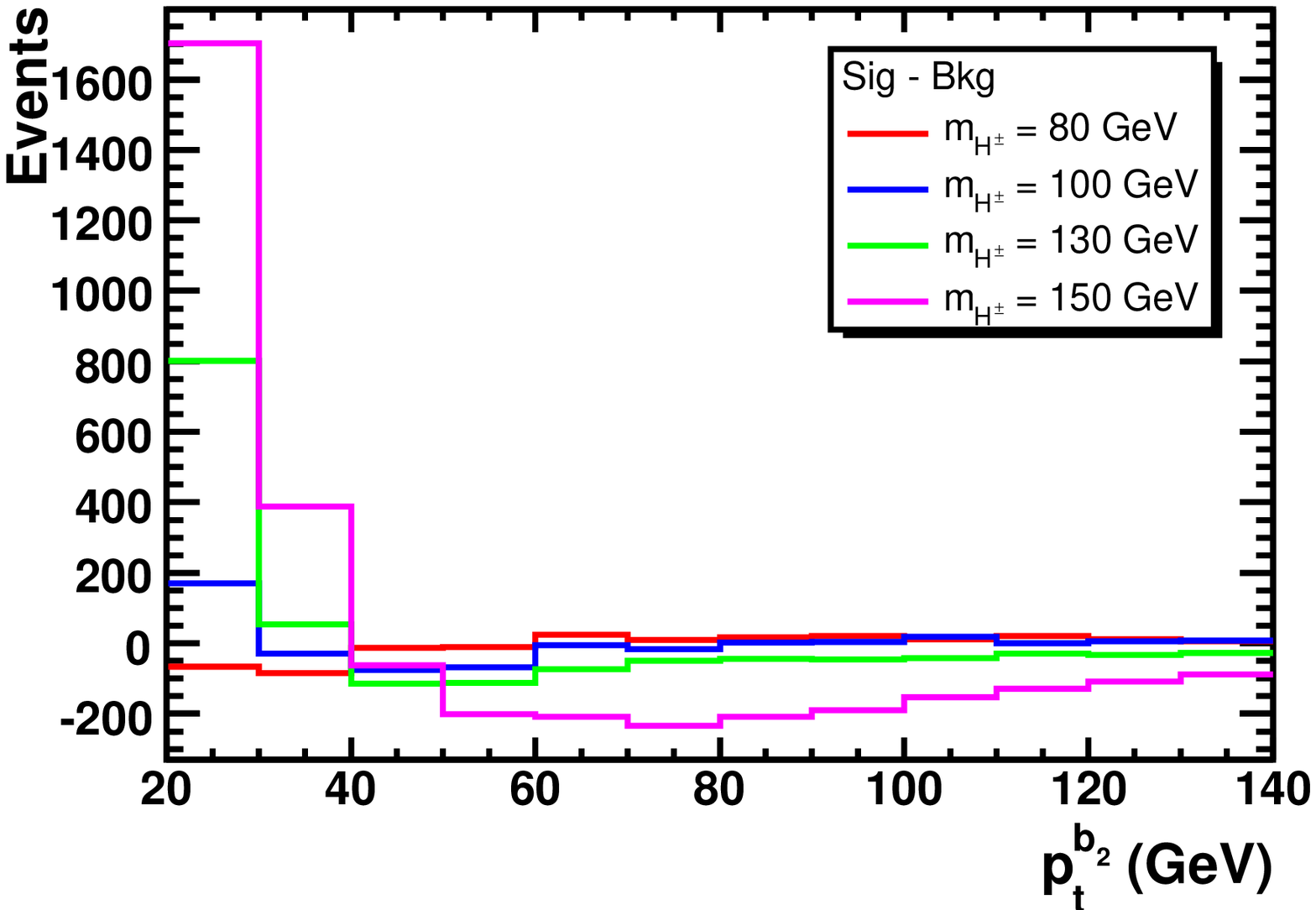, width=0.5\textwidth}
\caption{
$p_t$ distributions of the second (least energetic) $b$ quark jet
for the $tbH^\pm$ signal
and the $t\bar{t}$ background for $\sqrt{s}=14$~TeV (left)
and the respective differences between signal and background (right).
}
\label{fig:lhc_ptb2}
\end{figure}

\begin{figure}[htbp]
\epsfig{file=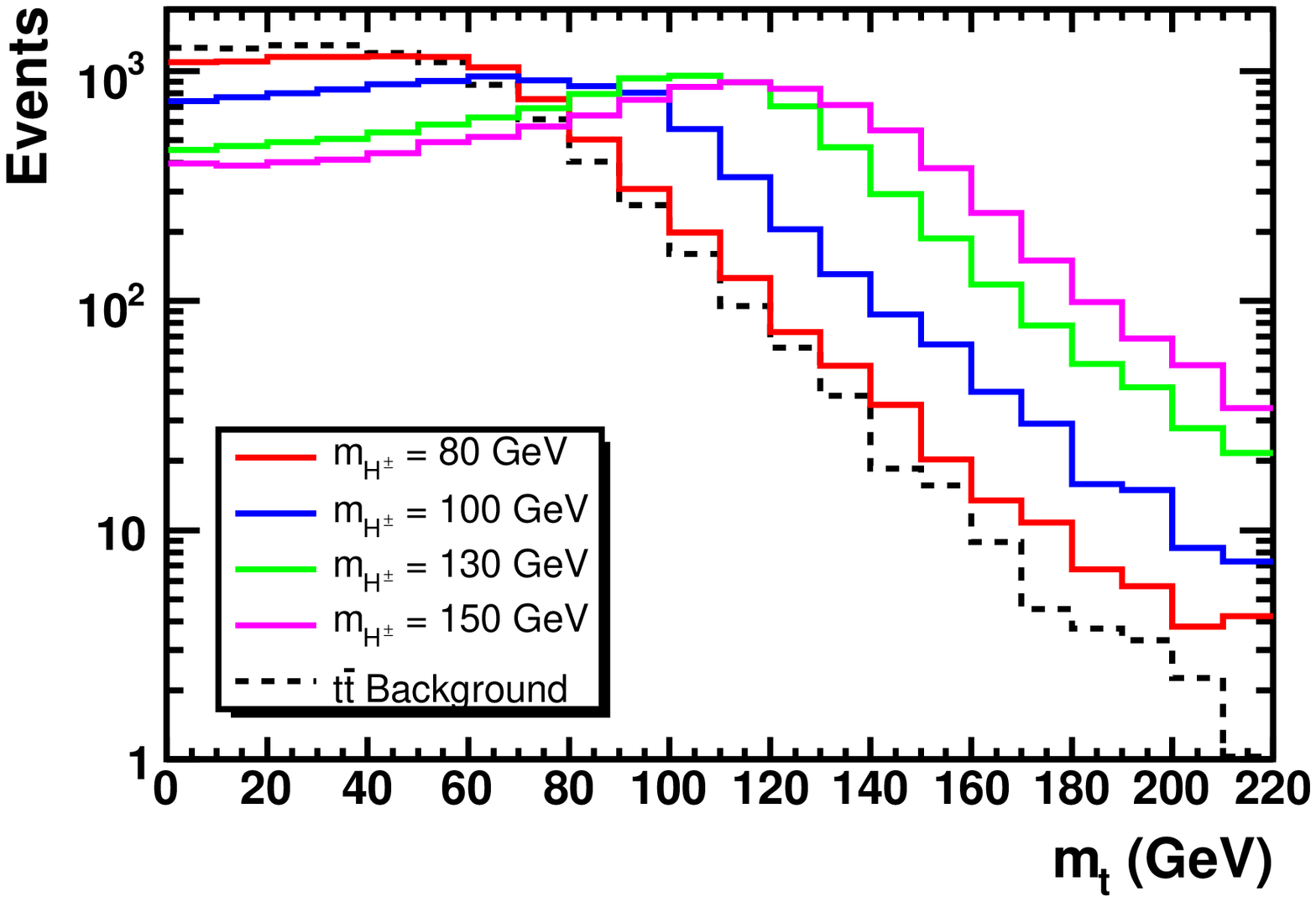, width=0.5\textwidth}  \hfill
\epsfig{file=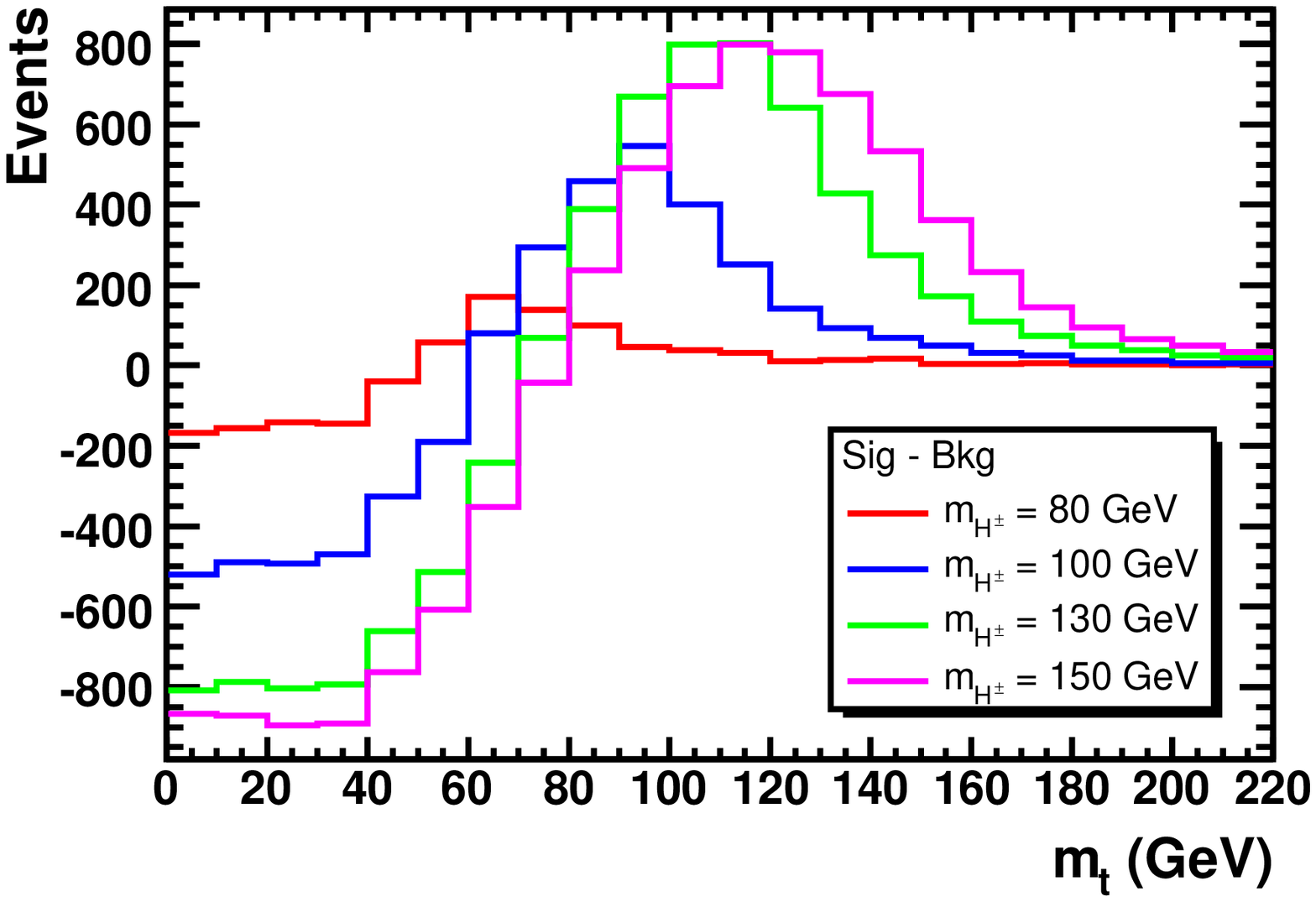, width=0.5\textwidth}
\caption{
Transverse mass
$m_t = \sqrt{2 p_t^{\tau_\mathrm{jet}} p_t^{\rm miss}
  [1-\cos(\Delta\phi)]}$
distributions of the
$\tau_{\rm jet} + p_t^{\rm miss}$ system
($\Delta\phi$ is the azimuthal angle
between $p_t^{\tau_\mathrm{jet}}$ and $p_t^{\rm miss}$)
for the $tbH^\pm$ signal
and the $t\bar{t}$ background for $\sqrt{s}=14$~TeV (left)
and the respective differences between signal and background (right).
}
\label{fig:lhc_mtransverse}
\end{figure}

\begin{figure}[htbp]
\epsfig{file=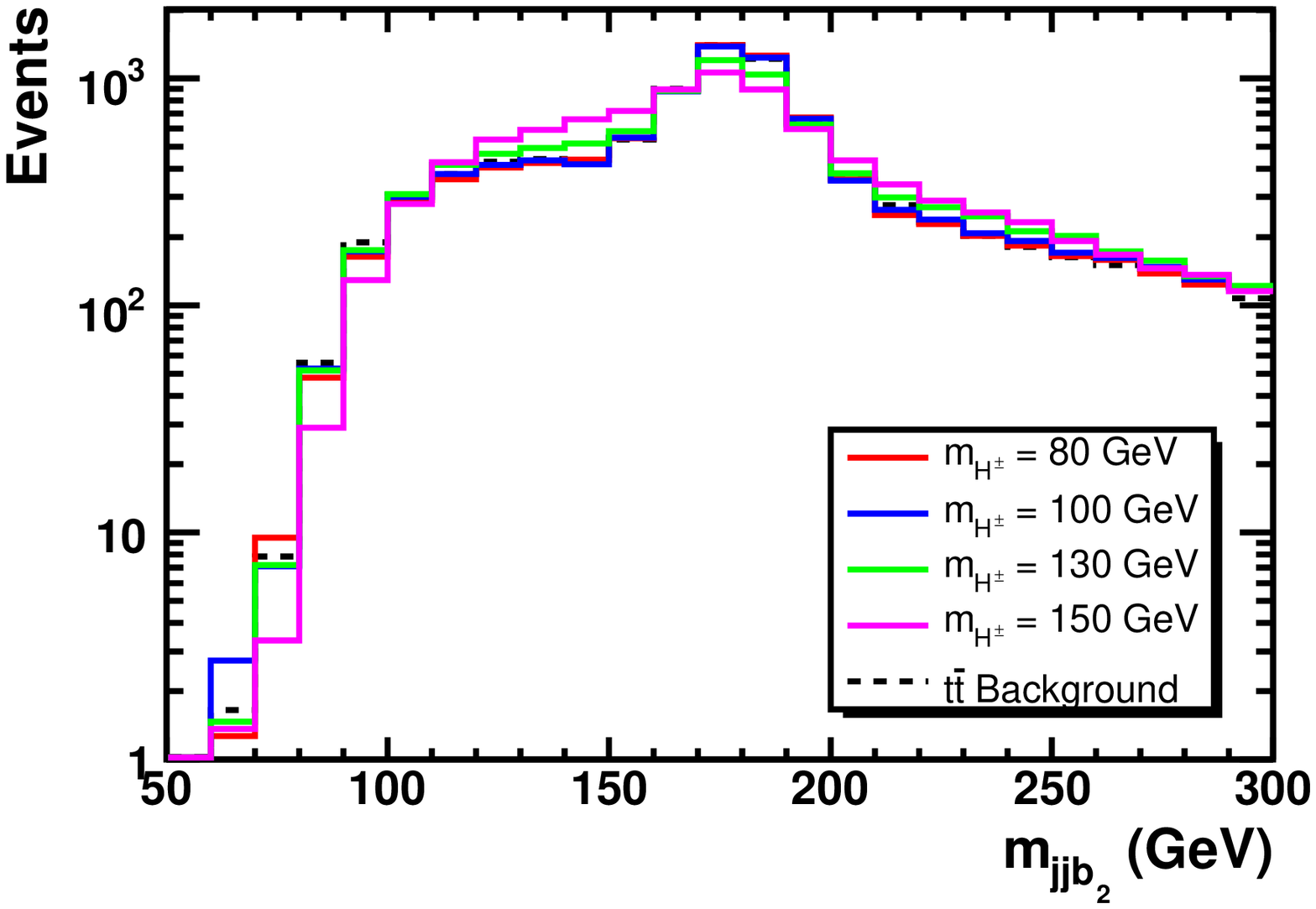, width=0.5\textwidth}  \hfill
\epsfig{file=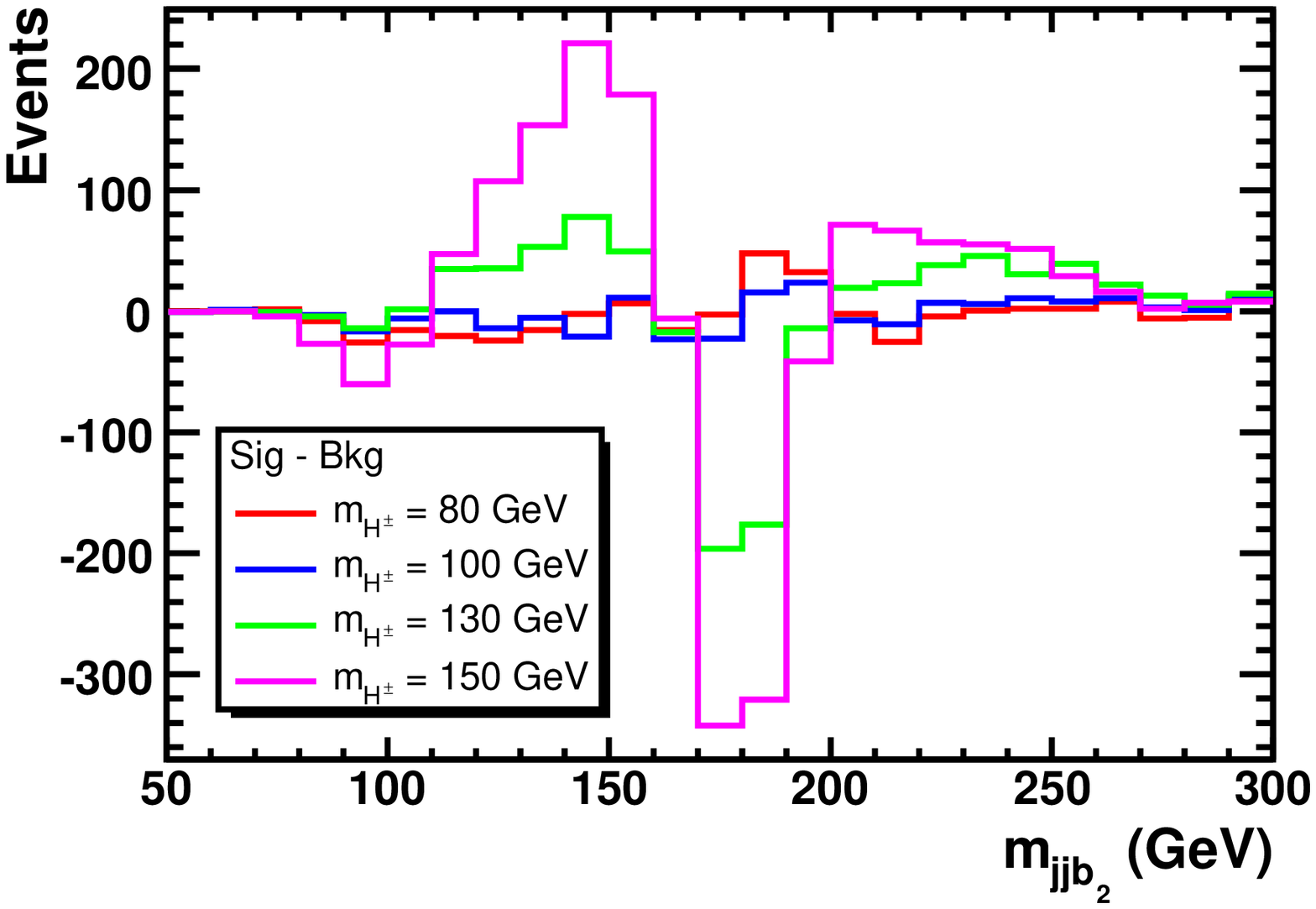, width=0.5\textwidth}
\caption{
Invariant mass distributions of the two light quark jets and the 
second (least energetic) $b$ quark jet
for the $tbH^\pm$ signal
and the $t\bar{t}$ background for $\sqrt{s}=14$~TeV (left)
and the respective differences between signal and background (right).
}
\label{fig:lhc_mjjb2}
\end{figure}

\begin{figure}[htbp]
\epsfig{file=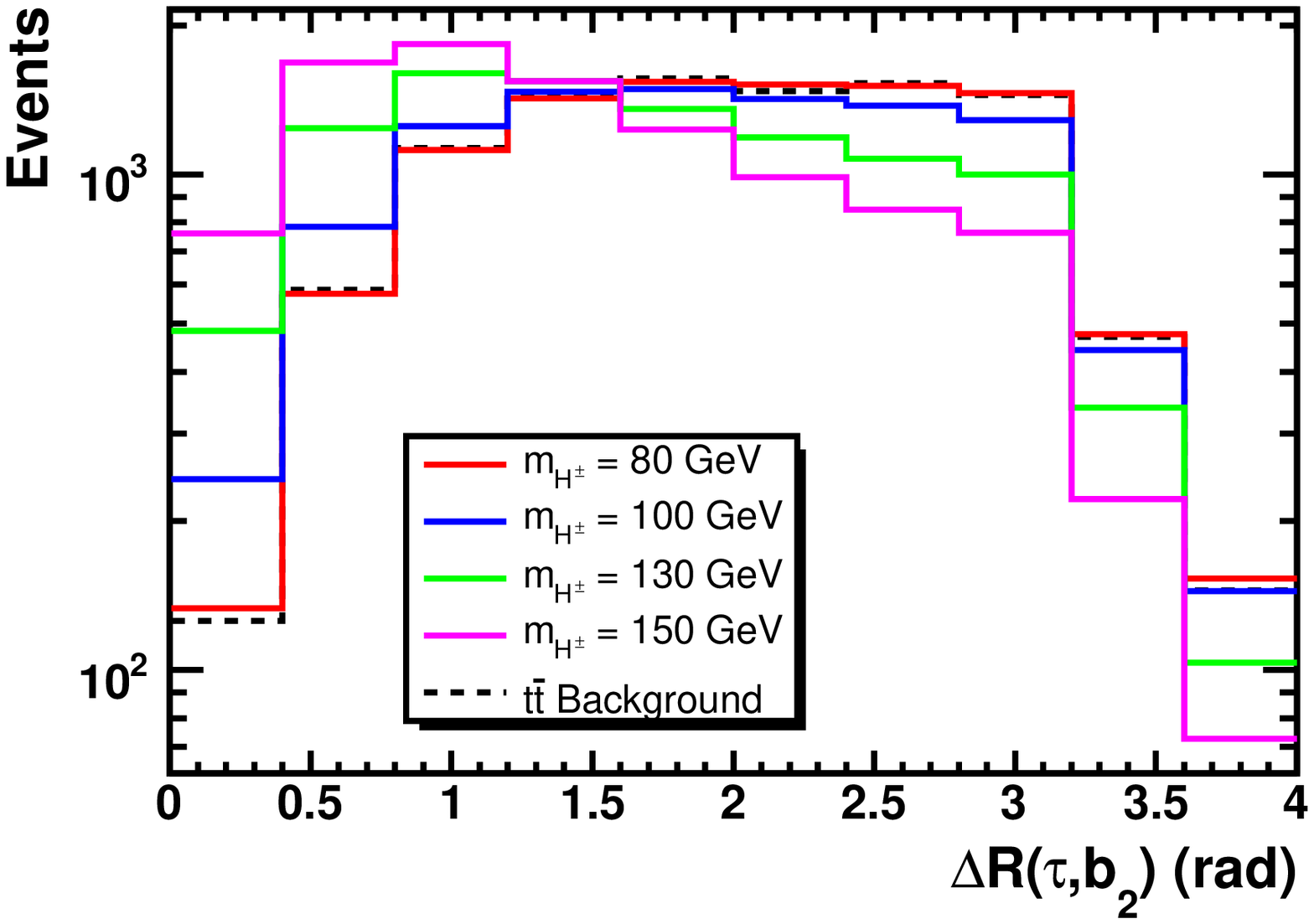, width=0.5\textwidth}  \hfill
\epsfig{file=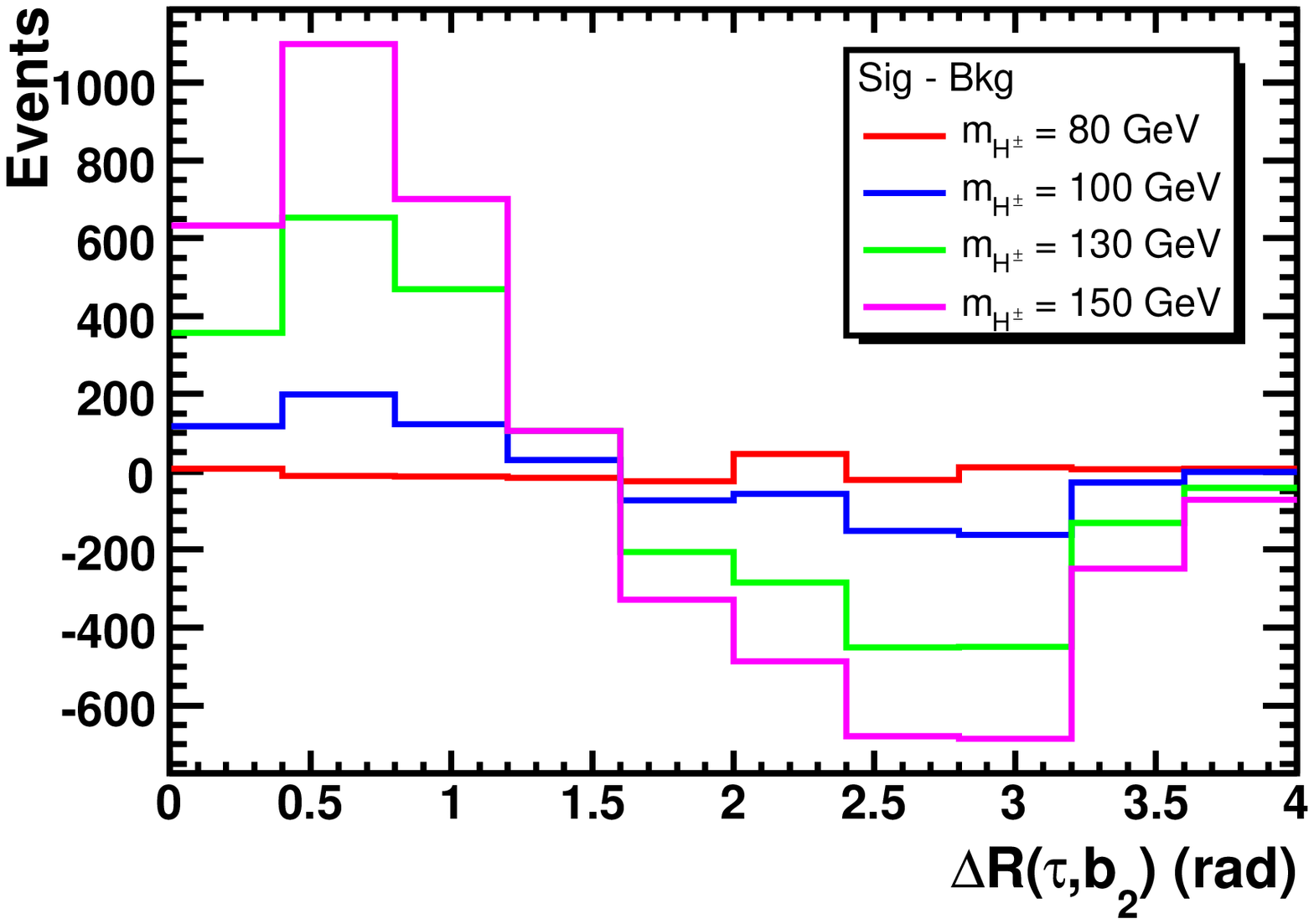, width=0.5\textwidth}
\caption{
Spatial distance
$\Delta R(\tau,b_2) = \sqrt{(\Delta\phi)^2 + (\Delta\eta)^2}$
distributions
(where $\Delta\phi$ is the azimuthal angle in rad between
the $\tau$ and $b$ jet)
for the $tbH^\pm$ signal
and the $t\bar{t}$ background for $\sqrt{s}=14$~TeV (left)
and the respective differences between signal and background (right).
}
\label{fig:lhc_distance-tau-b}
\end{figure}

\begin{figure}[htbp]
\epsfig{file=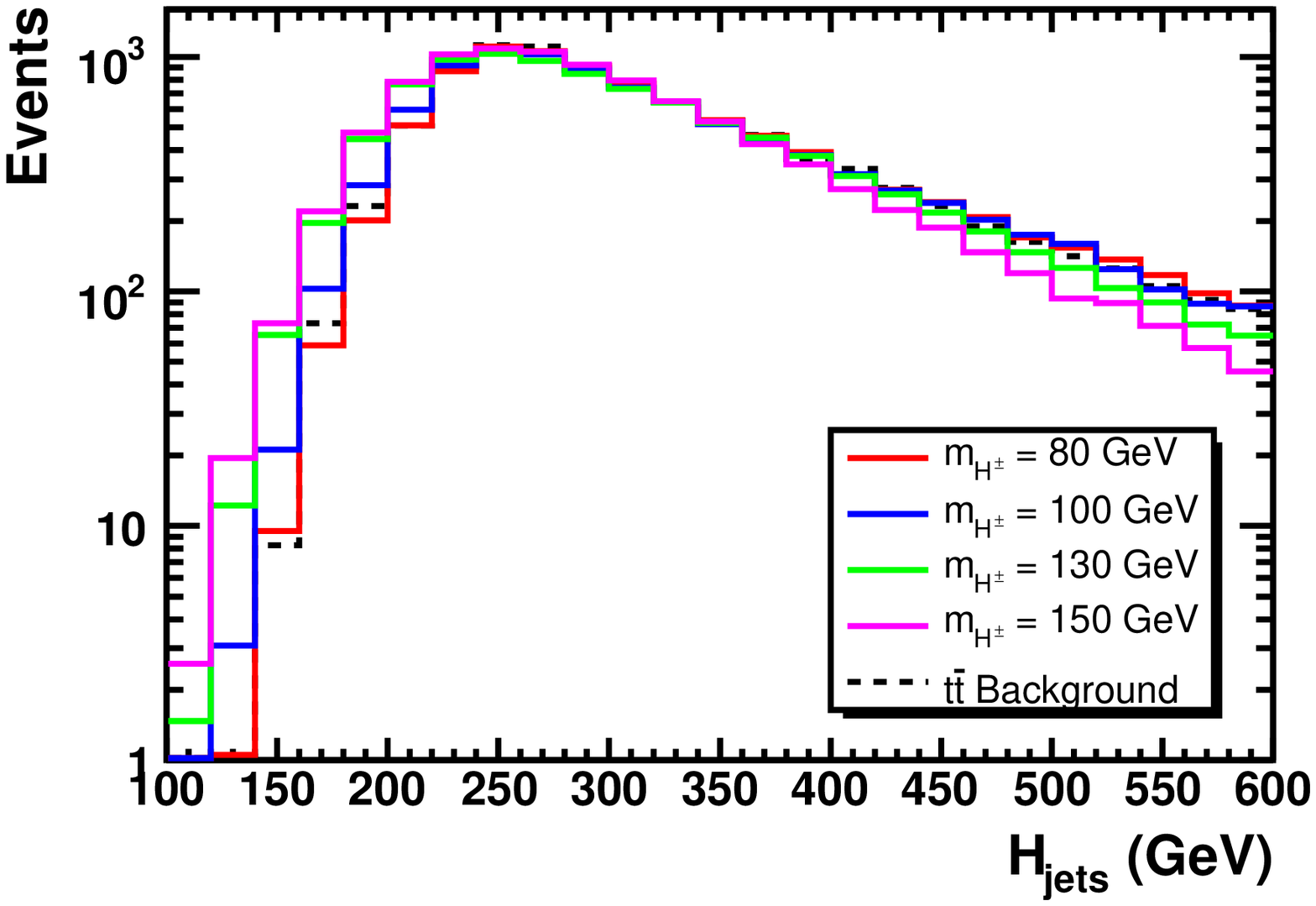, width=0.4\textwidth}  \hfill
\epsfig{file=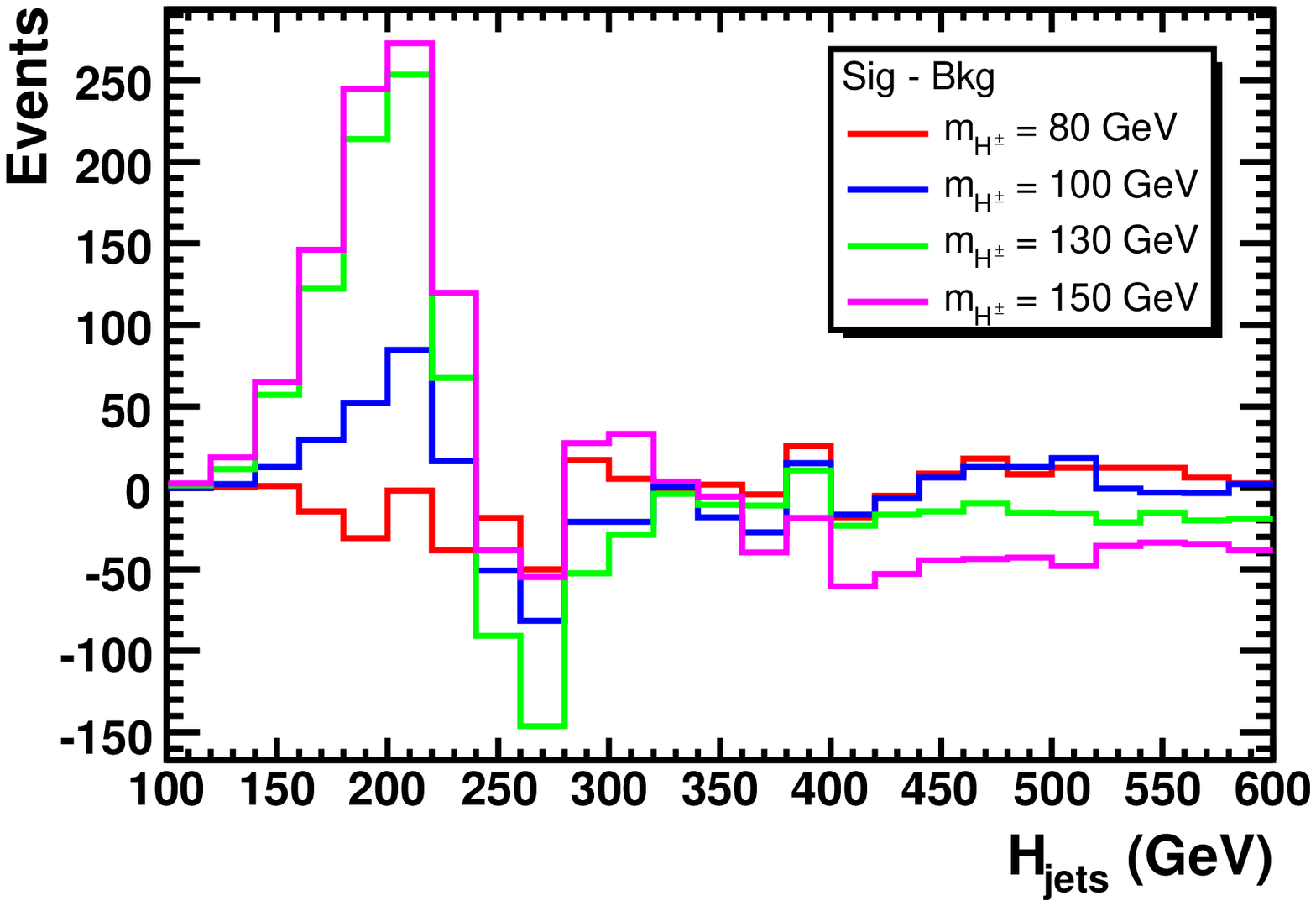, width=0.4\textwidth}
\vspace*{-3mm}
\caption{
Distributions of the
total transverse momentum of all quark jets,
$H_{\rm jets} = p_t^{j_1} + p_t^{j_2} + p_t^{b_1} + p_t^{b_2}$,
for the $tbH^\pm$ signal
and the $t\bar{t}$ background for $\sqrt{s}=14$~TeV (left)
and the respective differences between signal and background (right). 
}
\label{fig:lhc_hjet}
\end{figure}

\begin{figure}[htbp]
\epsfig{file=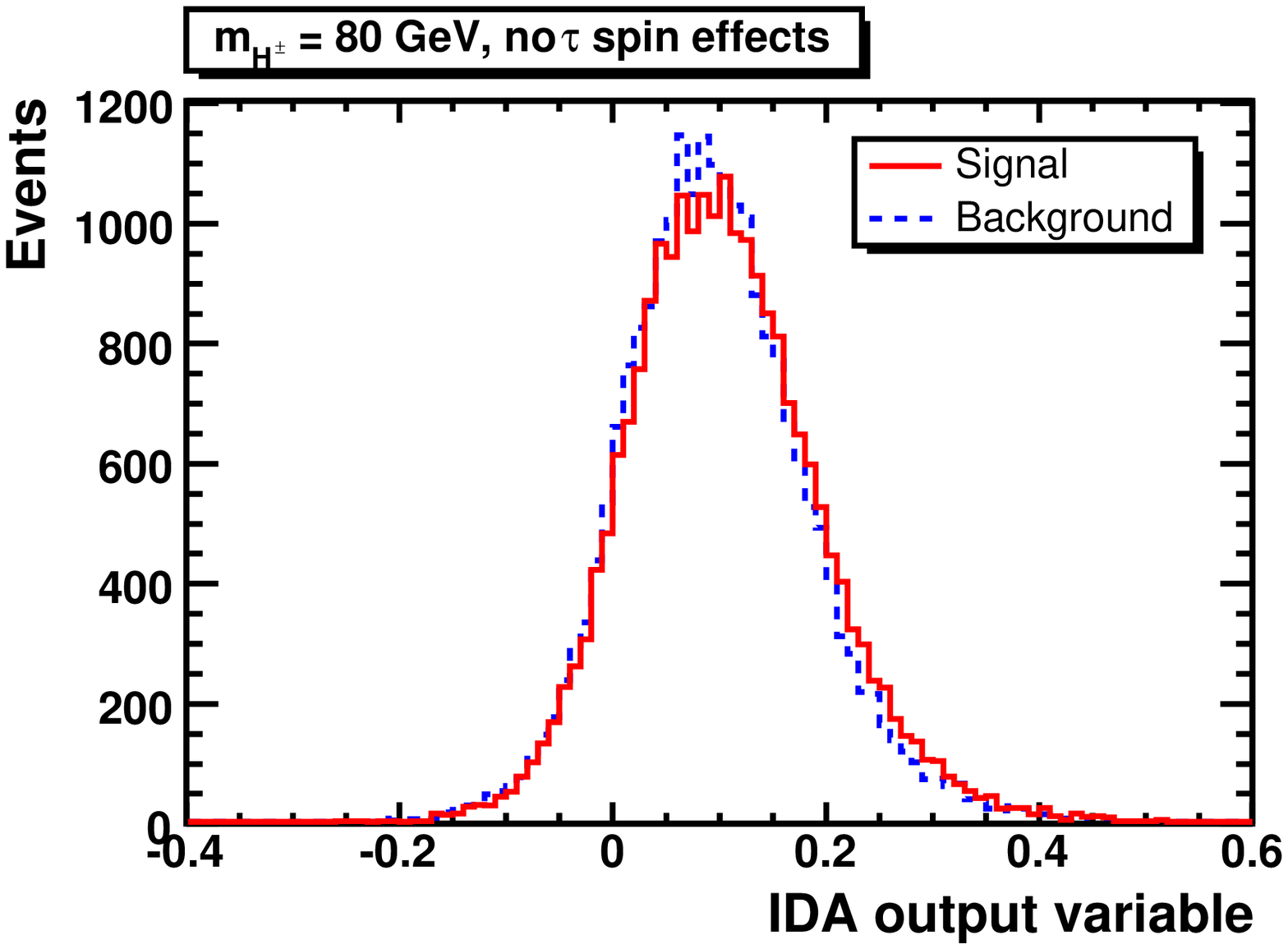, width=0.4\textwidth}  \hfill
\epsfig{file=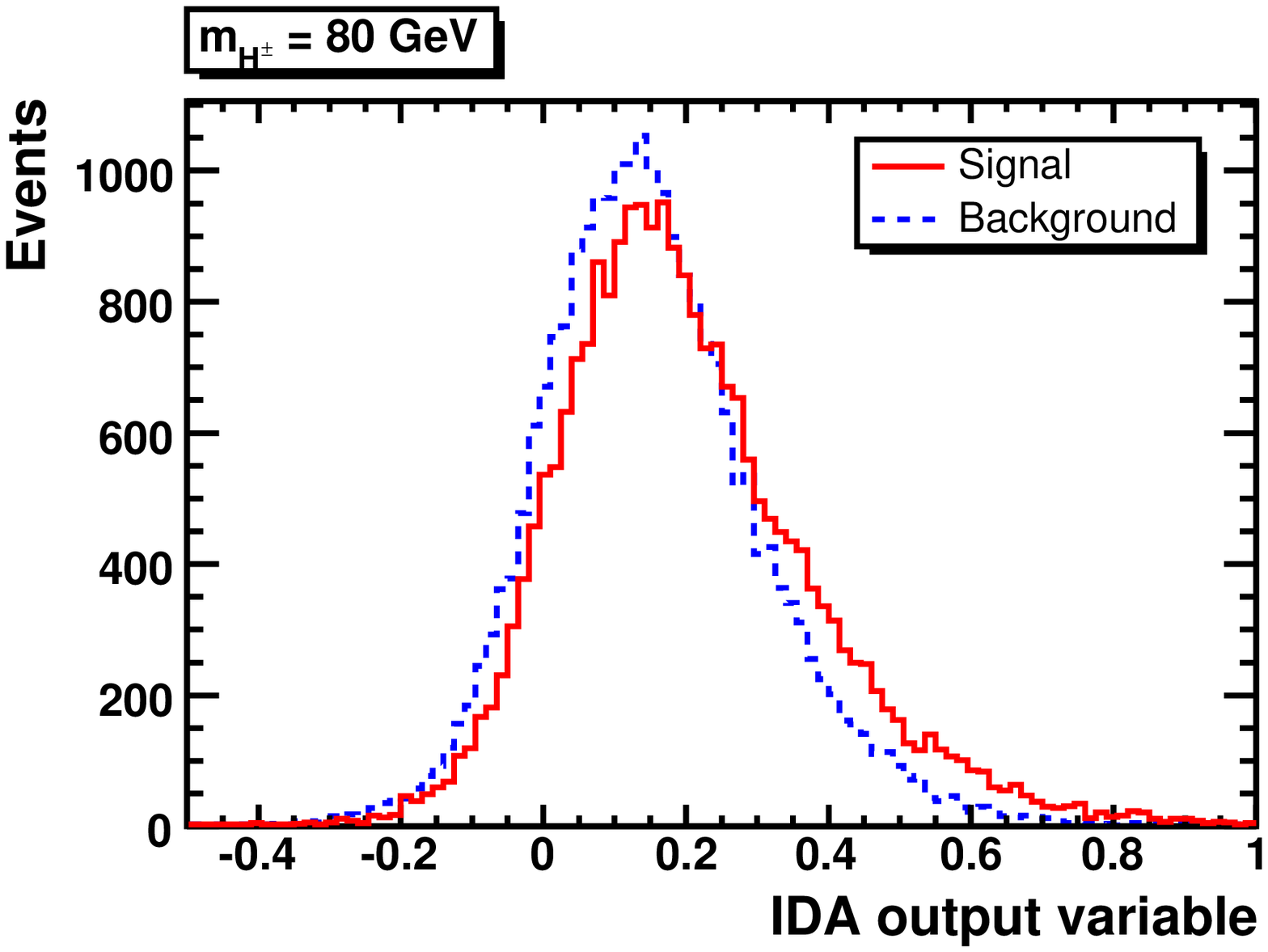, width=0.4\textwidth}
\epsfig{file=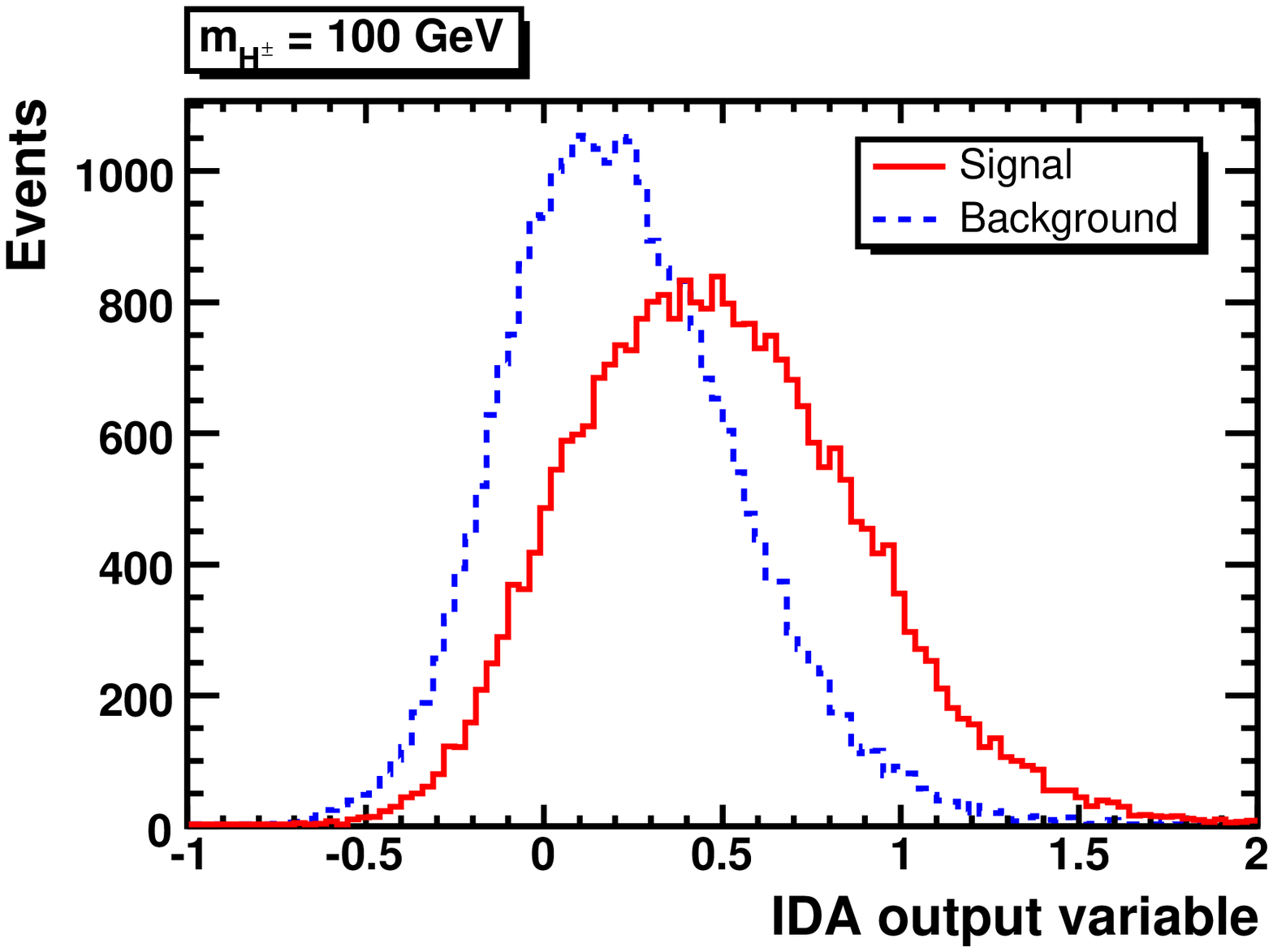, width=0.4\textwidth}  \hfill
\epsfig{file=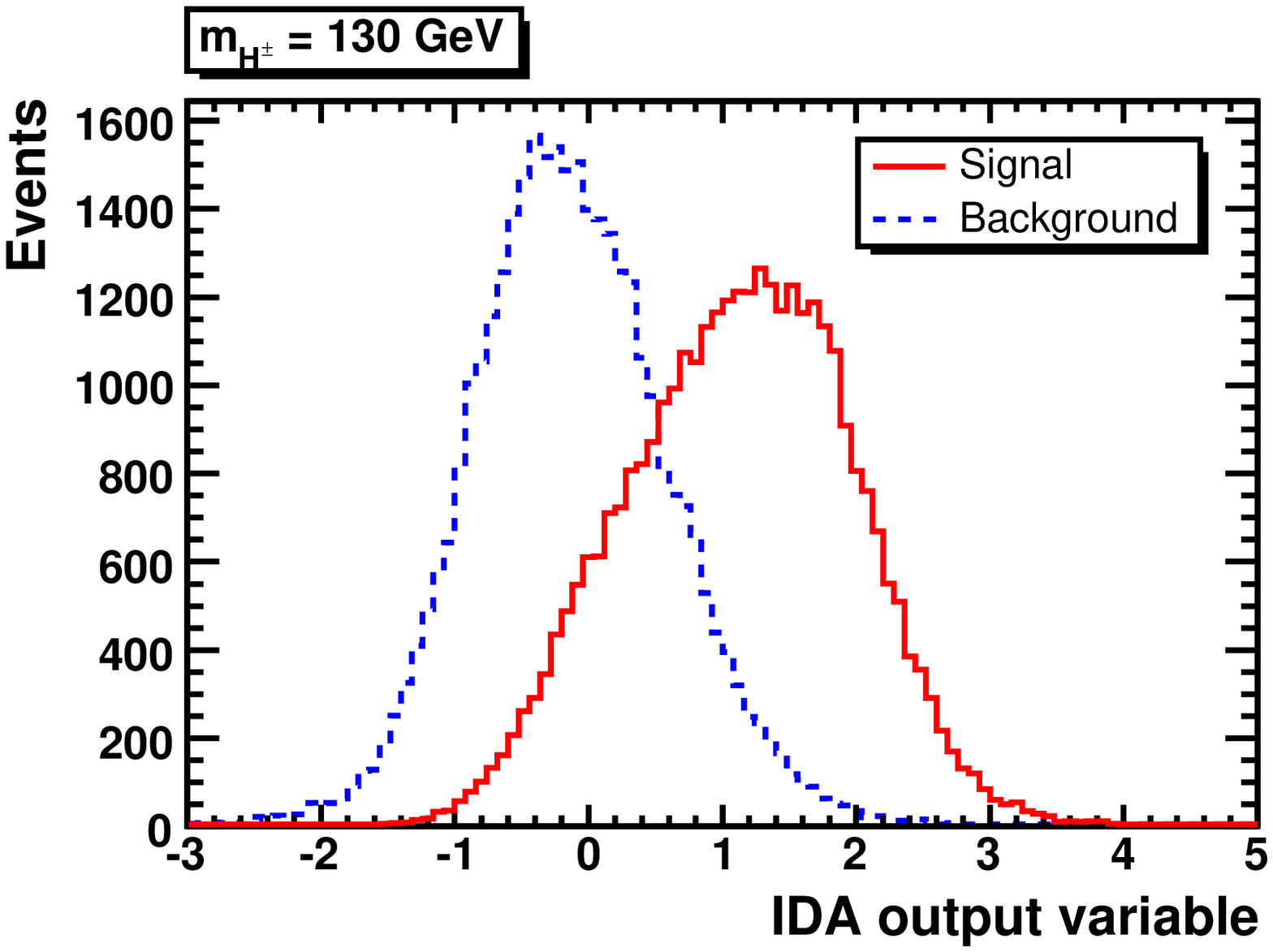, width=0.4\textwidth}
\epsfig{file=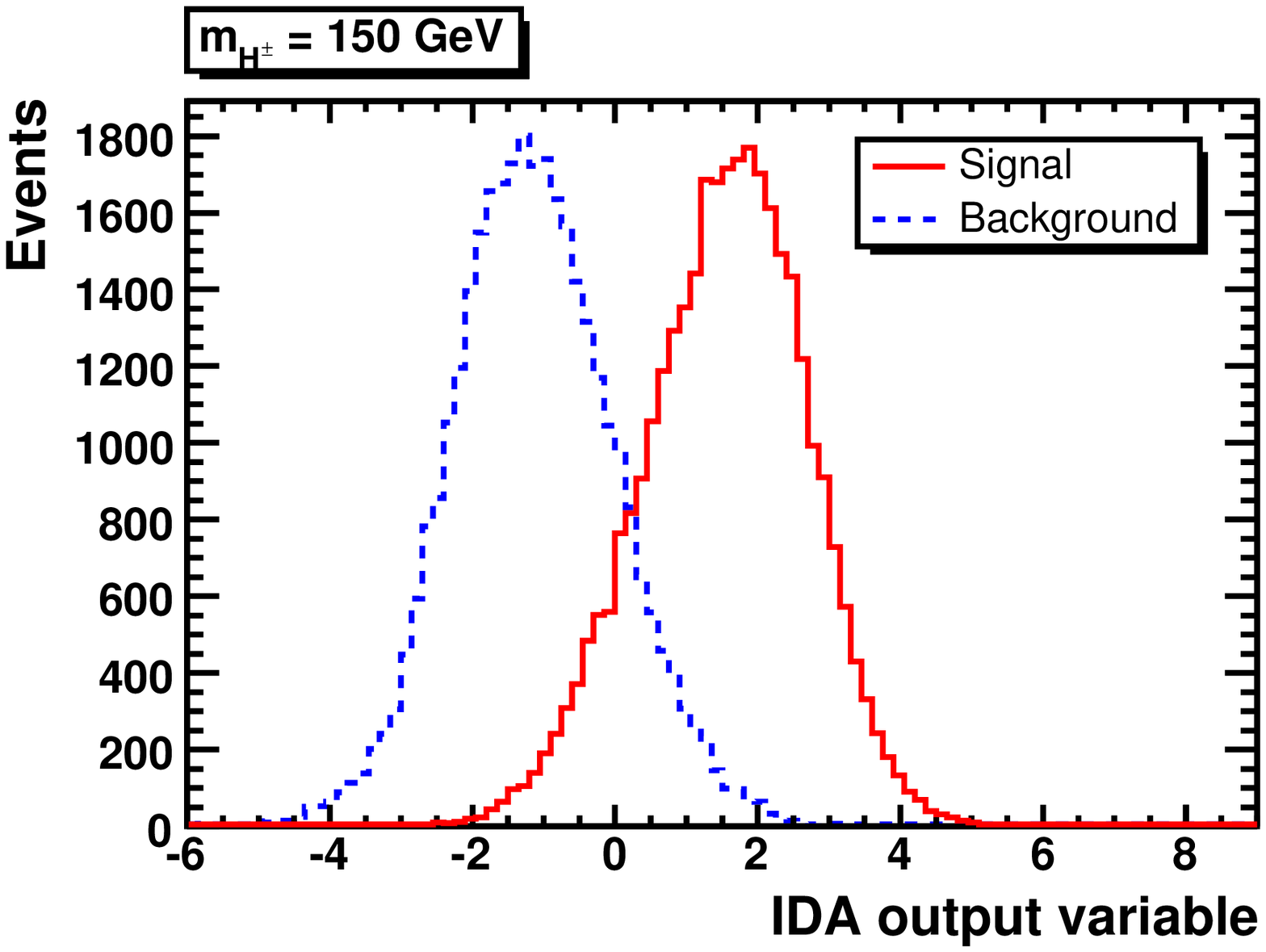, width=0.4\textwidth}  \hfill
\vspace*{-3mm}
\caption{
Distributions of the IDA output variable in the first IDA
step for the $tbH^\pm$ signal (solid, red)
and the $t\bar{t}$ background (dashed, blue) for $\sqrt{s}=14$~TeV.
}
\label{fig:lhc_ida1}
\end{figure}

\begin{figure}[htbp]
\epsfig{file=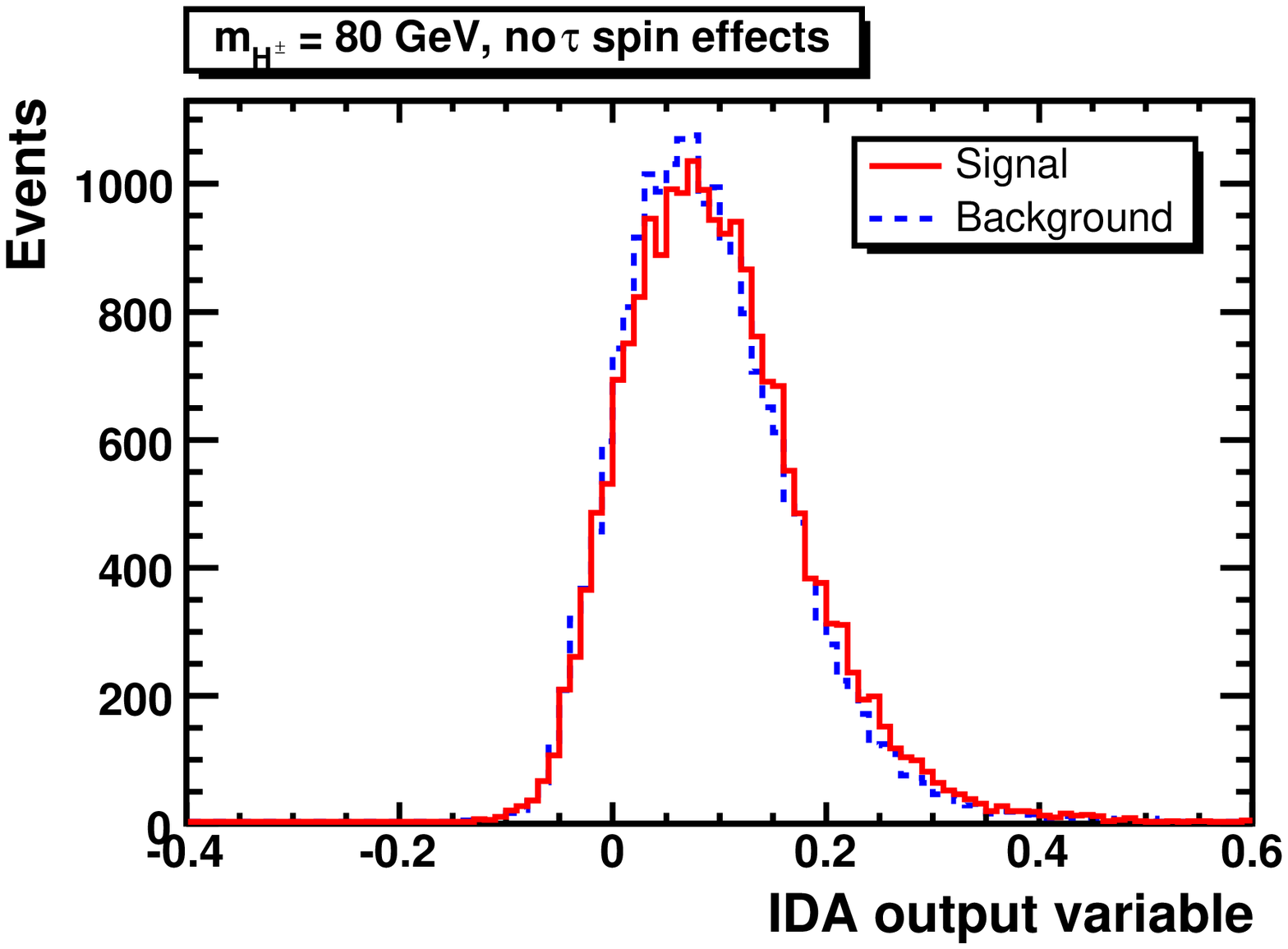, width=0.5\textwidth}  \hfill
\epsfig{file=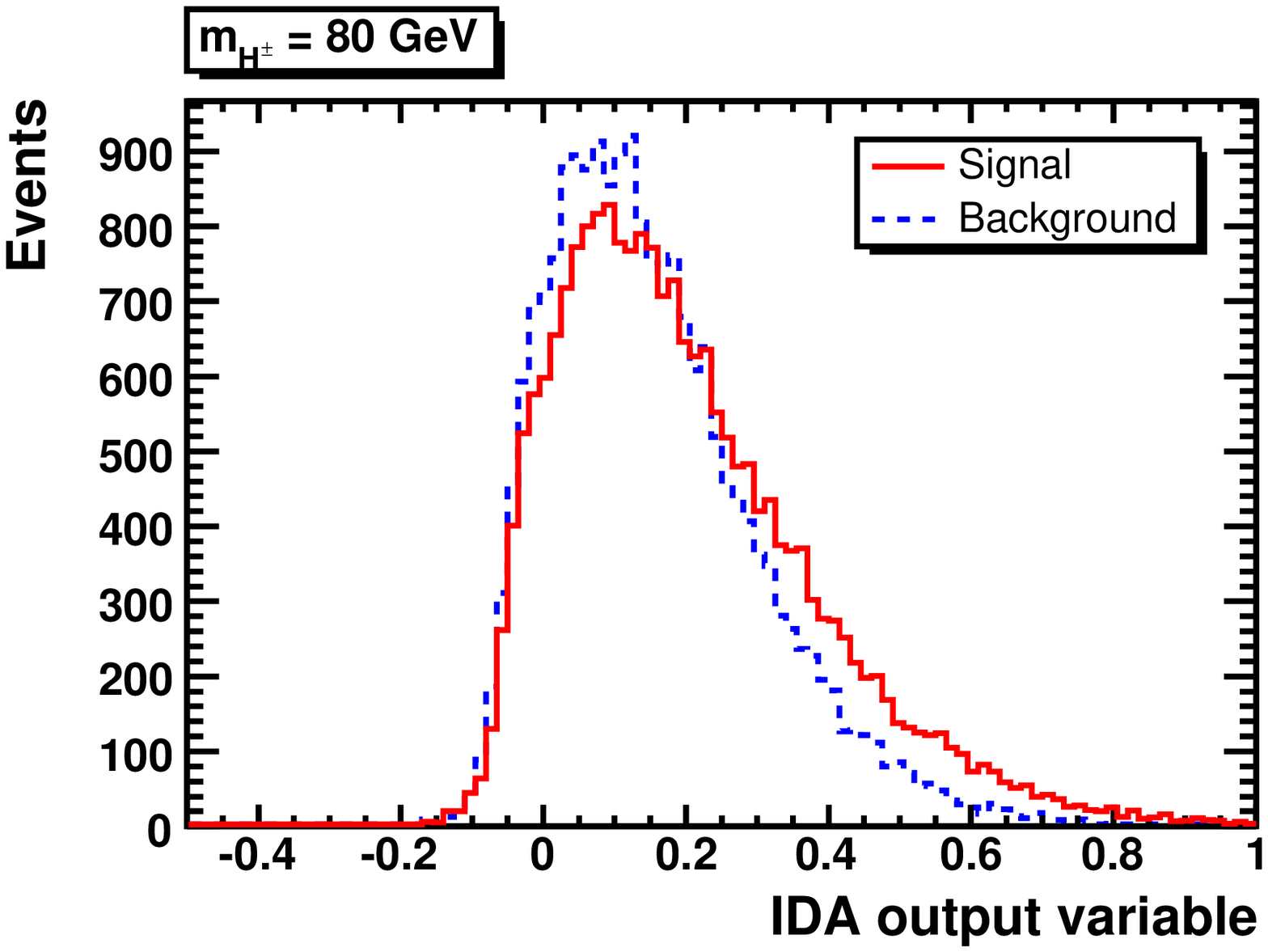, width=0.5\textwidth}
\epsfig{file=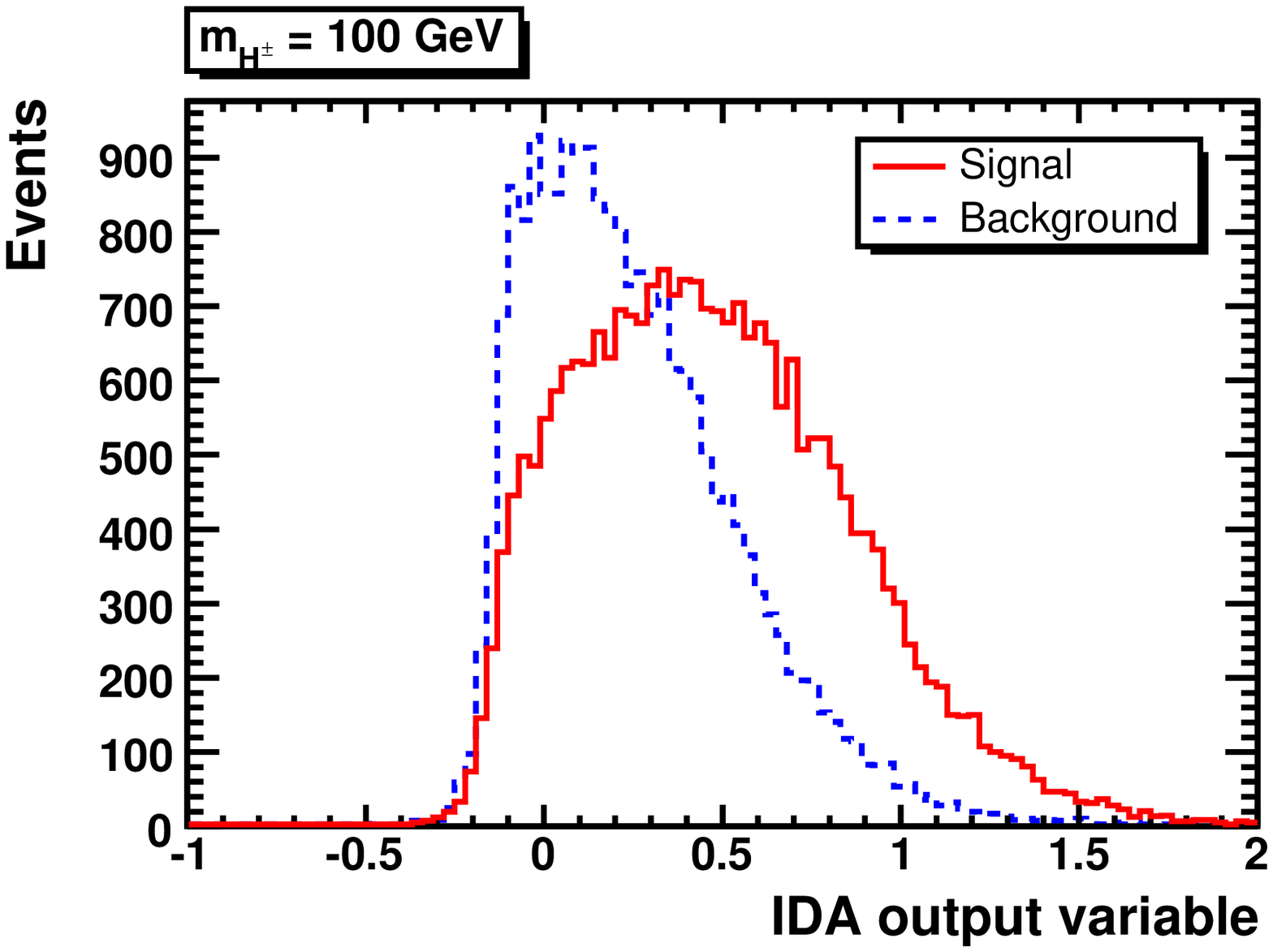, width=0.5\textwidth}  \hfill
\epsfig{file=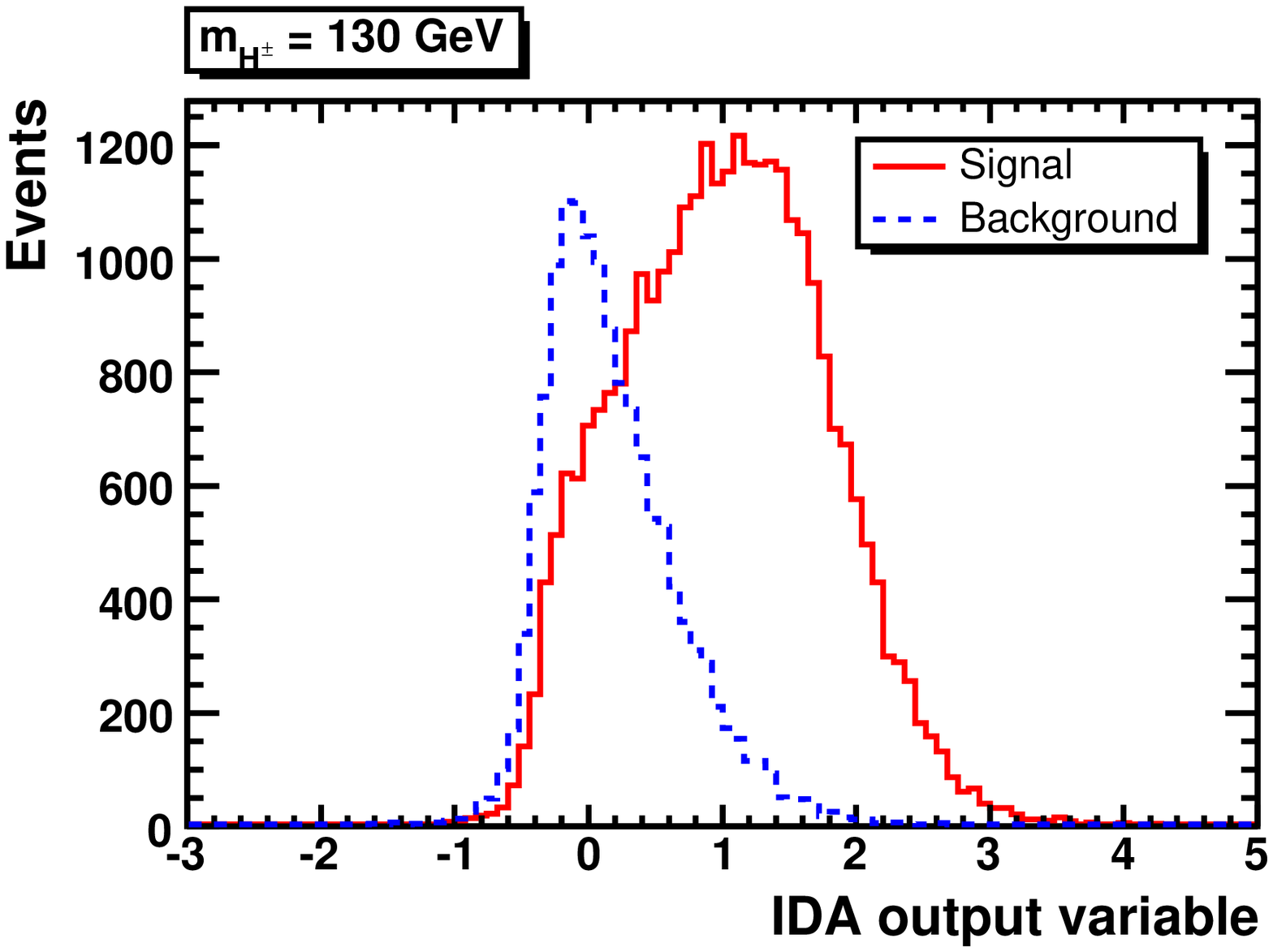, width=0.5\textwidth}
\epsfig{file=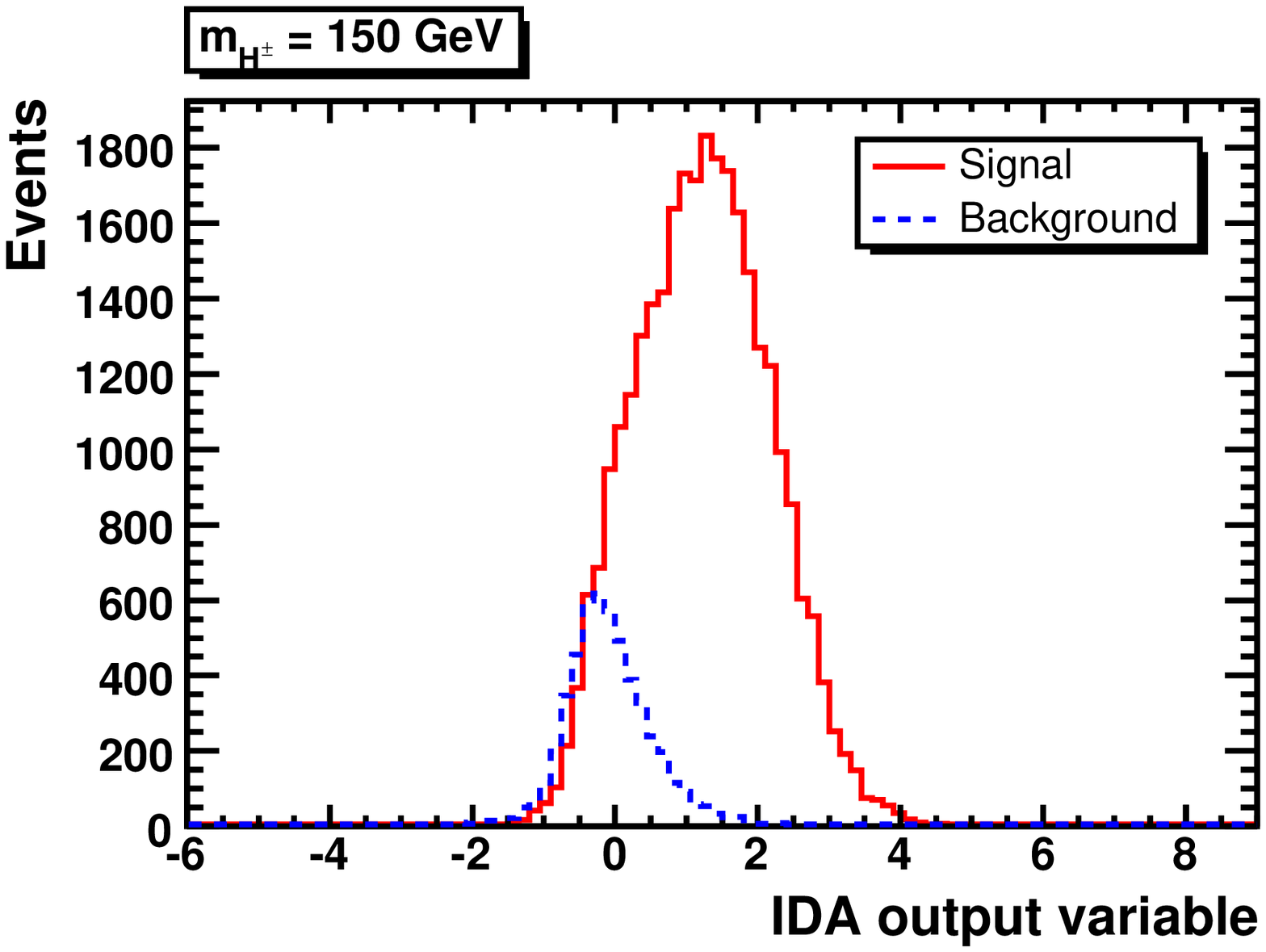, width=0.5\textwidth}  \hfill
\epsfig{file=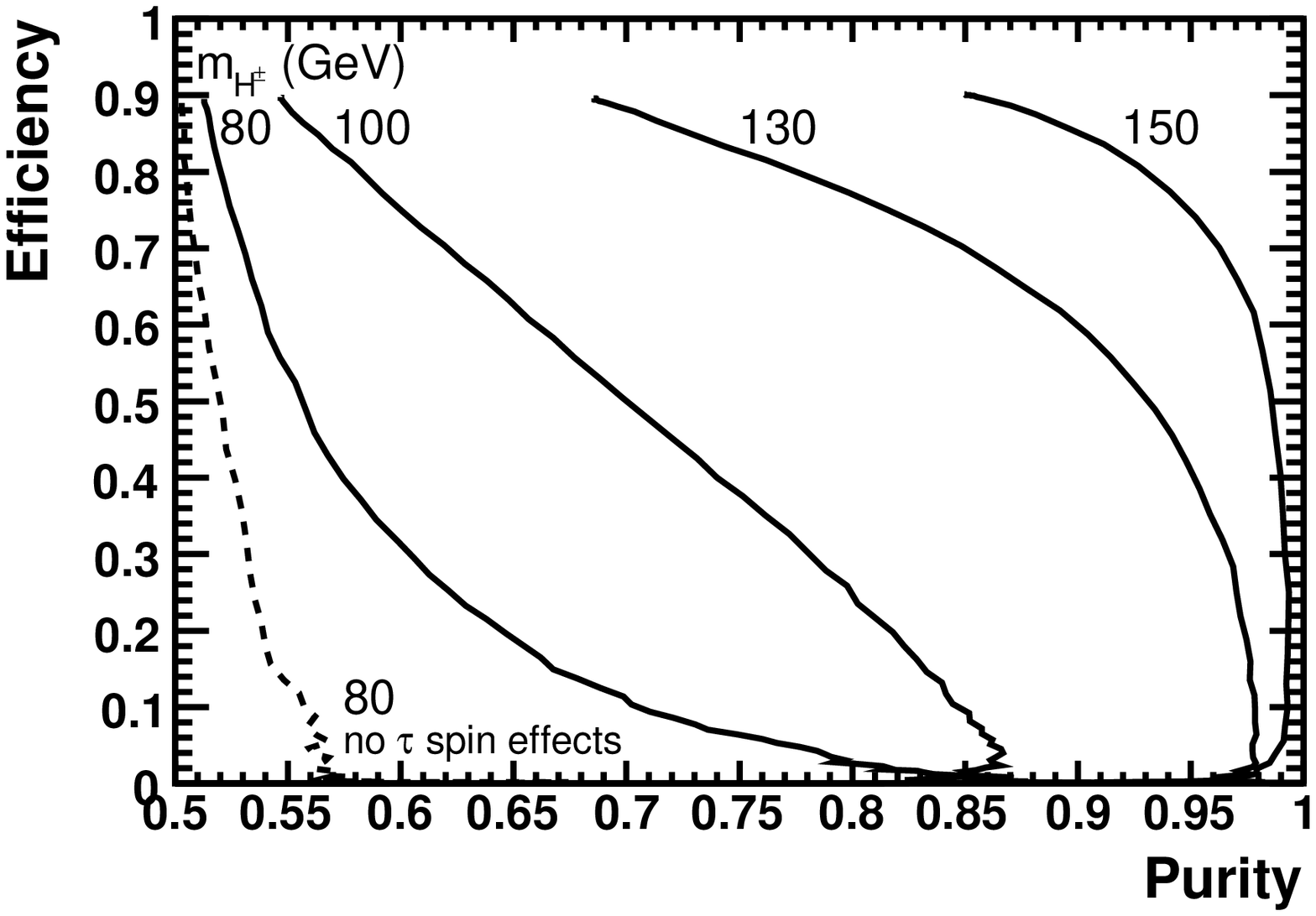, width=0.5\textwidth}
\caption{
Upper row, middle row and lower left figure:
distributions of the IDA output variable 
in the second IDA step
for 90\% efficiency in the first IDA
step (corresponding to a cut at 0 in Fig.~\ref{fig:lhc_ida1})
for the $tbH^\pm$ signal (solid, red)
and the $t\bar{t}$ background (dashed, blue).
Lower right figure: efficiency as a function of the purity
when not taking the spin effects in the $\tau$ decay into account for
$m_{H^\pm}=80$~GeV (dashed) and with spin effects in the $\tau$ decay for
$m_{H^\pm}=80,100,130,150$~GeV (solid, from left to right).
Results are for the LHC.
}
\label{fig:lhc_ida}
\end{figure}

\clearpage
\subsection{Conclusions}
The discovery of charged Higgs bosons 
would be a clear sign of physics beyond the SM.
In this case study we have investigated charged Higgs boson topologies 
produced at the current Tevatron and LHC energies and compared
them against the irreducible SM background due to top-antitop production
and decay. 
While sizable differences between signal and background
are expected whenever $m_{H^\pm}\ne m_{W^\pm}$, 
near the current mass limit of about $m_{H^\pm}\approx 80$ GeV the
kinematic spectra are very similar between SM decays and
those involving charged Higgs bosons. In this case,
spin information will significantly
distinguish between signal and irreducible SM background. In fact,
we have considered hadronic $\tau\nu_\tau$ decays of charged Higgs bosons, 
wherein the $\tau$ polarization induced by a decaying (pseudo)scalar object
is significantly different from those emerging in the vector ($W^\pm$) decays
onsetting in the top-antitop case. 
For a realistic analysis which is not specific for a particular detector, 
a dedicated Monte Carlo event generation and a simplified multipurpose 
detector response approximation have been applied.
The identification of a hadronic tau-lepton will be an experimental challenge 
in an environment with typically four jets being present. 
We have demonstrated how an IDA method can be an applied to separate signal and background
when the differences between the signal and background distributions are small. 
Our results show that the IDA method 
will be equally effective at both the Tevatron and LHC. 
While only the dominant irreducible $t\bar{t}$ background has been dealt 
with in detail, we have also specifically addressed the QCD background.
A suitably hard missing transverse momentum cut has been applied to
reject such jet activity 
and we have demonstrated that 
although the discriminative power is reduced by such a cut, the reduction is
small compared to the gain from including the $\tau$ polarization effects.
Using the differences in $\tau$
polarization between the signal and the dominant SM irreducible  $t\bar{t}$
background is crucial for disentangling the former
from the latter.

\clearpage
\section{Energy scale for b jets in \DO}
\label{sec:D0-bJES}
%
%
\textbf{Contributed by: J. Cammin}
\vspace{0.25in}

This section describes the determination of the energy
response of b jets in the \DO calorimeter. Since this measurement is
work in progress, and no final results are available yet, we discuss
only the concept of the measurement.

\subsection{Introduction}
In \DO, jet energies measured in the
calorimeter~\cite{Abazov:2005pn}, are corrected for offset,
response, and showering~\cite{Abbott:1998xw}. The response is the
largest single correction factor and is measured from the energy
balance in \gj events using the ``Missing \ET Projection Fraction
Method'' (MPF)~\cite{Abbott:1998xw}. The response is measured in
bins of an energy estimator,
$E'=E_{T_\gamma}\cosh(\eta_{\mathrm{jet}})$, which later is mapped to
the raw measured jet energy, $E_{\mathrm{jet}}^{\mathrm{meas}}$ in
order to get an energy-dependent measurement of the jet response.

The above mentioned corrections are derived for light jets and do not
take into account peculiarities of heavy flavor jets, such as
different fragmentation and hadronization and the presence of
semileptonic decays. The latter leads to an energy response
considerably smaller than that of light jets because of the neutrino
involved in the decay. It is therefore crucial to derive special
energy corrections for b jets, so that particle masses measured in
decays containing b quarks, such as $\mathrm{Z\to b\bar{b}}$,
$\mathrm{t\to bW}$, or $\mathrm{H\to b\bar{b}}$, are reconstructed at
the correct energy scale.

In \DO special b jet corrections exist only for semi-muonic decays, which
are applied if the jet contains a ``soft'' muon. The following section
describes the response measurement to compensate for the remaining
effects (mainly due to semi-electronic decays, for which no dedicated
corrections exist).

\subsection{The concept of the measurement}
The response for b jets that do \emph{not} have a soft muon tag is
measured in b-tagged \gj events using the same method (MPF) as for the
light jet response described above. However, b-tagged jets are a
mixture of true b and c jets and mistagged light jets which have
different energy responses. The situation is sketched in
Figure~\ref{fig:sketch-response}, where the measured mean value of the
response distribution in a particular energy bin is the weighted sum
of the mean values of response distributions for light, b, and c jets.
The response of b jets thus needs to be disentangled from the response
measured in a tagged \gj{} sample. Since there are three unknows, the
mean values \Rj, \Rb, and \Rc of the responses, this can be
accomplished by algebraically solving a system of three equations,
where each equation corresponds to a measurement of the response in a
sample with different flavor composition of the jets. In \DO, the
first measurement is performed on an untagged sample. With good
approximation, the measured response is that of light jets, \Rj,
Equation~(\ref{eq:Rut}). The second measurement is obtained from a
b-tagged \gj sample using the ``counting signed impact parameter''
(\CSIP) algorithm\footnote{With this algorithm a jet qualifies as a b
  jet if it has at least two tracks with impact parameter significance
  $\sigma$ above three, or at least three tracks with $\sigma$ above
  two.} \cite{Khanov:2004ze}, Equation~(\ref{eq:Rt}). The third
equation is taken from a sample with a tighter b-tagging criterion:
The jets must be tagged with the \CSIP algorithm and have a ``track
mass'' \mtrk above 1.93~GeV, Equation~(\ref{eq:Rmt}). \mtrk is the
invariant mass of the tracks that tag the jet, \emph{i.e.}, tracks
with impact parameter significance greater than two.

\begin{figure}[t]
  \begin{minipage}[t]{0.46\linewidth}
    \includegraphics[width=\linewidth]{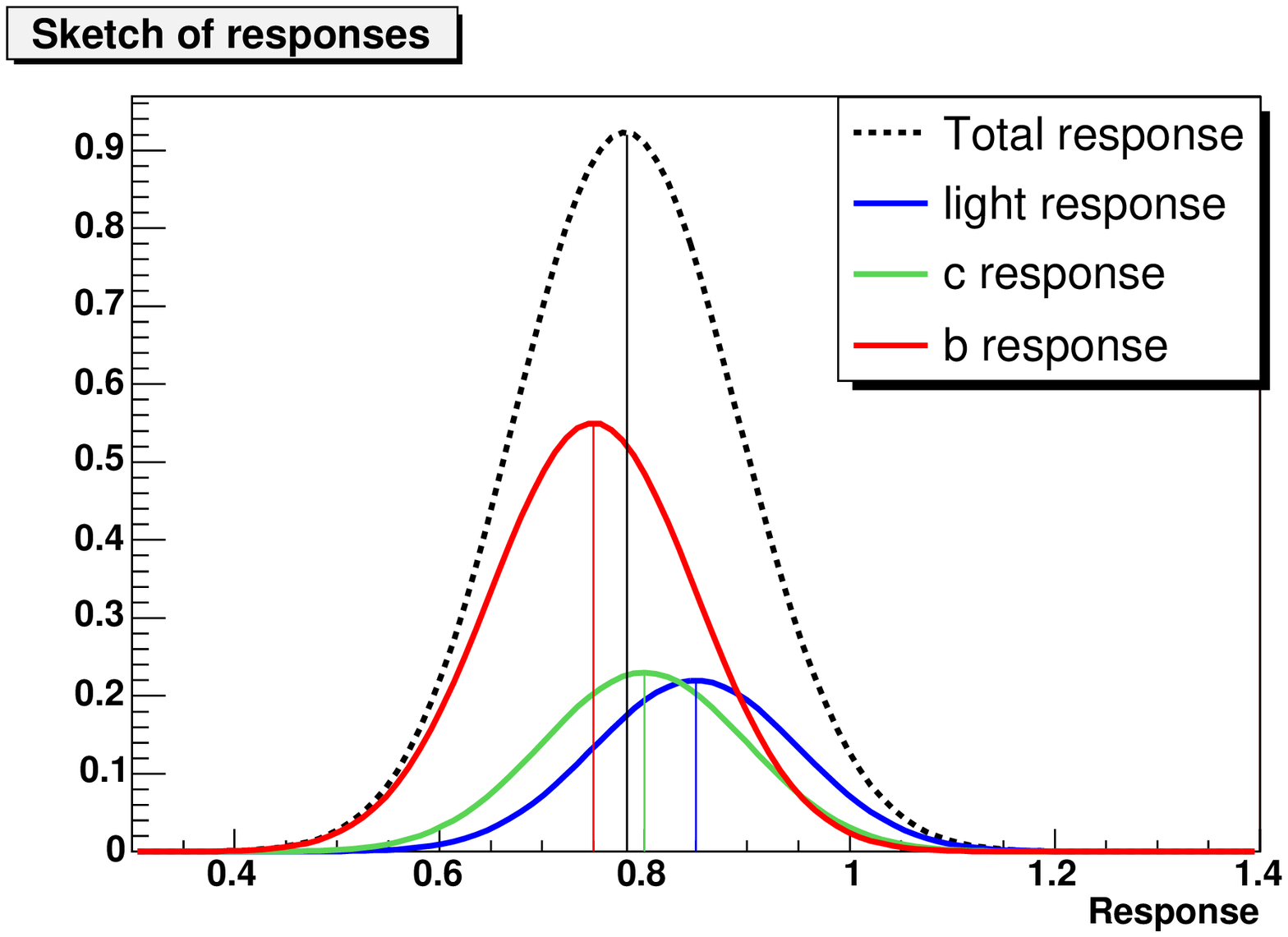}
    \caption{Sketch of response distribution in a particular energy bin.}
    \label{fig:sketch-response}
  \end{minipage}
  \hfill
  \begin{minipage}[t]{0.46\linewidth}
    \includegraphics[width=\linewidth]{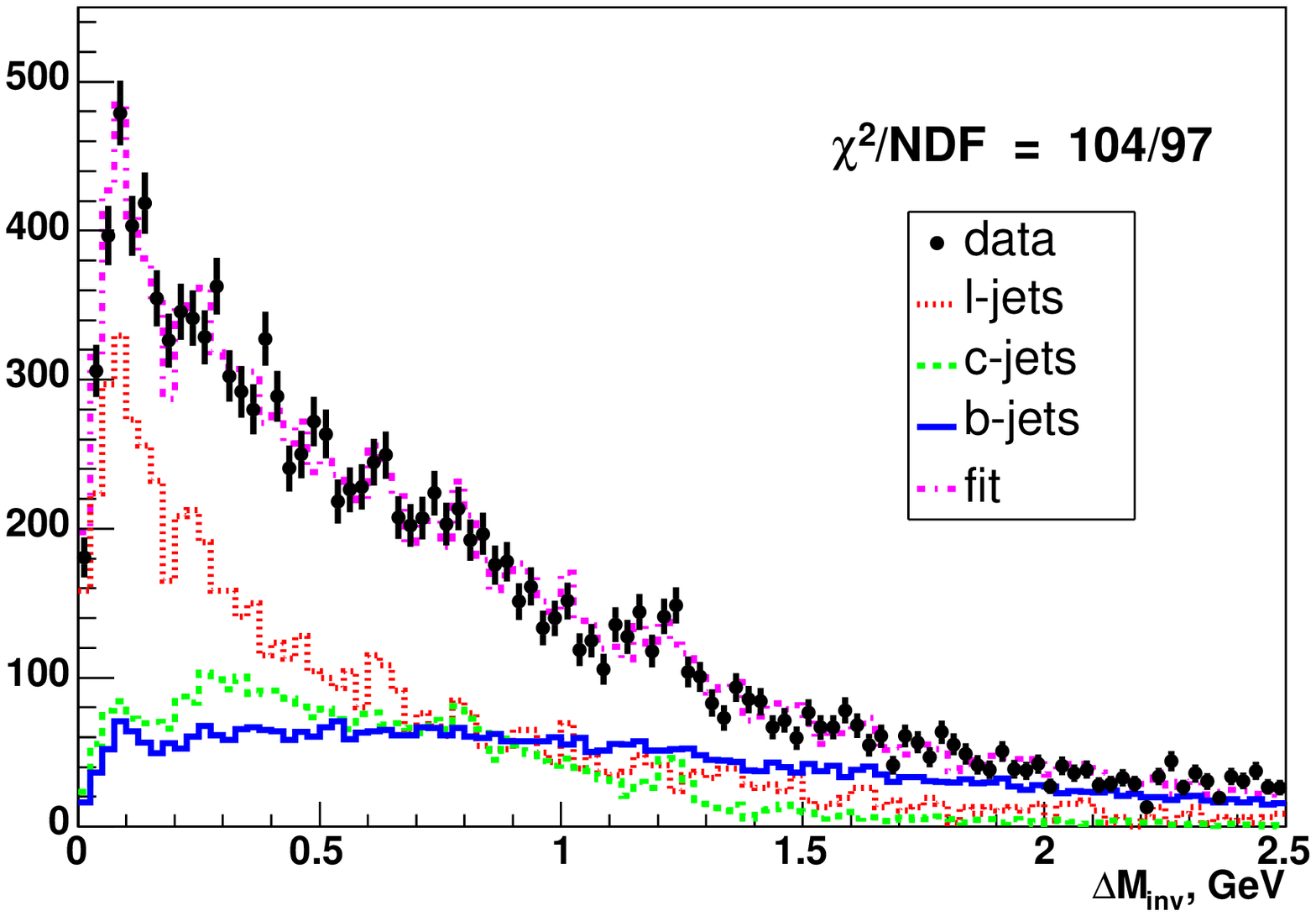}
    \caption{``Invariant track mass'' minus an offset of
      $m_\pi+m_K$.} \label{fig:csip-massfit}
  \end{minipage}
\end{figure}

\begin{eqnarray}
\mathrm{untagged:}\qquad R_{ut} &\approx& R_l\label{eq:Rut}\\
\mathrm{tagged:}\qquad   R_{t} &=& f_l R_l + f_b R_b + f_c R_c\label{eq:Rt}\\
\mathrm{tagged:}\qquad   R_{mt} &=& f'_l R_l + f'_b R_b + f'_c R_c\label{eq:Rmt}
\end{eqnarray}

The distribution of the invariant mass \mtrk is depicted in
Figure~\ref{fig:csip-massfit} for light, b, and c jets. The energy response
of b jets is then obtained by solving the system of
Equations~(\ref{eq:Rut})--~(\ref{eq:Rmt}) and measuring $R_{ut}$,
$R_{t}$, and $R_{mt}$ in various energy bins:

\begin{eqnarray}
  R_b =  \frac{1}{f_b-\frac{f_c}{f_c'}f_b'}
  \left[R_t - R_{ut}\left(f_l-\frac{f_c}{f_c'}f_l'\right)
 - R_{mt}\frac{f_c}{f_c'}\right]\label{eq:Rb_full}.
\end{eqnarray}

The flavor fractions $f^{(')}_l, f^{(')}_b, f^{(')}_c$ can be obtained from
fits of the mass templates to the data distribution as shown in
Figure~\ref{fig:csip-massfit}, or from a similar distribution that
discriminates between the jet flavors. The flavor composition is also
a function of the energy and must be measured separately in each
energy bin.

The advantage of this method is that the energy response of b jets is
measured directly in data and relies only very little on Monte Carlo
simulations (template distributions for the fit to the flavor
fractions). However, an inclusive \gj sample contains only a few
percent of \gb events, hence a large data sample is needed in order to
keep the statistical uncertainties at a reasonable level. 

Since jets in \DO are already corrected for the light jet energy
scale, the resulting b response will be provided as a residual scale factor
\Rj/\Rb. This scale factor derived from tagged \gj data can also be
compared to Monte Carlo simulations of \gj and \gb events, a
preliminary result of which is shown in Figure~\ref{fig:MCRjRb}.
The study suggests that b jets need additional energy corrections of
as much as 10\% at energies around 20~GeV and about 5\% at energies
of 150~GeV.

\begin{figure}[tbh]
  \centering
  \includegraphics[width=0.6\linewidth]{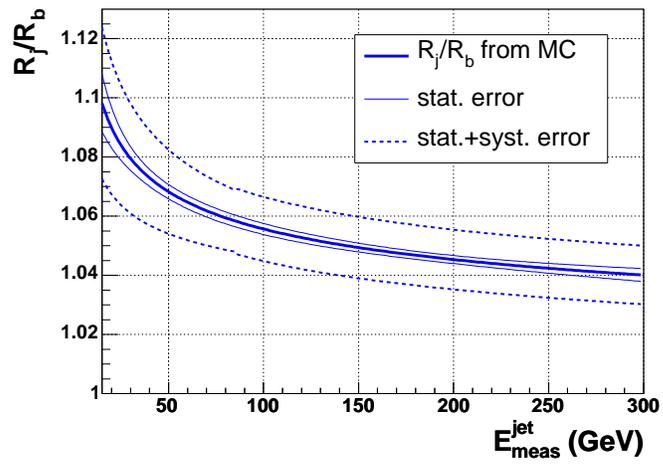}
  \caption{Response of b jets relative to the light jet response as a
  function of the raw jet energy.}
  \label{fig:MCRjRb}
\end{figure}


\clearpage
\section{Insights into $H\rightarrow\gamma\gamma$ from CDF Searches}
%
%
%
%
%
%
%
%
\textbf{Contributed by: S.-W. Lee}
\vspace{0.25in}

We describe how all of the diphoton measurements at CDF provide important
insights into the Higgs search at the LHC, $H \rightarrow \gamma\gamma$.  A
brief review of diphoton physics at CDF is also given here.
%

\subsection{Introduction}

The study of photon production at a hadron collider is important for many
reasons. As the photon energy is well-measured, compared to jets, it can be a
good tool to further our understanding of Quantum Chromodynamics (QCD). One of
the important reason to study photons at a hadron collider, as well as all QCD
measurements, is that they are the backgrounds to new physics. The most famous
of these is the Higgs search at the LHC, where diphoton backgrounds are the
most serious experimental difficulty. In the next section we will illustrate
how all of the diphoton measurements at CDF provide important insights into
backgrounds for new physics, specifically the diphoton backgrounds to the Higgs
search at the LHC, $H \rightarrow \gamma\gamma$. 

In addition, there are a large number of important and well-motivated
theoretical models which make a strong case for looking for new physics in
events with two photons in the final state. These theories include
Supersymmetry~(SUSY), Extra Dimensions~(ED), Grand Unified Theories, Composite
models of quarks and leptons, and Technicolor models. Therefore it is important
to understand diphoton production at Tevatron experiments in order to reliably
search for the Standard Model Higgs and new physics at LHC.

The aim of this talk is to present the recent measurement of diphoton
production at Tevatron experiment, CDF, to lead us to a deeper understanding of
new physics signatures at LHC experiments.

\subsection{Diphoton physics at CDF}
A brief review of physics with diphoton final states using the CDF detector at
the Tevatron is given here. These include searches for supersymmetry, extra 
dimensions and bosophilic Higgs, as well as QCD diphoton cross section 
measurement. Recent results from CDF Run II experiment are presented, but some 
result from Run I is also reviewed.

Diphoton final states are a signature of many interesting processes. For
example, at the LHC, one of the main discovery channels for the Higgs boson
search is the $\gamma\gamma$ final state. An excess of  $\gamma\gamma$
production at high invariant mass could be a signature of large extra
dimensions, and in many theories involved physics beyond the standard model,
cascade decays of heavy new particles generate a  $\gamma\gamma$ signature in
the final state. However, the QCD production rate is large compared to most new
physics, so an understanding of the QCD production mechanism is a prerequisite
to searching reliably for new physics in this channel. 

CDF has good analysis tool to identify the photon signal from the mixture of
photons and a neutral meson background. For the CDF measurement the fraction of
photon candidate events that have an observed conversion in the materials just
in front of the calorimeter is used, along with the transverse shower shape
measured in a proportional chamber at shower maximum in the calorimeter itself.
In the end one of the two methods is used to evaluate point-by-point the
fraction of photons in the data sample.~\protect\cite{Lee:2003dy}.

\subsubsection*{Diphoton Cross Section}

Recently CDF has performed pure QCD test with prompt diphotons using a data
sample of 207 $pb^{-1}$ in Run II~\protect\cite{Acosta:2004sn}. The analysis required two photon candidates
with $E_T >$ 14 GeV (13 GeV) for the leading (softer) photon candidate in the
event. The background from non-prompt photon sources is determined using a
statistical method based on differences in the electromagnetic showers. 

\begin{figure}[ht]
\begin{centering}
\includegraphics[width=8.0cm]{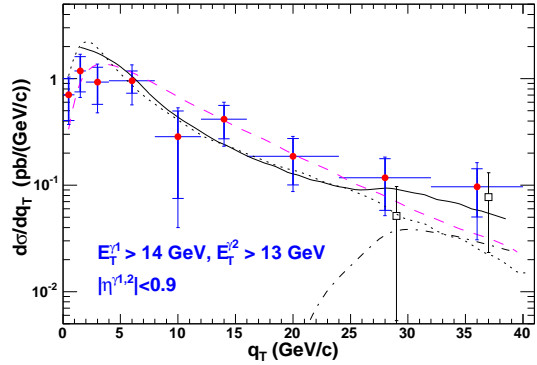}
\caption{
\label{fig:qcd}
The $\gamma\gamma$  $q_{T}$ distribution, along with predictions from  DIPHOX
(solid), RESBOS (dashed), and PYTHIA (dotted). The PYTHIA predictions have been
scaled by a factor of 2. Also shown, at larger $q_T$, are the DIPHOX prediction
(dot-dashed) and the CDF data (open squares) for the configuration where the
two photons are required  to have $\Delta\phi < \pi/2$.
}
\end{centering}
\end{figure}

CDF has measured the cross section for  prompt diphoton production as a
function of three kinematic variables - diphoton mass, the transverse momentum
of the diphoton system ($q_{T}$), and the azimuthal angle between the two
photons, $\Delta\phi$. Comparisons have been made with predictions from DIPHOX,
RESBOS and PYTHIA.  The data are in good agreement with the predictions for the
mass distribution.   At low to moderate $q_{T}$ and $\Delta\phi$  greater than
$\pi/2$, where the effect of soft  gluon emissions are important, the data
agree better with RESBOS than DIPHOX. By contrast, in the regions where the
2$\rightarrow$3  fragmentation contribution becomes important, large $q_{T}$, 
$\Delta\phi$  less than $\pi/2$ and low diphoton mass, the data agree better
with DIPHOX. The $q_{T}$ distribution is shown in Fig.~1. 

This result would appear to indicate a need to have a full theoretical
calculation of diphoton production; a resummed full NLO calculation will be
necessary. Again, an understanding of the QCD diphoton production mechanism  is
a prerequisite to searching reliably for new physics in this channel. 

\begin{figure}[ht]
\begin{centering}
\includegraphics[width=8.0cm]{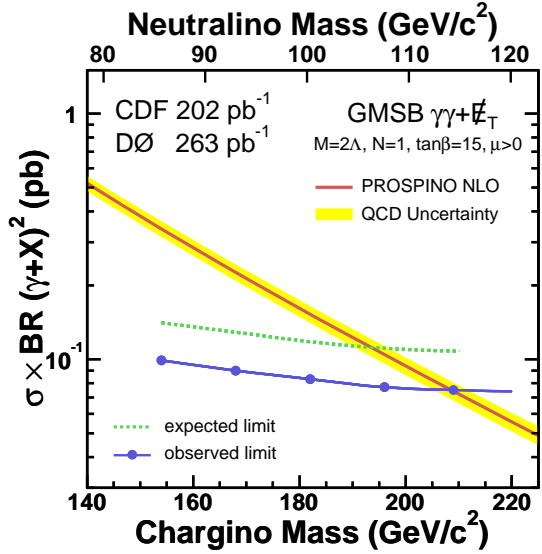}
\caption{
\label{fig:xsec}
The NLO cross section and combined experimental limits 
as a function of chargino and neutralino mass in GMSB model.
}
\end{centering}
\end{figure}

\subsubsection*{Search for Supersymmetry}
Among various SUSY models, two SUSY breaking mechanism are interesting,  which
predict photons in the final states. Supergravity models can produce events
which decay down to the second lightest neutralino via a loop into  the
lightest neutralino~($\NONE$) and a photon, where $\NONE$ is  the lightest
supersymmetric particle (LSP). Gauge-Mediated SUSY breaking  models (GMSB) with
the $\NONE$ decaying into a photon and gravitino can  produce a final state of
two photons and large missing transverse energy~($\met$). $\met$ is often used
as a pointer to possible SUSY signals because it indicates the escape of a
non-interacting SUSY particle from the detector. The LSP  signals are of
particular interest as they provide a natural explanation for  the dark
matter. 

CDF has searches 202 $pb^{-1}$ of inclusive diphoton events of Run II data  for
anomalous production of $\met$ as evidence of new physics. Events are selected
as having two photon candidates with $E_T >$ 13 GeV in the  central. CDF
observe no candidate events, with an expected standard model  background of
0.27 $\pm$ 0.07 (stat) $\pm$ 0.10 (syst) events. Using these  results, CDF has
set limit on the lightest chargino \mbox{M$_{\CONE}>167$ GeV/$c^2$},  and the
lightest neutralino \mbox{M$_{\NONE}>93$~GeV/$c^2$} at 95\% C.L. in a GMSB
scenario with a light gravitino~\protect\cite{Acosta:2004sb}.

Fig.~2 shows the combined CDF and D0 result for the observed cross section  as
a function of $\CONE$ and $\NONE$ along with the theoretical LO and NLO 
production cross sections. The final mass limit for the lightest chargino is 
209~GeV$/c^2$ which translates to a mass limit of 114~GeV$/c^2$ on the 
lightest neutralino and a limit of 84.6~TeV on $\Lambda$. This result
significantly extends the individual experimental limits~\protect\cite{Buescher:2005he}.

\subsubsection*{Search for Extra Dimensions}

Recent theories postulate the existence of new space-time dimensions. Such
extra dimensions might be found by looking for graviton exchange  processes in
the diphoton final state. For example in the Randall-Sandrum  model with a
warped extra dimension, diphoton resonances can be produced  via the graviton.

CDF has searches for diphoton mass resonance with a data sample of  345
$pb^{-1}$. Two isolated photons , each with $E_T >$ 25 GeV, are required  in
the analysis. The main background comes from standard model diphoton production
which  accounts for 30\% of the events, and from jets which fake photons. No
deviation from standard model expectations is observed, and set upper limit on
the cross section  times branching ratio of the Randall-Sandrum graviton
production and decay to diphotons. The lower mass bounds obtained for the first
excited states of the Randall-Sandrum graviton  are 690 and 220~GeV$/c^2$ for
coupling, k/M$_{pl}$ = 0.1 and 0.01,  respectively.~\protect\cite{CDF-PUB-EXOTIC-PUBLIC-7098} 

\begin{figure}[ht]
\begin{centering}
\includegraphics[width=8.0cm]{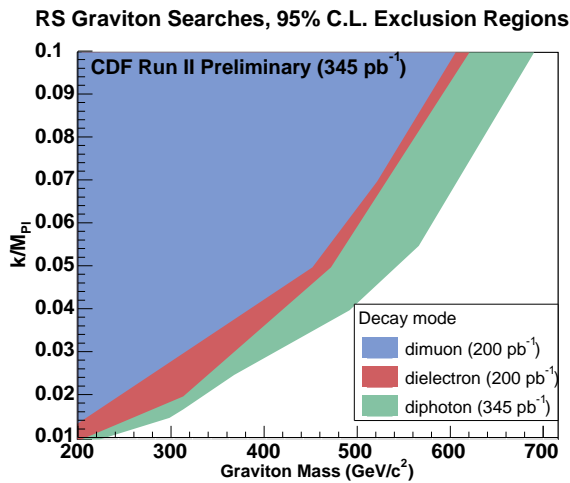}
\caption{
\label{fig:xdim}
Combined 95\% C.L. RS gaviton mass limit of the diphoton and dilepton
searches}
\end{centering}
\end{figure}

Fig.~3 shows the combined 95\% C.L. RS graviton mass limit of the diphoton and 
dilepton searches in the graviton mass versus k/M$_{pl}$ plane.

\subsubsection*{Search for Bosophilic Higgs}

The signature of high mass photon pairs is attractive for searches for new
physics as the photon is the lightest gauge boson, and hence might be more
easily produced in decays of new physics. There are models in which a Higgs
boson could decay into two photons with a branching ratio much larger than
predicted in the standard model; bosophilic Higgs boson.

In Run I CDF has searched for departures from standard model expectations for
inclusive high mass diphoton production in association with a W or Z boson~\protect\cite{Affolder:2001hx}.
This analysis is complimentary to the diphoton cross section analysis, in which
very strict photon selection requirements are used to reduce the large jet fake
backgrounds maximizing signal significance, but which become progressively less
efficient with $E_T$ for high energy photons. It is also complimentary to the
recent diphoton+X search analysis which was focused on non-resonant diphoton
signatures such as GMSB SUSY.

\begin{figure}[ht]
\begin{centering}
\includegraphics[width=8.0cm]{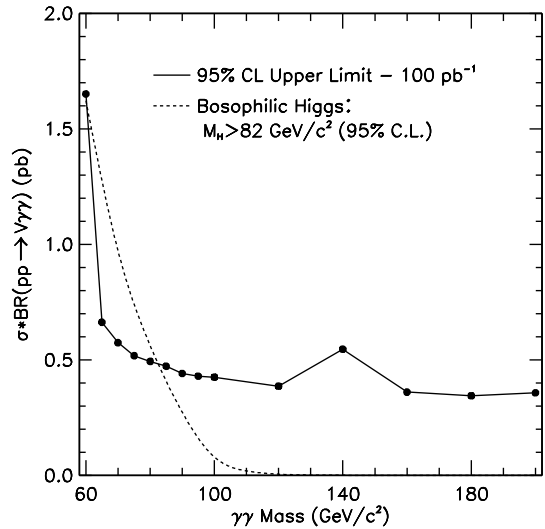}
\caption{
\label{fig:fh}
Upper limit at 95\% C.L. on the $\gamma\gamma + W/Z$\  cross  section as a
function of $\gamma\gamma$\ mass.  The dashed curve  shows the prediction for
cross section times branching  fraction for a bosophilic $H\to \gamma\gamma$\
with branching fraction from reference~\protect\cite{Stange:1993ya} and the cross
section for associa ted  Higgs production is a standard model NLO calculation
from  reference~\protect\cite{Han:1991ia}.} 
\end{centering}
\end{figure}

CDF found no evidence for a resonant structure and set an upper limit on the
cross section times branching ratio for $\ppbar \rightarrow H \rightarrow
\gamma\gamma$  between 60 and 200 GeV/$c^2$ (see Fig.~4).  A 95\% C.L. lower
limit on the mass of a bosophilic Higgs boson (one which couples only to
$\gamma$, W and Z with standard model couplings) is set at 82 GeV/c$^2$.

\subsection{Conclusion}

In this article we summarize the current CDF experimental results of diphoton
physics, test of standard model and searches for new physics in final states
containing energetic photons, at Tevatron. we also describe how all of the
diphoton measurements at CDF provide important insights into backgrounds for
new physics, specifically the diphoton backgrounds to the Higgs search at the
LHC.  As we learned it is important to understand diphoton production in order
to reliably search for the standard model Higgs and new physics. There may be
an interesting connection between Tevatron and LHC.

%





\bibliographystyle{tev4lhc}
\bibliography{tev4lhc}

\end{document}